\documentclass[11pt,twoside]{ThesisStyle}

\newcommand\titoltesi{Contributions to Context-Aware Smart Healthcare: A Security and Privacy Perspective}
\newcommand\nomautortesi{Edgar}
\newcommand\cognomautortesi{Batista}
\newcommand\autortesi{\nomautortesi~\cognomautortesi}
\newcommand\nomdirectortesi{Agusti}
\newcommand\cognomdirectortesi{Solanas}
\newcommand\directortesi{\nomdirectortesi~\cognomdirectortesi}
\newcommand\depttesi{Department of Computer Engineering and Mathematics}
\newcommand\ciutattesi{Tarragona}
\newcommand\diatesi{16th}
\newcommand\mestesi{March}
\newcommand\anytesi{2022}

\usepackage[utf8]{inputenc}
\usepackage[T1]{fontenc}
\usepackage[paper=b5paper,bindingoffset=10mm,top=27mm,bottom=20mm,left=20mm,right=20mm,headheight=13.6pt]{geometry}

\usepackage{booktabs}
\usepackage{longtable}
\usepackage{multirow}
\usepackage{tabularx}

\newcolumntype{C}{>{\centering\arraybackslash}X} 
\usepackage{colortbl}
\arrayrulecolor{black}
\usepackage{rotating}

\usepackage{graphicx}
\usepackage[font=footnotesize]{caption}
\usepackage{subcaption}

\usepackage{algorithm,algpseudocode}
\counterwithin{algorithm}{chapter}

\usepackage{color}
\definecolor{linkcol}{rgb}{0,0,0} 
\definecolor{black}{rgb}{0,0,0}
\definecolor{_mygreen}{rgb}{0,0.5,0}
\definecolor{_myred}{rgb}{0.89,0.0,0.13}
\newcommand{\gr}[1]{\textcolor{_mygreen}{#1}}
\newcommand{\re}[1]{\textcolor{_myred}{#1}}

\usepackage{amsmath,amssymb}
\usepackage[hyperindex=true]{hyperref}
\usepackage{url}
\usepackage{pifont}
\usepackage{afterpage}
\usepackage{pdfpages}

\usepackage[nottoc,notlof,notlot]{tocbibind}
\usepackage{minitoc}
\let\minitocORIG\minitoc
\renewcommand{\minitoc}{
    \begingroup
    \hypersetup{linkcolor=linkcol}
    \minitocORIG 
    \endgroup 
    \vspace{1.5em}
}

\usepackage{fancyhdr}                    
\let\headruleORIG\headrule
\renewcommand{\headrule}{\color{black} \headruleORIG}

\fancypagestyle{plain}{
    \fancyhead{}
    \fancyfoot{}
    
}

\newcommand\blankpage{%
    \null
    \thispagestyle{empty}%
    \addtocounter{page}{-1}%
    \newpage}

\makeatletter
\def\cleardoublepage{\clearpage\if@twoside \ifodd\c@page\else%
  \hbox{}%
  \thispagestyle{empty}
  \newpage%
  \if@twocolumn\hbox{}\newpage\fi\fi\fi}
\makeatother

\hypersetup{
    pdftitle=\titoltesi,
    pdfauthor=\autortesi,
    pdfsubject=PhD Thesis,
    pdfhighlight=/O, 
    colorlinks=true, 
    linkcolor=black, 
    citecolor=black, 
    urlcolor=black 
}

\newcommand{\ie}{\textit{i.e., }}
\newcommand{\eg}{\textit{e.g., }}

\usepackage{hyphenat}

\begin{document}

\includepdf[pages=-]{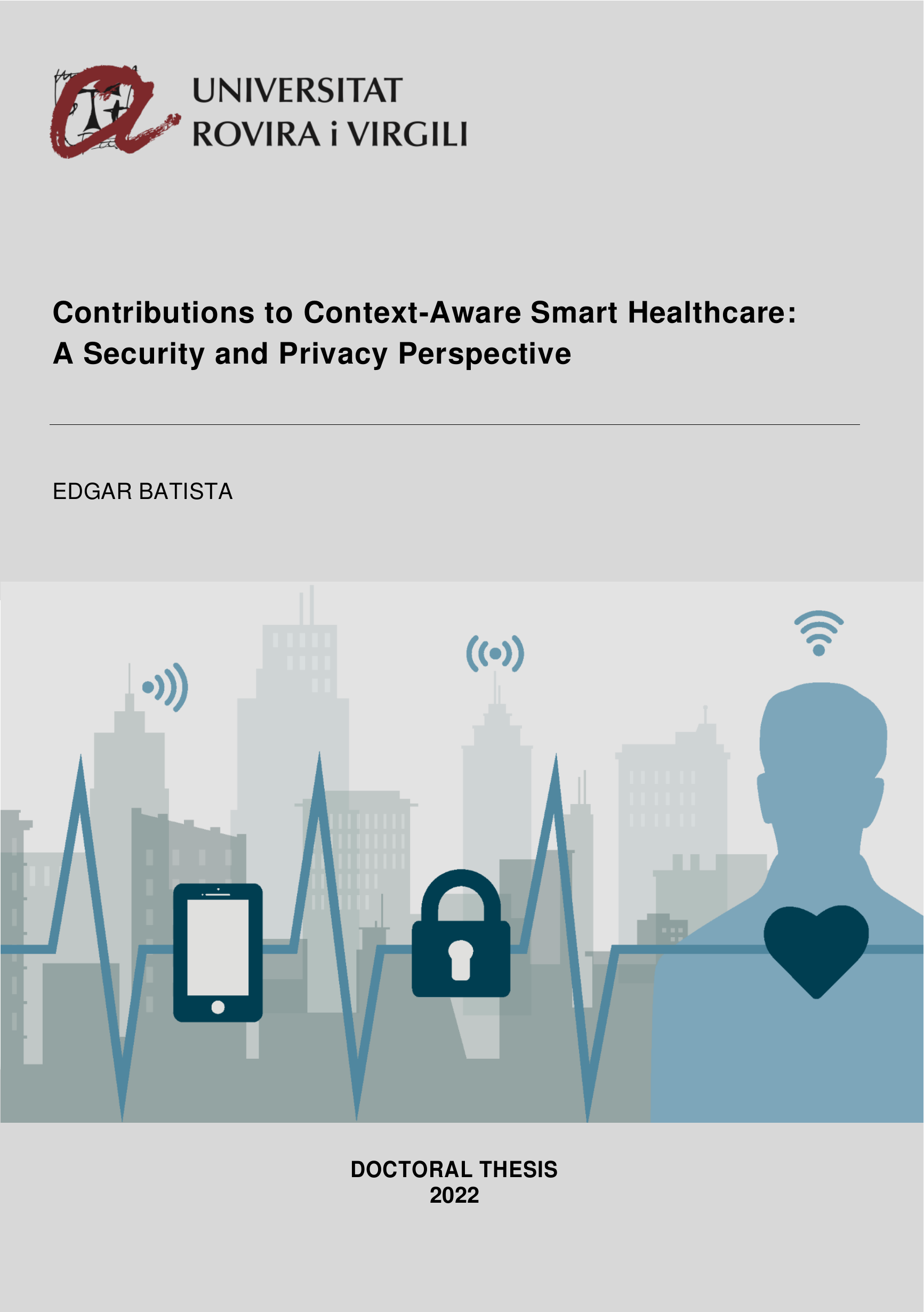}\cleardoublepage
\thispagestyle{empty}

\vspace*{16cm}

\noindent \small{Amb el suport del Pla de Doctorats Industrials del Departament de Recerca i Universitats de la Generalitat de Catalunya (projecte 2017-DI-002).}

\vspace*{0.5cm}

\begin{figure}[h!]
\centering
\begin{minipage}{.45\textwidth}
  \centering
  \includegraphics[width=0.8\linewidth]{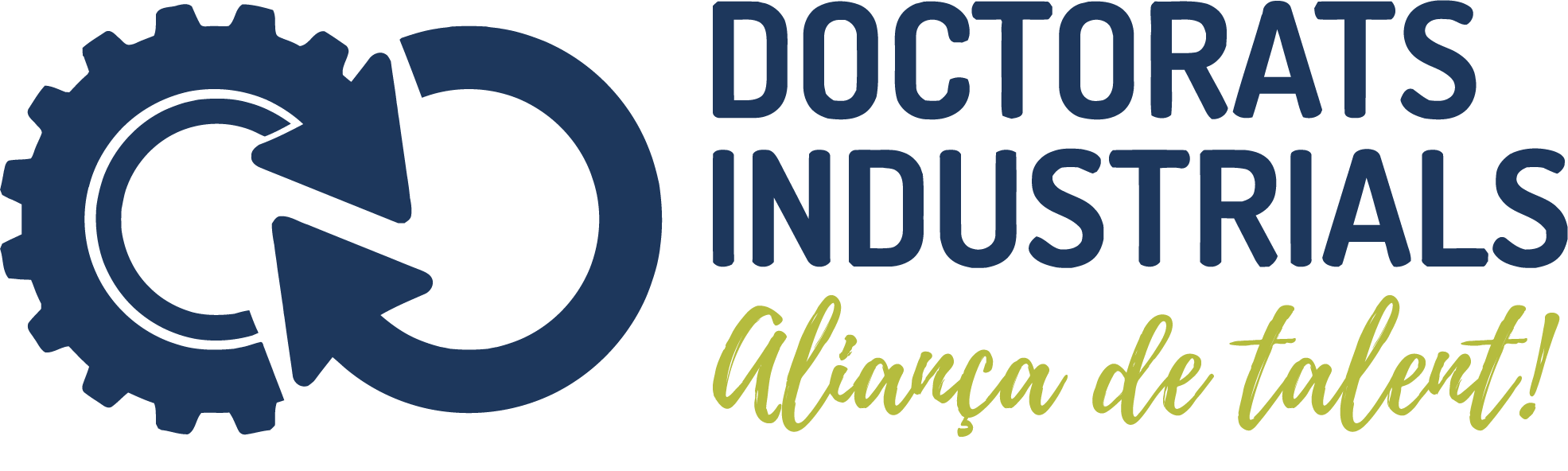}
\end{minipage}%
\begin{minipage}{.5\textwidth}
  \centering
  \includegraphics[width=\linewidth]{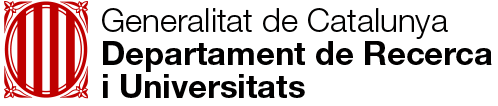}
\end{minipage}
\end{figure}
\cleardoublepage
\thispagestyle{empty}

\begin{center}

\vspace*{0.5cm}
\noindent \LARGE{\nomautortesi~\textsc{\cognomautortesi}} \\
\vspace*{1.7cm}
\noindent \LARGE{\textbf{\titoltesi}} \\
\vspace*{1.7cm}
\noindent \Large{\textbf{DOCTORAL THESIS}} \\
\vspace*{0.35cm}
\noindent \large{supervised by Dr. \nomdirectortesi~\textsc{\cognomdirectortesi}} \\
\vspace*{2.1cm}
\noindent \large{\depttesi} \\
\vspace*{3.2cm}

\begin{figure}[h]
	\centering
	\includegraphics[width=0.35\columnwidth]{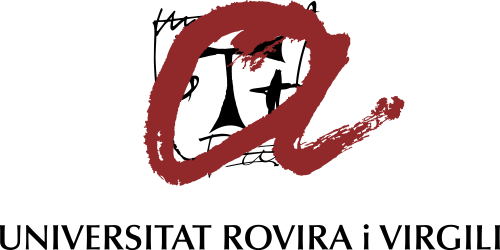}
\end{figure}

\noindent \ciutattesi \\
\vspace*{0.3cm}
\noindent \anytesi

\end{center}\cleardoublepage
\thispagestyle{empty}

\vspace*{0.6cm}

\begin{figure}[h!]	
\includegraphics[width=0.35\columnwidth]{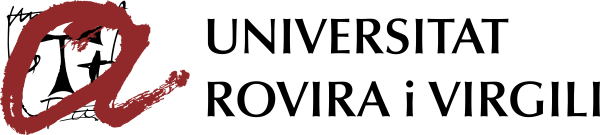}
\end{figure}

\vspace*{0.2cm}

\noindent\small{\textbf{\depttesi}\\
\noindent\small{Campus Sescelades}\\
\noindent\small{Av. Pa\"isos Catalans, 26}\\
\noindent\small{43007 Tarragona}\\

\vspace{2cm}

\noindent I STATE that the present study, entitled ``\titoltesi'', presented by \autortesi~for the award of the degree of Doctor, has been carried out under my supervision at the \depttesi~of this university, and that it fulfils all the requirements to be eligible for the Industrial Doctorate Distinction.

\vspace{3cm}

\noindent \ciutattesi,~\diatesi~\mestesi~\anytesi

\vspace{1.5cm}

\noindent Doctoral Thesis Supervisor

\vspace{0.15cm}

\begin{figure}[h!]	
\includegraphics[width=0.35\columnwidth]{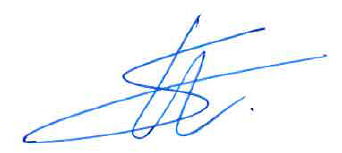}
\end{figure}

\vspace{0.1cm}

Dr.~\directortesi
\cleardoublepage

\dominitoc

\pagenumbering{roman}
\section*{Abstract}

Information and communication technologies have irreversibly changed our lives.
The healthcare industry, one of the world’s largest and fastest-growing industries, is dedicating many efforts in adopting the latest technologies into daily medical practice.
It is not therefore surprising that healthcare paradigms are constantly evolving seeking for more efficient, effective and sustainable services.
In this context, the potential of ubiquitous computing through smartphones, smartwatches, wearables and IoT devices has become fundamental to collect large volumes of data, including people's health status and people's location.
The enhanced sensing capabilities together with the emergence of high-speed telecommunication networks have facilitated the implementation of context-aware environments, such as smart homes and smart cities, able to adapt themselves to the citizens needs.
The interplay between ubiquitous computing and context-aware environments opened the door to the so-called smart health paradigm, focused on the provision of added-value personalised health services by meaningfully exploiting vast amounts of health, mobility and contextual data.

The management of health data, from their gathering to their analysis, arises a number of challenging issues due to their highly confidential nature.
In particular, this dissertation contributes to several security and privacy challenges within the smart health paradigm.
More concretely, we firstly develop some contributions to context-aware environments enabling smart health scenarios.
We present an extensive analysis on the security aspects of the underlying sensors and networks deployed in such environments, a novel user-centred privacy framework for analysing ubiquitous computing systems, and a complete analysis on the security and privacy challenges that need to be faced to implement cognitive cities properly.
Second, we contribute to process mining, a popular analytical field that helps analyse business processes within organisations.
Despite its popularity within the healthcare industry, we address two major issues: the high complexity of healthcare processes and the scarce research on privacy aspects.
Regarding the first issue, we present a novel process discovery algorithm with a built-in heuristic that simplifies complex processes and, regarding the second, we propose two novel privacy-preserving process mining methods, which achieve a remarkable trade-off between accuracy and privacy.
Last but not least, we present some smart health applications, namely a context-aware recommender system for routes, a platform supporting early mobilisation programmes in hospital settings, and a health-oriented geographic information system.

The results of this dissertation are intended to help the research community to enhance the security of the intelligent environments of the future as well as the privacy of the citizens regarding their personal and health data.
\cleardoublepage
\section*{Resumen}

Las tecnologías de la información y la comunicación han cambiado nuestras vidas de forma irreversible.
La industria sanitaria, una de las industrias más grandes y de mayor crecimiento, está dedicando muchos esfuerzos por adoptar las últimas tecnologías en la práctica médica diaria.
Por tanto, no es sorprendente que los paradigmas sanitarios estén en constante evolución en busca de servicios más eficientes, eficaces y sostenibles.
En este contexto, el potencial de la computación ubicua mediante teléfonos inteligentes, relojes inteligentes,  dispositivos \textit{wearables} e IoT ha sido fundamental para recopilar grandes volúmenes de datos, especialmente relacionados con el estado de salud y la localización de las personas.
Las mejoras en las capacidades de detección junto con la aparición de redes de telecomunicaciones de alta velocidad han facilitado la implementación de entornos sensibles al contexto, como las casas y las ciudades inteligentes, capaces de adaptarse a las necesidades de los ciudadanos.
La interacción entre la computación ubicua y los entornos sensibles al contexto abrió la puerta al paradigma de la \textit{salud inteligente} (del inglés, smart health), centrado en la prestación de servicios de salud personalizados y de valor añadido mediante la explotación significativa de grandes cantidades de datos sanitarios, de movilidad y contextuales.

La gestión de datos sanitarios, desde su recogida hasta su análisis, plantea una serie de cuestiones desafiantes debido a su naturaleza altamente confidencial.
En particular, esta tesis contribuye a varios retos de seguridad y privacidad dentro del paradigma de la salud inteligente.
Más concretamente, en primer lugar, hacemos contribuciones a los entornos sensibles al contexto que habilitan escenarios de salud inteligente.
Para ello, presentamos un amplio análisis sobre los aspectos de seguridad de los sensores y redes desplegados en dichos entornos, un novedoso marco de privacidad centrado en el usuario para analizar sistemas de computación ubicuos, y un análisis completo acerca de los retos de seguridad y privacidad que deben abordarse para implementar adecuadamente las ciudades cognitivas.
En segundo lugar, también contribuimos a la minería de procesos, un popular campo que ayuda a analizar procesos de negocio dentro de las organizaciones.
A pesar de su popularidad en la industria sanitaria, abordamos dos de sus grandes problemas: la alta complejidad de los procesos sanitarios y la poca investigación sobre los aspectos de privacidad.
En relación al primer tema, presentamos un algoritmo de descubrimiento de procesos con una heurística que simplifica su complejidad y, en relación al segundo tema, proponemos dos nuevos métodos de minería de procesos con privacidad, que logran un notable equilibrio entre precisión y privacidad.
Por último, pero no menos importante, presentamos algunas aplicaciones de salud inteligente, a saber, un sistema de recomendación sensible al contexto para rutas, una plataforma de apoyo a programas de movilización temprana en entornos hospitalarios, y un sistema de información geográfica orientado a la salud.

Los resultados de esta tesis pretenden ayudar a la comunidad científica a mejorar la seguridad de los entornos inteligentes del futuro, así como la privacidad de los ciudadanos con respecto a sus datos personales y sanitarios.\cleardoublepage
\section*{Resum}

Les tecnologies de la informació i la comunicació han canviat les nostres vides de manera irreversible.
La indústria sanitària, una de les indústries més grans i de major creixement, està dedicant molts esforços per adoptar les últimes tecnologies en la pràctica mèdica diària.
Per tant, no és sorprenent que els paradigmes sanitaris estiguin en constant evolució cercant serveis més eficients, eficaços i sostenibles.
En aquest context, el potencial de la computació ubiqua mitjançant telèfons intel·ligents, rellotges intel·ligents, dispositius \textit{wearables} i IoT ha esdevingut fonamental per recopilar grans volums de dades, especialment relacionats amb l'estat de salut i la ubicació de les persones.
Les millores en les capacitats de detecció juntament amb l'aparició de xarxes de telecomunicacions d'alta velocitat han facilitat la implementació d'entorns sensibles al context, com les cases i les ciutats intel·ligents, capaços d'adaptar-se a les necessitats dels ciutadans.
La interacció entre la computació ubiqua i els entorns sensibles al context va obrir la porta al paradigma de la \textit{salut intel·ligent} (de l'anglès, smart health), centrat en la prestació de serveis de salut personalitzats i de valor afegit mitjançant l'explotació significativa de grans quantitats de dades sanitàries, de mobilitat i contextuals.

La gestió de dades sanitàries, des de la seva recollida fins a la seva anàlisi, planteja una sèrie de problemes desafiants a causa del seu caràcter altament confidencial.
En particular, aquesta tesi contribueix a diversos reptes de seguretat i privadesa dins del paradigma de la salut intel·ligent.
Més concretament, en primer lloc, fem contribucions als entorns sensibles al context que habiliten escenaris de salut intel·ligent.
Per a això, presentem una àmplia anàlisi sobre els aspectes de seguretat dels sensors i xarxes desplegats en aquests entorns, un nou marc de privadesa centrat en l'usuari per analitzar sistemes de computació ubics, i una anàlisi completa dels reptes de seguretat i privadesa que s'han d'abordar per implementar les ciutat cognitives correctament.
En segon lloc, també contribuïm a la mineria de processos, un camp popular que ajuda a analitzar processos de negoci dins de les organitzacions.
Malgrat la seva popularitat dins de la indústria sanitària, abordem dos dels seus grans problemes: l'elevada complexitat dels processos sanitaris i l'escassa investigació sobre els aspectes de privadesa.
En relació al primer tema, presentem un nou algorisme de descobriment de processos amb una heurística que simplifica la seva complexitat i, en relació al segon tema, proposem dos nous mètodes de mineria de processos amb privadesa, que aconsegueixen un notable equilibri entre precisió i privadesa.
Per últim, però no menys important, presentem algunes aplicacions de salut intel·ligent, com són un sistema de recomanació de rutes sensible al context, una plataforma de suport a programes de mobilització primerenca en entorns hospitalaris, i un sistema d'informació geogràfica orientat a la salut.

Els resultats d'aquesta tesi pretenen ajudar a la comunitat científica a millorar la seguretat dels entorns intel·ligents del futur, així com la privadesa dels ciutadans respecte a les seves dades personals i sanitàries.
\cleardoublepage

\begingroup
    \hypersetup{linkcolor=linkcol}
    \tableofcontents
\endgroup

\mainmatter

\part{Introduction} \label{part:intro}
\chapter{Introduction}
\label{chap:intro}

\emph{This chapter introduces the issues faced in this doctoral thesis. In addition, it briefly describes the solutions proposed to tackle those issues. First, Section \ref{sec:intro:motivation} motivates the research conducted in this dissertation. More specifically, the main contributions of this thesis are listed in Section \ref{sec:intro:contributions}. Finally, the structure and organisation of the thesis are outlined in Section \ref{sec:intro:organisation}.}

\minitoc

\section{Motivation} \label{sec:intro:motivation}

\textit{Smart health} is a relatively new context-aware healthcare paradigm \cite{solanas2014shealth} influenced by several fields of knowledge, including medical informatics, telecommunications, electronics and bioengineering, among others.
The ever-changing healthcare industry found in information and communication technologies (ICT) an unconditional ally to achieve concrete objectives, such as increasing efficiency, improving patients' treatments, shortening waiting times and reducing healthcare expenditure.
From a global perspective, there is little doubt that ICT have deeply influenced our daily lives: they have changed how we live, how we learn, how we interact with each other, and how we take care of our health.

ICT systems have suffered a tremendous evolution in the last few decades: from monolithic servers and workstations, to mobile, pervasive and embedded devices. 
This transformation has enabled \textit{ubiquitous computing}, founded on the principles of bringing computation and communication capabilities to everything, everywhere and anytime \cite{krumm2018ubiquitous}.
The synergies between ICT and ubiquitous computing ushered in the era of the \textit{Internet of Things} (IoT).
The main idea of IoT is to connect all sorts of ``things'' (\eg sensors, phones, watches, wearables, TVs, appliances, home automation devices, augmented/virtual reality glasses, voice assistant devices\ldots) to the Internet so as to create smart environments with self-aware capabilities \cite{nivzetic2020iot}.
Medicine and healthcare encountered in IoT great opportunities to streamline health services.
Consequently, the \textit{Internet of Medical Things} (IoMT) emerged as a practical application of IoT for medical care \cite{al2020intelligence}.
The IoMT has indeed become a key enabler of smart health since both fields have interrelated objectives: to provide efficient, accurate, personalised and real-time health services to the citizenship.
Feasible smart health services exploiting the benefits of IoMT include, but are not limited to, personalised patient care, real-time patient monitoring, remote patient monitoring, self-monitoring, disease management and early emergency response systems.
All in all, the IoT industry, in addition to revolutionise the entire technological market and the society as a whole, is leading to an ever more connected world.

Advancements in the miniaturisation of electronic devices together with the deployment of cheaper and faster data networks have propelled environments augmented with vast amounts of contextual and real-time data.
This kind of environments, known as \textit{context-aware environments}, are capable of adapting themselves to the citizens needs as well as to the requirements of their underlying systems \cite{temdee2018context}.
Well-known examples of context-aware environments are smart homes, smart buildings, smart hospitals and smart cities at a larger scale.
These environments are emerging in response to the increasingly urbanisation concern dealing with larger and older populations and scarcity of resources worldwide \cite{un2019population}.
In this sense, the paradigm of context-aware environments provides the required resource sustainability by means of handling multiple systems and resources in a coordinated fashion, such as transportation systems, energy management, waste management, e-governance and, obviously, healthcare systems.

Complementing the above, there is a general agreement on the fact that the rampant growth in sensor technologies, communications and information systems in the last years has enabled an unprecedented capacity to collect huge amounts of data, \ie infobesity.
The amount of data in the digital universe is growing at a dizzying pace, and it is expected to reach the 175 ZB by 2025, IDC forecasts \cite{idc2018data}.
However, these data are no longer well-structured and homogeneous but, on the contrary, they are unstructured and heterogeneous.
Hence, despite the ability to gather data effortlessly, the real aim is to transform these data into wisdom and extract value and meaningful knowledge from them.
This journey becomes even tougher in the healthcare domain, where very sensitive data are managed and human lives could be at stake.
Consequently, plenty of business opportunities in the field of data analytics are emerging, namely artificial intelligence, big data, collaborative filtering, data mining and process mining, to name a few.

All in all, smart health is built upon many interwoven fields: IoT/IoMT, ubiquitous computing, data networks, context-aware environments, communication technologies, data science, artificial intelligence, computer vision and so on.
Consequently, many challenging problems are related to smart health but, in many cases, they are explored individually in their respective fields of study and are unknown by the smart health research community working in more specific domains.
In this dissertation, security and privacy challenges of some of these underlying fields enabling smart health are explored.
More specifically, we concentrate these efforts on two directions: (i) context-aware environments enablers of smart health, and (ii) process mining, a trendy data analytics discipline within the healthcare domain \cite{aalst2011book,aalst2016book}.

Nowadays, security and privacy are some of the major concerns of the healthcare domain due to the humongous amount of health data collected and transmitted over the Internet.
The research community is struggling to securing context-aware environments and smart health systems from malicious actors, who are continuously seeking novel mechanisms to bypass the established security barriers.
These tireless efforts made by attackers generally respond to economical motivations because of the high value of personal data, especially financial and health-related data \cite{blackmarketprice}.
Hence, to prevent jeopardising people's privacy, the development of privacy-enhancing techniques for conducting healthcare-oriented analyses are of utmost importance.
Unfortunately, some fields of study do not put much emphasis on these aspects, such as the case of process mining, whose privacy-preserving process mining research is still in an embryonic stage.

All things considered, the ultimate goal of this dissertation is to pave the way to pursue the best knowledge-security-privacy equilibrium in smart health scenarios, by securing the underlying infrastructures and enhancing the quality of life and privacy of citizens.

\section{Contributions} \label{sec:intro:contributions}

The main contributions of this dissertation are the following:

\begin{enumerate}
\item \textit{Security and privacy contributions to context-aware environments}:
As previously stated, context-aware environments are the interplay between multiple cyber-physical systems and ubiquitous computing to provide added-value personalised services to citizens.
To enable intelligent services in them, these environments and their systems need to be secure and private, especially when providing smart health services.
First, we conduct a throughout analysis on the security aspects of the sensors and networks deployed in context-aware environments.
Second, we propose a novel approach to analyse the privacy risks of ubiquitous computing systems using a novel taxonomy.
And third, with an eye on the connectivism theory, we also provide an exhaustive analysis on the security and privacy aspects to consider once implementing cognitive cities, the next evolution of smart cities.

\item \textit{Contributions to process mining}:
Process mining in healthcare has become one of the most actively researched fields within the area over the recent years.
In order to deal with the high complexity of healthcare processes, we propose a novel process discovery algorithm with a simplification heuristic by design so as to facilitate knowledge acquisition.
Moreover, we explore the challenges and opportunities of conducting process mining analyses by considering, not only event data, but also all sorts of data collected from context-aware environments.

\item \textit{Contributions to privacy-preserving process mining}:
Privacy is one of the most relevant concerns when conducting process mining analysis.
This research direction is still in its infancy though.
To this end, we propose two new privacy-preserving process mining techniques able to limit the knowledge of the discovered process models according to a privacy threshold under different attacker models.
These approaches, evaluated with different real-life event logs, achieve a notable trade-off between data utility and privacy.

\item \textit{Applications to smart health}:
In addition to the security and privacy concerns reviewed along the thesis, we also provide some contributions to smart health applications.
First, we present a smart route recommender application to foster healthier lifestyles considering the health status of citizens, contextual data and crowdsourced-based information.
Second, we propose a novel platform to support the management of early mobilisation programmes in patients hospitalised in intensive care units.
And third, we describe an integral system to support healthcare decisions using geographic information systems.

\end{enumerate}

\section{Organisation} \label{sec:intro:organisation}

The thesis is organised into the following five parts as follows.

\begin{itemize}
\item Part \ref{part:intro}: \nameref{part:intro}
\begin{itemize}
    \item Chapter \ref{chap:intro} introduces the rationale of the research conducted in this dissertation and enumerates the main contributions.
    \item Chapter \ref{chap:background} provides an extensive background on the different key topics of this dissertation, this is, smart health, context-aware environments, privacy protection, statistical disclosure control, process mining and privacy-preserving process mining.
\end{itemize}
\item Part \ref{part:cae}: \nameref{part:cae}
\begin{itemize}
    \item Chapter \ref{chap:sensors} provides a security analysis on the technical deployment of more secure and private context-aware environments enabling smart health services. In particular, the most prominent sensors and communication networks are detailed and classified. Moreover, the main security attacks, threats and vulnerabilities in these environments are studied, and a number of countermeasures are proposed to overcome them. Moreover, key future research directions are briefly outlined.
    \item Chapter \ref{chap:ubicomp} presents a novel framework to analyse the privacy aspects of ubiquitous computing systems from an individual-centred perspective using five dimensions, namely identity privacy, query privacy, location privacy, footprint privacy and intelligence privacy. Fundamental challenges and the role of new technologies in ubiquitous computing are briefly discussed too.
    \item Chapter \ref{chap:cogcities} explores cognitive cities, and classifies their main challenges in the field of security and privacy into three categories, namely technical challenges, societal challenges and regulatory challenges. Besides, to make cognitive cities a near reality, key opportunities and future research directions are enumerated.
\end{itemize}
\item Part \ref{part:pm}: \nameref{part:pm}
\begin{itemize}
    \item Chapter \ref{chap:pmcae} explores the opportunities of conducting on-line and proactive process analyses in context-aware environments, and identifies and classifies the main challenges into four categories, namely data challenges, ecosystem challenges, computational challenges and automation challenges.
    \item Chapter \ref{chap:skipminer} presents a novel process discovery algorithm, called Skip Miner, with a probabilistic built-in heuristic able to bring structure to the already complex healthcare processes.
    \item Chapter \ref{chap:upppm} proposes a novel privacy-preserving process mining method, called \textit{u}-PPPM, based on the uniformisation of sensitive distributions of event data attributes, which are prone to be exploited to re-identify people by means of distribution-based attacks in combination with targeted location-oriented attacks.
    \item Chapter \ref{chap:kpppm} presents a novel privacy-preserving process mining method, called \textit{k}-PPPM, based on microaggregation techniques to increase people's privacy through the \textit{k}-anonymity protection model during process mining analyses.
\end{itemize}
\item Part \ref{part:apps}: \nameref{part:apps}
\begin{itemize}
    \item Chapter \ref{chap:apps} proposes three applications to smart health aiming to improve the quality of life of citizens. First, a context-aware recommender system for healthy routes, called SmartRoute, is presented. Second, a platform supporting early mobilisation programmes, called e-PEMICU, is presented with the aim to improve the recovery process of critically ill patients hospitalised in intensive care units. Finally, HGIS, a geographic information system supporting the integration, analysis and visualisation of heterogeneous health data with spatio-temporal features, is presented.
\end{itemize}
\item Part \ref{part:concl}: \nameref{part:concl}
\begin{itemize}
    \item Chapter \ref{chap:conclusions} summarises the contributions of this dissertation and points out some future research lines.
    \item Chapter \ref{chap:publications} enumerates the scientific publications supporting the content of this dissertation.
\end{itemize}
\end{itemize}
\chapter{Background}
\label{chap:background}

\emph{This chapter provides the reader with the necessary context to understand the topics discussed in this dissertation. First, the evolution of the provision of healthcare services by means of ICT-based paradigms, from electronic health to smart health, is discussed in Section \ref{sec:background:evolhealth}. Next, Section \ref{sec:background:privprot} provides background on statistical disclosure control, a primary field within privacy protection research. The most common privacy-preserving data transformation techniques and the main privacy models are presented in Sections \ref{subsec:background:dtt} and \ref{subsec:background:priv_mod}, respectively. Finally, background on process mining is provided in Section \ref{sec:background:procmin}. More specifically, in addition to the conceptualisation of this research field in Sections \ref{subsec:background:proc_models} to \ref{subsec:background:pm}, the applicability of process mining in the healthcare domain is demonstrated in \ref{subsec:background:pm_health}, and a throughout review of the state of the art on the novel privacy-preserving process mining research direction is provided in \ref{subsec:background:pppm}.}

\minitoc

\section{The Evolution of Healthcare} \label{sec:background:evolhealth}

The social, economical, political and medical developments of the latest century have reshaped the world's demography.
The steady growth of the global population along with the increase of the life expectancy are increasingly contributing to the ageing of our society.
Over the next three decades, predictions estimate a 25\% increase of the world's population --from today's 7.8 billion to 9.7 billion in 2050, being the highest growth rates in African and Asian countries--, and a 5-years increase of the life expectancy at birth --from today's 72.6 years to 77.1 years in 2050, being this increment more significant in high-income countries-- \cite{un2019population,kontis2017future}.
As a result, people over age 65 are the fastest-growing age group, and they have already outnumbered, for the first time in history, children under the age of five in 2018.
Prospects estimate that one in six people in the world will be over age 65 by 2050 \cite{un2020ageing}.
This global phenomenon is accompanied by the acceleration of the urbanisation process, wherein people are increasingly moving from rural areas to urban areas, thus leading to highly-dense cities or ``mega-cities''.
Since 2007, more people are living in urban areas than in rural areas, reaching more than 4.2 billion people (55\% of the world's population) in urban areas in 2018.
Over the coming decades, it is expected that the urbanisation level will reach 70\%, due to the fast urbanisation process that Africa and Asia --still mostly rural-- will experience \cite{un2019urba}.

Countries all over the world must be prepared for this demographic shift, so governments face tremendous pressure to effectively adapt their cities to a growing, older and more demanding population.
Modernising urban infrastructures, managing natural resources sustainably, and protecting the environment from pollution are just some of the many challenges to face \cite{shahidehpour2018smart,khanna2015urbanisation}.
In particular, the healthcare sector will gain even more relevance in the future, as older people are prone to suffer from multiple (and sometimes chronic) illnesses with expensive and long-term treatments.
Nowadays, cardiovascular diseases are the first death cause globally, mainly due to heart attacks and strokes, taking the life of approximately 18 million people annually, representing a third of all global deaths.
These diseases, generally influenced by risk factors such as hypertension or diabetes, require early detection and management using counselling and medicines to prevent fatal consequences \cite{who2021cvd}.
Significantly enough, respiratory diseases are an immense worldwide health burden due to their generally chronic nature.
Preventing respiratory diseases are among the most important cost-effective health interventions, so that decision-makers should consider promoting environmental-friendly regulations as top priority \cite{firs2019respiratory}.
Asthma, a common respiratory disease with high prevalence among children, affects around 260 million people worldwide \cite{who2021asthma}, and Chronic Obstructive Pulmonary Disease (COPD), a progressive life-threatening lung disease, causes the death of 3 million people every year, making it the third leading cause of death worldwide \cite{who2021copd}.
Moreover, as an inevitable consequence of population ageing, dementia syndrome, responsible for progressively impairing the cognitive abilities, has gained importance in the latest years among elderly.
The most common form of dementia is Alzheimer disease.
Whereas around 50 million people have dementia nowadays, nearly 10 million new cases are diagnosed every year --one case every 3 seconds!--, and estimations predict that these figures will triple by 2050 \cite{un2017dementia,adi2018alzheimer}.
Unlike other diseases, dementia has a psychological and social impact, not only on the people with dementia, but also on their relatives and caregivers, who must support them in their daily activities.

External factors also influence public health and can accelerate the appearance of diseases and even cause deaths.
First, poor environment conditions and climate change must be seriously handled to enhance worldwide health.
For instance, an early exposure to air pollutants, chemicals and dust increases the probability to suffer from cardiovascular or chronic respiratory diseases throughout people's life \cite{kim2018air}, and overexposure to ultraviolet radiation can lead to various types of skin cancer \cite{watson2016uv}.
Healthier environments could prevent more than 12 million deaths (the 23\% of all global deaths) every year \cite{who2017environment}.
Second, the overcrowded areas resulting from the urbanisation process can accelerate the spreading of infectious diseases and cause epidemics, as recently observed with the COVID-19 disease \cite{li2020propagation}.
Finally, negative behavioural habits, such as unhealthy diets, sedentarism, tobacco use and alcohol and drugs consumption, are risk factors associated with disease outbreaks (generally cardiovascular-related) and unhealthy lifestyles.
With the aim to raise global awareness on healthy ageing, the 73rd World Health Assembly adopted the 2020-2030 as the ``Decade of Healthy Ageing'' as a pivotal period to join forces and align actions towards this direction \cite{who2020decade}.

Healthcare systems are gradually adapting to the socio-economical needs to achieve objectives, such as reducing costs and times, increasing efficiency, improving treatments and care protocols and enhancing the quality of life of patients.
These objectives are required to achieve the sustainability of the entire healthcare sector.
To this end, the adoption of ICT has played a fundamental role and opened the door to the appearance of novel healthcare paradigms integrating the latest technological advancements (\eg smartphones and other mobile devices, sensors, IoT devices\ldots) to medical daily practice.
Also, ICT have deeply changed the conception of healthcare among citizenship, involving the patients as active elements of the system that can proactively contribute to their own health.
Far from classic healthcare paradigms comprising handwritten records, modern ICT-based paradigms have evolved towards \textit{electronic health} considering the use of ICT devices, later to \textit{mobile health} using mobile technology, and currently to \textit{smart health} that complements health data with sensing information gathered from context-aware environments.
Figure \ref{fig:background:health_paradigms} illustrates the relationship between the most popular healthcare paradigms explained below.

\begin{figure}[b!]
\centering
\includegraphics[width=0.85\columnwidth]{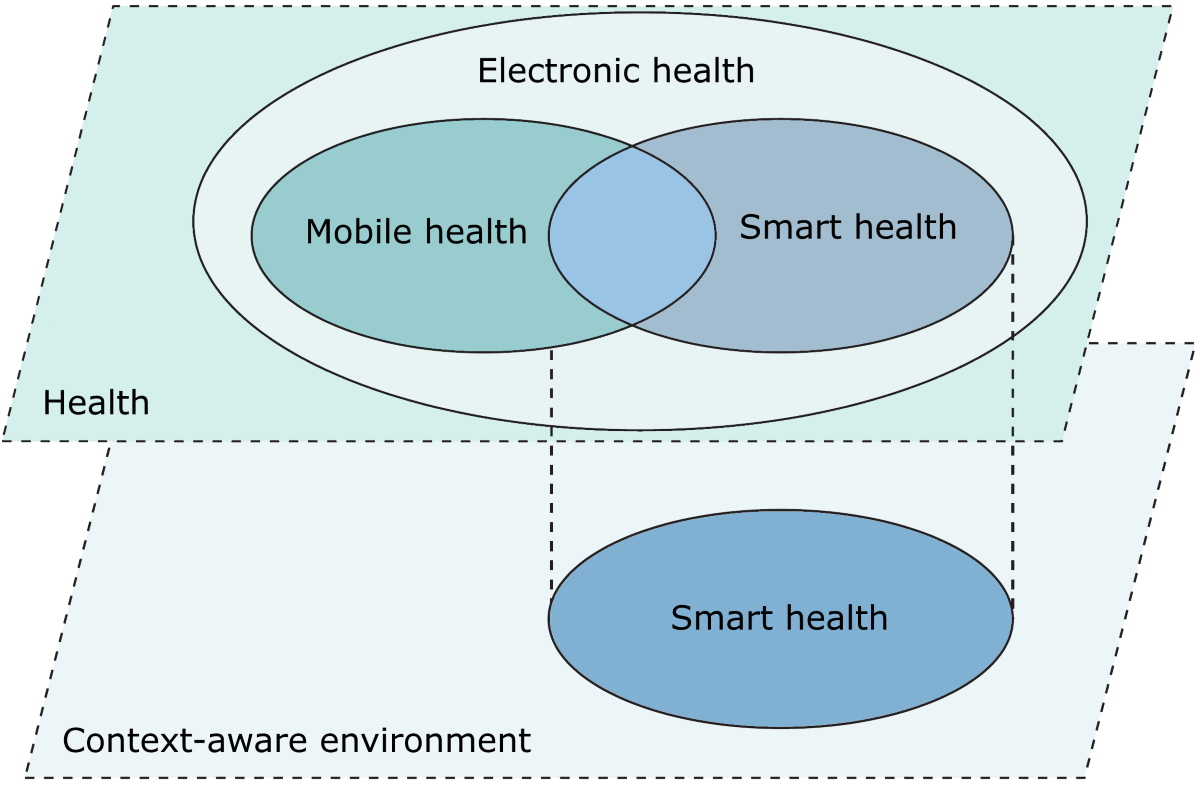}
\caption{Set diagram with the relationship between the electronic health, mobile health and smart health paradigms (adapted from \cite{solanas2014shealth}).}
\label{fig:background:health_paradigms}
\end{figure}

\subsection{Electronic Health}

With the adoption of ICT within the healthcare sector, the classic health paradigm evolved to embrace the so-called \textit{electronic health} (or \textit{e-health}), introduced by Eysenbach as:

\begin{quotation}
  \noindent \textit{``an emerging field in the intersection of medical informatics, public health and business, referring to health services and information delivered or enhanced through the Internet and related technologies.''}
  \begin{flushright}G. Eysenbach (2001) \cite{eysenbach2001ehealth}\end{flushright}
\end{quotation}

Even though e-health is no longer an emerging field, its conception represented a revolution in this area.
Thanks to the use of complex systems considering computers, electronic devices, databases and network infrastructures, e-health helped to re-design, improve and optimise numerous healthcare processes.
Among these, it enabled the provision of on-line medical services and care treatments, the exchange of electronic health records in a standardised way between healthcare providers, and remote communications between patients and medical personnel.
Properly implemented, e-health can shorten waiting times and reduce costs (\eg avoiding duplicate or unnecessary diagnostics or therapeutic interventions), thereby increasing efficiency and improving the comfortability and quality of life of patients \cite{ball2001health}.

\subsection{Mobile Health}

Following the consolidation of e-health, mobile devices (\eg smartphones, tablets) became commonplace, and their use generalised rapidly thanks to their remarkable computing and communication capabilities at low costs.
This fact led to the \textit{mobile health} (or \textit{m-health}) paradigm, by delivering healthcare services through mobile communication devices.
More formally, the first definition for m-health was proposed by Istepanian \textit{et al.} as:

\begin{quotation}
  \noindent \textit{``an emerging mobile communications and network technologies for healthcare systems.''}
  \begin{flushright}R. Istepanian \textit{et al.} (2006) \cite{istepanian2006mhealth}\end{flushright}
\end{quotation}

Clearly, m-health has a much broader potential than e-health and incorporates novel capabilities to the provision of healthcare services, namely 
(i) ubiquity, allowing devices with global monitoring capabilities having access to an unprecedented number of services, data and knowledge;
(ii) mobility, collecting data and accessing to services from everywhere using the wireless communication networks;
and (iii) user-centric, patients receive customised added-value services tailored to their needs (\eg medical data, location\ldots).
The use of mobile devices enables the collection of patients data in real-time so providing effective and early access to health services, while reducing times and costs \cite{steinhubl2013can}.
Moreover, the popularity of mobile devices, especially among youth, may engage citizenship to have healthier habits by controlling their own health (\ie self-monitoring) and foster adherence to treatments and medication plans in already-diagnosed patients \cite{anderson2016mobile,cechetti2019developing}.
The range of m-health applications vary from remote monitoring and rehabilitation, to the early detection of emergency situations (\eg falls or getting lost) or abnormal changes in health conditions (\eg hypertension, hypoglycemia or tachycardia) that can prevent at-risk patients from fatal outcomes.
For instance, the work in \cite{batista2015monitoring} proposes a remote monitoring system using low-cost smartphones for people with mild cognitive impairments (MCI) who can experience wandering behaviour \cite{solanas2015wandering}.
With the aim to overcome unhealthy addictions, such as tobacco, alcohol and drugs, \cite{carpenter2020developments} reviews the latest m-health developments to intervene and support addicted or recovered patients.
Also, patients suffering from chronic respiratory diseases have a large number of self-monitoring and feedback m-health tools to manage and monitor their symptoms \cite{sleurs2019mobile}.
Like the e-health paradigm, although m-health can no longer be considered an emerging field, it still stands as a consolidated area of application and research.

\subsection{Context-Aware Environments}

Along with the irreversible implementation of ICT in the healthcare domain, cities already started to equip their infrastructures with ICT to face the urbanisation challenges.
In this context, ICT empower the capabilities of the cities, and enable gathering real-time data to help local governments to make better decisions, manage resources more sustainably and enhance the quality of life of their citizens.
This phenomenon allows the transformation of cities into \textit{smart cities}.
Despite the many definitions, one of the first accepted definitions for smart cities was suggested by Caragliu \textit{et al.} \cite{caragliu2011smart}, which was later extended by P{\'e}rez-Mart{\'i}nez \textit{et al.} as:

\begin{quotation}
  \noindent \textit{``cities strongly founded on information and communication technologies that invest in human and social capital to improve the quality of life of their citizens by fostering economic growth, participatory governance, wise  management of resources, sustainability, and efficient mobility, whilst they guaranty the privacy and security of the citizens.''}
  \begin{flushright}P. P{\'e}rez-Mart{\'i}nez \textit{et al.} (2013) \cite{perez2013cities}\end{flushright}
\end{quotation}

Smart cities, as an example of smart regions \cite{roth2013smart}, take into account the interactivity between people with their immediate environment.
Therefore, citizens act as intelligent sensors and enrich the city with their knowledge and sensing capabilities about their nearby context, thus becoming information producers instead of mere information consumers.
By generalising the concept, environments being capable of adapting themselves to the users needs as well as to the requirements of such systems are commonly known as \textit{context-aware environments} \cite{lopez2017challenges}.
Smart homes, smart buildings, smart hospitals and smart cities are well-known examples of context-aware environments.

\subsection{Smart Health}

Rapidly, the natural synergy between the e-health and m-health paradigms converged with the rising of smart cities and context-aware environments, and inspired the next healthcare paradigm: \textit{smart health} (or \textit{s-health}), first defined by Solanas \textit{et al.} as:

\begin{quotation}
  \noindent \textit{``the provision of health services by using the context-aware network and sensing infrastructure of smart cities.''}
  \begin{flushright}A. Solanas \textit{et al.} (2014) \cite{solanas2014shealth}\end{flushright}
\end{quotation}

Although the provided definition was inspired by smart cities, the concept goes beyond the boundaries of smart cities, and it could generalise to context-aware environments.
Considering the above definition, a smart health application or service is truly ``smart'' only if it provides healthcare services or promotes healthy habits by using the potential of context-aware environments.
Like the m-health paradigm, the use of patient data and mobility data is considered in the smart health, however the later surpasses the potential of m-health and augments its capabilities in two ways.
On the one hand, smart health also considers data coming from the sensing infrastructure of the context-aware environment, a completely different source that is independent from the patient.
On the other hand, whereas user-centrality is considered in both paradigms, smart health also introduces an environment-centric approach, since the information gathered by the users also modifies the behaviour of the environment \cite{solanas2014shealth,patsakis2014personalized}.
For instance, in emergency situations, such as road accidents where people's lives could be at stake, the behaviour of the environment could be modified (\eg by changing the traffic lights in a smart city) so as to fasten the arrival of ambulances \cite{patsakis2015emergency}.
Another s-health application can provide real-time health recommendations to high-risk population considering the environmental conditions \cite{casino2015context}, especially significant for people suffering from COPD or other respiratory diseases \cite{riano2014copd}.
The widespread use of the IoT, together with their inherent capabilities to gather contextual data, can ease the transition from environments into context-aware environments, and therefore contribute to the massive deployment of smart health solutions at large scale.

\section{Privacy Protection} \label{sec:background:privprot}

Digital services have become an essential part of our daily lives.
The seamlessly integration of these services with ubiquitous devices entails endless possibilities to collect and analyse all sorts of (personal) data.
However, this excessive level of integration arises a number of challenges, including privacy, one of the most fundamental human rights \cite{stewart2017comment}.
To limit the disclosure of excessive information to unauthorised parties, \textit{information privacy} considers the privacy aspects associated to personally identifiable individuals with regards to their personal data and communications \cite{torra2017privacy}.
Due to the myriad connotations and perspectives of privacy \cite{cockcroft2016relationship}, numerous definitions for ``information privacy'' have been suggested \cite{belanger2011privacy,martin2017role}.
One of the first definitions was given by Clarke, who conceptualised information privacy as:

\begin{quotation}
  \noindent \textit{``the interest an individual has in controlling, or at least significantly influencing, the handling of data about themselves.''}
  \begin{flushright}R. Clarke (1997) \cite{clarke1997introduction}\end{flushright}
\end{quotation}

The value of digital data, especially people's confidential information like health records and banking details, has significantly increased over the past decade \cite{blackmarketprice}.
Consequently, espionage, data misuse and data abuse have become a sad reality in some services providers, sacrificing user's privacy in exchange of economic compensations or further benefits \cite{buck2017appip}.
The Cambridge Analytica \cite{isaak2018user} and the Snowden cases \cite{fuchs2017internet} are some of the most recent scandals against privacy that have had a worldwide impact.
Indeed, many m-health applications fail at properly addressing the privacy aspects and confidentiality of users medical data, despite their high sensitivity \cite{papageorgiou2018security,sunyaev2015availability}.
Nevertheless, it is worth noting that attacks against privacy are not always intentional.
For instance, data breaches may be unintentional in some cases (\eg accidental online publications, hacks or system glitches), although they equally jeopardise users privacy \cite{leakhiv,myfitnesspalhack,yahoohack}.
All in all, ensuring information privacy to preserve people's privacy is one of the main challenges and a very active research field today.

The rapid changes in the digital data landscape led to the necessity for adapting the legislative domain to the new era.
With the aim to harmonise the current data protection laws across the European Union, the General Data Protection Regulation (GDPR) became enforceable in all the Member States on May 25th 2018 \cite{eugdpr}.
This regulation responded to people's worries about the inadequate protection of their personal data with the obsolete 1995's Data Protection Directive (DPD) 95/46/EC \cite{eudpd}, which left room for interpretation into the individual national laws of each Member State and did not consider most of the technological advancements.
The GDPR strengthens the already-established data protection principles defined by the DPD, such as consent and purpose limitation, and attempts to return the control of personal data back to the citizens by granting them extensive rights, such as the right to be forgotten and the right to data portability, among others \cite{politou2018forgetting,de2016new}.
Significantly enough, GDPR builds upon the \textit{privacy by design} principles, enforcing organisations to consider all the privacy requirements and implement appropriate measures at the early stages of the design of a service.
These principles, defined by Cavoukian a decade ago \cite{cavoukian20107principles}, encapsulate concepts like data minimisation, purpose specification, disclosure limitation, transparency and end-to-end security through pseudonymisation or encryption, among others.
Controversially, the application of these principles may go against the nature of the big data, machine learning and the IoT technologies, thus arising a number of serious challenges to face \cite{edwards2016privacy,zarsky2017incompatible}.
To foster its implementation, the failure in the compliance of the GDPR entails severe economic sanctions, which have already been suffered by some companies \cite{tambou2019fines}.
Law experts agree on the fact that GDPR has supposed the biggest overhaul on data protection laws across EU.

\subsection{Foundations} \label{subsec:background:sdc}

\textit{Statistical Disclosure Control} (SDC) \cite{hundepool2012statistical} comprises a set of data-driven techniques used to anonymise data in microdata sets, \ie datasets consisting of multiple records corresponding to individual respondents.
The anonymisation is conducted before the publication in such a way that it prevents (i) \textit{identity disclosure}, \ie a respondent can not be re-identified from any particular record in the published microdata set, and (ii) \textit{attribute disclosure}, \ie the value of a confidential attribute (\eg health condition, salary\ldots) can not be discovered for a specific respondent.

A microdata set $\mathcal{T}$ with $r$ respondents (records) and $s$ attributes (dimensions) is a $r \times s$ matrix where $\mathcal{T}_{ij}$ is the value of attribute $j$ for respondent $i$.
Attributes can be classified into four categories in accordance to their disclosure potential~\cite{hundepool2012statistical}.

\begin{itemize}
    \item \textit{Direct Identifier}: Attributes that can unambiguously and uniquely identify the respondent, such as the social security number, passport number and full name.
    \item \textit{Quasi-Identifier} or \textit{Key Attribute}: Attributes that identify the respondent with some ambiguity, but not uniquely. However, the combination of multiple quasi-identifiers might re-identify the respondent uniquely. Common quasi-identifier attributes refer to publicly known characteristics of the respondents, such as the gender, age, birth data and ZIP code.
    \item \textit{Sensitive Attribute}: Attributes that contain confidential respondent information, including the salary, disease, race, religion belief and political affiliation. Adversaries aim to link these attributes to direct identifiers: when succeeded, respondents privacy is compromised.
    \item \textit{Non-Sensitive Attribute}: Attributes that contain non-confidential respondent information, \ie other attributes than direct identifiers, quasi-identifiers and sensitive attributes, such as the job.
\end{itemize}

Table \ref{tbl:background:mds} shows a toy example of a microdata set $\mathcal{T}$.
This example discloses medical information of twelve respondents from a fictitious hospital institution.
With regards to the five attributes, one of them is a direct identifier (\ie full name, in blue), three are quasi-identifiers (\ie age, gender and ZIP code, in green), and the latest is a sensitive attribute (\ie disease, in red).
Clearly, $\mathcal{T}$ can not be published as such, because it seamlessly links confidential information to its respondents.
With the objective to create a protected microdata set $\mathcal{T}'$ for release, direct identifiers are removed in the first place.
Although this process helps reducing disclosure risks, it is often not sufficient.
More specifically, disclosure risks come from the feasibility to re-identify respondents by cross-correlating multiple quasi-identifiers shared between external (and non-anonymous) datasources.
This is exemplified in Figure \ref{fig:background:qid_shared}: confidential medical information might be correlated to direct identifiers by linking the quasi-identifier attributes shared between the assumingly anonymous medical microdata set and a public voter list datasource.
The high analytical value of quasi-identifiers invalidates the possibility to suppress them, so they must be anonymised through a transformation technique \cite{willenborg2012elements} (see Section \ref{subsec:background:dtt}).

\begin{table}[b!]
\centering
\caption{Example of a microdata set $\mathcal{T}$. Direct identifiers are highlighted in blue, quasi-identifiers in green, and sensitive attributes in red.}
\label{tbl:background:mds}
\resizebox{0.8\textwidth}{!}{%
\renewcommand{\arraystretch}{0.85}
\begin{tabular}{|c|>{\columncolor[RGB]{237,247,255}}c||>{\columncolor[RGB]{232,246,243}}c>{\columncolor[RGB]{232,246,243}}c>{\columncolor[RGB]{232,246,243}}c||>{\columncolor[RGB]{253,240,239}}c|} \hline
{\bf \#} & {\bf Full name} & {\bf Age} & {\bf Gender} & {\bf ZIP code} & {\bf Disease} \\ \hline
1 & Drew Jones & 34 & Male & 43053 & Asthma \\
2 & William Jones & 39 & Male & 43068 & Heart Disease  \\
3 & John Blake & 37 & Male & 43053 & Heart Disease \\
4 & Martha Terry & 31 & Female & 43068 & Asthma  \\
5 & Joseph Greene & 41 & Male & 43053 & Cancer  \\
6 & Debbie Dougan & 47 & Female & 43053 & Cancer  \\
7 & Nancy Wagner & 46 & Female & 43068 & Cancer \\
8 & Mario Bryden & 49 & Male & 43068 & Cancer \\
9 & Marie Collins & 65 & Female & 44853 & Heart Disease  \\
10 & Vanessa Williamson & 60 & Female & 44853 & Cancer \\
11 & Elizabeth Rio & 57 & Female & 44850 & Asthma  \\
12 & Anthony Boswell & 59 & Male & 44850 & Asthma \\
\hline
\end{tabular}%
}
\end{table}

\begin{figure}[t!]
\centering
\includegraphics[width=0.59\columnwidth]{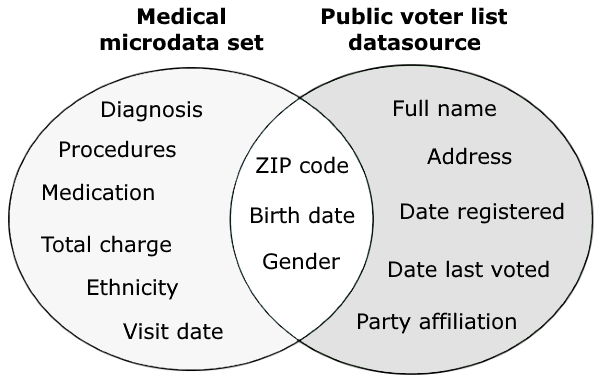}
\caption{Quasi-identifiers shared between datasources that enable respondent re-identification (adapted from \cite{sweeney2002kanonymity}).}
\label{fig:background:qid_shared}
\end{figure}

\begin{figure}[b!]
\centering
\includegraphics[width=0.74\columnwidth]{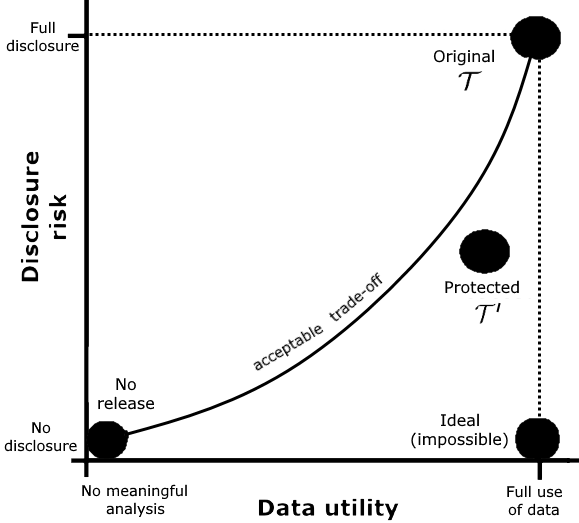}
\caption{Trade-off between data utility and disclosure risk during anonymisation using SDC techniques (adapted from \cite{shlomo2018sdc}).}
\label{fig:background:utility_vs_disclosure}
\end{figure}

It is worth noting that there is no single anonymity procedure, and the transformation techniques to be used can vary according to the microdata sets and the privacy requirements.
By anonymising the data, the released microdata set $\mathcal{T}'$ will lose information and, therefore, its usefulness diminishes.
From a statistical perspective, the main challenge of SDC is to maintain an appropriate balance between the privacy level and the utility of the data \cite{templ2014sdc}.
This well-known trade-off has been extensively reported in the literature \cite{domingo2009measuring,li2009tradeoff,shlomo2018sdc}.
As observed in Figure \ref{fig:background:utility_vs_disclosure}, the original microdata set $\mathcal{T}$ has the maximum data utility, but also maximum disclosure risk, because data is not distorted and re-identification is trivial (\ie no privacy considered).
On the contrary, perfect privacy (\ie no disclosure risk) would only be guaranteed if the protected microdata set is not released, so the utility of the data is zero.
Ideally, although impossible, the achievement of a perfect privacy by maintaining the utility of the data in the original microdata set would represent a perfect anonymisation.
Generally, when an anonymisation technique is applied, the disclosure risk of the protected microdata set $\mathcal{T}'$ lowers along with the data utility.

\subsection{Data Transformation Techniques} \label{subsec:background:dtt}

Anonymisation techniques are required to hide information in $\mathcal{T}$ with the aim to preserve privacy.
Many techniques have been proposed in the literature to transform a microdata set $\mathcal{T}$ into a new protected microdata set $\mathcal{T}'$ that fulfils a set of privacy requirements.
Despite the many taxonomies, these techniques can generally be classified into \textit{perturbative} and \textit{non-perturbative} \cite{domingo2001quantitative,matthews2011data,zigomitros2020survey,domingo2005ordinal}.
Roughly speaking, perturbative methods transform $\mathcal{T}$ into $\mathcal{T}'$ with statistically meaningful values so that certain statistical properties are maintained.
On the other hand, non-perturbative methods transform $\mathcal{T}$ into $\mathcal{T}'$ through partial suppressions or reduction of detail, although the data in $\mathcal{T}'$ is still true.
Next, the most prevalent data transformation techniques are introduced.

\subsubsection{Perturbative Techniques}

\textit{Noise addition} is a classical perturbative technique typically applied to introduce random noise to the numerical values in $\mathcal{T}$.
Taking a Gaussian distribution with zero mean and standard deviation $\sigma$ (\ie $\mathcal{N}$(0, $\sigma$)), values can be masked: the higher the $\sigma$ value, the greater the range of the generated values in $\mathcal{T}'$.
Although original values are disguised, computations are still statistically sound if there is interest in the aggregated data rather than the very individual data.
The scalar product and random sum are widely used methods that can estimate with decent accuracy aggregated data.

Widely used to anonymise sensitive attributes, the \textit{data swapping} technique consists in the exchange of sensitive attributes values among individual records.
As the values are not modified, statistical distributions and frequencies are not affected.
To maximise the utility of the data, swaps should not be random due to the possible inter-dependencies between the quasi-identifiers and sensitive attributes.
For instance, in a protected medical microdata set $\mathcal{T}'$, ``prostate cancer'' should not be assigned to a woman.
Although data swapping can be used for both numerical and categorical values, there is also the \textit{rank swapping} variant that can only be applied on numerical values.
In this variant, after sorting the numerical values in ascending order, a value can be swapped with another randomly chosen value within a restricted range.

Another perturbative technique is \textit{synthetic data generation}, whereby a synthetic microdata set $\mathcal{T}'$ is created from a mathematical model that preserves the basic statistical measures from the original data $\mathcal{T}$.
In general, this technique virtually eliminates the possibility of re-identification because data is synthetic.
However, it limits the utility of the data for analysis on random subdomains.

Significantly popular, \textit{microaggregation} is a family of perturbative SDC techniques used to reduce re-identification risk.
Originally defined for numerical data, it was later extended for categorical data as well.
This technique works in two different steps.
The first step partitions the set of records of the microdata set $\mathcal{T}$ into clusters in such a way that (i) each record must be assign to one cluster, (ii) each cluster contains, at least, $k$ records, and (iii) the records within a cluster are as similar as possible.
Although the latter condition is not mandatory, it contributes to maximise the within-group homogeneity and lower the information loss in the resulting protected microdata set $\mathcal{T}'$.
In the second step, each record in a cluster is replaced by the representative or centroid of the cluster, typically the mean for numerical data or the median for categorical data.
When microaggregation is applied to the projection of records on their quasi-identifiers, the protected microdata set $\mathcal{T}'$ achieves the \textit{k}-anonymous property (explained in Section \ref{subsec:background:priv_mod}).
Exact optimal microaggregation is \textit{NP}-hard, so heuristics are used to approximate the solution to the optimal one in a reasonable time.
The Maximum Distance to Average Vector (MDAV) is a standard microaggregation algorithm for clustering records based on classical Euclidean distances in a multivariate space \cite{domingo2005ordinal}.
The \textit{k}-member \cite{byun2007kmember} and One-pass K-means Algorithm (OKA) \cite{lin2008oka} are other microaggregation algorithms from the literature.
Whereas the aforementioned algorithms force the creation of clusters with size $k$, other algorithms, such as Variable-MDAV (V-MDAV) \cite{solanas2006vmdav}, allow groups with variable sizes between $k$ and $2k-1$.

\subsubsection{Non-Perturbative Techniques}

A straightforward non-perturbative technique is \textit{sampling}.
This technique consists in creating a protected microdata set $\mathcal{T}'$ by selecting a portion (\ie a sample) of the records from the microdata set $\mathcal{T}$.
The sampled microdata set $\mathcal{T}'$, which is expected to recreate the average behaviour of $\mathcal{T}$, is used to draw similar conclusions as if working with the original $\mathcal{T}$.
Although $\mathcal{T}'$ contains real data, it reduces the disclosure risks due to the difficulties to demonstrate that a unique record in $\mathcal{T}'$ is also unique in $\mathcal{T}$.

Another approach is \textit{top/bottom coding}, consisting in limiting the highest/lowest values for a given attribute.
Particularly, \textit{top coding} sets an upper limit on all the values of an attribute and replaces any value greater than this limit by the established upper limit.
Contrary, \textit{bottom coding} replaces any value below a pre-specific lower limit by the lower limit.
This technique can reduce the disclosure risk of individuals with extreme values.
For instance, if an individual has an extremely high salary (\eg the CEO of a large company), instead of reporting the exact amount, which would make the observation vulnerable to disclosure, it is reported with the non-identifiable upper limit.
A more robust variant consists in selecting the $p$ highest (lowest) values of an attribute, and replacing them by the upper (lower) limit, so that such limits appear in $p$ records.

Generalising the above technique, \textit{generalisation} is a non-perturbative technique aiming to replace the value of an attribute with an abstraction of the original value.
These abstractions can be user-defined using a taxonomy.
Generalisation in categorical attributes is possible by combining several categories into one less specific category but with higher frequency count.
For instance, multiple education levels in $\mathcal{T}$ (\eg ``high school'', ``undergraduate'', ``graduate'' and ``doctoral'') can be combined in $\mathcal{T}'$ (\eg ``high school'', ``undergraduate'' and ``graduate and above'').
As the number of doctoral students is generally smaller, their disclosure might be feasible.
The combination of this category with a broader one diminishes the re-identification risk.
Similarly, attributes corresponding to categorical geographical locations are prone to generalisation to prevent disclosing too much details (\eg generalising the towns into provinces or countries).
For numerical attributes, generalisation can be achieved through discretisation.
For example, the accuracy of ZIP codes can be generalised from 5 digits to less than 5 digits, \eg generalising 43068 in $\mathcal{T}$ as 4306*, 430**, etc., in $\mathcal{T}'$.
Also, ages can be generalised with intervals, \eg instead of reporting ages 25, 27 and 23 in $\mathcal{T}$, one could report $[20-30]$, 2* or $<30$ in $\mathcal{T}'$.

\textit{Suppression} techniques appear as a necessity to prevent releasing too specific records or values that could lead to disclosure.
Records containing unique or rare values that facilitate re-identification are removed.
Similarly, records whereby the combination of attributes is unique or nearly unique should also be suppressed.
Instead of suppressing the entire record, local suppression could be applied to suppress only the value in question.

\subsection{Privacy Models} \label{subsec:background:priv_mod}

Data anonymisation is performed with the aim to achieve some clearly-defined privacy requirements that reduce disclosure risks.
The definition of the privacy requirements comes determined by privacy models, which establish the level of protection of the protected microdata set $\mathcal{T}'$ \cite{zigomitros2020survey}.
Once a privacy model is chosen, then one can apply diverse data transformation techniques in $\mathcal{T}$ to maximise the utility of the resulting $\mathcal{T}'$ as long as it fulfils the requirements of the privacy model.

Introduced by Samarati and Sweeney two decades ago, \textit{k-anonymity} is one of the most well-known privacy model \cite{samarati1998kanonymity,sweeney2002kanonymity}.
By definition, a microdata set $\mathcal{T}'$ is \textit{k}-anonymous if it includes $k$ or more records (for $k>1$) for any combination of quasi-identifiers.
Therefore, each record in $\mathcal{T}'$ is indistinguishable from at least $k-1$ other records with respect to the quasi-identifiers.
The group of records sharing the same quasi-identifiers form an equivalence class.
If adversaries attempt to link records from $\mathcal{T}'$ with external datasources using shared quasi-identifiers, a unique match in $\mathcal{T}'$ will never be obtained, but a group of $k$ or more records.
Therefore, the probability to successfully associate a record to an individual is never greater than $1/k$.
Although the model counteracts identity disclosure, it is not able to prevent attribute disclosure.
This situation is depicted in Table~\ref{tbl:background:k-anonymity}, which shows a 4-anonymous version of the microdata set in Table~\ref{tbl:background:mds} by means of generalisation.
It can be observed that the microdata set $\mathcal{T}_1'$ is 4-anonymous because for every record there exist three others with the same quasi-identifier values.
Despite of that, the sensitive value of the second equivalence class (\#5--\#8) coincides in all its records.
The lack of diversity in the sensitive attribute may enable attribute disclosure when combined with background knowledge attacks.

\begin{table}[b!]
\parbox{.5\linewidth}{
\centering
\caption{A microdata set $\mathcal{T}_1'$ fulfilling \textit{k}-anonymity ($k=4$).}
\label{tbl:background:k-anonymity}
\resizebox{0.5\textwidth}{!}{%
\begin{tabular}{|c||>{\columncolor[RGB]{232,246,243}}c>{\columncolor[RGB]{232,246,243}}c>{\columncolor[RGB]{232,246,243}}c||>{\columncolor[RGB]{253,240,239}}c|} \hline
{\bf \#} & {\bf Age} & {\bf Gender} & {\bf ZIP code} & {\bf Disease} \\ \hline
1 & $<$40 & * & 430** & Asthma \\
2 & $<$40 & * & 430** & Heart Disease  \\
3 & $<$40 & * & 430** & Heart Disease \\
4 & $<$40 & * & 430** & Asthma  \\ \hline
5 & 4* & * & 430** & Cancer  \\
6 & 4* & * & 430** & Cancer  \\
7 & 4* & * & 430** & Cancer \\
8 & 4* & * & 430** & Cancer \\ \hline
9 & $>$50 & * & 4485* & Heart Disease  \\
10 & $>$50 & * & 4485* & Cancer \\
11 & $>$50 & * & 4485* & Asthma  \\
12 & $>$50 & * & 4485* & Asthma \\
\hline
\end{tabular}%
}
}
\hfill
\parbox{.5\linewidth}{
\centering
\caption{A microdata set $\mathcal{T}_2'$ fulfilling \textit{l}-diversity ($l=3$).}
\label{tbl:background:l-diversity}
\resizebox{0.5\textwidth}{!}{%
\begin{tabular}{|c||>{\columncolor[RGB]{232,246,243}}c>{\columncolor[RGB]{232,246,243}}c>{\columncolor[RGB]{232,246,243}}c||>{\columncolor[RGB]{253,240,239}}c|} \hline
{\bf \#} & {\bf Age} & {\bf Gender} & {\bf ZIP code} & {\bf Disease} \\
\hline
3 & $<$50 & * & 4305* & Heart Disease \\
1 & $<$50 & * & 4305* & Asthma \\
5 & $<$50 & * & 4305* & Cancer  \\
6 & $<$50 & * & 4305* & Cancer  \\ \hline
2 & $<$50 & * & 4306* & Heart Disease  \\
4 & $<$50 & * & 4306* & Asthma  \\
7 & $<$50 & * & 4306* & Cancer \\
8 & $<$50 & * & 4306* & Cancer \\ \hline
9 & $>$50 & * & 4485* & Heart Disease  \\
10 & $>$50 & * & 4485* & Cancer \\
11 & $>$50 & * & 4485* & Asthma  \\
12 & $>$50 & * & 4485* & Asthma \\ 
\hline
\end{tabular}%
}
}
\end{table}

To counter the limitations of \textit{k}-anonymity, the \textit{l-diversity} model introduced by Machanavajjhala \textit{et al.} \cite{machanavajjhala2007ldiversity} requires that the distribution of a sensitive attribute in each equivalence class contains, at least, $l$ well-represented values.
According to the definition of the privacy model, the term ``well-represented'' denotes that there are at least $l$ distinct values for the sensitive attributes in each equivalence class.
Following the example from Table \ref{tbl:background:mds}, a 3-diverse microdata set $\mathcal{T}_2'$ is illustrated in Table \ref{tbl:background:l-diversity}.
In this example, each equivalence class is represented by three different sensitive values, thus reducing disclosure risks.
Even though \textit{l}-diversity solves many weaknesses of \textit{k}-anonymity, it still presents two major shortcomings.
On the one hand, \textit{l}-diversity does not consider the semantic closeness of the sensitive values, thus $l$ different, but similar from a semantic perspective, can disclose some information.
For instance, assuming an equivalent class which is 3-diverse and the three values of this class are ``gastritis'', ``gastric ulcer'' and ``stomach cancer'', adversaries can deduce that the respondent has a stomach-related issue, regardless of the specific disease.
On the other hand, this model does not prevent attribute disclosure in case of skewed distributions, \ie the sensitive values have different degrees of sensitivity (\eg the probability of HIV+ is 1\% and HIV-- is 99\%).
For instance, privacy could be jeopardised if (i) an equivalence class contains half of the records with HIV+ and the other half HIV-- (\ie the probability of HIV+ would be of 50\%, rather than the original 1\%), or (ii) an equivalence class with many records contains one HIV-- and the rest HIV+ (\ie the probability of HIV+ would be close to 100\%).

With the aim to overcome the drawbacks of \textit{l}-diversity, Li \textit{et al.} \cite{li2007tcloseness} proposed \textit{t-closeness}, establishing that the distribution of a sensitive attribute in any equivalence class is close to the distribution of the attribute in the overall microdata set.
More specifically, the distance (computed using the Earth Mover Distance function \cite{rubner2000emd}) between both distributions should not exceed a threshold $t$.
Although the aforementioned privacy models are the most popular, the literature is plenty of further privacy models, such as $\delta$-presence \cite{nergiz2007hiding} and $\beta$-likeness \cite{cao2012publishing}, among others.
Moreover, some schemes attempt to guarantee multiple privacy models simultaneously, such as \textit{p}-sensitive \textit{k}-anonymity or \textit{t}-closeness \textit{k}-anonymity \cite{sun2008enhanced,truta2006privacy}.
Last, Dwork \cite{dwork2006differential} proposed the notion of \textit{differential privacy} (or $\varepsilon$-privacy).
Unlike the above models, the privacy protection of this model is not considered a property of the microdata set, but a property of a data process method.
It states that the risk to someone's privacy should not substantially increase (bounded by a parameter $\varepsilon$) as a result of participating in a statistical database.
Adversaries should not be able to learn any information about any individual that they could not learn if the participant had opted out of the microdata set.
Therefore, differential privacy ensures that the removal or addition of a single microdata set item does not significantly affect the outcome of any analysis.

\section{Process Mining} \label{sec:background:procmin}

Organisations operate in increasingly competitive markets, considering not only local and regional competitors, but also international competitors.
Moreover, the expansion of organisational operations together with complex interactions between organisations, suppliers and customers result into more complex organisations, whose management becomes increasingly harder.
With the aim to enhance efficiency and reduce costs, it is crucial to understand and properly manage business processes, which specify and enforce how organisations operate.
Many approaches to improve performance within organisations, such as Six Sigma \cite{pyzdek2003six} and Business Process Management \cite{dumas2013bpm}, show that efficient business processes contribute to improve the overall performance of organisations.
Nowadays, many organisations already use specific information systems to support the execution of their business processes \cite{dumas2005pais,aalst2009pais}.
Such systems typically manage, guide and support the execution of business processes by logging all information that occurred during their executions in the form of events.
These records provide a detailed and objective view of the processes executions and enable process-related analysis at different detail levels and from different perspectives.

Figure \ref{fig:background:pm_overview} contextualises the role played by process mining within organisations.
From a general view, process models describe the behaviour of business processes in the ``real world''.
With the specifications of process models, information systems are configured to support such business processes.
While business processes are being executed, information systems record all the event data generated in the form of event logs.
As a result, event logs contain historical facts about the actual executions of business processes.
Process mining \cite{aalst2011book,aalst2016book} establishes links between the actual observed process executions (\ie event logs) and the modelled process behaviour (\ie process models).

\begin{figure}[t!]
\centering
\includegraphics[width=0.95\columnwidth]{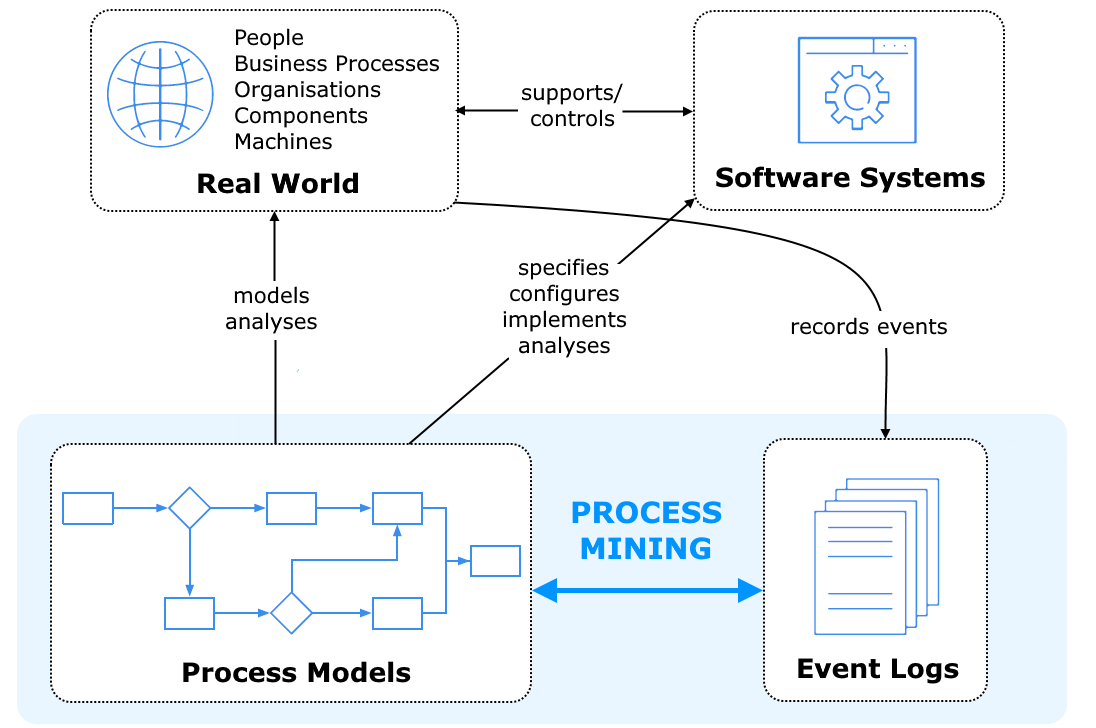}
\caption{High-level conceptual model of process mining (adapted from \cite{aalst2011book}).}
\label{fig:background:pm_overview}
\end{figure}

\subsection{Business Processes and Process Models} \label{subsec:background:proc_models}

A \textit{business process} (or \textit{process} in short) consists of a set of activities that are performed in coordination in an organisational and technical environment with the aim to achieve a business goal \cite{weske2007bpm}.
Despite the numerous definitions \cite{dumas2013bpm,hammer2009reengineering,havey2005essential,aalst2004workflow}, all pinpoint two key concepts: (i) the business process describes a finite collection of activities with some logical relationship and ordering between them, and (ii) the outcome of the business process provides value to the customers or the market.
This value can either be external or internal to the organisation \cite{dumas2013bpm}.
On the one hand, business processes with internal value could refer to the proper organisation's financial management, the inter-departmental communications protocols, or the hiring procedures from the human resources department.
On the other hand, business processes with external value are generally associated to the provision of services to customers, the production of products, or the management of customers complaints.
A business process can be as simple as two activities performed in a sequence, or it can be as complex as hundreds of activities to be performed considering loops, dependencies, conditionals and concurrency.
For example, when a patient enters a hospital, there exists a business process detailing the procedure to treat him/her according to the observed symptoms; when an applicant applies for a loan from a bank, the bank follows a business process to assess the application; and when a customer orders goods to a wholesale company, it initiates a predefined business process to deliver and charge these goods (see Figure~\ref{fig:background:ex_process}).

\begin{figure}[b!]
\centering
\includegraphics[width=0.83\columnwidth]{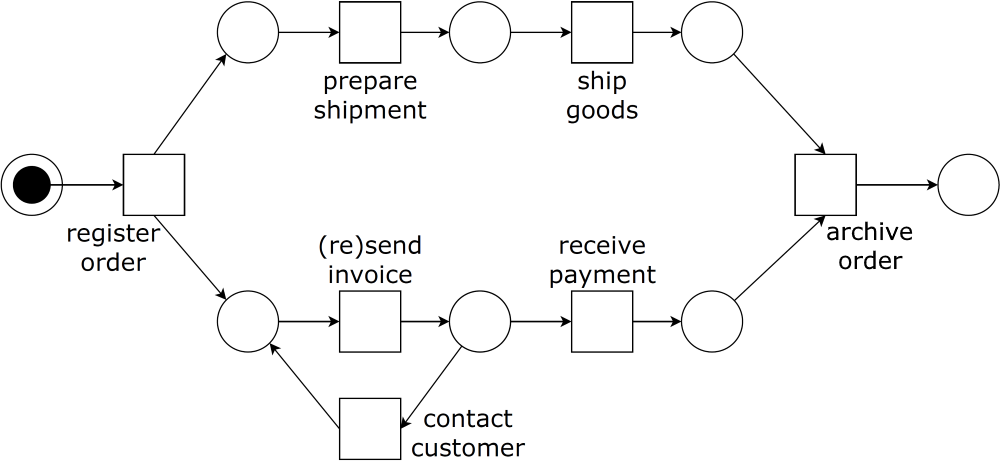}  
\caption{Model of a business process from a wholesale company for goods delivery.}
\label{fig:background:ex_process}
\end{figure}

Business processes are graphically represented using a certain notation in the form of \textit{process models} (or \textit{models}).
Process models can be used for a variety of reasons \cite{aalst2011book}.
Significantly important in large or complex organisations, process models serve to document business processes for management, certification (\eg ISO 9000 quality standards) or regulation compliance purposes (\eg the Sarbanes-Oxley Act \cite{karagiannis2007sox}).
Such documentation also provides an effective and efficient way to specify the functioning of a task that software developers need before its implementation.
Process models are also often useful to analyse business processes from various angles and find potential errors, deadlocks or inconsistencies.
Typically, for a given business process, two types of process models can be created, either formal or informal.
Formal models describe unambiguously all the details explicitly, which results into large and complex models that are difficult for stakeholders or non-expert users to understand.
In contrast, informal models ignore many details and show processes from a high-level view, thus becoming much simpler.
However, this simplicity may sometimes yield misleading insights.

Designing good process models is not straightforward.
In many cases, models tend to describe an idealised version of the reality, and ignore situations that, despite being undesired or uncommon, may happen.
Therefore, it is important to align the observed behaviour occurring in reality to the ones described in process models.
In consequence, process models should be flexible to changes and be resilient to deviations without operationally affecting the organisations outcomes as much.
Complex and dynamic environments, such as the healthcare sector, are prone to this issue.
For example, the patient handling process in an emergency department normally starts with the registration of an incoming patient followed by the assignment of the patient's priority before any further action is performed.
However, patients with life-threatening emergencies (\eg heart attacks, strokes or road accidents) may go directly to reanimation or surgery, skipping both registration and priority assignment activities.

The importance of modelling business processes is reflected by the plethora of modelling notations available in the literature \cite{aalst2011book}.
Process models are usually represented in terms of graphs, such as D/F-graphs, which are directed and labelled \cite{weijters2003rediscovering}.
Hereafter, we formalise a graph-based model $\mathcal{M} = \{N,E\}$ as a non-empty set $N$ of nodes and a set $E$ of edges, corresponding to ordered pairs of elements of $N$ with a certain weight.
Representing process models as graphs helps facing processes-related problems by means of graph techniques, such as calculating distances, similarities, paths and connectivity.
Further advanced graph-derived notations, such as Petri nets \cite{murata1989petri,peterson1977petri}, BPMN \cite{wohed2006bpmn,dijkman2008bpmn}, YAWL \cite{aalst2005yawl} and C-nets \cite{aalst2011cnets}, are also used to model business processes by introducing additional logic for managing concurrency, priorities, decision making or semantics.
For instance, the process model from Figure \ref{fig:background:ex_process} is represented using a petri net.

\subsection{Event Data and Event Logs}

Business processes are set in the information systems to leave traces in the form of \textit{events}, represented as records describing well-defined steps (\ie activities) of processes executions.
These events represent run-time information about how processes were actually executed.
All generated events are stored sequentially in the so-called \textit{event logs}.
For the sake of clarity, each event log contains events related to a single process.
Table \ref{tbl:background:ex_event_log} shows an excerpt of the event log related to the business process depicted in Figure \ref{fig:background:ex_process}.
Next, the notions related to event logs are formalised \cite{aalst2011book,suriadi2017event}.

\begin{table}[b!]
\centering
\caption{Excerpt of some event log. Each record corresponds to an event.}
\label{tbl:background:ex_event_log}
\resizebox{0.99\textwidth}{!}{%
\renewcommand{\arraystretch}{0.85}
\begin{tabular}{ccccccc} \toprule
\multirow{2}{*}{\bf Case ID} & \multirow{2}{*}{{\bf Event ID}} & \multicolumn{5}{c}{{\bf Properties}} \\ \cmidrule{3-7}
& & {\bf Activity} & {\bf Timestamp} & {\bf Resource} & {\bf Cost} & {\bf \ldots} \\ \midrule
1 & 54245 & Register order & 2017-03-15 08:03 & Lewis Khan & 15 & \ldots \\
1 & 54246 & Prepare shipment & 2017-03-16 11:27 & Cory Dixon & 200 & \ldots \\
1 & 54247 & Ship goods & 2017-03-16 23:14 & Bryan Benson & 100 & \ldots \\
1 & 54248 & Send invoice & 2017-03-17 04:34 & Freddie Jean & 50 & \ldots \\
1 & 54249 & Receive payment & 2017-03-19 10:52 & Annie Leslie & 20 & \ldots \\
1 & 54250 & Archive order & 2017-03-19 12:05 & Annie Leslie & 5 & \ldots \\ \\
2 & 60141 & Register order & 2017-10-23 11:55 & Lewis Khan & 15 & \ldots \\
2 & 60142 & Prepare shipment & 2017-10-25 06:26 & Bryan Benson & 200 & \ldots \\
2 & 60143 & Send invoice & 2017-10-26 17:08 & Mary Hudson & 50 & \ldots \\
2 & 60144 & Ship goods & 2017-10-27 14:32 & Bryan Benson & 100 & \ldots \\
2 & 60145 & Contact customer & 2017-11-05 09:41 & Mary Hudson & 40 & \ldots \\
2 & 60146 & Resend invoice & 2017-11-05 19:12 & Mary Hudson & 50 & \ldots \\
2 & 60147 & Receive payment & 2017-11-08 12:03 & Anthony Monroe & 20 & \ldots \\
2 & 60148 & Archive order & 2017-11-08 17:16 & Annie Leslie & 5 & \ldots \\ \\
3 & 61209 & Register order & 2017-12-10 15:36 & Karen Johnson & 15 & \ldots \\
3 & 61210 & Send invoice & 2017-12-13 09:07 & Anthony Monroe & 50 & \ldots \\
3 & 61211 & Prepare shipment & 2017-12-13 18:56 & Cory Dixon & 200 & \ldots \\
3 & 61212 & Ship goods & 2017-12-14 11:50 & Bryan Benson & 100 & \ldots  \\
3 & 61213 & Receive payment & 2017-12-17 02:03 & Freddie Jean & 20 & \ldots \\
3 & 61214 & Archive order & 2017-12-18 17:52 & Annie Leslie & 5 & \ldots  \\ \\
\ldots & \ldots & \ldots & \ldots & \ldots & \ldots & \ldots \\
\bottomrule
\end{tabular}%
}
\end{table}

Let $\mathcal{E}$ be the event universe, \ie the set of all possible event identifiers.
Events may have many \textit{properties} (also referred to as \textit{attributes}), \eg an event has a timestamp, corresponds to an activity, is executed by a particular resource, has associated costs, etc.
Being $\mathcal{N}$ the attribute names universe, for any event $e \in \mathcal{E}$ and attribute name $n \in \mathcal{N}$: $\#_n(e)$ is the value of attribute $n$ for event $e$.
Thus, $\#_{activity}(e)$ is the activity associated to event $e$,  $\#_{time}(e)$ is the timestamp of event $e$, $\#_{resource}(e)$ is the resource associated to event $e$, and $\#_{cost}(e)$ is the cost to perform event $e$.
For convention, timestamps should be non-descending in the event log.
For instance, the event in the first row in Table \ref{tbl:background:ex_event_log}, with identifier 54245, corresponds to the ``Register order'' activity, has the timestamp 15-03-2017 08:03, and the resource executing the activity was Lewis Khan with a cost of 15 units.

Each event is associated to a \textit{case}, \ie a process execution or process instance.
The events referring to the same case are represented in the form of a \textit{trace}.
Formally, a trace over some event universe $\mathcal{E}$ is a finite sequence of events $\sigma \in \mathcal{E}'$, such that each event appears only once, \ie for $1 \leq i < j \leq \vert \sigma \vert: \sigma(i) \neq \sigma(j)$.
Let $\mathcal{C}$ be the case universe, \ie the set of all possible case identifiers.
For any case $c \in \mathcal{C}$, $\hat{c} \in \mathcal{E}'$ is a shorthand for referring to the trace of $c$, such that (i) all events of the same case $c$ are in sequence $\hat{c}$, and (ii) all events are ordered chronologically based on their timestamps.
It is worth emphasising that cases, although belonging to the same business process, are not intertwined, \eg the way a patient is treated in a hospital does not influence the way that other patients are treated.
Finally, an event log $\mathcal{L}$ is represented as a set of cases $\mathcal{L} \subseteq \mathcal{C}$, such that each event appears at most once in the entire event log.
For example, the event log in Table \ref{fig:background:ex_process} is formalised as $\mathcal{L} = \{1,2,3,\ldots\}$.
For case $c = 3$, $\hat{c} = \langle 61209, 61210,\ldots,61214\rangle$.
Then, $\#_{activity}(61212) = $ ``Ship goods'' is the activity associated with event 61212, and so on.

To store or exchange event logs, the current standard is the XES format, an XML-based format.
XES \cite{verbeek2010xes,xeswww}, which stands for eXtensible Event Stream, was adopted in 2010 by the IEEE Task Force on Process Mining as the standard format for logging events.
On November 2016, XES became the 1849-2016 IEEE Standard \cite{ieee1849}.
A XES document contains an event log consisting of a number of traces.
Each trace is described as a sequential list of events corresponding to a particular case.
The event log, its traces and its events may have a number of attributes and extensions to give semantics to such attributes if needed.
Other formats that may be used for event logs are MXML \cite{dongen2005mxml}, the predecessor of XES, and CSV, a general-purpose file format.

Drawing convincing conclusions from artificial data is risky, hence, real-life event logs are more desirable for process mining related research.
For example, the 4TU.ResearchData \cite{logrepository} is a prominent data repository for science and engineering containing many open-source event logs widely used in process mining.
With the aim to assess the quality of the process mining methods proposed in this dissertation, several real-life event logs with different characteristics and domains have been selected.
Table \ref{tbl:background:event_logs} describes the characteristics of these event logs, in terms of the number of events, cases, activities and resources in them.
Besides, all event logs except one are publicly available, and the only proprietary event log was collected by the author in a hospital institution in the area of Tarragona for the purposes of this research.
This event log describes the activities conducted by healthcare practitioners during the ongoing of patients medical treatments.

\begin{table}[t!]
\centering
\caption{Characteristics of the event logs used in this dissertation.}
\label{tbl:background:event_logs}
\resizebox{0.99\textwidth}{!}{%
\renewcommand{\arraystretch}{0.85}
\begin{tabular}{cccccc} \toprule
\multirow{2}{*}{\bf Name} & \multirow{2}{*}{{\bf Availability}} & \multicolumn{4}{c}{{\bf Characteristics}} \\ \cmidrule{3-6}
& & {\bf N. events} & {\bf N. cases} & {\bf N. activities} & {\bf N. resources} \\ \midrule
BPI12 & Public \cite{bpi12} & 262.200 & 13.087 & 23 & 68 \\
BPI13 & Public \cite{bpi13} & 6.660 & 1.487 & 7 & 584 \\
BPI14 & Public \cite{bpi14} & 466.155 & 46.507 & 39 & 242 \\
BPI15 & Public \cite{bpi15} & 262.628 & 5.649 & 356 & 72 \\
CoSeLoG & Public \cite{coselog} & 8.577 & 1.434 & 27 & 48 \\
TGN-Hospital & Propietary & 122.179 & 58.836 & 36 & 280 \\
\bottomrule
\end{tabular}%
}
\end{table}

\subsection{Foundations of Process Mining} \label{subsec:background:pm}

Lots of events are recorded by today's information systems due to the concurrent execution of numerous business processes.
However, extracting value from vast amounts of data is not a trivial task.
Consequently, managers have difficulty getting a bird's-eye view of the actual execution of processes within the organisation.
Also, this hinders the ability of an organisation to improve their business processes and become more efficient, effective and sustainable.
Introduced at the beginning of the 2000s, \textit{process mining} \cite{aalst2011book,maita2018systematic} is a research discipline focused on the interplay between people, data, processes and technology.
It sits between the data-oriented nature of machine learning and data mining on the one hand, and the process-oriented nature of process modelling and analysis on the other hand.
Generally accepted, van der Aalst defined the concept of process mining as follows:

\begin{quotation}
  \noindent \textit{``discover, monitor and improve real processes (\ie not assumed processes) by extracting knowledge from event logs readily available in today's systems.''}
  \begin{flushright}W. M. P. van der Aalst (2011) \cite{aalst2011book}\end{flushright}
\end{quotation}

The ultimate goal of process mining is to exploit event data meaningfully to extract process-related knowledge.
As process mining is based on facts (\ie event logs), its analysis can provide valuable insights into process executions, recommend measures to improve business processes in terms of compliance and performance, identify bottlenecks, record deviations and violations, measure service levels, monitor resources utilisation, and predict times and costs, among others.
Besides, instead of creating a single process models, process mining provides various views on the same reality from different perspectives and at different abstraction levels.

\begin{figure}[t!]
\centering
\includegraphics[width=0.84\columnwidth]{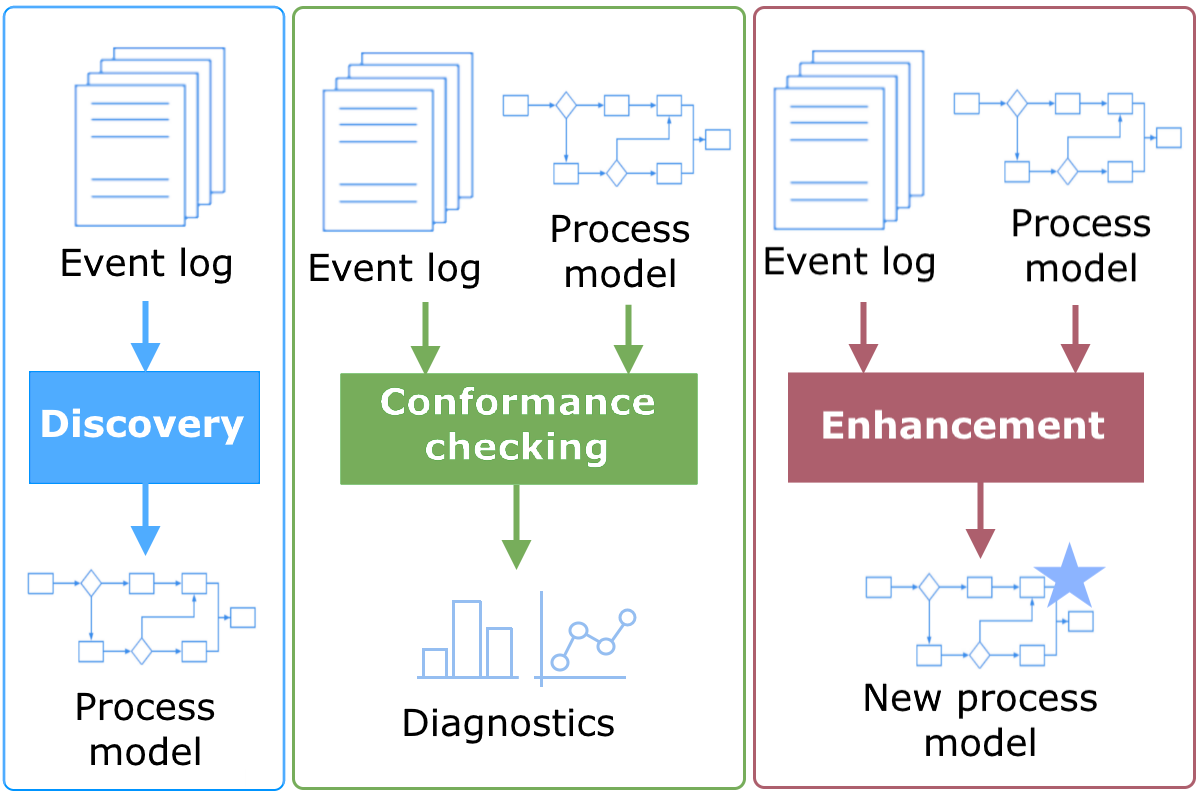}
\caption{Three types of process mining techniques: discovery, conformance and enhancement.}
\label{fig:background:pm_techniques}
\end{figure}

Three main types of process mining techniques are identified, illustrated in Figure \ref{fig:background:pm_techniques}.
The first and most prominent type of process mining is \textit{discovery}.
A process discovery technique takes an event log as an input and produces a model that best describes the behaviour observed in the log without using any a-priori information.
The purpose of discovery techniques is to provide insights into what happens in reality.
Besides, these techniques are not limited to producing models representing the control-flow of the activities, but also other perspectives (explained later) such as a social network between the resources that perform activities.
Some of the most popular process discovery algorithms are the heuristics miner \cite{weijters2003rediscovering,weijters2006heuristic}, $\alpha$-algorithm \cite{aalst2004alpha} (with subsequent versions overcoming flaws \cite{alves2004alpha1,li2007process}), fuzzy miner \cite{gunther2007fuzzy}, genetic process miner \cite{aalst2005genetic} and inductive miner \cite{leemans2013inductive}.
The second type of process mining is \textit{conformance} (also known as \textit{conformance checking}).
These techniques compare an existing process model (representing the ideal process behaviour) with an event log of the same process (representing the actual observed process behaviour).
In short, conformance techniques are used to check if reality, as recorded in the event log, conforms to the model and vice versa.
Hence, they are ideal for identifying and explaining process deviations, and measuring the severity of such deviations.
Fraud detection is a classical use case of conformance techniques.
Last, the third type of process mining is \textit{enhancement}.
In this case, the idea is to take both an event log and a process model as inputs, and extend or improve the existing process model using information extracted from the event log.
There exist various types of enhancement techniques, such as \textit{repair}, consisting in modifying the process model to better reflect reality, and \textit{extension}, consisting in adding new information or perspectives to the process model (\eg resources, durations, decision rules, quality metrics\ldots) by cross-correlating it with the event log.

The representation of process models goes beyond the control-flow of process activities.
Hence, orthogonal to the three types of process mining, different perspectives can be identified: control-flow, organisational, case and time.
The \textit{control-flow} perspective of a process focuses on the activities and their ordering.
Essentially, the goal is to find the best characterisation of all possible paths expressed in terms of some modelling notation.
The \textit{organisational} perspective concentrates on information about resources in the log, \ie which actors (\eg people, roles, departments\ldots) are involved and how are they related.
This perspective serves to classify people in terms of roles and organisational units, or to show a social network.
Next, the \textit{case} perspective is concerned with the properties or characteristics of cases \ie the value of the corresponding data elements.
For example, in the aforementioned example in Table \ref{tbl:background:ex_event_log}, it may be interesting to know the buyer or the number of goods ordered.
Finally, the \textit{time} perspective focuses on the timing and frequency of events, so as to discover bottlenecks, measure service levels, monitor the utilisation of resources, and predict the remaining processing times of running cases.
These perspectives are partially overlapping and non-exhaustive.

\subsection{Process Mining in Healthcare} \label{subsec:background:pm_health}

In the light of the cross-cutting nature of process mining, many sectors saw opportunities to improve the management of their business processes and started adopting process mining into practice \cite{maita2018systematic,dos2019process}.
%
%
Healthcare is one of the most popular sectors where process mining analyses are conducted \cite{mans2015process,mans2012process,gatta2019opportunities}.
The healthcare sector comprises large and complex business processes that usually take place in very dynamic contexts \cite{helfert2009challenges}.
Healthcare business processes comprise a set of medical and non-medical activities supported by medical professionals aiming to diagnose, treat and prevent diseases in order to improve patients health and quality of life \cite{mans2015process}.
In general, these processes can be classified as medical treatment processes or organisational processes \cite{lenz2007support}.
On the one hand, \textit{medical treatment processes} (or \textit{clinical processes}) refer to diagnostic and therapeutic procedures carried out for particular patients through the observation, the collection of patients data and the interpretation of patients-specific information according to medical knowledge and experience.
On the other hand, \textit{organisational processes} are generic process procedures supporting medical treatment processes, such as patient scheduling, test requests and result reporting.
Whereas medical treatment processes are linked to patients, organisational processes help coordinate interoperating healthcare professionals and organisational units.
Unlike other environments, healthcare processes entail a number of challenging characteristics \cite{rebuge2012business}.
It is well-known that healthcare processes are 
(i) highly dynamic, \ie processes are subject to variation over time due to the different conditions of patients and the appearance of new treatment procedures and technological developments,
(ii) highly complex, \ie the medical decision process requires interpreting large amounts of patient-specific information and treatment outcomes may be unpredictable due to patient-specific reactions, 
(iii) increasingly multi-disciplinary, \ie processes are executed according to a wide range of distributed activities performed by professionals from different departments and disciplines, 
and (iv) ad-hoc, \ie specific situations could require from particular procedures resulting from the knowledge and expertise of healthcare professionals.

Beyond process mining, for traceability purposes, healthcare information systems already record all actions performed to patients as a way to have a global vision on the procedures realised.
These records, in the form of event logs, are suitable for process mining too.
Due to the aforementioned challenges, it is likely that the discovery of the actual execution of business processes does not conform with the theoretical process model.
By exploiting process mining techniques, one could measure and understand these misalignments, analyse organisational performance, identify bottlenecks and reduce waiting times and costs, with the aim to enhance such business processes and achieve a more effective, cost-efficient and sustainable healthcare model.
All in all, the ultimate objective of enhancing healthcare processes is to provide better procedures for a better quality of life of patients.

The interest of process mining for healthcare is not new, whose research has multiplied over the last years \cite{batista2018process,rojas2016review,erdogan2018systematic,ghasemi2016review}.
Many studies are devoted to validate the applicability of process mining in different medical specialities and facilities, through the discovery of healthcare processes from different perspectives.
For example, this has been applied for oncological processes \cite{mans2008application,baker2017process,kurniati2018process,caron2013healthcare,caron2014monitoring}, emergency care processes \cite{alvarez2018discovering,duma2020ad,delias2014applying}, sepsis care \cite{vries2017towards,mannhardt2017analyzing}, stroke processes \cite{mans2008stroke}, surgery interventions \cite{blum2008workflow,fernandez2015process}, nursing care \cite{fernandez2013process}, and general hospitalary or ambulatory care processes \cite{kim2013discovery,zhou2014process}, among others.
As discovery techniques by themselves might not be enough to gather knowledge, some investigations concentrated on the comparison of processes to either (i) detect and understand the causes of processes deviations, or (ii) assess the compliance with guidelines and standards \cite{partington2015process,poelmans2010combining,basole2015visual,suriadi2014measuring,dunkl2011assessing,yoo2016assessment}.
In addition, several studies exploited process mining to obtain medical key performance indicators, such as hospitalisation length of stay or resources utilisation rates, and validate with experts the suitability of the actual processes executions \cite{perimal2014health,perimal2012gaining,poelmans2010combining,partington2015process,zhou2014process}.
More theoretically, several articles highlighted the importance of introducing semantics and ontologies throughout process mining to achieve consistent analyses and prevent misinterpretations or ambiguous results \cite{helm2019adopting,helm2020towards,grando2011semantic,detro2016enhancing}.
From a broader perspective, some studies have also proposed a number of methodologies for applying process mining in the particular case of healthcare: from the extraction of data from healthcare information systems to the interpretation of results and the improvement of processes if needed \cite{rebuge2012business,rojas2017question,kurniati2019assessment,caron2014process}.

\subsection{Privacy-Preserving Process Mining} \label{subsec:background:pppm}

Given the need for high confidentiality in all sorts of data managed in the healthcare domain, privacy aspects are of major importance, including the event logs and the business processes.
Whereas privacy aspects have been extensively studied in many computer fields, such as data mining, collaborative filtering or big data, they have barely been addressed in process mining.
However, the consideration of privacy issues in process mining has recently attracted the attention of researchers in the last few years, and it has opened the door to \textit{privacy-preserving process mining}, a young and promising direction within process mining research \cite{mannhardt2019pppm}.
More specifically, this research falls within the scope of the challenging initiative of ``Responsible Process Mining'', being fairness, accuracy, confidentiality and transparency its core principles \cite{aalst2016responsible}.
Privacy-preserving process mining aims to apply the classical privacy protection principles into process mining in order to develop privacy-enhancing techniques for preventing both identity and attribute disclosure to unauthorised parties during process mining analyses.
These techniques normally imply the transformation (\ie distortion) of event data, which directly affect the utility of the event data and, therefore, the process mining results, mainly process models.
Hence, the main challenge of privacy-preserving process mining is to determine the best approach to distort event data to counteract specific attacks, so that the utility of the process mining results maximises, while people's privacy is protected.

There are many use cases wherein the adoption of privacy within process mining is essential \cite{mannhardt2018privacy}.
The first and most prominent case is the need for introducing privacy-preserving process mining analyses in sensitive domains, such as healthcare and banking, which manage lots of confidential data about personally identifiable information.
The unauthorised disclosure of medical or financial knowledge inferred from process mining could clearly jeopardise people's privacy: for instance, insurance companies could use this knowledge in their daily operations to accept or deny policies.
Also, as process mining is less popular than other analytical techniques, such as data mining, some organisations might not have enough resources to apply these techniques and understand the results accordingly.
In such cases, organisations could opt to externalise process mining services to third parties.
However, instead of analysing the original event logs, it might be interesting to analyse a privacy-preserved version of these event logs so as to limit the third-parties' ability to retrieve privacy information beyond the very process mining results.
Privacy issues can also emerge during cross-organisational process mining analyses, in which the execution of process instances is not conducted by a single organisation, but by a group of independent (and possible competitor) organisations, thus requiring the sharing of their own event logs.
Although all organisations could benefit and gain knowledge from the aggregated process mining results, some of them may be reluctant to sharing their event logs with potential competitors unless they are properly preserved from a confidentiality and privacy perspective.
Globally, the worldwide trend towards open data models for transparency purposes also prompts the need for privacy-preserving techniques on event data to be released.
In this case, organisations must consider privacy techniques paramount due to the appearance of possible re-identification risks due to the observed individual uniqueness in event logs \cite{von2020quantifying}.
Last but not least, privacy-preserving techniques could be seen as safeguards or preventive countermeasures against data breaches too.

Initial studies proposed the use of pseudonymisation and encryption techniques to achieve confidentiality within event logs.
Considering a cross-organisational context, Liu \textit{et al.} \cite{liu2016towards} presented a cooperative trusted-third-party scheme dealing with public and private process models.
To mitigate the risks associated with the externalisation of process mining services, Burattin \textit{et al.} \cite{burattin2015toward} took advantage of symmetric and homomorphic cryptosystems to obfuscate confidential data in events logs.
Also, Tillem \textit{et al.} \cite{tillem2016privalpha} proposed a simple protocol to generate process models in a privacy-preserved fashion from encrypted event logs using a modified version of the $\alpha$-algorithm.
However, more recently and after analysing the weaknesses and open challenges of event data encryption in \cite{rafiei2018ensuring}, Rafiei \textit{et al.} \cite{rafiei2018supporting} stated that confidentiality could not be achieved by merely encrypting all data.
To this end, a confidentiality framework based on the encryption and abstraction of event logs to discover process models was presented.

A GDPR-friendly ecosystem for conducting process mining in a privacy by design manner was developed in Michael \textit{et al.} \cite{michael2019user}.
Within an IoT context, authors designed an user-centred privacy-driven system supporting different privacy policies on event logs (\eg data capture, data storage, data use\ldots) using an ABAC-based authorisation model.
Significantly enough, the system allowed tracking who does what, when, why, where and how, with personal data during process mining analyses.

Recently, Pika \textit{et al.} \cite{pika2020privacy} analysed the data privacy and utility requirements for process models in the healthcare domain.
To do this, a variety of existing data transformation techniques, such as noise addition, suppression, microaggregation, data swapping or encryption, were assessed with regards to their suitability on the anonymisation of event log attributes.
The impact of the anonymisation was observed to be dependent on the very characteristics of the event logs and that it varies among process mining analyses.
Despite of that, the proper selection of anonymisation techniques could help interpret and improve the accuracy of the process mining results.
To address the above challenges, a theoretical privacy-preserving process mining framework for supporting healthcare process mining analyses was proposed, along with a privacy metadata model to record all the privacy-preserving transformations applied on the event logs.
Similarly, Rafiei and van der Aalst \cite{rafiei2020privacy} also explored common data transformation techniques to anonymise event log attributes, such as the suppression of events according to the activity attribute or the generalisation of temporal attributes, and presented a XES-based extension for defining the privacy metadata model used in privacy-preserved event logs.

More generally, Fahrenkrog-Petersen \cite{fahrenkrog2019providing} highlighted the main privacy challenges in process mining and outlined the privacy guarantees to be achieved.
This preliminary work proposed two different approaches to achieve privacy, namely event log sanitisation, \ie preprocessing event logs until they guarantee a certain level of privacy, and privatised process mining, \ie developing novel process mining algorithms that generate process mining artifacts (\eg process models) guaranteeing a certain level of privacy.
Both approaches were later extended in further works.
With regards to event log sanitisation, Fahrenkrog-Petersen \textit{et al.} \cite{fahrenkrog2019pretsa} proposed PRETSA, an algorithm providing privacy guarantees in event logs in terms of \textit{k}-anonymity and \textit{t}-closeness.
This algorithm aims to maximise the utility of the discovered process models through step-wise transformations of a prefix-tree representation of event logs.
Besides, to break the link between personally identifiable information and confidential data, personal data (\ie resource information) is removed from event logs, and infrequent (but potentially identifiable) process behaviour is discarded too.
With regards to privatised process mining, Mannhardt \textit{et al.} \cite{mannhardt2019pppm} presented a holistic privacy model for process discovery based on differential privacy.
The resulting process models safeguard the privacy of individuals by introducing a level of noise to each process mining analysis.
A correlation between the utility of the process models and the structure of event logs (\eg in terms of trace variants) was observed.
Both solutions have been integrated in ELPaaS (Event Log Privacy as a Service), a publicly available web application \cite{bauer2019elpaas}.

Although most studies concentrate on the control-flow perspective of processes, some of them also consider the privacy issues related to other perspectives.
For example, Rafiei and van der Aalst \cite{rafiei2019mining} highlighted the privacy issues associated to the organisational perspective of process models, useful for resource behaviour, performance and bottleneck analyses.
Authors presented a decomposition algorithm for discovering people's roles in a privatised fashion against frequency-based attacks, in which attackers could infer the most/least frequent activities or the first/last activities.
Concerning the case perspective, Rafiei \textit{et al.} \cite{rafiei2020tlkc} presented the \textit{TLKC}-privacy model using group-based anonymisation.
More comprehensively, an in-depth analysis of group-based anonymisation techniques considering different attacker models is reported in \cite{rafiei2021group}.
To quantify the effectiveness of the presented models, Rafiei and van der Aalst \cite{rafiei2020towards} also proposed some measures for quantifying the disclosure risks and the data utility preservation.
Similar to ELPaaS, another open-source web-based applications, called PC4PM and PPDP-PM, have integrated the aforementioned privacy-preserving techniques \cite{rafiei2021pc4pm,rafiei2020practical}.

To incorporate the latest privacy contributions within the cross-organisational process mining context, a secure multi-party computation tool, called Shareprom \cite{elkoumy2020shareprom}, enabled multiple parties conducting basic process mining analyses over event logs without disclosing sensitive information from/to the other parties.
However, to overcome potential scalability issues, Elkoumy \textit{et al.} \cite{elkoumy2020secure} proposed a distributed divide-and-conquer scheme enabling parallel processing of event logs.
Moreover, differential privacy is another fundamental aspect in the tool to ensure that the results are protected against possible privacy leakages by, among others, oversampling cases or adding noise to temporal attributes \cite{elkoumy2021mine,elkoumy2020privacy}.

Last but not least, Elkoumy \textit{et al.} \cite{elkoumy2021privacy} presented a conceptual model for threats and requirements that must be fulfilled by all privacy-preserving methods for process mining.
After synthesising existing literature, it was observed that none of these techniques address all those threats, including re-identification, reconstruction or cryptanalysis, and requirements, including anonymity, unlinkability or accountability.
Therefore, this investigation highlighted the importance for future techniques to comprehensively address these issues, and set the ground for future challenges to be considered, such as the quantification of the privacy disclosure, balance between risk and utility, and traceability, among others.


\part{Context-Aware Environments} \label{part:cae}
\chapter{Security Analysis of Sensors Enabling Smart Health}
\chaptermark{Security Analysis of Sensors for Smart Health}
\label{chap:sensors}

\emph{The provision of smart health services require a large number of heterogeneous and complex sensing devices, communications networks and systems, which need to be operating uninterruptedly. The high sensitivity of the information handled, such as medical records, arises numerous challenges from a security and privacy perspective. Hence, attackers can attempt to exploit vulnerabilities in these systems to obtain valuable information and pursuit economic benefits. This chapter deepens on these challenges from a technical side, and sets the ground to deploy more secure and private context-aware environments enabling smart health. First, Section \ref{sec:sensors:intro} overviews the rationale of the proposed study. Then, Section \ref{sec:sensors:sensors} describes the most widely used sensors to gather user attributes and contextual attributes in s-health scenarios. Next, Section \ref{sec:sensors:communication} details the wireless communication technologies available for deploying complex s-health services. After that, Section \ref{sec:sensors:security} discusses the main security requirements to fulfil in smart health systems, by analysing the main threats and vulnerabilities to be considered and proposing several countermeasures. Considering the previous, Section \ref{sec:sensors:future} provides an extensive discussion on open issues and research opportunities to be faced in the future. Finally, the chapter concludes in Section \ref{sec:sensors:concl}.}

\minitoc

\section{Introduction} \label{sec:sensors:intro}

Nowadays, there exists an overwhelming number and variety of devices with sensing capabilities, with different features, technologies, dimensions and costs.
These sensing devices, required to enable context-aware environments and provide smart health applications, are able to collect and transmit data from multiple physical locations.
More specifically, they are organised topologically as networks, thus arising wireless sensor networks (WSNs) and the wireless body area networks (WBANs).
Throughput, latency, capacity, reliability and security are some of the challenges involved once creating these networks.

Interestingly enough, data security stands as one of the most important features in the healthcare domain.
Medical data, such as electronic health records, biomedical signals and physiological parameters, are highly sensitive and must be handled with the highest security and privacy requirements.
This is, sensors data must be thoroughly managed during all their life cycle: from their gathering, through their transmission and analysis, to their final presentation.
Unfortunately, despite the many security safeguards, the history of communication networks as well as the Internet's encompasses countless security flaws, vulnerable cryptographic protocols and threatening data breaches.
Hence, every information system or communication network is virtually prone to be attacked by cybernetic criminals or to suffer from irreparable damages because of human errors.
Smart health systems, involving a large number of complex and diverse devices and parties, may present risks from a security and privacy perspective that deserve attention.

All in all, it is essential to provide a comprehensive approach to a realistic deployment of smart health services.
To that end, this chapter provides a down-to-earth landscape of sensors that could be used to deploy these scenarios.
More specifically, first we provide a thorough review and characterisation of user-centric and contextual sensors that enable these services.
Then, we discuss the most common wireless communication technologies that allow those sensors to interoperate and transfer the sensed information in a secure manner.
Finally, we also elaborate on the most serious vulnerabilities and threats in such settings, and suggest the corresponding countermeasures.

\section{Sensors: Definition and Taxonomy} \label{sec:sensors:sensors}

Advancements in microelectronics and manufacturing technologies have enabled the development of a large variety of sensors, embedded in electronic small-scale devices, with high sensitivity, low energy consumption and contained costs.
This section elaborates on the different sensors to be considered within the smart health paradigm.
Sensors are categorised into two groups according to the nature of the sensor data: (i) user-centric data, \ie referring to personally identifiable individuals, and (ii) contextual data, \ie referring to the context or the immediate environment. 


\subsection{User-centric Sensors}

User-centric sensors aim to acquire specific data related to individuals.
Within this context, most of it relates to medical data, such as physiological parameters, biosignals and health status.
Complementing these data, individuals' location or body movements could also be of great interest to contextualise individuals.
All these sensors have already been seamlessly integrated within wearable devices, whose popularity has grown during the last decade, as part of the IoMT technology \cite{yetisen2018wearables}.
In particular, wearable technology has revolutionised ubiquitous computing with low-cost yet powerful devices, including body-worn accessories, smart textiles, garments, on-skin tattoos, ingestible sensors and implantable appliances, among others.
This technology, capable of monitoring, analysing and transmitting individuals' data, opens the door to numerous healthcare opportunities, ranging from the remote or self-monitoring of patients' health to the early detection of medical complications.
The generalised use of wearables will contribute to reducing health expenditure and develop more sustainable healthcare models.
This section reviews some of the most common methods for the sensing of user-centric attributes (see Figure \ref{fig:sensors:user}), which are summarised in Table \ref{tbl:sensors:cardiovascular} to Table \ref{tbl:sensors:location_motion}.

\begin{figure}[b!]
\centering
\includegraphics[width=0.93\linewidth]{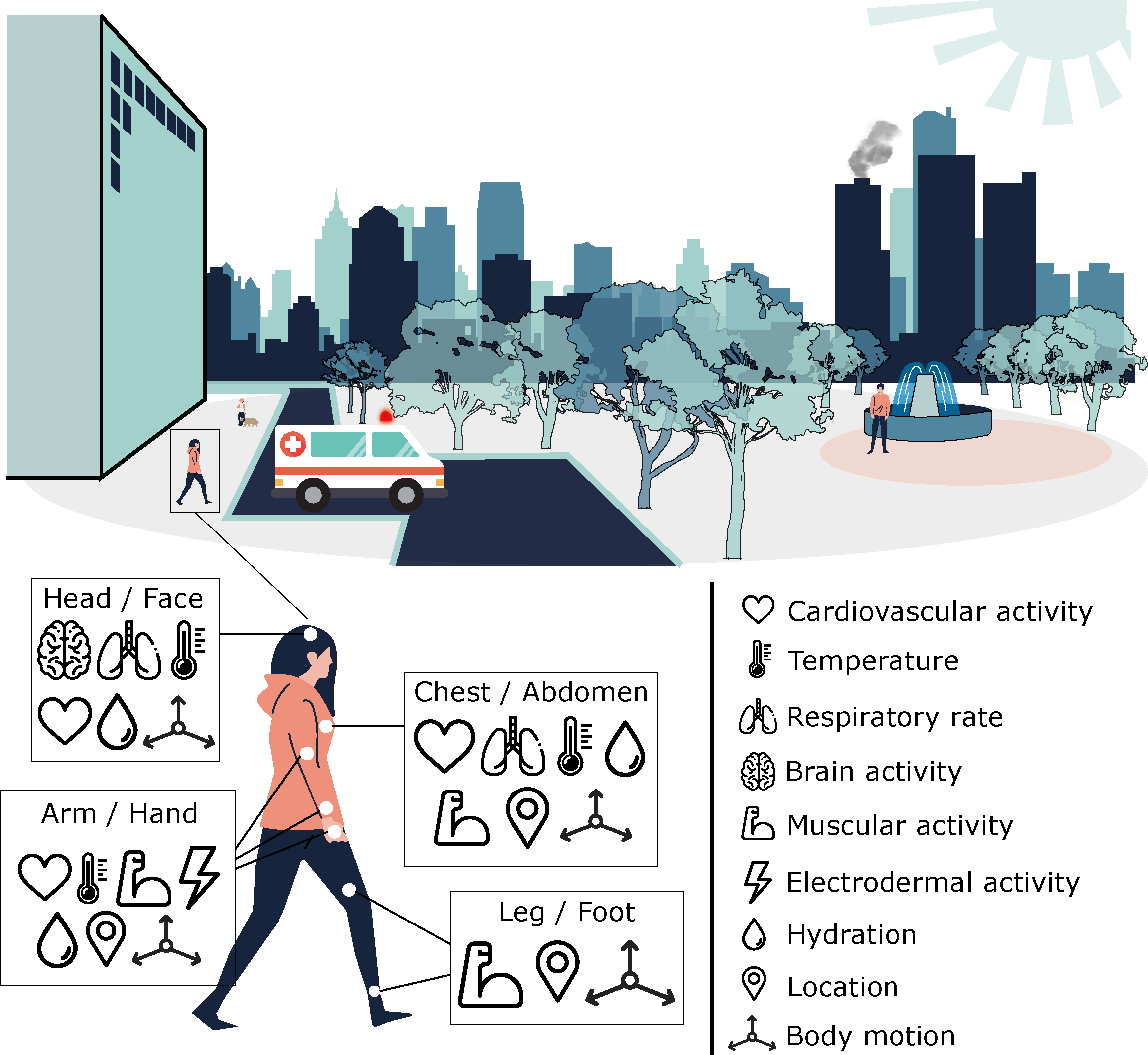}
\caption{User-centric attributes for smart health: each icon, representing the sensors of an attribute, is assigned to the parts of the body where that attribute can be collected (adapted from \cite{batista2021sensors}).}
\label{fig:sensors:user}
\end{figure}

\paragraph{Cardiovascular Activity}~

Many wearable and IoMT devices monitor cardiovascular parameters, such as heart rate, blood pressure, oxygen saturation and blood glucose concentration, to early detect abnormal conditions (\eg tachycardia, hypertension, stroke\ldots) or help cardiac rehabilitation recovery.

The \textit{heart rate} (or pulse) is one of the most commonly measured vital signs.
There are mainly four methods to sense this attribute: electrocardiography (ECG), photoplethysmography (PPG), ballistocardiography (BCG) and phonocardiography (PCG).
On the one hand, ECG sensors record the electrical activity and rhythm of the heart, in the form of electrocardiograms, by attaching series of electrodes to the skin.
Traditional ECG monitoring uses Holter monitors, well-known medical portable units, able to capture long recording periods in both home- and hospital-environments.
Despite the high reliability, they are quite obstructive, invasive and uncomfortable due to their wired architecture.
To overcome this, wireless ECG monitoring solutions embedded in wearable devices are emerging.
On the other hand, heart rate data can also be acquired using optical PPG methods by means of pulse oximeter sensors, which measure the intensity of a LED light reflected or transmitted through the skin.
Thanks to the low-cost and non-invasive nature of this method, most wearables and fitness trackers opt for PPG methods.
However, the accuracy of PPG measurements is not excellent, since they are strongly affected by a number of factors, including the measurement procedure, environmental factors and the skin pigmentation, and post-processing techniques are common to enhance quality.
Finally, body motion-based BCG methods and sound-based PCG methods are less prominent in wearable devices.

Another popular measurement is the \textit{blood oxygen level}, whose monitoring is fundamental for people suffering from blood disorders, circulatory problems or respiratory diseases.
The most accurate method to measure this attribute is the arterial blood gas (ABG) test, although it is invasive, painful and typically conducted in hospital settings only.
More suitable for smart health purposes is the use of PPG-based oximeter sensors, commonly integrated in many wearables, to measure the blood oxygen saturation.
Compared to ABG tests, this method is simpler, cheaper, non-invasive and non-painful, but slightly less accurate.

Cardiovascular activity monitoring also considers \textit{blood pressure} to early detect cardiovascular-related risks, such as hypertension.
Cuff-based sphygmomanometers are the traditional in-hospital devices to measure this attribute.
Although they are cheap, accurate and even available in portable electronic devices for home monitoring, they are invasive and unsuitable for outdoor usage or long-term monitoring.
To overcome this, accurate and non-invasive blood pressure values can be estimated from the pulse transit time (PTT) measure obtained from the combination of PPG and ECG signals.
Many smartwatches, wristbands, armbands and chest bands using similar approaches are emerging.

\begin{table}[t!]
\centering
\scriptsize
\caption{Characteristics of the methods for cardiovascular activity sensing.}
\label{tbl:sensors:cardiovascular}
\resizebox{\textwidth}{!}{
\renewcommand{\arraystretch}{0.7}
\begin{tabular}{ccccccc}
\toprule
\multirow{2}{*}{\textbf{Attribute}} & \multirow{2}{*}{\textbf{Method}} & \multirow{2}{*}{\textbf{Sensor}} & \textbf{Device /} & \multirow{2}{*}{\textbf{Location}} & \multirow{2}{*}{\textbf{Properties}} & \textbf{Smart} \\
& & & \textbf{Wearable} & & & \textbf{health} \\ \midrule
\multirow{4}{*}{Heart rate} & & & & \multirow{4}{*}{Chest} & $\checkmark$~Accuracy & \multirow{4}{*}{$\times$} \\
& Traditional & Skin & Holter & & $\sim$~Cost & \\
& ECG & electrodes & monitor & & $\sim$~Cont.~monit. & \\
& & & & & $\times$~Non-invasive & \\ \midrule
\multirow{4}{*}{Heart rate} & & & \multirow{2}{*}{Patch} & & $\checkmark$~Accuracy & \multirow{4}{*}{$\checkmark$} \\
& Wireless & Skin & \multirow{2}{*}{Band} & Chest & $\sim$~Cost & \\
& ECG & electrodes & \multirow{2}{*}{Textile} & Arm & $\checkmark$~Cont.~monit. & \\
& & & & & $\sim$~Non-invasive & \\ \midrule
\multirow{4}{*}{Heart rate} & \multirow{4}{*}{PPG} & & \multirow{2}{*}{Smartwatch} & & $\sim$~Accuracy & \multirow{4}{*}{$\checkmark$} \\
& & Pulse & \multirow{2}{*}{Wristband} & Wrist & $\checkmark$~Cost & \\
& & oximeter & \multirow{2}{*}{Ring} & Finger & $\checkmark$~Cont.~monit. & \\
& & & & & $\checkmark$~Non-invasive & \\ \midrule
\multirow{4}{*}{Heart rate} & \multirow{4}{*}{BCG} & \multirow{2}{*}{Tilt} & \multirow{4}{*}{Patch} & \multirow{4}{*}{Chest} & $\times$~Accuracy & \multirow{4}{*}{$\sim$} \\
& & \multirow{2}{*}{Force} & & & $\sim$~Cost & \\
& & \multirow{2}{*}{Pressure} & & & $\sim$~Cont.~monit. & \\
& & & & & $\checkmark$~Non-invasive & \\ \midrule
\multirow{4}{*}{Heart rate} & \multirow{4}{*}{PCG} & \multirow{4}{*}{Sound} & Microphone & \multirow{4}{*}{Chest} & $\times$~Accuracy & \multirow{4}{*}{$\sim$} \\
& & & Smartphone & & $\checkmark$~Cost & \\
& & & Electronic & & $\sim$~Cont.~monit. & \\
& & & stethoscope & & $\checkmark$~Non-invasive & \\ \midrule
& \multirow{4}{*}{ABG} & \multirow{4}{*}{Chemical} & & & $\checkmark$~Accuracy & \multirow{4}{*}{$\times$} \\
Blood & & & Chemical & Arm & $\times$~Cost & \\
oxygen & & & analyser & Wirst & $\times$~Cont.~monit. & \\
& & & & & $\times$~Non-invasive & \\ \midrule
& \multirow{4}{*}{PPG} & & Smartwatch & \multirow{2}{*}{Wrist} & $\sim$~Accuracy & \multirow{4}{*}{$\checkmark$} \\
Blood & & Pulse & Strap & \multirow{2}{*}{Earlobe} & $\checkmark$~Cost & \\
oxygen & & oximeter & Band & \multirow{2}{*}{Finger} & $\checkmark$~Cont.~monit. & \\
& & & Textile & & $\checkmark$~Non-invasive & \\ \midrule
& \multirow{4}{*}{Traditional} & \multirow{4}{*}{Pressure} & & \multirow{4}{*}{Arm} & $\checkmark$~Accuracy & \multirow{4}{*}{$\times$} \\
Blood & & & Sphygmo- & & $\checkmark$~Cost & \\
pressure & & & manometer & & $\times$~Cont.~monit. & \\
& & & & & $\times$~Non-invasive & \\ \midrule
& \multirow{2}{*}{PTT} & \multirow{2}{*}{Pulse} & \multirow{2}{*}{Smartwatch} & Wrist & $\checkmark$~Accuracy & \multirow{4}{*}{$\checkmark$} \\
Blood & \multirow{2}{*}{(ECG \&} & \multirow{2}{*}{oximeter} & \multirow{2}{*}{Band} & Arm & $\sim$~Cost & \\
pressure & \multirow{2}{*}{PPG)} & \multirow{2}{*}{Electrodes} & \multirow{2}{*}{Patch} & Ear & $\checkmark$~Cont.~monit. & \\
& & & & Chest & $\checkmark$~Non-invasive & \\ \midrule
& & & & \multirow{4}{*}{Finger} & $\checkmark$~Accuracy & \multirow{4}{*}{$\times$} \\
Blood & Traditional & Electro- & Glucose & & $\checkmark$~Cost & \\
glucose & (chemical) & chemical & meter & & $\times$~Cont.~monit. & \\
& & & & & $\times$~Non-invasive & \\ \midrule
& & & \multirow{2}{*}{Wristband} & & $\sim$~Accuracy & \multirow{4}{*}{$\checkmark$} \\
Blood & Epidermal & Electro- & \multirow{2}{*}{Patch} & Wrist & $\sim$~Cost & \\
glucose & chemical & chemical & \multirow{2}{*}{Tattoo} & Arm & $\checkmark$~Cont.~monit. & \\
& & & & & $\checkmark$~Non-invasive & \\ \midrule
& & \multirow{2}{*}{Photo-} & & \multirow{2}{*}{Wrist} & $\sim$~Accuracy & \multirow{4}{*}{$\checkmark$} \\
Blood & Optical & \multirow{2}{*}{sensor} & Wristband & \multirow{2}{*}{Finger} & $\sim$~Cost & \\
glucose & spectroscopy & \multirow{2}{*}{Infrared} & Patch & \multirow{2}{*}{Earlobe} & $\checkmark$~Cont.~monit. & \\
& & & & & $\checkmark$~Non-invasive & \\
\bottomrule
\end{tabular}
}
\end{table}

Last, \textit{blood glucose concentration} is crucial for the management of diabetes, and many m-health solutions emerged with the aim to monitor, suggest and engage diabetic people with their treatment.
Traditionally, blood glucose is measured using glucose meters, \ie portable devices with electrochemical sensors that chemically analyse a blood drop.
These devices are easy to use, cheap and accurate, but they are invasive and do not provide a continuous monitoring.
For a smart health approach, less invasive and obstructive technologies are required.
To this end, electrochemical sensors embedded in patches and tattoos and optical measurements using spectroscopy techniques have already been proposed.

All in all, monitoring cardiovascular activity is particularly important in at-risk patients, such as elderly, who may sometimes need immediate response in a proactive fashion.
Hence, actuators are likely to play a key role in the years to come.
Today, non-implantable cardioverter-defibrillators and in-body insulin pumps are already a reality \cite{cheung2021wearable}.

\paragraph{Temperature}~

The \textit{body temperature} is an indicator of the overall physiological status of an individual and helps determine illnesses, such as hypothermia or fever, among others.
Unfortunately, standard medical measurements using thermometers are invasive, obstructive and not aligned with s-health solutions.
Notwithstanding, these measurements can be estimated in a less-invasive way from the \textit{skin temperature}, usually acquired from the wrist, arm, armpit, chest or forehead.

\begin{table}[t!]
\centering
\scriptsize
\caption{Characteristics of the methods for temperature sensing.}
\label{tbl:sensors:temperature}
\resizebox{\textwidth}{!}{
\renewcommand{\arraystretch}{0.7}
\begin{tabular}{ccccccc}
\toprule
\multirow{2}{*}{\textbf{Attribute}} & \multirow{2}{*}{\textbf{Method}} & \multirow{2}{*}{\textbf{Sensor}} & \textbf{Device /} & \multirow{2}{*}{\textbf{Location}} & \multirow{2}{*}{\textbf{Properties}} & \textbf{Smart} \\
& & & \textbf{Wearable} & & & \textbf{health} \\ \midrule
& & \multirow{4}{*}{Mercury} & \multirow{2}{*}{Mercury-} & & $\checkmark$~Accuracy & \multirow{4}{*}{$\times$} \\
Body & Traditional & & \multirow{2}{*}{in-glass} & Oral & $\checkmark$~Cost & \\
temperature & (chemical) & & \multirow{2}{*}{thermometer} & Rectal & $\times$~Cont.~monit. & \\
& & & & & $\times$~Non-invasive & \\ \midrule
& \multirow{4}{*}{Electrical} & \multirow{4}{*}{Thermistor} & & Arm & $\sim$~Accuracy & \multirow{4}{*}{$\checkmark$} \\
Skin & & & Patch & Chest & $\checkmark$~Cost & \\
temperature & & & Band & Ear & $\checkmark$~Cont.~monit. & \\
& & & & Forehead & $\checkmark$~Non-invasive & \\ \midrule
& \multirow{4}{*}{Electrical} & & & Arm & $\times$~Accuracy & \multirow{4}{*}{$\sim$} \\
Skin & & Thermo- & Patch & Chest & $\checkmark$~Cost & \\
temperature & & couple & Band & Ear & $\checkmark$~Cont.~monit. & \\
& & & & Forehead & $\checkmark$~Non-invasive & \\ \midrule
& \multirow{4}{*}{Optical} & & Smartwatch & Wrist & $\sim$~Accuracy & \multirow{4}{*}{$\checkmark$} \\
Skin & & FBG & Band & Chest & $\checkmark$~Cost & \\
temperature & & Infrared & Patch & Ear & $\checkmark$~Cont.~monit. & \\
& & & Textile & Forehead & $\checkmark$~Non-invasive & \\
\bottomrule
\end{tabular}
}
\end{table}

Skin temperature is usually measured using thermistors, thermocouples or optical methods, among others.
In particular, thermistor sensors are generally popular, cost-efficient, and ideal for many wearables, yet their accuracy is influenced by a number of factors, including the measurement's location, ambient temperature or strain.
Optical methods, based on fiber Bragg grating (FBG) or infrared technology, are quite similar to thermistor sensors in terms of accuracy, comfortability and cost.
However, thermocouples sensors, although being cost-efficient too, have a worse accuracy than the previous methods.

\paragraph{Respiratory Rate}~

The number of breathing cycles per minute determines the \textit{respiratory rate}, one of the main vital signs of the human body and a clear indicator of overall health.
Monitoring this attribute helps identifying or keeping track of disorders, such as asthma attacks, panic attacks, sleep apnea, shortness of breath, COPD and pneumonia.
However, in contrast to other vital signs, the recording of the respiratory rate is less automated.
The gold standard technique for its measurement consists in counting the number of times that the chest or abdomen rises during one minute while the patient is resting.
This manual technique is inefficient for smart health purposes.

More automated, the most popular monitoring methods are contact-based, \ie the sensor directly contacts the skin. 
Breathing can be monitored considering the expansion and contraction movements of the chest walls using strain sensors (\eg resistive, capacitive and inductive sensors), transthoracic impedance sensors or motion sensors (\eg IMU, explained later).
This technique is generally accurate, even though body motion artefacts and environmental factors can diminish the quality, and the small dimensions, low-power consumption and reduced costs of these sensors facilitate their integration into wearable devices and textiles to be minimally intrusive.
There are further contact-based methods for respiratory rate sensing, although they might be less accurate or more obstructive for smart health, such as acoustic-based methods using microphones, air temperature-based methods using thermistors, thermocouples or pyroelectric sensors, air humidity-based methods using capacitive, resistive or nanocrystal sensors, and based on the modulation of cardiac activity recorded from ECG and PPG signals.

The other type of monitoring methods are contactless, \ie the sensor does not directly contact the skin, which are more comfortable solutions and ease long-term monitoring or during sleep. 
The main drawback of these methods is their susceptibility to environmental noise or motion artefacts, so they should only be considered under very constrained environments.
Most contactless methods are based on camera sensing, which estimate the respiration rate by tracking the chest movements.
However, further methods have proposed the use of infrared thermal imaging sensors to detect the temperature fluctuations during the respiration cycle, or ultrasonic proximity sensors.

\begin{table}[t!]
\centering
\scriptsize
\caption{Characteristics of the methods for respiratory rate sensing.}
\label{tbl:sensors:respiratory}
\resizebox{\textwidth}{!}{
\renewcommand{\arraystretch}{0.7}
\begin{tabular}{ccccccc}
\toprule
\multirow{2}{*}{\textbf{Attribute}} & \multirow{2}{*}{\textbf{Method}} & \multirow{2}{*}{\textbf{Sensor}} & \textbf{Device /} & \multirow{2}{*}{\textbf{Location}} & \multirow{2}{*}{\textbf{Properties}} & \textbf{Smart} \\
& & & \textbf{Wearable} & & & \textbf{health} \\ \midrule
& Traditional & \multirow{4}{*}{-} & \multirow{4}{*}{-} & \multirow{4}{*}{-} & $\checkmark$~Accuracy & \multirow{4}{*}{$\times$} \\
Respiratory & (observation & & & & $\checkmark$~Cost & \\
rate & of chest or & & & & $\times$~Cont.~monit. & \\
& abdomen) & & & & $\checkmark$~Non-invasive & \\ \midrule
& & \multirow{2}{*}{Resistive} & \multirow{2}{*}{Patch} & \multirow{4}{*}{Chest} & $\checkmark$~Accuracy & \multirow{4}{*}{$\checkmark$} \\
Respiratory & Chest wall & \multirow{2}{*}{Capacitive} & \multirow{2}{*}{Belt} & & $\checkmark$~Cost & \\
rate & strain & \multirow{2}{*}{Inductive} & \multirow{2}{*}{Textile} & & $\checkmark$~Cont.~monit. & \\
& & & & & $\checkmark$~Non-invasive & \\ \midrule
& & \multirow{4}{*}{Impedance} & \multirow{2}{*}{Patch} & \multirow{4}{*}{Chest} & $\checkmark$~Accuracy & \multirow{4}{*}{$\checkmark$} \\
Respiratory & Electrical & & \multirow{2}{*}{Belt} & & $\checkmark$~Cost & \\
rate & impedance & & \multirow{2}{*}{Textile} & & $\checkmark$~Cont.~monit. & \\
& & & & & $\checkmark$~Non-invasive & \\ \midrule
& & \multirow{4}{*}{IMU} & \multirow{2}{*}{Patch} & & $\checkmark$~Accuracy & \multirow{4}{*}{$\checkmark$} \\
Respiratory & Motion & & \multirow{2}{*}{Belt} & Chest & $\checkmark$~Cost & \\
rate & (contact) & & \multirow{2}{*}{Textile} & Abdomen & $\checkmark$~Cont.~monit. & \\
& & & & & $\checkmark$~Non-invasive & \\ \midrule
& \multirow{4}{*}{Acoustic} & \multirow{4}{*}{Microphone} & & \multirow{2}{*}{Nose} & $\times$~Accuracy & \multirow{4}{*}{$\sim$} \\
Respiratory & & & Microphone & \multirow{2}{*}{Mouth} & $\checkmark$~Cost & \\
rate & & & Headset & \multirow{2}{*}{Chest} & $\checkmark$~Cont.~monit. & \\
& & & & & $\sim$~Non-invasive & \\ \midrule
& & \multirow{2}{*}{Thermistor} & & & $\sim$~Accuracy & \multirow{4}{*}{$\sim$} \\
Respiratory & Air temp. & \multirow{2}{*}{Thermocouple} & Headset & Nose & $\checkmark$~Cost & \\
rate & (electrical) & \multirow{2}{*}{Pyroelectric} & Patch & Mouth & $\checkmark$~Cont.~monit. & \\
& & & & & $\sim$~Non-invasive & \\ \midrule
& & \multirow{2}{*}{Capacitive} & & & $\sim$~Accuracy & \multirow{4}{*}{$\sim$} \\
Respiratory & Air humid. & \multirow{2}{*}{Resistive} & Headset & Nose & $\checkmark$~Cost & \\
rate & (electrical) & \multirow{2}{*}{Nanocrystal} & Patch & Mouth & $\checkmark$~Cont.~monit. & \\
& & & & & $\sim$~Non-invasive & \\ \midrule
& \multirow{2}{*}{Cardiac} & \multirow{2}{*}{Pulse} & \multirow{2}{*}{Smartwatch} & & $\checkmark$~Accuracy & \multirow{4}{*}{$\checkmark$} \\
Respiratory & \multirow{2}{*}{activity} & \multirow{2}{*}{oximeter} & \multirow{2}{*}{Band} & Wrist & $\sim$~Cost & \\
rate & \multirow{2}{*}{modulation} & \multirow{2}{*}{Electrodes} & \multirow{2}{*}{Patch} & Chest & $\checkmark$~Cont.~monit. & \\
& & & & & $\checkmark$~Non-invasive & \\ \midrule
& & \multirow{4}{*}{Camera} & & \multirow{4}{*}{-} & $\sim$~Accuracy & \multirow{4}{*}{$\sim$} \\
Respiratory & Motion & & RGB camera & & $\checkmark$~Cost & \\
rate & (contactless) & & Smartphone & & $\times$~Cont.~monit. & \\
& & & & & $\checkmark$~Non-invasive & \\ \midrule
& & \multirow{4}{*}{Camera} & & \multirow{4}{*}{-} & $\sim$~Accuracy & \multirow{4}{*}{$\sim$} \\
Respiratory & Thermal & & Infrared & & $\times$~Cost & \\
rate & imaging & & camera & & $\times$~Cont.~monit. & \\
& & & & & $\checkmark$~Non-invasive & \\ \midrule
& \multirow{4}{*}{Ultrasonic} & \multirow{2}{*}{Ultrasonic} & & \multirow{4}{*}{-} & $\sim$~Accuracy & \multirow{4}{*}{$\sim$} \\
Respiratory & & \multirow{2}{*}{proximity} & Recording & & $\times$~Cost & \\
rate & & \multirow{2}{*}{Capacitive} & device & & $\times$~Cont.~monit. & \\
& & & & & $\checkmark$~Non-invasive & \\
\bottomrule
\end{tabular}
}
\end{table}

\paragraph{Brain Activity}~

Neurological disorders are one of the most prevalent disorders in our society, namely Alzheimer's disease or other forms of dementia, epilepsy or meningitis, among others.
Unfortunately, classical \textit{brain activity} monitoring methods are highly sophisticated and require big and expensive instrumentation.
Hence, developing novel methods considering the size, cost and power constraints so as to be embedded into portable and wearable devices is truly challenging.

The assessment of the quality of brain activity is popularly conducted through electroencephalography (EEG), a method that measures the electrical activity in the brain by placing small electrodes at multiple locations on the scalp.
Conventional EEG measurements, conducted in medical facilities, require a head cap with electrodes connected to a recording device through long wires.
This method is accurate, but significantly obstructive and invasive.
To address these shortcomings, the adoption of wireless technologies has enabled comfortable EEG monitoring using wearable devices, such as headset-based solutions, ear-based devices or temporary tattoos \cite{casson2019wearable}.
Besides EEG, there are other methods for brain activity monitoring, although their complexity complicates their integration into wearable devices.
These methods include functional near-infrared spectroscopy (fNIRS) for hemodynamic changes, magnetoencephalography (MEG) considering magnetic fields and positron-emission tomography (PET).

\begin{table}[b!]
\centering
\scriptsize
\caption{Characteristics of the methods for brain activity sensing and muscular activity sensing.}
\label{tbl:sensors:brain_muscular}
\resizebox{\textwidth}{!}{
\renewcommand{\arraystretch}{0.7}
\begin{tabular}{ccccccc}
\toprule
\multirow{2}{*}{\textbf{Attribute}} & \multirow{2}{*}{\textbf{Method}} & \multirow{2}{*}{\textbf{Sensor}} & \textbf{Device /} & \multirow{2}{*}{\textbf{Location}} & \multirow{2}{*}{\textbf{Properties}} & \textbf{Smart} \\
& & & \textbf{Wearable} & & & \textbf{health} \\ \midrule
& & & \multirow{4}{*}{Head cap} & \multirow{4}{*}{Scalp} & $\checkmark$~Accuracy & \multirow{4}{*}{$\times$} \\
Brain & Traditional & Skin & & & $\times$~Cost & \\
activity & EEG & electrodes & & & $\times$~Cont.~monit. & \\
& & & & & $\times$~Non-invasive & \\ \midrule
& & & \multirow{2}{*}{Headband} & Scalp & $\sim$~Accuracy & \multirow{4}{*}{$\checkmark$} \\
Brain & Wireless & Skin & \multirow{2}{*}{Headset} & Head & $\sim$~Cost & \\
activity & EEG & electrodes & \multirow{2}{*}{Tattoo} & Forehead & $\checkmark$~Cont.~monit. & \\
& & & & Ear & $\checkmark$~Non-invasive & \\ \midrule
& \multirow{4}{*}{fNIRS} & \multirow{4}{*}{Optodes} & \multirow{4}{*}{Head cap} & & $\checkmark$~Accuracy & \multirow{4}{*}{$\times$} \\
Brain & & & & Scalp & $\sim$~Cost & \\
activity & & & & Head & $\sim$~Cont.~monit. & \\
& & & & & $\sim$~Non-invasive & \\ \midrule
& \multirow{4}{*}{MEG} & Optically & \multirow{4}{*}{Head cap} & & $\checkmark$~Accuracy & \multirow{4}{*}{$\times$} \\
Brain & & pumped & & Scalp & $\times$~Cost & \\
activity & & magneto- & & Head & $\sim$~Cont.~monit. & \\
& & meteres & & & $\sim$~Non-invasive & \\ \midrule
& \multirow{4}{*}{PET} & & & \multirow{4}{*}{Head} & $\checkmark$~Accuracy & \multirow{4}{*}{$\times$} \\
Brain & & Photosensor & Head cap & & $\times$~Cost & \\
activity & & Photodiode & Helmet & & $\times$~Cont.~monit. & \\
& & & & & $\times$~Non-invasive & \\ \midrule
& \multirow{2}{*}{Intra-} & \multirow{2}{*}{Monopolar} & \multirow{2}{*}{Needle \&} & & $\checkmark$~Accuracy & \multirow{4}{*}{$\times$} \\
Muscular & \multirow{2}{*}{muscular} & \multirow{2}{*}{or concentric} & \multirow{2}{*}{recording} & Region of & $\times$~Cost & \\
activity & \multirow{2}{*}{EMG} & \multirow{2}{*}{electrodes} & \multirow{2}{*}{device} & interest & $\times$~Cont.~monit. & \\
& & & & & $\times$~Non-invasive & \\ \midrule
& & & Patch & & $\sim$~Accuracy & \multirow{4}{*}{$\checkmark$} \\
Muscular & Surface & Skin & Band & Region of & $\checkmark$~Cost & \\
activity & EMG & electrodes & Cap & interest & $\checkmark$~Cont.~monit. & \\
& & & Textile & & $\checkmark$~Non-invasive & \\ \midrule
& \multirow{4}{*}{MMG} & Accelerometer & & & $\sim$~Accuracy & \multirow{4}{*}{$\checkmark$} \\
Muscular & & Pressure & Patch & Region of & $\checkmark$~Cost & \\
activity & & Force- & Band & interest & $\checkmark$~Cont.~monit. & \\
& & sensitive & & & $\checkmark$~Non-invasive & \\
\bottomrule
\end{tabular}
}
\end{table}

\paragraph{Muscular Activity}~

Monitoring \textit{muscular activity} helps detecting and evaluating the severity of neurodegenerative disorders, such as Parkinson disease or bradykinesia and dyskinesia symptoms.
Wearable technology can benefit the early detection of these disorders in non-diagnosed patients, as well as to remotely monitor the evolution of these conditions in already-diagnosed patients.

The most popular diagnostic procedure to assess the functioning of the muscles and the nerve cells is the electromyography (EMG), which measures the electrical signals generated by the muscles during their movement.
There are two main methods for EMG recordings.
On the one hand, intramuscular EMG methods are invasive and somehow painful, and are not well aligned with smart health solutions.
On the other hand, surface EMG methods are non-invasive procedures that require placing some patch electrodes on the muscle's skin, so they could be integrated into wearable devices, such as wristbands, armbands, caps or even textiles to ease long-term, real-time monitoring.
Yet, it is noteworthy that the accuracy of these methods is affected by the skin's properties, tissue structure and external electromagnetic interference.
In addition to electrical measurements, muscular activity can also be measured from a mechanical perspective through mechanomyography (MMG), which measures the mechanical vibrations of muscles fibres using accelerometers, pressure sensors or force-sensitive resistors.

\paragraph{Electrodermal Activity}~

\textit{Electrodermal activity}, also known as skin conductance or galvanic skin response, aims to detect changes in the electrical properties of the skin, particularly due to sweating.
This property, highly valuable in behavioural medicine, helps detecting emotional states (\eg stress, anxiety, depression, fatigue\ldots), characterise sleep activity or manage the neurological status.

The instrumentation required to measure this attribute is quite simple, as it only needs a couple of electrodes placed on the skin surface, generally the wrist or fingertip.
In general, this method is little invasive, facilitates long-term monitoring and can be integrated into wearable solutions based on wristbands or finger straps.
Although initial devices were wired, many current solutions are already wireless with the goal of enhancing comfortability.

\begin{table}[b!]
\centering
\scriptsize
\caption{Characteristics of the methods for electrodermal activity sensing and hydration sensing.}
\label{tbl:sensors:eda_hydration}
\resizebox{\textwidth}{!}{
\renewcommand{\arraystretch}{0.7}
\begin{tabular}{ccccccc}
\toprule
\multirow{2}{*}{\textbf{Attribute}} & \multirow{2}{*}{\textbf{Method}} & \multirow{2}{*}{\textbf{Sensor}} & \textbf{Device /} & \multirow{2}{*}{\textbf{Location}} & \multirow{2}{*}{\textbf{Properties}} & \textbf{Smart} \\
& & & \textbf{Wearable} & & & \textbf{health} \\ \midrule
& \multirow{4}{*}{Electrical} & \multirow{2}{*}{Skin} & \multirow{2}{*}{Smartwatch} & & $\checkmark$~Accuracy & \multirow{4}{*}{$\sim$} \\
Electrodermal & & \multirow{2}{*}{electrodes} & \multirow{2}{*}{Band} & Wrist & $\checkmark$~Cost & \\
activity & & \multirow{2}{*}{(wired)} & \multirow{2}{*}{Strap} & Finger & $\checkmark$~Cont.~monit. & \\
& & & & & $\sim$~Non-invasive & \\ \midrule
& \multirow{4}{*}{Electrical} & \multirow{2}{*}{Skin} & \multirow{2}{*}{Smartwatch} & & $\checkmark$~Accuracy & \multirow{4}{*}{$\checkmark$} \\
Electrodermal & & \multirow{2}{*}{electrodes} & \multirow{2}{*}{Band} & Wrist & $\checkmark$~Cost & \\
activity & & \multirow{2}{*}{(wireless)} & \multirow{2}{*}{Strap} & Finger & $\checkmark$~Cont.~monit. & \\
& & & & & $\checkmark$~Non-invasive & \\ \midrule
\multirow{4}{*}{Hydration} & Traditional & \multirow{4}{*}{-} & \multirow{4}{*}{-} & \multirow{4}{*}{-} & $\checkmark$~Accuracy & \multirow{4}{*}{$\times$} \\
& (observation & & & & $\checkmark$~Cost & \\
& of eyes or & & & & $\times$~Cont.~monit. & \\
& lips) & & & & $\checkmark$~Non-invasive & \\ \midrule
\multirow{4}{*}{Hydration} & & \multirow{4}{*}{Infrared} & \multirow{2}{*}{Band} & \multirow{2}{*}{Wrist} & $\checkmark$~Accuracy & \multirow{4}{*}{$\checkmark$} \\
& Optical & & \multirow{2}{*}{Patch} & \multirow{2}{*}{Arm} & $\sim$~Cost & \\
& spectroscopy & & \multirow{2}{*}{Textile} & \multirow{2}{*}{Head} & $\sim$~Cont.~monit. & \\
& & & & & $\checkmark$~Non-invasive & \\ \midrule
\multirow{4}{*}{Hydration} & & & \multirow{2}{*}{Band} & \multirow{2}{*}{Wrist} & $\sim$~Accuracy & \multirow{4}{*}{$\checkmark$} \\
& Electro- & Impedance & \multirow{2}{*}{Patch} & \multirow{2}{*}{Arm} & $\checkmark$~Cost & \\
& magnetic & Capacitive & \multirow{2}{*}{Textile} & \multirow{2}{*}{Head} & $\sim$~Cont.~monit. & \\
& & & & & $\checkmark$~Non-invasive & \\ \midrule
\multirow{4}{*}{Hydration} & & & Band & & $\checkmark$~Accuracy & \multirow{4}{*}{$\checkmark$} \\
& Epidermal & Electro- & Patch & Wrist & $\sim$~Cost & \\
& chemical & chemical & Tattoo & Arm & $\sim$~Cont.~monit. & \\
& & & Textile & & $\checkmark$~Non-invasive & \\
\bottomrule
\end{tabular}
}
\end{table}

\paragraph{Hydration}~

Dehydration is a dangerous condition that leads to physical and cognitive performance loss and, in the long term, to more serious diseases, such as kidney disease, heart diseases or respiratory infections. 
Although less prominent, the management of the \textit{hydration level} contributes to people's health.

Traditionally, the method to assess dehydration is qualitative, \ie looking directly at the patient's eyes or lips.
In order to measure the hydration level in a quantitative way, several techniques based on optical spectroscopic, electromagnetic or electrochemical measurements have been proposed.
These sensing methods are being integrated in stretchable devices, such as wristbands, patches, headbands and smart textiles.

\paragraph{Location}~

Since health services are no longer provided in healthcare facilities only, other user-related information might be valuable to complement medical data and, hence, contextualise users.
In particular, thanks to the self-location capabilities integrated in most smartphones and smartwatches, \textit{location} data is an extremely valuable asset to assist at-risk patients (\eg elderly, children and people with some medical conditions) by finding medical assistance nearby, notifying to the emergency services the exact location of the emergency, and communicating rapidly with caregivers.

People's location is usually determined using GPS, integrated in most smartphones, wearables and IoT devices.
This satellite-based technology is highly accurate in outdoors for deploying smart health services, although it worsens in indoors or upon bad weather conditions.
Further satellite-based positioning solutions are GLONASS and Galileo, yet their availability in mobile device is less common.
With regards to indoors, there exist several technologies for precise indoor positioning and proximity-based systems, such as BLE beacons, WPS, RFID and UWB (details in Section \ref{sec:sensors:communication}).
These technologies help locate and keep track of people's trajectories in indoor environments, such as elderly in nursing homes or patients in a smart hospital.

\begin{table}[t!]
\centering
\scriptsize
\caption{Characteristics of the methods for location sensing and body motion sensing.}
\label{tbl:sensors:location_motion}
\resizebox{\textwidth}{!}{
\renewcommand{\arraystretch}{0.7}
\begin{tabular}{ccccccc}
\toprule
\multirow{2}{*}{\textbf{Attribute}} & \multirow{2}{*}{\textbf{Method}} & \multirow{2}{*}{\textbf{Sensor}} & \textbf{Device /} & \multirow{2}{*}{\textbf{Location}} & \multirow{2}{*}{\textbf{Properties}} & \textbf{Smart} \\
& & & \textbf{Wearable} & & & \textbf{health} \\ \midrule
\multirow{4}{*}{Location} & & \multirow{2}{*}{GPS} & \multirow{2}{*}{Smartphone} & \multirow{4}{*}{Any} & $\checkmark$~Accuracy & \multirow{4}{*}{$\checkmark$} \\
& Satellite & \multirow{2}{*}{GLONASS} & \multirow{2}{*}{Smartwatch} & & $\checkmark$~Cost & \\
& (outdoor) & \multirow{2}{*}{Galileo} & \multirow{2}{*}{Band} & & $\checkmark$~Cont.~monit. & \\
& & & & & $\checkmark$~Non-invasive & \\ \midrule
\multirow{4}{*}{Location} & & BLE beacon & \multirow{2}{*}{IoT} & \multirow{4}{*}{-} & $\checkmark$~Accuracy & \multirow{4}{*}{$\checkmark$} \\
& Proximity & WPS & \multirow{2}{*}{Access point} & & $\checkmark$~Cost & \\
& (indoor) & RFID & \multirow{2}{*}{Tag} & & $\sim$~Cont.~monit. & \\
& & UWB & & & $\checkmark$~Non-invasive & \\ \midrule
& \multirow{2}{*}{Optical} & \multirow{4}{*}{Camera} & & \multirow{2}{*}{Markers} & $\checkmark$~Accuracy & \multirow{4}{*}{$\times$} \\
Body & \multirow{2}{*}{motion} & & Camera & \multirow{2}{*}{distributed} & $\times$~Cost & \\
motion & \multirow{2}{*}{capture} & & Marker & \multirow{2}{*}{in the body} & $\times$~Cont.~monit. & \\
& & & & & $\times$~Non-invasive & \\ \midrule
& \multirow{4}{*}{Optical} & \multirow{4}{*}{Camera} & & \multirow{4}{*}{-} & $\sim$~Accuracy & \multirow{4}{*}{$\sim$} \\
Body & & & RGB-depth & & $\sim$~Cost & \\
motion & & & camera & & $\sim$~Cont.~monit. & \\
& & & & & $\checkmark$~Non-invasive & \\ \midrule
& \multirow{4}{*}{Kinematic} & \multirow{4}{*}{IMU} & \multirow{2}{*}{Band} & & $\sim$~Accuracy & \multirow{4}{*}{$\checkmark$} \\
Body & & & \multirow{2}{*}{Patch} & Region of & $\checkmark$~Cost & \\
motion & & & \multirow{2}{*}{Textile} & interest & $\checkmark$~Cont.~monit. & \\
& & & & & $\checkmark$~Non-invasive & \\
\bottomrule
\end{tabular}
}
\end{table}

\paragraph{Body Motion}~

Human motion analysis helps physicians and physiotherapists to identify abnormal movements and plan and assess the correctness of rehabilitation programs.
More specifically, applications of \textit{body motion} measurements in healthcare are diverse, including the assessment of gait abnormalities, the development of corrective posture systems for rehabilitation purposes, the detection of falls (especially for elderly) and the recognition of gestures and activities.

Many solutions are based on optical motion capture methods, aiming to track human motions in a 3D space using multiple cameras triangulating markers attached to different parts of the body.
However, these systems are very complex, require expensive time-consuming operations and are only feasible in indoor settings under very controlled conditions.
Similarly, there are alternative optical methods that use RGB-depth cameras that do not require attaching markers to the body.
These methods are more practical and less invasive, but they do not provide spatio-temporal information and are only viable under controlled conditions too.
More interesting is the use of inertial measurement units (IMU), which consider a triaxial accelerometer, a triaxial gyroscope and a non-inertial triaxial magnetometer, to recreate the motion of the movements.
This method, which works indoors and outdoors, can be integrated in wearable devices for enhancing people's comfort while enabling long-term monitoring.

\subsection{Contextual sensors}

Large networks of sensors and IoT devices collecting contextual parameters in real-time enable context-aware environments, where smart health applications can be deployed.
Although this information is rarely exploited in other healthcare paradigms, it can provide more efficient and effective health services to enhance people's health status and welfare.
This section reviews some of the most popular methods for the sensing of contextual attributes (see Figure \ref{fig:sensors:context}), which are summarised in Table \ref{tbl:sensors:air_temp_hum_pres} to Table \ref{tbl:sensors:em_seismic}.

\begin{figure}[t!]
\centering
\includegraphics[width=0.95\linewidth]{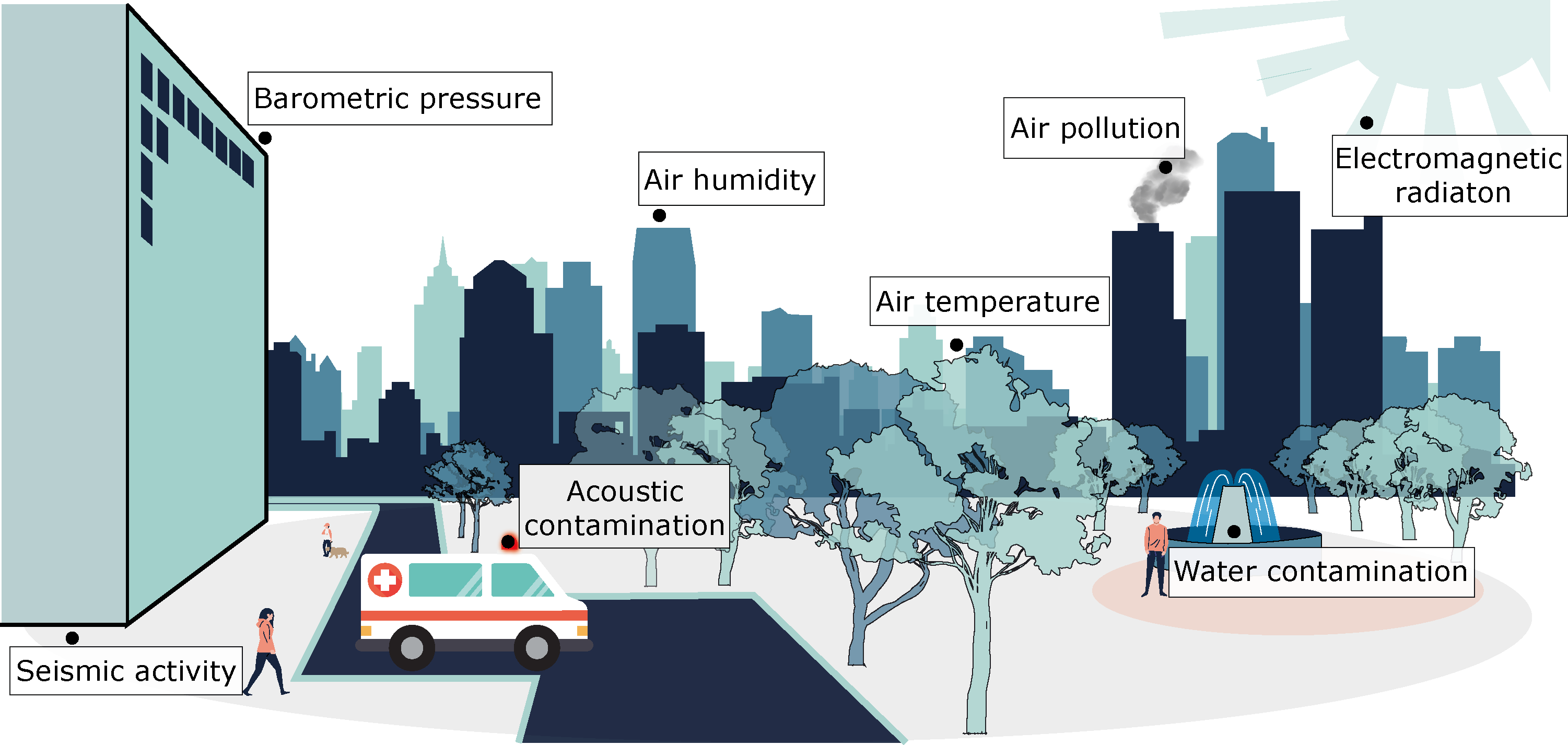}
\caption{Contextual attributes for smart health that can be sensed from context-aware environments (adapted from \cite{batista2021sensors}).}
\label{fig:sensors:context}
\end{figure} 

\paragraph{Air Temperature}~

\textit{Air temperature} is one of the main contextual parameters contributing to the comfort, welfare and health of people.
Studies on the relationship between air temperature and mortality rate or the appearance of diseases or disorders are on the rise, especially important with the incoming climate change effects \cite{xu2018association}.

There are different types of temperature sensors integrated in a large number of IoT devices.
Thermocouples are popular sensing solutions because of their reduced costs, rapid responses to temperature changes and large range of temperature detection values.
But, their accuracy is not excellent, so they should not be used when very precise measurements are needed.
Resistance temperature detectors, with a metal core, are more accurate, but their response to temperature changes is slower and their cost is higher compared to other sensors.
Similarly, thermistors are low-cost sensors with a ceramic or polymer core that lead to faster responses to temperature changes with notable accuracy.
Last but not least, several devices prefer semiconductor-based sensors using integrated circuits because of their low cost, low energy consumption and fair accuracy.

\paragraph{Air Humidity}~

Another popular contextual parameter is \textit{air humidity}.
Abnormal humidity values can cause physical discomfort that can aggravate to serious health outcomes, especially related to infectious diseases or respiratory complications.

Similar to air temperature sensing, many IoT devices already integrate mechanisms for air humidity monitoring, which can be measured using different types of sensors.
Capacitive humidity sensors are commonly used in the market due to their accuracy, small dimensions, low-power consumption and wide measurement range.
Resistive sensors are inexpensive and more practical for smart health applications that do not require extremely precise results.
More recently, optical methods based on fiber-optic sensors are being currently researched as an alternative solution providing more durability, higher accuracy and electromagnetic immunity.
This method however is less implemented in commercial devices, although its future is encouraging.

\paragraph{Barometric pressure}~

Differences in the \textit{barometric pressure} because of weather shifts or altitude changes may harmfully affect the human's body, resulting in headache, migraine or joint pain, such as arthritis.
Current barometric pressure sensors are MEMS (microelectromechanical systems), based on the piezo-resistive effect, which offer high accuracy, low power consumption, low cost and can be manufactured at low cost and small dimensions so that they could be seamlessly integrated into portable IoT devices.

\begin{table}[b!]
\centering
\scriptsize
\caption{Characteristics of the methods for air temperature sensing, air humidity sensing and barometric pressure sensing.}
\label{tbl:sensors:air_temp_hum_pres}
\resizebox{0.9\textwidth}{!}{
\renewcommand{\arraystretch}{0.7}
\begin{tabular}{ccccccc}
\toprule
\textbf{Attribute} & \textbf{Method} & \textbf{Sensor / Device} & \textbf{Properties} & \textbf{Smart health} \\ \midrule
& \multirow{4}{*}{Electrical} & \multirow{4}{*}{Thermocouple} & $\times$~Accuracy & \multirow{4}{*}{$\sim$} \\
Air & & & $\checkmark$~Cost & \\
temperature & & & $\checkmark$~Resp.~time & \\
& & & $\checkmark$~Energy~cons. & \\ \midrule
& \multirow{4}{*}{Electrical} & \multirow{2}{*}{Resistance} & $\checkmark$~Accuracy & \multirow{4}{*}{$\sim$} \\
Air & & \multirow{2}{*}{temperature} & $\sim$~Cost & \\
temperature & & \multirow{2}{*}{detector} & $\times$~Resp.~time & \\
& & & $\checkmark$~Energy~cons. & \\ \midrule
& \multirow{4}{*}{Electrical} & \multirow{4}{*}{Thermistor} & $\checkmark$~Accuracy & \multirow{4}{*}{$\checkmark$} \\
Air & & & $\sim$~Cost & \\
temperature & & & $\checkmark$~Resp.~time & \\
& & & $\checkmark$~Energy~cons. & \\ \midrule
& \multirow{4}{*}{Electrical} & \multirow{2}{*}{Semiconductor} & $\sim$~Accuracy & \multirow{4}{*}{$\checkmark$} \\
Air & & \multirow{2}{*}{integrated} & $\checkmark$~Cost & \\
temperature & & \multirow{2}{*}{circuit} & $\checkmark$~Resp.~time & \\
& & & $\checkmark$~Energy~cons. & \\ \midrule
& \multirow{4}{*}{Electrical} & \multirow{4}{*}{Capacitive} & $\checkmark$~Accuracy & \multirow{4}{*}{$\sim$} \\
Air & & & $\times$~Cost & \\
humidity & & & $\checkmark$~Resp.~time & \\
& & & $\checkmark$~Energy~cons. & \\ \midrule
& \multirow{4}{*}{Electrical} & \multirow{4}{*}{Resistive} & $\sim$~Accuracy & \multirow{4}{*}{$\checkmark$} \\
Air & & & $\checkmark$~Cost & \\
humidity & & & $\sim$~Resp.~time & \\
& & & $\checkmark$~Energy~cons. & \\ \midrule
& \multirow{4}{*}{Optical} & \multirow{4}{*}{Fiber-optic} & $\checkmark$~Accuracy & \multirow{4}{*}{$\sim$} \\
Air & & & $\times$~Cost & \\
humidity & & & $\sim$~Resp.~time & \\
& & & $\checkmark$~Energy~cons. & \\ \midrule
& \multirow{4}{*}{MEMS} & & $\checkmark$~Accuracy & \multirow{4}{*}{$\checkmark$} \\
Barometric & & Piezoresistive & $\checkmark$~Cost & \\
pressure & & pressure & $\checkmark$~Resp.~time & \\
& & & $\checkmark$~Energy~cons. & \\ 
\bottomrule
\end{tabular}
}
\end{table}

\paragraph{Air Pollution}~

Carbon monoxide, particulate matter, sulphur dioxide, lead and nitrogen oxide are some of the most hazardous air pollutants.
Collecting and managing \textit{air pollution} information is highly valuable in smart health applications due to the number of adverse effect that may appear from breathing polluted air, such as respiratory problems, cardiovascular diseases, neurological disorders and even certain cancers after long-term exposures, among others \cite{manisalidis2020environmental}.

Optical spectroscopy methods are standard analytical techniques to detect gas pollutants in the air, although they are time-consuming, expensive and cannot be used in real-time.
More interestingly, there are two main sensing methods for measuring this information more efficiently.
The metal oxide semiconductor (MOS) sensors are the most commonly used sensors in the industry thanks to their low cost, reduced dimensions, fast response times, low power consumption and high durability.
However, they are highly sensitive to changes in environmental conditions and interfering gases.
These limitations are overcome in electrochemical sensors, but their dimension and cost are significantly higher.

\paragraph{Water Contamination}~

In addition to air, water is another vital resource that may contain microbiological or chemical contamination.
\textit{Polluted water}, especially abundant in low-income countries, can lead to waterborne parasitic infections, such as cholera and dysentery, chronic diseases or reproductive complications.

\begin{table}[b!]
\centering
\scriptsize
\caption{Characteristics of the methods for air pollution sensing, water contamination sensing and acoustic contamination sensing.}
\label{tbl:sensors:air_water_acous_contam}
\resizebox{0.98\textwidth}{!}{
\renewcommand{\arraystretch}{0.7}
\begin{tabular}{ccccccc}
\toprule
\textbf{Attribute} & \textbf{Method} & \textbf{Sensor / Device} & \textbf{Properties} & \textbf{Smart health} \\ \midrule
& & & $\checkmark$~Accuracy & \multirow{4}{*}{$\times$} \\
Air & Optical & Infrared & $\times$~Cost & \\
pollution & spectroscopy & Fluorescence & $\times$~Resp.~time & \\
& & & $\sim$~Energy~cons. & \\ \midrule
& \multirow{4}{*}{Chemiresistive} & \multirow{4}{*}{MOS} & $\sim$~Accuracy & \multirow{4}{*}{$\checkmark$} \\
Air & & & $\checkmark$~Cost & \\
pollution & & & $\checkmark$~Resp.~time & \\
& & & $\sim$~Energy~cons. & \\ \midrule
& \multirow{4}{*}{Electrochemical} & \multirow{4}{*}{Electrochemical} & $\checkmark$~Accuracy & \multirow{4}{*}{$\sim$} \\
Air & & & $\times$~Cost & \\
pollution & & & $\checkmark$~Resp.~time & \\
& & & $\checkmark$~Energy~cons. & \\ \midrule
& & & $\checkmark$~Accuracy & \multirow{4}{*}{$\times$} \\
Water & Traditional & In-lab & $\times$~Cost & \\
contamination & (chemical) & instrumentation & $\times$~Resp.~time & \\
& & & $\times$~Energy~cons. & \\ \midrule
& \multirow{4}{*}{Electrochemical} & \multirow{2}{*}{Resistive} & $\sim$~Accuracy & \multirow{4}{*}{$\checkmark$} \\
Water & & \multirow{2}{*}{Capacitive} & $\sim$~Cost & \\
contamination & & \multirow{2}{*}{Conductance} & $\checkmark$~Resp.~time & \\
& & & $\checkmark$~Energy~cons. & \\ \midrule
& \multirow{4}{*}{Optical} & \multirow{4}{*}{CMOS camera} & $\sim$~Accuracy & \multirow{4}{*}{$\checkmark$} \\
Water & & & $\checkmark$~Cost & \\
contamination & & & $\checkmark$~Resp.~time & \\
& & & $\checkmark$~Energy~cons. & \\ \midrule
& \multirow{4}{*}{Acoustic} & \multirow{4}{*}{Microphone} & $\checkmark$~Accuracy & \multirow{4}{*}{$\checkmark$} \\
Acoustic & & & $\checkmark$~Cost & \\
contamination & & & $\checkmark$~Resp.~time & \\
& & & $\checkmark$~Energy~cons. & \\
\bottomrule
\end{tabular}
}
\end{table}

Traditional water quality analyses are laboratory-based, \ie by chemically analysing water samples collected manually at various locations and at different time periods.
Despite accurate, this procedure is highly inefficient, resource-consuming and off-line because no real-time information is provided, an essential characteristic to detect outbreaks of contaminated water rapidly.
To evaluate the water quality in real-time, specific electrochemical sensors can monitor changes in different water parameters that become affected by chemical and biological pollutants, namely turbidity, free/total chlorine, oxidation-reduction potential, electrical conductivity, pH, nitrates level or temperature.
Also, further approaches proposed the use of aquatic sensors embedding a CMOS camera to detect floating debris in the water.

\paragraph{Acoustic Contamination}~

Regular exposure to \textit{acoustic contamination}, this is, elevated sound levels and environmental noise, can lead to adverse health outcomes, such as hearing impairments, sleep disturbance and chronic stress.
Today, continuous and real-time monitoring of noise levels is feasible with low-cost and small microphones embedded in IoT devices.

\paragraph{Electromagnetic Radiation}~

Electromagnetic radiation has become a popular form of pollution due to the omnipresent telecommunication equipment.
In short, there exist two types of radiation: non-ionising radiation and ionising radiation.
Whereas non-ionising radiation (\eg radio-frequency, microwaves or infrared) does not have a strong impact in human health, prolonged exposures to ionising radiation (\eg Gamma rays or X-rays) can result to serious adverse health outcomes, such as cancer.

To protect against radiation exposure, portable and affordable devices are being developed to detect Gamma radiation using Geiger-M{\"u}ller tubes or fiber-optic radiation sensors, and infrared and ultraviolet lights with optical sensors as well.

\paragraph{Seismic Activity}~

Surface vibrations are common, but they are generally imperceptible for humans due to their low intensity and do not suppose an apparent risk.
However, sudden high-intensity shakes can produce seismic waves able to collapse buildings and trigger catastrophic consequences for people, including death.
Continuous seismic monitoring does not contribute to people's health status \textit{per se} under normal conditions, but it can be essential for alerting or predicting seismic events so to guarantee people's safety.

Traditional sensing solutions use seismometers, reliable and accurate instruments but bulky, expensive and sensitive to electromagnetic interference.
More recently, this activity is being monitored using inexpensive tri-axial acceleterometer sensors that, in combination with machine learning techniques, enables detecting and predicting seismic events.
Additionally, opto-mechanical sensors based on optical fiber technology have also been assessed to monitor ground motions.

\begin{table}[t!]
\centering
\scriptsize
\caption{Characteristics of the methods for electromagnetic radiation sensing and seismic activity sensing.}
\label{tbl:sensors:em_seismic}
\resizebox{0.94\textwidth}{!}{
\renewcommand{\arraystretch}{0.7}
\begin{tabular}{ccccccc}
\toprule
\textbf{Attribute} & \textbf{Method} & \textbf{Sensor / Device} & \textbf{Properties} & \textbf{Smart health} \\ \midrule
& \multirow{4}{*}{Electrical} & & $\checkmark$~Accuracy & \multirow{4}{*}{$\sim$} \\
Electromagnetic & & Geiger-M{\"u}ller & $\times$~Cost & \\
radiation & & tubes & $\sim$~Resp.~time & \\
& & & $\sim$~Energy~cons. & \\ \midrule
& \multirow{4}{*}{Optical} & \multirow{4}{*}{Fiber-optic} & $\checkmark$~Accuracy & \multirow{4}{*}{$\checkmark$} \\
Electromagnetic & & & $\sim$~Cost & \\
radiation & & & $\checkmark$~Resp.~time & \\
& & & $\sim$~Energy~cons. & \\ \midrule
& & \multirow{4}{*}{Seismometer} & $\checkmark$~Accuracy & \multirow{4}{*}{$\times$} \\
Seismic & Traditional & & $\times$~Cost & \\
activity & (motion) & & $\sim$~Resp.~time & \\
& & & $\sim$~Energy~cons. & \\ \midrule
& \multirow{4}{*}{Kinematic} & \multirow{4}{*}{Accelerometer} & $\sim$~Accuracy & \multirow{4}{*}{$\checkmark$} \\
Seismic & & & $\checkmark$~Cost & \\
activity & & & $\sim$~Resp.~time & \\
& & & $\checkmark$~Energy~cons. & \\ \midrule
& \multirow{4}{*}{Optical} & \multirow{4}{*}{Opto-mechanical} & $\sim$~Accuracy & \multirow{4}{*}{$\checkmark$} \\
Seismic & & & $\sim$~Cost & \\
activity & & & $\sim$~Resp.~time & \\
& & & $\checkmark$~Energy~cons. & \\
\bottomrule
\end{tabular}
}
\end{table}

\section{Communication Networks} \label{sec:sensors:communication}

Smart health services are not fed from a single sensing device, but from many.
To ease their management, devices are structured logically as networks, mostly wireless to endow the system with a major flexibility and cost effectiveness.
Wireless sensor networks (WSNs) provide a contextualisation-enabling infrastructure within physical environments.
For instance, the coverage of a certain region with a WSN comprising multiple temperature, humidity and air pollution sensors enables the transmission of real-time data to a smart health service aiming to alert nearby patients with respiratory diseases.
With the rise of wearable technology, WSNs evolved towards a more user-centric approach, and the wireless body area networks (WBANs) appeared with the aim to collect user-centric attributes and provide efficient, personalised and real-time health services.
For example, a diabetic-oriented WBAN enables continuous blood glucose monitoring and, when abnormal values are detected, it alerts an external smart health service and/or activates an insulin pump actuator.
However, in addition to the technical challenges inherited from WSNs, such as latency, throughput or energy consumption, WBANs have to face additional obstacles for their practical adoption, including reliability, fault tolerance, interoperability and security, among others.

The devices involved in WBANs are commonly resource-constrained in terms of computation, memory, storage and battery capacity.
Hence, designing the proper communication architecture to transmit data in a time- and energy-efficient manner is of great interest.
WBAN communication architectures are generally based on three tiers (see Figure \ref{fig:sensors:architecture}).

\begin{itemize}
\item \textit{Intra-WBAN communications (Tier 1)}:
This tier considers the communications between the sensors and actuators (\ie nodes) placed in, on and around the human body, in a range of approximately two meters.
Besides the direct communication among these nodes, they can also communicate with a sink node, a portable device attached to the body, to transmit the user-centric data.
The sink node, which usually refers to a smartphone in the s-health context, is the WBAN coordinator and gateway to the next tier.
Short-range communication technologies, such as Bluetooth, are used in this tier.

\item \textit{Inter-WBAN communications (Tier 2)}:
This tier aims to connect the users' WBANs with other networks that are easily accessible to other users, such as the Internet and cellular networks.
Therefore, the communications in this tier take place between the sink node and one or more access points, which are the gateways to those networks.
Large-range communication technologies, such as ZigBee, BLE, Wi-Fi and cellular, are adopted in Tier 2.

\item \textit{Beyond-WBAN communications (Tier 3)}:
The communications in this tier refer to those from the health provider.
The received user-centric data from Tier 2 is stored in the healthcare information system (HIS), and then analysed by medical practitioners or automated systems.
With the medical records and the patients' profile, smart health systems can automate real-time diagnosis, adjust medical treatments or alert the emergency services if needed.
\end{itemize}

Conceptually, four main actors are involved in this architecture.
First, as reviewed in the previous section, nodes are a primary information source in s-health systems.
Second, these systems are supported by the HIS, responsible for the storage, retrieval, analysis and presentation of the data in accordance to the services provided.
Third, the users of the s-health services, either patients, physicians and caregivers, must also be considered.
And fourth, all these actors can interact among them thanks to the deployment of communications networks.

\begin{figure}[t!]
\centering
\includegraphics[width=0.93\linewidth]{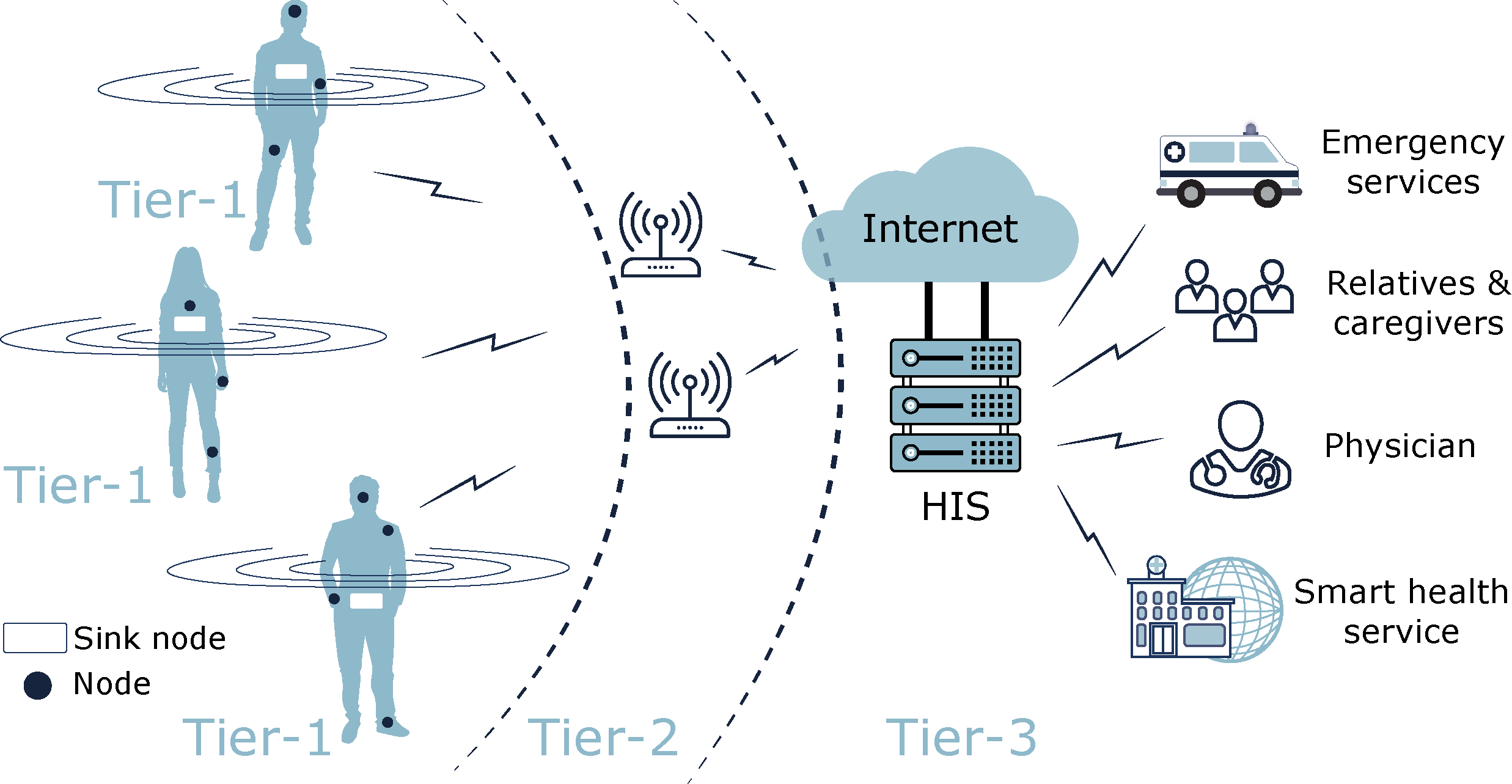}
\caption{Component-based representation of the 3-tier communication architecture for WBANs (adapted from \cite{batista2021sensors}).}
\label{fig:sensors:architecture}
\end{figure} 

There exist plenty of wireless technologies allowing deploying smart health systems, each one with its own properties in terms of radio coverage, data rate, latency, power consumption, etc.
All these features must be considered when designing any smart health solution.
The landscape of wireless communication technologies for s-health is described as follows, and a comparison between these technologies is summarised in Table \ref{tbl:sensors:communications_1} and Table \ref{tbl:sensors:communications_2}.

\paragraph{Bluetooth and BLE}~

\textit{Bluetooth} is one of the most popular wireless technologies for short-range communications.
Data can be transmitted between two wireless devices, one of them acting as a master (commonly, the sink node) and the other as a slave, in a range up to 100 meters at a data rate of 3 Mbps at most.
Like Wi-Fi and ZigBee, Bluetooth operates in the 2.4 GHz ISM band, and applies frequency hopping-related techniques to reduce possible interference.
This technology reached its popularity in the early 2000s with the emergence of mobile devices, and even today is extensively used in numerous general-purpose portable devices, including smartphones, smartwatches, fitness trackers, laptops, computer peripherals and alike.
But, with the advent of resource-constrained devices, a very low-power Bluetooth specification was developed: the \textit{Bluetooth Low Energy} (BLE), able to transmit data in a range up to 400 meters at a very low latency and energy consumption.
The BLE characteristics are well aligned with smart health applications, such as critical emergency response, so to communicate with wearables, IoT and other devices deployed in WSNs and WBANs.
Hence, it would be unsurprising to see BLE as an excellent technology for the next-generation medical purpose oriented devices.

\begin{table}[b!]
\centering
\scriptsize
\caption{Comparison of the main wireless communication technologies for smart health (I).}
\label{tbl:sensors:communications_1}
\resizebox{\textwidth}{!}{
\renewcommand{\arraystretch}{0.8}
\begin{tabular}{lccccc}
\toprule
& \textbf{Bluetooth} & \textbf{BLE} & \textbf{ZigBee} & \textbf{IEEE 802.15.6} & \textbf{Wi-Fi} \\ \midrule
\multirow{2}{*}{Frequency} & \multirow{3}{*}{2.4 GHz} & \multirow{3}{*}{2.4 GHz} & \multirow{2}{*}{868/915 MHz} & 14--29 MHz &  \\
\multirow{2}{*}{bands} & & & \multirow{2}{*}{2.4 GHz} & 400--2400 MHz & 2.4/5 GHz\\
& & & & 3.2--10.3 GHz \\ \midrule
\multirow{2}{*}{Coverage} & Short/Medium & Medium & Short/Medium & Short & Medium \\ \cmidrule{2-6}
& 10--100 m & 400 m & 10--100 m & 2 m & 50--100 m\\ \midrule
\multirow{3}{*}{Data rate} & Moderate & Moderate & Low & Low/Moderate & High \\ \cmidrule{2-6}
& \multirow{2}{*}{1--3 Mbps} & \multirow{2}{*}{1--2 Mbps} & \multirow{2}{*}{20--250 kbps} & 10 kbps & 400 Mbps \\
& & & & to 15 Mbps & to 10 Gbps \\ \midrule
\multirow{2}{*}{Latency} & Moderate & Very low & Very low/Low & Low/Moderate & Low \\ \cmidrule{2-6}
& 100 ms & 10 ms & 10--30 ms & 125 ms & 50 ms\\ \midrule
\multirow{2}{*}{Power} & Moderate & Very low & Very low/Low & Very low & High \\ \cmidrule{2-6}
& 0.2--0.5 W & 10 mW & 1--60 mW & 0.1--3 mW & 0.8--1 W\\ \midrule
Size & 8 & 32.000 & 65.000 & 256 & 250 \\ \midrule
\multirow{2}{*}{Topology} & \multirow{2}{*}{Scatternet} & Star & Star, Tree, & Star, & Star, Mesh, \\ 
& & Mesh & Mesh & Multi-hop & Ad-hoc \\ \midrule
\multirow{2}{*}{Security} & 56,64,128-bit & 128-bit & 128-bit & Level 1 \& & 128,256-bit \\ 
& AES & AES & AES & Level 2 & AES \\ \midrule
Cost & Medium & Low & Low & Low & High \\ \midrule
\multirow{2}{*}{WBAN tier} & Tier 1 & Tier 1 & \multirow{2}{*}{Tier 2} & \multirow{2}{*}{Tier 1} & Tier 2 \\
& Tier 2 & Tier 2 & & & Tier 3 \\ \midrule
Smart health & $\sim$ & $\checkmark$ & $\sim$ & $\checkmark$ & $\checkmark$ \\
\bottomrule
\end{tabular}
}
\end{table}

\paragraph{ZigBee}~ 

Built on the IEEE 802.15.4 standard, \textit{ZigBee} is another low-power technology suitable for resource-constrained devices.
Depending on the ZigBee module, different characteristics in terms of coverage, data rate, power consumption and operational frequency bands are offered.
Data can be generally transmitted through ZigBee up to 100 meters (similar to Bluetooth's), but at a low data rate up to 250 kbps.
Hence, this technology might not be adequate for transmitting real-time user-centric data that requires immediate action.
Yet, it can be suitable for battery-powered IoT devices oriented to contextual sensing, which could be operational for years before batteries depletion.

\paragraph{IEEE 802.15.6}~

\textit{IEEE 802.15.6}, the latest international standard for WBANs communications, enables short-distance communications between devices operating on, in or around the human body.
Three different types of communications are supported in the standard, each of them operating at different frequency bands for different purposes.
First, the narrowband (NB) between the 400 MHz and 2.4 GHz with data rates up to 900 kbps that is used for medical telemetry or implant communication, among other applications.
Second, the ultra wideband (UWB) operating at higher frequencies enable higher data rates of several Mbps between on-body devices and on/off-body devices, such as entertainment systems.
And third, the human body communication (HBC) that operates in a low frequency band and transmits data at a maximum data rate of 2 Mbps in an energy-efficient manner. 

In addition to the low energy consumption requirements, reliability is another fundamental characteristic of this technology because devices are continuously changing their location due to humans movements.
Also, to secure these communications, three security levels are defined:
(i) level 0 does not provide any security mechanisms and unsecured communications are established,
(ii) level 1 provides message authentication and integrity, but no encryption mechanisms,
and (iii) level 2 provides message authentication, integrity and encryption.
This standard should certainly be adopted by miniaturised and resource-constrained medical devices to properly communicate the user-centric data.

\begin{table}[t!]
\centering
\scriptsize
\caption{Comparison of the main wireless communication technologies for smart health (and II).}
\label{tbl:sensors:communications_2}
\resizebox{0.97\textwidth}{!}{
\renewcommand{\arraystretch}{0.8}
\begin{tabular}{lccccc}
\toprule
& \textbf{4G/LTE} & \textbf{5G} & \textbf{LoRa} & \textbf{SigFox} & \textbf{NB-IoT} \\ \midrule
\multirow{2}{*}{Frequency} & \multirow{3}{*}{0.7--2.6 GHz} & 600--700 MHz & \multirow{2}{*}{863--928} & \multirow{2}{*}{868/915} & \multirow{2}{*}{800--900} \\
\multirow{2}{*}{bands} & & 2.5--3.8 GHz & \multirow{2}{*}{MHz} & \multirow{2}{*}{MHz} & \multirow{2}{*}{MHz} \\
& & 25--100 GHz & & & \\ \midrule
\multirow{2}{*}{Coverage} & High & Medium/High & High & High & High  \\ \cmidrule{2-6}
& 10 km & 300m--1 km & 5--20 km & 10--50 km & 15 km  \\ \midrule
\multirow{2}{*}{Data rate} & High & Very High & Very low & Very low & Low \\ \cmidrule{2-6}
& 10--300 Mbps & 1--20 Gbps & 37.5 kbps & 100--600 bps & 250 kbps  \\ \midrule
\multirow{2}{*}{Latency} & Low & Very low & High & High & High  \\ \cmidrule{2-6}
& 50--70 ms & 1--10 ms & 3 s & 10 s & 1 s \\ \midrule
\multirow{2}{*}{Power} & Moderate & Low & Low & Low & Low \\ \cmidrule{2-6}
& 250--700 mW & N/A & 25 mW & 10--100 mW & 20--200 mW \\ \midrule
\multirow{2}{*}{Size} & Thounsands & 1 Million & \multirow{2}{*}{1.000} & \multirow{2}{*}{1.000.000} & \multirow{2}{*}{50.000} \\ 
& per km$^2$ & per km$^2$ \\ \midrule
Topology & Cellular & Cellular & Star of stars & Star & Star \\ \midrule
Security & 128-bit & 256-bit & 128-bit AES & Optional & 128,256-bit  \\ \midrule
Cost & Medium & High & Low & Low & Low \\ \midrule
\multirow{2}{*}{WBAN tier} & Tier 2 & Tier 2 & Tier 2 & Tier 2 & Tier 2 \\ 
& Tier 3 & Tier 3 & & & \\ \midrule
Smart health & $\checkmark$ & $\checkmark$ & $\sim$ & $\times$ & $\sim$  \\
\bottomrule
\end{tabular}
}
\end{table}

\paragraph{Wi-Fi}~

One of the most extensively used wireless technologies is \textit{Wi-Fi}, available in most devices from the digital ecosystem, namely computers, smartphones, smartwatches, TVs and so on.
Wi-Fi, encompassed within the IEEE 802.11 standards family, is suitable for transmitting large volumes of data in a range of tens of meters at very high data rates, where energy consumption is not a critical concern.
For instance, the latest specification IEEE 802.11ax (Wi-Fi 6) reaches data rates of up to 10 Gbps, strengthens the security requirements by introducing WPA3, and even is more energy-efficient than its predecessors, which may open the door to its possible use in some resource-constrained devices in the years to come.
These characteristics make Wi-Fi technology suitable for large-scale and real-time smart health services.

\paragraph{Cellular Networks}~

The unprecedented growth of smartphones in the last decade motivated the evolution of cellular networks, originally devoted to the provision of telephony services, towards high-bit rate transmissions of data.
Today, the LTE-based \textit{4G} technology is available in many smartphones and other mobile devices.
This technology is able to transfer data at high data rates of hundreds of Mbps at a relatively low latency.
Like Wi-Fi, the main limitation of 4G is its high power consumption, which hinders its implementation in resource-constrained devices, although it fits perfectly for long-range communications in outdoors, where secure Wi-Fi access points are less available.

The fifth generation of mobile networks, \textit{5G}, is expected to enable the massive deployment of IoT in a truly connected world comprising billions of devices.
This emerging technology promises very high data rates up to several Gbps in an almost negligible latency (1 ms ideally), using only a fraction of the energy consumption of 4G.
To make 5G a reality, lots of antennas will need to be installed in order to manage an unprecedented coverage density.
This requires a substantial investment in infrastructure though.
5G will certainly arise numerous s-health related opportunities, even though some of them could sound futuristic today, such as augmented/virtual reality assistance for blind people, remote collaboration in surgical interventions, or video-enabled medication adherence.

\paragraph{Low Power Wide Area Networks}~

Long-range communications can hardly be implemented in sensors and IoT devices due to its aggressive power consumption.
To fill this gap, \textit{Low Power Wide Area Networks} (LPWANs) emerged.
This kind of technologies are able to transmit data along large distances (of several kilometres) at a very low-power consumption.
Yet, this communication is conducted at a low data rate and with a high latency.
All in all, these technologies are not suitable for real-time applications, although they could be adopted for contextual sensing, where real-time constraints might be relaxed, or for non-critical healthcare monitoring, such as rehabilitation.

Some of the most prominent LPWAN technologies are \textit{LoRa}, \textit{SigFox} and \textit{NB-IoT}.
In short, SigFox is an easy-to-deploy technology enabling very large network connectivity at low infrastructure costs, although the transmission rate is very low (in the order of bps) and the latency is higher compared to similar technologies.
More interestingly, LoRa offers an excellent trade-off between distance coverage, data rate and energy consumption, and its popularity in the IoT arena has significantly grown in the latest years and it is expected to grow further.
NB-IoT, although enhancing the LoRa's properties in terms of latency and data rate, has a scarce adoption in IoT devices and lacks from deployment readiness.

\paragraph{Miscellaneous}~

In addition to the aforementioned technologies, we cannot forget other existing solutions that could be well suited for s-health purposes.
RFID and NFC are popular solutions for very short-range communications, particularly interesting for indoors.
Further promising low-power technologies that could complement or even replace ZigBee or Bluetooth in the incoming years are, among others, Z-Wave, ANT and RuBee.
Within the LPWAN standards, Weightless could be an interesting solution for communicating devices in the industrial and medical field.
Finally, WiMAX (IEEE 802.16) might be another suitable technology for establishing long-range transmissions of several kilometres where energy consumption is not critical.

\section{Security Analysis}  \label{sec:sensors:security}

Current information systems must face numerous concerns, but security and privacy issues are essential.
Given the high confidentiality of the information managed, these issues become foremost in smart health systems.
Security and privacy aspects must be considered throughout the entire system: from the very sensing devices where data is gathered, going through the network where data is transmitted, to the HIS where data is stored, analysed and visualised.
The impact of attacks against s-health systems may go beyond the leakage of medical records and the loss of privacy, and life-threatening situations may arise in case of exploiting vulnerabilities in implantable devices, such as pacemakers or insulin pumps \cite{kuehn2018pacemaker}.
Moreover, wireless communications and the proper HIS may also entail a number of security flaws depending on the design and implementation of the system: insecure programming practices, vulnerable communication protocols or obsolete technologies open the door to numerous security attacks \cite{papageorgiou2018security}.

This section addresses the information security aspects in smart health systems from a global scope, by describing the security requirements that all s-health systems should fulfil, categorising the most common attacks in these systems, and proposing appropriate solutions to avert them.

\subsection{Requirements}

Smart health systems, like any other information system, must pursue a number of security requirements and put in place the appropriate mechanisms to guarantee them.
The main security requirements considered in these systems are briefly discussed next.

\begin{itemize}
\item \textit{Data confidentiality}:
Confidentiality ensures that data is only disclosed to authorised entities (\eg people, devices\ldots), whilst remaining unintelligible to unauthorised entities.
User-centric data, but especially medical data, should be kept confidential during their transmission over the network and storage periods.
Encryption is a widely used technique to achieve this property, since only authorised entities can decode the data.

\item \textit{Data integrity}:
Integrity ensures the accuracy, trustworthiness and completeness of data, hence guaranteeing that data has not been modified or destroyed by unauthorised entities.
Unless detected, smart health systems would react to users upon faux data, and potentially endanger their health.
Cryptographic hashes help detect data modifications, and redundancy and backup policies enable restoring affected data if necessary.

\item \textit{Data availability}:
Availability is a property that guarantees that authorised entities have constant access to the data.
Since medical decisions could be made anytime and anywhere, smart health systems must be resilient to service disruptions, either intentional or accidental.
Availability issues could be avoided with redundancy, recovery policies and fail-over strategies.

\item \textit{Non-repudiation}: 
Non-repudiation is the guarantee that a particular interaction between two entities actually occurred.
Given the communication of a message between two authorised entities, this property ensures that the sender cannot deny having sent a message to the receiver in the future, and the receiver cannot deny having received the message from the sender in the future.

\item \textit{Authentication and Authorisation}:
Authentication refers to the process of confirming the identity of an entity, \ie determining whether the entity is who it claims to be.
Authorisation refers to the process of determining whether the authenticated entity has access to the particular resources and services of the system.
Authentication procedures, such as credential-based (\eg passwords, biometrics, digital certificates\ldots), are mandatory to establish communications only with properly authenticated entities.
If succeeded, then systems must ensure whether the entities have permission to do the requested actions (\eg access or delete medical information).

\item \textit{Privacy}:
Smart health systems must process personal data in a lawful, fair and transparent manner for a specific and legitimate purpose.
These systems must be compliant with the current data protection regulations and adopt the appropriate safeguards to reduce disclosure risks, both identity and attribute.
To this end, anonymisation techniques are desirable.

\end{itemize}

\subsection{Attacks, Threats and Vulnerabilities}

Different types of security attacks can be conducted in smart health systems.
Next, these attacks are classified according to the target actor, \ie nodes, communications, HIS and users.
However, other different taxonomies have been proposed:
(i) based on the attack's nature, either passive attack (\ie monitor the system to launch further attacks) or active attack (\ie inject, alter or destroy the system),
(ii) based on the attack's origin, either internal attack (\ie initiated by insiders within the system) or external attack (\ie initiated by outsiders located outside the system),
(iii) based on the attack's launch method, either physical (\ie have physical access to the cyber-physical system), logical (\ie exploit vulnerabilities in logical systems) or side-channel (\ie indirect physical effects of systems),
or (iv) based on the TCP/IP model layer, either application layer, transport layer, network layer or network interface layer.
A summary of the security attacks discussed below is provided in Table \ref{tbl:sensors:attacks}.

\subsubsection{Attacks against Nodes}

The constrained nature of most of the nodes deployed in WSNs and WBANs limits the implementation of robust security mechanisms.
Hence, they are often primary targets for attackers.

\paragraph{Node capture attacks}~

\textit{Node capture attacks} occur when attackers take control of a node after successfully exploiting a vulnerability in it.
These potentially devastating attacks need to be rapidly detected to disconnect the compromised node from the network as soon as possible.
Otherwise, attackers may seek for further vulnerabilities within the system to elevate their privileges and, eventually, have control over the entire system.
These attacks threaten confidentiality because attackers could be able to extract private information from the captured node, such as sensitive user-centric data or the cryptographic keys used to encrypt and decrypt the communications.
Besides, within the s-health context, people's privacy could also be compromised.

\begin{table}[b!]
\centering
\scriptsize
\caption{Summary and classification of security attacks in smart health systems.}
\label{tbl:sensors:attacks}
\resizebox{\textwidth}{!}{
\renewcommand{\arraystretch}{0.9}
\begin{tabular}{lcccccc}
\toprule
\multirow{2}{*}{\textbf{Attack}} & \textbf{Target} & \multirow{2}{*}{\textbf{Nature}} & \multirow{2}{*}{\textbf{Origin}} & \textbf{Launch} & \multirow{2}{*}{\textbf{TCP/IP layer}} & \textbf{Requirements} \\ 
& \textbf{actor} & & & \textbf{method} & & \textbf{threats} \\ \midrule
\multirow{4}{*}{Node capture} & \multirow{4}{*}{Nodes} & \multirow{4}{*}{Active} & \multirow{4}{*}{External} & \multirow{4}{*}{Physical} & \multirow{4}{*}{Netw. interface} & Confidentiality \\
& & & & & & Non-repudiation \\
& & & & & & Authentication \\
& & & & & & Privacy \\ \midrule
False data & \multirow{2}{*}{Nodes} & \multirow{2}{*}{Active} & \multirow{2}{*}{Internal} & \multirow{2}{*}{Physical} & \multirow{2}{*}{Netw. interface} & \multirow{2}{*}{Integrity} \\
injection & & & & & & \\ \midrule
Sleep & \multirow{2}{*}{Nodes} & \multirow{2}{*}{Active} & \multirow{2}{*}{External} & Physical & \multirow{2}{*}{Netw. interface} & \multirow{2}{*}{Availability} \\
deprivation & & & & Logical & & \\ \midrule
\multirow{2}{*}{Side-channel} & \multirow{2}{*}{Nodes} & Passive & \multirow{2}{*}{External} & \multirow{2}{*}{Side-channel} & \multirow{2}{*}{Netw. interface} & Confidentiality \\ 
& & Active & & & & Availability \\ \midrule
& \multirow{4}{*}{Nodes} & \multirow{4}{*}{Active} & \multirow{4}{*}{External} & \multirow{4}{*}{Logical} & \multirow{4}{*}{Netw. interface} & Confidentiality \\
Firmware & & & & & & Non-repudiation \\
update & & & & & & Authentication \\
& & & & & & Authorisation \\ \midrule
\multirow{2}{*}{Eavesdropping} & Communi- & \multirow{2}{*}{Passive} & \multirow{2}{*}{External} & \multirow{2}{*}{Logical} & Netw. interface & Confidentiality \\
& cations & & & & Network & Privacy \\ \midrule
Data & Communi- & \multirow{2}{*}{Active} & \multirow{2}{*}{Internal} & \multirow{2}{*}{Physical} & \multirow{2}{*}{Netw. interface} & \multirow{2}{*}{Integrity} \\
tampering & cations & & & & & \\ \midrule
& \multirow{2}{*}{Communi-} & & & & & Integrity \\
Replay & \multirow{2}{*}{cations}  & Active & Internal & Physical & Network & Authentication \\
& & & & & & Authorisation \\ \midrule
\multirow{4}{*}{Spoofing} & & \multirow{4}{*}{Active} & & \multirow{4}{*}{Physical} & Netw. interface & \multirow{4}{*}{Integrity} \\
& Communi- & & Internal & & Network & \\
& cations & & External & & Transport & \\
& & & & & Application &  \\ \midrule
& & \multirow{4}{*}{Active} & & \multirow{4}{*}{Logical} & & Confidentiality \\
Man-in-& Communi- & & Internal & & Network & Integrity \\
the-Middle & cations & & External & & Transport & Authentication \\ 
& & & & & & Privacy \\ \midrule
& \multirow{2}{*}{Communi-} & & \multirow{2}{*}{Internal} & & Network & \\
Flooding & \multirow{2}{*}{cations} & Active & \multirow{2}{*}{External} & Logical & Transport & Availability \\
& & & & & Application & \\ \midrule
\multirow{2}{*}{Jamming} & Communi- & \multirow{2}{*}{Active} & \multirow{2}{*}{External} & \multirow{2}{*}{Physical} & \multirow{2}{*}{Netw. interface} & \multirow{2}{*}{Availability} \\
& cations & & & & &  \\ \midrule
\multirow{2}{*}{Black hole} & Communi- & \multirow{2}{*}{Active} & \multirow{2}{*}{Internal} & \multirow{2}{*}{Physical} & \multirow{2}{*}{Network} & \multirow{2}{*}{Availability} \\
& cations & & & & &  \\ \midrule
& & & & & & Confidentiality \\
& & & & & & Integrity \\
& \multirow{2}{*}{HIS} & & & & & Availability \\
Malware & \multirow{2}{*}{Nodes} & Active & External & Logical & Application & Non-repudiation \\
& & & & & & Authentication \\
& & & & & & Authorisation \\
& & & & & & Privacy \\ \midrule
\multirow{2}{*}{Data leakage} & \multirow{2}{*}{HIS} & \multirow{2}{*}{Passive} & \multirow{2}{*}{External} & \multirow{2}{*}{Logical} & \multirow{2}{*}{Application} & Confidentiality \\
& & & & & & Privacy \\ \midrule
\multirow{4}{*}{Phishing} & \multirow{4}{*}{Users} & \multirow{4}{*}{Active} & \multirow{4}{*}{External} & \multirow{4}{*}{Logical} & \multirow{4}{*}{Application} & Confidentiality \\
& & & & & & Authentication \\
& & & & & & Authorisation \\
& & & & & & Privacy \\
\bottomrule
\end{tabular}
}
\end{table}

\paragraph{False data injection attacks}~

Once nodes are compromised, attackers can inject malicious code and redeploy them in the network (as if they were legitimate nodes) with the aim to perform unintended actions.
Usually, attackers use the captured node to conduct \textit{false data injections attacks}, \ie fabricating erroneous data as if it was true or preventing passing true data.
These attacks break systems integrity, since they would naively react to fake data and might take unsuitable health decisions for people's health \cite{ahmed2017false}.
For example, injecting false physiological attributes could lead to wrong medical diagnosis or treatments.
More significantly, these attacks become severe during critical missions, such as surgeries, where the injection of false values might result in a loss of life.

\paragraph{Sleep deprivation attacks}~

\textit{Sleep deprivation attacks} (also known as energy drain attacks) are more aggressive attacks consisting in damaging, either physically or logically, the sensors so as to disrupt networks.
To this end, these attacks increase the power consumption of the captured nodes with useless tasks, such as running infinite loops, to accelerate the battery draining of the devices.
Once these nodes are disconnected, the system's availability is affected.
Within the healthcare domain, this attack could disconnect life-assistance devices, namely pacemakers \cite{hei2010defending} or cardiac defibrillators \cite{marin2016security}, and threaten people's lives.

\paragraph{Side-channel attacks}~

Whereas most attacks focus on exploiting the weaknesses of algorithms and protocols, \textit{side-channel attacks} aim to exploit the physical effects of computing devices during their normal functioning to infer sensitive information, such as cryptographic keys or passwords.
There exist different analyses of physical side signals, such as timing analyses, power analyses, electromagnetic analyses and acoustic analyses, among others.
In general, these attacks are difficult to handle due to their non-invasive nature and passive mode together with the fact that they target the physics, rather than the implementation, of the nodes.

The numerous nodes deployed in smart health systems open the door to side-channel attacks \cite{spence2020side}.
For instance, by means of body motion sensors, key-based security systems of smartphones or smartwatches could be bypassed, which gives attackers an additional advantage to break into the system more easily.
More active side-channel attacks might pose people's life at stake, such as in case of injecting electromagnetic signals that would bogus the legitimate signals of cardiac implantable devices \cite{kune2013ghost}.

\paragraph{Firmware update attacks}~

Updates on the nodes firmware is crucial to support latest technological developments and improve performance.
Nowadays, these updates are performed remotely, in which the devices download the latest firmware version available and upgrade themselves automatically.
However, this procedure usually lacks encryption and/or authentication mechanisms, and opens the door to \textit{firmware update attacks}.

These attacks aim to inject a malicious firmware into a vulnerable node so as to grant attackers total control over it.
The severity of these attacks lay on the feasibility to affect entire families of nodes: if a manufacturer uses the same firmware update mechanism, all its devices would be vulnerable.
Vulnerabilities in the firmware update procedure of fitness trackers and even implantable devices have been reported \cite{classen2018anatomy,liebowitz2015biological}.

\subsubsection{Attacks against Communications}

Communications networks are a key enabler of smart health systems.
These communications are prone to be hijacked by attackers to capture the information transmitted in them and, hence, they need to be properly secured.

\paragraph{Eavesdropping}~

In \textit{eavesdropping} (or sniffing) attacks, attackers capture and listen secretly the data packets transiting the communications without the knowledge of the legitimate entities.
All messages are compromised and eavesdroppers may acquire private information from the whole system, undermining the system's confidentiality.
For this reason, messages must never be transmitted in plain-text or encrypted with vulnerable algorithms.

Some Bluetooth, BLE, Wi-Fi and ZigBee communications have already been proven to be vulnerable to eavesdroppers, who are able to bypass their encryption and extract private information.
In this sense, a number of medical devices have been compromised using these attacks, including hospital equipment, wearables and implantable devices, among others \cite{goyal2016mind,marin2016security,rahman2015secure}.

\paragraph{Data tampering}~

Attackers can also conduct more active attacks aiming to deliberately alter or destroy data transiting the network.
This attack, commonly known as \textit{data tampering}, compromises the integrity of the data in order to manipulate the system's functioning.
For example, systems would malfunction in case of modifying properties of the data packets, such as their timestamps (\ie the flow of events would be erroneous) or their destination address (\ie packets would be redirected to illegitimate destinations).

Even worse, the unauthorised modification of more sensitive data, such as medical information, would make systems react upon malicious data.
This situation may harm people's health, leading to mistreatment or even death.
Tampering attacks have been successfully conducted on medical equipment, fitness trackers and implantable devices \cite{classen2018anatomy,halperin2008pacemakers,liebowitz2015biological,morgner2018malicious,rahman2015secure}.
Besides medical data, tampering contextual data may also negatively impact the lifestyle of large populations, since s-health systems would adapt themselves to false contextual settings.
Hence, whereas tampering user-centric data leads to individual damages, tampering contextual data might lead to large-scale damages.

\paragraph{Replay attacks}~

During eavesdropping, valid data packets are captured.
Although messages are encrypted (and hence unreadable for attackers), they have an effect on the recipient entity.
Attackers can exploit this to mislead legitimate entities and acquire the trust of the system with the aim to duplicate transactions, impersonate entities or raise confusion.
Therefore, \textit{replay attacks} occur when unauthorised entities re-send legitimate captured data packets at a later time while acting as the original sender, hoping to repeat some action that benefits the attacker.
For example, if attackers intercept messages related to a valid login procedure, they could try to replay them later on to get access into the system without even knowing the actual credentials.

Replay attacks can be tragic in smart health systems, particularly in case of replaying messages describing old medical data of users.
In such cases, systems would react to old physiological attributes and, in consequence, leading to mistreatment.
Among others, research has shown the possibility of launching replay attacks to target diabetic people, implantable devices or fitness trackers due to the lack of authentication and integrity validation techniques \cite{classen2018anatomy,halperin2008pacemakers,marin2016security,radcliffe2011hacking}.

\paragraph{Spoofing attacks}~

\textit{Spoofing attacks} consist in masquerading attackers who act as legitimate entities by using forged data.
If legitimate entities trust the incoming (malign) entity, attackers would gain access to once inaccessible resources and conduct further insider attacks.
These attacks can be conducted at different layers, namely ARP spoofing, IP spoofing, DNS spoofing and DCHP spoofing, among others.

Spoofing countermeasures can barely be implemented in most nodes due to their computational and storage limitations.
Hence, these devices may be vulnerable to spoofing attacks.
Successful attacks enable attackers retrieve medical data gathered from the nodes and even trigger life-threatening situations in case of spoofing insulin pumps \cite{park2016ain}.
Beyond medical data, users location can also be forged by spoofing GPS data.

\paragraph{Man-in-the-Middle attacks}~

One of the most popular and devastating attacks in networks are \textit{Man-in-the-Middle attacks}.
These attacks aim to intercept the communication between two legitimate entities, who believe that they are communicating directly with each other.
Once a communication is hijacked, attackers can eavesdrop the data packets seeking for private data, manipulate data packets by tampering their data or injecting false data, redirect traffic to malicious entities or even spread malware through the network.

The lack of robust security mechanisms of many nodes enables Man-in-the-Middle attacks in sensor networks.
For instance, the lack of encryption of some sensing devices and implantable devices enables attackers to seamlessly hijack communications and capture private data, such as session identifiers, passwords and health data \cite{classen2018anatomy,goyal2016mind,kazlouski2021still,rostami2013balancing}.

\paragraph{Distributed denial of service attacks}~

Availability-threatening attacks prevent the normal functioning and responsiveness of the systems.
In particular, \textit{distributed denial of service attacks} (DDoS) require multiple and coordinated sources controlled by an attacker targeting a victim with the aim to make its resources unavailable for legitimate entities.
In particular, the aforementioned sleep deprivation attacks are a type of DDoS attack at the node level.
Notwithstanding, most DDoS attacks are at the communication level.
\textit{Flooding}, a prominent DDoS attack, consists in overwhelming legitimate resources with purposeless requests, in such a way that they are not able to handle all the incoming packets (even the legitimate ones) and then collapse.
Other DDoS-oriented attacks are \textit{jamming attacks}, which use specific devices to generate random radio-frequency signals to cause interference, and \textit{black hole attacks} (also called packet drop attacks), which exploit vulnerabilities in routing protocols to redirect and block the traffic.

Unfortunately, many DDoS attacks have been launched during the latest years with the Mirai botnet, specialised at exploiting the low-security implementations of IoT devices to disrupt services \cite{de2018ddos}.
Smart health systems need to be always-on to provide real-time monitoring and act immediately in case of emergency \cite{classen2018anatomy,goyal2016mind}.
Moreover, availability becomes particularly crucial in certain critical-mission medical devices, such as implantable devices, wherein these threatening attacks can lead to the loss of people's lives \cite{halperin2008pacemakers,marin2016security}.

\subsubsection{Attacks against HIS}

Some security attacks aim to target the HIS infrastructure of the health service providers, \ie the servers, databases, routers, firewalls and computers that manage the system's applications, data and flows.
These attacks are likely to be more sophisticated since these systems are not resource-constraint and, hence, are able to implement more robust countermeasures.

\paragraph{Malware}~

\textit{Malware} encompasses different types of unwanted, hostile and malicious programs used to invade, damage, disrupt or disable computer systems and networks by exploiting their vulnerabilities.
Infected systems are partly or completely under the control of attackers and, therefore, prone to data theft, hijacks and propagate further malware into other systems.
Among the different categories of malware, including worms, trojans, viruses or rootkits, among others, ransomware attacks have become widely extended in the latest years.
In these attacks, attackers encrypt the files of an hijacked system with a cryptographic key, and ask for some payment in cryptocurrencies to the system's owners to restore them.

For decades, malware was developed to target conventional computing devices.
For example, the popular ransomware attack in 2017, WannaCry, hijacked multiple healthcare systems, including the British National Health Service.
With the rise of large sensor networks and IoT devices, different malware variants emerged to target these more vulnerable devices \cite{classen2018anatomy,de2018ddos,vignau201910}.
Smart health systems must monitor the functioning of their entities, seeking for abnormal malware-derived conditions and, once detected, apply the proper countermeasures to prevent the malware spread and mitigate the impact on the entire system.

\paragraph{Data leakage}~

\textit{Data leakages} (also called data breaches) occur when personal and/or confidential data from an organisation is released by unauthorised entities.
Medical records, financial information, trade secrets and intellectual property are profitable assets that can be leaked \cite{leakhiv,myfitnesspalhack,yahoohack,blackmarketprice}.
These attacks clearly constitute a threat to the confidentiality and privacy of the people involved in the leak.
Yet, it is noteworthy that data leakages might not be due to malicious attacks, but due to unintentional human actions, such as unintentional emailing to wrong recipients, or system glitches.
In any case, this does not prevent organisations from legal liabilities and economical sanctions, in addition to the considerable reputational damage.

\subsubsection{Attacks against Users}

It could be said that the weakest link in the security chain of a system is the human factor \cite{mitnick2003art}.
Consequently, attackers can attempt to infiltrate into target systems by taking advantage of the lack of knowledge of most users regarding computer security, rather than seeking for vulnerabilities in their infrastructure, network and nodes.

\paragraph{Phishing attacks}~

\textit{Phishing attacks} aim to deceive users to obtain sensitive information, in which attackers disguise themselves as legitimate entities.
Email is the most common channel to perform phishing attacks, where attackers forge the sender's address to seem legitimate, hoping that users rely on it and introduce private data (typically, the credentials) or download some malware.
Phishing has evolved towards more sophisticated attacks, such as spear phishing (\ie targeted phishing), whaling (\ie targeted phishing to high-privilege users), vishing (\ie voice phishing) and smishing (\ie SMS phishing).

The COVID-19 pandemic provided an excellent opportunity for attackers to intensity phishing campaigns for deceiving users with fraudulent messages \cite{apwg2020trends}.
Within the healthcare domain, attackers could impersonate legitimate users to try gain access into the HIS and, then, launch further insider attacks to acquire confidential data \cite{jalali2020employees,priestman2019phishing,wright2016big}.
Systems should be able to detect abnormal accesses (\eg unusual source IP address or country) and prevent the entrance of users unless performing a second authentication.

\subsection{Countermeasures}

Smart health systems need to adopt the proper security measures to protect them from security attacks.
Next, the most popular security solutions and safeguards to be considered in these systems are classified into four groups, namely secure communications, always-on systems, trust management and data protection.
A summary of the countermeasures discussed below is provided in Table \ref{tbl:sensors:solutions}.

\subsubsection{Secure Communications}

Making communications and networks secure is of utmost importance in s-health systems.
To this end, all entities must transmit encrypted data.

\paragraph{Lightweight cryptography}~

Servers, desktop computers, tablets or smartphones are powerful enough to implement state-of-the-art cryptographic primitives.
However, conventional cryptography is unsuitable for resource-constrained devices, which rely on simpler and faster primitives within the arena of \textit{lightweight cryptography}.

Confidentiality is ensured with encryption.
Regarding symmetric key solutions, different block cipher algorithms have been proposed in lightweight cryptography, such as PRESENT, CLEFIA and LEA, using smaller block sizes and key sizes in comparison to conventional cryptography, where AES is the current standard.
Stream ciphers, although less prominent than block ciphers, have also been proposed in lightweight cryptography, such as Enocoro and Trivium.
Regarding public-key solutions, elliptic-curve lightweight cryptography enhances classical approaches because they can use shorter key lengths and perform less computationally demanding operations to reach the same level of security \cite{lara2018elliptic}.
In parallel and with an eye on the post-quantum era, lattice-based cryptography could gain its momentum in resource-constrained devices \cite{nejatollahi2019post}.

Cryptographic primitives can also provide data integrity assurance by means of hash functions.
Conventional hash functions, such as SHA-2 and SHA-3, might not be efficient enough for constrained devices, and lightweight hash functions using shorter messages and shorter outputs, namely PHOTON, SPONGENT and Lesamanta-LW, have been proposed.
Also, lightweight message authentication codes (MAC), used to verify the authenticity and integrity of messages, have also been defined, such as LightMAC and Chaskey.
All the aforementioned lightweight cryptographic algorithms are standardised under the ISO/IEC 29192.

\begin{table}[t!]
\centering
\scriptsize
\caption{Summary and classification of security countermeasures in smart healthcare systems.}
\label{tbl:sensors:solutions}
\resizebox{\textwidth}{!}{
\renewcommand{\arraystretch}{0.9}
\begin{tabular}{clccc}
\toprule
\multirow{2}{*}{\textbf{Type}} & \multirow{2}{*}{\textbf{Solution}} & \multirow{2}{*}{\textbf{Actor}} & \multirow{2}{*}{\textbf{TCP/IP layer}} & \textbf{Requirements} \\ 
& & & & \textbf{protected} \\ \midrule
\multirow{5}{*}{Secure} & & \multirow{2}{*}{Nodes} & \multirow{4}{*}{Netw. interface} & Confidentiality \\
\multirow{5}{*}{communications} & Lightweight & \multirow{2}{*}{Communications} & & Integrity \\
& cryptography & \multirow{2}{*}{HIS} & & Non-repudiation \\
& & & & Authentication \\ \cmidrule(lr){2-5}
& \multirow{2}{*}{Key management} & Nodes & \multirow{2}{*}{Netw. interface} & Confidentiality \\
& & HIS & & Authentication \\ \midrule
& Secure routing & Communications & Network & Availability \\ \cmidrule(lr){2-5}
Always-on & \multirow{2}{*}{DDoS} & Nodes & & \\
systems & \multirow{2}{*}{countermeasures} & Communications & Network & Availability \\
& & HIS & & \\ \midrule
\multirow{13}{*}{Trust} & \multirow{2}{*}{Authentication} & \multirow{2}{*}{Nodes} & \multirow{2}{*}{Transport} & Authentication \\ 
\multirow{13}{*}{management} & \multirow{2}{*}{protocols} & \multirow{2}{*}{HIS} & \multirow{2}{*}{Application} & Confidentiality \\ 
& & & & Privacy \\ \cmidrule(lr){2-5}
& \multirow{2}{*}{Access control} & & & Authentication \\ 
& \multirow{2}{*}{mechanisms} & HIS & Application & Confidentiality \\ 
& & & & Privacy \\ \cmidrule(lr){2-5}
& & \multirow{4}{*}{Communications} & & Confidentiality \\ 
& \multirow{2}{*}{Intrusion detection} & \multirow{4}{*}{HIS} & Network & Integrity \\ 
& \multirow{2}{*}{systems} & & Transport & Availability \\ 
& & & Application & Authentication \\ 
& & & & Privacy \\ \cmidrule(lr){2-5}
& Traceability of & \multirow{2}{*}{HIS} & \multirow{2}{*}{Application} & \multirow{2}{*}{Integrity} \\ 
& digital evidences & & & \\ \midrule
\multirow{3}{*}{Data} & Privacy protection & \multirow{2}{*}{HIS} & \multirow{2}{*}{Application} & \multirow{2}{*}{Privacy} \\ 
\multirow{3}{*}{protection} & models & & & \\ \cmidrule(lr){2-5}
& Awareness & \multirow{2}{*}{Users} & \multirow{2}{*}{-} & \multirow{2}{*}{Privacy} \\
& programmes & & & \\
\bottomrule
\end{tabular}
}
\end{table}

\paragraph{Key management}~

The security of the communications is guaranteed as long as the cryptographic keys are safe.
These keys need to be generated, exchanged, stored, used and revoked within any distributed system in an efficient and secure way.
Hence, smart health systems must implement robust \textit{key management} services and policies to safeguard the confidentiality of such keys.

More specifically, to generate cryptographic keys, randomness is an important factor to ensure the uniqueness and unpredictability of its calculation.
The implementation of random number generators in resource-constrained environments is challenging due to the hardware and software limitations.
To this end, the implementation of lightweight algorithms is on the rise at the moment \cite{bakiri2018hardware}.
Within the wearables context, several private key generation schemes have been proposed based on the randomness of the proper user-centric attributes, such as the heart rate or the body motion \cite{revadigar2017accelerometer,xu2017gait}.

\subsubsection{Always-on Systems}

Although smart health systems must be uninterruptedly available, increasingly sophisticated availability-oriented attacks make this fulfilment challenging.

\paragraph{Secure routing}~

Smart health systems require communicating the routing information through the network in a time- and energy-efficient way.
Besides, these routing solutions must be autonomous, scalable and adapt dynamically to the network topology since nodes can join and leave the network on-the-fly, \eg a new sensor is introduced, a sensor has crashed, or a sensors has been compromised and removed \cite{anwar2018green}.
\textit{Routing protocols} must also be resilient to attackers who can inject malicious routing information into the network to cause routing inconsistencies and disrupt communications.
Securing routing protocols in WSNs and WBANs could be based on the trustworthiness of neighbour nodes, clustering methods, hierarchical methods or genetic evolutionary techniques \cite{nidhya2019end}.

\paragraph{DDoS countermeasures}~

There exist different types of \textit{DDoS countermeasures}, usually categorised into (i) preventive measures, (ii) detection measures, and (iii) responsive measures.

First, prevention measures are intended to decrease the probability of suffering DDoS attacks.
For instance, filtering-based mechanisms aim to decrease the network traffic by distinguishing between legitimate traffic and attacking traffic, which is dropped.
To do this, the IP traceback method helps determine the true IP origin address of a data packet, rather than its spoofed IP address.
Other methods to filter data packets using different criteria include probabilistic/deterministic packet marking, route-based packet filtering, history-based IP filtering or ingress/egress filtering.

Second, detecting abnormal behaviour can be done by monitoring the systems metrics.
In the event of a DDoS attack, these metrics worsen.
Traditionally, these detection methods were based on statistical analysis, aiming at monitoring the incoming traffic at different time periods.
More recently, many methods based on machine learning techniques, such as neural networks and support vector machines, aim to identify DDoS attacks considering the characteristics of the data packets, namely their size, origin and destination addresses, ports or time intervals \cite{doshi2018machine}.

An third, the detection of DDoS attacks must be immediately followed by a proactive response.
Smart health systems must be designed to be fault-tolerant and limit the DDoS damage.
Scaling hardware resources, queuing techniques or migration-enabling services are some strategies that shall be considered.

\subsubsection{Trust Management}

Smart health systems need mechanisms to be trustworthy, \ie produce reliable and authentic data and communications as well as provide accountability for it.

\paragraph{Authentication protocols}~

Both users and devices must be \textit{authenticated} in s-health systems to prevent disclosing information to unauthorised entities.
For decades, passwords have been the most extensively used method of user authentication.
However, weak passwords or the systematic reuse of passwords are common malpractices that facilitate the task of attackers.
To overcome this, current authentication mechanisms strengthen this procedure by combining knowledge factors (\eg passwords, PINs), possession factors (\eg smart cards, one-time password tokens) and biometric factors (\eg fingerprint, iris scan, facial recognition).
With the advent of smartphones and wearables, users can be continuously being authenticated with the real-time collection of user-centric attributes, namely heart rate, ECG signals or body motion \cite{wu2018continuous}.
Regarding devices authentication, energy and time efficient lightweight protocols are used to overcome the limitations of traditional protocols \cite{bilal2017authentication}.

\paragraph{Access control mechanisms}~

\textit{Access control} establishes a subject-to-object segregation, this is, limits the access of users or devices (\ie subjects) to the resources (\ie objects) of the system.
For instance, patients can only access their own information, physicians can only access the medical information of their patients, and nodes can only have access to the services associated to their task.
To provide scalable and flexible mechanisms, fine-grained access policies are defined to enforce different access privileges to the system's subjects.

Role-based access control (RBAC) models associate each subject to a role, and each role has a set of access permissions.
Therefore, a subject has as many permissions as the role indicates \cite{de2018health}.
Alternatively, cryptography can also help define access control mechanisms in a more secure way with the attribute-based encryption (ABE) based fine-grained access control.
In this scheme, information is encrypted with a set of attributes of the user (\eg department, age, gender\ldots), and only those users fulfilling those conditions are able to decrypt the information and gain access.
Decentralised and private access control mechanisms within an IoT ecosystem can also be achieved with the use of blockchain technology \cite{ouaddah2017towards}.

\paragraph{Intrusion detection systems}~

\textit{Intrusion detection systems} (IDS) are continuously monitoring and analysing the activities occurring within a system in order to alert upon unknown or suspicious malicious behaviour.
These systems can either be network-based, \ie monitoring data packets across the network, or host-based, \ie monitoring all the activities in the end device.
Although they are a mature technology in traditional environments, they might not be adequate at all in context-aware environments yet.
In this context, IDS must be as lightweight as possible to minimise the overhead introduced in the system's infrastructure, in such a way that they do not impact with the proper functioning of the system \cite{zarpelao2017survey}.

The most popular method implemented in IDS is based on anomalies detection.
By properly defining the normal behaviour of the system (captured during a training phase), the real-time activity can be compared against the normal behaviour.
When the distance between the normal and the real-time behaviour exceeds a predefined threshold, an alarm is raised.
This method enables detecting new attacks, but it is susceptible to high false alarm rates because the method's accuracy depends on the completeness of the behaviour captured during the training phase.
Another method is based on signatures, referring to the effects and patterns suffered by a system due to an attack.
By storing all the signatures of known attacks in a database, IDS are able to detect whether the real-time behaviour of the system matches with any stored signature.
Although accurate, unknown attacks remain undetectable in this method.
Hybrid specification-based IDS methods combine the advantage of these two methods.

\paragraph{Traceability of digital evidences}~

When systems are compromised, some attackers footprint might remain in the system in the form of digital evidence.
To report the incident to the judicial authorities and initiate an investigation to prosecute the criminals, the gathering, preservation, integrity, sharing and \textit{traceability} of these evidences must be unquestionable and accountable.
Unfortunately, this procedure is not standardised due to the disparities among jurisdictions, hence endangering the recognition of evidences in courts of justice, especially involving cross-border crimes.
Despite the complexity, a series of international ISO/IEC standards, such as the ISO/IEC 27037, 27041, 27042 and 27043, can help provide guidance about the management of digital evidences.
The correct implementation of these standards may be helpful to build trust and, therefore, foster cross-border cooperation.
Related technological tools, such as LOCARD \cite{lopez2021effective}, will gain momentum in the years to come.

\subsubsection{Data Protection}

Managing personal information arises a number of privacy concerns throughout the entire data life cycle.
Smart health systems must carefully implement privacy by design principles and adopt the adequate privacy-preserving safeguards.

\paragraph{Privacy protection models}~

People's identities cannot be seamlessly associated to confidential data in case of data leakages.
Pseudonymisation, a GDPR-friendly practice, can reduce privacy risks by masking individuals identities with artificial identifiers, called pseudonyms.
This strategy should be considered when communicating user-centric data between two entities across the network, in such a way that eavesdropper attacks are not able to identify to whom the captured data belongs to and only legitimate entities can correlate the pseudonym with the individual's identity.
Besides, once information is stored in the HIS for secondary use, it should undergo an \textit{anonymisation} process, ensuring that third-parties are not able to re-identify individuals identities from the data stored.
To this end, the data transformation techniques and the privacy models explained in Section \ref{sec:background:privprot} shall be considered.

\paragraph{Awareness programmes}~

Organisations should foster \textit{awareness programmes} on cybersecurity and data protection to educate non-expert users.
These programmes must provide high-quality updated information, tips, recommendations and campaigns that users could easily remember and apply in their daily digital habits to prevent or mitigate user-oriented attacks.
For instance, these programmes could be related to phishing, work computer security, mobile device security, best practices on strong passwords or Wi-Fi security.
Unfortunately, these actions are often not applied and, when conducted, they are considered from a very generic perspective.
Latest research is exploring how psychological traits affect users susceptibility to fall into attackers traps, hence, allowing the creation of personalised awareness campaigns \cite{lopez2021human}.

\section{Future Research Directions}  \label{sec:sensors:future}

Despite the steadily adoption of context-aware environments, smart health applications are still constrained to very specific scenarios and cannot exploit all their potential.
Thanks to the latest developments in the manufacturing of IoT and high-speed communications networks, the implementation of these environments will be accelerated in the years to come and, when they are a reality, the ecosystem of smart health will reach a higher dimension.

Significant advancements in the miniaturisation of sensors have opened the door to nanotechnology, able to revolutionise myriad aspects of healthcare and open the door to new frontiers and research opportunities, including disease diagnostics and monitoring, surgical devices, drug delivery and vaccine development, among others.
Although initial nanotechnology-based devices have already been set in the form of ingestible sensors and textile-based wearables, their use is still not generalised due to their manufacturing cost.
Next-generation nanotechnology-based devices may consider smart pills with sensing, imaging and drug delivery capabilities for nanomedicine purposes, nanobots working as miniature surgeons with repairing capabilities of cellular structures, and nanofibers for regenerative medicine.
The optimism about nanotechnology has fostered the Internet of Nano-Things (IoNT) \cite{miraz2018internet}, whose success in smart health will mostly depend on the success to address its security concerns, not only regarding the safety in human lives, but also from the technological side.
The development of security solutions in such technologies will be a major technical challenge.

The increase of mobile devices integrating wireless communication capabilities is adding complexity to the already complex electromagnetic spectrum of context-aware environments.
In these scenarios, the placement of the different devices from WSNs and WBANs can determine the correct performance of the entire system.
Inadequate configurations can dramatically decrease the quality of service of devices operating in context-aware environments, arising severe consequences in sensitive contexts, such as smart hospitals or smart ICUs.
However, the continuous movement of both humans and wireless devices in these settings hinders the analysis of such communications systems.
To this end, radio-planning analyses in terms of coverage/capacity relations, power distribution, potential interference, power delay profiles and delay spread can play a key role \cite{lopez2018implementation}.
These analyses do not only help evaluate appropriate configurations within context-aware environments, but also anticipate and protect networks from malicious interference-based attacks, such as jamming attacks.
Another challenge of next-generation networks is their evolution towards green models \cite{yu2016green}, this is, decrease the energy consumption of current communication technologies, whose intensive demand contributes to an alarming carbon footprint.
Energy-efficient lightweight security solutions will hence be a mandatory requirement in the wireless communication technologies of the future.

There are multiple emerging architectures that can enhance the security robustness of current systems, even though their adoption in the smart health domain is still in an embryonic stage.
On the one hand, zero trust architectures \cite{rose2020zero} offer a security model based on the premise of not trusting any entity until passing a validation, legitimation and authorisation process.
Hence, this model supports the implementation of least privileged access and continually requires the identification of actors who have gained access to the network.
On the other hand, blockchain technology \cite{dasaklis2018blockchain} can make electronic health records more efficient, transparent and secure, and secure information through smart contracts by removing the need for a mediator.
In a near future, this technology could allow organisations creating a secure system to store patient records enabling, therefore, faster diagnosis and interventions to each patient.

Equivalently impressive is the evolution of artificial intelligence (AI) in the last decade, in particular machine learning and deep learning.
Today, with the fast-evolving security threats and attacks, AI-based applications for cybersecurity offer a strategic advantage to thwart malicious endeavours of attackers at contained costs \cite{kroll2021enhancing}.
In particular, many systems, including HIS and other components involved in smart health environments, can enhance their robustness, response and resilience through AI.
For instance, security attacks could be mitigated or defeated autonomously, including zero-day attacks, security countermeasures could be launched on-the-fly according to the attacks severity, honeypots could be generated dynamically, and sophisticated malware and phishing attacks could be efficiently detected.
Although AI can significantly improve security solutions, it is a double-edged sword because it could also facilitate novel sorts of attacks that adversaries might exploit to generate new categories of vulnerabilities and arise unforeseen security threats.
With the aim to make AI reliable for cybersecurity, some development and monitoring practices should be followed, and ethical and legal challenges must be properly addressed too.

\section{Conclusions} \label{sec:sensors:concl}

Information security is a major challenge within smart health.
The high confidentiality of the information together with the large number and variety of actors, devices and resources involved in smart health systems enable the appearance of potential security threats that attackers can exploit for their own benefit.
Moreover, these issues strengthen evermore in sensing devices (\ie nodes), whose constrained nature hinders the possibility to integrate robust security measures.

In this chapter, an exhaustive and technical overview of the deployment of secure smart health services is provided.
First, we have assess the suitability of sensing devices, able to collect user-centric attributes and/or contextual attributes, for smart health according to their accuracy, cost, size and energy consumption.
To convey all these data, specific sensors networks like WBANs and WSNs have been briefly introduced, and the most prominent wireless technology for this domain, such as BLE, ZigBee or 5G networks, have been described in terms of radio coverage, data rate, latency and energy consumption.
Moreover, we have also disserted the information security requirements for these systems, and we have classified the security attacks according to the target actor, this is, nodes, communications, information systems and users.
Security countermeasures, considering the capabilities of the constrained resources and the wireless communications networks have also been proposed.
Finally, we have pointed out some promising future research directions concerning the security issues of smart health systems, namely nanotechnology, green communications, zero trust architectures, blockchain technology and artificial intelligence.

\chapter{Privacy Analysis of Ubiquitous Computing Systems}
\chaptermark{Privacy Analysis of Ubiquitous Computing Syst.}
\label{chap:ubicomp}

\emph{The relevance of ubiquitous computing systems is increasingly growing. Many everyday objects are already ``smart'', being capable of sensing their nearby environment, processing data and communicating outputs to people or other intelligent devices. However, to provide profitable services, they often require information about people, sometimes personal data. In this context, privacy issues may emerge. This chapter presents a novel approach to analyse the privacy risks of ubiquitous computing systems into five dimensions: identity privacy, query privacy, location privacy, footprint privacy and intelligence privacy. Section \ref{sec:ubicomp:intro} justifies the importance of considering privacy aspects within ubiquitous computing systems. The privacy dimensions of the proposed privacy model are described in Section \ref{sec:ubicomp:model}, emphasising the main risks associated to these systems and some privacy-enhancing techniques to avert them. Section \ref{sec:ubicomp:disc} discusses some fundamental challenges in the future of privacy protection in the ubiquitous computing field, by analysing the impact of new technologies and services. Finally, Section \ref{sec:ubicomp:concl} concludes the chapter with some final remarks.}

\minitoc

\section{Introduction} \label{sec:ubicomp:intro}

Ubiquitous computing systems are commonplace in context-aware environments.
Envisaged many years ago \cite{weiser1991computer}, \textit{ubiquitous computing} (also called \textit{pervasive computing}) refers to the provision of computational and communication capabilities to many different (or all) objects in an environment, such as televisions, fridges, cars and watches, among others.
Environments can range from small scales, such as homes, to larger scales, such as hospitals or even cities.
In this context, interactions are not only between people and machines, but also among machines.
These machine-to-machine interactions strengthen the overall sensing capabilities of these objects environment to provide services more effectively.
When these objects are connected to the Internet, IoT emerge.
In a nutshell, ubiquitous computing systems, a generalisation of IoT systems, aim to provide computation to everything and everywhere.

These systems make our lives easier with personalised services and improved user experience, but in exchange for providing (consciously or not) large amounts of personal data, \eg locations, preferences, etc.
Consequently, the storage and processing of these data could jeopardise people's privacy.
Although privacy is not new in the ubiquitous computing field \cite{langheinrich2001privacy,langheinrich2002privacy,duan2004protecting}, determining whether a given ubiquitous computing system is privacy-friendly is not straightforward, since current techniques are based on individual analyses.
Understanding the rationality of the systems, how they work, and why they work that way are key questions that raise from the analysis of their privacy features \cite{schaub2015context}.
Besides, the increasing number and variety of ubiquitous computing systems, together with the new advances in the privacy and security fields, hinder the assessment of the proper management of personal data in all these devices using a unified criteria.

This chapter describes a holistic privacy model to analyse the privacy risks of ubiquitous computing systems from a new individual-centred perspective based on five privacy dimensions, namely identity, location, footprint, query and intelligence.
The compartmental explanation of our model using different dimensions helps to comprehensively understand the nuances of privacy protection in this field.

\section{The 5-D Conceptual Model of Privacy Risks} \label{sec:ubicomp:model}

The proposed privacy model results from the combination of two simpler privacy models, the 3-D conceptual framework for database privacy \cite{domingo20073d} and the W$^3$-privacy model for location-based services \cite{perez2011w3}.
The first addresses database privacy issues into three dimensions related to the main actors involved, namely (i) respondent privacy, \ie preventing the re-identification of individuals, (ii) owner privacy, \ie accessing only to specific subsets of data and revealing only the results of the query, and (iii) user privacy, \ie ensuring the privacy of the queries and preventing user profiling.
The latter describes privacy issues of location-based services according to three independent dimensions, namely (i) who, \ie preventing the disclosure of the identity of individuals, (ii) where, \ie obfuscating the location of individuals, and (iii) what, \ie guaranteeing the privacy of the queries and preventing profiling.
Indeed, the proposed privacy model is an adapted version of the one proposed in \cite{martinez2013pursuit}, which was introduced within the context of smart cities.
Further, more insights regarding the scope of each dimension with regard to individuals privacy are provided, in opposition to organisations’ privacy, which in the original model was called ``owner privacy'' and we have renamed it for the sake of clarity as ``intelligence privacy''.
In a nutshell, our privacy model builds upon five dimensions: (i) identity privacy, (ii) query privacy, (iii) location privacy, (iv) footprint privacy and (v) intelligence privacy.
Next, for each dimension, the definition, risks, countermeasures and practical scenarios within the context of ubiquitous computing systems are provided.
Also, Table \ref{tbl:ubicomp:summary} summarises the dimensions of the proposed model.

\begin{table}[b!]
\centering
\scriptsize
\caption{Summary of the proposed 5-D privacy model for ubiquitous computing systems.}
\label{tbl:ubicomp:summary}
\resizebox{\textwidth}{!}{
\renewcommand{\arraystretch}{0.7}
\begin{tabularx}{\textwidth}{ccccCC}
  \toprule
   \textbf{Our 5-D model} & \multicolumn{3}{c}{\textbf{Mapping to existing models}} & \multirow{2}{*}{\textbf{Definition}} & \multirow{2}{*}{\textbf{Solutions}} \\ \cmidrule{2-4}
   \cite{solanas2021privacy} & \cite{domingo20073d} & \cite{perez2011w3} & \cite{martinez2013pursuit} \\ \midrule
   Identity & - & Who & Identity & Non-disclosure of individuals identities & Pseudonymisers, Attribute based credentials \\  \midrule
   Query & User & What & Query & Privacy preservation of formulating queries and retrieving information & Private information retrieval \\ \midrule
   Location & - & Where & Location & Non-disclosure of the physical location of individuals & Collaboration masking mechanisms, cloaking, proximity-based approaches \\ \midrule
   Footprint & Respondent & - & Footprint & Limit the information retrieved from microdata sets & Statistical disclosure control \\ \midrule
   Intelligence & Owner & - & Owner & Disclosure of joint data when different parties collaborate & Privacy-preserving data mining \\
   \bottomrule
\end{tabularx}
}
\end{table}

\subsection{Identity Privacy} \label{subsec:ubicomp:identity}

Within the context of ubiquitous computing systems, service providers cover people's needs through a variety of added-value ubiquitous-based services.
Generally, to access these services, users must identify themselves to their providers using some identification mechanism that controls who is accessing the services.
Although reasonable from the providers perspective, this may not always be convenient from the users perspective, who may prefer to avoid disclosing their identities.
The identification of users (\eg using direct identifiers) helps improve their experience since it enables to personalise the outcomes of the services in accordance with their preferences.
However, identification mechanisms based on this kind of personal data allow providers to uniquely identify their users and track their use of the provided services.
With user profiling concerns in mind, \textit{identity privacy} refers to the non-disclosure of individuals identities to ubiquitous computing providers when using their services.

When disclosing real identities to service providers, these can create digital profiles with personal data that, in combination with data from multiple providers (or from multiple services offered by the same provider), could infer advanced personal information, such as daily activities, habits and routines.
The more information providers collect and the more services deployed, the more realistic and accurate these digital profiles will become.
In addition to the privacy risks emerged from user profiling, there exist other concerns, such as the trustworthiness of the providers, the purposes of the gathered data and the potential privacy impact in case of data misuse or theft.

One mechanism to preserve identity privacy is by using pseudonyms, which aim to link a certain pseudonym(s) to an individual's identity in a secret, unique and non-trivial way to deduce.
Commonly, the management of pseudonyms is handed over to pseudonymisers (\ie third parties), whose trustworthiness is paramount to ensure identity privacy.
To prevent a single point of failure and therefore improve the privacy-resistance, multiple and geographically distributed pseudonymisers could be considered \cite{perez2009location}.

Increasingly popular, people might be identified by means of the devices they use.
Different risk levels could be presented in certain scenarios.
The riskier situation represents those ubiquitous computing devices that normally belong to a unique individual (\eg smartwatches, fitness trackers, smart glasses\ldots).
To preserve the identity privacy of individuals in this situation, the relationship between each individual and his/her device must be unknown.
This threat relaxes when an ubiquitous device provides services to a controlled group of people, for instance, devices deployed in a smart home or in an autonomous vehicle.
The relationship between individuals and devices should remain unknown though.
However, in this case, if the service identifies the devices, it cannot identify a single individual, since he/she is anonymised within the group.
Therefore, the more people using the same device, the more preserved their identities.
This could be extended to larger systems, such as smart cities in which services are provided to the entire population and the identity privacy of their individuals is practically guaranteed.

Last but not least, attribute based credentials (ABC) are also a suitable solution to protect identity privacy in the ubiquitous computing field \cite{fuentes2018attribute}, specially when a single device is considered.

\subsection{Query Privacy} \label{subsec:ubicomp:query}

Normally, ubiquitous computing systems provide services on demand, this is upon the reception of requests from individuals.
These requests can be seen as queries that users perform to obtain specific services.
Although no direct identifiers are necessarily sent in these queries, they could disclose lots of personal information.
To prevent this, \textit{query privacy} refers to the privacy preservation of formulating queries and retrieving information from ubiquitous computing service providers.

Queries and their results can be easily analysed by service providers unless proper countermeasures are put in place.
By collecting queries from anonymous users, providers could acquire knowledge and infer their habits and preferences, and even be able to re-identify them \cite{adar2007user}.
To avoid having to trust providers, which is sub-optimal, queries should contain the least information possible, following the data minimisation principle.
By doing so, users hinder service providers to learn information, but it can be difficult for them because they are not trained to tune their queries that way.
To solve this, query privacy concerns can be mitigated by using private information retrieval (PIR) techniques.
PIR-based schemes are cryptographic protocols that retrieve records from databases while masking the identity of the retrieved records from the database owners (\ie providers) \cite{yekhanin2010pir}.
By using PIR tools, the correlation between queries and individuals could be broken and profiling becomes much more difficult.
For instance, if an individual wants to query a service provider, a PIR tool could generate a number of $k$ different queries $\{q_1,q_2,\ldots,q_k\}$, in such a way that the provider is not able to determine which is the real query of the user \cite{domingo2009h,wang2018protecting}.

Let's contextualise the query privacy risks in the ubiquitous computing field.
Providers of fitness services could infer the health status, habits and routines of individuals when interacting with their fitness trackers or smartwatches since they collect a large variety of health-related data (\eg physiologic data, exercises, calories intake, etc.).
Furthermore, service providers from smart homes and autonomous vehicles may extract information about daily habits, such as work schedules or sleep routines, of their owners.
More challenging, voice recognition enabling ubiquitous services (\eg voice assistants) could endanger query privacy since devices listen and record the exact query of individuals.
In this case, it would be necessary to guarantee that the signal processing is done on the device, which currently is not the case for most services.

\subsection{Location Privacy} \label{subsec:ubicomp:location}

The ubiquitous computing systems' ability to bring computation anywhere was groundbreaking.
However, this situation may arise some privacy concerns because the physical location of users of such devices (owners) could be known and, then, other sensitive information could be inferred, \eg health-related data, social interactions and religious or political beliefs.
The importance of preserving individuals’ location privacy in the context of ubiquitous computing justifies its addition as an independent dimension to be analysed.
\textit{Location privacy} aims to guarantee the preservation of the physical location of individuals interacting with ubiquitous computing services.

One of the most popular services offered in ubiquitous computing devices is location-based services (LBS), which require location data to provide their services, such as roadside assistance, real-time traffic information or proximity-based marketing \cite{huang2018lbs}.
Besides, many devices, such as smartphones, smartwatches or autonomous vehicles, already incorporate built-in sensors with self-location capabilities, commonly GPS.
Notwithstanding, some service providers often receive location data directly from the individuals.
For example, individuals must disclose their location explicitly to require the weather forecast information or the best route to go a specific location according to real-time traffic.
In other cases, providers could infer the people's locations through proximity data.
For instance, video surveillance systems could identify individuals through face recognition techniques and associate their location to that of the video camera, without the intervention of the users.
Moreover, in the case of autonomous vehicles or smart homes, the location of users is indirectly disclosed since it coincides with the location of the vehicle and the home, respectively.

The sensitiveness of location data fosters the search for solutions that obfuscate this information while preserving functionality.
It is worth emphasising that these solutions might degrade the quality of the results obtained though.
Hence, there is a trade-off between location disclosure and results quality \cite{wang2018privacy}.
Scenarios where the location of the ubiquitous computing systems changes over time, collaboration mechanisms between nearby ubiquitous devices (\ie users) could mask exact locations, so that the location data sent to providers would not directly disclose the real locations.
In addition to collaboration protocols, real locations could also be protected by means of cloaking services or by determining the proximity to entities without revealing their whereabouts \cite{kotzanikolaou2016lightweight,niu2014fine}.

\subsection{Footprint Privacy} \label{subsec:ubicomp:footprint}

Service providers collect and store large amounts of information about users.
This information, mainly used for traceability and analytical purposes, can be seen as microdata sets.
Roughly speaking, these microdata sets contain detailed data about the use and traffic of individuals on their services, this is, the footprint left by the users on the services.
If the microdata sets are published or released to external third parties, they could be able to retrieve meaningful information about such individuals, and privacy concerns may appear.
Even worse, if third parties obtain microdata sets from several service providers used by the same user, further knowledge could be inferred about the individuals' actions.
To address this, \textit{footprint privacy} guarantees that microdata sets collected from individuals' interactions with ubiquitous services are properly sanitised before their publishing or release so as to control the amount of information retrieved or inferred from them.

Whereas users played a major role in protecting their privacy in the aforementioned dimensions, most of the effort to guarantee footprint privacy is handed over to the service providers.
To prevent malpractices, service providers must comply with data protection laws (\eg GDPR), and adopt the proper measures and safeguards when releasing/sharing their microdata sets.
Otherwise, the privacy of these individuals, whose data has been collected for a long time, would be jeopardised.
The adoption of SDC techniques (see Section \ref{sec:background:privprot}) is usually necessary to preserve footprint privacy.

\subsection{Intelligence Privacy} \label{subsec:ubicomp:intelligence}

In the current globalised world, numerous service providers are competing offering similar products and services.
The data collected by each provider is, in most cases, a very valuable asset for these organisations, which enable extracting knowledge and providing user-oriented, personalised and added-value services.
Hence, sharing or releasing these data is not a common practice, especially if competitors could have a chance of taking advantage from them.
However, under certain circumstances, organisations (not necessarily competitors) could take mutual benefit by collaborating, even though they do not want to share their data.
To cover this situation, \textit{intelligence privacy} focuses on the disclosure of information when different organisations collaborate, whilst the actual information of each organisation is not shared or revealed.

Let's exemplify the concept of intelligence privacy using manufacturers of autonomous and intelligent vehicles.
Each manufacturer integrates a lot of built-in sensors around vehicles to gather, store and analyse the status of the vehicle, the nearby environment and further driving-related parameters.
Due to the high sensitivity of these data, manufacturers do not want to share them with their competitors.
However, the collaboration among manufacturers by sharing their data could be extremely important to enhance roads safety and reduce the number of collisions.
Therefore, each manufacturer (even being competitors) would benefit from this collaboration (this is, to obtain joint results), even though they are reluctant to share their intelligence data.
It is worth emphasising that intelligence privacy considers data belonging to companies (\eg the heat dissipated by wheels when breaking).
Thus, data collected by organisations but belonging to individuals should not be considered under this dimension because they belong to the users, and not to the companies, so they should be managed as such.

In this situation of mutual distrust, privacy-preserving data mining (PPDM) techniques emerge as the natural solution to protect intelligence privacy \cite{agrawal2000ppdm}.
PPDM methods are applicable once independent entities want to collaborate to obtain common results that benefit both of them, but without sharing their data.
By applying these methods to the queries submitted across several organisations' databases, the amount of information transferred to every party is controlled and does not pose risks on revealing the original data, but only the joint results.

\section{Discussion} \label{sec:ubicomp:disc}

Ubiquitous computing systems are so intricately fused with our daily lives that they must play a key role in the protection of people's privacy.
In these systems, privacy has been considered from a data-oriented perspective for a long time.
In this sense, data have been seen as a valuable asset that belong to who has the ability to collect, store and exploit them, so privacy protection has been mixed with related problems, such as access control, network and database security, and alike.
As privacy issues mainly affect people, the focus must be put on the people and, from there, rethink the whole architecture that aims at protecting it.
In the recent years, we are already witnessing a shift towards an individual-centred privacy, in which the focus of the technology and the privacy are the users.
Therefore, to protect privacy, the personal dimensions of users must be understood.
In this chapter, those dimensions respond to questions such as ``Who am I?'' (identity privacy), ``Where am I?'' (location privacy), ``What do I need?'' (query privacy) and ``What have I done?'' (footprint privacy).
This paradigm shift is particularly relevant when considering the impact of wearable devices, in which most the data generated are personal and sensitive.
Consequently, protecting the binding between these data and users considering the aforementioned dimensions is paramount.

With the recent enforcement of GDPR, several challenges related to privacy protection appear when managing ubiquitous computing systems.
GDPR reinforces the conditions for processing personal data, which is only allowed when individuals give explicit and informed consent for such processing according to some well-defined and unambiguous purposes and uses.
Contrary, individuals also have the right to withdraw this consent easily and at any time, thus denying the future processing of these data.
Within the ubiquitous computing field, these requirements are challenging, especially when deploying ubiquitous computing systems in public environments (\eg in a smart city), whose purposes of processing might not be already clear at the very beginning.
Also challenging is the right to data portability, in which individuals can obtain their data in a structured and interoperable format, due to the heterogeneity of ubiquitous computing devices as they can return a wide spectrum of information, ranging from health data, wearable trackers and opinions, to even biometric and financial data \cite{urquhart2018realising}.
Moreover, the widespread deployment of ubiquitous computing systems opens the door to attacks and data breaches wherein people's data is compromised (see Section \ref{sec:sensors:security}).
As organisations are enforced to communicate these breaches to supervisory authorities within 72 hours, service providers should permanently keep track of their systems seeking for abnormal behaviour that could compromise personal data \cite{davies2016data}.

Thanks to the privacy by design principle, already introduced in GDPR, the design process of (ubiquitous computing) systems needs to be redefined in most cases.
This is particularly important when building ubiquitous computing systems that surround us all the time.
Therefore, emerging technologies based on context-awareness must be designed with privacy in its core, and not as an additional layer out of the systems.
This can easily be observed with the evolution of face recognition technology.
In the early days of this technology (and similar ones), biometric information was sent to servers over the Internet, analysed, and the authentication results was sent back to the edge device.
Clearly, this procedure had many points of failure from a privacy perspective.
Currently, most of these technologies are directly executed on the devices (\eg smartphones) and, as a result, the data is privately stored by the user and privacy risks are relaxed.

Last, privacy risks related to ubiquitous computing systems affect us all.
Regardless the technological and legal solutions to reduce these issues, it is paramount that people foster awareness on their privacy when using these systems and realise about the importance of their privacy in the today's technological world.

\section{Conclusions} \label{sec:ubicomp:concl}

Privacy is one of the most fundamental human rights.
Unfortunately, privacy aspects are sometimes still too fragile.
In particular, this fragility is observed in ubiquitous computing systems, which have been intricately fused with our daily lives for years.

With the aim to improve people's awareness, this chapter has provided an overview of the privacy issues that might arise as a result of the generalisation of ubiquitous computing systems in context-aware environments.
We have proposed a novel five dimensional framework for the analysis of privacy protection from an individual-centred perspective, in opposition to older approaches centred on data.
In our model, we focused on four dimensions related to individuals, \ie identity, query, location and footprint privacy, and a fifth dimension related to companies, \ie intelligence privacy.
This high-level model of privacy dimensions will help researchers and practitioners to approach the difficult problem of analysing individuals privacy in a more comprehensive way.
Besides, we analysed some of the main trends affecting privacy that should be changed in the coming years, namely the consolidation of the privacy by design approach, the paradigm shift from data privacy to individuals privacy, and the growing importance of legislation.

\chapter{Security and Privacy Analysis of Cognitive Cities}
\label{chap:cogcities}

\emph{Cognitive cities are gaining momentum as a promising solution to the challenges of future mega-cities. This emerging paradigm augments smart cities with learning and behavioural change capabilities. Cognitive cities are built upon artificial learning and behavioural analysis techniques founded on the exploitation of human-machine collective intelligence. Hence, they rely on the sharing of citizens' daily-life data, which might be sensitive. In this context, information security and privacy become critical concerns that must be addressed to guarantee the proper deployment of cognitive cities along with the fundamental rights of people. This chapter identifies the main challenges, opportunities and actionable solutions in the field of information security and privacy for cognitive cities. First, Section \ref{sec:cogcities:intro} conceptualises the paradigm of cognitive cities. Then, the challenges and opportunities in the field of study are described in Section \ref{sec:cogcities:chall_opport}, by classifying them into three categories: technical, societal and regulatory. Then, Section \ref{sec:cogcities:future} foresees some of the most important research directions in the future. Finally, the chapter closes in Section \ref{sec:cogcities:concl} with some final remarks.}

\minitoc

\section{Introduction} \label{sec:cogcities:intro}

Smart cities, one of the most popular implementations of context-aware environments, rely on the assumption that gathering vast amounts of data would (hypothetically) enable better decision-making.
However, efficiency sometimes contradicts resilience, so ICT alone are not enough to build liveable cities.
The latest developments in artificial intelligence, IoT and ubiquitous computing, along with connected learning theories such as connectivism \cite{siemens2017connectivism}, have allowed the emergence of the new augmented urban paradigm of \textit{cognitive cities}.
Introduced in \cite{mostashari2011cognitive}, cognitive cities learn and adapt their behaviour based on past experiences, and they are able to sense, understand and respond to changes in their environment.
Hence, cognitive cities become more sustainable and resilient by exploiting the collective intelligence of the city, this is, learning from the information flows circulating among agents (either humans or devices) and the city, and adapting as the environment changes as well.
Unlike smart cities, cognitive cities use other sources of information, besides technological sources, to sense their conditions: for instance, cultural, behavioural, spatial and political information can be used to make up the personality of the city.
Moreover, whereas smart cities are mainly reactive, providing an efficient response to incoming data from nearby sensors, cognitive cities are proactive as they are capable of generating hypothesis about the future and evaluating the possible consequences of their decisions.

Figure \ref{fig:cogcities:concept} depicts a scenario that highlights the features of cognitive cities.
Assume that an intelligent vehicle has suffered a blowout, remaining stopped blocking the lane, and is unable to notify its state to nearby agents because the communication capabilities of the vehicle have been affected due to the incident (number \ding{192}).
A nearby citizen reports the incident by sending an alert with his/her smartphone, which is forwarded to other nearby agents, such as other citizens, vehicles and smart devices (number \ding{193}).
Upon the reception of the alert, the closest traffic light (\ie an agent) adapts its behaviour/role and coordinates with other nearby traffic lights to rearrange their lights patterns, so that traffic could continue to flow without major disturbances (number \ding{194}).
Moreover, the built-in camera of the traffic light records details about the incident (\eg location, pictures\ldots) and sends an alert to emergency services if needed.
In such case, the traffic lights adapt their light patterns to prioritise the arrival of the emergency services to the incident (number \ding{195}).
In the midterm, after several blowouts in the same location, agents send a collective report to the authorities so they could be aware of the numerous incidents in that area.
With this process, the cognitive city learns that the area is prone to vehicles blowouts, and recommends to asphalt the pavement again so as to prevent further incidents in the future (number \ding{196}).

\begin{figure}[b!]
\includegraphics[width=0.95\textwidth]{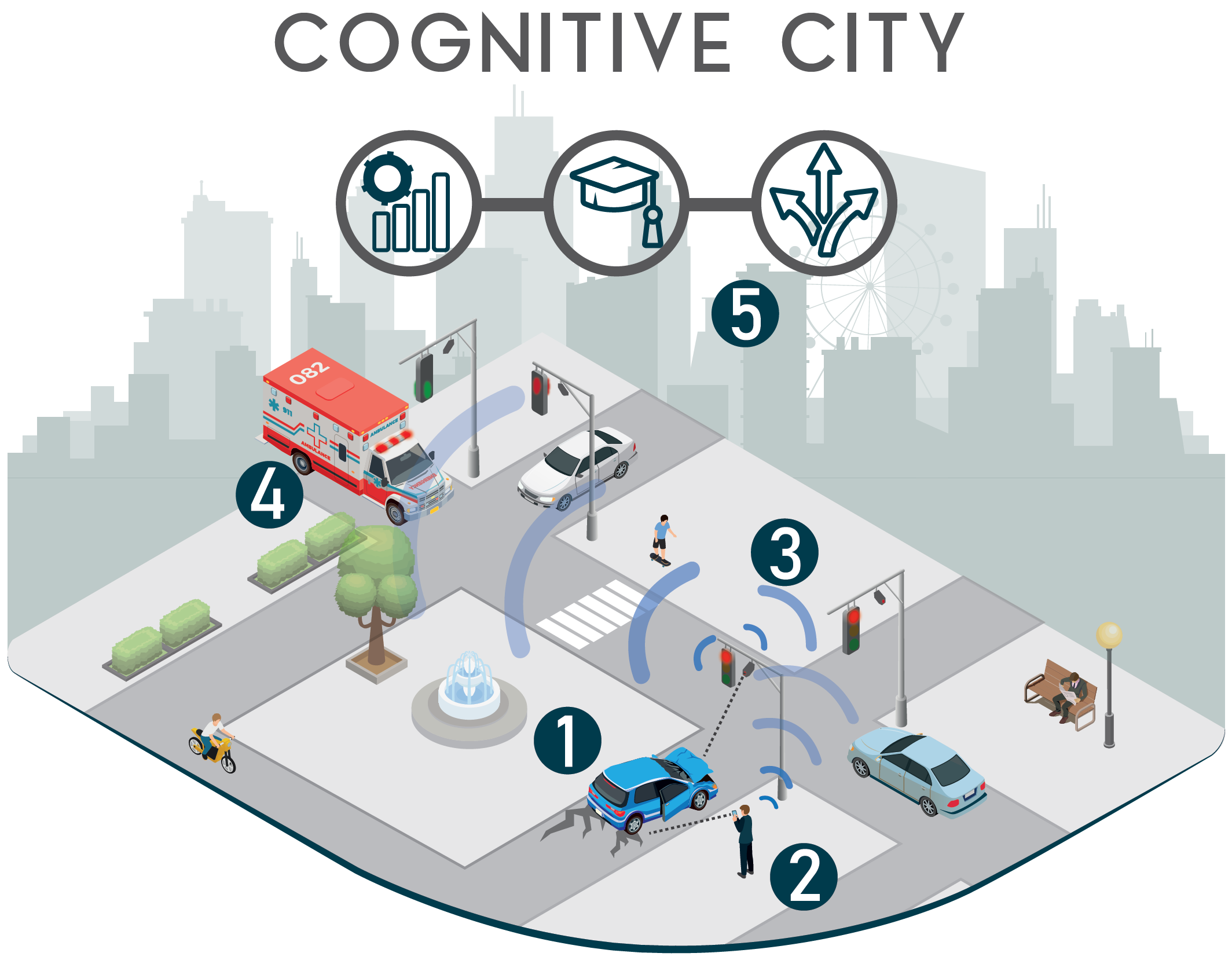}
\caption{Scenario of a cognitive city (reprinted from \cite{machin2021privacy}).}
\label{fig:cogcities:concept}
\end{figure}

Cognitive cities use citizens' data to make decisions and influence their daily lives.
The ubiquitous collection of these data, the availability of diverse urban-wide data sources (\eg transportation, healthcare, energy, surveillance cameras\ldots) together with the inferential capabilities of cognitive systems introduce serious threats from a privacy perspective, such as personal information disclosure, information appropriation and citizens' profiling.
Also, these consequences are not only at the individual level, but also at the societal level, opening the door to other threats, such as over-surveillance, power imbalance and citizens manipulation in favour of governments, corporations, or whoever is in control of the city data.
From a security perspective, failures or malfunctions in cognitive systems can paralyse the city activities, leading to loss of confidence, economical damages, or harm (or even lose) citizens' lives if mission-critical infrastructures, such as healthcare, become affected.
Therefore, infrastructures need to be prepared to face DDoS attacks, ransomware, integrity-threatening attacks and other disruptive attacks, and implement the proper security protection countermeasures.
Furthermore, since cognitive cities rest upon the active sharing of citizens' data, privacy challenges become paramount.
All these challenges must be addressed as soon as possible to make the development of cognitive cities a reality.

This chapter explores the current landscape of the field of privacy and information security in cognitive cities.
The identification and classification of the main challenges along with the discussion around the opportunities to address them help provide a holistic view of the current panorama in this field and foresee the most important future research directions.

\section{Challenges and Opportunities} \label{sec:cogcities:chall_opport}

In the urban environments of the future, cognitive and smart systems would become critical because they will make decisions on highly-valuable assets from key sectors.
Given that cognitive cities will be fed by a constant flow of information from cognitive agents, potential security and privacy issues could emerge.
In this section, we elaborate on the challenges and opportunities of cognitive cities from their security and privacy perspective.
For the sake of clarity, they are classified into three categories: (i) technical aspects, (ii) societal aspects, and (iii) regulatory aspects.

\subsection{Technical Aspects}

Most of the challenges identified are related to the \textit{technical} aspects of implementing and deploying cognitive cities, ranging from the very sensors and actuators to the analysis of aggregated data.
Below, a list of the main technical challenges and opportunities is provided.

\begin{itemize}
    \item Privacy and security are interwoven, and one can hardly be achieved without the other. If technology fails, then privacy might be compromised. In any case, although privacy by design principles are a necessary starting point, they are not enough if cognitive cities are insecure from the information security perspective. Hence, building strong security measures is first and foremost. Classical integrity checking mechanisms, access control systems, and authentication protocols, among others, might need to be redefined to fit with the numerous and heterogeneous cognitive agents surrounding cognitive cities. For instance, the distributed nature of blockchain technology could ease the management of credentials verification and legitimacy, in contrast to current public-key authentication infrastructures based on certificate chains, revocation lists and multiple-sourced trust.
    \item Citizens interact with cognitive cities through cyber-physical systems. These physical interfaces need to be protected from adverse environmental conditions (\eg air temperature, air humidity, wind, rain or air pollution) or malicious citizens actions (\eg vandalism, robbery or sabotage). Further, since information security issues could lead to physical consequences, cognitive systems should be equipped with remote deactivation capabilities (\eg secure kill-switches devices) to prevent hazardous situations for the citizenship. All in all, protecting the physical layer is not trivial, and numerous protection mechanisms to detect, prevent, resist and react to a wide range of physical threats should be considered. Besides, these mechanisms need to be periodically revisited over a continuous risk assessment procedure.
    \item Devices should comply with a number of strict requirements to develop the full potential of cognitive cities. First, on-device intelligence is required. Second, these agents need to be resilient and operate autonomously. Third, agents also have to be flexible and provide adaptive responses to address the ever-changing conditions of the real world. Fourth, self-awareness, \ie the ability to sense their own state, is another property that infrastructures supporting cognitive cities will require to achieve real resiliency. This characteristic will allow agents to prevent, detect and react autonomously to any event that may affect their proper functioning and alert, if needed, external meta-cognitive agents that could manage these disturbances accordingly. Similarly, self-reconfiguration and self-healing are further desirable properties to recover from destructive events. However, all these requirements imply a computational and power capabilities that most devices cannot afford nowadays.
    \item Citizens and cognitive systems in the city are expected to learn from aggregated streams of information. Hence, it is fundamental to guarantee the integrity and the authenticity of the data. Machine learning algorithms imply a number of security-related challenges. On the one hand, cognitive systems must be able to detect and resist adversarial machine learning attacks, in which attackers supply deceptive inputs \cite{joseph2018adversarial}. On the other hand, these algorithms should also be thoroughly tested to avoid the so-called ``illusion inertial thinking'' problem, consisting in applying concepts learned from one problem (\ie testing samples) to another significantly different problem due to poor generalisations and leading to wrong conclusions \cite{li2020modeling}. This issue is important because the accuracy of machine learning methods depends on large number of training samples that may not always be available. Even worse, if these wrong conclusions are passed onto another system as inputs for a new cascading learning process, the deviation from the desired behaviour can be considerable.
    \item Lots of different technologies converge and interact with each other within cognitive cities. Hence, legacy software and hardware might be a problem unless they are properly maintained. Security vulnerability management processes must be thoroughly planned before deploying any technology, and periodically executed through all its life-cycle to detect security or privacy risks. Moreover, managing obsolescence is important to identify discontinued or simply obsolete ICT components beforehand, and replace them by newer and/or patched equivalent solutions. Mixing older and newer components will also require in-depth studies to guarantee interoperability, avoid leakages and uncover hidden security risks or privacy threats.
    \item The distributed and heterogeneous nature of cognitive cities creates a huge attack surface that is hard to protect. Efforts must be devoted to coordinate and efficiently harmonise the functioning of diverse technologies and devices, by fostering the creation of international standards and certifications on the basis of real-life urban scenarios. This certification process would be required along the life-cycle of any cognitive system, and would complement law enforcement by transnational regulations. To this end, a global Security by Default framework particularly designed for cognitive cities is encouraged, so that it (i) provides a formal description of entities (\ie ontology) that constitute the cognitive cities, (ii) is based on a multi-layered model from technological layers to social city-level layers, (iii) encompasses all the stages in the life-cycle of cognitive products, processes and services for the cognitive city (from design to monitoring stages), (iv) considers the potential security and privacy risks that might arise in every phase and architecture layer, (v) includes new threat models for cognitive cities, and (vi) includes metrics and risk assessment models to evaluate security and privacy at several levels according to the multiple stakeholders.
\end{itemize}

\subsection{Societal Aspects}

In addition to the aforementioned technical challenges, there are a number of \textit{societal} aspects that also play a key role in the success of cognitive cities.
A list of the main societal challenges and opportunities is provided below.

\begin{itemize}
    \item Citizens may be reluctant to share their information with cognitive cities. Given the growing awareness of citizens on preserving their privacy, service providers may incorporate privacy as an added valuable feature and a competitive business advantage. Hence, to make cognitive cities real, citizens should perceive fairness in the collection and processing of their personal data, and understand the benefits that technology could provide to them. Hence, automatic data processing must strictly adhere to clear transparent practices so as to foster trust. As a starting point, it would be ideal that citizens could make real-time interactive control over their personal data. However, the ubiquitous nature of data acquisition in cognitive cities hinders this situation. All in all, efforts should be made to narrow the digital divide, so citizens understand how they are interacting with the always-on technologies of cognitive cities.
    \item The unprecedented intelligent mechanisms of cognitive systems could be perceived as threats to the civil liberties and privacy. Citizens profiling, the potential erosion of freedom of choice, the resemblance with surveillance systems (``big brother'') and the potential development of manipulation tools are issues that all stakeholders should properly clarify.
    \item Cognitive cities should adapt their physical components and their interactive capabilities to all members of the society, including elderly, unskilled people and impaired citizens.
    \item Information security incidents may have serious consequences in cognitive cities. Cognitive cities should be equipped with monitoring capabilities to detect and respond to security threats and anomalies while protecting their infrastructures. Besides the technological and procedural aspects, attention should be paid to the (often undervalued) human factor \cite{nobles2018human} since talent shortage, human errors or copying behaviour can affect the mission of any organisation. Continuous risk management processes of human behaviour-based risks should be assessed. Moreover, new promising security concepts, such as cognitive security \cite{andrade2019cognitive}, might be considered.
    \item Both the number and cost of incidents caused by insiders is increasingly growing \cite{ponemon2020insider}. Besides external attacks, cognitive cities' stakeholders should also consider the chance of infractions committed by employees or contractual personnel responsible for the design, management, operation or maintenance of the infrastructures and services. Insider threats could be mitigated under the framework of behavioural theories, by privately monitoring human interactions, detecting tendencies shifts and developing accountable practices. By lowering the insiders' perceived benefit or increasing the perceived cost of violation, many risks would be minimised.
    \item Beyond technical aspects, human factors leading to attacks should be addressed. To this end, fostering engagement and educating citizens on cybersecurity, as well as raising awareness on good security and privacy practices is primary.
\end{itemize}

\subsection{Regulatory Aspects}

Last but not least, there are some \textit{regulatory} aspects that must be taken into account once implementing any cognitive system.
A list of the main regulatory challenges and opportunities is provided next.

\begin{itemize}
    \item Data protection regulations (\eg GDPR) make consent and awareness mandatory prior to data processing. This safeguard does not only apply to citizen-generated data, but also to derived secondary data inferred from prior citizens' information. In cognitive cities, as data might be gathered opportunistically, it might be hard to balance the fulfilment of legal obligations with the development of innovative services. To this end, technical and regulatory mechanisms should be developed to boost user awareness and guarantee consent across the ubiquitous landscape of cognitive environments.
    \item Legal complexity could become intractable due to the vast diversity of technologies and scenarios that may arise in cognitive cities. For instance, a supplier of a cognitive city will be accountable under the laws of that city country, but these laws may be in conflict with those of the supplier's country. To improve law enforcement across different jurisdictions, although complex, it could be interesting to develop inter-operable markup policy languages and protocols, so devices could automatically exchange data protection policies.
    \item Cognitive cities may become targets of cybercrime and terrorism, so liability emerges as a serious concern. Capturing, storing and processing digital evidences of criminal activities conducted in cognitive systems might be feasible by means of international standardised procedures and techniques. To this end, blockchain could become a promising technology to build a transnational trust system for digital evidences at the same time that ensures the chain of custody along the forensic workflow \cite{lopez2021effective}.
\end{itemize}

\section{Future Research Directions} \label{sec:cogcities:future}

The previous analysis demonstrates that there is still a difficult stretch of the road ahead to have secure and private cognitive cities.
Hence, some of the most promising research lines that deserve the attention from the scientific community are enumerated below.

\begin{itemize}
    \item Define and balance open and comprehensible data policies, which are needed for fostering citizens' participation, with strong and state-of-the-art privacy-preserving mechanisms.
    \item Develop techniques to distinguish AI-generated contents from human-genera-ted contents in order to build reliable data-driven policies for government innovation.
    \item Balance the fulfilment of user consent and the ubiquitous nature of cognitive environments in the city.
    \item Introduce self-awareness capabilities into autonomous agents to make them able to detect and recover from outages.
    \item Develop information security standards on cognitive cities.
    \item Enhance existing models to achieve citizens' privacy in the cognitive city environment.
    \item Develop analytical and summarising techniques based on machine learning techniques to detect and respond to events impacting on the cognitive city.
    \item Develop locking mechanisms to guarantee the integrity of the information, either original data or derived secondary data inferred from prior information, such as based on blockchain technology.
    \item Standardise mechanisms and procedures to provide worldwide incontrovertible digital evidences, whilst ensuring their chain of custody, in order to prosecute criminal offences.
\end{itemize}

\section{Conclusions} \label{sec:cogcities:concl}

Information security and privacy protection are open issues yet to be solved in the field of cognitive cities, an emerging urban paradigm evolved from the well-known smart cities but augmenting them with connected learning theories.
Unlike smart cities, cognitive cities are proactive, so they are capable of learning from past events and automatically adapting their future behaviour to become more efficient, resilient and sustainable.
Unfortunately, insufficient efforts have been devoted to this promising and young field of study.

In this chapter, we have outlined the most relevant challenges and research lines related to security and privacy that still remain open.
More specifically, these challenges have been classified into three categories: technical, societal and regulatory.
This taxonomy will help researchers to rapidly identify the main concerns and propose actionable solutions to them.
Besides, information security and privacy are dynamic ever-changing issues and, since cognitive cities are still ahead in the future, it is hard to imagine what new challenges will bring to our cognitive cities in the future.
Nevertheless, by already addressing the concerns highlighted in this chapter, when cognitive systems begin to interact with citizens on a daily basis, new unforeseen privacy and security problems will arise.
Only acting in advance, we will be able to avert these threats and guide the development of sustainable and resilient cities where technology is at the service of citizens.

\part{Process Mining} \label{part:pm}
\chapter{Process Mining Meets Context-Aware Environments} 
\chaptermark{Process Mining in Context-Aware Environments} 
\label{chap:pmcae}

\emph{With the unstoppable deployment of context-aware environments, massive amounts of events are generated and stored by distributed, heterogeneous and highly uncoupled devices able to sense their surroundings and report precise and timely updates on the environment. By exploiting recently available contextual information, automatised process mining analyses could generate novel opportunities to increase the efficiency of systems operating in context-aware environments. This chapter discusses the benefits, opportunities and challenges of bringing process mining in this new, events-rich, context-aware environments. Section \ref{sec:pmcae:intro} motivates the reader with the importance of considering contextual data in today's analyses. The benefits and opportunities arising from the combination of process mining and contextual data gathered from context-aware environments is described in Section \ref{sec:pmcae:opportunities}, by coining the paradigm of context-aware process mining. Next, Section \ref{sec:pmcae:challenges} enumerates the main difficulties and challenges to be faced. Finally, Section \ref{sec:pmcae:concl} closes with some concluding remarks and future directions.}

\minitoc

\section{Introduction} \label{sec:pmcae:intro}

Classical process mining analyses are fed by event logs automatically generated by information systems within organisations, such as enterprise resource planning systems, expert systems, management information systems and systems alike.
With these data, organisations apply process mining techniques periodically to monitor the execution of their business processes.
When unexpected behaviours or deviations are found, managers apply corrective measures on the processes to redress their execution towards the desired behaviour.
Since event logs describe historical facts, the time lapse between the anomalies and their correction could be high though.
During this period of time, the provision of services might have been far from optimal, service times might have increased and resources might have been wasted.
Due to the high competitiveness in today's market, this fact could lead to an irreparable loss of customers, revenues and reputational damage.

Fortunately, events are no longer generated by this kind of systems only, but also by most of the computerised systems that we interact with everyday.
With the advent of the IoT, more and more devices form part of our daily lives, and all these interactions generate events that could be exploited as well.
Despite data heterogeneity, the growing significance of events has allowed coining the term of the \textit{Internet of Events} \cite{aalst2014scientist}.
This term categorises all available event data into four dimensions:
(i) the Internet of Content, referring to all information generated by humans to increase knowledge on particular subjects, including webpages, articles, e-books and multimedia,
(ii) the Internet of People, related to people's social interactions, such as e-mails, forums and social networks,
(iii) the Internet of Things, considering data generated by physical devices connected to the network,
and (iv) the Internet of Locations, referring to all data with a geospatial dimension, such as those data created from mobile devices with self-location capabilities (see Figure~\ref{fig:pmcae:ioe}).

\begin{figure}[t!]
\centering
\includegraphics[width=0.72\textwidth]{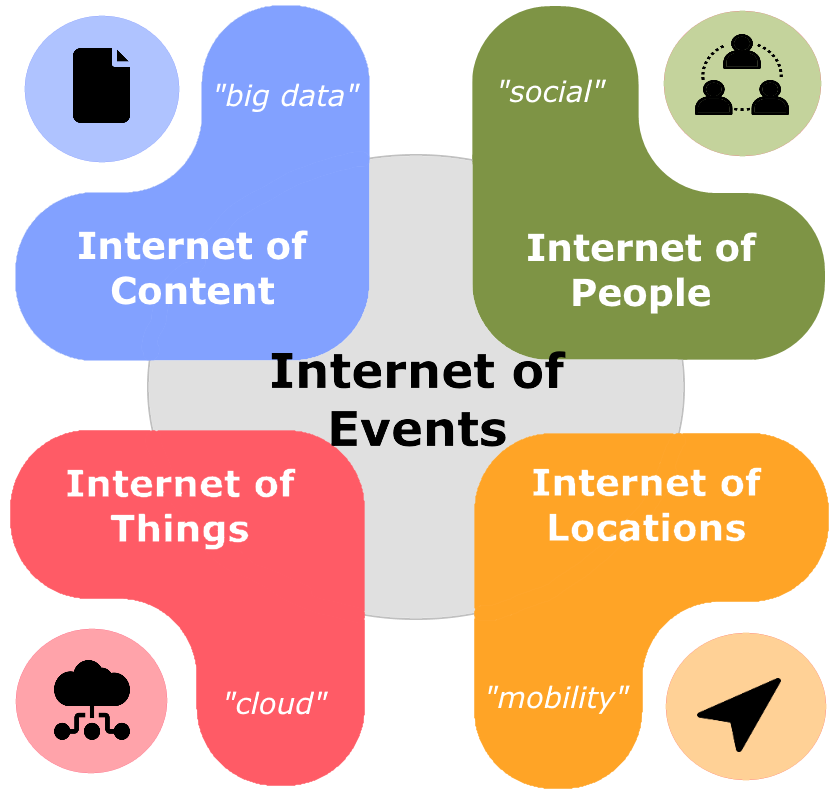}
\caption{The Internet of Events is based on the Internet of Content, the Internet of People, the Internet of Things and the Internet of Locations.}
\label{fig:pmcae:ioe}
\end{figure}

The adoption of sensor-equipped intelligent environments, such as context-aware environments, raises the potential of process mining to a higher dimension.
In this sense, process mining would not only consider event data generated by the execution of processes, but also by all sorts of event data generated by external devices surrounding the execution of such processes.
Real-time contextual event data enables \textit{process contextualisation}, this is, understanding processes in the context where they are being executed.
This information could be used to assess the impact of the environment on the processes.
Highly dynamic and complex environments, such as healthcare-related environments, can benefit from this contextualisation to acquire hidden knowledge.

With the imminent adoption of the Internet of Events, this chapter envisages the benefits that process mining could play in these context-aware environments.
Moreover, the chapter describes the main challenges to overcome and provides the fundamental food-for-though about future research opportunities in the field.

\section{Context-Aware Process Mining} \label{sec:pmcae:opportunities}

Business processes are generally designed to operate optimally under ideal or controlled settings.
However, upon certain changes, processes may not operate as efficiently as they should, hence worsening their functioning and resulting into undesired outcomes.
Without further ado, the recent COVID-19 pandemic has definitely changed the patients care processes executed in hospitals, emergency rooms, intensive care units and primary care facilities.
The increment of the hospitalisation pressure together with longer hospitalisation stays required realigning the care processes before reaching overcrowded situations.
Process mining analyses played a fundamental role to detect critical workflows, identify deviations and organise the resources available in the most efficient way \cite{chang2020impact,pegoraro2021analyzing}.
In this situation, decisions and countermeasure actions were taken using (historical) event data of the care processes themselves from the first days/weeks/months of the pandemic, which implies having a worrying time lapse.
Besides, the high variability of processes in these environments implies investing many efforts to analyse, monitor and redefine these processes periodically.
Unless properly detected, serious long-term consequences may arise.

The growing implementation of context-aware environments leads to the collection of massive amounts of real-time data, treated as event data, describing contextual settings.
These data could be generated by a wide range of devices with plenty of sensing capabilities, such as wearable devices (\eg smartwatches and fitness trackers), smartphones, beacons, voice assistants, smart appliances, video cameras and further IoT devices ranging from presence detectors to environmental sensing devices, among others.
Considering the event data generated from the execution of processes in context-aware environments, it could be augmented with contextual event data, thus opening the door to \textit{context-aware process mining}, a proactive paradigm for conducting process mining analyses within context-aware environments (see Figure \ref{fig:pmcae:concept}).
As a result, the provision of services could modify its own behaviour in real-time and adapt itself to the current contextual settings.
Service models would hence be more flexible, intelligent, efficient and sustainable.

\begin{figure}[t!]
\centering
\includegraphics[width=\textwidth]{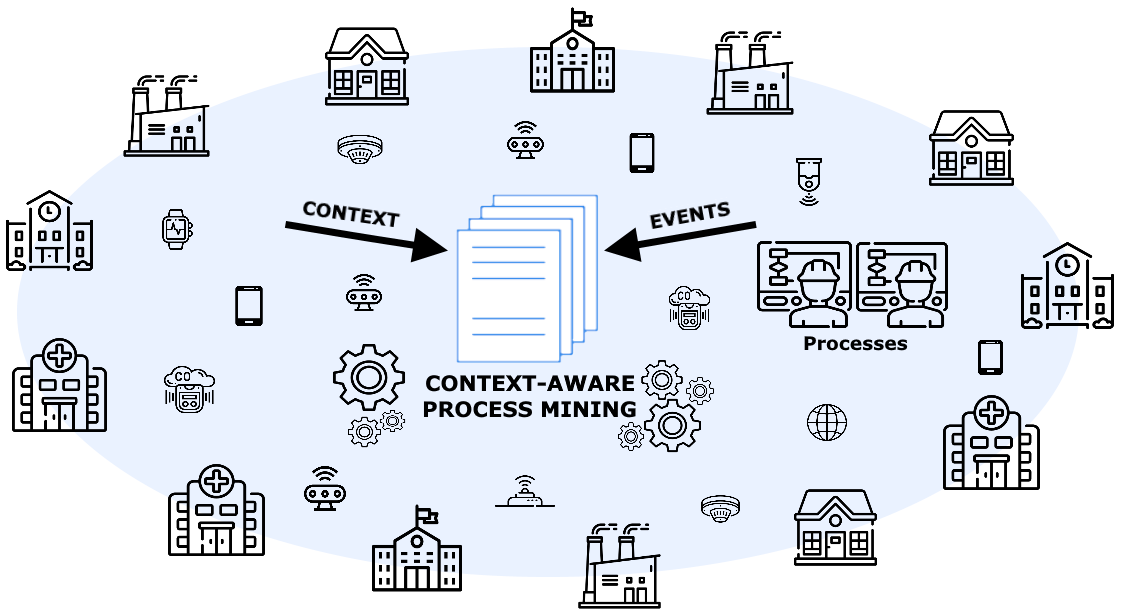}
\caption{The concept of context-aware process mining: proactive process mining analyses using event logs comprising contextual data from context-aware environments (adapted from \cite{batista2019pmcae}).}
\label{fig:pmcae:concept}
\end{figure}

Under this paradigm, the different types of process mining techniques entail promising opportunities.
Given an event log of a certain process executed in a context-aware environment, by means of process discovery techniques, one could retrieve, not only the actual behaviour of the process, but also augment it with context-related annotations, that describe under which conditions the process has been executed.
Process discovery techniques could also represent the behaviour of processes according to the contextual settings, as they could affect the processes executions.
This would ease the comprehension of the processes executions and, moreover, would enable discovering hidden relationships between processes and their immediate context.
From the conformance checking perspective, these techniques could study whether the context has any influence on how processes should behave and how processes do actually behave.
If so, the resulting diagnostics would be able to detail in which contexts the processes perform more (or less) similar to the ideal executions defined by the model.
More importantly, process enhancement techniques can play a major role to provide context-dependent up-to-date processes.
These techniques would be able to make real-time decisions to realign processes and be more efficient according to the current context.
Consider the provision of a service, defined by the same process, in two different environments.
Due to the influence of the context, it is likely that the process might behave better in one environment than in the other.
For example, a healthcare institution could find out that the (same) process responsible for treating asthma turns into better outcomes (\eg less costs) in facility A rather than in facility B, because the context of facility A has a more favourable weather and less pollution rates.
In this case, even though the service provided and the business process are exactly the same, different outcomes for the (theoretically) same service are obtained.
By automatically readjusting the processes to the contextual settings, service providers may increase their quality and sustainability.
The enhancement of processes can be benefited from the learning strategies adopted in cognitive environments so as to learn from previous experiences and, therefore, adapt these processes to novel contexts.

\section{Challenges} \label{sec:pmcae:challenges}

The proper contextualisation of business processes in context-aware environments by means of process mining implies a number of challenging considerations.
Some of the most important challenges are discussed as follows.
For the sake of clarity, they have been grouped into four categories, namely (i) data challenges, (ii) ecosystem challenges, (iii) computing challenges, and (iv) automation challenges.

\subsection{Data Challenges}

Process mining in context-aware environments does not only consider event data resulting from the historical execution of business processes, but also new sets of contextual event data from the sensing infrastructure of context-aware environments.
As these new data sets are external to the processes executions, several \textit{data} challenges may emerge.

\begin{itemize}
    \item \textit{Data quality}: The event data generated by the sensing infrastructure of context-aware environments must be true, reliable and trustful. Erroneous and noisy data (\eg from broken or wrongly calibrated sensors) must be detected and discarded prior to their analysis with process mining.
    \item \textit{Data authenticity}: The development of methods to ensure the integrity of the event data as well as detect whether data have been malicious tampered is fundamental. The detection of fraudulent data is important to guarantee the robustness and reliability of process mining results.
    \item \textit{Data privacy}: Contextual event data might contain sensitive information, such as people's health conditions or locations. It is important to properly define procedures and protocols to deal with this kind of data, so as to prevent the disclosure of personal information that could jeopardise people's privacy. Defining clear and transparent policies is a necessary first step.
\end{itemize}

\subsection{Ecosystem Challenges}

Context-aware environments involve complex management and coordination tasks due to the large and diverse amount of agents forming part of this \textit{ecosystem}.
Among others, two of the most important challenges are as follows.

\begin{itemize}
    \item \textit{Timestamp synchronisation}: Within context-aware environments, event data are generated by a variety of sources, and each of them has its own internal clock. Multi-sourced timestamp synchronisation techniques are required to aggregate and chronologically order all these events. This step is needed to ensure the quality of context-aware process mining analyses.
    \item \textit{Systems interoperability}: To connect all the agents in context-aware environments and enable the exchange of data, a series of standards must be defined at different levels. On the one hand, at a technical level, wireless communication technologies (\eg ZigBee, IEEE 802.11, BLE\ldots) and application-level protocols (\eg SOAP, REST\ldots) must be defined. On the other hand, at a syntactical level, it is also important to agree on the encoding, either at a symbol level (\eg UTF-8, ASCII\ldots) or at a message level (\eg JSON, XML, CSV, HL7\ldots).
\end{itemize}

\subsection{Computational Challenges}

The gathering of contextual events in context-aware environments increases the number of events to process. This fact arises tremendous \textit{computational} challenges that, whereas omitted in traditional environments, should be considered in context-aware environments.

\begin{itemize}
    \item \textit{Big data}: The adoption of big data paradigms and technologies, such as Spark, Hadoop and Map-Reduce, within process mining analyses seems unavoidable in order to deal with very large sets of event data. Although classical process mining algorithms are not yet integrated with these technologies, the development of new algorithms considering these technologies by design opens to door to plenty of opportunities.
    \item \textit{Specific context-aware algorithms}: Current process mining algorithms are general-purpose-oriented, and do not consider neither the domain nor the contextual constraints where they are being executed. This solution might not be suitable when considering context-aware environments. To this end, \textit{ad-hoc} process mining algorithms could be developed, by taking into account the nature of the domain where they are going to be executed (\eg either a smart hospital or a smart home).
\end{itemize}

\subsection{Automation Challenges}

Process enhancement techniques conducted within context-aware environments open the door to improve the efficiency of service models thanks to the \textit{automation} of certain tasks. However, some challenges must be considered.

\begin{itemize}
    \item \textit{Evaluation and execution phase}: Deciding when a process needs to be readjusted to the current context is not trivial. This task must guarantee that the efficiency of the new process executions would improve the current one in the current context. The design of this proactive automation task must be carefully tested to prevent abnormal behaviours and undesired deviations.
    \item \textit{Learning}: Enhancement techniques should deal with and learn from positive deviants, \ie outlier process executions that perform better or more efficiently than a normal process execution within a given context, and from new contexts, \ie situations that have never occurred in the past, so these techniques do not have any previous knowledge on how to deal with them.
\end{itemize}

\section{Conclusions} \label{sec:pmcae:concl}

The IoT revolution has paved the way towards augmenting environments with plenty of sensing devices, named context-aware environments.
These devices are able to generate vast volumes of event data providing timely updates on the contextual settings.
With the Internet of Events paradigm in mind, this information could be exploited and complement classical event logs, detailing historical executions of processes, to provide added-value knowledge and deliver services more efficiently.

In this chapter, the opportunities and benefits of considering contextual information in conjunction with process mining analyses have been highlighted.
To this end, the paradigm of context-aware process mining has been coined.
Contrary to classical process mining analyses, the proactive nature of this paradigm enables the automatic realignment of business processes according to their immediate context in a timely efficient manner.
However, this arises a number of challenges to be faced, namely data challenges (related to data quality, authenticity and privacy), ecosystem challenges (related to the interoperability between devices and information systems, together with their appropriate synchronisation), computing challenges (related to the algorithms and technoloies to be used) and automation challenges (related to the automatic enhancement of processes).
Despite the infancy on context-aware process mining, there is no doubt that it will become a paramount research line in the near future.
\chapter{Skip Miner: Simplifying Spaghetti Processes}
\label{chap:skipminer}

\emph{The high complexity of today's business processes together with the large variability among process executions lead to the appearance of spaghetti process models, characterised by their high complexity and lack of structure that hinder their analysis. In order to acquire meaningful knowledge, the research community is struggling to apply several simplification methods to bring structure to this kind of processes. However, this task is challenging and opens the door to many opportunities yet to be explored. To this end, this chapter describes a novel process discovery algorithm, named Skip Miner, integrating a probabilistic-based simplification heuristic. First, Section \ref{sec:skipminer:spaghetti} provides an overview of spaghetti processes, and some of the most common simplification methods are explained in Section \ref{sec:skipminer:background}. Next, Section \ref{sec:skipminer:proposal} describes the details of the proposed Skip Miner method. Section \ref{sec:skipminer:expsetup} includes the methodology used to test the algorithm using a real-life medical event log, whose results are discussed and compared in Section \ref{sec:skipminer:discussion}. Finally, the chapter concludes in Section \ref{sec:skipminer:concl}.}

\minitoc

\section{Spaghetti Processes} \label{sec:skipminer:spaghetti}

Process discovery techniques are extensively used within organisations to visualise the actual execution of their processes.
To be effective, these techniques must represent the discovered process models in the most realistic and understandable way.
Unfortunately, sometimes this results into large and complex models with no (or very little) structure.
This type of processes, known as \textit{spaghetti processes}, are extremely hard to comprehend, and pose serious analytical challenges.
An example of a spaghetti process is depicted in Figure \ref{fig:skipminer:spaghetti}.
Spaghetti processes might be discovered from event logs for several reasons.
The first reason is because the discovery algorithm used represents the reality contained in the event logs very accurately, with plenty of details of all process instances, instead of showing only a representative behaviour of these instances.
Excessive details could jeopardise the visualisation of the processes.
Second, low quality event logs can also produce spaghetti processes, as models would contain noisy, incomplete or infrequent behaviour that distorts reality \cite{aalst2011manifesto}.
Last, processes executed within highly complex and dynamic contexts, \ie where the execution of process instances is likely to present multiple and changing variants, might be spaghetti-like too.
In this context, healthcare processes are prone to be discovered as spaghetti processes due to their inherent characteristics (refer to Section \ref{subsec:background:pm_health}).
For instance, understanding medical treatments as processes, the discovery of the execution of a certain treatment might produce a spaghetti model because the treatment might be followed differently for different patients.

\begin{figure}[b!]
 \centering
 \includegraphics[width=0.93\columnwidth]{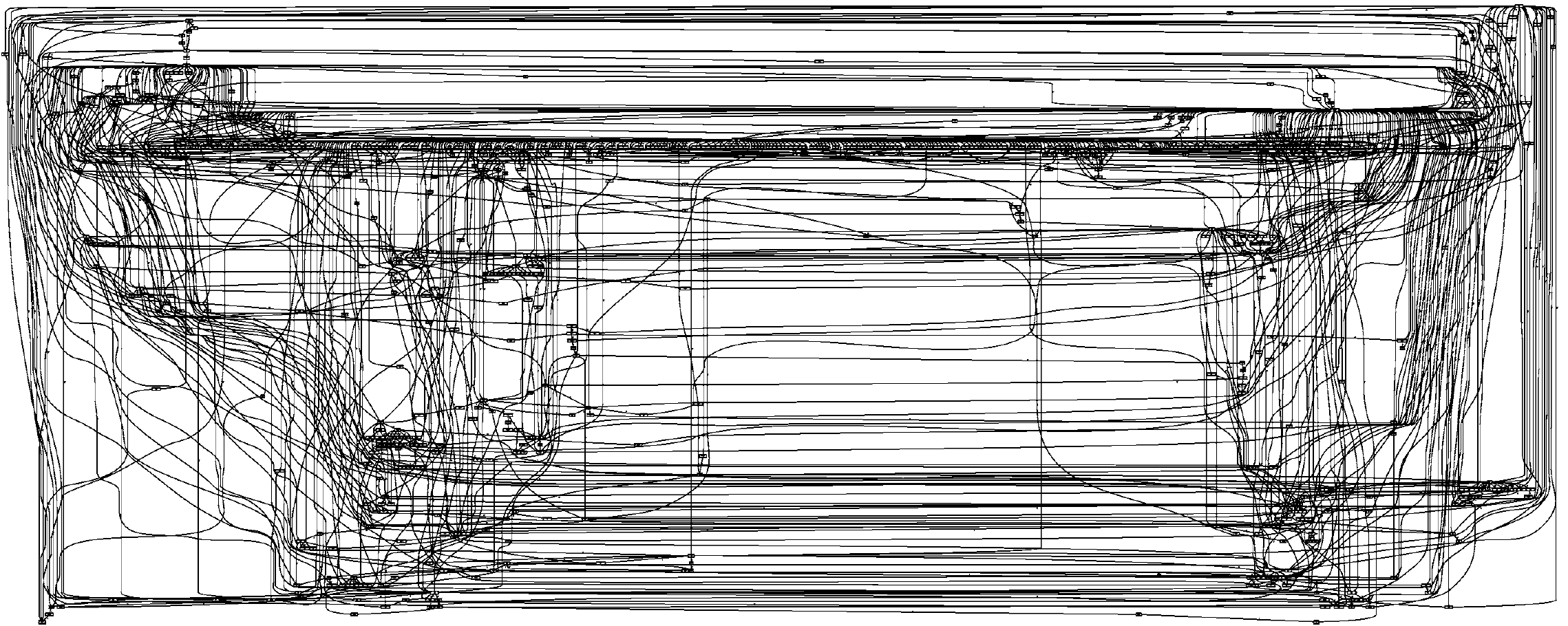}
 \caption{Example of a spaghetti process model describing 619 activities related to the diagnosis and treatment of 2765 patients in a Dutch hospital (reprinted from \cite{aalst2011book}).}
 \label{fig:skipminer:spaghetti}
\end{figure}

The discovery of process models from event logs in a spaghetti-like fashion is of little interest due to the limitations for further analysis and understanding.
Despite the many process discovery techniques, the challenges around spaghetti process models are still present.
This chapter presents Skip Miner, a novel process discovery algorithm with a probabilistic built-in heuristic to confer structure to process models.
The proposed method has been evaluated and compared against other methods from the literature, using a real-life medical event log from a healthcare institution containing spaghetti-like process models.

\section{Classical Process Simplification Methods} \label{sec:skipminer:background}

Dealing with spaghetti processes is not straightforward and raises a number of challenging opportunities at the time of designing process discovery techniques and visualisation strategies.
Conceptually, bringing structure to process models can be faced from many angles.
According to van der Aalst and G{\"u}nther \cite{aalst2007structure}, one could 
(i) \textit{aggregate} semantically similar information to minimise the number of process items, \ie creating clusters of low-level detail information,
(ii) represent high-level \textit{abstractions} of the information and discard insignificant or too detailed information for a given context,
(iii) make \textit{emphasis} on meaningful and relevant information within the model,
or (iv) \textit{customise} the models according to their specific purpose for a given context.
These strategies should be considered when attempting to simplify process models, especially those being spaghetti.

The simplification of spaghetti-like processes can be conducted at any stage of the discovery phase, generally before or after the discovery of the process model.
On the one hand, event data can be simplified prior to the execution of the discovery algorithm (\textit{a-priori} methods), with the assumption that a simpler event log will produce a simpler process model.
For instance, a generalised practice is to filter out events whose activity is too detailed and/or is not observed in the majority of process instances.
Therefore, only events whose activity determines a frequent behaviour are considered.
Thanks to the abstraction of the resulting process models from detailed information, the risk of discovering spaghetti processes minimises.
In this line, the investigations in \cite{conforti2017filtering,zelst2020detection} aim at sanitising event logs by filtering out noisy events that present infrequent behaviour within the process.
With the ``divide and rule'' principle in mind, trace clustering \cite{song2008trace} aims to divide the event log into groups of events whose process instances are relatively homogeneous and, for each of them, represent the behaviour in an independent process model.
This method is based on the assumption that each of the individual process models will be better structured, hence facilitating their comprehension, in comparison to a single process model representing the behaviour of the entire event log.

On the other hand, process models can be simplified after their discovery, \ie \textit{a-posteriori} methods.
A common practice is to remove irrelevant behaviour from the discovered process models, by eliminating those activities and/or transitions between activities (\ie nodes and/or edges from a graph) that have little weight in the overall process model.
One of the most popular process discovery techniques, fuzzy miner \cite{gunther2007fuzzy}, applies aggregation and abstraction techniques directly to the process models, thanks to the computation of a correlation and significance value to each node and transition of the model.
By means of thresholds, one is able to seamlessly fine tune the discovered process models, resulting in the aggregation of highly-correlated nodes into clusters and the filtering of irrelevant edges.
Last, inductive miner \cite{leemans2013inductive} also applies many filter throughout the entire process discovery phase, by dividing the event log, discarding infrequent instances and removing infrequent edges.

\section{The Skip Miner Algorithm} \label{sec:skipminer:proposal}

\begin{figure}[b!]
 \centering
 \includegraphics[width=0.93\columnwidth]{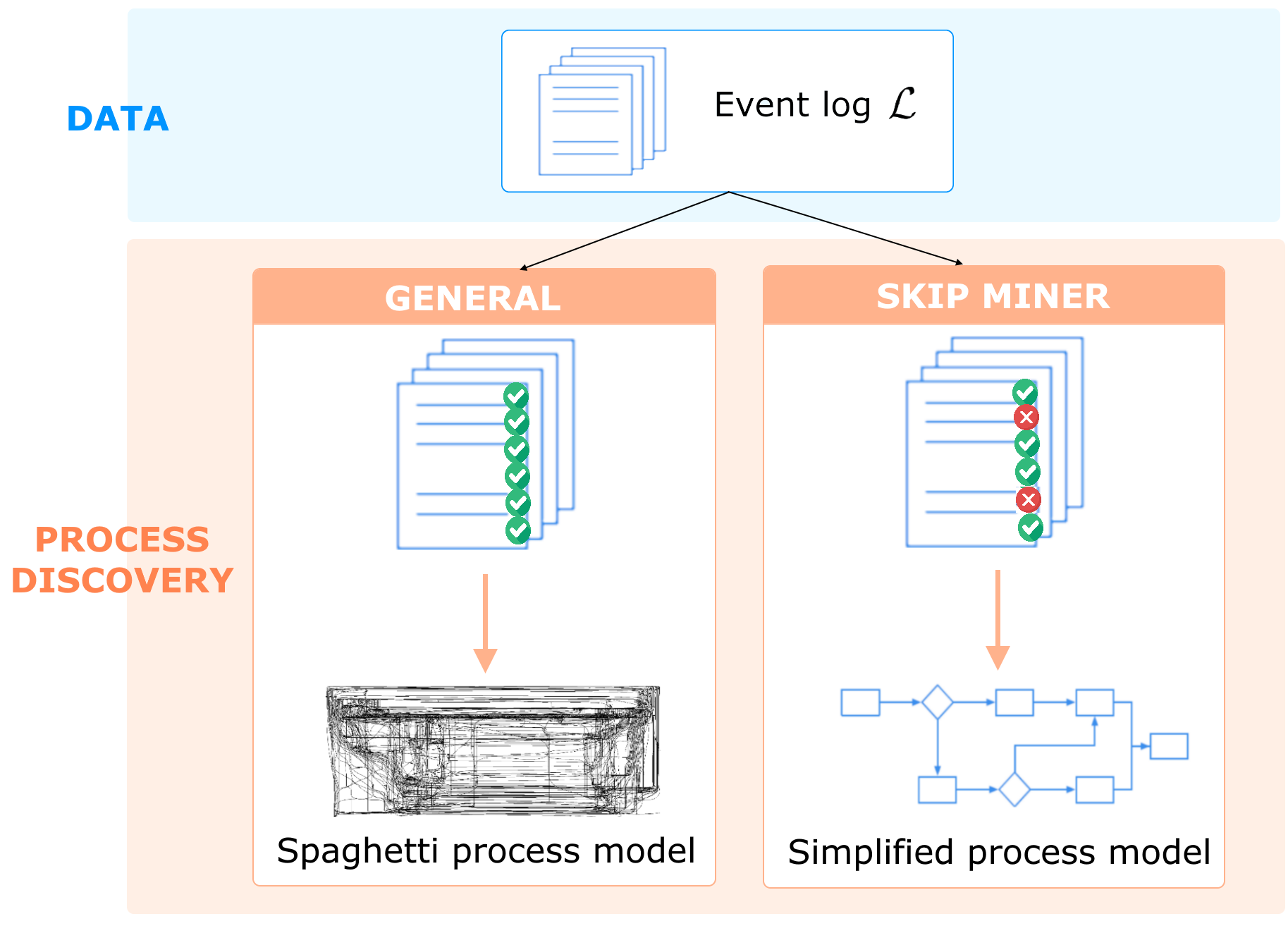}
 \caption{Conceptual overview of the Skip Miner algorithm.}
 \label{fig:skipminer:overview}
\end{figure}

In this section, we describe a novel process discovery algorithm, named Skip Miner, that represents the control-flow perspective of the processes as graphs.
This algorithm builds upon a \textit{simplification by design} approach, \ie the simplification heuristic is integrated within the design of the algorithm, thereby avoiding the appearance of spaghetti-like process models.
Skip Miner is based on the assumptions that (i) not all the events within an event log are equally important, and (ii) some events are more likely responsible than others for converting a process model into a spaghetti-like process model.
Therefore, the algorithm uses a probabilistic approach to decide whether to consider or discard (\ie ``skip'') each event during the discovery phase.
An overview of the rationale of the algorithm is depicted in Figure \ref{fig:skipminer:overview}.
Given an event log $\mathcal{L}$, a general algorithm would consider all the events at the time of discovering the process model.
If $\mathcal{L}$ contains a large amount of traces or these traces present very heterogeneous behaviour, then it is likely that the resulting process model becomes spaghetti-like.
To prevent this, Skip Miner skips some of these events and considers the rest of events for creating the process model, which is expected to be simpler.

The implementation details and the main design decisions of the Skip Miner algorithm are described as follows.

\subsection{Preliminaries} \label{subsec:skipminer:prelim}

Skip Miner focuses on the control-flow perspective of the processes.
Therefore, being the resulting process model a graph $\mathcal{M} = \{N, E\}$, the node set $N$ represents the activities of the event log (\ie $\#_{activity}(e), \forall e \in \mathcal{L}$), and the edge set $E$ represents the transitions between pairs of activities from $N$.
Besides, each edge has a weight $w \in [0,1)$ representing the probability of transiting from the origin node to the destination node.

To compute the weight $w$ of each edge, first we need to count the number of existing transitions between each pair of nodes from $N$.
Let $O$ be an co-occurrence matrix of length $n \times n$, where $n$ is the number of distinct activities in $\mathcal{L}$ (\ie the number of nodes in $N$).
Each item in $O$ indicates the number of times that a node $j$ has been reached from a node $i$ within each trace.
In this context, it is worth noting that $O$ is asymmetric (\ie $O[i][j] \neq O[j][i]$), because is not the same doing activity $j$ after activity $i$, than doing activity $i$ after activity $j$.
Figure \ref{fig:skipminer:matrices} illustrates the following toy example.
Assume an event log $\mathcal{L}$ containing activities from the set $\{$A, B, C, D$\}$.
A co-occurrence matrix $O$ can be computed by annotating the transitions between consecutive activities.
In this case, activity B was performed after activity A in six occasions (\ie $O[$A$][$B$] = 6$), and activity C was performed after activity A in two occasions (\ie $O[$A$][$C$] = 2$).
However, activity D was never performed after activity A (\ie $O[$A$][$D$] = 0$).

The main drawback of matrix $O$ is the lack of normalisation, \ie its values may range from 0 to $\infty$.
Consequently, we create a frequency matrix $F$, by conducting a row-wise normalisation of matrix $O$, \ie each value in $O$ is divided by the sum of values in its row.
This matrix transformation is also depicted in Figure \ref{fig:skipminer:matrices}.
As a result, the values in $F$ are bounded between 0 and 1, and the sum of the values of each row is 1.
Thus, each value in $F$ represents the probability of moving from one node to another (\ie the probability of doing an activity after doing another), which directly corresponds to the value $w$ of the edges in $E$.
Finally, the information in $F$ can be graphically represented in the form of a graph $\mathcal{M}$.

\begin{figure}[t!]
 \centering
 \includegraphics[width=0.97\columnwidth]{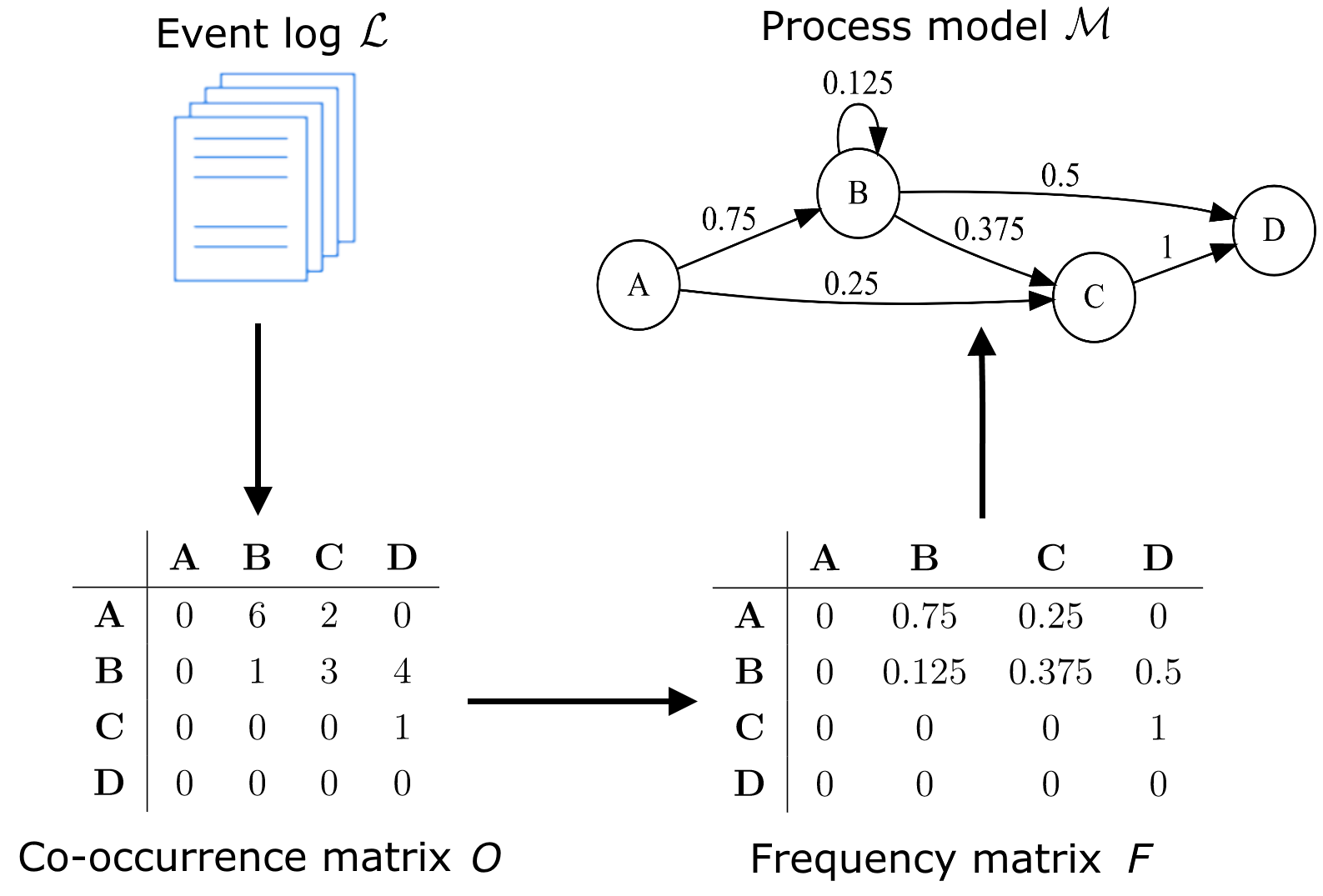}
 \caption{From an event log $\mathcal{L}$ to a process model $\mathcal{M}$ using auxiliary matrices.}
 \label{fig:skipminer:matrices}
\end{figure}

\subsection{Skipping Events: The \textit{LoS} Parameter}

The differentiating characteristic of this algorithm is the ability to skip events.
The number of events to skip comes defined by a numeric parameter called \textit{Length of Skip} (\textit{LoS}), where \textit{LoS}~$\geq 0$.
In practice, although \textit{LoS} could tend to the infinite, the maximum meaningful value of \textit{LoS} would be the length of the largest trace in the event log.
When \textit{LoS}~$= 0$, the algorithm would not skip any event, thus generalising to a discovery algorithm that does not simplify the resulting process models.

Be $\mathcal{L}$ an event log with three traces, and each of them containing events ordered chronologically associated to activities from the set $\{$A, B, C, D, E$\}$, as follows: 
$\mathcal{L} = \langle \langle $A, B, C, D, E$\rangle, \langle$A, B, C, E$\rangle, \langle$A, B, C, D, B, C, D, E$\rangle \rangle$.
Configuring \textit{LoS}~$= 1$, the activities that might be considered in Skip Miner are as follows (in bold): 
$\mathcal{L}' = \langle \langle $\textbf{A}, \textbf{B}, C, \textbf{D}, \textbf{E}$\rangle, \langle$\textbf{A}, \textbf{B}, C, \textbf{E}$\rangle, \langle$\textbf{A}, \textbf{B}, C, \textbf{D}, \textbf{B}, C, \textbf{D}, \textbf{E}$\rangle \rangle$.
In the first trace, we can notice that activity C has been skipped, and a transition between activity B and activity D would be created.
Indeed, in this example, activity C would not be discovered in the resulting process model because it was skipped in all traces.
It is worth emphasising that the above $\mathcal{L}'$ is only one of the solutions that may be returned by the algorithm, which works in a non-deterministic manner (explained later).
Configuring \textit{LoS}~$= 2$, Skip Miner might consider the following activities (in bold):
$\mathcal{L}'' = \langle \langle $\textbf{A}, B, C, \textbf{D}, \textbf{E}$\rangle, \langle$\textbf{A}, B, C, \textbf{E}$\rangle, \langle$\textbf{A}, B, C, \textbf{D}, B, C, \textbf{D}, \textbf{E}$\rangle \rangle$.
In this case, both activities B and C together with all their transitions would not be considered in the resulting process model.

The \textit{LoS} parameter has a direct relationship with the number of events used for the discovery of the algorithm: the larger the value of \textit{LoS}, the more events are skipped and fewer events are considered.
In this sense, the discovered process model will become simpler when increasing the \textit{LoS}.
However, largely simplified process models will not represent the original behaviour in an accurate and complete way, which implies an information loss.
This trade-off between simplicity and information loss needs to be considered when deciding the value of the \textit{LoS} parameter.

\subsection{The Probability to Skip Events}

To balance the trade-off between process simplification and information loss, it is important to define an heuristic for formalising when events should be skipped.
A naive heuristic could determine to always skip events.
For instance, be $\mathcal{L}$ an event log containing a single trace with seven events referring to different activities, $\mathcal{L} = \langle \langle$A, B, C, D, E, F, G$\rangle \rangle$.
If \textit{LoS}~$= 1$, the heuristic would consider the activities in bold: $\mathcal{L}' = \langle \langle$\textbf{A}, B, \textbf{C}, D, \textbf{E}, F, \textbf{G}$\rangle \rangle$, this is, a process model with four nodes and three edges.
Similarly, if \textit{LoS}~$= 2$, the heuristic would consider the activities in bold: $\mathcal{L}'' = \langle \langle$\textbf{A}, B, C, \textbf{D}, E, F, \textbf{G}$\rangle \rangle$, this is, a process model with only three nodes and two edges.
By contrast, the original process model (\textit{LoS}~$= 0$) would consider seven nodes and six edges.
It can be observed that this heuristic enables a rapid simplification of the process models, but also increases rapidly the information loss.
Moreover, the main drawback of this heuristic is the arbitrariness to skip events, because it does not consider neither which events are going to be skipped nor their relevance within the entire event log.
In the long run, despite the simplifications, this heuristic would significantly decrease the quality of the process models.

Next, we describe the heuristic to skip events implemented in Skip Miner.
The proposed heuristic uses a non-deterministic approach using probabilities: each activity within the event log, \ie $\forall \#_{activity}(e), e \in \mathcal{L}$, has a \textit{Skip Probability} (\textit{SP}), where \textit{SP}($\#_{activity}(e)$) $\in [0,1)$.
The probability of each activity determines how probable is to skip the following \textit{LoS} events after reaching an event of that activity.
If \textit{SP}($\#_{activity}(e)$)~$= 0$, no events are skipped when finding an event referring to that activity (\ie the next event is always considered).
On the contrary, if \textit{SP}($\#_{activity}(e)$) $\rightarrow 1$, the following \textit{LoS} events after that activity are skipped, thus the next event to consider is that after \textit{LoS} + 1 positions.
Let $\mathcal{A}$ be the activity universe of all activities within an event log, \ie $\mathcal{A} = \{\#_{activity}(e)\}, \forall e \in \mathcal{L}$, the \textit{SP} value of each activity $a \in \mathcal{A}$ comes defined by Equation~\ref{eq:skipminer:sp}.

\begin{equation}
  \textit{SP}(a) = 1 - \frac{1}{n}
  \label{eq:skipminer:sp}
\end{equation}
where $n$ is the number of different activities performed just right after activity $a$.

According to this heuristic, it is more likely to skip events when the number of different activities that follow that activity is high.
Let's illustrate the rationale of this heuristic.
Within spaghetti-like process models (remember Figure \ref{fig:skipminer:spaghetti}), it is easy to observe lots of paths (\ie edges) after visiting a specific node.
These nodes are more likely responsible for turning the model into a spaghetti-like model.
Our heuristic aims to minimise the complexity of this kind of nodes.
This is analogous as saying that the purpose of the heuristic is to reduce (i) the number of less significant nodes and (ii) the outdegree of the most significant nodes.
Next, we describe different scenarios, illustrated in Figure \ref{fig:skipminer:heuristics}, to emphasise the value of the heuristic proposed.

\begin{figure}[b!]
\centering
\begin{subfigure}{.36\textwidth}
  \centering\includegraphics[height=8.5cm]{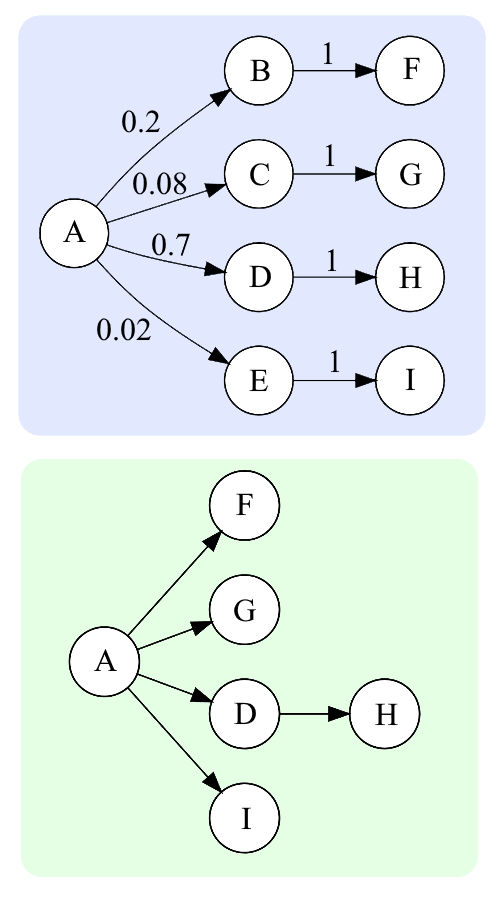}  
  \caption{Scenario 1}\label{subfig:skipminer:sce1}
\end{subfigure}
\hfill
\begin{subfigure}{.36\textwidth}
  \centering\includegraphics[height=8.5cm]{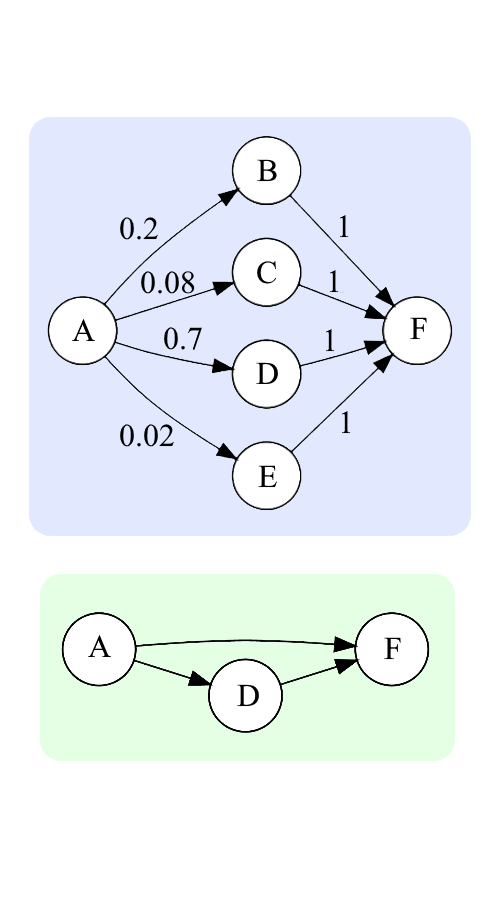}  
  \caption{Scenario 2}\label{subfig:skipminer:sce2}
\end{subfigure}
\hfill
\begin{subfigure}{.26\textwidth}
  \centering\includegraphics[height=8.5cm]{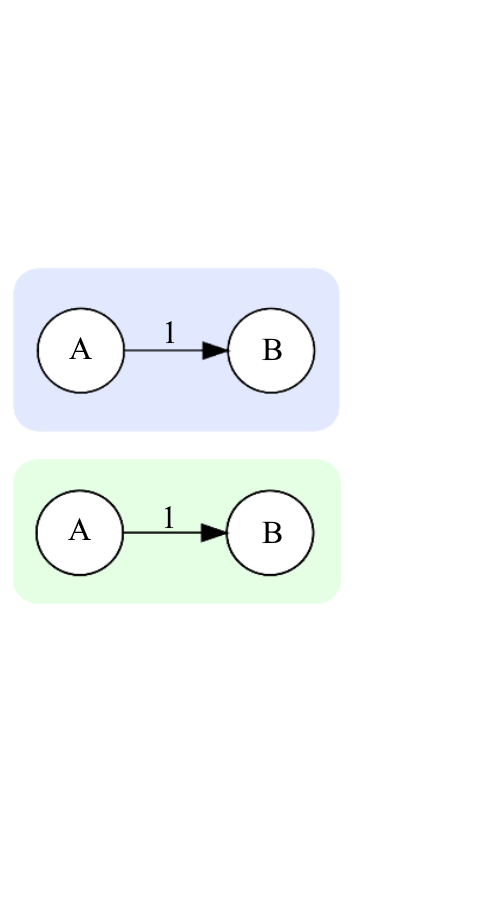}  
  \caption{Scenario 3}\label{subfig:skipminer:sce3}
\end{subfigure}
\caption{Toy examples about the simplification of the heuristic implemented in Skip Miner using an \textit{LoS}~$= 1$. The models in blue represent the scenario in a spaghetti process model, and the models in green represent a potential simplification of the Skip Miner heuristic (adapted from \cite{batista2019skip}).}
\label{fig:skipminer:heuristics}
\end{figure}

\begin{itemize}
    \item \textit{Scenario 1}: A node with a high outdegree, where each of the following nodes conduct to a different node (see Figure \ref{subfig:skipminer:sce1}). As node A leads to many other nodes (B, C, D and E), the heuristic assumes that these nodes might not very relevant and are candidates to be simplified. In this case, \textit{SP}(A)~$= 0.75$, which means that the algorithm will skip (with a large probability) the following \textit{LoS} events after finding an event referring to activity A.
    Even though it is likely to skip events in most cases (and creating new edges from A that did not appear in the original model), when not taken, the original edges will be maintained. However, from a probability perspective, it is more likely that only edges with large weight appear, and those with little weight will be omitted. In this scenario, simplification is achieved by discarding less significant nodes and edges.
    \item \textit{Scenario 2}: A node with a high outdegree, where each of the following nodes conduct to the same node (see Figure \ref{subfig:skipminer:sce2}). This scenario is similar to the previous one, although nodes B, C, D and E are merely drivers to reach a more important node (\ie F). With a large probability, the heuristic will create a new edge connecting A to F with a higher weight, as well as maintain also the original nodes and edges from A with higher weights. In contrast, less significant nodes and edges are likely to be discarded.
    \item \textit{Scenario 3}: A node with a low outdegree (see Figure \ref{subfig:skipminer:sce3}). The outdegree of node A is 1, \ie the node after A is always B. In this scenario, the heuristic preserves the transition and does not conduct any simplification: as \textit{SP}(A)~$= 0$, no events are skipped after A, and B will always be considered. It is apparent that it makes no sense to skip events in this scenario because it does not provide any benefit from the simplification point of view and, in fact, it would introduce unnecessary noise.
\end{itemize}

\subsection{Outlook}

For the sake of completeness, we provide the general scheme of the Skip Miner algorithm in Algorithm \ref{alg:skipminer:algorithm}.
The algorithm requires two parameters: (i) an event log $\mathcal{L}$ and (ii) a non-negative integer for the \textit{LoS}.
First, the skip probabilities (\textit{SP}) of each activity in $\mathcal{L}$ are calculated using Equation \ref{eq:skipminer:sp}, and the co-occurrence matrix \textit{O} is instantiated with zeros (lines 2--3).
Then, the algorithm starts iterating over each trace in $\mathcal{L}$ (lines 4--17).
Hereafter, transitions between activities are determined by establishing connections between origin activities (controlled by variable i) and destination activities (controlled by variable j).
By design, the first event of a trace is always considered, and immediately becomes controlled by i (line 5).
To seek the destination activity, the non-deterministic heuristic is executed considering the skip probability of the origin activity (line 8).
If taken, a total of \textit{LoS} events are skipped and the destination event becomes the one after \textit{LoS} positions (line 9).
Otherwise, no events are skipped and the destination event becomes the one immediately after the origin event (line 11).
When both origin and destination activities are chosen, a connection between them is accumulated in the co-occurrence matrix \textit{O} (line 14).
Finally, the pointers for the next connection are adjusted (\ie the destination event is the new origin event, and the new destination event is still unknown) for further iterations until the end of the trace (line 15).
Once all traces have been processed, the co-occurrence matrix \textit{O} contains all the information required to create a process model $\mathcal{M}$ in the form of a graph following the procedure explained in Section \ref{subsec:skipminer:prelim}.

\begin{algorithm}[t]
\footnotesize
\begin{algorithmic}[1]
\Require
  \Statex $\mathcal{L}$ is a non-empty event log.
  \Statex \textit{LoS} is a non-negative integer.
\Ensure
  \Statex $\mathcal{M}$ is a process model representing the control-flow perspective of $\mathcal{L}$.
\Function{SkipMiner}{\textbf{EventLog} $\mathcal{L}$, \textbf{Integer} \textit{LoS}}
\State \textbf{HashMap$\langle$String, Float$\rangle$} SP $\gets$ CalculateAllSkipProbabilities($\mathcal{L}$);
\State \textbf{Matrix2D$\langle$Integer$\rangle$} \textit{O} $\gets$ Initialise($\mathcal{L}$);
\ForAll {trace t in $\mathcal{L}$}
  \State \textbf{Integer} i, j $\gets$ 0;
  \While {i $<$ length(t)}
    \State \textbf{String} act\_i $\gets$ $\#_{activity}$(t[i]);
    \If {Random(0, 1) $<$ SP.get(act\_i)}  \Comment{Skip events}
      \State j $\gets$ i + 1 + \textit{LoS};
    \Else  \Comment{Do not skip events}
      \State j $\gets$ i + 1;
    \EndIf
    \State \textbf{String} act\_j $\gets$ $\#_{activity}$(t[j]);
    \State \textit{O}[act\_i][act\_j] += 1;
    \State i $\gets$ j; j $\gets$ null; 
  \EndWhile
\EndFor
\State \textbf{ProcessModel} $\mathcal{M} \gets$ CreateGraph(\textit{O});
\State \Return $\mathcal{M}$;
\EndFunction
\end{algorithmic}
\caption{Skip Miner algorithm}
\label{alg:skipminer:algorithm}
\end{algorithm}

\section{Experimental Setup} \label{sec:skipminer:expsetup}

This section describes the experimental setup of the proposed Skip Miner algorithm.
Experiments have been conducted using the TGN-Hospital event log, already described in Table \ref{tbl:background:event_logs}.
For managerial purposes, the workflow of healthcare practitioners could be analysed to retrieve performance and costs metrics.
These workflows, which can be represented as process models, can be quite complex, especially when evaluating practitioners responsible for more (and diverse) patients.
In the experiments, five of the most active practitioners (hereafter, referred to as $D_1$, $D_2$, $D_3$, $D_4$ and $D_5$, respectively), whose process model is spaghetti-like and difficult to understand, are evaluated.

\begin{figure}[t!]
 \centering
 \includegraphics[width=0.9\columnwidth]{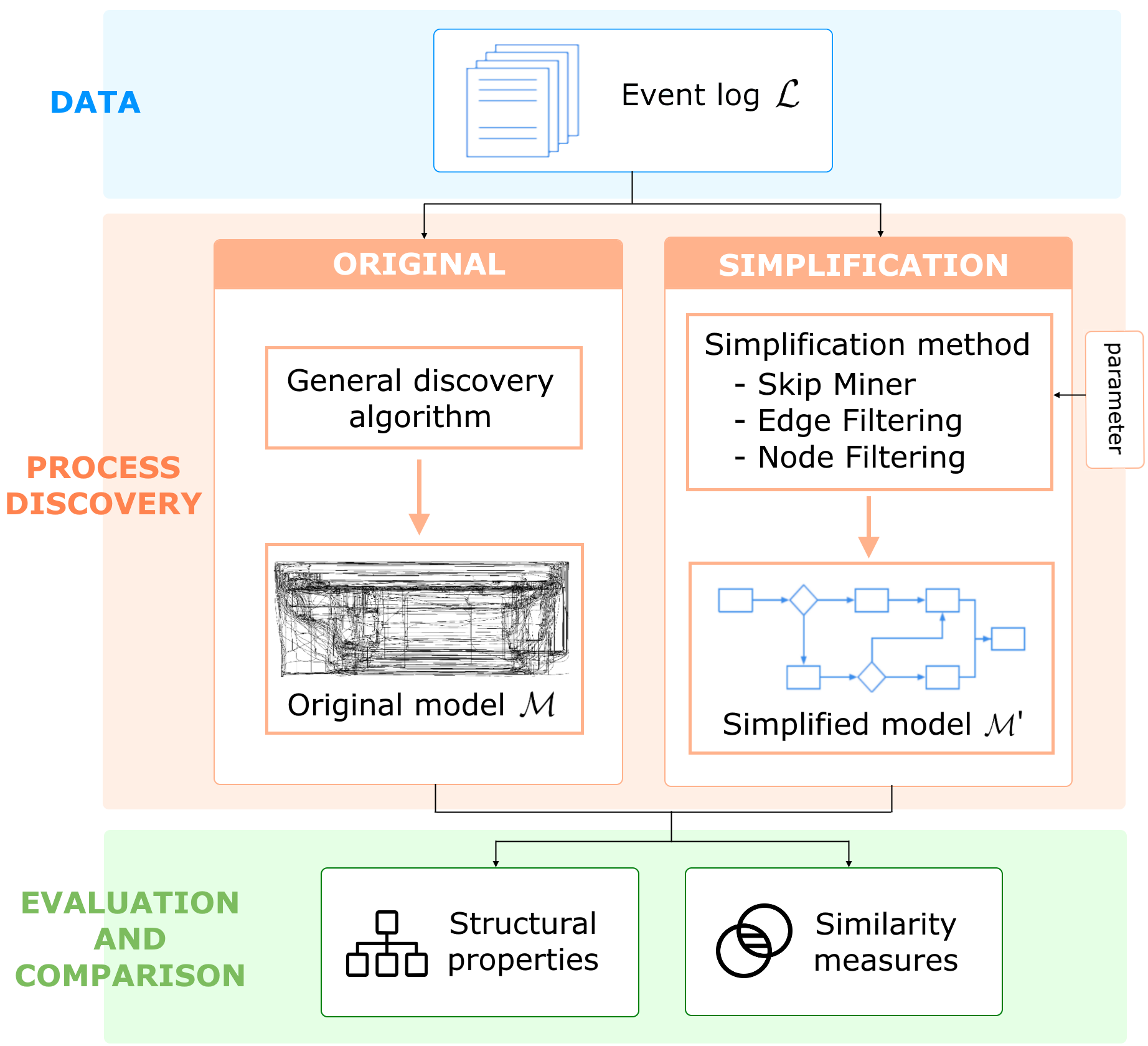}
 \caption{Experimental setup.}
 \label{fig:skipminer:expsetup}
\end{figure}

For each practitioner $D$, the procedure depicted in Figure \ref{fig:skipminer:expsetup} is followed.
First, the process model $\mathcal{M}$ is discovered by exploiting the events of the event log $\mathcal{L}$ associated to the practitioner (\ie $\#_{resource}(e) = D, \forall e \in \mathcal{L}$).
This process model represents the original (spaghetti) behaviour, because all the transitions between consecutive events within the same trace are considered.
This model sets a baseline to evaluate the effectiveness of the simplification methods.
Three simplification methods are evaluated:
(i) the proposed Skip Miner algorithm for different \textit{LoS} values, being \textit{LoS} $= \{1, 2, 3, 4, 5\}$;
(ii) an \textit{a-posteriori} method, hereafter EdgeFil, that removes edges from the process model whose weight is below a certain threshold $\alpha$, being $\alpha = \{0.01, 0.03, 0.05, 0.08, 0.10, 0.15\}$ (\eg if $\alpha = 0.05$, the resulting process model contains only edges whose weight is 0.05 or above);
and (iii) an \textit{a-priori} method, hereafter NodeFil, that discovers the process model by considering only those events whose activity appears in a specific proportion $\beta$ of traces from the event log, being $\beta = \{0.01, 0.03, 0.05, 0.08, 0.10, 0.15\}$ (\eg if $\beta = 0.1$, the resulting process models contains only activities that appear in, at least, the 10\% of traces).
The execution of any of these simplification methods with a certain parameter value results into a new process model $\mathcal{M}'$, which is a simplified version of the original process model $\mathcal{M}$.

Having a pair $\langle \mathcal{M}, \mathcal{M}' \rangle$, they are compared from two angles: from a structural perspective and from a quality (information loss) perspective.
On the one hand, the structural properties of the graphs can be compared according to their topological variations, this is, the number of nodes (nN), the number of edges (nE), the proportion of edges per node (EN) and the graph's density (gD).
On the other hand, the quality of the simplified process models with regards to the original process models can be measured using graph similarity measures, such as the Vertex Edge Overlap (VEO) \cite{papadimitriou2010web}, Vertex Ranking (VR) \cite{papadimitriou2010web}, DeltaCon (DC) \cite{koutra2013deltacon} and Weight Distance (WD) \cite{shoubridge2002detection}.
All measures are bounded between 0 and 1, where total similarity (\ie no information loss) is indicated by 1 in the first three measures and by 0 in the last measure.

\section{Results and Discussion} \label{sec:skipminer:discussion}

\begin{table}[t!]
\centering
\scriptsize
\caption{Results of the practitioner $D_1$.}
\label{tbl:skipminer:results_d1}  
\resizebox{0.965\textwidth}{!}{
\renewcommand{\arraystretch}{0.85}
\begin{tabular}{crcccccccc} \toprule
\multicolumn{2}{c}{} & \multicolumn{4}{c}{\textbf{Structural properties}} & \multicolumn{4}{c}{\textbf{Similarity measures}} \\ \cmidrule(lr){3-6} \cmidrule(lr){7-10}
\multicolumn{2}{c}{} & {\textbf{nN}} & {\textbf{nE}} & {\textbf{EN}} & {\textbf{gD}} & {\textbf{VEO}} & {\textbf{VR}} & {\textbf{DC}} & {\textbf{WD}} \\ \midrule
\textbf{Original $\mathcal{M}_1$} & & 27 & 117 & 4.333 & 0.167 & - & - & - & - \\ \midrule
\multirow{4}{*}{\textbf{Skip Miner}} & \textbf{1} & 25 & 107 & 4.280 & 0.178 & 0.725 & 0.863 & 0.460 & 0.687 \\
\multirow{4}{*}{(\textit{LoS})} & \textbf{2} & 25 & 86 & 3.440 & 0.143 & 0.698 & 0.846 & 0.428 & 0.734 \\
& \textbf{3} & 21 & 66 & 3.143 & 0.157 & 0.580 & 0.682 & 0.380 & 0.806 \\
& \textbf{4} & 19 & 46 & 2.421 & 0.135 & 0.498 & 0.616 & 0.349 & 0.864 \\
& \textbf{5} & 18 & 43 & 2.389 & 0.141 & 0.488 & 0.565 & 0.339 & 0.872 \\ \midrule
\multirow{5}{*}{\textbf{EdgeFil}} & \textbf{0.01} & 26 & 114 & 4.385 & 0.175 & 0.986 & 0.957 & 0.940 & 0.026 \\
\multirow{5}{*}{($\alpha$)} & \textbf{0.03} & 26 & 101 & 3.885 & 0.155 & 0.937 & 0.925 & 0.840 & 0.137 \\
& \textbf{0.05} & 26 & 83 & 3.192 & 0.128 & 0.862 & 0.925 & 0.721 & 0.291 \\
& \textbf{0.08} & 26 & 71 & 2.731 & 0.109 & 0.805 & 0.919 & 0.642 & 0.393 \\
& \textbf{0.10} & 26 & 55 & 2.115 & 0.085 & 0.720 & 0.870 & 0.565 & 0.530 \\
& \textbf{0.15} & 26 & 49 & 1.885 & 0.075 & 0.685 & 0.859 & 0.546 & 0.581 \\ \midrule
\multirow{5}{*}{\textbf{NodeFil}} & \textbf{0.01} & 19 & 105 & 5.526 & 0.307 & 0.910 & 0.643 & 0.445 & 0.146 \\
\multirow{5}{*}{($\beta$)} & \textbf{0.03} & 15 & 97 & 6.467 & 0.462 & 0.836 & 0.512 & 0.383 & 0.272 \\
& \textbf{0.05} & 11 & 80 & 7.273 & 0.727 & 0.715 & 0.391 & 0.346 & 0.466 \\
& \textbf{0.08} & 8 & 52 & 6.500 & 0.929 & 0.578 & 0.281 & 0.331 & 0.648 \\
& \textbf{0.10} & 7 & 40 & 5.714 & 0.952 & 0.492 & 0.247 & 0.327 & 0.733 \\
& \textbf{0.15} & 5 & 25 & 5.000 & 1.250 & 0.333 & 0.176 & 0.315 & 0.869 \\
\bottomrule
\end{tabular}
}
\end{table}

\begin{table}[t!]
\centering
\scriptsize
\caption{Results of the practitioner $D_2$.}
\label{tbl:skipminer:results_d2}  
\resizebox{0.965\textwidth}{!}{
\renewcommand{\arraystretch}{0.85}
\begin{tabular}{crcccccccc} \toprule
\multicolumn{2}{c}{} & \multicolumn{4}{c}{\textbf{Structural properties}} & \multicolumn{4}{c}{\textbf{Similarity measures}} \\ \cmidrule(lr){3-6} \cmidrule(lr){7-10}
\multicolumn{2}{c}{} & {\textbf{nN}} & {\textbf{nE}} & {\textbf{EN}} & {\textbf{gD}} & {\textbf{VEO}} & {\textbf{VR}} & {\textbf{DC}} & {\textbf{WD}} \\ \midrule
\textbf{Original $\mathcal{M}_2$} & & 33 & 195 & 5.909 & 0.185 & - & - & - & - \\ \midrule
\multirow{4}{*}{\textbf{Skip Miner}} & \textbf{1} & 33 & 162 & 4.909 & 0.153 & 0.605 & 0.930 & 0.355 & 0.800 \\
\multirow{4}{*}{(\textit{LoS})} & \textbf{2} & 31 & 122 & 3.935 & 0.131 & 0.546 & 0.859 & 0.330 & 0.831 \\
& \textbf{3} & 31 & 110 & 3.548 & 0.118 & 0.537 & 0.822 & 0.320 & 0.871 \\
& \textbf{4} & 31 & 81 & 2.613 & 0.087 & 0.471 & 0.817 & 0.311 & 0.894 \\
& \textbf{5} & 30 & 85 & 2.833 & 0.098 & 0.449 & 0.748 & 0.291 & 0.916 \\ \midrule
\multirow{5}{*}{\textbf{EdgeFil}} & \textbf{0.01} & 33 & 195 & 5.909 & 0.185 & 1.000 & 1.000 & 1.000 & 0.000 \\
\multirow{5}{*}{($\alpha$)} & \textbf{0.03} & 33 & 185 & 5.606 & 0.175 & 0.978 & 0.980 & 0.858 & 0.051 \\
& \textbf{0.05} & 33 & 132 & 4.000 & 0.125 & 0.840 & 0.936 & 0.618 & 0.323 \\
& \textbf{0.08} & 33 & 110 & 3.333 & 0.104 & 0.771 & 0.912 & 0.551 & 0.436 \\
& \textbf{0.10} & 33 & 85 & 2.576 & 0.080 & 0.682 & 0.901 & 0.481 & 0.564 \\
& \textbf{0.15} & 33 & 67 & 2.030 & 0.063 & 0.610 & 0.890 & 0.442 & 0.656 \\ \midrule
\multirow{5}{*}{\textbf{NodeFil}} & \textbf{0.01} & 33 & 195 & 5.909 & 0.185 & 1.000 & 1.000 & 1.000 & 0.000 \\
\multirow{5}{*}{($\beta$)} & \textbf{0.03} & 32 & 193 & 6.031 & 0.195 & 0.993 & 0.964 & 0.693 & 0.010 \\
& \textbf{0.05} & 28 & 180 & 6.429 & 0.238 & 0.931 & 0.820 & 0.479 & 0.139 \\
& \textbf{0.08} & 18 & 132 & 7.333 & 0.431 & 0.704 & 0.506 & 0.355 & 0.493 \\
& \textbf{0.10} & 18 & 132 & 7.333 & 0.431 & 0.704 & 0.506 & 0.355 & 0.493 \\
& \textbf{0.15} & 9 & 57 & 6.333 & 0.792 & 0.327 & 0.249 & 0.303 & 0.867 \\
\bottomrule
\end{tabular}
}
\end{table}
\begin{table}[t!]
\centering
\scriptsize
\caption{Results of the practitioner $D_3$.}
\label{tbl:skipminer:results_d3}  
\resizebox{0.965\textwidth}{!}{
\renewcommand{\arraystretch}{0.85}
\begin{tabular}{crcccccccc} \toprule
\multicolumn{2}{c}{} & \multicolumn{4}{c}{\textbf{Structural properties}} & \multicolumn{4}{c}{\textbf{Similarity measures}} \\ \cmidrule(lr){3-6} \cmidrule(lr){7-10}
\multicolumn{2}{c}{} & {\textbf{nN}} & {\textbf{nE}} & {\textbf{EN}} & {\textbf{gD}} & {\textbf{VEO}} & {\textbf{VR}} & {\textbf{DC}} & {\textbf{WD}} \\ \midrule
\textbf{Original $\mathcal{M}_3$} & & 25 & 112 & 4.480 & 0.187 & - & - & - & - \\ \midrule
\multirow{4}{*}{\textbf{Skip Miner}} & \textbf{1} & 25 & 111 & 4.440 & 0.185 & 0.725 & 0.843 & 0.468 & 0.716 \\
\multirow{4}{*}{(\textit{LoS})} & \textbf{2} & 23 & 90 & 3.913 & 0.178 & 0.672 & 0.819 & 0.440 & 0.726 \\
& \textbf{3} & 25 & 75 & 3.000 & 0.125 & 0.624 & 0.872 & 0.415 & 0.795 \\
& \textbf{4} & 22 & 62 & 2.818 & 0.134 & 0.561 & 0.651 & 0.382 & 0.801 \\
& \textbf{5} & 20 & 45 & 2.250 & 0.118 & 0.495 & 0.680 & 0.343 & 0.880 \\ \midrule
\multirow{5}{*}{\textbf{EdgeFil}} & \textbf{0.01} & 25 & 105 & 4.200 & 0.175 & 0.974 & 0.997 & 0.942 & 0.062 \\
\multirow{5}{*}{($\alpha$)} & \textbf{0.03} & 25 & 88 & 3.520 & 0.147 & 0.904 & 0.978 & 0.818 & 0.214 \\
& \textbf{0.05} & 25 & 70 & 2.800 & 0.117 & 0.819 & 0.962 & 0.714 & 0.375 \\
& \textbf{0.08} & 25 & 62 & 2.480 & 0.103 & 0.777 & 0.965 & 0.669 & 0.446 \\
& \textbf{0.10} & 25 & 52 & 2.080 & 0.087 & 0.720 & 0.955 & 0.630 & 0.536 \\
& \textbf{0.15} & 25 & 45 & 1.800 & 0.075 & 0.676 & 0.946 & 0.588 & 0.598 \\ \midrule
\multirow{5}{*}{\textbf{NodeFil}} & \textbf{0.01} & 18 & 99 & 5.500 & 0.324 & 0.921 & 0.656 & 0.452 & 0.133 \\
\multirow{5}{*}{($\beta$)} & \textbf{0.03} & 12 & 74 & 6.167 & 0.561 & 0.735 & 0.424 & 0.367 & 0.429 \\
& \textbf{0.05} & 10 & 61 & 6.100 & 0.678 & 0.644 & 0.352 & 0.346 & 0.566 \\
& \textbf{0.08} & 6 & 29 & 4.833 & 0.967 & 0.372 & 0.222 & 0.323 & 0.820 \\
& \textbf{0.10} & 4 & 16 & 4.000 & 1.333 & 0.255 & 0.144 & 0.310 & 0.925 \\
& \textbf{0.15} & 3 & 9 & 3.000 & 1.500 & 0.161 & 0.102 & 0.309 & 0.955 \\
\bottomrule
\end{tabular}
}
\end{table}

\begin{table}[t!]
\centering
\scriptsize
\caption{Results of the practitioner $D_4$.}
\label{tbl:skipminer:results_d4}  
\resizebox{0.965\textwidth}{!}{
\renewcommand{\arraystretch}{0.85}
\begin{tabular}{crcccccccc} \toprule
\multicolumn{2}{c}{} & \multicolumn{4}{c}{\textbf{Structural properties}} & \multicolumn{4}{c}{\textbf{Similarity measures}} \\ \cmidrule(lr){3-6} \cmidrule(lr){7-10}
\multicolumn{2}{c}{} & {\textbf{nN}} & {\textbf{nE}} & {\textbf{EN}} & {\textbf{gD}} & {\textbf{VEO}} & {\textbf{VR}} & {\textbf{DC}} & {\textbf{WD}} \\ \midrule
\textbf{Original $\mathcal{M}_4$} & & 23 & 84 & 3.652 & 0.166 & - & - & - & - \\ \midrule
\multirow{4}{*}{\textbf{Skip Miner}} & \textbf{1} & 22 & 89 & 4.045 & 0.193 & 0.706 & 0.898 & 0.464 & 0.760 \\
\multirow{4}{*}{(\textit{LoS})} & \textbf{2} & 21 & 82 & 3.905 & 0.195 & 0.695 & 0.771 & 0.389 & 0.719 \\
& \textbf{3} & 21 & 69 & 3.286 & 0.164 & 0.650 & 0.811 & 0.428 & 0.786 \\
& \textbf{4} & 18 & 58 & 3.222 & 0.190 & 0.645 & 0.690 & 0.397 & 0.766 \\
& \textbf{5} & 18 & 51 & 2.833 & 0.167 & 0.580 & 0.681 & 0.361 & 0.815 \\ \midrule
\multirow{5}{*}{\textbf{EdgeFil}} & \textbf{0.01} & 23 & 80 & 3.478 & 0.158 & 0.981 & 0.996 & 0.940 & 0.048 \\
\multirow{5}{*}{($\alpha$)} & \textbf{0.03} & 23 & 70 & 3.043 & 0.138 & 0.930 & 0.970 & 0.833 & 0.167 \\
& \textbf{0.05} & 23 & 49 & 2.130 & 0.097 & 0.804 & 0.981 & 0.725 & 0.417 \\
& \textbf{0.08} & 23 & 41 & 1.783 & 0.081 & 0.749 & 0.936 & 0.642 & 0.512 \\
& \textbf{0.10} & 23 & 39 & 1.696 & 0.077 & 0.734 & 0.932 & 0.634 & 0.536 \\
& \textbf{0.15} & 23 & 34 & 1.478 & 0.067 & 0.695 & 0.902 & 0.589 & 0.595 \\ \midrule
\multirow{5}{*}{\textbf{NodeFil}} & \textbf{0.01} & 19 & 79 & 4.158 & 0.231 & 0.946 & 0.779 & 0.479 & 0.091 \\
\multirow{5}{*}{($\beta$)} & \textbf{0.03} & 13 & 63 & 4.846 & 0.404 & 0.831 & 0.550 & 0.407 & 0.262 \\
& \textbf{0.05} & 11 & 52 & 4.727 & 0.473 & 0.741 & 0.463 & 0.378 & 0.411 \\
& \textbf{0.08} & 11 & 52 & 4.727 & 0.473 & 0.741 & 0.463 & 0.378 & 0.411 \\
& \textbf{0.10} & 8 & 42 & 5.250 & 0.750 & 0.586 & 0.318 & 0.333 & 0.645 \\
& \textbf{0.15} & 5 & 19 & 3.800 & 0.950 & 0.351 & 0.197 & 0.323 & 0.834 \\
\bottomrule
\end{tabular}
}
\end{table}

\begin{table}[t!]
\centering
\scriptsize
\caption{Results of the practitioner $D_5$.}
\label{tbl:skipminer:results_d5}  
\resizebox{0.965\textwidth}{!}{
\renewcommand{\arraystretch}{0.85}
\begin{tabular}{crcccccccc} \toprule
\multicolumn{2}{c}{} & \multicolumn{4}{c}{\textbf{Structural properties}} & \multicolumn{4}{c}{\textbf{Similarity measures}} \\ \cmidrule(lr){3-6} \cmidrule(lr){7-10}
\multicolumn{2}{c}{} & {\textbf{nN}} & {\textbf{nE}} & {\textbf{EN}} & {\textbf{gD}} & {\textbf{VEO}} & {\textbf{VR}} & {\textbf{DC}} & {\textbf{WD}} \\ \midrule
\textbf{Original $\mathcal{M}_5$} & & 29 & 94 & 3.241 & 0.116 & - & - & - & - \\ \midrule
\multirow{4}{*}{\textbf{Skip Miner}} & \textbf{1} & 29 & 93 & 3.207 & 0.115 & 0.759 & 0.926 & 0.478 & 0.655 \\
\multirow{4}{*}{(\textit{LoS})} & \textbf{2} & 29 & 69 & 2.379 & 0.085 & 0.706 & 0.917 & 0.448 & 0.690 \\
& \textbf{3} & 28 & 65 & 2.321 & 0.086 & 0.685 & 0.869 & 0.417 & 0.729 \\
& \textbf{4} & 27 & 51 & 1.889 & 0.073 & 0.687 & 0.784 & 0.407 & 0.712 \\
& \textbf{5} & 26 & 47 & 1.808 & 0.072 & 0.653 & 0.769 & 0.392 & 0.763 \\ \midrule
\multirow{5}{*}{\textbf{EdgeFil}} & \textbf{0.01} & 29 & 88 & 3.034 & 0.108 & 0.975 & 0.990 & 0.907 & 0.064 \\
\multirow{5}{*}{($\alpha$)} & \textbf{0.03} & 29 & 80 & 2.759 & 0.099 & 0.940 & 0.983 & 0.838 & 0.149 \\
& \textbf{0.05} & 29 & 73 & 2.517 & 0.090 & 0.907 & 0.974 & 0.797 & 0.223 \\
& \textbf{0.08} & 29 & 55 & 1.897 & 0.068 & 0.812 & 0.941 & 0.652 & 0.415 \\
& \textbf{0.10} & 29 & 55 & 1.897 & 0.068 & 0.812 & 0.941 & 0.652 & 0.415 \\
& \textbf{0.15} & 29 & 41 & 1.414 & 0.050 & 0.725 & 0.891 & 0.527 & 0.564 \\ \midrule
\multirow{5}{*}{\textbf{NodeFil}} & \textbf{0.01} & 20 & 81 & 4.050 & 0.213 & 0.893 & 0.653 & 0.450 & 0.161 \\
\multirow{5}{*}{($\beta$)} & \textbf{0.03} & 15 & 66 & 4.400 & 0.314 & 0.784 & 0.467 & 0.376 & 0.371 \\
& \textbf{0.05} & 7 & 31 & 4.429 & 0.738 & 0.447 & 0.224 & 0.312 & 0.776 \\
& \textbf{0.08} & 6 & 25 & 4.167 & 0.833 & 0.390 & 0.188 & 0.311 & 0.818 \\
& \textbf{0.10} & 4 & 15 & 3.750 & 1.250 & 0.254 & 0.114 & 0.301 & 0.930 \\
& \textbf{0.15} & 4 & 15 & 3.750 & 1.250 & 0.254 & 0.114 & 0.301 & 0.930 \\
\bottomrule
\end{tabular}
}
\end{table}

This section discusses the experimental outcomes of the proposed Skip Miner algorithm in comparison to other simplification methods.
Tables \ref{tbl:skipminer:results_d1} to \ref{tbl:skipminer:results_d5} show the results of the experiments, for each practitioner $D_1$ to $D_5$, respectively.
With regards to the results obtained, it can be observed that the increase of the \textit{LoS} value leads to a smooth simplification of the structural properties of the graphs.
As expected, both the number of nodes and edges decreases in comparison to the original model.
Besides, this phenomenon conducts to decreasing the proportion of edges per node and the graphs density.
This might enhance the readability of graphs.
Indirectly, as \textit{LoS} increases, the simplified models introduce more distortion since more events are skipped.
The behaviour of the four similarity measures is similar in all cases, thus showing a uniform tendency.
Considering \textit{LoS} $= 1$, it can be noticed that the distortion introduced might be already quite significant in some cases, but this distortion does not grow at the same pace when increasing the value of \textit{LoS}.
Hence, Skip Miner has a fast initial effect on the quality of the simplified process models, but it then stabilises when the models are more simplified.

The use of the EdgeFil method leads to a decrease of the number of edges in the simplified models when value $\alpha$ increases.
But, the number of nodes remains intact.
Accordingly, the proportion of edges per node and the graphs density decreases as well.
Although the importance of reducing less significant edges, the main drawback of this method is its inability to simplify nodes from the model, especially those less relevant.
As this method is designed to remove low-weight edges, the information loss in the simplified process models is not dramatic and the quality might be relatively similar to the original model.
However, similarity measures do not have the same tendency when $\alpha$ increases: for instance, the similarity results are better using the VR measure than the WD measure, this is because the first measure gives more emphasis on the nodes differences (less affected) and the second on the edges differences (more affected).

With regards to the NodeFil method, the main purpose of simplifying less relevant nodes leads to a rapid diminishing on the number of nodes (and edges, in consequence) in the simplified process models when increasing the value of $\beta$.
Indeed, even for small $\beta$ values (\eg $\beta = 0.03$), the number of nodes is halved in some cases.
This means that many activities are conducted in a small number of cases, which may not be relevant when representing the overall behaviour.
In contrast to the other simplification methods, the drastic removal of nodes increments the proportion of edges per node and the density of the graphs.
Thus, the resulting graphs are barely complete, with very few nodes and only the most relevant edges.
Although the quality results are quite notable for small $\beta$ values, as long as $\beta$ value increases, the extreme simplification leads to a significant loss of quality.
Even, in extreme cases, the similarity between the original and the simplified model is almost negligible.

Overall, the largest simplifications of process models are achieved using the NodeFil method, whose topology is largely affected, but at the price of sacrificing their quality.
In this sense, both Skip Miner and EdgeFil methods simplify the structure of the process models in a smoother way, even though EdgeFil is not capable of removing nodes.
From the quality perspective, the best results are achieved with the EdgeFil method, because the simplification, which is performed after the discovery of the process model (\textit{a-posteriori} method), is designed to remove those elements with less relevance, hence reducing its impact.
Comparing the Skip Miner and NodeFil methods, whereas results worsen steadily using Skip Miner, they worsen rapidly when using the NodeFil method.
According to the results obtained, we can easily observe that NodeFil is suitable to achieve simplicity and EdgeFil to achieve quality.
However, they both have problems in achieving good results in the other property.
In this sense, Skip Miner provides an acceptable trade-off between simplicity and quality.

\section{Conclusions} \label{sec:skipminer:concl}

The need for sustainable service models has fostered organisations to put more attention on the correct management, design, execution and monitoring of their business processes.
To conduct powerful and meaningful analysis, the use of process discovery techniques to visualise the actual execution of processes has emerged.
However, the process models discovered might sometimes be extremely complex and difficult to understand due to their lack of structure.
This kind of processes, called spaghetti processes, hinder the ability to make the proper strategic decisions for the organisations.

In this chapter, we have proposed a process discovery algorithm, called Skip Miner, that contains a novel simplification heuristic to bring structure to the process models discovered directly from the event logs.
In contrast to classical simplification methods, this algorithm uses a non-deterministic approach to predict which events are more likely to convert the process model into a spaghetti-like process model, so that they could be discarded from the discovery and visualisation phase.
This method has been evaluated with a real-life medical event log containing some spaghetti processes, and it has been compared with two simplification methods from the literature.
Experimental results demonstrated the ability of Skip Miner for finding a good balance between simplification and process models quality.

\chapter{A Uniformisation Method to Privacy-Preserving Process Mining} 
\chaptermark{A Uniformisation Method to PPPM}
\label{chap:upppm}

\emph{Privacy-preserving process mining (PPPM) is a young and emerging research direction within process mining. With the aim to prevent disclosing confidential information to unauthorised parties during process mining analyses, the use of privacy-enhancing techniques becomes paramount. This chapter presents a novel privacy-preserving process mining method, called \textit{u}-PPPM, based on the uniformisation of confidential distributions of event data attributes that might be exploited to re-identify people by means of distribution-based attacks in combination with location-oriented targeted attacks. First, in Section \ref{sec:upppm:attacks}, we present an attack able to re-identify individuals from event logs gathered in public areas that were previously pseudonymised or encrypted. Next, Section \ref{sec:upppm:method} details the proposed \textit{u}-PPPM method, which aims to produce privacy-preserved versions of event logs, in such a way that the statistical distributions of their attributes become uniform according to a privacy threshold. Then, Section \ref{sec:upppm:experiments} describes the evaluation methodology followed to assess the impact of our approach using six real-life event logs. Results, discussed in Section \ref{sec:upppm:disc}, measure the distortion introduced to the protected process models for different privacy thresholds. Finally, the chapter closes in Section \ref{sec:upppm:concl} with some final remarks.}

\minitoc

\section{Distribution-based Attacks} \label{sec:upppm:attacks}

GDPR recommends the use of pseudonymisation and encryption to obfuscate the link between personally identifiable information and confidential data.
Although common within the PPPM field, these techniques do not offer strong protection against certain attacks.
First, Section \ref{subsec:upppm:vuln} demonstrates the potential privacy issues that could emerge when releasing pseudonymised or encrypted event logs.
Then, in Section \ref{subsec:upppm:attacker}, we formalise an attacker model that enables the re-identification of individuals, by exploiting the distribution of attributes from this kind of event logs when conducting location-oriented attacks.
Institutions with free-public access (\eg hospitals, emergency units, banks, governmental institutions\ldots) are susceptible to this kind of attacks and, when attackers have enough background knowledge, the privacy of the individuals involved might be at stake.

\subsection{Vulnerabilities of Pseudonymised or Encrypted Event Logs} \label{subsec:upppm:vuln}

Let $\mathcal{L}$ be an event log describing the activities carried out by people in an institution with free-public access that manages confidential data (\eg employees from a public hospital, public administration\ldots).
It is apparent that $\mathcal{L}$ cannot be released or shared unmodified, because it associates confidential data to personally identifiable information.
In this situation, institutions can opt for different strategies to create a privacy-preserved event log $\mathcal{L}'$.
A straightforward solution would be the suppression of personally identifiable information and/or the confidential data from $\mathcal{L}$, as suggested in \cite{fahrenkrog2019pretsa,pika2020privacy}.
Notwithstanding, if this suppression is applied, the utility of the event data would decrease dramatically, and the number of process mining analyses would be limited.
For instance, resources-oriented analyses, performance analyses or organisational analyses would not be possible anymore.

To prevent this, in accordance to GDPR recommendations \cite{eugdpr,mannhardt2018privacy,spindler2016encryption} and previous PPPM literature \cite{burattin2015toward,pika2020privacy,rafiei2018supporting,tillem2016privalpha}, institutions might take advantage of pseudonymisation or encryption techniques to distort personally identifiable information (among other attributes), and therefore break the linkage between personally identifiable information and confidential data.
Hence, a protected event log $\mathcal{L}'$ with unreadable personally identifiable information associated to confidential data would be released.
However, this kind of strategies present a potential weakness that, if properly exploited, would enable the re-identification of individuals.

\begin{table}[b!]
\centering
\caption{Sample of an event log $\mathcal{L}$ considering confidential data from a medical institution, and the resulting event log $\mathcal{L}'$ with the personally identifiable information encrypted or pseudonymised.}
\label{tbl:upppm:log}
\resizebox{\textwidth}{!}{
\begin{tabular}{ccccccc}
\toprule
\multicolumn{7}{c}{\textbf{Original event log $\mathcal{L}$}} \\ \midrule
\textbf{Event ID} & \textbf{Case ID} & \textbf{Patient} & \textbf{Physician} & \textbf{Activity} & \textbf{Timestamp} & \textbf{Disease} \\ \midrule
25037 & 2021TGC365 & Tony Green & Seth Murray & Start surgery & 2021-04-25 17:03:54 & Fracture \\
25038 & 2021CZE652 & Alice Fisher & Tom Adams & Mammogram & 2021-04-25 17:12:05 & Cancer \\
25039 & 2021KPI825 & Peter Brown & Bob West & Admission & 2021-04-25 17:15:26 & Flu \\
25040 & 2021CZE652 & Alice Fisher & Tom Adams & Discharge & 2021-04-25 17:26:20 & Cancer \\
25041 & 2021KPI825 & Peter Brown & Bob West & Prescription & 2021-04-25 17:32:53 & Flu \\
25042 & 2021KPI825 & Peter Brown & Bob West & Discharge & 2021-04-25 17:34:17 & Flu \\
25043 & 2021TGC365 & Tony Green & Seth Murray & End surgery & 2021-04-25 17:58:10 & Fracture \\ \bottomrule \\ \toprule
\multicolumn{7}{c}{\textbf{Event log $\mathcal{L}'$ with the personally identifiable information distorted}} \\ \midrule
\textbf{Event ID} & \textbf{Case ID} & \textbf{Patient} & \textbf{Physician} & \textbf{Activity} & \textbf{Timestamp} & \textbf{Disease} \\ \midrule
25037 & 2021TGC365 & Ok4\&ff)785 & 38-!bC3\_El68 & Start surgery & 2021-04-25 17:03:54 & Fracture \\
25038 & 2021CZE652 & L5g62;b98 & 1Ba\_:5Dfa/2 & Mammogram & 2021-04-25 17:12:05 & Cancer \\
25039 & 2021KPI825 & B6*fMs3-fc & 36:\%7c2\_f!7E & Admission & 2021-04-25 17:15:26 & Flu \\
25040 & 2021CZE652 & L5g62;b98 & 1Ba\_:5Dfa/2 & Discharge & 2021-04-25 17:26:20 & Cancer \\
25041 & 2021KPI825 & B6*fMs3-fc & 36:\%7c2\_f!7E & Prescription & 2021-04-25 17:32:53 & Flu \\
25042 & 2021KPI825 & B6*fMs3-fc & 36:\%7c2\_f!7E & Discharge & 2021-04-25 17:34:17 & Flu \\
25043 & 2021TGC365 & Ok4\&ff)785 & 38-!bC3\_El68 & End surgery & 2021-04-25 17:58:10 & Fracture \\ \bottomrule
\end{tabular}
}
\end{table}

Despite the distortion of personally identifiable information in $\mathcal{L}'$, the appearance distribution of attributes' values is not affected (although the actual values in $\mathcal{L}'$ seem unreadable).
This is, a certain text $x$ in the original event log is always replaced by the same unreadable text $x'$ in the pseudonymised/encrypted event log.
So, anyone could figure out that all events with the same value $x'$ would belong to the same individual, even though his/her identity is unknown.
For instance, if \textit{Alice Fisher} appears in $\mathcal{L}$ a total of $n$ times, then the pseudonym or the ciphertext associated with \textit{Alice Fisher} appears a total of $n$ times in $\mathcal{L}'$ too.
Table \ref{tbl:upppm:log} exemplifies this fact in an event log sample.
Although appearance distributions might not seem to be harmful, attackers could exploit them meaningfully, opening the door to the re-identification of individuals.
Trying to break this direct correlation, one could use dynamic encryption or dynamic pseudonymisation, where multiple values $\{x'_1,x'_2,\ldots,x'_m\}$ are assigned in $\mathcal{L}'$ to the same value $x$ in $\mathcal{L}$.
From a privacy perspective, this approach is a clear improvement.
However, it complicates process mining extremely, especially when grouping events/traces from the same individual is required so as to discover specific individuals processes.
This is why solutions found in the literature proposing these mechanisms use static approaches.

\subsection{Attacker Model} \label{subsec:upppm:attacker}

Let's assume that an attacker has gained access to an organisation's event log $\mathcal{L}'$ with its personally identifiable information pseudonymised or encrypted, because either $\mathcal{L}'$ was released for transparency purposes, shared with another organisation, or obtained through malicious ways (\eg data theft).
The attacker aims to exploit such information for disclosing private information.

By analysing the event data in $\mathcal{L}'$ (\eg corresponding to the activity in a public hospital), it is likely that the attacker realises that some people do more activities or participate in more cases than others (\eg some physicians are likely to have more patients --cases-- or do more activities --events-- than other physicians).
Knowing that the distribution of the values remains unaltered between $\mathcal{L}$ and $\mathcal{L}'$, the attacker could model the frequency of activity of each person $I'$ in $\mathcal{L}'$.
Therefore, he/she is able to know which are the pseudonyms/ciphertexts with more and less activity in~$\mathcal{L}'$.

With this knowledge, then the attacker is able to conduct a location-oriented targeted attack based on restricted space identification (RSI) and object identification (OI), \ie well-known attacks (especially against Location-Based Services) that imply the direct or approximate contact with the targets.
The attacker stays physically in the institution (\eg the waiting room of the hospital) and annotates the frequency of activity of each physician (\eg how many patients each physician has received).
By extending this attack for a reasonable period of time, the attacker is able to know which physicians are more active within the institution.
Finally, the attacker could infer from his/her observations that the physician with more activity corresponds to the pseudonym/ciphertext from $\mathcal{L}'$ with more appearances, and so on.
By following this strategy, the attacker is able to infer the correlation between the pseudonyms/ciphertexts associated to people in $\mathcal{L}'$ to real people (\ie identity disclosure).
In addition, the attacker could also infer their confidential data: those confidential data associated with the corresponding pseudonym/ciphertext from $\mathcal{L}'$ (\ie attribute disclosure).

This attack, illustrated in Figure \ref{fig:upppm:attacker}, demonstrates that static data transformations, although popular in practice, might not be enough to preserve people's privacy.

\begin{figure}[t]
\centering\includegraphics[width=0.87\linewidth]{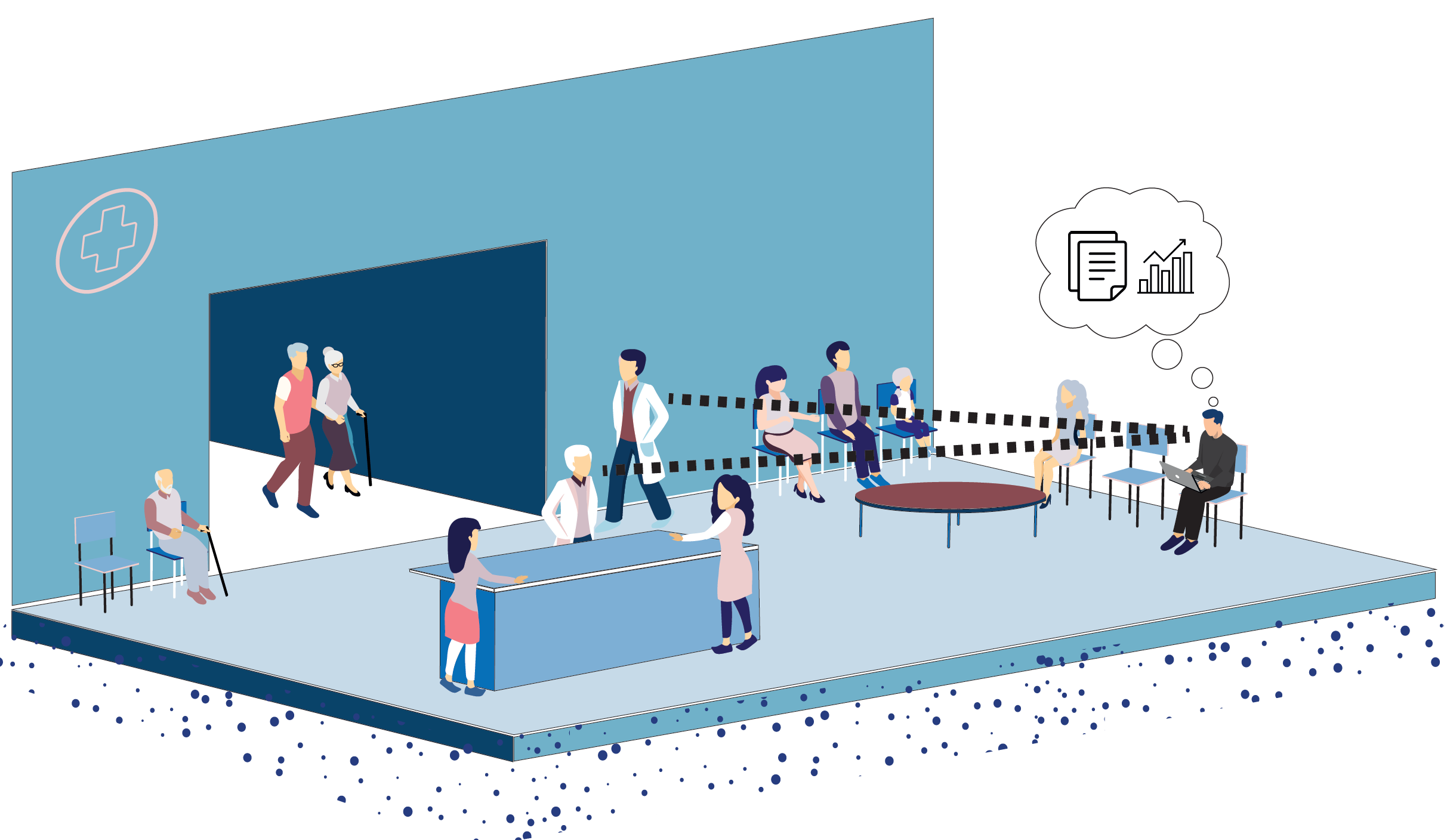}
\caption{Scenario of an RSI/OI attack in the waiting room of a public hospital (adapted from \cite{batista2021uniformization}).}
\label{fig:upppm:attacker}
\end{figure}

\section{\textit{u}-PPPM: The Uniformisation Method} \label{sec:upppm:method}

The distribution of attributes' values in an event log is a key factor to consider at the time of protecting individuals' privacy.
As a countermeasure against possible attacks in which classical encryption and pseudonymisation techniques fail, in this section we propose \textit{u}-PPPM, a novel PPPM method based on the uniformisation of potentially identifiable distributions from an event log.
Assuming that attackers are capable of inferring the frequency/distribution of individuals' activities (\eg the number of cases an individual participates on, or the number of activities an individual performs), the proposed method distorts the event data in $\mathcal{L}$ to create a privacy-preserved event log $\mathcal{L}'$, in such a way that renders the distribution knowledge acquired by the attackers useless.
Hence, the event data in $\mathcal{L}'$ prevents the direct re-identification of individuals and averts distribution-based attacks.

The basis of the proposed method is to group similar individuals from $\mathcal{L}$ in groups of size $k$, being $k$ a privacy threshold, and exchange events among the individuals in the same group, until all individuals within the same group are uniform from a distribution perspective.
This exchange procedure is repeated for all the groups.
As a result, given a certain group, all their $k$ individuals are indistinguishable to an attacker.
Finding a good balance between information loss (caused by distortion/exchanges of event data) and privacy (achieved by uniformisation) is of utmost importance to preserve the quality of the process models.

The implementation details and the main design decisions of the \textit{u}-PPPM method are described in Section \ref{subsec:upppm:algorithm}.
Then, an illustrative example of the functioning of the method is explained in Section \ref{subsec:upppm:example}.
Finally, Section \ref{subsec:upppm:security} discusses the security and privacy enhancements of the event logs protected using the \textit{u}-PPPM method in comparison to classical approaches.

\subsection{Algorithm Details} \label{subsec:upppm:algorithm}

\begin{algorithm*}[b!]
\footnotesize
\begin{algorithmic}[1]
\Require
\Statex $\mathcal{L}$ is a non-empty event log describing the activities of $p$ individuals, where $p > 0$.
\Statex $k$ is a privacy level in the range from 2 to $p$.
\Statex $sel$ is a selection strategy.
\Ensure
\Statex $\mathcal{L}'$ is a non-empty event log, whose distribution (\ie frequency of activity) is indistinguishable among $k$ individuals.
\Function{\textit{u}-PPPM}{\textbf{EventLog} $\mathcal{L}$, \textbf{Integer} k, \textbf{Strategy} $sel$}
\State $\mathcal{L} \gets$ orderByFrequencyOfActivity($\mathcal{L}$);
\State \textbf{List$\langle$List$\langle$Individual$\rangle\rangle$} groups $\gets$ createGroups($\mathcal{L}, k$);
\ForAll{group in groups}
 \While{group \textbf{is not} uniformed}
  \State \textbf{Individual} $I_{pv}, I_{rc} \gets$ selectIndividuals(group, $sel$);
  \State \textbf{List$\langle$Event$\rangle$} events\_to\_exchange $\gets$ selectRandomEvents($I_{pv}$);
  \ForAll{event in events\_to\_exchange}
   \State $\#_{resource}($event$) \gets I_{rc}$;
  \EndFor
 \EndWhile
 \State $\mathcal{L} \gets$ updateGroup(group);
\EndFor
\State \textbf{EventLog} $\mathcal{L}' \gets$ obfuscatePII($\mathcal{L}$);
\State \Return $\mathcal{L}'$;
\EndFunction
\end{algorithmic}
\caption{\textit{u}-PPPM algorithm}
\label{alg:upppm:algorithm}
\end{algorithm*}

Let $\mathcal{L}$ be an event log describing the activities of $p$ individuals, $\mathcal{I} = \{I_1,I_2,\ldots,I_p\}$, where $\mathcal{I}$ is the set of all individuals in $\mathcal{L}$.
Each individual $I \in \mathcal{I}$ participates in a number of cases in $\mathcal{L}$, namely $c_I$, where $c_I > 0$.
Therefore, the frequency of activity per individual, namely $D$, can be known as $D = \{c_{I_1},c_{I_2},\ldots,c_{I_p}\}$.
It is worth noting that $D$ is the distribution knowledge that attackers could infer from location-oriented attacks and distribution-based attacks.
The privacy-preserved event log $\mathcal{L}'$, created by \textit{u}-PPPM, is characterised for having a distorted distribution $D'$ in such a way that follows a uniform distribution within groups of $k$ individuals, this is, groups of $k$ individuals will have the same frequency of activity.
The outline of the algorithm is provided in Algorithm \ref{alg:upppm:algorithm}.

\begin{figure*}[b!]
 \centering
 \begin{subfigure}{0.49\textwidth}
   \includegraphics[width=\textwidth]{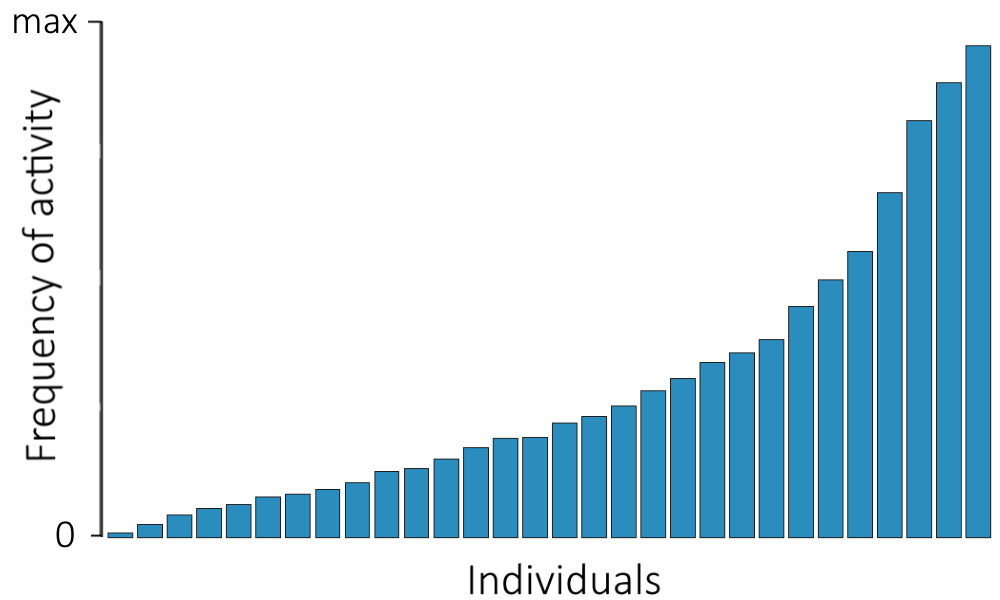}
   \caption{Original distribution in $\mathcal{L}$}
 \end{subfigure}
 \begin{subfigure}{0.49\textwidth}
   \includegraphics[width=\textwidth]{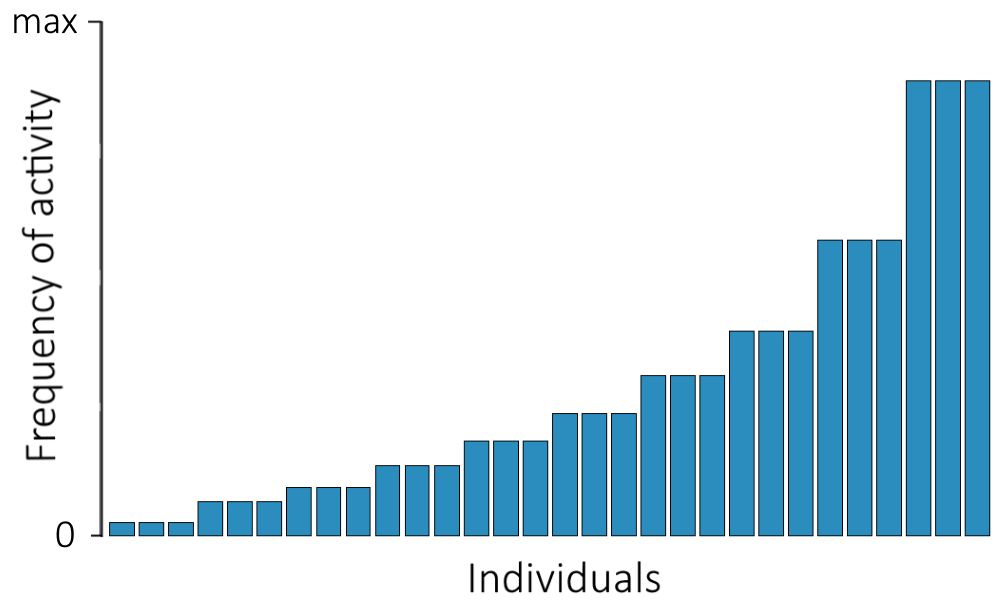}
   \caption{Distorted distribution in $\mathcal{L}'$ for $k = 3$} 
 \end{subfigure}
 \par\medskip 
  \begin{subfigure}{0.49\textwidth}
   \includegraphics[width=\textwidth]{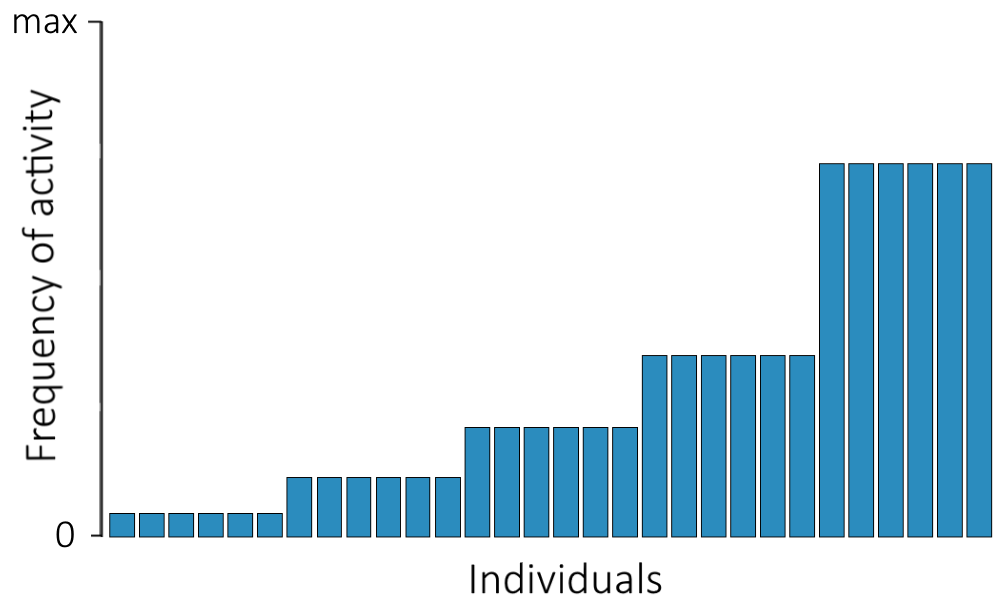}
   \caption{Distorted distribution in $\mathcal{L}'$ for $k = 6$}
 \end{subfigure}
 \begin{subfigure}{0.49\textwidth}
   \includegraphics[width=\textwidth]{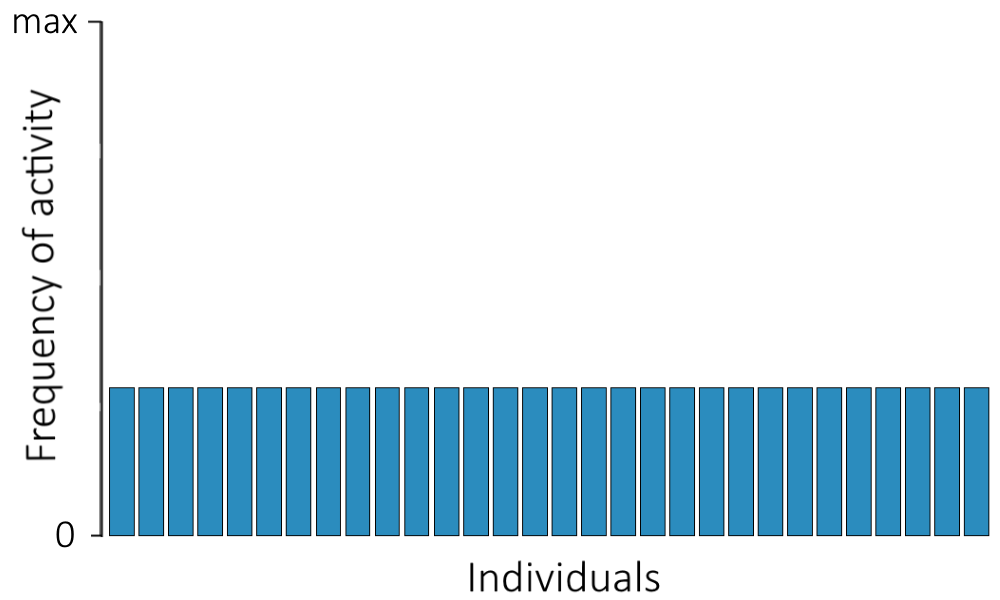}
   \caption{Distorted distribution in $\mathcal{L}'$ for $k = p$}
 \end{subfigure}
 \caption{Distortion of the frequency of activity per individual according to $k$. Each bar represents the frequency of activity of a certain individual.}
 \label{fig:upppm:distribution}
\end{figure*}

The anonymisation procedure conducted in \textit{u}-PPPM requires setting a \textit{privacy level}.
This privacy level is directly related to the individuals' group size $k$, where $k \geq 2$: the higher $k$, the more privacy is achieved.
Indeed, the maximum privacy level is achieved when $k = p$, this is, all individuals are encompassed within a single group.
From a distribution perspective, all the individuals share the same frequency of activity and, hence, are indistinguishable from the attacker's perspective.
However, higher privacy levels imply more exchanges of events and a higher distortion of the original distribution, which negatively affect the quality of the process models to be discovered.
It can be easily observed that the number of groups created is equal to $\lfloor p/k \rfloor$.
Regarding the creation of these groups, it is well-known that grouping similar individuals allows the creation of homogeneous groups and, hence, helps to reduce information loss.
In our method, the similarity between individuals is measured according to the individual's frequency of activity in $\mathcal{L}$,\ie the knowledge attackers try to exploit.
Hence, from this perspective, we assume that $k$ individuals are similar if their frequency of activity are similar.
Since privacy is guaranteed by making this distribution uniform, this criteria allows minimising the distortion of the event data, since fewer exchanges of events are required to reach a uniform state within each group.
Figure \ref{fig:upppm:distribution} depicts the impact of $k$ to the distribution knowledge that attackers could acquire.
As aforementioned, although larger groups contribute to preserve individuals' privacy because re-identification is more challenging, the event log suffers a higher distortion that worsens the quality of the process mining results.

When all groups are determined, the distribution of each group must be uniformed.
To do this, events/cases are interchanged between the individuals belonging to the same group until these $k$ individuals share the same frequency of activity.
To this end, first a \textit{selection strategy} ($sel$) for choosing two individuals within the same group must be defined.
In particular, an individual is selected as a \textit{provider} ($I_{pv}$) of events and another individual is selected as a \textit{receiver} ($I_{rc}$) of events.
Efficient selections of individuals allow minimising the number of events exchanges, which helps reduce the distortion introduced to the event log $\mathcal{L}'$.
Although any selection strategy could be used in \textit{u}-PPPM, four different strategies, namely $S_1, S_2, S_3$ and $S_4$, are suggested below.

\begin{itemize}
    \item \textit{$S_1$ -- Roulette-wheel}: Each individual has a probability of being selected, according to their need for providing events to or receiving events from the other members of the group. Individuals with a higher-than-average activity within the group have a larger probability of being selected as providers, while individuals with a lower-than-average activity within the group have a higher probability of being selected as receivers.
    \item \textit{$S_2$ -- Max-Min}: The individual with the highest frequency of activity in the group is chosen as the provider, and the individual with the lowest frequency of activity in the group is chosen as the receiver.
    \item \textit{$S_3$ -- Random}: Individuals are randomly selected without considering their needs to be providers or receivers. This is a blind strategy that does not consider the actual distribution/frequency of activity.
    \item \textit{$S_4$ -- Lateral}: Individuals within the group are sorted in descending order according to their frequency of activity. The first individual (\ie the one with the highest activity) provides events to the second individual until their frequency of activity is uniform. Next, the third individual acts as receiver from the previous two, until the three individuals are uniform. This procedure is repeated for all the $k$ individuals within the group until all individuals activity distributions are uniform.
\end{itemize}

Once the two individuals, $I_{pv}$ and $I_{rc}$, are selected by following any of the above strategies, a number of events are exchanged from $I_{pv}$ to $I_{rc}$ so that their frequency of activity becomes uniform.
For instance, consider a group of $k = 2$ comprised by individuals $I_{8}$ and $I_{13}$ that are responsible for 35 and 49 cases, respectively, in a given event log $\mathcal{L}$.
In this case, $I_{13}$ would be the provider $I_{pv}$, and $I_{8}$ would be the receiver $I_{rc}$.
Then, the events associated to 7 cases from $I_{13}$ would be assigned to $I_{8}$, so both individuals would have 42 cases each in $\mathcal{L}'$.
Therefore, $I_{8}$ and $I_{13}$ would be indistinguishable from the attacker's distribution point of view in $\mathcal{L}'$.
With regards to the events exchange, the very \textit{selection of the events} to be exchanged from $I_{pv}$ to $I_{rc}$ is done at random.
This results in \textit{u}-PPPM being a non-deterministic method.
In a nutshell, the exchange of events among individuals results in the modification of their individuals' information in the event data (\ie the information stored in $\#_{resource}(e), \forall e \in \mathcal{L}$ to be exchanged).
Once all groups are made uniform, \textit{u}-PPPM obfuscates the personally identifiable information associated to the event data by means of pseudonymisation or encryption (like other methods in the literature).
Finally, the method returns the created privacy-preserved event log $\mathcal{L}'$.

The distortion introduced to the protected event logs $\mathcal{L}'$ implies a worsening on the quality of the process models to be discovered with process mining.
More specifically, \textit{u}-PPPM directly affects the quality of the process models associated to each individual, which is important in some analyses used to evaluate the performance or behaviour of specific people.
Therefore, given an individual $I_{i}$ ($1 \leq i \leq p$), the original process model $\mathcal{M}_{I_i}$ discovered from $\mathcal{L}$ will be different to the process model $\mathcal{M}'_{I_i}$ discovered from $\mathcal{L}'$ after applying \textit{u}-PPPM, because $\mathcal{M}'_{I_i}$ is affected by 
(i) the loss of events that $I_{i}$ has provided to the rest of the $k - 1$ individuals within the same group,
and (ii) the events that $I_{i}$ has received from the rest of the $k - 1$ individuals within the same group.
Indeed, to the best of our knowledge, in the area of process mining, this is the very first method that focuses on the quality of the discovered individuals' process models once applying a privacy-preserving technique on event data.

\subsection{Illustrative Example} \label{subsec:upppm:example}

Figure \ref{fig:upppm:example_orig} provides an illustrative example of an event log $\mathcal{L}$ describing the activities performed by four individuals (Bob, Pete, Marie and Sam) in five independent cases ($t_1,\ldots,t_5$) in chronological order.
For the sake of brevity, $\mathcal{L}$ only shows the most important attributes for this example and other attributes, such as the event identifier, the timestamp and the confidential data, have been omitted.
It can be observed that some individuals are responsible for more cases than others: Bob participates in the five cases, Pete and Marie participate in three cases each, and Sam participates in only one case.
If their names were only replaced by pseudonyms or encrypted, attackers could exploit this distribution knowledge to re-identify them through location-oriented attacks.

The impact of \textit{u}-PPPM on the privacy-preserved event logs $\mathcal{L}'$ and their distribution knowledge according to the privacy level $k$ can be observed in Figure \ref{fig:upppm:example_prot}.
More specifically, the example illustrates the impact for $k = 2$ and $k = 4$.
The differences produced by \textit{u}-PPPM with regards to the original event log $\mathcal{L}$ are highlighted in blue.
By applying \textit{u}-PPPM with a privacy level $k = 2$, the method creates two groups of two individuals each according to the similarity of their frequencies of activity.
Hence, Bob and Pete belong to a group, and Marie and Sam belong to the other group.
Then, to uniform each group, the events corresponding to a (random) case is exchanged from Bob to Pete (for instance, the three events from $t_3$), and the events corresponding to a (random) case is exchanged from Marie to Sam (for instance, the two events from $t_5$).
As a result, Bob and Pete are now responsible for four cases each, and Marie and Sam for two cases each, thus being indistinguishable in $\mathcal{L}'$ for the distribution perspective.
If applying \textit{u}-PPPM with a privacy level $k = 4$, then a single group with all four individuals would be created.
In this scenario, it is likely that Bob provides the events corresponding to two (random) cases to Sam (for instance, the events from $t_1$ and $t_5$).
As a result, all four individuals appear to be responsible for three cases each in $\mathcal{L}'$, thus preventing their distribution-based re-identification.

\begin{figure}[t]
\centering\includegraphics[width=0.99\linewidth]{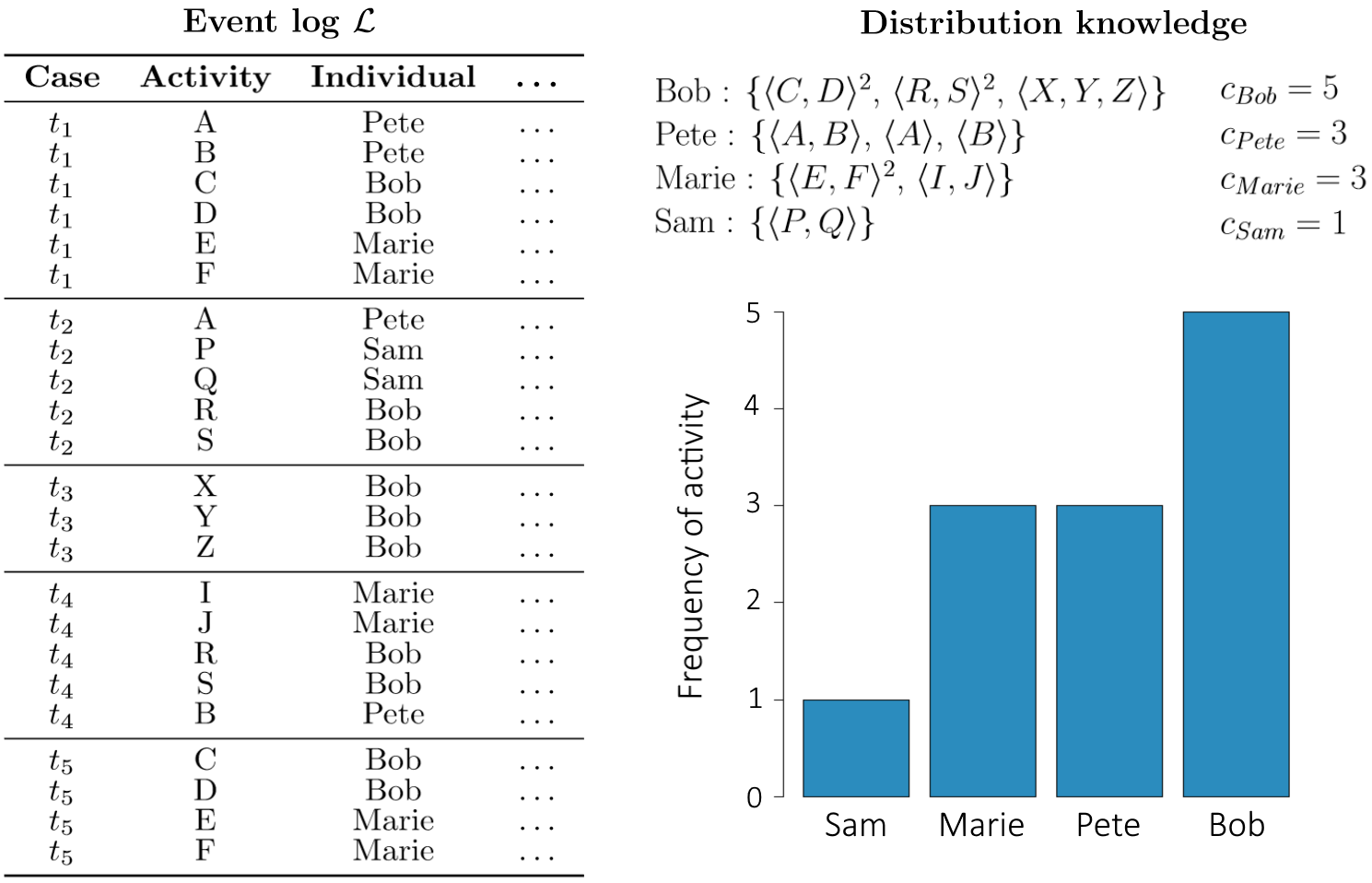}
\caption{Toy example of an event log $\mathcal{L}$ with its distribution information (adapted from~\cite{batista2021uniformization}).}
\label{fig:upppm:example_orig}
\end{figure}

\begin{figure}[t!]
\centering\includegraphics[width=0.99\linewidth]{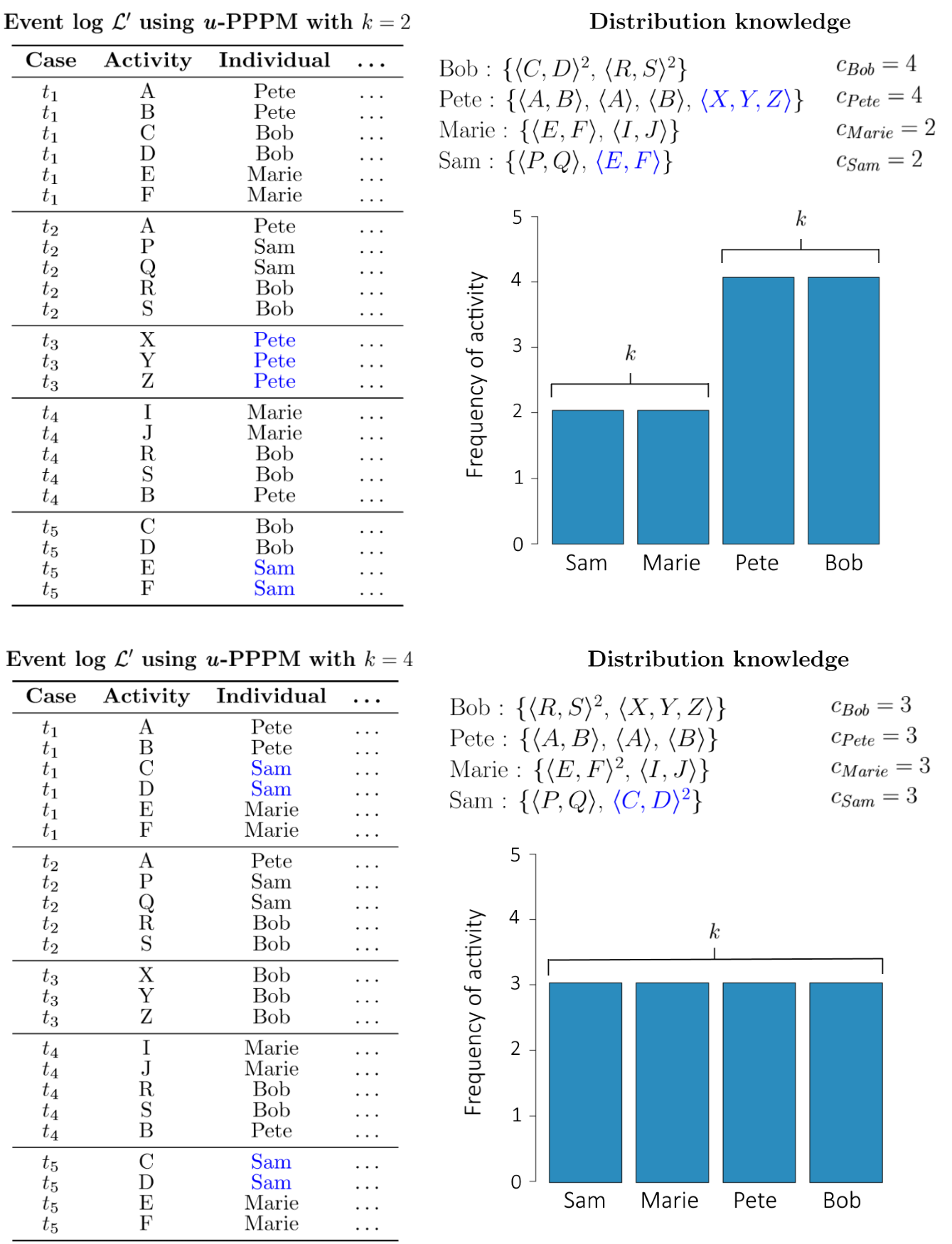}
\caption{Impact of \textit{u}-PPPM on the privacy-preserved event logs $\mathcal{L}'$ and their distribution knowledge for $k = 2$ and $k = 4$ (adapted from \cite{batista2021uniformization}).}
\label{fig:upppm:example_prot}
\end{figure}

\subsection{Security and Privacy Analysis} \label{subsec:upppm:security}

This section provides a security analysis of the proposed \textit{u}-PPPM method in terms of confidentiality and privacy enhancements.
Other security requirements, such as integrity, availability and authentication, are beyond the scope of this method.
Thanks to the offline nature of the method, no online communication with external parties is required to conduct the anonymisation procedure.
In this sense, the avoidance of communications with further computer systems relaxes the potential security threats, and the security of the method resides in the distortion of the statistical properties of the event data introduced by the very method.
For the sake of completeness, we compare the security of the privacy-preserved event logs obtained from the proposed method in comparison to the event logs that can be obtained by means of pseudonymisation or encryption mechanisms.
Hence, let $\mathcal{L}$ be the original (unprotected) event log, we define $\mathcal{L}'$ as a privacy-preserved event log version of $\mathcal{L}$ obtained from \textit{u}-PPPM, and $\mathcal{L}''$ as a protected event log version of $\mathcal{L}$ with the personally identifiable information pseudonymised or encrypted.

The confidentiality of \textit{u}-PPPM is supported by the obfuscation of personally identifiable information by means of pseudonymisation or state-of-the-art encryption techniques at the last stage of the algorithm.
These techniques ensure the confidentiality of $\mathcal{L}'$ as long as the private cryptographic keys remain secret.
Hence, users exploiting the data in $\mathcal{L}'$ are not able to decrypt the personally identifiable information and, therefore, it cannot be directly associated to the confidential data in $\mathcal{L}'$.
As both $\mathcal{L}'$ and $\mathcal{L}''$ rely on pseudonymisation or encryption, the confidentiality level of both approaches is the same.

The main differentiating factor of the proposed method is the added privacy guarantees that are not supported when using pseudonymisation or encryption only.
Comparing the event data in $\mathcal{L}'$ and $\mathcal{L}''$, the knowledge that attackers gain from the distribution is significantly different.
Whereas the attributes' distribution has not been altered in $\mathcal{L}''$, the attackers distribution knowledge has been limited in $\mathcal{L}'$ and re-identification is constrained to groups of $k$ individuals.
Formally, whereas $\mathcal{L}''$ describes its attributes with an unknown distribution (\eg normal, Poisson\ldots) that might be susceptible to location-oriented attacks, \textit{u}-PPPM reshapes this distribution in $\mathcal{L}'$ towards a uniform distribution.
The main advantage of using a uniform distribution is that it renders the distribution knowledge that attackers could gain useless, because groups of $k$ individuals are indistinguishable among them.
In case that an attacker would be able to infer the probability distribution of a given individual in $\mathcal{L}'$, the re-identification risk is upper-bounded by $1/k$.
Figure \ref{fig:upppm:upppm_vs_encryption} depicts this privacy enhancement of \textit{u}-PPPM.

\begin{figure}[b!]
\centering\includegraphics[width=\linewidth]{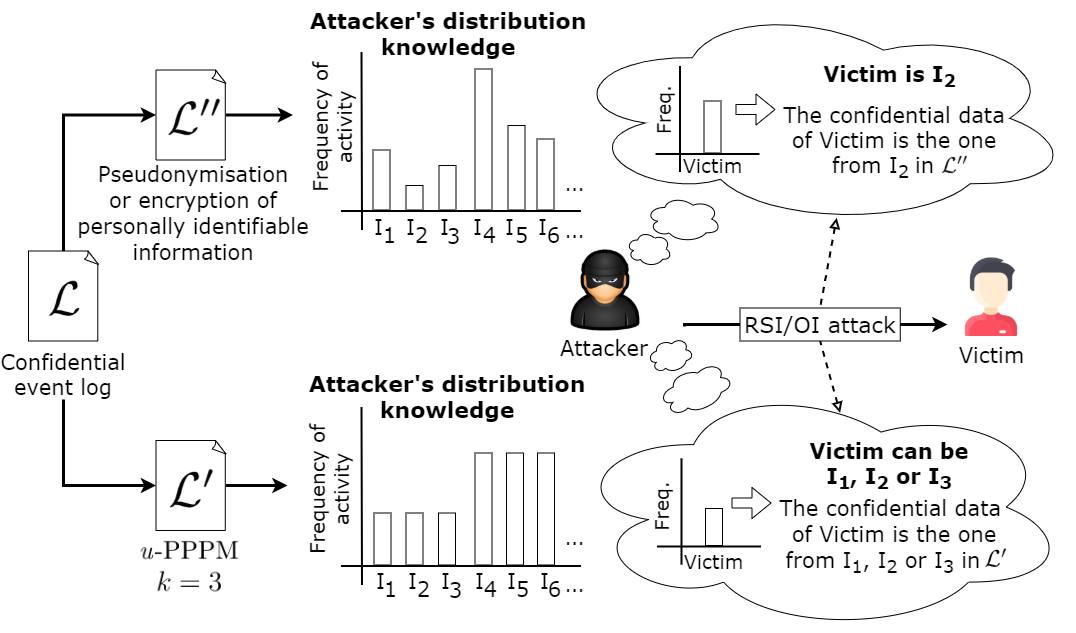}
\caption{Differences, from the attacker's perspective, between the privacy enhancements when using \textit{u}-PPPM or using pseudonymisation or encryption only (adapted from~\cite{batista2021uniformization}).}
\label{fig:upppm:upppm_vs_encryption}
\end{figure}

Although there might be some resemblance between the proposed method and those studied in related privacy fields (\eg SDC), it is worth emphasising that \textit{u}-PPPM can hardly be compared to the existing SDC privacy-preserving models, namely \textit{k}-anonymity or \textit{l}-diversity.
First, SDC-related protection models aim to protect microdata sets by modifying their statistical properties with an eye on the utility of the data sets themselves.
However, \textit{u}-PPPM focuses on minimising the disclosure risks against location-oriented attacks, and evaluates the utility of the discovered process models.
Second, data sources in SDC protection models and \textit{u}-PPPM are different: while the records in a microdata set belong to different individuals, the records in an event log could eventually belong to the same individual.
Besides, in an event log, records have a relationship among them, but records are independent in a microdata set.
Certainly, similarities between \textit{u}-PPPM and the \textit{k}-anonymity model could be found.
Although both approaches create clusters of $k$ individuals/records as homogeneous as possible, \textit{k}-anonymity uses distance functions to create these groups, while \textit{u}-PPPM does not consider the distance among events, and creates the groups based on the activity distribution knowledge that attackers could gain.
Finally, \textit{k}-anonymity results in $k$ indistinguishable records in the protected microdata set, while \textit{u}-PPPM results in $k$ indistinguishable individuals from a distribution perspective.

\section{Experimental Setup} \label{sec:upppm:experiments}

This section provides the details of the methodology followed to evaluate the proposed \textit{u}-PPPM method.
The main purpose of this evaluation is to test the impact of \textit{u}-PPPM on the quality of the process models discovered from the protected event logs $\mathcal{L}'$, by comparing them with the corresponding process models discovered from the original (and unprotected) event logs $\mathcal{L}$.
More specifically, as \textit{u}-PPPM distorts the individuals' information in the event data, the process models to be evaluated are those associated to each individual, represented from a control-flow perspective.
Therefore, the impact of \textit{u}-PPPM on the quality of the process models has been measured with the aim to answer the following three research questions.

\begin{itemize}
\item \textit{Q1 -- Individual distortion}: How similar are the process models of an individual $I$ when discovered from the original event log (\ie $\mathcal{M}_I$), and when discovered from the protected event log (\ie $\mathcal{M}'_I$)?
\item \textit{Q2 -- Inter-individual distortion}: Are the differences among the individuals' process models from the original event log (\ie $\mathcal{M}_{I_1}, \mathcal{M}_{I_2},\ldots,\mathcal{M}_{I_p}$) also maintained among the individuals' process models from the protected event log (\ie $\mathcal{M}'_{I_1}, \mathcal{M}'_{I_2},\ldots,\mathcal{M}'_{I_p}$)?
\item \textit{Q3 -- Conformance}: How well the event data of an individual $I$ in the original event log (\ie $\mathcal{L}$) conforms with the process model of that individual when discovered from the protected event log (\ie $\mathcal{M}'_I$)?
\end{itemize}

Whereas many evaluation methodologies only focus on the individual distortion of the process models or the conformance, this evaluation methodology aims to provide a holistic view on the impact of \textit{u}-PPPM with regards to the quality of the process models from three independent perspectives.
A graphical summary of the evaluation methodology in accordance with the previous questions is depicted in Figure \ref{fig:upppm:methodology}.

\begin{figure}[t!]
\centering\includegraphics[width=0.81\linewidth]{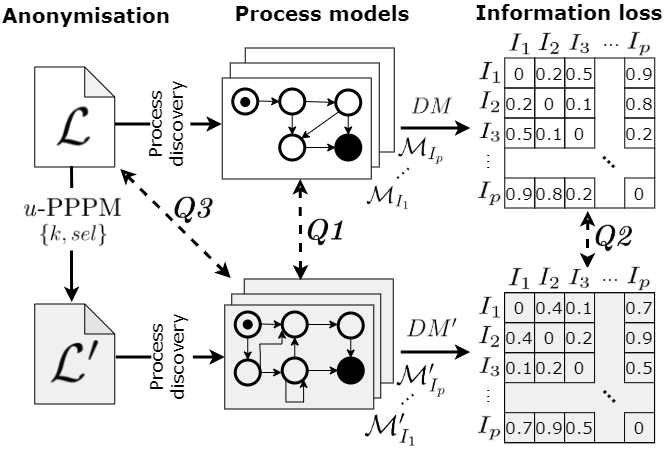}
\caption{Methodology to evaluate the \textit{u}-PPPM method (adapted from~\cite{batista2021uniformization}).}
\label{fig:upppm:methodology}
\end{figure}

Since the evaluation of \textit{u}-PPPM is conducted at the level of the very process models, first they need to be discovered.
In particular, they are represented as D/F-graphs.
This generic and less-restrictive notation enables observing the very impact of our method on the quality of the process models, and avoids the modelling constraints and restrictions introduced by subsequent modelling notations.

The parameters in which \textit{u}-PPPM is executed affect the quality of the obtained process models.
Therefore, our method is tested with different combinations of its two parameters: (i) the size of the groups ($k$), being $k = \{2, 3, 4, 5, 8, 10\}$, and (ii) the selection strategy ($sel$) that corresponds to the aforementioned $S_1, S_2, S_3$ and $S_4$ strategies.
With the aim to observe the impact of these parameters, \textit{u}-PPPM is executed for all combinations of these parameters: ($k_{i}, sel_{i}$), $\forall k_i \in k, \forall sel_{i} \in sel$.
Therefore, each event log $\mathcal{L}$ is anonymised 24 (6 $\times$ 4) times, resulting in 24 different privacy-preserved event log versions $\mathcal{L}'$, achieved from the execution of \textit{u}-PPPM with (2, $S_1$), (2, $S_2$),\ldots,(10, $S_4$).
Besides, experiments have been conducted using the six event logs described in Table \ref{tbl:background:event_logs}.
It is noteworthy that the number of resources in each event log corresponds to $p$.

First, the individual quality of the process models (\textit{Q1}) is evaluated through a model-by-model comparison.
This evaluation determines the distortion that \textit{u}-PPPM introduces to the process model of each individual.
It is useful to estimate how different a given process model is with regards to its original version.
This evaluation procedure works as follows.
For each \textit{u}-PPPM execution with a certain combination of parameters, first the $p$ original process models from $\mathcal{L}$ (\ie $\mathcal{M}_{I_1}, \mathcal{M}_{I_2},\ldots,\mathcal{M}_{I_p}$) and the $p$ protected process models from $\mathcal{L}'$ (\ie $\mathcal{M}'_{I_1}, \mathcal{M}'_{I_2},\ldots,\mathcal{M}'_{I_p}$) are discovered.
Next, the pairs of process models belonging to the same individual (\ie $\{\mathcal{M}_{I_1},\mathcal{M}'_{I_1}\},\{\mathcal{M}_{I_2},\mathcal{M}'_{I_2}\},\ldots,\{\mathcal{M}_{I_p},\mathcal{M}'_{I_p}\}$) are compared using a similarity measure.
To prevent bias in this measure, process models are compared using four independent similarity measures: VEO \cite{papadimitriou2010web}, VR \cite{papadimitriou2010web}, WD \cite{shoubridge2002detection} and DC \cite{koutra2013deltacon}.
Since all measures are bounded between 0 and 1, for the sake of consistency, it has been standardised that total similarity (\ie no distortion) is indicated by 0.
As a result, each comparison of a pair of process models results in four similarity measures, thus obtaining a total of $4 \times p$ similarity values for the complete set of process models to evaluate.
Taking the average of all these values, a \textit{quality score} (QS) can be associated to the \textit{u}-PPPM execution.
This QS value, bounded between 0 and 1 (the lower, the more similar), indicates the averaged distortion that a given \textit{u}-PPPM execution introduces to the protected process models.

Assuming the unavoidable distortion introduced by \textit{u}-PPPM in the process model of each individual, it is also paramount to evaluate how this distortion affects the entire event data and the process models as a whole, this is the inter-individual distortion (\textit{Q2}).
For example, if the process models of two individuals, $\mathcal{M}_{I_1}$ and $\mathcal{M}_{I_2}$, are similar in the original event log $\mathcal{L}$, it is desirable to preserve this similarity in the protected process models, $\mathcal{M}'_{I_1}$ and $\mathcal{M}'_{I_2}$, too.
Achieving a small inter-individual distortion allows gathering similar insights when comparing protected process models from different individuals (\eg for performance analysis) as if using the original process models.
This evaluation procedure works as follows.
For each \textit{u}-PPPM execution with a certain combination of parameters, the distance matrices $DM$ and $DM'$ (both with size $p \times p$) are computed.
These matrices contain the similarity between all pairs of process models in $\mathcal{L}$ and $\mathcal{L}'$, respectively.
As before, the similarity values in $DM$ and $DM'$ are computed using the aforementioned four similarity measures.
Therefore, from an event log $\mathcal{L}$ and a protected event log $\mathcal{L}'$, four distance matrices can be obtained, respectively (\ie $\{DM_\textrm{VEO},\ldots,DM_\textrm{DC}\}$ and $\{DM'_\textrm{VEO},\ldots,DM'_\textrm{DC}\}$).
Then, pairs of matrices calculated with the same similarity measure (\ie $\{DM_\textrm{VEO},DM'_\textrm{VEO}\},\ldots,\{DM_\textrm{DC},DM'_\textrm{DC}\}$) are compared using the well-known Mean Absolute Error (MAE).
The MAE, bounded between 0 and 1, where lower values indicate more similarity, serves as an indicator of the information loss.
Taking the average of the four MAE values, an \textit{information loss score} (ILS) can be associated to the \textit{u}-PPPM execution.
This ILS value, bounded between 0 and 1 (the lower, the more similar), indicates the averaged information loss that a given \textit{u}-PPPM execution introduces to the protected event log in comparison to the original event log.

Last but not least, the quality of the protected process models can also be measured against the original event data (\textit{Q3}), which resembles the conformance checking techniques.
This evaluation shows whether the behaviour of a given individual $I$ in the original event data $\mathcal{L}$ can be replayed in the distorted process model $M'_I$ of that individual.
From a quality perspective, it would be ideal that all (or most) of the original behaviours could be preserved, despite the unavoidable distortion introduced by \textit{u}-PPPM.
This evaluation procedure works as follows.
For each \textit{u}-PPPM execution with a certain combination of parameters, first the $p$ protected process models from $\mathcal{L}'$ (\ie $\mathcal{M}'_{I_1}, \mathcal{M}'_{I_2},\ldots,\mathcal{M}'_{I_p}$) are discovered.
Next, from the original event log $\mathcal{L}$, the different traces of each individual are extracted.
Then, the traces of each individual $I_1,I_2,\ldots,I_p$ are replayed in their corresponding process model $\mathcal{M}'_{I_1}, \mathcal{M}'_{I_2},\ldots,\mathcal{M}'_{I_p}$.
Since processes are modelled as D/F-graphs, a case is replayed only if there exists a graph connection between all the consecutive events in the trace.
By replaying all the traces of an individual, a replay score can be associated to an individual by dividing the number of cases that have been successfully replayed by the total number of traces.
This value, bounded between 0 and 1, shows the degree of conformance of a given individual: high values indicate that the original behaviour is preserved and, therefore, can be replayed in the protected process model.
By taking the average of the replay scores of all the $p$ individuals in $\mathcal{L}$, a \textit{conformance score} (CS) can be associated to the \textit{u}-PPPM execution.
This CS value, bounded between 0 and 1 (the higher, the more conformance), shows the averaged conformance level of the original event data with regard to the protected process models.

\section{Results and Discussion} \label{sec:upppm:disc}

This section discusses the experimental results for assessing the impact of the proposed \textit{u}-PPPM method on the anonymisation of event logs for process mining.
It is worth mentioning that, since the execution of \textit{u}-PPPM is non-deterministic, all the results reported correspond to the average of five executions for each combination of parameters.

All the quantitative results are detailed in Appendix \ref{app:upppm}.
Tables \ref{tbl:app_upppm:results_qs}, \ref{tbl:app_upppm:results_ils} and \ref{tbl:app_upppm:results_cs} describe the QS, ILS and CS results obtained, respectively, for each execution of \textit{u}-PPPM with a certain combination of parameters.
In addition to the six event logs evaluated, an average result from all the event logs is also provided (last column in the tables).
For the sake of comprehensiveness, results are grouped by $k$ and, within each group, the best and the worst QS/ILS/CS results are highlighted in green and red, respectively.
Stated in summary form, Tables \ref{tbl:upppm:summary_qs}, \ref{tbl:upppm:summary_ils} and \ref{tbl:upppm:summary_cs} help evaluate the impact of each parameter in the results.
These tables group and average the QS, ILS and CS results, respectively, according to each parameter value.
As before, the best and the worst results are highlighted in green and red, respectively, within each parameter group.

\begin{sidewaystable}[htbp]
\renewcommand{\tabcolsep}{0.2cm}
\renewcommand{\arraystretch}{0.93}
\centering
\scriptsize  
\caption{Summary of the QS results grouped by parameter.} \label{tbl:upppm:summary_qs}
\begin{tabular}{ccrccccccc}
   \toprule
   &&& \multicolumn{6}{c}{\textbf{Event logs}} & \multirow{2}{*}{\textbf{Avg.}} \\
   \cmidrule(lr){4-9}
   &&& \textbf{BPI12} & \textbf{BPI13} & \textbf{BPI14} & \textbf{BPI15} & \textbf{CoSeLoG} & \textbf{TGN-Hospital} \\
   \midrule
   \multirow{10}{*}{\rotatebox{90}{\textbf{Grouping criteria}}} & \multirow{5}{*}{\textbf{Priv.}} & \textbf{2} & \gr{0.0637 $\pm$ 0.1053} & \gr{0.0027 $\pm$ 0.0299} & \gr{0.1348 $\pm$ 0.1285} & \gr{0.1764 $\pm$ 0.1366} & \gr{0.1112 $\pm$ 0.1389} & \gr{0.0148 $\pm$ 0.048} & \gr{0.084 $\pm$ 0.0979} \\
   & \multirow{5}{*}{($k$)} & \textbf{3} & 0.151 $\pm$ 0.1615 & 0.0046 $\pm$ 0.0427 & 0.192 $\pm$ 0.1474 & 0.2501 $\pm$ 0.1545 & 0.1784 $\pm$ 0.1749 & 0.0242 $\pm$ 0.0674 & 0.1337 $\pm$ 0.1249 \\
   & & \textbf{4} & 0.1879 $\pm$ 0.1763 & 0.0049 $\pm$ 0.0423 & 0.2091 $\pm$ 0.1505 & 0.2899 $\pm$ 0.1685 & 0.2084 $\pm$ 0.1828 & 0.0398 $\pm$ 0.0887 & 0.1563 $\pm$ 0.1347 \\
   & & \textbf{5} & 0.2254 $\pm$ 0.1799 & 0.0079 $\pm$ 0.056 & 0.2342 $\pm$ 0.1531 & 0.312 $\pm$ 0.1822 & 0.2244 $\pm$ 0.1906 & 0.0583 $\pm$ 0.1052 & 0.177 $\pm$ 0.1445 \\
   & & \textbf{8} & 0.3023 $\pm$ 0.1765 & 0.0085 $\pm$ 0.0561 & 0.2622 $\pm$ 0.1586 & 0.3491 $\pm$ 0.1881 & 0.2774 $\pm$ 0.1944 & 0.0862 $\pm$ 0.123 & 0.2143 $\pm$ 0.1494 \\
   & & \textbf{10} & \re{0.363 $\pm$ 0.1882} & \re{0.0103 $\pm$ 0.0612} & \re{0.2733 $\pm$ 0.1631} & \re{0.3944 $\pm$ 0.1937} & \re{0.2971 $\pm$ 0.2054} & \re{0.1033 $\pm$ 0.1314} & \re{0.2402 $\pm$ 0.1572} \\
   \cmidrule{2-10}
   & \multirow{3}{*}{\textbf{Strat.}} & $\pmb{S_1}$ & 0.2205 $\pm$ 0.166 & 0.0064 $\pm$ 0.048 & 0.2214 $\pm$ 0.1501 & 0.299 $\pm$ 0.1717 & 0.22 $\pm$ 0.1817 & 0.0571 $\pm$ 0.0963 & 0.1707 $\pm$ 0.1356 \\
   & \multirow{3}{*}{($sel$)} & $\pmb{S_2}$ & \gr{0.166 $\pm$ 0.1646} & \gr{0.0059 $\pm$ 0.0449} & \gr{0.1775 $\pm$ 0.1469} & \gr{0.2372 $\pm$ 0.1627} & \gr{0.1741 $\pm$ 0.1716} & \gr{0.0387 $\pm$ 0.0824} & \gr{0.1332 $\pm$ 0.1289} \\
   & & $\pmb{S_3}$ & 0.2215 $\pm$ 0.1681 & 0.0066 $\pm$ 0.0483 & 0.2224 $\pm$ 0.1516 & 0.3054 $\pm$ 0.1725 & 0.2243 $\pm$ 0.1824 & 0.0552 $\pm$ 0.0953 & 0.1726 $\pm$ 0.1364 \\
   & & $\pmb{S_4}$ & \re{0.2542 $\pm$ 0.1598} & \re{0.007 $\pm$ 0.0509} & \re{0.249 $\pm$ 0.1522} & \re{0.3396 $\pm$ 0.1756} & \re{0.2462 $\pm$ 0.1889} & \re{0.0667 $\pm$ 0.1018} & \re{0.1938 $\pm$ 0.1382} \\ 
   \bottomrule
\end{tabular}
\vspace{1.1\baselineskip}
\caption{Summary of the ILS results grouped by parameter.} \label{tbl:upppm:summary_ils}
\begin{tabular}{ccrccccccc}
   \toprule
   &&& \multicolumn{6}{c}{\textbf{Event logs}} & \multirow{2}{*}{\textbf{Avg.}} \\
   \cmidrule(lr){4-9}
   &&& \textbf{BPI12} & \textbf{BPI13} & \textbf{BPI14} & \textbf{BPI15} & \textbf{CoSeLoG} & \textbf{TGN-Hospital} \\
   \midrule
   \multirow{10}{*}{\rotatebox{90}{\textbf{Grouping criteria}}} & \multirow{5}{*}{\textbf{Priv.}} & \textbf{2} & \gr{0.0256 $\pm$ 0.0069} & \gr{0.0014 $\pm$ 0.0004} & \gr{0.0275 $\pm$ 0.0111} & \gr{0.0148 $\pm$ 0.0069} & \gr{0.0237 $\pm$ 0.0108} & \gr{0.0051 $\pm$ 0.002} & \gr{0.0167 $\pm$ 0.0126} \\
   & \multirow{5}{*}{($k$)} & \textbf{3} & 0.0623 $\pm$ 0.0161 & 0.0019 $\pm$ 0.0005 & 0.0371 $\pm$ 0.0143 & 0.0248 $\pm$ 0.0119 & 0.0387 $\pm$ 0.018 & 0.0081 $\pm$ 0.0053 & 0.0295 $\pm$ 0.0237 \\
   & & \textbf{4} & 0.072 $\pm$ 0.0206 & 0.0021 $\pm$ 0.0007 & 0.0394 $\pm$ 0.0149 & 0.0311 $\pm$ 0.0145 & 0.0455 $\pm$ 0.0206 & 0.0132 $\pm$ 0.0048 & 0.0344 $\pm$ 0.0271 \\
   & & \textbf{5} & 0.0915 $\pm$ 0.0247 & 0.0032 $\pm$ 0.0007 & 0.0446 $\pm$ 0.0167 & 0.0332 $\pm$ 0.0153 & 0.0469 $\pm$ 0.0205 & 0.019 $\pm$ 0.0069 & 0.04 $\pm$ 0.0316 \\
   & & \textbf{8} & 0.1065 $\pm$ 0.025 & 0.0033 $\pm$ 0.0007 & 0.0492 $\pm$ 0.0186 & 0.0434 $\pm$ 0.0194 & 0.0597 $\pm$ 0.0258 & 0.028 $\pm$ 0.0102 & 0.0489 $\pm$ 0.0365 \\
   & & \textbf{10} & \re{0.1346 $\pm$ 0.0346} & \re{0.0039 $\pm$ 0.0008} & \re{0.0522 $\pm$ 0.0196} & \re{0.0513 $\pm$ 0.0228} & \re{0.0662 $\pm$ 0.027} & \re{0.0328 $\pm$ 0.0118} & \re{0.0573 $\pm$ 0.0451} \\
   \cmidrule{2-10}
   & \multirow{3}{*}{\textbf{Strat.}} & $\pmb{S_1}$ & 0.0839 $\pm$ 0.0225 & \gr{0.0026 $\pm$ 0.0005} & 0.0422 $\pm$ 0.0163 & 0.0333 $\pm$ 0.0155 & 0.0477 $\pm$ 0.0211 & 0.0188 $\pm$ 0.0073 & 0.0386 $\pm$ 0.0302 \\
   & \multirow{3}{*}{($sel$)} & $\pmb{S_2}$ & \gr{0.0631 $\pm$ 0.0174} & 0.0026 $\pm$ 0.0006 & \gr{0.0356 $\pm$ 0.0137} & \gr{0.0233 $\pm$ 0.0111} & \gr{0.0373 $\pm$ 0.0168} & \gr{0.013 $\pm$ 0.0049} & \gr{0.0296 $\pm$ 0.0233} \\
   & & $\pmb{S_3}$ & 0.0852 $\pm$ 0.0227 & 0.0026 $\pm$ 0.0006 & 0.0427 $\pm$ 0.0165 & 0.0344 $\pm$ 0.0159 & 0.0474 $\pm$ 0.021 & 0.0181 $\pm$ 0.007 & 0.0389 $\pm$ 0.0305 \\
   & & $\pmb{S_4}$ & \re{0.0962 $\pm$ 0.0227} & \re{0.0027 $\pm$ 0.0006} & \re{0.0462 $\pm$ 0.017} & \re{0.0414 $\pm$ 0.018} & \re{0.0547 $\pm$ 0.0228} & \re{0.021 $\pm$ 0.008} & \re{0.0443 $\pm$ 0.0337} \\ 
   \bottomrule
\end{tabular}
\end{sidewaystable}

\begin{sidewaystable}[htbp]
\renewcommand{\tabcolsep}{0.2cm}
\renewcommand{\arraystretch}{0.92}
\centering
\scriptsize  
\caption{Summary of the CS results grouped by parameter.} \label{tbl:upppm:summary_cs}
\begin{tabular}{ccrccccccc}
   \toprule
   &&& \multicolumn{6}{c}{\textbf{Event logs}} & \multirow{2}{*}{\textbf{Avg.}} \\
   \cmidrule(lr){4-9}
   &&& \textbf{BPI12} & \textbf{BPI13} & \textbf{BPI14} & \textbf{BPI15} & \textbf{CoSeLoG} & \textbf{TGN-Hospital} \\
   \midrule
   \multirow{10}{*}{\rotatebox{90}{\textbf{Grouping criteria}}} & \multirow{5}{*}{\textbf{Priv.}} & \textbf{2} & \gr{0.9996 $\pm$ 0.002} & \gr{0.9999 $\pm$ 0.0012} & \gr{0.8356 $\pm$ 0.2286} & \gr{0.9553 $\pm$ 0.0872} & \gr{0.9897 $\pm$ 0.0302} & \gr{0.9999 $\pm$ 0.0007} & \gr{0.9633 $\pm$ 0.0583}  \\
   & \multirow{5}{*}{($k$)} & \textbf{3} & 0.996 $\pm$ 0.021 & 0.9999 $\pm$ 0.0022 & 0.8317 $\pm$ 0.2266 & 0.9311 $\pm$ 0.0977 & 0.9677 $\pm$ 0.0791 & 0.9998 $\pm$ 0.0011 & 0.9544 $\pm$ 0.0713 \\
   & & \textbf{4} & 0.9897 $\pm$ 0.0465 & 0.9998 $\pm$ 0.0027 & 0.8308 $\pm$ 0.229 & 0.9086 $\pm$ 0.1136 & 0.9595 $\pm$ 0.097 & 0.9997 $\pm$ 0.0017 & 0.948 $\pm$ 0.0818 \\
   & & \textbf{5} & 0.9871 $\pm$ 0.0567 & 0.9996 $\pm$ 0.0047 & 0.8289 $\pm$ 0.2284 & 0.8936 $\pm$ 0.1384 & 0.9493 $\pm$ 0.1264 & 0.9995 $\pm$ 0.0021 & 0.943 $\pm$ 0.0928 \\
   & & \textbf{8} & 0.9757 $\pm$ 0.0909 & 0.9994 $\pm$ 0.0059 & 0.8249 $\pm$ 0.2292 & 0.8686 $\pm$ 0.1409 & 0.9338 $\pm$ 0.12 & 0.9991 $\pm$ 0.0027 & 0.9336 $\pm$ 0.0983 \\
   & & \textbf{10} & \re{0.9735 $\pm$ 0.0919} & \re{0.9989 $\pm$ 0.0106} & \re{0.8241 $\pm$ 0.2287} & \re{0.8364 $\pm$ 0.1821} & \re{0.9205 $\pm$ 0.1683} & \re{0.9989 $\pm$ 0.0033} & \re{0.9254 $\pm$ 0.1141} \\
   \cmidrule{2-10}
   & \multirow{3}{*}{\textbf{Strat.}} & $\pmb{S_1}$ & 0.9866 $\pm$ 0.0548 & 0.9996 $\pm$ 0.0044 & 0.8292 $\pm$ 0.2284 & 0.899 $\pm$ 0.1254 & 0.952 $\pm$ 0.1108 & 0.9995 $\pm$ 0.0019 & 0.9443 $\pm$ 0.0876 \\
   & \multirow{3}{*}{($sel$)} & $\pmb{S_2}$ & \gr{0.9919 $\pm$ 0.0369} & \gr{0.9996 $\pm$ 0.0041} & \gr{0.833 $\pm$ 0.2273} & \gr{0.9361 $\pm$ 0.1103} & \gr{0.9675 $\pm$ 0.0857} & \gr{0.9997 $\pm$ 0.0017} & \gr{0.9546 $\pm$ 0.0776}  \\
   & & $\pmb{S_3}$ & 0.9867 $\pm$ 0.0531 & 0.9996 $\pm$ 0.0044 & 0.8283 $\pm$ 0.23 & 0.8921 $\pm$ 0.1287 & 0.9479 $\pm$ 0.1112 & 0.9995 $\pm$ 0.0019 & 0.9423 $\pm$ 0.0882 \\
   & & $\pmb{S_4}$ & \re{0.9826 $\pm$ 0.0613} & \re{0.9995 $\pm$ 0.0052} & \re{0.8268 $\pm$ 0.228} & \re{0.8685 $\pm$ 0.1423} & \re{0.9463 $\pm$ 0.1063} & \re{0.9993 $\pm$ 0.0022} & \re{0.9372 $\pm$ 0.0909} \\ 
   \bottomrule
\end{tabular}
\end{sidewaystable}

Experimental results show that \textit{u}-PPPM behaves consistently regardless the event log to which it is applied, and similar trends can be observed when the \textit{u}-PPPM's parameters vary.
However, the data-dependence nature of the proposed method makes that the quality of the protected process models (either evaluated using QS, ILS or CS) depends on the event data.
This is, although executing the method with the same parameters values, the results from QS, ILS or CS vary within each event log.
For example, by using $k = 5$ and $sel = S_1$, the QS results in BPI12 and BPI15 event logs are 0.232 and 0.31, respectively.
Hence, to evaluate the high-level impact of the \textit{u}-PPPM's parameters on the quality of the process models, the averaged results are particularly useful.
Despite the above, Tables \ref{tbl:upppm:summary_qs} to \ref{tbl:upppm:summary_cs} show similar tendencies regarding the effect of the \textit{u}-PPPM's parameters to the quality of the protected process models.
The impact of each of these parameters is discussed next.

The selection of the \textit{u}-PPPM's parameters values clearly affects the quality of the protected process models.
It can be noticed that increasing the privacy level $k$ affects negatively the process models quality.
The larger $k$, the more individuals share the same distribution, so the more difficult for the attacker.
However, this introduces more distortion in the protected process models, as observed.
This observation is, indeed, aligned with those from other privacy-related fields, such as SDC, in which data utility decreases at higher privacy levels.
Notwithstanding, it is worth noting that this is not always true in our method, as the selection strategy can slightly contribute to maximise the quality at a given privacy level.
In our results, $S_2$ stands as the best selection strategy in all event logs.
The reason why this happens is because this strategy selects the two most appropriate individuals (from a distribution perspective) for the exchange, and avoids using probabilistic or random guesses.
In opposition to $S_2$, the worst strategy is $S_4$, hence, demonstrating that its iterative design propagates the distortion among all individuals within the group, which results in a quality decrease.
Thus, better results are achieved at higher privacy levels if using the appropriate selection strategy.
For instance, using the BPI15 event log, all QS, ILS and CS results are better using $k = 8$ and $sel = S_2$ rather than $k = 4$ and $sel = S_3$.

\subsection{Relationship between QS, ILS and CS Results} \label{subsec:upppm:correlation}

To contextualise the previous results, Figures \ref{fig:app_upppm:results_bpi12_bpi13}, \ref{fig:app_upppm:results_bpi14_bpi15} and \ref{fig:app_upppm:results_coselog_tgn} from Appendix \ref{app:upppm} depict the relationship between the QS, ILS and CS results for each combination of parameters for all the event logs.
Each plot, which represents the results of an event log, consists of four charts:
the first indicates the correlation between ILS and CS results,
the second represents the correlation between QS and CS results, 
the third represents the correlation between QS and ILS results, 
and the fourth combines the aforementioned planes to build a 3D space and shows the correlation between QS, ILS, and CS results.
The background colour of each projection plane helps to compose the planes and visualise the 3D perspective.
Every point in the chart corresponds to a specific \textit{u}-PPPM execution with a certain combination of parameters: 
the dot's colour indicates the group size $k$ and the dot's shape indicates the selection strategy.
In addition, to evaluate the generic impact of \textit{u}-PPPM from a global perspective (regardless the particularities of each event log), Figure \ref{fig:upppm:results_avg} shows the averaged results obtained from all event logs.

Interestingly enough, there exists an apparent direct correlation among the results, although they have been obtained from different perspectives.
This suggests that the protection of event logs using \textit{u}-PPPM introduces a uniform/homogeneous distortion in the process models: the more distortion in the process models individually (QS), the more distant the relationships among them (ILS), and less conformed with the original event data (CS).
In particular, despite the individual distortion of the protected process models (\ie QS values grow rapidly at high privacy levels), the relationships between the different process models are notably preserved (\ie ILS results grow at a much slower pace), and the original event data mostly conforms with the protected process models (\ie CS results slightly decrease at high privacy levels).
In this regard, the resulting protected event logs, in addition to anonymise the individual process models (\ie individuals' privacy), also preserve most of the patterns and tendencies from the original event logs.
From a privacy perspective, this property enables gathering similar insights and conclusions from the protected event logs (and the discovered protected process models) as if they were obtained from the original event logs (and the original process models too), but without releasing confidential data.

\begin{figure}[t!]
\centering
\includegraphics[width=0.95\linewidth]{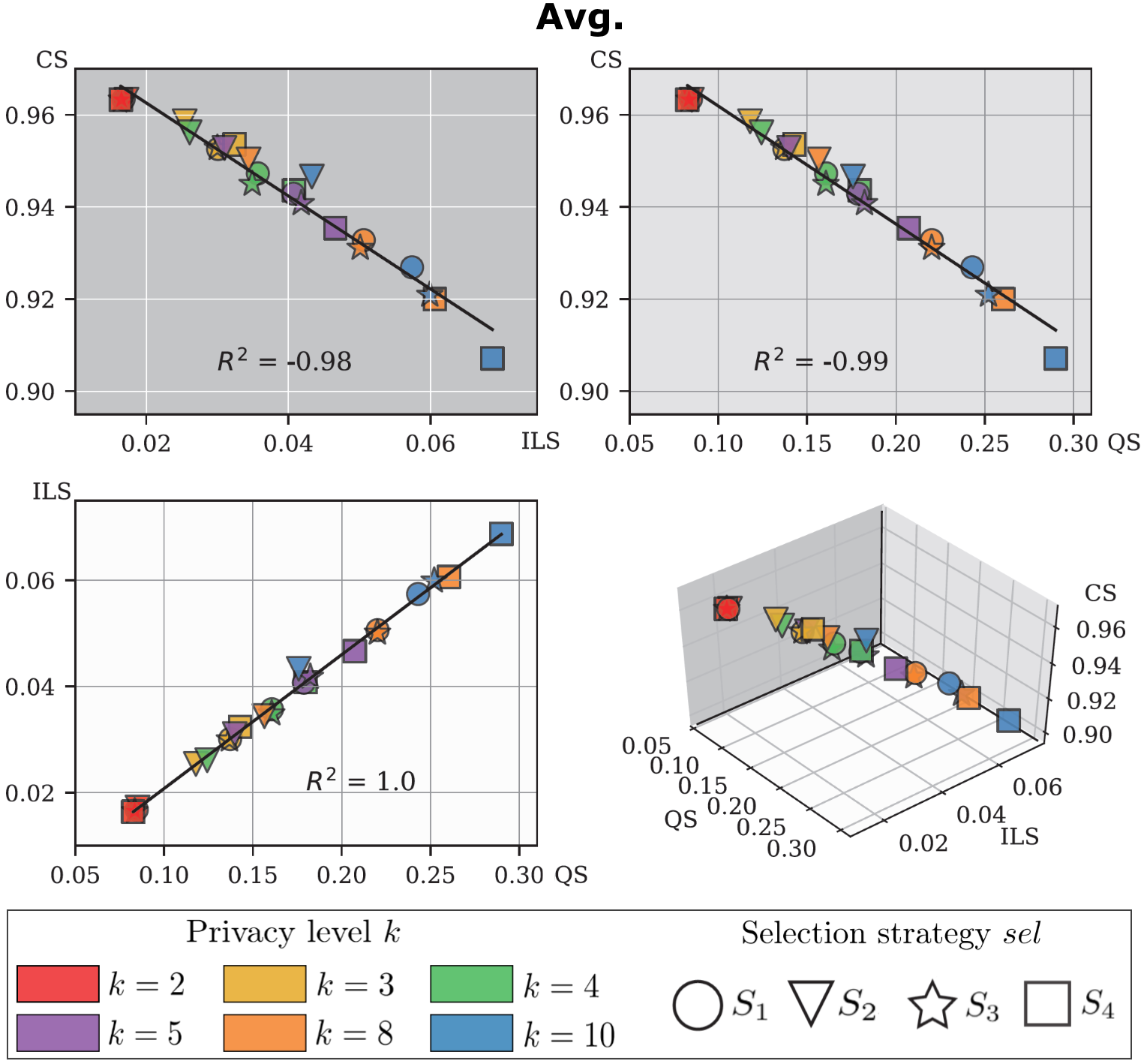}
\caption{Correlation between the averaged QS, ILS and CS results from all the event logs (adapted from \cite{batista2021uniformization}).}
\label{fig:upppm:results_avg}
\end{figure}

\begin{figure}[b!]
\centering
\includegraphics[width=0.79\linewidth]{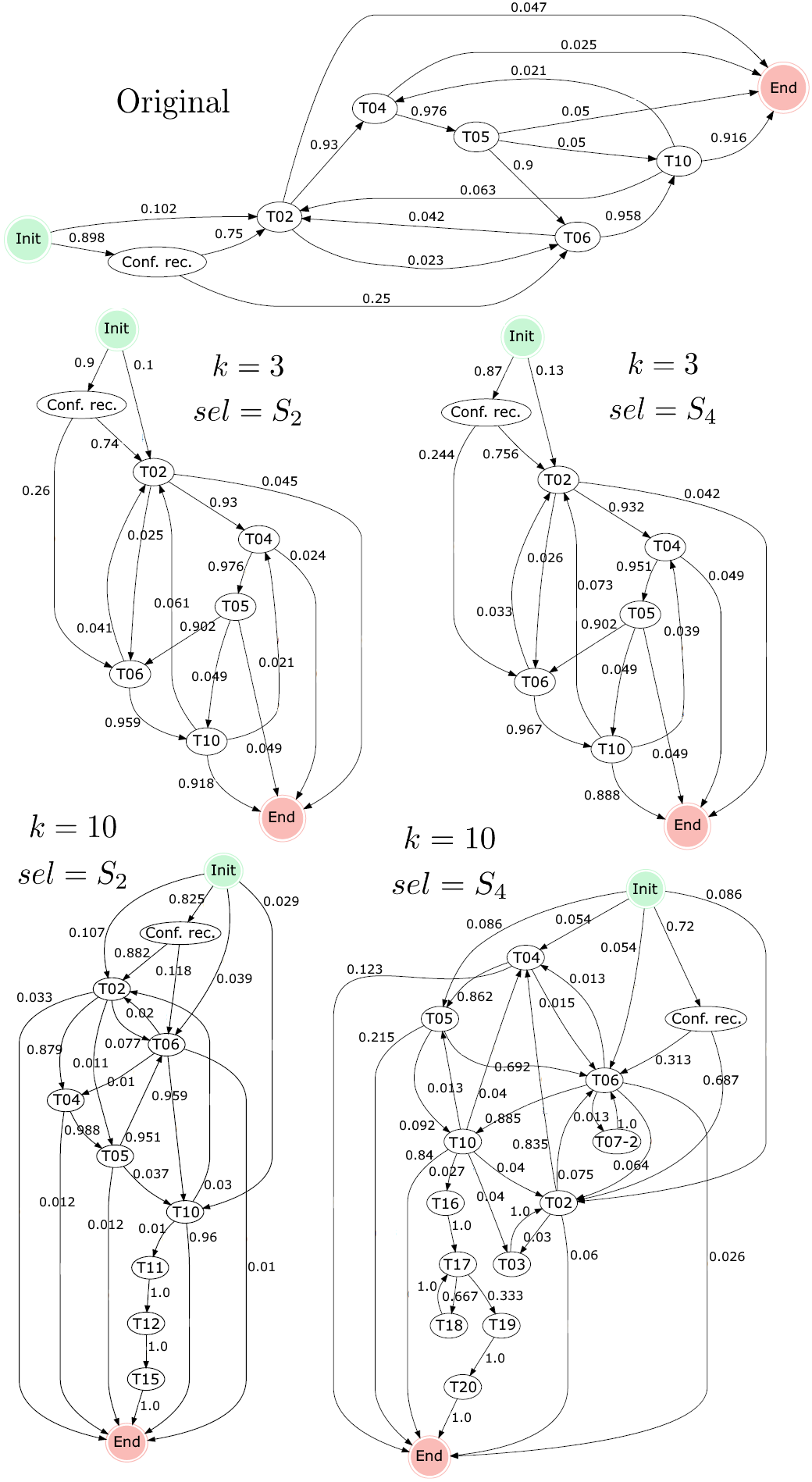}
\caption{Versions of the process model of a certain individual from the CoSeLoG event log after applying \textit{u}-PPPM with different parameters (adapted from \cite{batista2021uniformization}).}
\label{fig:upppm:models}
\end{figure}

As an illustrative example, with the aim to visually analyse the distortion of the process models from a qualitative perspective, Figure \ref{fig:upppm:models} depicts an example of the resulting process models of a given individual within the CoSeLoG event log after applying \textit{u}-PPPM for different parameters values.
Protected process models, using either deterministic or probabilistic selection strategies (\eg $S_2$ and $S_4$) and different privacy levels (\eg $k = 3$ and $k = 10$), can be compared to the original process model.
The qualitative differences in the process models are aligned with the quantitative results obtained in the experiments.

\subsection{Significance of the Anonymisation Parameters} \label{subsec:upppm:t-tests}

Experimental results are clearly affected by the choice of the parameters values.
To verify the significance of the results differences, statistical analyses of two sampled t-Tests are performed.
This analysis, widely used in statistics, compares the means of two independent groups (\ie populations) as a way to determine whether there is statistical evidence that the two means are significantly different.
If statistical differences are found, it can be inferred that using a certain parameter value instead of another affects significantly the quality of the process models.

Both \textit{u}-PPPM's parameters are analysed individually.
For each of them, the population of QS results obtained when using a certain value is respectively compared to the population of QS results obtained when using another value.
Then, the t-Test will evaluate whether the means of the two populations are significantly different.
This procedure is repeated for the ILS results and the CS results individually.
The settings of the t-Tests conducted follow the standard form.
The \textit{null hypothesis} states that there is no statistically significant difference in the mean of two populations (\ie means are equal).
Contrary, the \textit{alternative hypothesis} states that there is statistically significant difference in the mean of the two populations (\ie means are different).
Finally, the \textit{significance level} $\alpha$ is set to $\alpha = 0.05$ indicating the probability of rejecting the null hypothesis when it is true.
The t-Tests, which return a $p$-value, indicate that the null hypothesis is rejected (and the alternative hypothesis is accepted) if the $p$-value is lower than $\alpha$; otherwise, the null hypothesis cannot be rejected.
Figures \ref{fig:app_upppm:ttest_qs_k} to \ref{fig:app_upppm:ttest_cs_strategy} from Appendix \ref{app:upppm} depict the $p$-value results obtained from the t-Tests between each pair of parameter's values for the QS, ILS and CS results.

The t-Tests results confirm that the privacy level affects, as expected, the quality of the process models.
From a statistical point of view, as $p$-values are generally lower than $\alpha$, the means of two populations obtained with two different privacy level are different enough.
However, some t-Tests demonstrate that significant differences are not always detected, especially if two privacy levels are close: for instance, between $k = 3$ and $k = 4$, or between $k = 8$ and $k = 10$, in the majority of event logs.
However, by considering the averaged results, t-Tests show that there exist significant differences when increasing the privacy level.
So, this confirms that the quality of the process models improves at lower privacy levels.

Regarding the selection strategy, t-Tests agree that differences between using a certain strategy or another are not significant enough from a statistical perspective (\ie $p$-values are higher than $\alpha$).
Despite of that, these differences become more insignificant when comparing a certain strategy with $S_2$.
For example, focusing on the averaged QS results, $p$-values are lower than $0.2$ when $S_2$ is compared with another strategy, while $p$-values are higher than $0.55$ when comparing any other pair of strategies.
This pattern is mostly observed in the different event logs (except BPI13, in which the selection of the strategy does not seem relevant) and in all the QS, ILS and CS results.
Hence, although statistical results do not find a significant difference, experimental results have shown that strategy $S_2$ produces better outcomes than the other strategies and, hence, contributes to maximise the quality of the process models.

\section{Conclusions} \label{sec:upppm:concl}

Process mining in an emerging discipline that is attracting increasing attention.
Among the many challenges that are still to be met, considering privacy aspects throughout the entire process mining analysis is of utmost interest.
The modelling and discovery of process models may entail serious privacy risks that may allow disclosing confidential data, such as healthcare information or banking information, to unauthorised parties.
To lessen risks for people's privacy, the application of privacy-preserving methodologies, encompassed within the novel PPPM research direction, is paramount.

In this chapter, we have demonstrated the inability of pseudonymisation and encryption techniques, highly popular in practice, to avert distribution-based attacks.
These attacks, feasible in public places (\eg waiting room in a healthcare facility or public administration\ldots), in combination with location-oriented targeted attacks (such as RSI and OI attacks), may open the door to attackers for acquiring background knowledge based on the distribution of events' attributes in confidential event logs.
As a countermeasure, this chapter has presented \textit{u}-PPPM, a uniformisation-based PPPM method aiming to distort attributes distributions in event logs, thus rending the attackers background knowledge useless, according to a given privacy threshold.
We have evaluated this method with six real-life event logs from multiple domains.
To quantify the quality degradation introduced by our method in the protected process models, they have been analysed from three perspectives: individual distortion, inter-individual distortion and conformance.
Experimental results showed that increasing the privacy level increases the distortion, so the utility of the process models diminishes, and that properly choosing the selection strategy to conduct the uniformisation helps to keep the distortion under control.
Correlations between the results have suggested that our method introduces an homogeneous distortion on the process models.

\chapter{A Microaggregation Method to Privacy-Preserving Process Mining}
\chaptermark{A Microaggregation Method to PPPM}
\label{chap:kpppm}

\emph{Continuing with the research on PPPM reported in the preceding chapter, this chapter presents an advanced privacy protection method on the same conceptual basis. Deepening on the threats related to location-oriented targeted attacks, further attacks, such as the analysis and modelling of activities, could also be exploited to jeopardise people's privacy and disclose confidential information. This chapter proposes \textit{k}-PPPM, a novel PPPM method based on microaggregation techniques to increase privacy through \textit{k}-anonymity during process mining analyses. To the best of our knowledge, this is the first approach that uses microaggregation techniques to protect people's privacy in process mining. First, Section \ref{sec:kpppm:attacks} conceptualises the attacker model based on the inference and analysis of process models that might lead to people re-identification. To counteract it, Section \ref{sec:kpppm:method} describes the proposed \textit{k}-PPPM method aiming to produce privacy-preserved versions of event logs constrained by the \textit{k}-anonymity privacy model. Next, in Section \ref{sec:kpppm:experiments}, we detail the evaluation methodology followed to assess the impact of our method using six real-life event logs. Then, Section \ref{sec:kpppm:disc} discusses the results obtained in terms of data quality and information loss for different privacy thresholds. Finally, the chapter concludes in Section \ref{sec:kpppm:concl} with a summary of the main contributions.}

\minitoc

\section{Modelling-based Attacks} \label{sec:kpppm:attacks}

Location-oriented attacks thwart the use of pseudonymisation and encryption techniques \textit{per se} to protect people's privacy in event logs, as discussed in Section \ref{sec:upppm:attacks}.
These attacks, in combination with additional knowledge, lead to potential disclosure risks.
Whereas the previous chapter explored the significance of distributions within event logs, this chapter goes one step beyond and explores the feasibility to re-identify individuals through the analysis and modelling of their activities as process models.

Next, we formalise an attacker model able to re-identify targeted individuals and disclose their confidential data from pseudonymised or encrypted event logs.
In particular, serious privacy concerns might arise when analysing their activities and modelling inferred process models in institutions with free-public access when combined with location-oriented attacks.

\subsection{Attacker Model}

Let's establish the bases of the attacker model with the same premises as those from the previous chapter.
Be $\mathcal{L}$ an event log describing the activities carried out by people in an institution with free-public access that manages confidential data, such as employees from a public hospital.
To enable the attack, the attacker must have gained access (somehow) to the organisation's event log $\mathcal{L}'$ with its personally identifiable information pseudonymised or encrypted.
Assuming that $\mathcal{L}$ cannot be recovered from $\mathcal{L}'$, the attacker starts the attack with the information in $\mathcal{L}'$.
The step-by-step scheme of the attack is graphically depicted in Figure \ref{fig:kpppm:attacker_model}.

First, all the events in $\mathcal{L}'$ are separated according to the value of their $resource$ attribute (\ie $\#_{resource}(e), \forall e \in \mathcal{L}'$), which contains the obfuscated value of the individual's identity.
Hence, if $\mathcal{L}'$ contains information about $p$ individuals, $\mathcal{I}' = \{I'_1,I'_2,\ldots,I'_p\}$, where $\mathcal{I}'$ is the set of all obfuscated individuals in $\mathcal{L}'$, then $\mathcal{L}'$ is decomposed into $p$ sub-event logs, $\mathcal{L}'_{I'_1},\mathcal{L}'_{I'_2},\ldots,\mathcal{L}'_{I'_p}$, each of them with the events of the same individual (step \ding{192}).
For each of these sub-event logs, the attacker discovers its corresponding process model $\mathcal{M}'_{I'_1},\mathcal{M}'_{I'_2},\ldots,\mathcal{M}'_{I'_p}$, respectively, from a control-flow perspective using a specific process discovery algorithm.
Each process model represents the workflow and ordering of activities performed by an unknown individual (step \ding{193}).
With all this, the attacker has acquired a background knowledge that associates each individual's obfuscated identifier to an individual's process model that is representing his/her behaviour (step \ding{194}).
Since people's real identities cannot be retrieved from the obfuscated identifiers, the attacker takes advantage of the discovered process models to breach confidentiality.

\begin{figure}[b!]
\centering
\includegraphics[width=0.83\linewidth]{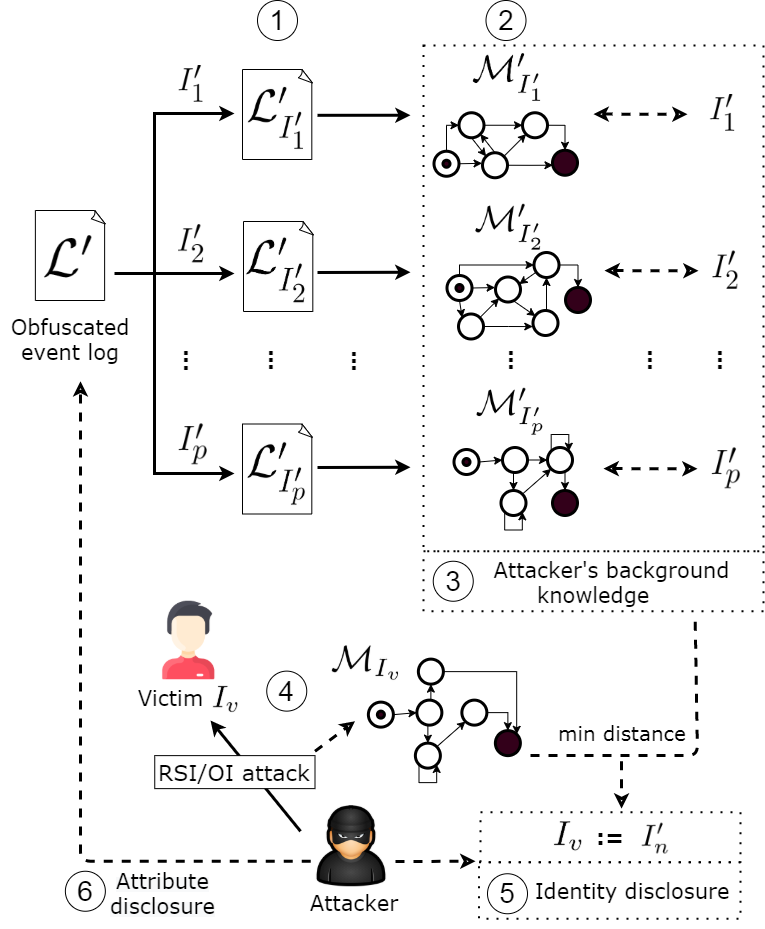}
\caption{Attacker model able to re-identify individuals through location-oriented attacks and modelling-based attacks in pseudonymised or encrypted event logs (adapted from \cite{kpppm}).}
\label{fig:kpppm:attacker_model}
\end{figure}

Then, the attacker can start the location-oriented attack, based on RSI and OI, against a targeted individual (victim) $I_v$.
The identity of $I_v$ is obfuscated in $\mathcal{L}'$ and, hence, unknown for the attacker.
In this attack, the attacker stays physically close to the victim and tracks all the activities that he/she is performing in the organisation.
After a reasonable period of time, this information will be helpful to cross-correlate it with his/her background knowledge.
This scenario is similar to the one already depicted in Figure \ref{fig:upppm:attacker}.
As a result, the attacker would be able to describe the behaviour of the victim $I_v$ as a process model $\mathcal{M}_{I_v}$ (step \ding{195}).

Finally, the attacker compares the observed process model $\mathcal{M}_{I_v}$ against all the process models from his/her background knowledge $\mathcal{M}'_{I'_1},\mathcal{M}'_{I'_2},\ldots,\mathcal{M}'_{I'_p}$ using some distance or similarity function.
The attacker can infer the obfuscated identifier of victim $I_v$, namely $I'_n$ (for $1 \leq n \leq p$), by identifying the process model $\mathcal{M}'_{I'_n}$ most similar to $\mathcal{M}_{I_v}$.
Hence, the attacker has been able to infer the identity of a victim $I_v$ to an obfuscated identifier $I'_n$ from $\mathcal{L}'$, \ie identity disclosure (step \ding{196}).
Indirectly, the attacker can also infer the confidential data of $I_v$ as those information associated to $I'_n$ in $\mathcal{L}'$, \ie attribute disclosure (step \ding{197}).

\section{\textit{k}-PPPM: The Microaggregation Method} \label{sec:kpppm:method}

In this section, we present \textit{k}-PPPM, a novel PPPM method based on microaggregation techniques to guarantee \textit{k}-anonymity.
Due to the feasibility to infer process models when conducting location-oriented attacks, the proposed method distorts the event data in $\mathcal{L}$ to create a privacy-preserved event log $\mathcal{L}'$ that renders the background knowledge acquired by the attackers useless.
Hence, $\mathcal{L}'$ prevents the direct re-identification of individuals and averts these attacks.

More specifically, given an event log $\mathcal{L}$, the proposed method aims to cluster similar individuals according to their process models into groups of size $k$, where $k$ is a privacy threshold.
Then, a representative process model of each cluster will be selected and associated to each individual within the cluster in $\mathcal{L}'$.
All the individuals within the same cluster will be represented by the same process model and, hence, indistinguishable to an attacker conducting modelling-based attacks.
This procedure, inspired by the classical microaggregation techniques, leverages the properties of the \textit{k}-anonymity privacy model.
Although microaggregation techniques have been extensively reported in the privacy literature, this is the very first PPPM method founded on them to the best of our knowledge.
The usefulness of \textit{k}-PPPM lies in the ability to balance the trade-off between information loss (caused by both the clustering and representative's selection phases) and privacy (achieved by \textit{k}-anonymity).

The details and the main design decisions of the proposed \textit{k}-PPPM method are detailed in Section \ref{subsec:kpppm:algorithm}.
Then, Section \ref{subsec:kpppm:security} discusses the security and privacy enhancements of the event logs protected through \textit{k}-PPPM compared to event logs protected through pseudonymisation or encryption.

\subsection{Algorithm Details} \label{subsec:kpppm:algorithm}

This section describes the implementation details of \textit{k}-PPPM, which are summarised in Algorithm \ref{alg:kpppm:algorithm}.
For the sake of completeness, this explanation is also supported by Figure \ref{fig:kpppm:algorithm}, which illustrates the step-by-step transformations applied to the event logs.

Be $\mathcal{L}$ an event log describing the activities of $p$ individuals, $\mathcal{I} = \{I_1,I_2,\ldots,I_p\}$, where $\mathcal{I}$ is the set of all individuals in $\mathcal{L}$ (step \ding{192}).
First of all, $\mathcal{L}$ is decomposed into $p$ sub-event logs, $\mathcal{L}_{I_1},\mathcal{L}_{I_2},\ldots,\mathcal{L}_{I_p}$, each of them containing the events of each individual (step \ding{193}).
Next, for each sub-event log, a process model $\mathcal{M}_{I_1},\mathcal{M}_{I_2},\ldots,\mathcal{M}_{I_p}$ is discovered, respectively, from a control-flow perspective using a specific process discovery algorithm.
Each process model represents the behaviour of each individual as a workflow of activities (step \ding{194}).
It can be noticed that these initial steps are similar to the ones from the attacker model so as to decrease the ability to acquire this background knowledge.

Following the microaggregation's principles, \textit{k}-PPPM determines that two individuals are similar if their process models are similar.
To this end, a similarity matrix $S$ of length $p \times p$ is built, by comparing all the pairs of process models between them using a similarity measure $sim$ (step \ding{195}).
Matrix $S$ fulfils two properties: it is
(i) symmetric, \ie $S[I_a][I_b] = S[I_b][I_a]$ because $sim(\mathcal{M}_{I_a},\mathcal{M}_{I_b}) = sim(\mathcal{M}_{I_b},\mathcal{M}_{I_a})$,
and (ii) hollow, \ie $S[I_a][I_a] = 0$, because $sim(\mathcal{M}_{I_a},\mathcal{M}_{I_a}) = 0$, in case that $0$ means total similarity.

Next, the microaggregation's clustering phase begins.
This phase aims at grouping individuals into clusters, in such a way that the individuals within the same cluster are as homogeneous as possible.
This homogeneity helps maximise within-cluster similarity and minimise information loss.
To do this, considering the information in $S$, a microaggregation-oriented clustering algorithm ($clus$) is executed by specifying the number of individuals per cluster $k$ (step \ding{196}).
The privacy level of \textit{k}-PPPM is directly related to $k$, where $k \geq 2$: the higher $k$, the more privacy.
Indeed, maximum privacy is achieved when $k = p$, as all individuals are grouped within a single cluster.
Also, it is noteworthy that any of the microaggregation clustering algorithms from the literature could be used.
As a result, a set $C$ of $n$ clusters, $C = \{c_1,c_2,\ldots,c_n\}$, is obtained.
It can be noticed that the number of clusters $n$ is equal to $\lfloor p/k \rfloor$, and that each cluster is a group of $k$ to $2k-1$ individuals.
In addition, all $p$ individuals must be assigned to one and only one cluster (step \ding{197}).

\begin{figure}[b!]
\centering
\includegraphics[width=0.97\linewidth]{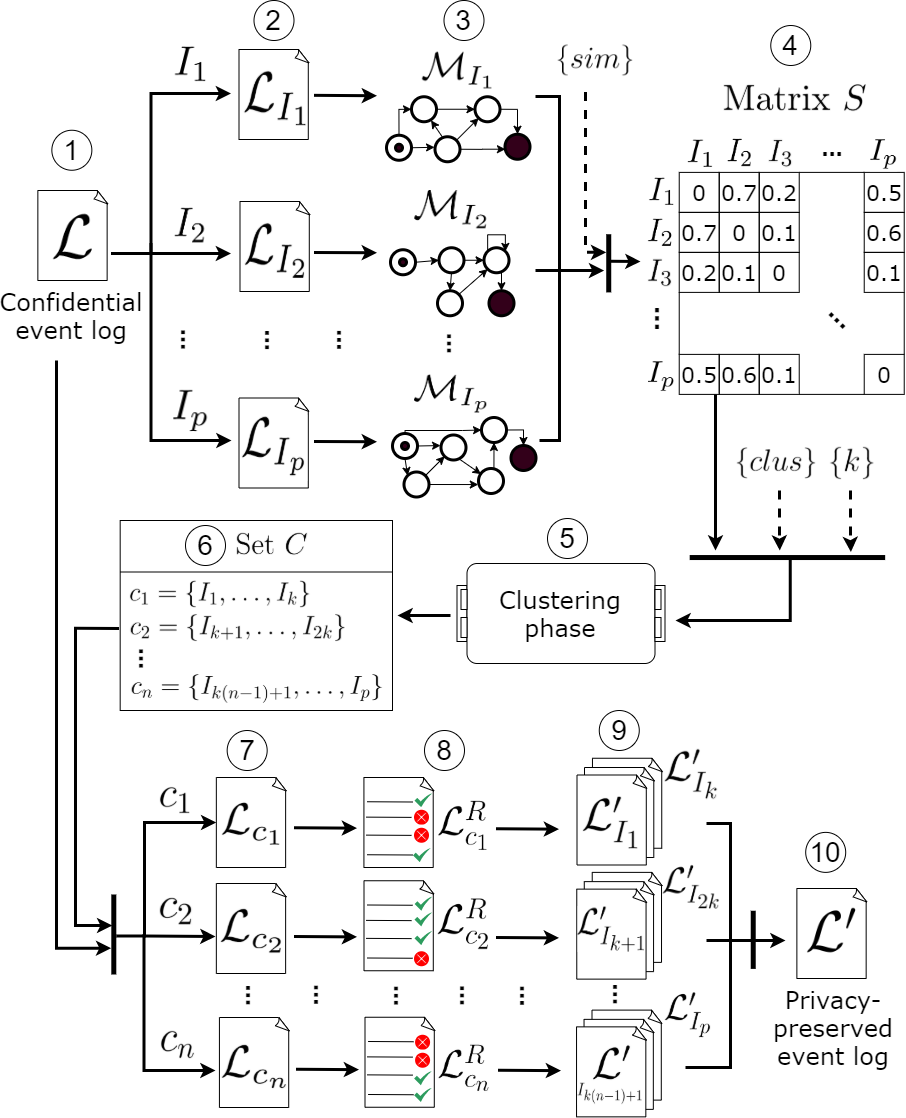}
\caption{Step-by-step scheme of the \textit{k}-PPPM method (adapted from \cite{kpppm}).}
\label{fig:kpppm:algorithm}
\end{figure}

Once clusters have been created, each cluster must select its representative (or centroid), this is, a \textit{virtual} individual that averagely represents all the individuals within the same cluster.
In \textit{k}-PPPM, this representation is observed from the process model perspective: the process model of the representative must averagely represent all the process models from all the individuals of the cluster.
To create this averaged process model, the original event data from $\mathcal{L}$ is used.
This phase requires two main steps: (i) aggregation, and (ii) sampling, as described below.

In the first place, $\mathcal{L}$ is decomposed into $n$ sub-event logs, $\mathcal{L}_{c_1},\mathcal{L}_{c_2},\ldots,\mathcal{L}_{c_n}$, each of them containing the events associated to all the individuals within each cluster $c_1,c_2,\ldots,c_n$, respectively.
Each sub-event log $\mathcal{L}_{c_i}$ (for $1 \leq i \leq n$) can be seen as an aggregator of events and traces of all the individuals within each cluster $c_i$ (step \ding{198}).
If these sub-event logs would be discovered as process models, they would show an aggregated behaviour of the entire cluster's individuals.
Next, using the event data in $\mathcal{L}_{c_i}$, the representative of each cluster $c_i$ is selected.
To do so, a number of traces from $\mathcal{L}_{c_i}$ are randomly sampled and stored in $\mathcal{L}^{R}_{c_i}$, \ie $\mathcal{L}^{R}_{c_i} \subset \mathcal{L}_{c_i}$ (step \ding{199}).
The number of traces to be sampled corresponds to the average of traces per individual within the cluster.
For example, considering that a certain cluster $c_i$ has three individuals ($k = 3$), and each of them has 24, 14 and 16 traces in $\mathcal{L}$, then $\mathcal{L}_{c_i}$ would contain 54 traces, but only 18 of them would be chosen for $\mathcal{L}^{R}_{c_i}$.
Thus, the representative individual, described as the event data in $\mathcal{L}^{R}_{c_i}$, is composed of partial behaviours of $k$ individuals.
The sampling of the traces is done at random, so resulting \textit{k}-PPPM a non-deterministic method.

With the representative selected, all the individuals within each cluster need to be replaced by their corresponding representative.
To do this, the event data in $\mathcal{L}^{R}_{c_i}$ is replicated $k$ times, one for each individual within the cluster $c_i$.
This step is fundamental to break the uniqueness of event data (and the process models to be discovered accordingly), as it duplicates the same information to $k$ different individuals to prevent direct identity disclosure.
In addition, instead of assigning each event to the personally identifiable information of the individual (as it is in $\mathcal{L}$), it is assigned to an obfuscated identifier (\eg pseudonymised or encrypted) of the individual.
The event data associated to each obfuscated individual $I_i$ (for $1 \leq i \leq p$) is stored in $\mathcal{L}'_{I_i}$ (step \ding{200}).
Formally, $\mathcal{L}'_{I_i}$ does no longer associate personally identifiable information to confidential data because its resource attribute is obfuscated.
Finally, \textit{k}-PPPM returns a privacy-preserved event log $\mathcal{L}'$ resulting from the union of the $p$ sub-event logs $\mathcal{L}'_{I_i}$, this is, $\mathcal{L}' = \mathcal{L}'_{I_1} \cup \mathcal{L}'_{I_2} \cup \ldots \cup \mathcal{L}'_{I_p}$ (step~\ding{201}).

The transformations applied by \textit{k}-PPPM on $\mathcal{L}$ to create $\mathcal{L}'$ degradate the quality of the individuals' process models to be discovered.
Given a certain individual $I_i$ ($1 \leq i \leq p$), the original process model $\mathcal{M}_{I_i}$ discovered from $\mathcal{L}$ and the protected process model $\mathcal{M}'_{I_i}$ discovered from $\mathcal{L}'$ will differ, because the latter is affected by 
(i) the loss of events/traces that have not been chosen to be in the cluster's representative, 
and (ii) the new events/traces (originally from the other $k - 1$ individuals within the same cluster) that have been chosen to be in the cluster's representative.

\begin{algorithm*}[t!]
\footnotesize
\begin{algorithmic}[1]
\Require
\Statex $\mathcal{L}$ is a non-empty event log describing the activities of $p$ individuals, where $p > 0$.
\Statex $k$ is a privacy level in the range from 2 to $p$.
\Statex $clus$ is a microaggregation-oriented clustering algorithm.
\Statex $sim$ is a measure to assess the similarity between two process models.
\Ensure
\Statex $\mathcal{L}'$ is a non-empty event log, whose events and individuals' process models are indistinguishable among $k$ individuals.
\Function{\textit{k}-PPPM}{\textbf{EventLog} $\mathcal{L}$, \textbf{Integer} $k$, \textbf{Algorithm} $clus$, \textbf{Measure} $sim$}
\State \textbf{List$\langle$Individual$\rangle$} individuals $\gets$ getAllIndividuals($\mathcal{L}$);
\State \textbf{HashMap$\langle$Individual, ProcessModel$\rangle$} models;
\ForAll{$I_i$ in individuals} \Comment{Steps \ding{193} and \ding{194}}
  \State \textbf{EventLog} $\mathcal{L}_{I_i} \gets$ getEventsFromIndividual($\mathcal{L}, I_i$);
  \State \textbf{ProcessModel} $\mathcal{M}_{I_i} \gets$ discoverModel($\mathcal{L}_{I_i}$);
  \State models.put($I_i, \mathcal{M}_{I_i}$);
\EndFor
\State \textbf{Matrix2D$\langle$Double$\rangle$} $S$;
\For{$i \gets 1$ to $p$} \Comment{Step \ding{195}}
  \For{$j \gets i$ to $p$}
    \State Individual $I_i$, $I_j \gets$ individuals[$i$], individuals[$j$];
    \State \textbf{Double} dist $\gets$ compare(models.get($I_i$), models.get($I_j$), $sim$);
    \State $S$[$I_i$][$I_j$] $\gets$ dist; $S$[$I_j$][$I_i$] $\gets$ dist;
  \EndFor
\EndFor
\State \textbf{List$\langle$List$\langle$Individual$\rangle\rangle$} $C \gets$ clustering($S, k, clus$); \Comment{Steps \ding{196} and \ding{197}}
\State \textbf{EventLog} $\mathcal{L}'$;
\For{$i \gets 1$ to $n$}
  \State \textbf{List$\langle$Individual$\rangle$} $c_i \gets C$[i]
  \State \textbf{EventLog} $\mathcal{L}_{c_i} \gets$ getEventsFromCluster($\mathcal{L}, c_i$); \Comment{Step \ding{198}}
  \State \textbf{EventLog} $\mathcal{L}^R_{c_i} \gets$ selectRepresentative($\mathcal{L}_{c_i}$);  \Comment{Step \ding{199}}
  \For{$j \gets 1$ to $k$}  \Comment{Step \ding{200}}
    \State \textbf{Individual} $I_j \gets c_i$[$j$];
    \State \textbf{EventLog} $\mathcal{L}'_{I_j} \gets \mathcal{L}^R_{c_i}$;
    \ForAll{event in $\mathcal{L}'_{I_j}$}
      \State $\#_{resource}($event$) \gets $obfuscatePII($I_j$);
    \EndFor
    \State $\mathcal{L}'$.append($\mathcal{L}'_{I_j}$);
  \EndFor
\EndFor
\State \Return $\mathcal{L}'$; \Comment{Step \ding{201}}
\EndFunction
\end{algorithmic}
\caption{\textit{k}-PPPM algorithm}
\label{alg:kpppm:algorithm}
\end{algorithm*}

\subsection{Security and Privacy Analysis} \label{subsec:kpppm:security}

This section provides a throughout analysis of \textit{k}-PPPM in terms of security and privacy enhancements.
For the sake of completeness, these properties are compared between the original (unprotected) event log $\mathcal{L}$, a protected event log version $\mathcal{L}'$ obtained after applying \textit{k}-PPPM, and a protected event log version $\mathcal{L}''$ obtained through pseudonymisation or encryption.

Regarding security aspects, \textit{k}-PPPM contributes to confidentiality by obfuscating direct identifiers and breaking their direct linkage with the confidential attributes in the event logs.
However, other security requirements, such as integrity, availability and authentication, are beyond the scope of this method.
Since the anonymisation procedure conducted by \textit{k}-PPPM does not require external parties nor network communications, the potential security threats are relaxed, and the security of the method lies in the distortion of the statistical properties of the event data.
Recalling confidentiality, \textit{k}-PPPM is supported by the obfuscation of personally identifiable information through pseudonymisation or state-of-the-art encryption techniques at the last stage of the algorithm.
These techniques guarantee the confidentiality of $\mathcal{L}'$ as long as the private cryptographic keys remain secret.
As both $\mathcal{L}'$ and $\mathcal{L}''$ rely on pseudonymisation or encryption mechanisms, the confidentiality level of both approaches is the same.

As a distinguishing factor, \textit{k}-PPPM provides privacy guarantees to the individuals appearing in $\mathcal{L}'$ in order to minimise the impact of modelling-based attacks in combination with location-oriented attacks, where classical obfuscation techniques fail.
Given an event log $\mathcal{L}'$, the event data and the individuals' process models that can be discovered are constrained by the \textit{k}-anonymity model:
(i) each trace within the event log is assigned to, at least, $k$ different individuals, each of them with their own confidential data, \ie minimising the attribute disclosure risk,
(ii) the same process model would be discovered in, at least, $k$ different individuals, \ie minimising the identity disclosure risk,
and (iii) each process model represents the behaviour of a group of, at least, $k$ individuals, instead of a single (and potentially identifiable) individual, \ie minimising the identity disclosure risk as well.

By comparing the data in $\mathcal{L}'$ and $\mathcal{L}''$, the knowledge that attackers gain from the modelling of processes is significantly different.
Whereas $p$ different process models can be discovered in $\mathcal{L}''$ (and $\mathcal{L}$ as well), only $n$ different process models would be discovered in $\mathcal{L}'$.
Despite this information loss, \textit{k}-PPPM is resilient to the attacker model described in Section \ref{sec:kpppm:attacks}, as attackers are not able to link an observed process model of the victim to their background knowledge, but to a group of $k$ indistinguishable individuals.
The re-identification risk is hence upper-bounded by $1/k$.
This situation is depicted in Figure \ref{fig:kpppm:kpppm_vs_encryption}.

\begin{figure}[b!]
\centering\includegraphics[width=0.82\linewidth]{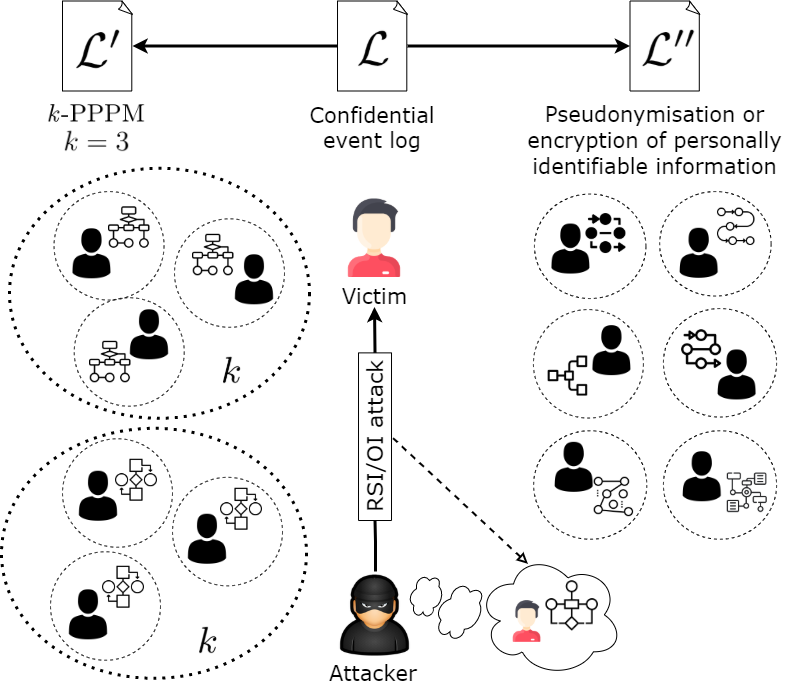}
\caption{Differences, from the attacker's view, between the privacy enhancements when using \textit{k}-PPPM or using pseudonymisation or encryption only (adapted from \cite{kpppm}).}
\label{fig:kpppm:kpppm_vs_encryption}
\end{figure}

Last but not least, the microaggregation strategy applied in \textit{k}-PPPM is slightly different compared to classical microaggregation approaches used in SDC.
On the one hand, there are some differences from a data perspective: event logs (used in \textit{k}-PPPM) cannot be directly understood as microdata sets (used in SDC).
For example, SDC techniques suppress the personally identifiable information in microdata sets, and the privacy guarantees reside on the ability to generalise or aggregate quasi-identifier attributes.
However, event logs rarely contain quasi-identifiers, and we have already noticed the limitations of process mining if personally identifiable information is completely removed from event logs (see Section \ref{subsec:upppm:vuln}).
On the other hand, there are also some discrepancies from an analytical perspective: the knowledge that can be acquired from event logs is different to that of microdata sets.
In microdata sets, each record corresponds to an independent individual, so there is no relationship between records.
However, events do have a relationship between them (mainly defined in the $case$ and $time$ attributes), and these relationships are exploited in process mining, such as when discovering process models from the control-flow perspective.
Consequently, in contrast to classical approaches, \textit{k}-PPPM must preserve people's privacy at the same time that preserves the relationships among events so as to maximise data utility.
A comparison between the two main microaggregation phases in classical SDC and \textit{k}-PPPM is provided in Table \ref{tbl:kpppm:comparison_sdc_kpppm}.

\begin{table}[t!]
\centering
\caption{Comparison between classical SDC and \textit{k}-PPPM.}
\label{tbl:kpppm:comparison_sdc_kpppm}
{\scriptsize
\renewcommand{\arraystretch}{0.75}
\begin{tabularx}{\textwidth}{clCC}
   \toprule
   & & \textbf{Classical SDC} & \textbf{\textit{k}-PPPM} \\
   \midrule
   \multirow{11}{*}{\rotatebox{90}{\textbf{Clustering phase}}} & \textbf{Input} & Microdata sets & Event logs \\ \cmidrule{2-4}
   & \textbf{Objective} & Minimise within-cluster records distance & Maximise within-cluster models similarity \\ \cmidrule{2-4}
   & \textbf{Elements} & Independent records & Process models, \ie individuals \\ \cmidrule{2-4}
   & \textbf{Criteria} & Euclidean, Manhattan, Minkowski, Chebyshev\ldots & Process model similarity, \eg VEO, VR, WD, DC\ldots \\ \cmidrule{2-4}
   & \textbf{Output} & Clusters of $k$ to $2k-1$ records & Clusters of $k$ to $2k-1$ individuals \\
   \midrule
   \multirow{23}{*}{\rotatebox{90}{\textbf{Representative phase}}} & \textbf{Input} & Clusters of $k$ to $2k-1$ records & Clusters of $k$ to $2k-1$ individuals \\ \cmidrule{2-4}
   & \textbf{Objective} & Compute a virtual centroid record, a within-cluster representative record & Compute a virtual centroid individual, a within-cluster representative process model \\ \cmidrule{2-4}
   & \textbf{Selection} & Average or median of all the records within the same cluster & Random sampling of the traces associated to all the individuals within the same cluster \\ \cmidrule{2-4}
   & \textbf{Replacement} & Replace each record's value by the representative's value & Replicate the traces of the representative's individual for all the individuals within the cluster \\ \cmidrule{2-4}
   & \textbf{Obfuscation} & Remove direct identifiers and aggregate quasi-identifiers & Obfuscate personally identifiable information of individuals (attribute $resource$) \\ \cmidrule{2-4}
   & \textbf{Output} & A microdata set containing, at least, $k$ records for each combination of quasi-identifiers & An event log containing, at least, $k$ individuals with the same events and traces, \ie process models \\
   \bottomrule
\end{tabularx}
}
\end{table}

\section{Experimental Setup} \label{sec:kpppm:experiments}

This section describes the experimental setup of the proposed \textit{k}-PPPM method.
Since the purpose of the evaluation is analogous to the one from the previous chapter, the evaluation methodology is identical as well.
For the sake of completeness, next we provide an overview of this methodology, also depicted in Figure \ref{fig:kpppm:methodology}, and we refer the reader to Section \ref{sec:upppm:experiments} for further details.

The experimental setup aims to assess the impact of \textit{k}-PPPM by focusing on the quality of the process models discovered from the protected event logs $\mathcal{L}'$ in comparison to the corresponding process models discovered from the original event logs $\mathcal{L}$.
The user-centric nature of the proposed method leads to evaluate the process models (from a control-flow perspective) associated to each individual.
The quality of the results obtained after applying \textit{k}-PPPM is measured according to the following two research questions.

\begin{itemize}
    \item \textit{Q1 -- Individual distortion}: How similar are the process models of an individual $I$ when discovered from the original event log (\ie $\mathcal{M}_I$), and when discovered from the protected event log (\ie $\mathcal{M}'_I$)?
    \item \textit{Q2 -- Inter-individual distortion}: Are the differences among the individuals' process models from the original event log (\ie $\mathcal{M}_{I_1}, \mathcal{M}_{I_2},\ldots,\mathcal{M}_{I_p}$) also maintained among the individuals' process models from the protected event log (\ie $\mathcal{M}'_{I_1}, \mathcal{M}'_{I_2},\ldots,\mathcal{M}'_{I_p}$)?
\end{itemize}

\begin{figure}[b!]
\centering\includegraphics[width=0.85\linewidth]{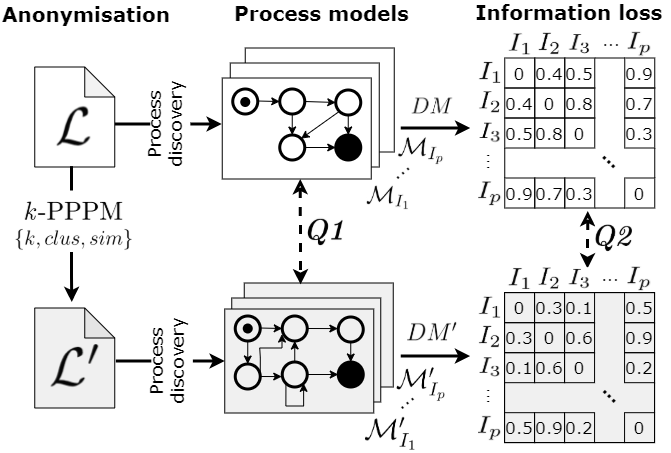}
\caption{Methodology to evaluate the \textit{k}-PPPM method (adapted from \cite{kpppm}).}
\label{fig:kpppm:methodology}
\end{figure}

As explained in the previous chapter, process models are represented using D/F-graphs, and experiments have been conducted using the six real-life event logs described in Table \ref{tbl:background:event_logs}, where the number of resources in each event log corresponds to the number of individuals $p$.
Also, the procedure to calculate the QS and ILS results associated to answer \textit{Q1} and \textit{Q2}, respectively, has been maintained too.

As aforementioned, \textit{k}-PPPM requires specifying three parameters to anonymise event logs, which directly affect the quality of the resulting process models.
To assess this impact, our method is tested with different combinations of its three parameters.
First, regarding the privacy level $k$, we use common values that are used in the \textit{k}-anonymity literature, namely $k = \{2, 3, 4, 5, 10, 20\}$.
Second, regarding the clustering algorithm $clus$, we use four strategies, in which three of them are popular heuristics from the literature, namely MDAV \cite{domingo2005ordinal}, \textit{k}-member (KM) \cite{byun2007kmember} and OKA \cite{lin2008oka}.
Additionally, we also use a fourth naive algorithm as a baseline (BL) that groups individuals according to their number of traces (\ie two individuals are similar if they have a similar number of traces in the event log).
The objective of BL is to evaluate whether there is a significant difference between using process-oriented clustering algorithms and event log-oriented clustering algorithms.
And third, regarding the similarity measure $sim$, we have chosen four popular graph similarity measures from the literature, namely VEO \cite{papadimitriou2010web}, VR \cite{papadimitriou2010web}, WD \cite{shoubridge2002detection} and DC \cite{koutra2013deltacon}.
With the aim to observe the impact of these parameters, the proposed method is executed for all combinations of these parameters: ($k_i,c_i,s_i$), $\forall k_i \in k, \forall c_i \in clus, \forall s_i \in sim$.
Therefore, each event log $\mathcal{L}$ is anonymised 96 times (6 $\times$ 4 $\times$ 4), leading to 96 different privacy-preserved event log versions $\mathcal{L}'$, achieved from the execution of \textit{k}-PPPM with (2, MDAV, VEO), (3, MDAV, VEO),\ldots,(20, BL, DC).

\section{Results and Discussion} \label{sec:kpppm:disc}

This section discusses the experimental results for assessing the impact of \textit{k}-PPPM on the protected process models in accordance to the aforementioned methodology.
Due to the non-deterministic nature of the method, all the results reported correspond to the average of five executions for each combination of parameters.

Appendix \ref{app:kpppm} provides all the details of the quantitative results obtained.
More specifically, Tables \ref{tbl:app_kpppm:results_qs_1} to \ref{tbl:app_kpppm:results_ils_4} describe the averaged QS and ILS results, respectively, for each \textit{k}-PPPM execution with a combination of its parameters for all the event logs.
Besides the six event logs evaluated, an averaged result from all of them is also provided in the last column of the tables.
For the sake of comprehensiveness, results are grouped by $k$ and, within each group, the best and the worst QS/ILS values are highlighted in green and red, respectively.
To ease evaluating the contribution of each parameter, Tables \ref{tbl:kpppm:summary_cs} and \ref{tbl:kpppm:summary_ils} group and average the QS and ILS results, respectively, according to each parameter value.
As before, the best and the worst results are highlighted in green and red, respectively, within each parameter group.

\begin{sidewaystable}[htbp]
\renewcommand{\tabcolsep}{0.2cm}
\renewcommand{\arraystretch}{0.92}
\centering
\scriptsize
\caption{Summary of the QS results grouped by parameter.} \label{tbl:kpppm:summary_cs}
\begin{tabular}{ccrccccccc}
   \toprule
   &&& \multicolumn{6}{c}{\textbf{Event logs}} & \multirow{2}{*}{\textbf{Avg.}} \\
   \cmidrule(lr){4-9}
   &&& \textbf{BPI12} & \textbf{BPI13} & \textbf{BPI14} & \textbf{BPI15} & \textbf{CoSeLoG} & \textbf{TGN-Hospital} \\
   \midrule
   \multirow{15}{*}{\rotatebox{90}{\textbf{Grouping criteria}}} & \multirow{5}{*}{\textbf{Privacy}} & \textbf{2} & \gr{0.268 $\pm$ 0.161} & \gr{0.169 $\pm$ 0.183} & \gr{0.407 $\pm$ 0.129} & \gr{0.496 $\pm$ 0.193} & \gr{0.337 $\pm$ 0.188} & \gr{0.324 $\pm$ 0.128} & \gr{0.334 $\pm$ 0.164} \\
   & \multirow{5}{*}{($k$)} & \textbf{3} & 0.335 $\pm$ 0.181 & 0.212 $\pm$ 0.195 & 0.459 $\pm$ 0.122 & 0.587 $\pm$ 0.17 & 0.424 $\pm$ 0.175 & 0.389 $\pm$ 0.126 & 0.401 $\pm$ 0.162 \\
   & & \textbf{4} & 0.365 $\pm$ 0.187 & 0.243 $\pm$ 0.199 & 0.485 $\pm$ 0.12 & 0.628 $\pm$ 0.162 & 0.46 $\pm$ 0.18 & 0.426 $\pm$ 0.125 & 0.435 $\pm$ 0.162 \\
   & & \textbf{5} & 0.389 $\pm$ 0.186 & 0.257 $\pm$ 0.205 & 0.502 $\pm$ 0.117 & 0.654 $\pm$ 0.149 & 0.5 $\pm$ 0.167 & 0.448 $\pm$ 0.126 & 0.459 $\pm$ 0.158 \\
   & & \textbf{10} & 0.476 $\pm$ 0.192 & 0.298 $\pm$ 0.208 & 0.544 $\pm$ 0.112 & 0.704 $\pm$ 0.127 & 0.543 $\pm$ 0.149 & 0.494 $\pm$ 0.124 & 0.51 $\pm$ 0.152 \\
   & & \textbf{20} & \re{0.534 $\pm$ 0.187} & \re{0.337 $\pm$ 0.211} & \re{0.576 $\pm$ 0.102} & \re{0.736 $\pm$ 0.12} & \re{0.588 $\pm$ 0.132} & \re{0.533 $\pm$ 0.121} & \re{0.551 $\pm$ 0.146} \\
   \cmidrule{2-10}
   & \multirow{3}{*}{\textbf{Algorithm}} & \textbf{MDAV} & \gr{0.363 $\pm$ 0.182} & 0.195 $\pm$ 0.188 & \gr{0.479 $\pm$ 0.114} & \gr{0.631 $\pm$ 0.151} & 0.466 $\pm$ 0.15 & \gr{0.399 $\pm$ 0.121} & \gr{0.422 $\pm$ 0.151} \\
   & \multirow{3}{*}{($clus$)} & \textbf{KM} & 0.368 $\pm$ 0.176 & \gr{0.195 $\pm$ 0.186} & 0.482 $\pm$ 0.111 & 0.632 $\pm$ 0.149 & \gr{0.465 $\pm$ 0.154} & 0.404 $\pm$ 0.113 & 0.424 $\pm$ 0.148 \\
   & & \textbf{OKA} & 0.373 $\pm$ 0.204 & 0.235 $\pm$ 0.208 & 0.496 $\pm$ 0.127 & 0.634 $\pm$ 0.153 & 0.474 $\pm$ 0.176 & 0.412 $\pm$ 0.142 & 0.437 $\pm$ 0.168 \\
   & & \textbf{BL} & \re{0.474 $\pm$ 0.168} & \re{0.386 $\pm$ 0.219} & \re{0.525 $\pm$ 0.116} & \re{0.642 $\pm$ 0.161} & \re{0.498 $\pm$ 0.181} & \re{0.529 $\pm$ 0.124} & \re{0.509 $\pm$ 0.162} \\
   \cmidrule{2-10}
   & \multirow{3}{*}{\textbf{Measure}} & \textbf{VEO} & \re{0.4 $\pm$ 0.177} & 0.248 $\pm$ 0.198 & \gr{0.491 $\pm$ 0.113} & 0.626 $\pm$ 0.149 & 0.469 $\pm$ 0.164 & \gr{0.431 $\pm$ 0.123} & \gr{0.444 $\pm$ 0.153} \\
   & \multirow{3}{*}{($sim$)} & \textbf{VR} & 0.393 $\pm$ 0.183 & \re{0.26 $\pm$ 0.202} & 0.495 $\pm$ 0.117 & \gr{0.624 $\pm$ 0.154} & \gr{0.467 $\pm$ 0.166} & 0.438 $\pm$ 0.12 & 0.446 $\pm$ 0.157 \\
   & & \textbf{WD} & 0.392 $\pm$ 0.189 & 0.257 $\pm$ 0.202 & 0.496 $\pm$ 0.118 & 0.639 $\pm$ 0.157 & 0.48 $\pm$ 0.162 & \re{0.439 $\pm$ 0.129} & \re{0.452 $\pm$ 0.155} \\
   & & \textbf{DC} & \gr{0.387 $\pm$ 0.182} & \gr{0.245 $\pm$ 0.198} & \re{0.5 $\pm$ 0.119} & \re{0.649 $\pm$ 0.153} & \re{0.485 $\pm$ 0.169} & 0.435 $\pm$ 0.128 & 0.45 $\pm$ 0.158 \\
   \bottomrule
\end{tabular}
\end{sidewaystable}

\begin{sidewaystable}[htbp]
\renewcommand{\tabcolsep}{0.2cm}
\renewcommand{\arraystretch}{0.92}
\centering
\scriptsize  
\caption{Summary of the ILS results grouped by parameter.} \label{tbl:kpppm:summary_ils}
\begin{tabular}{ccrccccccc}
   \toprule
   &&& \multicolumn{6}{c}{\textbf{Event logs}} & \multirow{2}{*}{\textbf{Avg.}} \\
   \cmidrule(lr){4-9}
   &&& \textbf{BPI12} & \textbf{BPI13} & \textbf{BPI14} & \textbf{BPI15} & \textbf{CoSeLoG} & \textbf{TGN-Hospital} \\
   \midrule
   \multirow{15}{*}{\rotatebox{90}{\textbf{Grouping criteria}}} & \multirow{5}{*}{\textbf{Privacy}} & \textbf{2} & \gr{0.071 $\pm$ 0.016} & \gr{0.069 $\pm$ 0.017} & \gr{0.047 $\pm$ 0.022} & \gr{0.044 $\pm$ 0.023} & \gr{0.075 $\pm$ 0.025} & \gr{0.058 $\pm$ 0.023} & \gr{0.061 $\pm$ 0.026} \\
   & \multirow{5}{*}{($k$)} & \textbf{3} & 0.103 $\pm$ 0.025 & 0.081 $\pm$ 0.019 & 0.058 $\pm$ 0.026 & 0.064 $\pm$ 0.028 & 0.096 $\pm$ 0.029 & 0.071 $\pm$ 0.028 & 0.08 $\pm$ 0.034 \\
   & & \textbf{4} & 0.116 $\pm$ 0.027 & 0.093 $\pm$ 0.021 & 0.066 $\pm$ 0.028 & 0.077 $\pm$ 0.032 & 0.116 $\pm$ 0.03 & 0.082 $\pm$ 0.032 & 0.093 $\pm$ 0.037 \\
   & & \textbf{5} & 0.131 $\pm$ 0.029 & 0.095 $\pm$ 0.025 & 0.071 $\pm$ 0.029 & 0.118 $\pm$ 0.033 & 0.131 $\pm$ 0.028 & 0.089 $\pm$ 0.034 & 0.108 $\pm$ 0.04 \\
   & & \textbf{10} & 0.201 $\pm$ 0.037 & 0.11 $\pm$ 0.025 & 0.087 $\pm$ 0.032 & 0.173 $\pm$ 0.041 & 0.236 $\pm$ 0.031 & 0.112 $\pm$ 0.038 & 0.155 $\pm$ 0.063 \\
   & & \textbf{20} & \re{0.275 $\pm$ 0.037} & \re{0.128 $\pm$ 0.029} & \re{0.123 $\pm$ 0.03} & \re{0.328 $\pm$ 0.035} & \re{0.371 $\pm$ 0.054} & \re{0.137 $\pm$ 0.039} & \re{0.229 $\pm$ 0.107} \\
   \cmidrule{2-10}
   & \multirow{3}{*}{\textbf{Algorithm}} & \textbf{MDAV} & 0.136 $\pm$ 0.025 & 0.072 $\pm$ 0.02 & \gr{0.071 $\pm$ 0.025} & \gr{0.125 $\pm$ 0.033} & \gr{0.158 $\pm$ 0.031} & \gr{0.076 $\pm$ 0.025} & \gr{0.107 $\pm$ 0.049} \\
   & \multirow{3}{*}{($clus$)} & \textbf{KM} & \gr{0.136 $\pm$ 0.024} & \gr{0.072 $\pm$ 0.019} & 0.072 $\pm$ 0.026 & 0.127 $\pm$ 0.036 & 0.162 $\pm$ 0.032 & 0.077 $\pm$ 0.024 & 0.108 $\pm$ 0.05 \\
   & & \textbf{OKA} & 0.136 $\pm$ 0.028 & 0.086 $\pm$ 0.021 & 0.075 $\pm$ 0.03 & 0.126 $\pm$ 0.031 & 0.164 $\pm$ 0.033 & 0.081 $\pm$ 0.028 & 0.112 $\pm$ 0.048 \\
   & & \textbf{BL} & \re{0.19 $\pm$ 0.037} & \re{0.154 $\pm$ 0.031} & \re{0.083 $\pm$ 0.031} & \re{0.159 $\pm$ 0.028} & \re{0.197 $\pm$ 0.035} & \re{0.133 $\pm$ 0.051} & \re{0.154 $\pm$ 0.058} \\ 
   \cmidrule{2-10}
   & \multirow{3}{*}{\textbf{Measure}} & \textbf{VEO} & \re{0.154 $\pm$ 0.03} & 0.094 $\pm$ 0.022 & 0.074 $\pm$ 0.027 & 0.13 $\pm$ 0.026 & 0.172 $\pm$ 0.032 & \gr{0.089 $\pm$ 0.031} & 0.119 $\pm$ 0.051 \\
   & \multirow{3}{*}{($sim$)} & \textbf{VR} & 0.145 $\pm$ 0.026 & 0.097 $\pm$ 0.024 & \gr{0.074 $\pm$ 0.026} & \gr{0.13 $\pm$ 0.025} & \gr{0.166 $\pm$ 0.031} & 0.091 $\pm$ 0.031 & \gr{0.118 $\pm$ 0.047} \\
   & & \textbf{WD} & 0.156 $\pm$ 0.033 & \re{0.1 $\pm$ 0.025} & 0.075 $\pm$ 0.029 & 0.136 $\pm$ 0.037 & \re{0.173 $\pm$ 0.034} & \re{0.094 $\pm$ 0.035} & \re{0.123 $\pm$ 0.055} \\
   & & \textbf{DC} & \gr{0.143 $\pm$ 0.025} & \gr{0.093 $\pm$ 0.021} & \re{0.078 $\pm$ 0.029} & \re{0.141 $\pm$ 0.039} & 0.172 $\pm$ 0.034 & 0.092 $\pm$ 0.033 & 0.12 $\pm$ 0.052 \\ 
   \bottomrule
\end{tabular}
\end{sidewaystable}

The anonymisation conducted in \textit{k}-PPPM is data-dependence, this is, the quality of the protected event logs and process models depends on the event data.
More specifically, due to the clustering nature of the method, this quality highly depends on the profile of the individuals within the event logs.
If the behaviour of the individuals is very heterogeneous and distant, the quality of the anonymisation will worsen significantly.
In contrast, if the event log contains individuals who behave similarly (\ie there are individuals from the same department or with the same role), the loss of quality will not be that severe because clusters will maintain certain similarity.
For this reason, although executing \textit{k}-PPPM with the same combination of parameters, the QS or ILS results can differ among different event logs.
For example, executing \textit{k}-PPPM with $k = 4$, $clus =$ MDAV and $sim =$ VR in event logs BPI13, BPI15 and CoSeLoG, the QS results are 0.184, 0.601 and 0.414, respectively, and the ILS results are 0.066, 0.072 and 0.1, respectively.
For this reason, computing the average quality of the \textit{k}-PPPM executions (available in the last column of the tables) can be meaningful to evaluate the high-level behaviour of \textit{k}-PPPM anonymisation.
Despite the above, experimental results show a similar tendency as the \textit{k}-PPPM's parameters vary, as reflected in Tables \ref{tbl:kpppm:summary_cs} and \ref{tbl:kpppm:summary_ils}.
The contribution of each of these parameters is discussed below.

First of all, results show that increasing the privacy level $k$ contributes significantly to decreasing the quality of the protected process models.
The larger $k$, the more individuals share the same process model, so the more difficult for the attacker to re-identify them.
But, the quality of the protected process model associated to each individual, influenced by the group-based anonymisation, worsens.
Unsurprisingly, this development is not new, as it has already been reported in privacy protection literature.
In average, both QS and ILS results steadily increase together with the privacy level.
However, note that this is not always true in the experimental results, and the other two parameters can slightly contribute to increase the quality at the same privacy level.

Our experiments also show that the proper selection of the clustering algorithm helps reducing the distortion introduced to the process models for a certain privacy level.
Better results can be achieved at higher privacy levels if using heuristic clustering algorithms, such as MDAV, KM or OKA, rather than using non-optimal clustering algorithms, such as BL, at lower privacy levels.
For instance, both QS and ILS results obtained using $k =$ 4, $clus =$ MDAV and $sim =$ DC are considerably better in comparison to the ones obtained using $k =$ 3, $clus =$ BL and $sim =$ DC.
In our experiments, the BL algorithm has demonstrated to be inefficient as the clustering criteria is based on an event log property, \ie the number of traces per individual.
Although process models are discovered from the event logs, it seems that forming the clusters according to an event log criteria is not optimal.
Contrary, the three heuristic algorithms, which use a process models similarity criteria, are more efficient to this end.
Despite the small differences between them, note that MDAV and KM behave slightly better than the OKA algorithm.
We believe that this aspect derives from the very design of the algorithm, as both MDAV and KM algorithms start clustering the most distant records, which are more likely to be part of less cohesive clusters.
The aim of starting clustering the most distant records is precisely to find the most appropriate clusters for these records to maximise the within-cluster similarity, already low by default.
Hence, clustering the most abnormal records first leads to better anonymisation results.

Besides, experiments demonstrate that the quality of the process models is not strongly affected by the similarity measure used during the anonymisation.
Although some measures could particularly behave better in certain executions or in certain event logs, the general differences between the quality results are relatively insignificant in average.
Therefore, the similarity measure does not significantly influence the anonymisation results, and users are free to use the measure that best fit with their interests, such as according to the modelling notation of the process models or based on computational criteria.

\subsection{Relationship between QS and ILS Results}

To put the previous results in context, the relationship between the QS and ILS results for each combination of parameters and for all the event logs is depicted in Figures \ref{fig:app_kpppm:results_bpi12_bpi13}, \ref{fig:app_kpppm:results_bpi14_bpi15} and \ref{fig:app_kpppm:results_coselog_tgn} from Appendix \ref{app:kpppm}.
Each point in the chart represents the execution of \textit{k}-PPPM with a certain combination of parameters.
To ease evaluating the direct impact of the privacy level in the quality of the process models, each dot has been coloured according to its $k$ value.
Moreover, executions obtained using the BL clustering have been discarded to prevent the appearance of naive and non-optimal solutions.
Finally, to evaluate the general impact of the proposed method from a high-level perspective (regardless the particularities of each event log), Figure \ref{fig:kpppm:results_avg} illustrates the averaged results obtained from all event logs.

\begin{figure}[t!]
\centering
\includegraphics[width=0.7\linewidth]{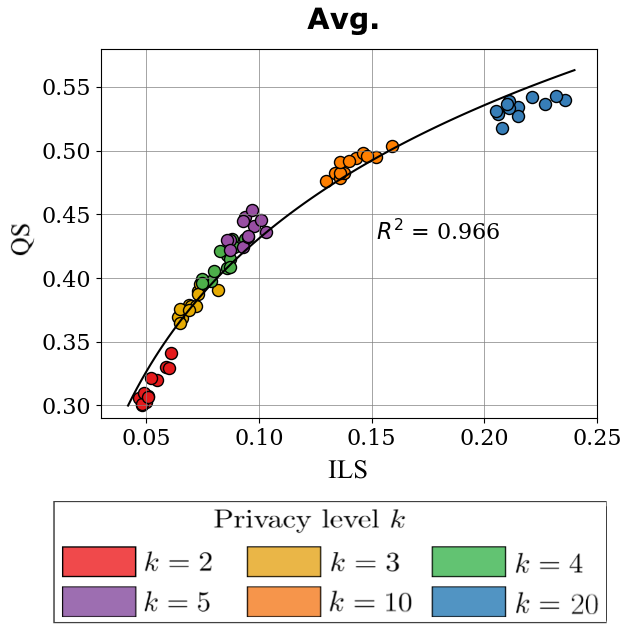}
\caption{Correlation between the averaged QS and ILS results from all the event logs (reprinted from \cite{kpppm}).}
\label{fig:kpppm:results_avg}
\end{figure}

Interestingly enough, although evaluating the process models from different perspectives, there exists an apparent direct correlation between the results: the more individual distortion in the process models (QS), the more distant the relationships among them (ILS).
The microaggregation nature of \textit{k}-PPPM leads to a non-negligible individual distortion at the very beginning, \ie QS values for $k = 2$ are relatively high.
But, as the privacy level increases, these individual distortions increments are not that severe.
In contrast, inter-individual distortions are greatly preserved, \ie ILS results are generally low.
However, increasing the privacy level has a larger impact on these inter-individual distortions rather than the individual distortions.
This phenomena occurs because little, but continuous, distortions of all the process models individually have a higher effect in all the relationships among all process models.
Notwithstanding, the low ILS results suggests that the method preserves most of the patterns from the original event logs, while protecting individuals' privacy (\ie QS) at the same time.
This property is indeed interesting from a privacy perspective, as it enables gathering similar insights and knowledge from the protected event logs (and the protected process models) as if they were obtained from the original event logs (and the original process models), but without disclosing confidential information.

In addition to the quantitative results, Figure \ref{fig:kpppm:models} depicts different versions of the process model corresponding to a certain individual from the CoSeLoG event log obtained from different executions of \textit{k}-PPPM.
More specifically, the original process model can be qualitatively compared to protected process models obtained using heuristic or non-heuristic algorithms (MDAV or BL, respectively) and for different privacy levels ($k = 2$ and $k = 5$).

\begin{figure}[b!]
\centering
\includegraphics[width=0.9\linewidth]{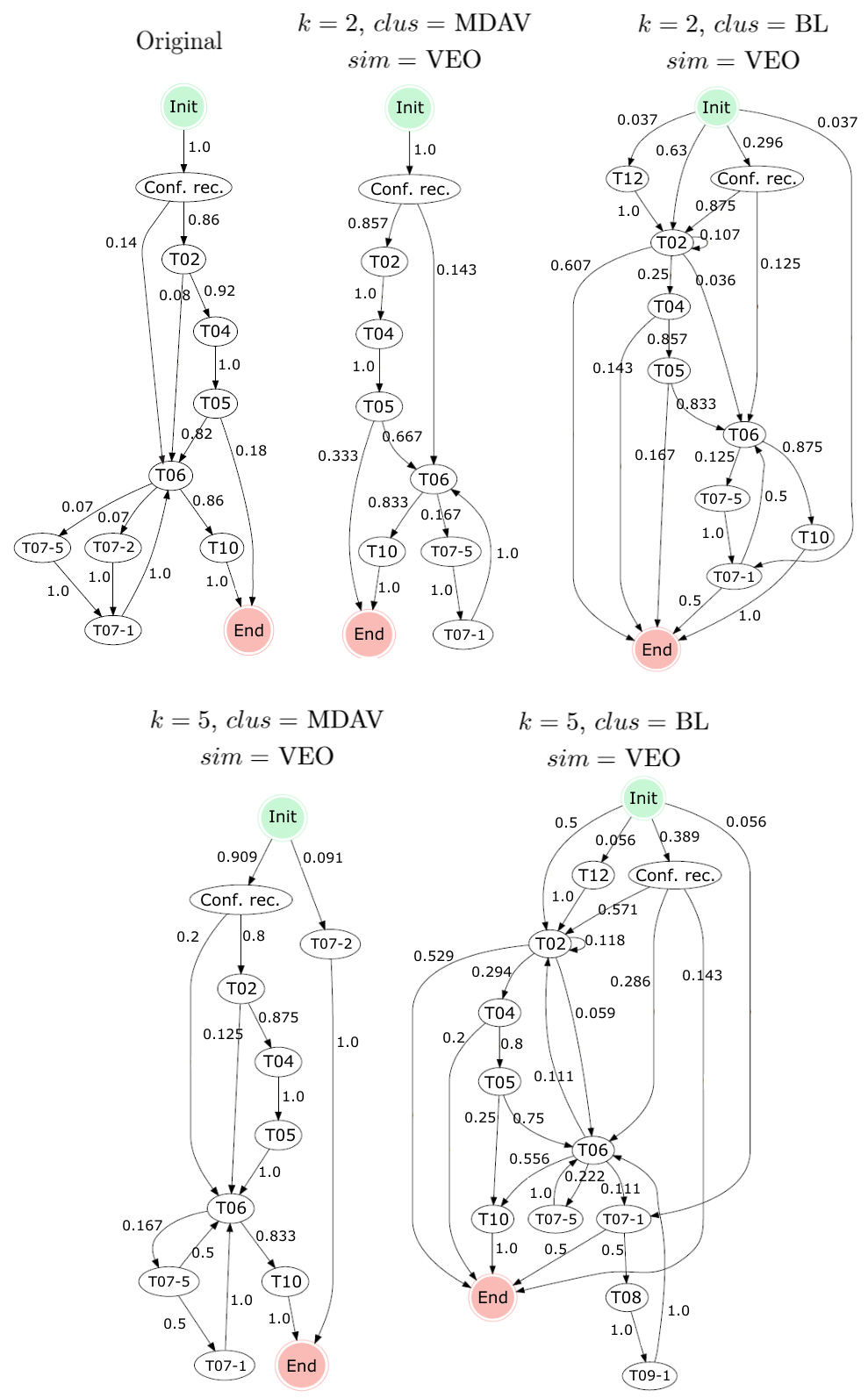}
\caption{Versions of the process model of a certain individual from the CoSeLoG event log after applying \textit{k}-PPPM with different parameters (reprinted from \cite{kpppm}).}
\label{fig:kpppm:models}
\end{figure}

\subsection{Significance of the Anonymisation Parameters}

Experimental results have shown that the proper selection of the parameters used in \textit{k}-PPPM can significantly influence the quality of the anonymisation results.
To verify the significance of the results differences, statistical analyses of two sampled t-Tests have been performed.
Since of evaluation procedure is analogous from the previous chapter, please refer to Section \ref{subsec:upppm:t-tests} for further details.
Figures \ref{fig:app_kpppm:ttest_qs_k} to \ref{fig:app_kpppm:ttest_ils_sim} from Appendix \ref{app:kpppm} depict the $p$-value results obtained from the t-Tests between each pair of parameter's values for the QS and ILS results.

With regards to the privacy level, in most cases, $p$-values are lower than the significance level $\alpha$ (set to $\alpha = 0.05$).
Therefore, t-Tests confirm that the privacy level truly affects the quality of the process models from a statistical perspective.
However, it is also noteworthy that t-Tests do not sometimes detect significant differences between consecutive privacy levels, such as between $k = 4$ and $k = 5$ in BPI12 and TGN-Hospital event logs, or for multiple combinations in BPI13 event log, to name a few.
All in all, averaged results determine that the quality of the process models significantly improves by lowering the privacy level.

Regarding the clustering algorithm, t-Tests generally agree that the differences between heuristic and non-heuristic algorithms are statistically significant enough.
Hence, we can ensure that selecting the proper clustering algorithm contributes to achieve better anonymisation results.
But, note that the particularities of the event logs can also affect to its significance.
For instance, in event logs BPI15 and CoSeLoG, t-Tests have not been able to detect significant differences between these two kinds of clustering algorithms.
Also, in the case of BPI13, significant differences were identified within the heuristic algorithms themselves, wherein results obtained using MDAV and KM are significantly better than the ones obtained using OKA.
Hence, although the use of heuristic algorithms contributes to minimise information loss, they are not a silver bullet to guarantee the best possible results.

Finally, our t-Tests confirm that the differences, for both the QS and ILS results, between using a similarity measure or another during the anonymisation are not statistically significant to ensure that they contribute to the quality of the results.
Hence, although it does not provide significant differences, experimental results show that some measures lead to better outcomes than other measures in particular event logs.

\section{Conclusions} \label{sec:kpppm:concl}

Location-oriented targeted attacks are threats to people's privacy in institutions with public access, such as public hospitals, in which attackers might be able to link confidential information to individuals with enough background knowledge.
Whereas the previous chapter concentrated on the ability to re-identify individuals through distribution-based attacks, this chapter has successfully demonstrated the risks surrounding modelling-based attacks.
Unfortunately, the use of pseudonymisation or encryption, classical solutions in process mining to protect individuals' privacy, are insufficient to protect event logs.

To face this issue, this chapter has presented a novel PPPM method, called \textit{k}-PPPM, aiming at distorting event data so as to affect the individuals' process models, and rendering the background knowledge of attackers useless, according to a certain privacy threshold.
More specifically, the proposed method, inspired by \textit{k}-anonymity, is pioneering in adopting microaggregation techniques within the PPPM field.
Experimental results, tested with six real-life event logs and for different combination of \textit{k}-PPPM's parameters, have demonstrated the ability to preserve most of the relationships and patterns among the different process models (\ie inter-individual distortion, ILS), while protecting each process model individually (\ie individual distortion, QS).
We have also demonstrated the impact of such \textit{k}-PPPM's parameters in the quality of the anonymised results conducting two sampled t-Tests statistical analyses.
In short, whereas the privacy level has a direct impact on the data utility, we have observed the importance of using heuristic-based clustering algorithms to reduce information loss.
In contrast, there is no evidence that the similarity measure improves the anonymised results quality.

Although the contributions of the last two chapters are a step forward in the field of PPPM, there is still room for improvement.
Indeed, gathering ideas from the classical privacy protection techniques and applying them into process mining may bring great research opportunities.
Future work will focus on the development of novel PPPM techniques able to cope with even more complex attacker models, which can gain advanced knowledge.
Also, we foresee the creation and application of more robust privacy-preserving models, which incorporate properties, such as \textit{l}-diversity or \textit{t}-closeness.
With regards to the proposed methods, it could be valuable to eliminate the non-deterministic nature of the methods and prevent the results randomness that may affect the quality results.
Besides, evaluating the suitability of the proposed methods using advanced modelling notations, such as BPMN or petri nets, would be of interest.
Last but not least, the differences between the original and the protected process models could also be assessed qualitatively, by asking experts and practitioners whether those differences could change their understanding and the decisions they would make.

\part{Applications} \label{part:apps}
\chapter{Specific Applications to Smart Health}
\label{chap:apps}

\emph{Smart health applications are the icing of the cake, this is, the final embodiment of solutions to problems solved by applying algorithms to all sorts of data so as to improve people's quality of life. This chapter presents three applications to smart healthcare. First, Section \ref{sec:apps:rs} describes a smart route recommendation system, which uses crowdsourcing-based information along with data gathered from the smart city infrastructure to provide recommendations to citizens. Then, Section \ref{sec:apps:mob} presents a platform founded on smartphones and motion sensors to support early mobilisation programmes of critically ill patients hospitalised in intensive care units. Next, Section \ref{sec:apps:hgis} concentrates on a healthcare-oriented geographic information system to support the integration of heterogeneous spatio-temporal data collected within a large healthcare provider. Finally, a summary of the main applications is explained in Section \ref{sec:apps:concl}.}

\minitoc

\section{SmartRoute: A Context-Aware Recommender System} \label{sec:apps:rs}

Within the context of smart cities, next-generation systems will be founded on the exploitation of context-aware data (using the sensing infrastructure of the environment) and collective behaviours (using crowdsourcing-based information) to improve the quality of life of their citizens.
In this section, we combine the aforementioned ingredients and present a novel context-aware recommender system for fostering healthier lifestyles amongst citizens within a smart city.
Section \ref{subsec:apps:rs:ratio} motives the reader with the rationale of our approach.
Next, Section \ref{subsec:apps:rs:arc} describes the architecture of the approach, this is, the main actors and general functioning.
Finally, the implementation details of our scheme are explained in Section \ref{subsec:apps:rs:impl}.

\subsection{Rationale} \label{subsec:apps:rs:ratio}

Many citizens perform physical activities in the city, namely walking, running and cycling, because of medical recommendations based on health issues.
However, many people do not receive the proper guidance on how, where or when to perform these activities, and basing their decisions on the distance of the route or the meteorological conditions, but neglecting other factors such as the route's difficulty or the air pollution.
With the aim to foster healthy habits and lifestyles in smart cities, healthcare services should be redefined and adapt dynamically to the citizens' needs.
In this approach, we show how recommendation systems could be used to provide healthcare services within the context of a smart city in which citizens collaborate with the city to improve their quality of life.
In particular, we describe a system able to provide citizens with real-time recommendations about the routes that better fit their capacities.
These recommendations consider diverse information from multiple sources: (i) citizens' health conditions, (ii) citizen's preferences, and (iii) contextual information from the smart city sensing infrastructure.
Below, the basic characteristics of our approach are enumerated.

\begin{itemize}
    \item The system provides a personalised healthcare service: Citizens can get recommendations of routes according to their own needs and preferences.
    \item The system is context-aware: The system is dynamic and adapts automatically and in real-time to contextual changes affecting variables monitored by the smart city sensors.
    \item The system is collaborative: Citizens are not only information consumers, but also information producers. Citizens are able to inform about unexpected situations that could affect other citizens (\eg bad pavement, obstacles in the route, etc.), create new routes, and provide feedback and rate routes after finishing their physical activities, among other functionalities.
    \item The system is inexpensive: Citizens only need a smartphone with self-location and communication capabilities to interact with the system.
\end{itemize}

\subsection{Architecture} \label{subsec:apps:rs:arc}

Our system comprises a number of interacting actors and resources, described as follows.

\begin{itemize}
    \item \textit{Citizens}: They are the inhabitants of a smart city. They could suffer from health conditions (then we refer to them as patients) or could be healthy. It is assumed that they have a mobile phone (generally a smartphone) that allows them to be connected to the Internet and exchange data with the system.
    \item \textit{Smart city sensors}: The smart city has a large number of sensors continuously monitoring several contextual variables in real time, such as temperature, humidity, wind, precipitation, pollution, luminosity, etc.
    \item \textit{Databases}: Data repositories containing information from citizens. We distinguish two kinds of information:
    \begin{itemize}
        \item \textit{Health-related information}: Healthcare information from citizens like electronic health records. To overcome privacy issues, these records do not have to be complete. The system only requires a general overview of the health issues of the citizens, such as visual impairments, respiratory problems, reduced mobility and health diseases.
        \item \textit{Preferences information}: Information about the preferences of the citizens with regard to the routes performed. At the completion of the routes, they are allowed to rate them according to their preferences.
    \end{itemize}
    \item \textit{Context-aware recommender system (CARS)}: Together with citizens, CARS is the most important actor of our system. Embodied in a software running on a server, it is responsible for recommending routes to citizens. These routes are recommended using collaborative filtering techniques and considering the health conditions and preferences of citizens as well as the real-time contextual information coming from the city sensors.
    \item \textit{Communications infrastructure}: It is presumed that the smart city is already equipped with the proper infrastructure to all the exchange and communication of data amongst all the actors. This communication infrastructure is likely to consist of wireless networks of different nature, such as IEEE 802.11, ZigBee, Bluetooth or 4G/5G.
\end{itemize}

\begin{figure}[t!]
\centering
\includegraphics[width=0.73\textwidth]{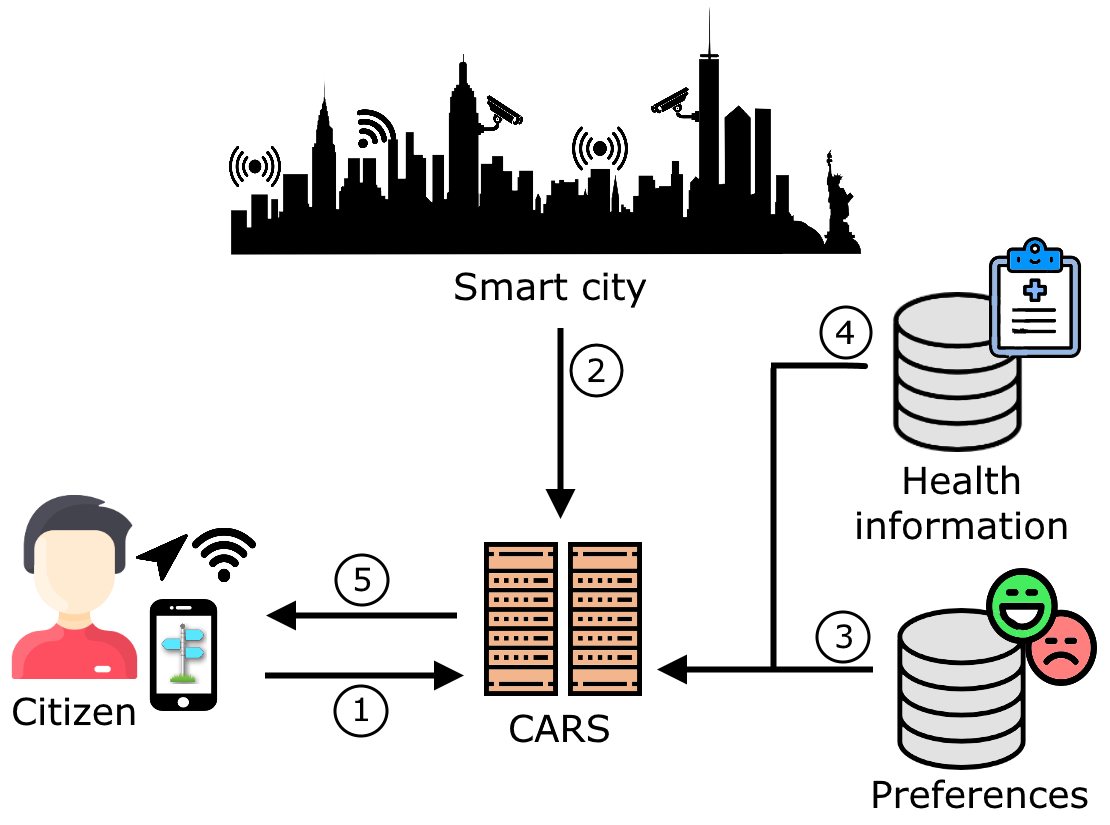}
\caption{Architecture and main actors of the recommender system (adapted from~\cite{casino2017healthy}).}
\label{fig:apps:rs:arch}
\end{figure}

Figure \ref{fig:apps:rs:arch} illustrates the general scheme of our system architecture and the interaction between the main actors.
According to this scheme, the main steps of the recommendation procedure are described as follows.
First, a citizen sends a query to the CARS (using the mobile application named SmartRoute) asking for a route recommendation.
In this query, extra information, such as location of the citizen, is also provided (step \ding{192}).
Second, the CARS checks the real-time environmental information coming from the smart city's sensing infrastructure (step \ding{193}).
Subsequently, the system evaluates the environmental information from the nearest sensor from each route using well-known metrics (air quality, ultraviolet radiation, wind speed, temperature and rain) \cite{elshout2008comparing}, and assigns a status to each route: ``danger'' if one or more measures indicate that the route may be hazardous for the user's health, ``caution'' if there are values that indicate possible risks (\eg users who are more sensitive to environmental factors), and ``idle'' if there is no evidence of health's risk.
Moreover, crowdsourcing-based information from other citizens about problems in routes is also considered when determining the route's status.
Third, the CARS checks the preferences of the citizen and applies a collaboration filtering method to the complete database of routes to produce the top-N recommendations with the N most promising routes for the citizen (step \ding{194}).
Fourth, CARS uses the healthcare information about the citizen to discard previously recommended routes that might endanger the citizen's health status (step \ding{195}).
For instance, for citizens with high impaired mobility problems, CARS discards routes that are not well paved or, for citizens with respiratory problems, routes with a lot of greenery are set to ``caution''.
After applying this filter, if no route satisfies the criteria, the CARS goes back to the third step and computes a new set of top-N recommendations that could satisfy the health and environmental constraints of the citizen.
Fifth and finally, CARS forwards the list of the N routes to the user, who can check these recommendations in the SmartRoute application (step \ding{196}).

\begin{table}[b!]
\centering
\caption{Excerpt of citizens' ratings in $[0,10]$. The higher the better.}
\label{tbl:apps:rs:ratings}
{\scriptsize
\renewcommand{\arraystretch}{0.9}
\begin{tabular}{ccccc}
  \toprule
  & \textit{Route}$_a$ & \textit{Route}$_b$ & \ldots & \textit{Route}$_m$ \\ \midrule
  $u_1$ & 5 & 8 & \ldots & 2  \\
  \vdots & \vdots & \vdots & $\ddots$ & \vdots \\
  $u_i$ & 9 & 4 & \ldots & 7 \\
  \vdots & \vdots & \vdots & $\ddots$ & \vdots \\
  $u_{n}$ & 8 & 5 & \ldots & 3 \\
  \bottomrule
\end{tabular}
}
\end{table}

\subsection{Implementation} \label{subsec:apps:rs:impl}

The data retrieved from users and sensors are stored in several tables in a relational database.
To illustrate the structure of these tables, Table \ref{tbl:apps:rs:ratings} depicts an excerpt of the database that stores the users' ratings about the real routes performed.
This table is used by the CARS (by means of the collaborative filtering technique) to determine the most similar users when computing the top-N route recommendations.
The quality/accuracy of the recommendations of the system is beyond the scope of this dissertation; for further details, please refer to \cite{casino2018smart}.
Moreover, the CARS uses a table containing information about the health condition of the citizens to avoid inappropriate recommendations.
An excerpt of this data structure is shown in Table \ref{tbl:apps:rs:health}.
The information about the routes is also stored.
At the time, the system contains real routes from the city of Tarragona and Barcelona.
Depicted in Figure \ref{fig:apps:rs:routes}, each route is defined by an initial point, a final point, and a number of intermediate points (\ie checkpoints) which add dynamism to routes and clarity possible loops.
An excerpt of this data structure is shown in Table \ref{tbl:apps:rs:routes}.
Finally, the contextual information, both weather data and air quality measurements, is updated hourly from the sensors deployed throughout the territory by the Catalan air quality monitoring network \cite{meteo_gencat}.
In the smart cities of the future, these data is expected to be more complete and fresh.

\begin{table}[b!]
\centering
\caption{Excerpt of citizens' health information assessed in $[0,1]$. Higher values indicate worse health condition.}
\label{tbl:apps:rs:health}
{\scriptsize
\renewcommand{\arraystretch}{0.9}
\begin{tabular}{cccccc}
  \toprule
  & \multirow{2}{*}{\textbf{Age}} & \textbf{Visual} & \textbf{Respiratory} & \textbf{Reduced} & \textbf{Heart} \\ 
  & & \textbf{Impairment} & \textbf{Problems} & \textbf{Mobility} & \textbf{Diseases} \\ \midrule
  $u_1$ & 50 & 0.3 & 0 & 0.7 & 0.1  \\
  \vdots & \vdots & \vdots & \vdots & \vdots & \vdots \\
  $u_i$ & 32 & 0 & 0.9 & 0.2 & 0 \\
  \vdots & \vdots & \vdots & \vdots & \vdots & \vdots \\
  $u_n$ & 67 & 0.1 & 0.3 & 0.6 & 0.8  \\
  \bottomrule
\end{tabular}
}
\end{table}

\begin{table}[t!]
\centering
\caption{Excerpt of real routes data collected in the city of Tarragona.}
\label{tbl:apps:rs:routes}
{\scriptsize
\renewcommand{\arraystretch}{0.9}
\begin{tabular}{cccc}
  \toprule
  & \textit{Route}$_a$ & \ldots & \textit{Route}$_k$ \\ \midrule
 \textbf{Start Coor.} & 41$^{\circ}$4'44.54"N, 1$^{\circ}$12'49.58"E & \ldots & 41$^{\circ}$7'45.65"N, 1$^{\circ}$14'32.90"E \\
 \textbf{End Coor.} & 41$^{\circ}$6'32.82"N, 1$^{\circ}$14'58.55"E  & \ldots & 41$^{\circ}$8'8.21"N, 1$^{\circ}$14'59.02"E \\
 \textbf{Distance (km)} & 9.6 & \ldots & 2.32 \\
 \textbf{Elev. Gain (m)} & 0 & \ldots & 55 \\
 \textbf{Pavement Quality} & Very good & \ldots & Average \\
 \textbf{Status} & Idle & \ldots & Caution \\
 \bottomrule
\end{tabular}
}
\end{table}

\begin{figure}[b!]
\centering
\begin{subfigure}{.45\textwidth}
  \centering\includegraphics[width=\linewidth]{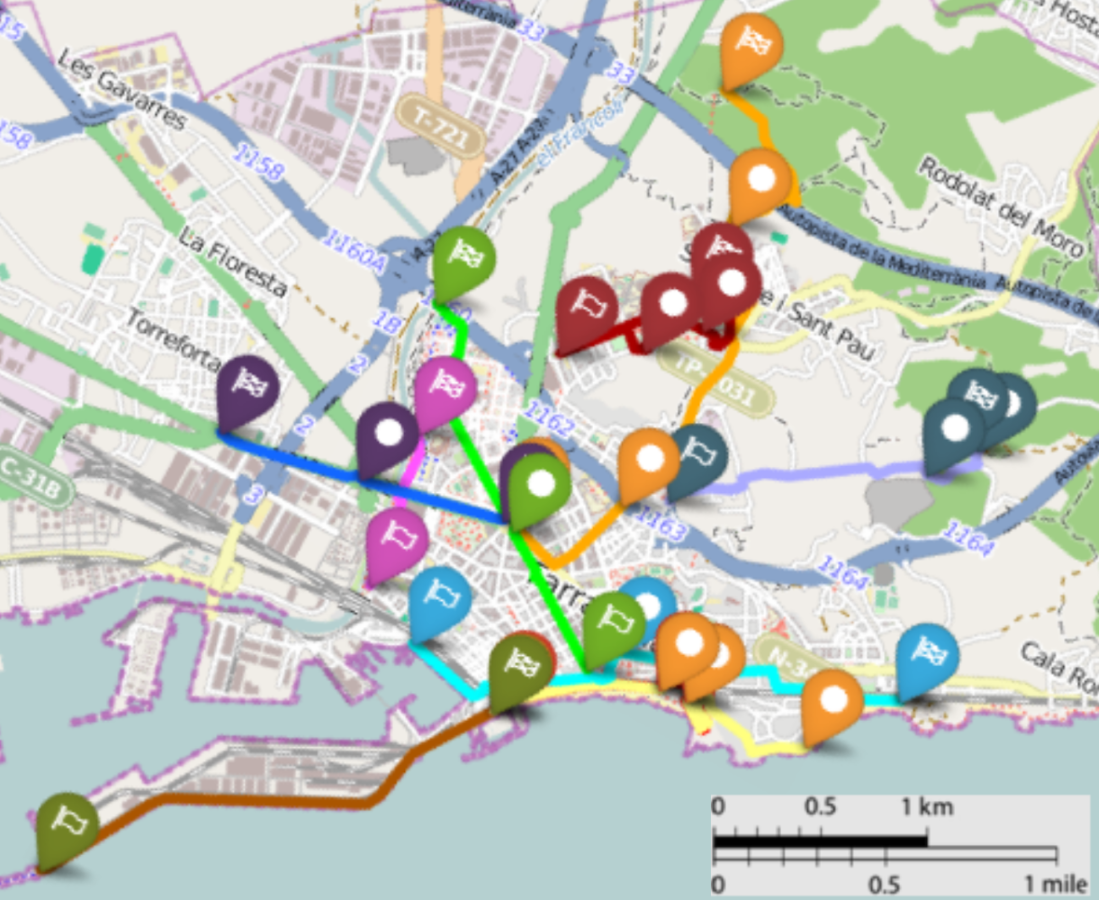}  
  \caption{Routes in Tarragona}
\end{subfigure}
\hfill
\begin{subfigure}{.53\textwidth}
  \centering\includegraphics[width=\linewidth]{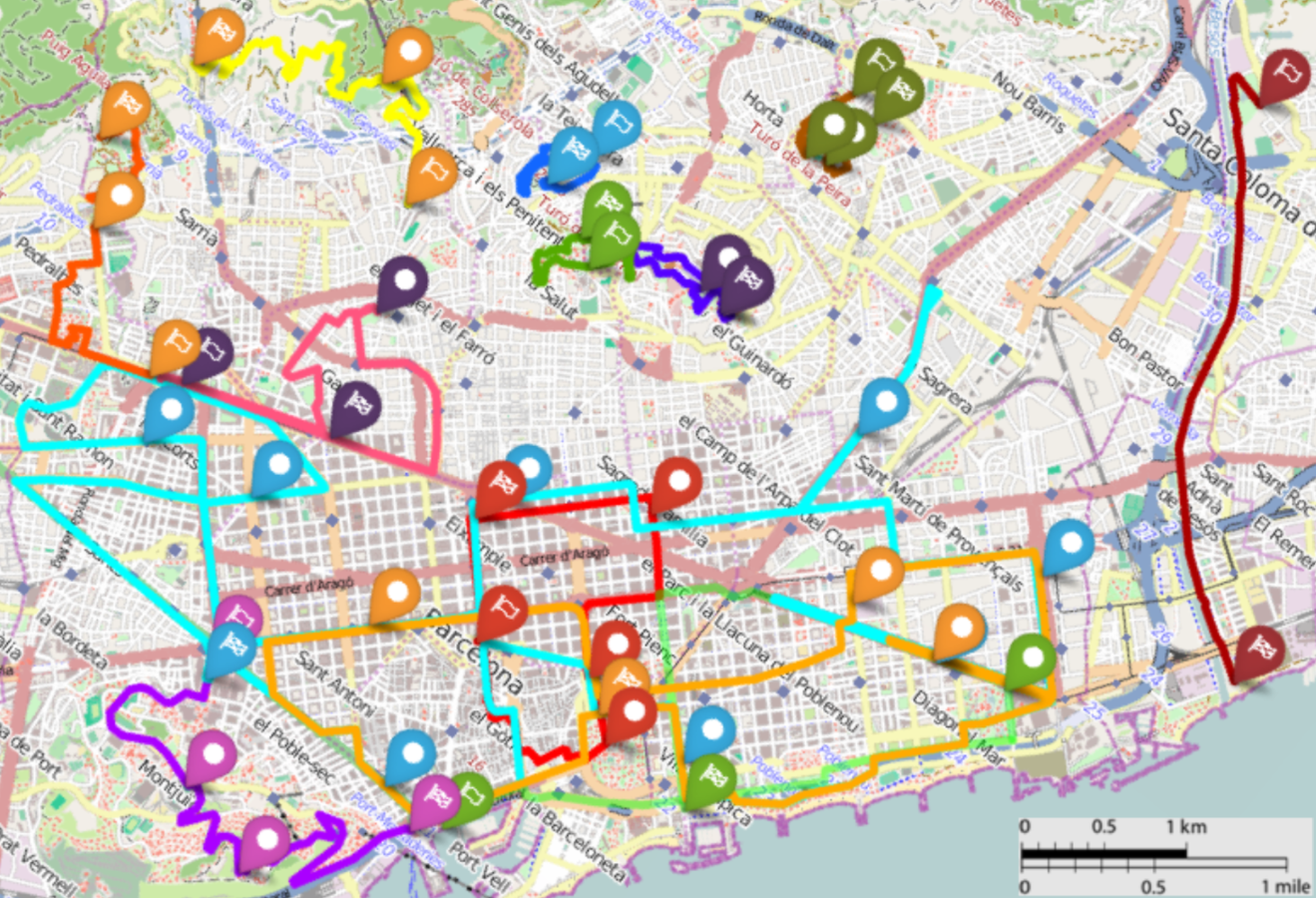}  
  \caption{Routes in Barcelona}
\end{subfigure}
\caption{Graphic representation of the routes in the system. Each route is represented by a colour. Markers with a blank flag indicate the start of a route, and markers with a chequered flag indicate the end. Markers with a white dot correspond to checkpoints (reprinted from \cite{casino2018smart}).}
\label{fig:apps:rs:routes}
\end{figure}

 \begin{figure*}[t!]
 \centering
 \begin{subfigure}{0.24\textwidth}
   \includegraphics[width=\textwidth]{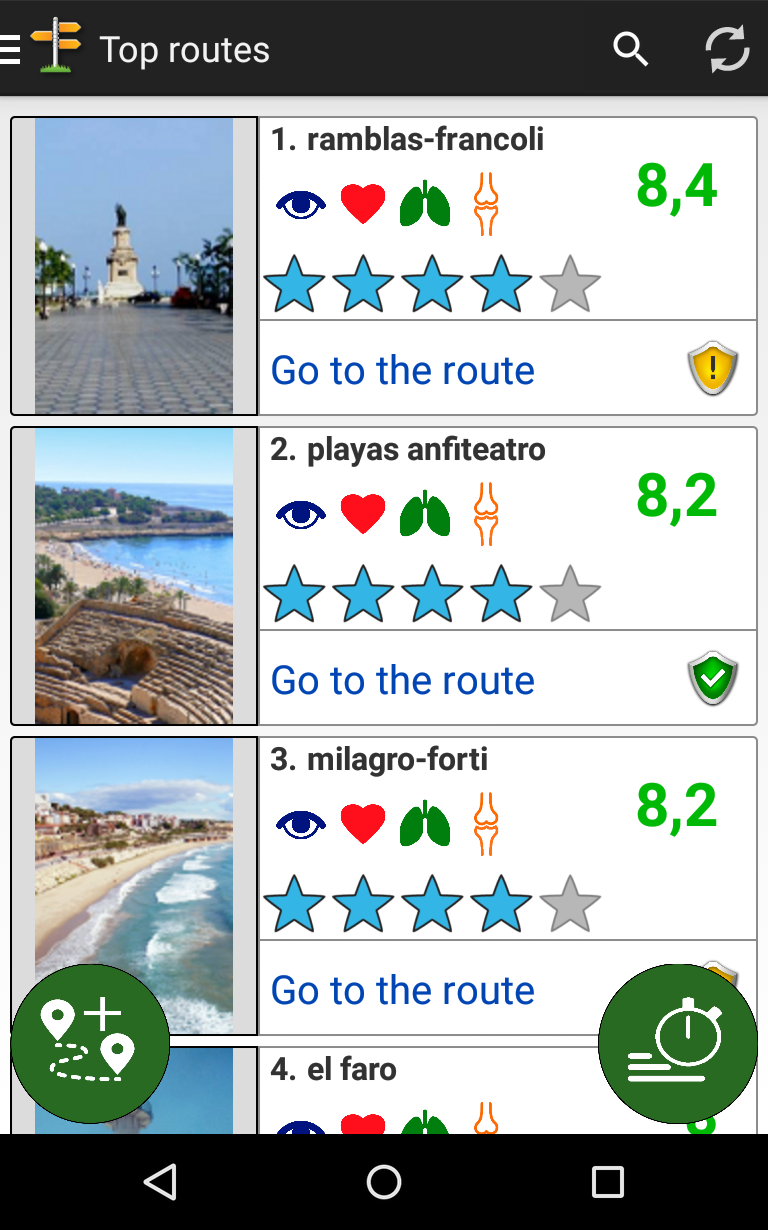}
   \caption{Home} \label{subfig:apps:rs:home}
 \end{subfigure}
 \begin{subfigure}{0.24\textwidth}
   \includegraphics[width=\textwidth]{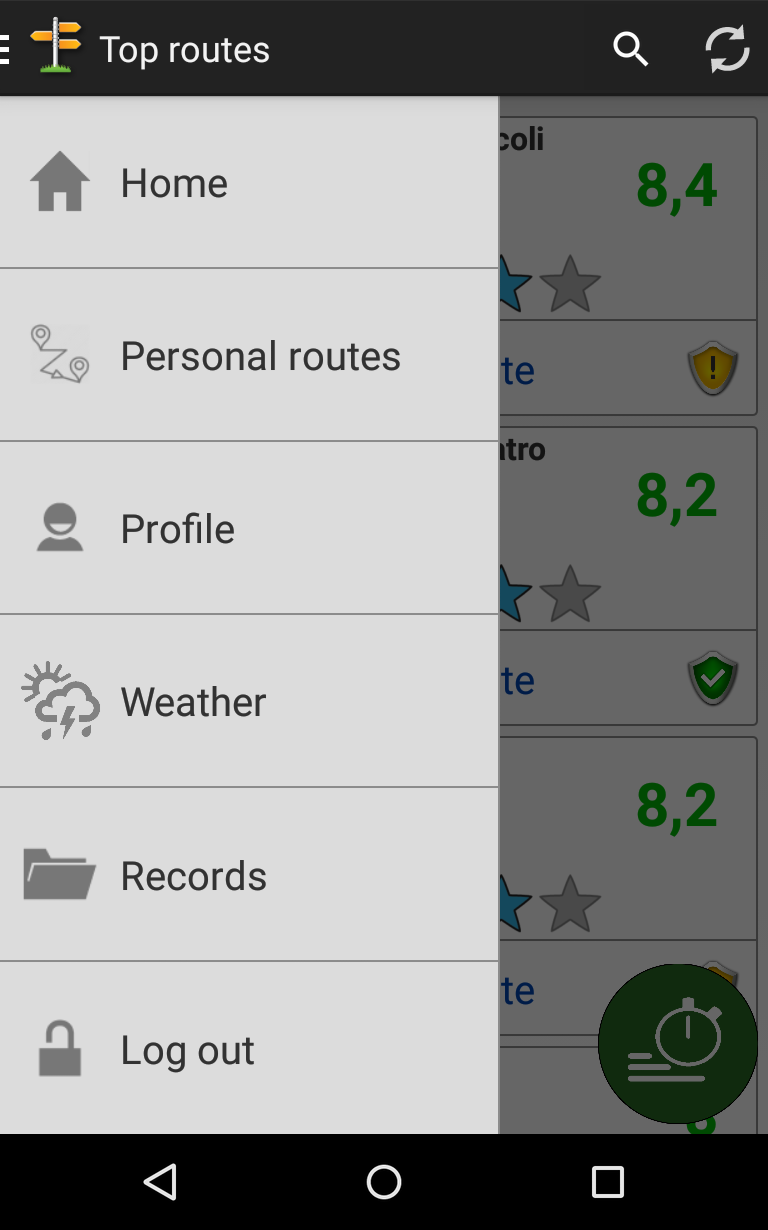}
   \caption{Application menu} \label{subfig:apps:rs:menu}
 \end{subfigure}
 \begin{subfigure}{0.24\textwidth}
   \includegraphics[width=\textwidth]{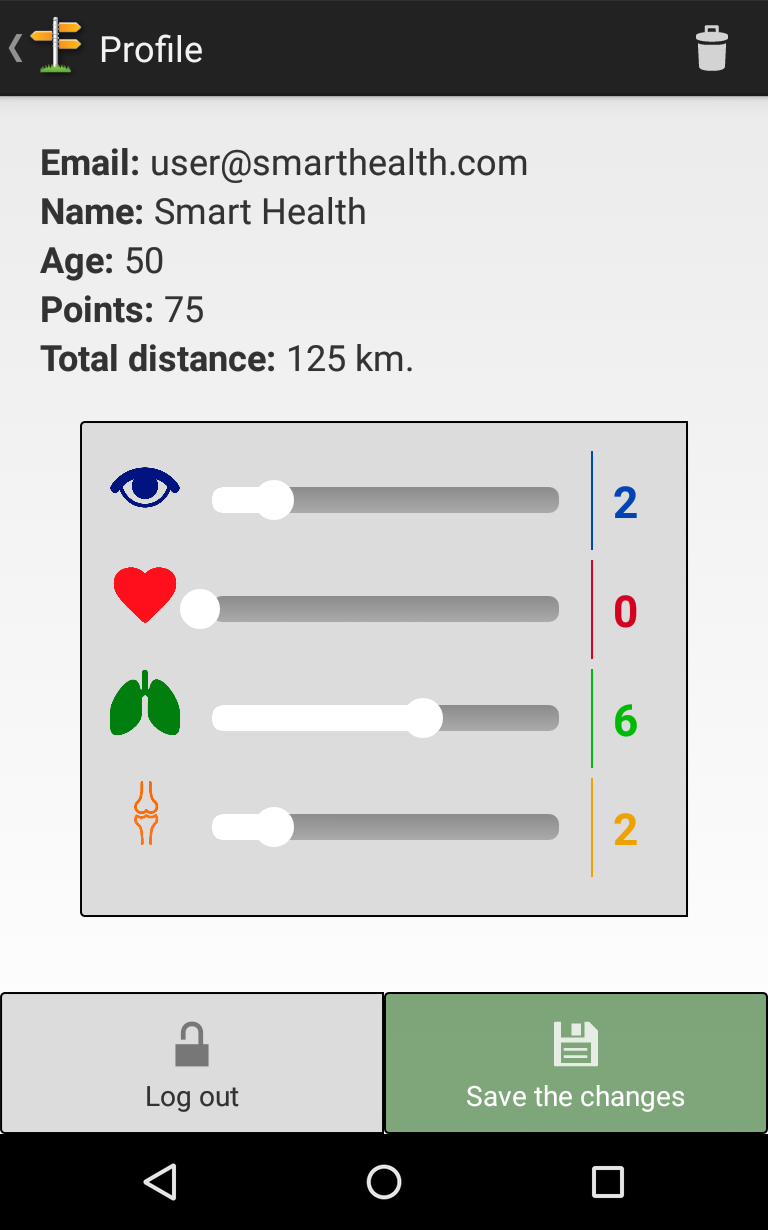}
   \caption{User profile} \label{subfig:apps:rs:profile}
 \end{subfigure}
 \begin{subfigure}{0.24\textwidth}
   \includegraphics[width=\textwidth]{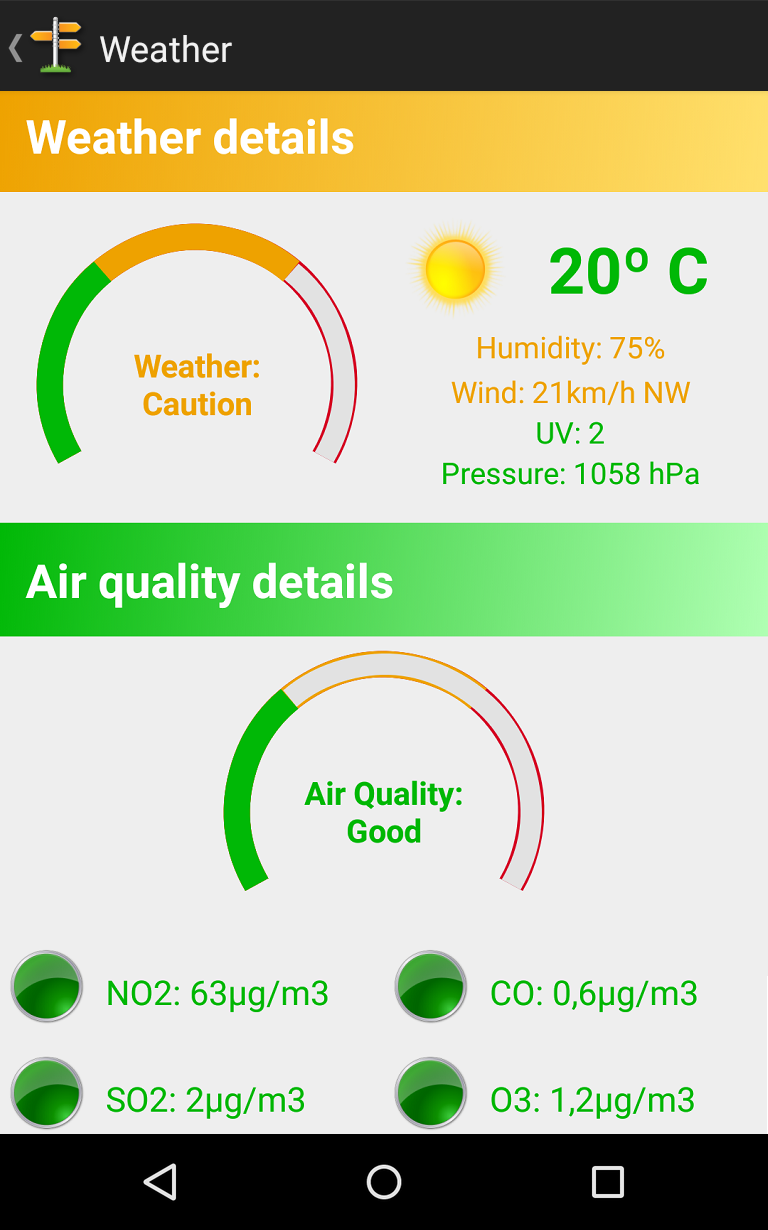}
   \caption{Contextual data} \label{subfig:apps:rs:environmental}
 \end{subfigure}
 \par\medskip 
 \begin{subfigure}{0.24\textwidth}
   \includegraphics[width=\textwidth]{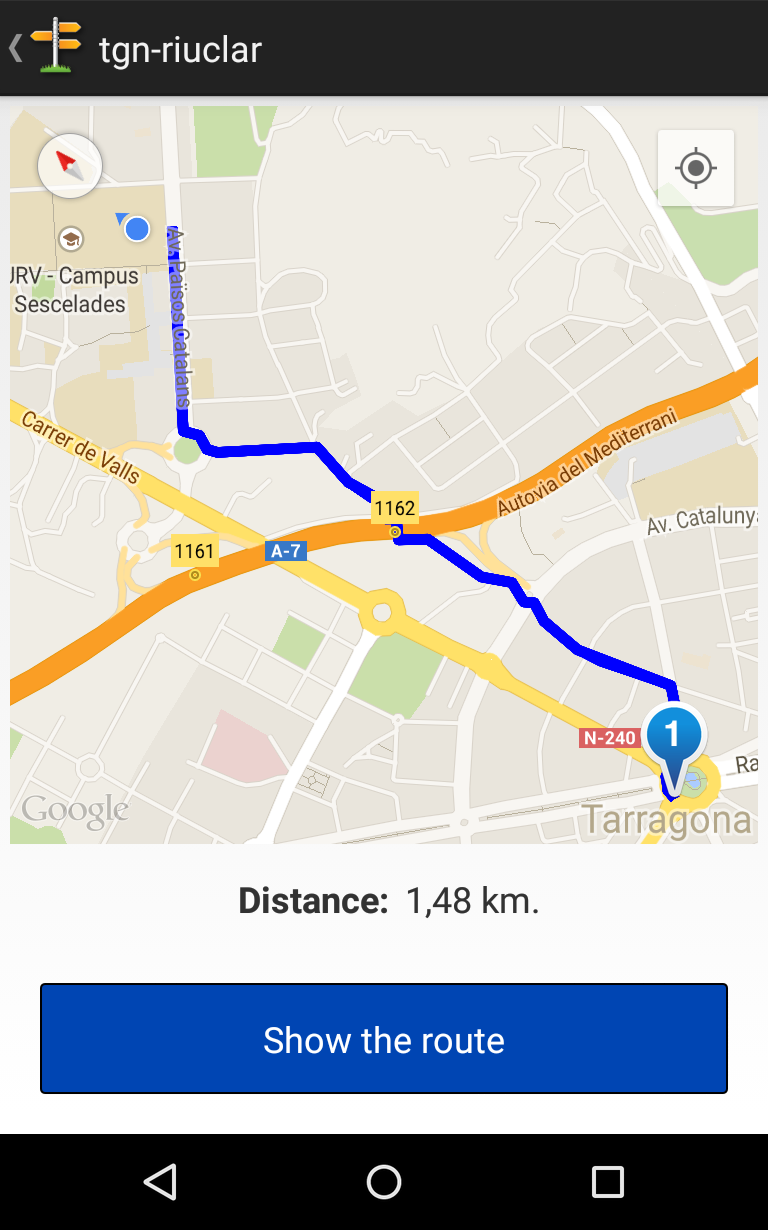}
   \caption{Route information} \label{subfig:apps:rs:info}
 \end{subfigure}
 \begin{subfigure}{0.24\textwidth}
   \includegraphics[width=\textwidth]{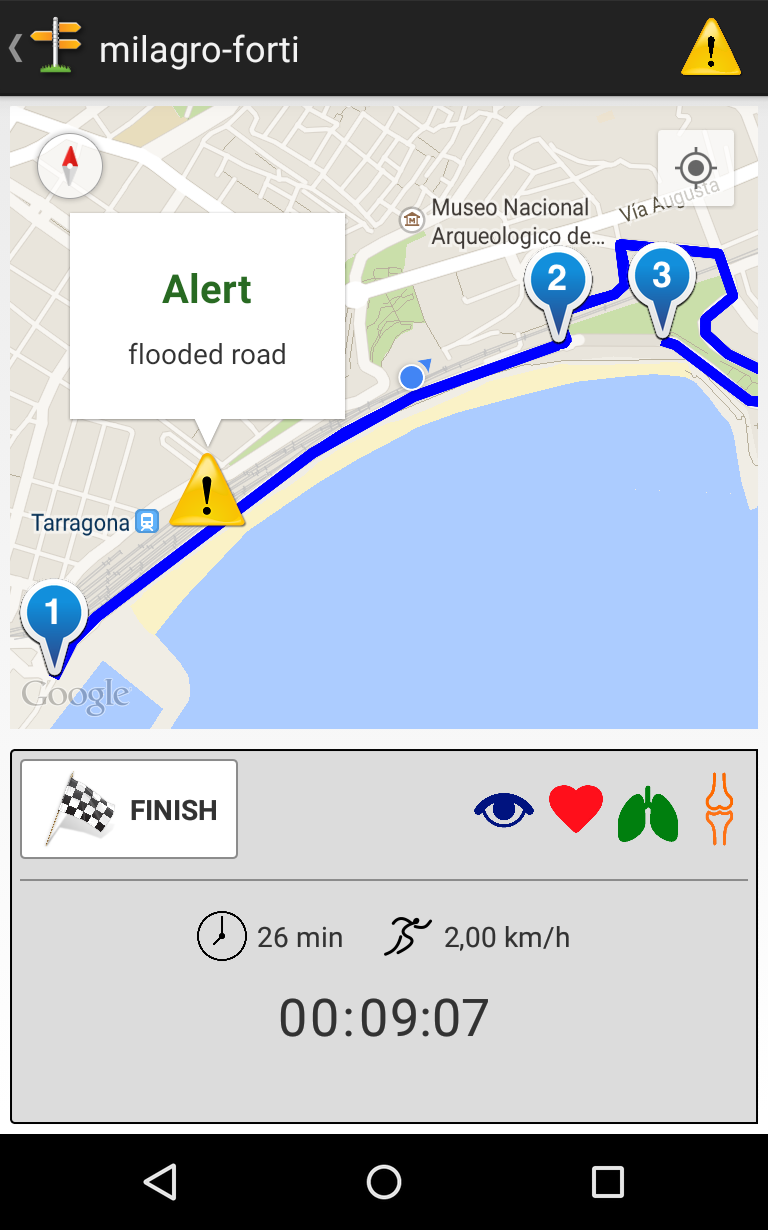}
   \caption{Route session} \label{subfig:apps:rs:session}
 \end{subfigure}
 \begin{subfigure}{0.24\textwidth}
   \includegraphics[width=\textwidth]{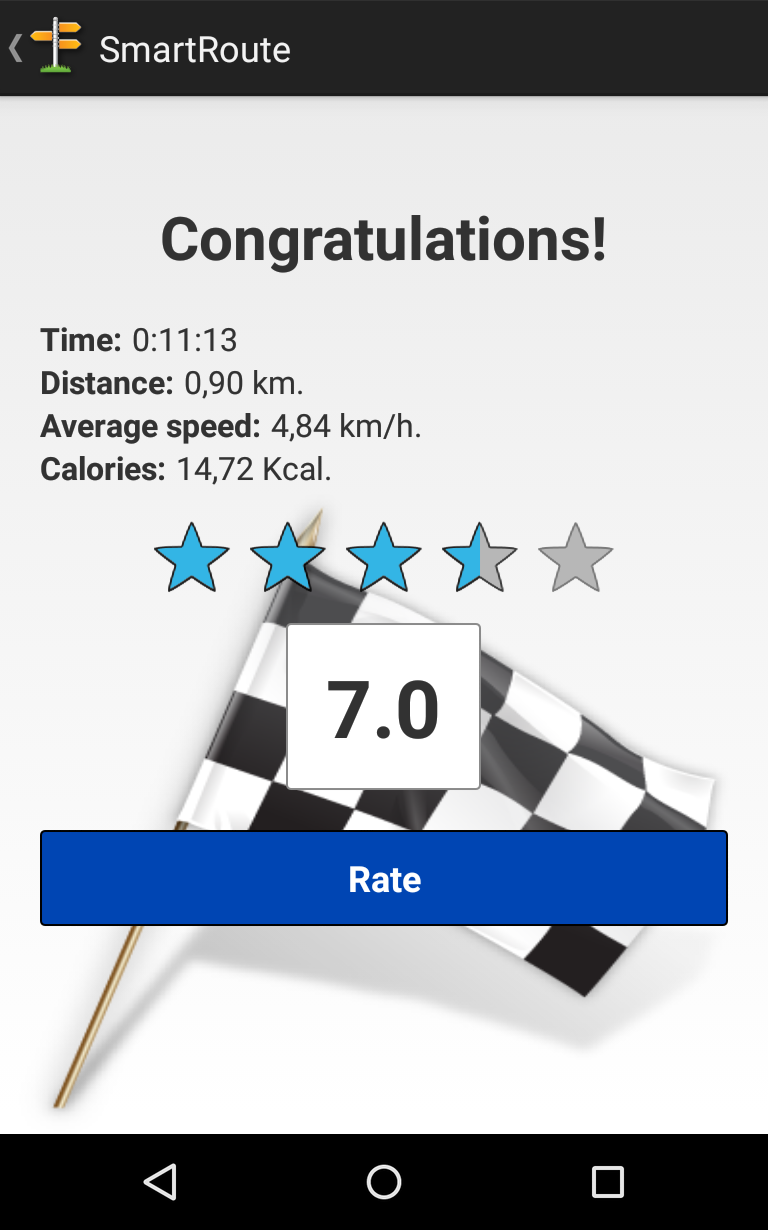}
   \caption{Finish \& Rate} \label{subfig:apps:rs:rate}
 \end{subfigure}
 \caption{Screenshots of the SmartRoute mobile application (reprinted from \cite{casino2017healthy}).}
 \label{fig:apps:rs:app}
\end{figure*}
 
Citizens interact with CARS through the SmartRoute mobile application, developed in Android.
Next, we describe the functioning of the application following the images from Figure \ref{fig:apps:rs:app}.
Citizens log onto the application and, from the home screen (see Figure \ref{subfig:apps:rs:home}), they are able to (i) go to the closest route and start, (ii) look for a specific route, (iii) get the list of the most rated routes, or (iv) create a new route by adding checkpoints during their activity.
The application menu (see Figure \ref{subfig:apps:rs:menu}) allows to get a list of personalised recommended routes, update their personal information, check the environmental information and check and manage records of different route sessions.
The healthcare information can be updated from the profile screen (see Figure \ref{subfig:apps:rs:profile}) using sliders that modify the degree of severity, as showed in Table \ref{tbl:apps:rs:health}.
Additionally, citizens can check the weather and air quality information from their closest sensor (see Figure \ref{subfig:apps:rs:environmental}) and check the feedback provided by the application (for example, the background colour changes according to the quality of dangerousness of such information).
Once citizens select a route, the application shows the path to reach the starting point from their current location, as well as additional information, such as a description of the route and possible warnings (see Figure \ref{subfig:apps:rs:info}).
When citizens are at the starting point, they can initiate the route and a new route session starts.
During a route session, the application shows statistics, such as the distance, approximate duration, speed and possible alerts or changes in the route in real time (see Figure \ref{subfig:apps:rs:session}).
In addition, citizens can send notifications to warn others of issues with that route.
At the completion of the route, the application shows the statistics collected during the session and asks for a rate (see Figure \ref{subfig:apps:rs:rate}).
This information is important to provide feedback to the CARS for further recommendations, as showed in Table~\ref{tbl:apps:rs:ratings}.

\section[e-PEMICU: Towards Smart Health Enhanced Early Mobilisation]{e-PEMICU: Towards Smart Health Enhanced Early Mobilisation%
  \sectionmark{e-PEMICU: Smart Health Enhanced Early Mobil.}}
\sectionmark{e-PEMICU: Smart Health Enhanced Early Mobil.} \label{sec:apps:mob}

Early mobilisation routines aim to improve the recovery process of critically ill patients that are hospitalised in intensive care units (ICUs).
These routines, which consist of passive and active progressive mobilisations, help reduce the side-effects derived from ICU stays.
With the aim to monitor the achievements in these programmes, in this section, we present e-PEMICU, a health platform that supports early mobilisation programmes in ICUs by means of motion sensors and smartphones.
First, we introduce the rationale of this research in Section \ref{subsec:apps:mob:ratio}.
Then, the architecture and the functioning of the system are described in Section \ref{subsec:apps:mob:arc}.
Finally, the implementation details are explained in Section \ref{subsec:apps:mob:impl}.

\subsection{Rationale} \label{subsec:apps:mob:ratio}

Physicians and physiotherapists struggle to shorten the length of stay of patients in ICUs in order to decrease the risks of suffering from post-intensive care syndrome.
This syndrome includes physical issues, such as muscle weakness and other muscular disorders, and mental and cognitive alterations \cite{bautista2019executive}.
These consequences, which may last for years, clearly jeopardise patients' quality of life.
For instance, half of the ICU surviving patients are not able to work for one year after being hospitalised, and 30\% might suffer from life-long impairments with no chances of returning back to their normal life \cite{phelan2018implementing}.
The application of early mobilisation routines into daily clinical practice can reduce these side-effects.
These routines comprise two types of mobilisations: (i) passive mobilisations, \ie passive exercises in which the movements are done by the therapists while the patients are relaxed, and (ii) active mobilisations, \ie active exercises in which the movements are done by the patients while sitting on a bed or chair, and includes cycling, tilting up and ambulation (with or without assistance).
Despite the benefits of early mobilisation, there is a lack of standardised frameworks.
For instance, the optimal level of physical therapy (\eg intensity, duration, frequency,\ldots) must be determined at different times (\eg for fully sedated patients, first awakening patients,\ldots) and it has to be adapted to each patient.
Additional barriers, including patients' symptoms and conditions and technical resources, must be considered during the design of these routines.

Therapists generally take notes during the exercises on paper forms or by using some basic word processor software.
To the best of our knowledge, there are no technological solutions to streamline the management and monitoring of early mobilisation programmes neither to analyse their patients, exercises and outcomes from a holistic perspective.
Consequently, it is cumbersome to detect similarities and trends among patients and, hence, to infer knowledge on the evolution of patients.
In this approach, we propose a platform, named e-PEMICU, to support the management of early mobilisation programmes in patients hospitalised in ICUs.
With the smart health paradigm in mind, this platform augments the ICU (and patients in it) with the proper sensing capabilities by using mobile devices and low footprint motion sensors.
Besides, the solution has been designed to be affordable, minimally intrusive, portable and easy to deploy in real environments.

\subsection{Architecture} \label{subsec:apps:mob:arc}

The proposed system comprises several interacting actors and resources, which are described as follows.

\begin{itemize}
    \item \textit{Hardware node}: Comprised of accelerometers, a smartphone and a tablet, the hardware node is the core of the system. It is the elementary actor for augmenting the patients with the proper sensing capabilities. Note that the system may consider multiple hardware nodes.
    \begin{itemize}
        \item \textit{Accelerometer}: The accelerometer (\ie motion sensor) is embedded in a wearable device and wrapped in a wristband. Placed in the part of the body where the movement is performed (\eg arm, leg,\ldots), it detects the acceleration changes in the X, Y and Z axes. Each hardware node can have multiple accelerometers.
        \item \textit{Smartphone}: The purpose of the smartphone is twofold: (i) obtain the sensor data from the accelerometers through a BLE connection, and (ii) forward the sensor data received to the back-end server over a Wi-Fi connection.
        \item \textit{Tablet}: This device is used by the users to access the platform's web front-end and interact with the tools provided.
    \end{itemize}
    \item \textit{Databases}: Data repositories containing all the information about the system. In particular, we distinguish two kinds of databases:
    \begin{itemize}
        \item \textit{Management database}: It contains information about the general management of the system, this is, information about the users and their roles, the configuration of the hardware nodes, the description of the different early mobilisation routines for each patient, etc.
        \item \textit{Motion database}: It contains the raw sensor data collected by the hardware nodes during the realisation of early mobilisation routines.
    \end{itemize}
    \item \textit{Users}: They are the individuals intended to use our platform. We distinguish four kinds of users (\ie roles):
    \begin{itemize}
        \item \textit{Patients}: The individuals hospitalised (in the ICU or in one of the hospital's room after their stay in the ICU) to whom the early mobilisation exercises are performed.
        \item \textit{Managers}: Typically the doctor responsible for the overall management of the platform. Managers are able to add, edit and remove patients and setup other users as well as to manage and configure the hardware nodes. Moreover, they configure and schedule routines for each patient.
        \item \textit{Professionals}: Intended for physiotherapists and nurses of the ICU who manage and assist the early mobilisation sessions with the patients. Also, they can assess the progress and evolution of their patients.
        \item \textit{Sporadic}: Intended for those users supervising the early mobilisation sessions in regular hospital rooms, such as physiotherapists, relatives, caregivers or even the very patients.
    \end{itemize}
    \item \textit{Back-end server}: It manages the workflow of the entire system, ranging from the storage of information to the databases to the rendering of the platform's web front-end to the users' devices. As the server is deployed using web technologies, note that it can be accessed (with the proper credentials) from virtually any device with Internet browsing capabilities.
    \item \textit{Communications infrastructure}: The hospital scenario is augmented with the proper communication infrastructure, responsible for the data exchange and communication among all the actors within the system. It mainly consists of wireless networks, such as IEEE 802.11 and BLE.
\end{itemize}

The portable and wireless nature of the proposed system enables its functioning in different hospital stays (\eg ICU boxes, regular hospital rooms, specialised rehabilitation rooms,\ldots) for multiple patients at the same time.
Therefore, we count with a number of portable versions of hardware nodes, one for each patient.
Figure \ref{fig:apps:mob:arch} illustrates a feasible scenario of a hospital using the e-PEMICU platform, which is described as follows.

Managers (coloured in white) use the management module of the web front-end to set up the e-PEMICU system.
This module enables the management of users as well as their roles in the system.
In addition, it considers the registration and identification of the devices of hardware nodes, \ie adding new motion sensors to the system, setting up new smartphones or disabling devices that are no longer being used.
Through this module, managers can check whether the sensors work properly, \ie the acceleration changes are detected, and their battery level (number \ding{192}).
Another functionality of the management module is to schedule early mobilisation routines to patients with a number of exercises.

\begin{figure}[b!]
\centering
\includegraphics[width=\textwidth]{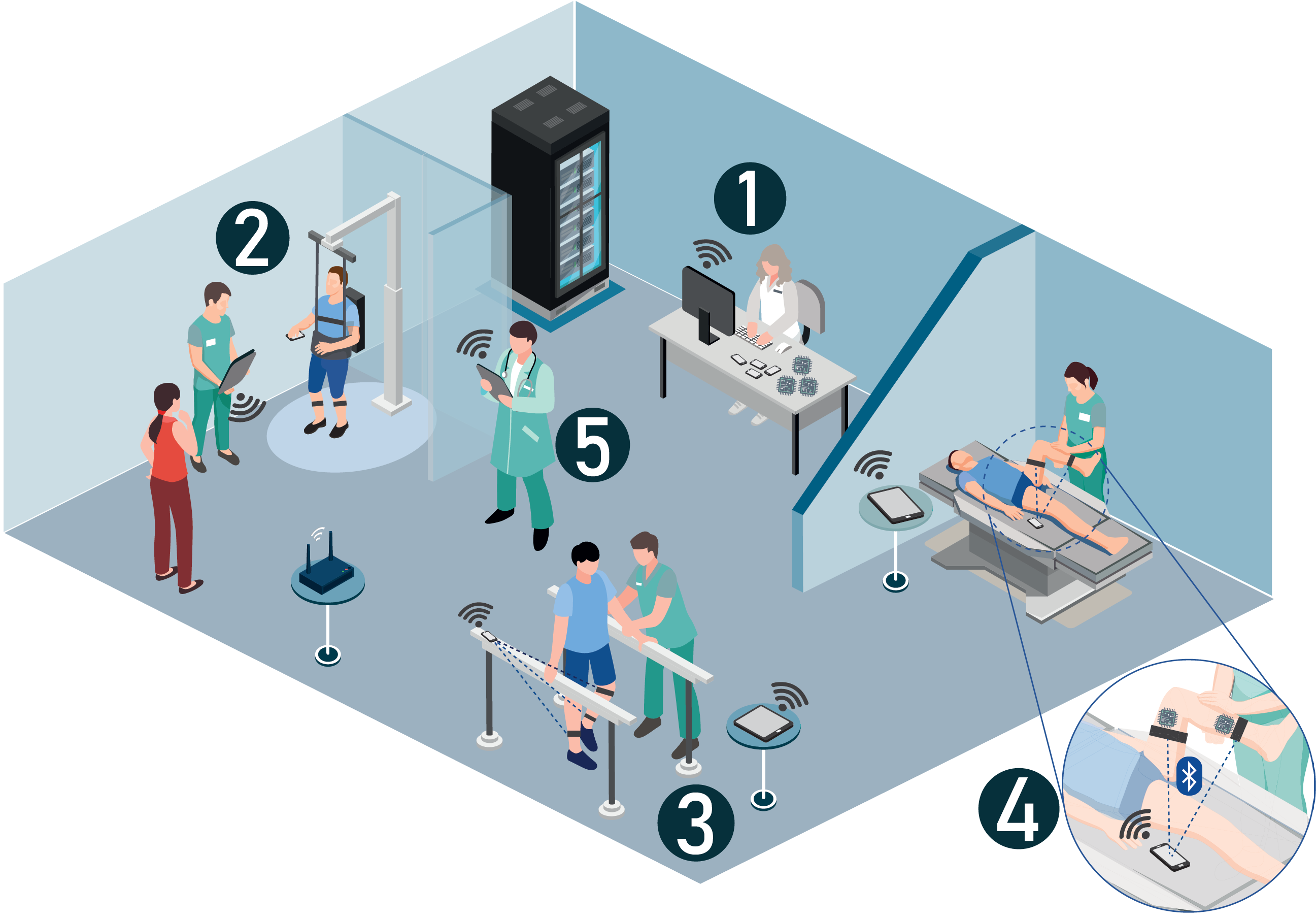}
\caption{Scenario of a hospital using the e-PEMICU platform (adapted from \cite{ferre2021smart}).}
\label{fig:apps:mob:arch}
\end{figure}

Whenever an early mobilisation routine is scheduled, professional or sporadic users (coloured in green and red, respectively) assist the patients during the exercises, either by performing passive mobilisations to them or, in case of conscious and recovered patients, just making sure that they follow the routines correctly.
For example, a professional performs a passive mobilisation to the patient placed in the right of Figure \ref{fig:apps:mob:arch}, while the other two patients perform active mobilisations with the supervision of professional or sporadic users.
During the session, these users interact with the session module of the web front-end using a tablet (number \ding{193}).
The main purpose of this module is to record the motion data from the early mobilisation sessions.
To do this, the details of the exercises to be performed during the session are provided together with setup information: the number of motion sensors needed, the specific place (\eg wrist, elbow, ankle, etc.) and orientation of these sensors, the duration and intensity of the exercises, the number of repetitions, and a short video depicting the correct execution of the exercises.
Once the sensors are properly placed on the patient's body, the exercise initiates when the professional or the sporadic user presses the ``start exercise'' button.
From that moment onward, movements detected by the motion sensors are stored in the motion database, until the ``end exercise'' button is pressed (number \ding{194}).
Note that, while the exercise is not active, motion data is not stored in the motion database, thus preventing the storage of faux movements.
Since sensors are resource-constrained, the movement data recorded is transmitted to the smartphone through a BLE connection.
Then, the smartphone forwards this data to the back-end server using a Wi-Fi connection (number \ding{195}).
The early mobilisation session finishes once the patient has performed all the exercises.

All the early mobilisation routines (executed and pending) can be visualised from the assessment module of the web front-end (number \ding{196}).
This module is intended for both professionals and managers, who can assess the evolution of the patients and measure the degree of achievement of the mobilisation routines performed.
Besides, for each performed session, the module shows an assessment indicator.
Since a session consists of a series of exercises, each one is assigned a colour (red, yellow and green) that indicates how far the real exercise is from the intended one.
For example, if an exercise consists in moving the hand up and down without moving the elbow (\ie thus requiring two sensors, one in the hand and another in the elbow), a red colour is shown if movement in the elbow is detected.

Finally, for the sake of completeness, Figure \ref{fig:apps:mob:modules} depicts a diagram with the relationship between the modules of the web application and the main actors.

\begin{figure}[t!]
\centering
\includegraphics[width=\textwidth]{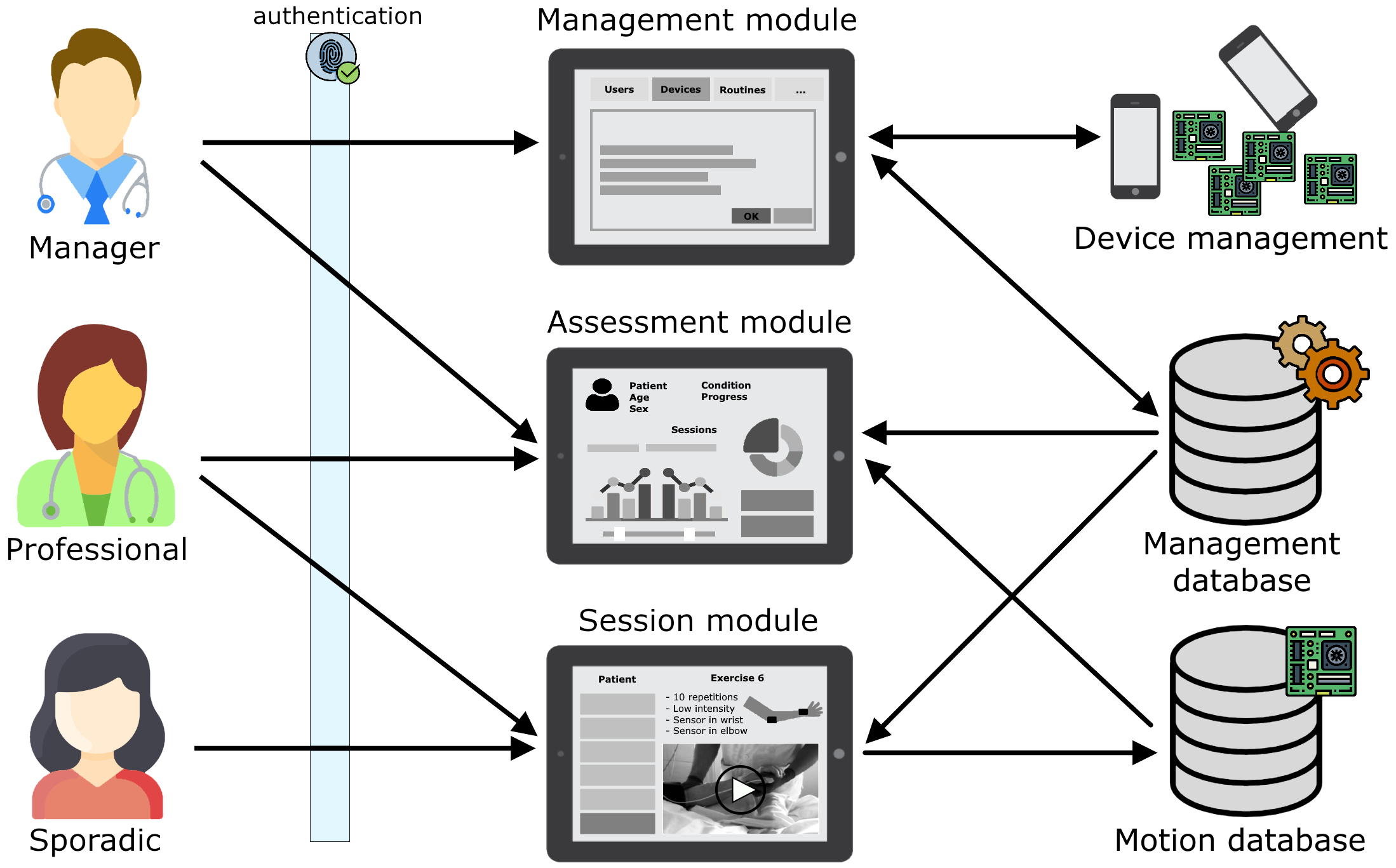}
\caption{Relationship between the modules of the e-PEMICU web application and the main actors involved (adapted from \cite{martinez2019epemicu}).}
\label{fig:apps:mob:modules}
\end{figure}

\subsection{Implementation} \label{subsec:apps:mob:impl}

A prototype of the platform has been implemented as a proof of concept to assess the feasibility of the proposed system.
This prototype (see Figure \ref{fig:apps:mob:proto}) has been tested with real patients hospitalised in Hospital Joan XXIII (Tarragona) that required early mobilisations.
The implementation details of the prototype are described next.

As motion sensors, we use micro:bit devices \cite{microbit}, low-cost programmable embedded systems, widely used for educational purposes among youth.
Each device counts with an accelerometer and a magnetometer sensor, a display LED matrix and two programmable buttons.
Besides, they fit perfectly into our system thanks to their pocket-size dimensions, so they can be placed on the patients' body in an almost non-intrusive manner, and their low-power consumption that allows operating for weeks without recharging the battery.
Within the same hardware node, these sensors are paired with a smartphone through BLE.
In the current prototype, for design decisions, up to three micro:bit sensors can be paired in a hardware node.
Nevertheless, the system is flexible in this sense, and more sensors could be introduced in the future to perform more complex movements if needed.
These sensors have been programmed to send a message over the BLE link as soon as a change in the acceleration in any of the axes is detected.
The message contains the sensor's UDN (Unique Derived Name, a five-characters code that is unique to each device), the acceleration values in the X, Y and Z axes, and the current timestamp.

The smartphones used in the prototype are low-cost, as they do not need to execute task-intensive processes, and their only mission is to forward the sensors' messages to the web API REST of the back-end server over a Wi-Fi connection.
This back-end server is a basic laptop with an Intel I5 dual-core processor and 8 GB RAM, which has installed a Linux Ubuntu, Apache as a web server and MySQL as a relational database management system.
Besides, it hosts a web REST API application implemented in PHP.
Note that this server is portable in the current prototype, but it could easily be extended in the future if needed.
Also, users use low-cost tablets to interact with the web front-end through a Wi-Fi connection.
Finally, to avoid connecting our devices to the wireless communication infrastructure of the hospital, we set up a Wi-Fi network with an 802.11ac access point.
This network enables the connection of the smartphones and tablets to the back-end server.
Hence, all the e-PEMICU actors are isolated from the hospital resources.

\begin{figure}[t!]
\centering
\includegraphics[width=0.63\columnwidth]{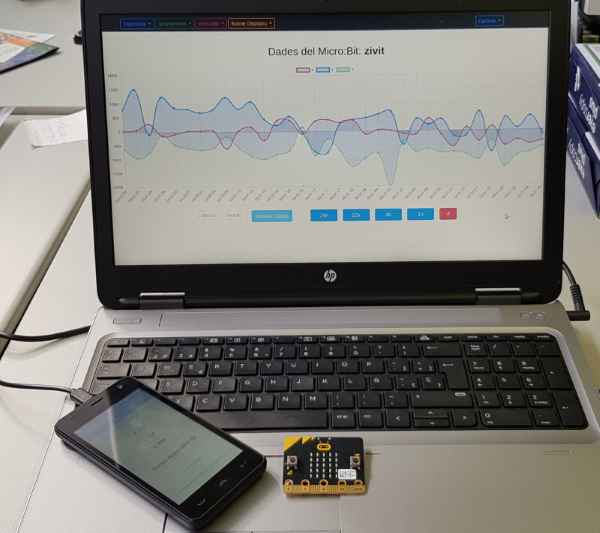}
\caption{Prototype of the e-PEMICU platform, showing the motion data collected by a micro:bit sensor (reprinted from \cite{martinez2019epemicu}).}
\label{fig:apps:mob:proto}
\end{figure}

\section{HGIS: A Healthcare Geographic Information System} \label{sec:apps:hgis}

Geographic Information Systems (GIS) are well-known information systems used to manage, analyse and visualise spatial and geographic data by representing them as maps.
Current GIS solutions are general-purpose, so they can easily be adopted by many domains for their applications.
However, due to the idiosyncrasies of the healthcare domain, these generic solutions may neglect insightful knowledge due to their lack of contextualisation.
With the aim to address the inherent requirements of the healthcare domain, this section presents HGIS, a GIS platform to support the integration, analysis and visualisation of heterogeneous healthcare-related data with spatio-temporal features.
First, Section \ref{subsec:apps:hgis:ratio} describes the rationale and motivation of the proposed platform.
Next, Section \ref{subsec:apps:hgis:arc} explains its architecture and functionalities.
Finally, implementation details of the platform are given in Section \ref{subsec:apps:hgis:impl}.

\subsection{Rationale} \label{subsec:apps:hgis:ratio}

Organisations count with numerous information systems that are continuously collecting vast amounts of heterogeneous data from a variety of sources.
Apart from operational purposes, these data are also used for strategic purposes and decision support when combined with advanced data analysis and visualisation techniques.
Since the generalisation of mobile devices and ubiquitous computing systems, more and more of the data gathered include spatio-temporal dimensions.
To work with this kind of data, GIS systems provide tools and methods to store, analyse, transform and visualise geolocated data.
GIS technologies are known for decades, contributing to the development and commercialisation of many applications from many domains, including territory planning, telecommunications, transportation and public health, among others \cite{waters2016gis}.

In particular, GIS technologies are well suited to play a key role in healthcare, since they could contribute to the analysis and improvement of its sustainability and efficiency from a geographical perspective.
Besides, due to the importance of the context, health conditions and spatio-temporal features are generally intertwined \cite{riano2014copd}.
For instance, disease propagation, environmental factors, socio-economic data, demographic status and the distribution of healthcare services are just a few examples of geographic, spatio-temporal features that affect to health outcomes in certain areas at particular times \cite{graves2008integrative}.
Following the spirit of the smart health paradigm, GIS could be coupled with the historical healthcare records and the processes stored in the data warehouses of medical institutions and integrate contextual information gathered from wearable and IoT devices from patients.
Unfortunately, this combination has not been considered yet, and potentially beneficial knowledge for both healthcare providers and citizens is neglecting.
To fill this gap, in this section, we propose HGIS (acronym for Healthcare Geographic Information System), a comprehensive, healthcare-oriented, GIS platform to exploit healthcare records and contextual data with spatio-temporal dimensions to provide strategic knowledge.
The proposed platform aims to assist healthcare managers in making better decisions on the planning and delivery of healthcare services.

\subsection{Architecture} \label{subsec:apps:hgis:arc}

The proposed platform comprises a number of interacting actors and resources, which are described below.

\begin{itemize}
\item \textit{Databases}: Data repositories containing all the information about the platform. In particular, we distinguish three kinds of information:
\begin{itemize}
    \item \textit{Medical information}: It refers to information about patients profiles, electronic health records, clinical history and clinical guidelines, among others. This information is generally stored in the data warehouse of the institution, a persistent data repository with the (medical) information imported from the operational information systems through ETL procedures.
    \item \textit{Geographical information}: Geospatial information can be visualised at different aggregation levels, depending on the granularity of the data. To cope with the needs of each healthcare organisation, we implemented an abstraction layer, called Locations Look Up Table (LLUT), for dealing with the geographical information. The LLUT is a tree-based data structure specially tailored to store and associate geographical constructs with their corresponding geographical shapes, defined by sets of customised latitude-longitude pairs. Therefore, the platform can dynamically translate each construct, identified with a unique code, into a geographical shape. Besides, the hierarchy of the LLUT enables the dynamic definition of different aggregation levels, according to the institution needs and the data granularity. For instance, at the lowest level (\ie the leafs of the tree), primary healthcare centres can be defined. One level above, hospitals can be defined. One level above, cities, and so on.
    \item \textit{Contextual information}: It may refer to environmental data, socio-economic data or demographic data, to name a few. It is used to augment medical information with cause-effect explanations. The combination of medical information in conjunction with contextual data enables personalised analyses and helps discover patterns or trends. Contextual information can be gathered from a variety of wearable and IoT devices as well as from context-aware environments.
\end{itemize}
\item \textit{Back-end server}: It manages the workflow of the entire platform, ranging from the analysis of heterogeneous spatio-temporal data to the rendering of the platform's web front-end to the users' devices. Being this server deployed using web technologies, users can interact with the platform from any device with Internet browsing capabilities.
\item \textit{Users}: They are the managers or decision-makers of a healthcare institution. They interact with the platform's web front-end to analyse sets of data and obtain strategic geographical visualisations that assist them in the decision-making process.
\item \textit{Communication infrastructure}: It consists of IEEE 802.11 wireless networks to exchange all the data amongst the different actors of the platform.
\end{itemize}

The general scheme of our platform and the interaction between its actors is depicted in Figure \ref{fig:apps:hgis:arch}.
As observed, the proposed platform considers three well-defined tiers: (i) a data tier consisting of different sets of information organised in databases, (ii) a processing tier with several analysis methods aiming at providing knowledge and decision-making support, and (iii) a visualisation layer to represent the information using maps.
Through the platform's web front-end, users interact with the system to obtain strategic knowledge by exploiting the organisation's data.
First, users query the back-end server asking for a certain analysis \ding{192}.
This request contains information about the kind of analysis to perform, such as the medical condition/disease to explore (\eg flu, cancer, COVID-19,\ldots), the time interval (\ie the initial date and the final date of the medical observations), the geographical level (according to the LLUT hierarchy) and the analysis method to apply.
Then, the back-end server retrieves the data needed from the different databases.
It first retrieves from the data warehouse the medical information related to the health condition or disease to study happened during the corresponding time range \ding{193}.
This information is then complemented with spatio-temporal contextual information related to the analysis to perform \ding{194}.
Next, the geographical shapes of the corresponding LLUT hierarchy level are retrieved and, if needed, the medical and contextual information are aggregated to achieve the desired level \ding{195}.
Note that both medical and contextual information are associated to geographical constructs defined in the LLUT.

\begin{figure}[b!]
\centering
\includegraphics[width=0.90\columnwidth]{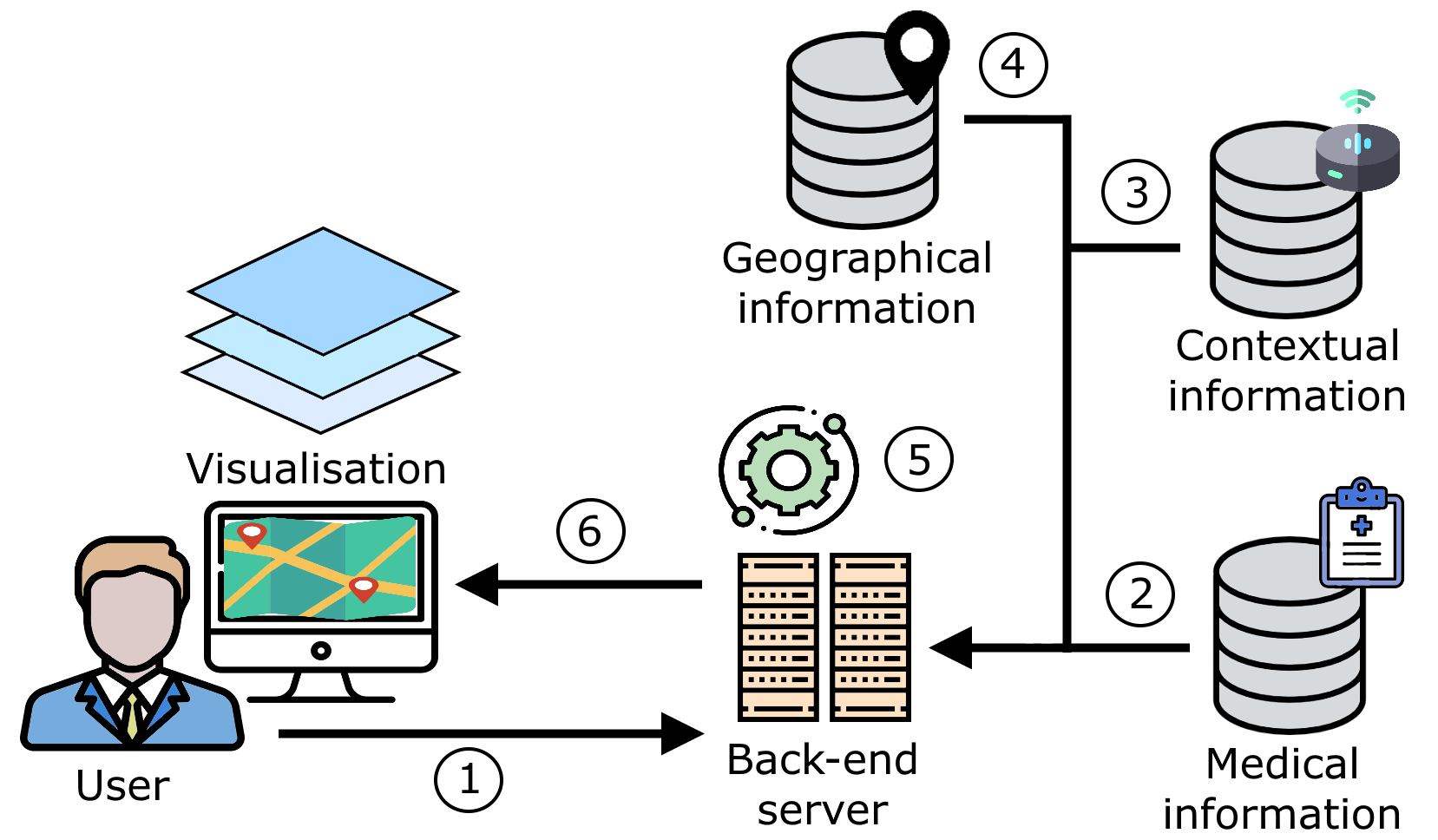}
\caption{Architecture and interaction between the main actors of HGIS.}
\label{fig:apps:hgis:arch}
\end{figure}

With all the data collected from the data tier, the back-end server applies the analysis method requested \ding{196}.
Initially, these methods can be fairly simple computations to calculate healthcare indicators, \ie numerical and categorical values resulting from the join evaluation of several medical and contextual related facts, such as the number of patients' relapses and their prescribed medication.
These indicators can be then statistically analysed by using data mining techniques to compute correlations, identify clusters, classify observations or detect outliers.
Also, thanks to the addition of timestamped data, process mining techniques can also be applied to discover processes, check processes adherence and optimise processes.
Thanks to process mining techniques, patients' flows could be discovered and analysed in terms of efficiency and performance from a spatio-temporal perspective.
To the best of our knowledge, the combination of process mining techniques with GIS technologies to study healthcare processes is yet to be researched.
Finally, according to the analysis method, the back-end server renders a specific map visualisation \ding{197}.
For instance, numerical indicators are shown with markers and choropleth maps, densities are represented with heat maps, flows are described with migration maps, and statistical data are shown with pie and bar charts.
Besides, the results of each analysis are depicted as a map layer, enabling stacking multiple layers from different analysis in the same map visualisation to find cross-analysis tendencies.

\subsection{Implementation} \label{subsec:apps:hgis:impl}

In order to obtain an initial pre-evaluation of HGIS, a proof of concept, web-based prototype has been developed by Xarxa Sanit{\`a}ria i Social Santa Tecla, a large healthcare provider in the south of Catalonia that comprises several hospitals and primary care centres.
The implementation details of the prototype are explained below.

The different databases of the data tier are built using the geographical PostGIS extension for PostgreSQL database management system.
This extension supports the management and mapping of geographic objects (\ie points, polygons,\ldots).
For the sake of patients' privacy, in this prototype, the data warehouse has been populated with synthetic data mimicking the real data from the operational information systems of the institution.
With regards to the LLUT structure, a five-level hierarchy was defined, in accordance to the geographic organisation of healthcare services in Catalonia; from a higher level to a lower level: province, healthcare region, municipality, basic health area and healthcare centre.
The data in the LLUT was populated with geographical information related to the places where the healthcare services of the institution are provided.

In the processing tier, we use server-side implementations in Java.
Some data analysis methods have been implemented to compute different indicators and further statistics (\eg aggregations, averages,\ldots).
Also, a process discovery algorithm has been considered to explore the flows (from a spatio-temporal perspective) during medical processes, such as the mobility of patients amongst healthcare facilities, which enables the identification of bottlenecks or hidden patterns.
Finally, for the visualisation tier, we use Javascript libraries to represent the information using map visualisations.
These visualisations depend on the hierarchy level of the data defined in the LLUT.
For instance, markers are used to visualise data at the lowest level (\ie healthcare centres) because they correspond to locations, rather than regions.
On the contrary, when data refers to higher levels of the LLUT entailing areas, choropleth visualisations are used.
In addition, processes are represented using traffic maps visualisations (see Figure \ref{fig:apps:hgis:maps}).

The initial testing of the proof of concept implementation shows that the platform is fully functional and helps representing information that, until now, was not accessible to decision makers.
Thanks to the use of open and scalable technologies, HGIS has proven to be an extensible, cost-efficient and practical tool for managers and data scientists in the healthcare organisation.

\begin{figure}[t!]
\centering
\begin{subfigure}{.48\textwidth}
  \centering\includegraphics[width=\linewidth]{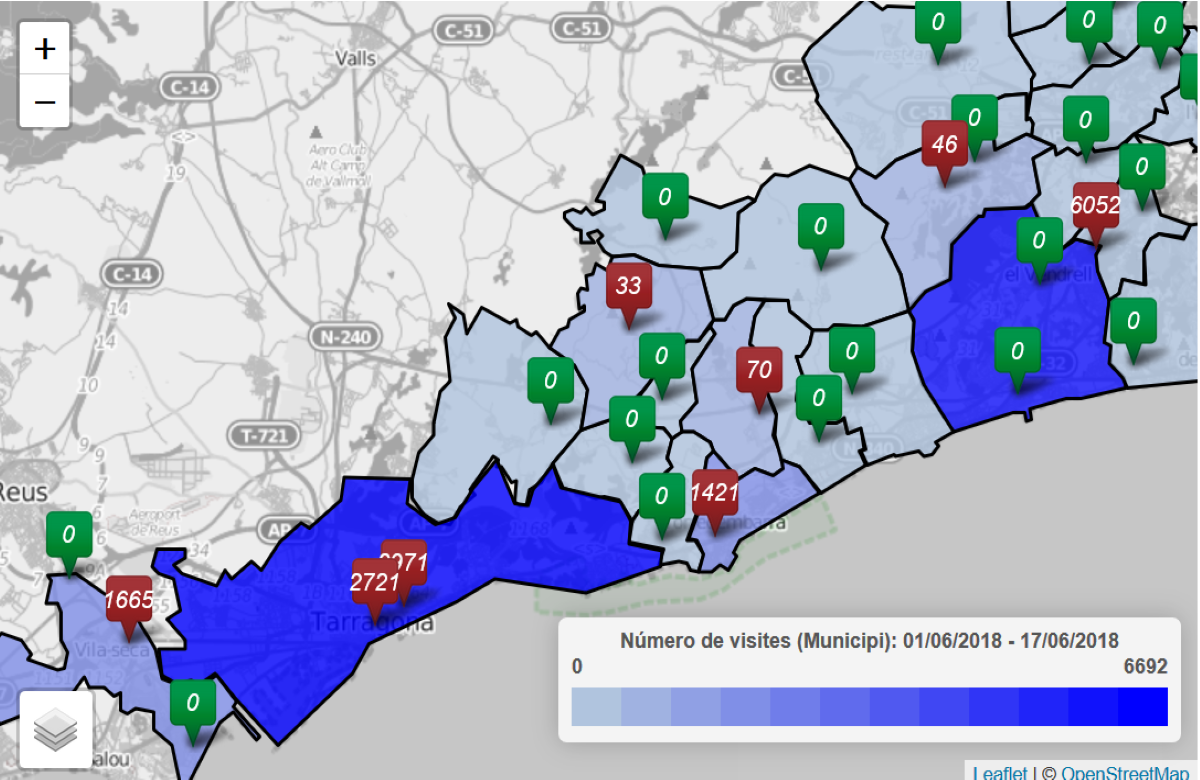}  
  \caption{Markers and choropleth map layer}
\end{subfigure}
\hfill
\begin{subfigure}{.48\textwidth}
  \centering\includegraphics[width=\linewidth]{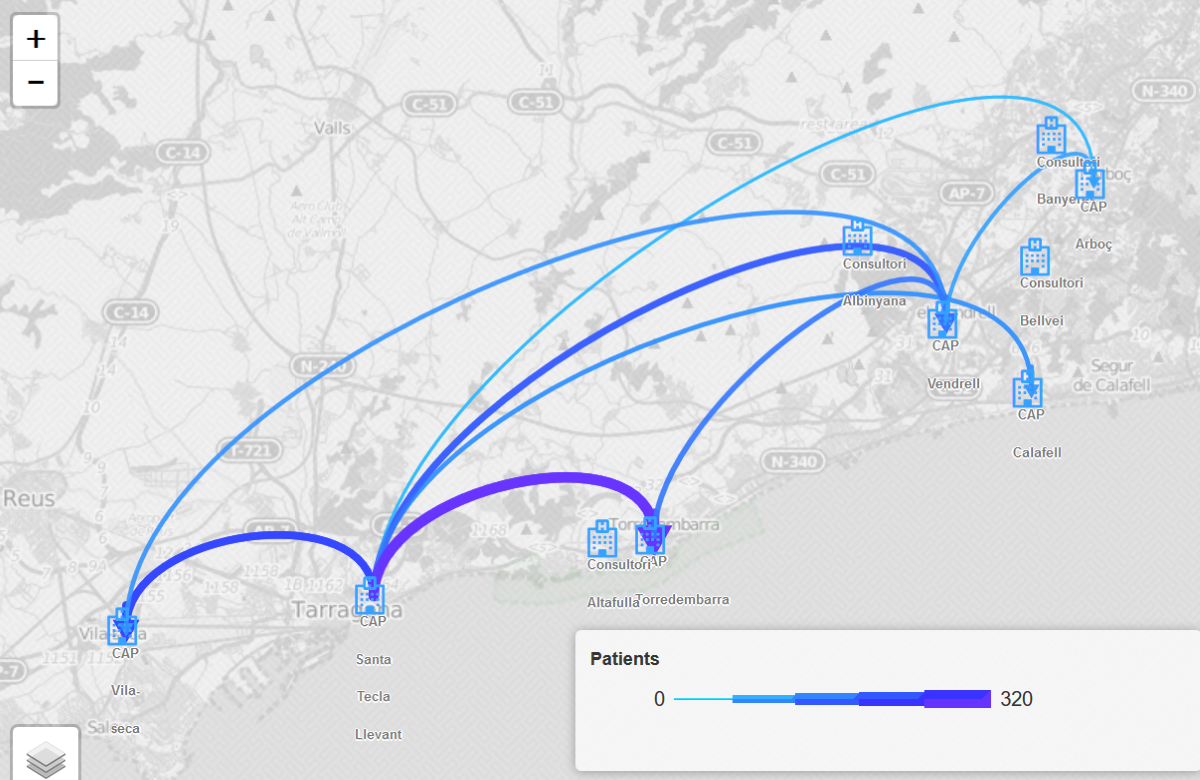}  
  \caption{Traffic map layer}
\end{subfigure}
\caption{Example of two map visualisations obtained in the HGIS platform (reprinted from \cite{batista2018hgis}).}
\label{fig:apps:hgis:maps}
\end{figure}

\section{Conclusions} \label{sec:apps:concl}

The number of applications to smart health is rapidly growing.
Whereas mobile health applications only consider data collected from mobile devices, such as healthcare information or location data, smart health applications augment this information with contextual data gathered from the sensing infrastructure of context-aware environments, such as smart cities.
All in all, the main goal of these applications is to improve the quality of life of citizens.
In this chapter, three applications have been proposed.

First, in Section \ref{sec:apps:rs}, we have presented the idea of integrating real-time recommender systems with the sensing infrastructure of smart cities to promote healthy lifestyles amongst citizens.
This recommender system provides citizens with routes recommendations for the realisation of physical exercises (\eg walking, running, cycling,\ldots) by considering the health conditions and preferences of citizens along with contextual data.
The collaborative nature of our approach empowers the figure of citizens, who can create new routes and rate routes performed, so that the recommender system could exploit this crowdsourcing-based information to provide more accurate recommendations to other citizens.
An overview of the proposed architecture, the main components and the implementation details have been provided.
Further work in this system will consider incorporating wearables and body sensors to determine the citizens' health conditions in real time while following a route. Further, gamification techniques to engage the participation of citizens could be considered.

Second, in Section \ref{sec:apps:mob}, we have focused on early mobilisation programmes, aiming to enhance the recovery process of critically ill patients that are hospitalised in ICUs.
Despite their benefits, there is a lack of technological solutions to evaluate the evolution and success of these programmes on patients.
To fill this gap, we have proposed e-PEMICU, a novel platform to support early mobilisation programmes in hospital settings.
The architecture, actors and functioning of the entire platform have been described.
Moreover, we have implemented a working prototype of the platform using non-invasive and affordable technology, which has been successfully tested with real patients from a hospital in Tarragona.
Future work will concentrate on the maturity of further smart health-based early mobilisation solutions from a technical perspective to address subsequent issues, such as interoperability, wireless channel characterisation, ethics, information security and data privacy.

Third, and last, in Section \ref{sec:apps:hgis}, we have sustained that the fusion of context-aware analyses with GIS technologies will reshape the understanding of business intelligence tools in the healthcare sector.
However, due to the general-purpose characteristics of current GIS technologies, advanced analyses of heterogeneous data are not straightforward.
To this end, we have presented HGIS, a comprehensive, healthcare-oriented, GIS platform to exploit healthcare records in conjunction with contextual data with spatio-temporal dimensions to provide strategic knowledge to decision-makers.
An overview of the proposed architecture and the interaction between its actors has been described.
Besides, a proof of concept prototype has been developed and tested within a large healthcare provider in the area of Tarragona.
Further work in this platform will concentrate on fully implementing the proposed tools and thoroughly testing it in real medical practice.

\part{Conclusions} \label{part:concl}
\chapter{Concluding Remarks}
\label{chap:conclusions}

\emph{This chapter summarises the main concluding remarks and findings of this dissertation. Moreover, it also sketches some future work in the studied research fields arising from either partially reached goals or expected improvements. More concretely, Section \ref{sec:conclusions:summary} provides a summary of the contributions presented in this dissertation, and Section \ref{sec:conclusions:future} describes future research lines of the topics studied.}

\minitoc

\section{Summary of Contributions} \label{sec:conclusions:summary}

In this dissertation, we have addressed security and privacy issues of the smart healthcare paradigm.
All these issues are especially relevant nowadays because of the large volume of personal and confidential data managed in these scenarios.
Further, given the growing popularity of context-aware environments, these issues will strengthen in a near future.
Even more challenging, being smart health a very complex system-of-systems, these issues cannot be addressed holistically from a single angle, but specific actions need to be carried out at different stages of the data life-cycle, ranging from its very collection, going through its transmission and storage, to its final exploitation.
Hence, this dissertation presents security and privacy contributions to smart health in two directions: on the one hand, to context-aware environments enablers of smart health (Part \ref{part:cae}) and, on the other hand, to the trendy analytical technique of process mining (Part \ref{part:pm}).
Moreover, since this paradigm is becoming popular, there is an increasing need for novel smart-health-oriented applications.
For this reason, the last contribution in this dissertation is devoted to applications (Part \ref{part:apps}).

In more detail, our contributions are summarised as follows.

\begin{enumerate}
\item In Chapter \ref{chap:sensors}, we have presented a neutral, high-level comprehensive analysis on the security aspects of context-aware environments that enable smart health services, from the very collection of data to its final analysis.
To characterise these environments, first we have presented a taxonomy to classify the sensors into user-centric sensors and contextual sensors.
Moreover, we have also analysed the most prominent wireless communication technologies to convey all the data handled in smart health services.
Considering the above, we have described and classified the main security attacks in these environments according to the target actor, followed by the description of common solutions to counteract them.
To the best of our knowledge, this is the first contribution providing an all-in-one security analysis contextualised to smart health scenarios.

\item In Chapter \ref{chap:ubicomp}, we have proposed a novel framework to analyse the privacy risks of ubiquitous computing systems from an individual-centred perspective.
To do this, we have identified five dimensions, namely identity privacy, query privacy, location privacy, footprint privacy and intelligence privacy, which have been contextualised within context-aware environments.
In contrast to classical approaches following a data-centred perspective, our framework enables developing (or enhancing) privacy-preserving ubiquitous computing systems from the individuals perspective.

\item In Chapter \ref{chap:cogcities}, we have provided an extensive analysis on the security and privacy challenges of cognitive cities, a specific implementation of cognitive environments that evolve from current context-aware environments.
Challenges have been classified into three categories, namely technical challenges, societal challenges and regulatory challenges.
To the best of our knowledge, this is the first contribution reviewing the open challenges of cognitive cities in terms of security and privacy aspects.

\item In Chapter \ref{chap:pmcae}, we have explored the opportunities of context-aware process mining by considering all sorts of data collected from context-aware environments, such as contextual data, in process mining analyses.
Under this paradigm, we have shown the benefits of conducting on-line and proactive process mining analyses, and we have also identified and classified the main challenges into four categories, namely data challenges, ecosystem challenges, computational challenges and automation challenges.

\item In Chapter \ref{chap:skipminer}, we have proposed a novel process discovery algorithm, called Skip Miner, aiming to confer structure to the process models discovered directly from the event logs.
In contrast to classical simplification methods, our algorithm uses a probabilistic built-in heuristic that diminishes the complexity of process models so as to prevent the appearance of spaghetti processes.
Using a real-life event log from a healthcare institution, where spaghetti processes are commonplace, we have shown that our method provides a remarkable balance between the simplification and the quality of the process models discovered in comparison to other simplification methods from the literature.

\item In Chapter \ref{chap:upppm} and Chapter \ref{chap:kpppm}, we have contributed to the young research field of privacy-preserving process mining.
Despite the popularity of pseudonymisation and encryption techniques in this field, we have shown their inability to avert either distribution-based attacks or modelling-based attacks when combined with location-oriented targeted attacks, such as RSI and OI attacks.
To protect people's privacy, we have proposed two novel privacy-preserving process mining methods, called \textit{u}-PPPM and \textit{k}-PPPM, able to create protected event logs that limit the knowledge of the discovered process models according to a privacy threshold.
On the one hand, \textit{u}-PPPM is based on the uniformisation of sensitive distributions of event data attributes and, on the other hand, \textit{k}-PPPM is based on microaggregation techniques through \textit{k}-anonymity.
To the best of our knowledge, \textit{k}-PPPM is the first method that uses microaggregation techniques to protect people's privacy in process mining.
By testing both methods with six real-life event logs, we have shown that they achieve a remarkable trade-off between data utility and privacy by introducing an homogeneous distortion on the entire set of process models.

\item In Chapter \ref{chap:apps}, we have provided three contributions to specific smart health applications.
First, we have presented a context-aware recommender system for healthy routes, called SmartRoute, by considering the health status and the preferences of citizens, contextual data from the city and crowdsourced-based information.
Second, we have presented a platform supporting the management of early mobilisation programmes, called e-PEMICU, oriented to enhance the recovery process of critically ill patients hospitalised in intensive care units.
And third, we have presented a holistic geographical information system, called HGIS, that cross-correlates healthcare data and contextual data in order to assist the decision-making process of healthcare institutions.

\end{enumerate}

\section{Future Research Lines} \label{sec:conclusions:future}

Next, we sketch possible research lines for future work in the same order in which we have presented our main contributions.

\begin{enumerate}
\item The security of smart healthcare systems partly depends on the security of context-aware environments.
As observed in Chapter \ref{chap:sensors}, these environments may integrate many different technologies, systems and actors, each of them with their own security threats.
The rapid evolution of these environments prompts to novel opportunities, but also new security threats and vulnerabilities yet to be known.
Among others, we highlight the need for lightweight and low-power security solutions in the nanotechnology-based devices and green communication technologies of the future.
Moreover, zero-trust architectures, blockchain technology and the use of artificial intelligence are promising tools to make smart healthcare even safer and more private.
To enhance the cyber resilience of these systems, educating citizens on the right use of technology through awareness programmes about cybersecurity and data protection is another not less important action.

\item Despite the generalisation of ubiquitous computing systems, privacy considerations are too often an afterthought.
Aligned with the growing importance of legislation on data protection, future research will focus on the consolidation of a paradigm shift from data privacy to individuals privacy, similar to the framework presented in Chapter \ref{chap:ubicomp}, and the development of ubiquitous computing systems with a privacy by design approach.

\item Although the field of cognitive cities is still in its infancy, it may open the door to the conceptualisation of new healthcare paradigms, such as cognitive healthcare.
To help success the applications operating on cognitive cities (or cognitive systems, in general), future endeavours are required to minimise the security and privacy risks of such systems, as they will be inherently be inherited by those applications.
In this sense, the work presented in Chapter \ref{chap:cogcities} can set the ground to face this matter.

\item Context-aware environments enable richer and online process mining analyses, as described in Chapter \ref{chap:pmcae}, able to readjust and enhance the execution of processes in almost real-time constraints.
In this context, future work will concentrate on quantitatively evaluating the performance of applying these analyses in these environments, in terms of efficiency, resource allocation, time and costs.

\item Dealing with spaghetti process models is not easy, as observed in Chapter \ref{chap:skipminer}.
Unfortunately, they are far too common during process mining analyses.
Future work in this topic will focus on developing further strategies and heuristics aiming to simplify and bring structure to this kind of processes.
For example, simplification strategies might consider the semantics of the models to conduct this task.
Moreover, defining high-level metrics for assessing the simplicity of process models would streamline the evaluation and comparison of different simplification strategies.

\item Research on privacy-preserving process mining is on the rise, and it is expected to continue growing in the coming years.
In this sense, applying classical privacy protection techniques to process mining will enable new privacy-preserving process mining methods, as those described in Chapter \ref{chap:upppm} and Chapter \ref{chap:kpppm}.
Future work will focus on three main avenues:
(i) the development of novel methods facing more complex attacker models able to acquire other sorts of knowledge,
(ii) the formalisation of more robust privacy protection models based on advanced privacy properties, such as \textit{t}-closeness, \textit{p}-sensitive or differential privacy,
and (iii) the definition of frameworks or metrics to evaluate and compare the suitability of privacy-preserving process mining methods.

\item Thanks to the widespread deployment of IoT devices and the generalisation of context-aware environments, it is expected that the number of smart healthcare applications will grow in the next years.
Although some applications have been presented in Chapter \ref{chap:apps}, numerous applications and services are yet to be developed in order to enhance the health, comfort and quality of life of citizens.

\end{enumerate}

\chapter{Publications}
\label{chap:publications}

\emph{This chapter enumerates all the research publications supporting the content of this dissertation, including articles published in ISI-JCR journals, book chapters and international conferences.}

\paragraph{Journals}~

\begin{longtable}{r|p{10cm}}

\textsc{2017} & \footnotesize{Fran Casino, Constantinos Patsakis, \underline{Edgar Batista}, Frederic Borr{\`a}s and Antoni Mart{\'i}nez-Ballest{\'e}, \textbf{``Healthy Routes in the Smart City: A Context-Aware Mobile Recommender''}, IEEE Software, 34(6):42--47, IEEE, 2017, ISSN 0740-7459. doi: \href{http://doi.org/10.1109/MS.2017.4121209}{10.1109/MS.2017.4121209}} \newline
\hspace*{0.5cm} --- ISI-JCR 2017: Q1 (CS, 13/104), IF 2.879\\

\multicolumn{2}{c}{} \\

\textsc{2021} & \footnotesize{\underline{Edgar Batista}, M. Angels Moncusi, Pablo L{\'o}pez-Aguilar, Antoni Mart{\'i}nez-Ballest{\'e} and Agusti Solanas, \textbf{``Sensors for Context-Aware Smart Healthcare: A Security Perspective''}, Sensors, 21(20):6886, MDPI, 2021, ISSN 1424-8220. doi: \href{https://doi.org/10.3390/s21206886}{10.3390/s21206886}} \newline
\hspace*{0.5cm} --- ISI-JCR 2020: Q1 (II, 14/64), IF 3.576\\

\multicolumn{2}{c}{} \\

\textsc{2021} & \footnotesize{Juvenal Machin, \underline{Edgar Batista}, Antoni Mart{\'i}nez-Ballest{\'e} and Agusti Solanas, \textbf{``Privacy and Security in Cognitive Cities: A Systematic Review''}, Applied Sciences, 11(10):4471, MDPI, 2021, ISSN 2076-3417. doi: \href{http://doi.org/10.3390/app11104471}{10.3390/app11104471}} \newline
\hspace*{0.5cm} --- ISI-JCR 2020: Q2 (EM, 38/91), IF 2.679\\

\multicolumn{2}{c}{} \\

\textsc{2021} & \footnotesize{\underline{Edgar Batista} and Agusti Solanas, \textbf{``A uniformization-based approach to preserve individuals’ privacy during process mining analyses''}, Peer-to-Peer Networking and Applications, 14(3):1500--1519, Springer, 2021, ISSN 1936-6450. doi: \href{https://doi.org/10.1007/s12083-020-01059-1}{10.1007/s12083-020-01059-1}} \newline
\hspace*{0.5cm} --- ISI-JCR 2020: Q2 (CS, 67/162), IF 3.307\\

\multicolumn{2}{c}{} \\

\textsc{2021} & \footnotesize{Maria Ferre, \underline{Edgar Batista}, Agusti Solanas and Antoni Mart{\'i}nez-Ballest{\'e}, \textbf{``Smart Health-Enhanced Early Mobilisation in Intensive Care Units''}, Sensors, 21(16):5408, MDPI, 2021, ISSN 1424-8220. doi: \href{https://doi.org/10.3390/s21165408}{10.3390/s21165408}} \newline
\hspace*{0.5cm} --- ISI-JCR 2020: Q1 (II, 14/64), IF 3.576\\

\multicolumn{2}{c}{} \\

\textsc{Submitted} & \footnotesize{\underline{Edgar Batista}, Antoni Mart{\'i}nez-Ballest{\'e} and Agusti Solanas, \textbf{``Privacy-Preserving Process Mining: A Microaggregation-based Approach''}, Journal of Information Security and Applications -- In Review.} \newline
\hspace*{0.5cm} --- ISI-JCR 2020: Q2 (CS, 56/161), IF 3.872\\

\end{longtable}

\paragraph{Book Chapters}~

\begin{longtable}{r|p{11cm}}

\textsc{2021} & \footnotesize{Agusti Solanas, \underline{Edgar Batista}, Fran Casino, Achilleas Papageorgiou and Constantinos Patsakis, \textbf{``Privacy-Oriented Analysis of Ubiquitous Computing Systems: A 5-D Approach''}, In: G. Avoine and J. Hernandez-Castro (eds), Security of Ubiquitous Computing Systems, pp. 201--213, Springer, 2021. doi: \href{https://doi.org/10.1007/978-3-030-10591-4\_12}{10.1007/978-3-030-10591-4\_12}}\\

\end{longtable}

\paragraph{Conferences}~

\begin{longtable}{r|p{11cm}}

\textsc{2017} & \footnotesize{Agusti Solanas, Fran Casino, \underline{Edgar Batista} and Robert Rallo, \textbf{``Trends and Challenges in Smart Healthcare Research: A Journey from Data to Wisdom''}, In Proceedings of the 3rd International Forum on Research and Technologies for Society and Industry (RTSI), pp. 1--6, Modena, Italy, IEEE, September 2017. doi: \href{https://doi.org/10.1109/RTSI.2017.8065986}{10.1109/RTSI.2017.8065986}}\\

\multicolumn{2}{c}{} \\

\textsc{2017} & \footnotesize{\underline{Edgar Batista}, \textbf{``Process Mining: Extracting Valuable Knowledge from Event Log Data''}, In Proceedings of the 4th URV Doctoral Workshop in Computer Science and Mathematics (DCSM), pp. 49--52, Tarragona, Spain, Publicacions URV, November 2017.}\\

\multicolumn{2}{c}{} \\

\textsc{2017} & \footnotesize{\underline{Edgar Batista} and Agusti Solanas, \textbf{``Privacy-Preserving Process Mining: Towards the new European General Data Protection Regulation''}, In Proceedings of the 2nd Cryptanalysis of Ubiquitous Computing Systems (CRYPTACUS) Workshop, pp. 22--24, Nijmegen, The Netherlands, November 2017.}\\

\multicolumn{2}{c}{} \\

\textsc{2018} & \footnotesize{\underline{Edgar Batista} and Agusti Solanas, \textbf{``Process Mining in Healthcare: A Systematic Review''}, In Proceedings of the 9th International Conference on Information, Intelligence, Systems and Applications (IISA), pp. 1--6, Zakynthos, Greece, IEEE, July 2018. doi: \href{https://doi.org/10.1109/IISA.2018.8633608}{10.1109/IISA.2018.8633608}}\\

\multicolumn{2}{c}{} \\

\textsc{2018} & \footnotesize{Fran Casino, Constantinos Patsakis, \underline{Edgar Batista}, Octavian Postolache, Antoni Mart{\'i}nez-Ballest{\'e} and Agusti Solanas, \textbf{``Smart Healthcare in the IoT Era: A Context-Aware Recommendation Example''}, In Proceedings of the 1st International Symposium on Sensing and Instrumentation in IoT Era (ISSI), pp. 1--6, Shanghai, China, IEEE, September 2018. doi: \href{https://doi.org/10.1109/ISSI.2018.8538106}{10.1109/ISSI.2018.8538106}}  \newline
\hspace*{0.5cm} --- Best Paper Application Award\\

\multicolumn{2}{c}{} \\

\textsc{2018} & \footnotesize{\underline{Edgar Batista}, Antoni Mart{\'i}nez-Ballest{\'e}, Marta Pe{\~n}a, Xavier Singla and Agusti Solanas, \textbf{``HGIS: A Healthcare-oriented Approach to Geographic Information Systems''}, In Proceedings of the 6th International Conference on Applications in Electronics Pervading Industry, Environment and Society (ApplePies), LNEE 573, pp. 59--65, Pisa, Italy, Springer, September 2018. doi: \href{https://doi.org/10.1007/978-3-030-11973-7\_8}{10.1007/978-3-030-11973-7\_8}}\\

\multicolumn{2}{c}{} \\

\textsc{2019} & \footnotesize{\underline{Edgar Batista}, \textbf{``Process Mining and Spaghetti Business Processes: An Uneasy Relationship''}, In Proceedings of the 5th URV Doctoral Workshop in Computer Science and Mathematics (DCSM), pp. 64--72, Tarragona, Spain, Publicacions URV, May 2019.}\\

\multicolumn{2}{c}{} \\

\textsc{2019} & \footnotesize{\underline{Edgar Batista} and Agusti Solanas, \textbf{``Skip Miner: Towards the Simplification of Spaghetti-like Business Process Models''}, In Proceedings of the 10th International Conference on Information, Intelligence, Systems and Applications (IISA), pp. 1--6, Patras, Greece, IEEE, July 2019. doi: \href{https://doi.org/10.1109/IISA.2019.8900713}{10.1109/IISA.2019.8900713}}\\

\multicolumn{2}{c}{} \\

\textsc{2019} & \footnotesize{Antoni Mart{\'i}nez-Ballest{\'e}, Pablo Gimeno, Aleix Marin{\'e}, \underline{Edgar Batista} and Agusti Solanas, \textbf{``e-PEMICU: an e-Health Platform to Support Early Mobilisation in Intensive Care Units''}, In Proceedings of the 10th International Conference on Information, Intelligence, Systems and Applications (IISA), pp. 1--6, Patras, Greece, IEEE, July 2019. doi: \href{https://doi.org/10.1109/IISA.2019.8900718}{10.1109/IISA.2019.8900718}}\\

\multicolumn{2}{c}{} \\

\textsc{2019} & \footnotesize{\underline{Edgar Batista}, Francisco Falcone, Antoni Mart{\'i}nez-Ballest{\'e} and Agusti Solanas, \textbf{``The Promising Future of Process Mining with the Internet of Events in Context-Aware Environments''}, In Proceedings of the 2nd International Conference on Sensing and Instrumentation in IoT Era (ISSI), pp. 1--6, Lisbon, Portugal, IEEE, August 2019. doi: \href{https://doi.org/10.1109/ISSI47111.2019.9043725}{10.1109/ISSI47111.2019.9043725}}\\

\end{longtable}

\paragraph{Online}~

\begin{longtable}{r|p{11cm}}

\textsc{2020} & \footnotesize{\underline{Edgar Batista}, \textbf{``Knowledge vs privacy: The healthcare dilemma''}, Tecnonews [Online], November 2020. Available at: \url{https://www.tecnonews.info/opiniones/knowledge_vs_privacy_the_healthcare_dilemma}}\\ 

\end{longtable}

\paragraph{Technical Reports}~

\begin{longtable}{r|p{11cm}}

\textsc{2017} & \footnotesize{Fran Casino, Constantinos Patsakis, Antoni Mart{\'i}nez-Ballest{\'e}, Frederic Borr{\`a}s and \underline{Edgar Batista}, \textbf{``Technical Report: Implementation and Validation of a Smart Health Application''}, arXiv preprint \href{https://arxiv.org/abs/1706.04109}{arXiv:1706.04109}, pp. 1--4, June 2017.}\\ 

\end{longtable}

\part{Appendices}
\appendix
\chapter{Results of \textit{u}-PPPM} \label{app:upppm}

This appendix presents the details of all the results obtained from the execution of the proposed \textit{u}-PPPM method in Chapter \ref{chap:upppm}.
More specifically, the following information can be found:

\begin{itemize}
\item Tables \ref{tbl:app_upppm:results_qs}, \ref{tbl:app_upppm:results_ils} and \ref{tbl:app_upppm:results_cs} list the QS, ILS and CS results obtained, respectively, for each execution of \textit{u}-PPPM with a certain combination of the parameters $k$ (privacy level) and $sel$ (selection strategy).
\item Figures \ref{fig:app_upppm:results_bpi12_bpi13}, \ref{fig:app_upppm:results_bpi14_bpi15} and \ref{fig:app_upppm:results_coselog_tgn} depict the correlation between the QS, ILS and CS results, for each execution of \textit{u}-PPPM with a certain combination of parameters.
\item Figures \ref{fig:app_upppm:ttest_qs_k} and \ref{fig:app_upppm:ttest_qs_strategy} illustrate the $p$-value results from the t-Tests according to the $k$ and $sel$ parameters values, respectively, for the QS results. The $p$-value results for the ILS results are illustrated in Figures \ref{fig:app_upppm:ttest_ils_k} and \ref{fig:app_upppm:ttest_ils_strategy}, and for the CS results in Figures \ref{fig:app_upppm:ttest_cs_k} and \ref{fig:app_upppm:ttest_cs_strategy}.
\end{itemize}

\begin{sidewaystable}[htbp]
\renewcommand{\tabcolsep}{0.2cm}
\renewcommand{\arraystretch}{0.96}
\centering
\scriptsize  
\caption{QS results for each combination of parameters.} \label{tbl:app_upppm:results_qs}
\begin{tabular}{ccccccccc} 
\toprule
\multicolumn{2}{c}{\bf Parameters} & \multicolumn{6}{c}{\bf Event logs}  & \multirow{2}{*}{\textbf{Avg.}} \\
\cmidrule(lr){1-2} \cmidrule(lr){3-8}
$\pmb{k}$ & $\pmb{sel}$ & {\bf BPI12} & {\bf BPI13} & {\bf BPI14} & {\bf BPI15} & {\bf CoSeLoG} & {\bf TGN-Hospital}  \\
\midrule
\textbf{2} & $\pmb{S_1}$ & \gr{0.0614 $\pm$ 0.1036} & \re{0.003 $\pm$ 0.0325} & \re{0.1356 $\pm$ 0.1296} & \re{0.1815 $\pm$ 0.1432} & 0.1115 $\pm$ 0.142 & \gr{0.0146 $\pm$ 0.0475} & 0.0846 $\pm$ 0.0997 \\
\textbf{2} & $\pmb{S_2}$ & 0.0645 $\pm$ 0.1066 & 0.0028 $\pm$ 0.0299 & \gr{0.1344 $\pm$ 0.1272} & 0.179 $\pm$ 0.1391 & \re{0.1147 $\pm$ 0.141} & 0.0149 $\pm$ 0.0482 & \re{0.0851 $\pm$ 0.0987} \\
\textbf{2} & $\pmb{S_3}$ & 0.0628 $\pm$ 0.1025 & \gr{0.0024 $\pm$ 0.028} & 0.1345 $\pm$ 0.1295 & 0.1744 $\pm$ 0.134 & 0.1118 $\pm$ 0.1386 & \re{0.0151 $\pm$ 0.0494} & 0.0835 $\pm$ 0.097 \\
\textbf{2} & $\pmb{S_4}$ & \re{0.0662 $\pm$ 0.1085} & 0.0026 $\pm$ 0.029 & 0.1347 $\pm$ 0.1278 & \gr{0.1705 $\pm$ 0.13} & \gr{0.1067 $\pm$ 0.1338} & 0.0147 $\pm$ 0.047 & \gr{0.0826 $\pm$ 0.096} \\
\midrule
\textbf{3} & $\pmb{S_1}$ & 0.1563 $\pm$ 0.1615 & 0.0045 $\pm$ 0.0436 & 0.1933 $\pm$ 0.1459 & 0.2614 $\pm$ 0.1595 & 0.183 $\pm$ 0.1731 & 0.0246 $\pm$ 0.0684 & 0.1372 $\pm$ 0.1253 \\
\textbf{3} & $\pmb{S_2}$ & \gr{0.1242 $\pm$ 0.1502} & \gr{0.0039 $\pm$ 0.0353} & \gr{0.168 $\pm$ 0.1468} & \gr{0.222 $\pm$ 0.1493} & \gr{0.1614 $\pm$ 0.1711} & \gr{0.0202 $\pm$ 0.0618} & \gr{0.1179 $\pm$ 0.1198} \\
\textbf{3} & $\pmb{S_3}$ & 0.155 $\pm$ 0.163 & 0.0048 $\pm$ 0.045 & 0.1991 $\pm$ 0.1507 & 0.2554 $\pm$ 0.1533 & 0.1815 $\pm$ 0.1762 & 0.0249 $\pm$ 0.068 & 0.1368 $\pm$ 0.126 \\
\textbf{3} & $\pmb{S_4}$ & \re{0.1683 $\pm$ 0.1713} & \re{0.0052 $\pm$ 0.0468} & \re{0.2074 $\pm$ 0.146} & \re{0.2615 $\pm$ 0.156} & \re{0.1878 $\pm$ 0.1792} & \re{0.0269 $\pm$ 0.0714} & \re{0.1428 $\pm$ 0.1285} \\
\midrule
\textbf{4} & $\pmb{S_1}$ & 0.1944 $\pm$ 0.1733 & 0.0049 $\pm$ 0.0412 & 0.2133 $\pm$ 0.1509 & 0.2889 $\pm$ 0.1676 & 0.2219 $\pm$ 0.1852 & 0.041 $\pm$ 0.0908 & 0.1607 $\pm$ 0.1348 \\
\textbf{4} & $\pmb{S_2}$ & \gr{0.1388 $\pm$ 0.1559} & \gr{0.0041 $\pm$ 0.04} & \gr{0.1757 $\pm$ 0.1464} & \gr{0.2334 $\pm$ 0.1555} & \gr{0.1705 $\pm$ 0.1664} & \gr{0.0312 $\pm$ 0.0796} & \gr{0.1243 $\pm$ 0.1233} \\
\textbf{4} & $\pmb{S_3}$ & 0.1928 $\pm$ 0.1864 & 0.0047 $\pm$ 0.0406 & 0.2148 $\pm$ 0.1512 & 0.3021 $\pm$ 0.1715 & 0.2084 $\pm$ 0.1887 & 0.0409 $\pm$ 0.0901 & 0.1606 $\pm$ 0.1381 \\
\textbf{4} & $\pmb{S_4}$ & \re{0.2257 $\pm$ 0.1895} & \re{0.006 $\pm$ 0.0475} & \re{0.2324 $\pm$ 0.1536} & \re{0.3352 $\pm$ 0.1793} & \re{0.2329 $\pm$ 0.1907} & \re{0.046 $\pm$ 0.0941} & \re{0.1797 $\pm$ 0.1425} \\
\midrule
\textbf{5} & $\pmb{S_1}$ & 0.2321 $\pm$ 0.1861 & 0.0077 $\pm$ 0.0549 & 0.2379 $\pm$ 0.1529 & 0.3101 $\pm$ 0.1813 & 0.2234 $\pm$ 0.1899 & 0.0608 $\pm$ 0.1071 & 0.1787 $\pm$ 0.1454 \\
\textbf{5} & $\pmb{S_2}$ & \gr{0.176 $\pm$ 0.1756} & \gr{0.0072 $\pm$ 0.0524} & \gr{0.1923 $\pm$ 0.1489} & \gr{0.2408 $\pm$ 0.1769} & \gr{0.1769 $\pm$ 0.177} & \gr{0.0455 $\pm$ 0.0958} & \gr{0.1398 $\pm$ 0.1378} \\
\textbf{5} & $\pmb{S_3}$ & 0.2287 $\pm$ 0.1874 & \re{0.0084 $\pm$ 0.0578} & 0.2377 $\pm$ 0.1532 & 0.3223 $\pm$ 0.1827 & 0.2383 $\pm$ 0.1911 & 0.0583 $\pm$ 0.1053 & 0.1823 $\pm$ 0.1462 \\
\textbf{5} & $\pmb{S_4}$ & \re{0.2647 $\pm$ 0.1703} & 0.0083 $\pm$ 0.0588 & \re{0.2687 $\pm$ 0.1573} & \re{0.3748 $\pm$ 0.188} & \re{0.2588 $\pm$ 0.2044} & \re{0.0687 $\pm$ 0.1124} & \re{0.2073 $\pm$ 0.1485} \\
\midrule
\textbf{8} & $\pmb{S_1}$ & 0.3045 $\pm$ 0.1826 & 0.0085 $\pm$ 0.056 & 0.2682 $\pm$ 0.161 & 0.3568 $\pm$ 0.1866 & 0.2908 $\pm$ 0.2024 & 0.0925 $\pm$ 0.1283 & 0.2202 $\pm$ 0.1528 \\
\textbf{8} & $\pmb{S_2}$ & \gr{0.2195 $\pm$ 0.183} & \gr{0.0081 $\pm$ 0.0543} & \gr{0.1962 $\pm$ 0.1527} & \gr{0.2582 $\pm$ 0.17} & \gr{0.2007 $\pm$ 0.1764} & \gr{0.056 $\pm$ 0.1006} & \gr{0.1565 $\pm$ 0.1395} \\
\textbf{8} & $\pmb{S_3}$ & 0.3127 $\pm$ 0.1774 & 0.0082 $\pm$ 0.0558 & 0.2665 $\pm$ 0.159 & 0.3592 $\pm$ 0.193 & 0.2874 $\pm$ 0.1931 & 0.0864 $\pm$ 0.1239 & 0.2201 $\pm$ 0.1504 \\
\textbf{8} & $\pmb{S_4}$ & \re{0.3726 $\pm$ 0.1628} & \re{0.0091 $\pm$ 0.0581} & \re{0.3179 $\pm$ 0.1617} & \re{0.4222 $\pm$ 0.2029} & \re{0.3307 $\pm$ 0.2055} & \re{0.1099 $\pm$ 0.1391} & \re{0.2604 $\pm$ 0.155} \\
\midrule
\textbf{10} & $\pmb{S_1}$ & 0.3744 $\pm$ 0.1886 & 0.0099 $\pm$ 0.0596 & 0.2802 $\pm$ 0.1604 & 0.3953 $\pm$ 0.1918 & 0.2891 $\pm$ 0.1973 & 0.1092 $\pm$ 0.1354 & 0.243 $\pm$ 0.1555 \\
\textbf{10} & $\pmb{S_2}$ & \gr{0.273 $\pm$ 0.2161} & \gr{0.0094 $\pm$ 0.0575} & \gr{0.1982 $\pm$ 0.1594} & \gr{0.2896 $\pm$ 0.1851} & \gr{0.2201 $\pm$ 0.1976} & \gr{0.0646 $\pm$ 0.1083} & \gr{0.1758 $\pm$ 0.154} \\
\textbf{10} & $\pmb{S_3}$ & 0.3767 $\pm$ 0.1917 & 0.0108 $\pm$ 0.0624 & 0.2819 $\pm$ 0.166 & 0.4192 $\pm$ 0.2003 & 0.3185 $\pm$ 0.2067 & 0.1053 $\pm$ 0.1351 & 0.2521 $\pm$ 0.1604 \\
\textbf{10} & $\pmb{S_4}$ & \re{0.4277 $\pm$ 0.1563} & \re{0.011 $\pm$ 0.0654} & \re{0.3329 $\pm$ 0.1666} & \re{0.4733 $\pm$ 0.1975} & \re{0.3605 $\pm$ 0.2198} & \re{0.134 $\pm$ 0.1469} & \re{0.2899 $\pm$ 0.1588} \\
\bottomrule
\end{tabular}
\end{sidewaystable}

\begin{sidewaystable}[htbp]
\renewcommand{\tabcolsep}{0.2cm}
\renewcommand{\arraystretch}{0.96}
\centering
\scriptsize
\caption{ILS results for each combination of parameters.} \label{tbl:app_upppm:results_ils}
\begin{tabular}{ccccccccc} 
\toprule
\multicolumn{2}{c}{\bf Parameters} & \multicolumn{6}{c}{\bf Event logs}  & \multirow{2}{*}{\textbf{Avg.}} \\
\cmidrule(lr){1-2} \cmidrule(lr){3-8}
$\pmb{k}$ & $\pmb{sel}$ & {\bf BPI12} & {\bf BPI13} & {\bf BPI14} & {\bf BPI15} & {\bf CoSeLoG} & {\bf TGN-Hospital}  \\
\midrule
\textbf{2} & $\pmb{S_1}$ & 0.0257 $\pm$ 0.0069 & \re{0.0015 $\pm$ 0.0004} & \re{0.0278 $\pm$ 0.0113} & \re{0.0157 $\pm$ 0.0076} & 0.023 $\pm$ 0.0104 & 0.0052 $\pm$ 0.0019 & 0.0168 $\pm$ 0.0126 \\
\textbf{2} & $\pmb{S_2}$ & \re{0.0264 $\pm$ 0.0072} & 0.0014 $\pm$ 0.0004 & \gr{0.0272 $\pm$ 0.0108} & 0.0148 $\pm$ 0.0068 & \re{0.0255 $\pm$ 0.0122} & 0.0051 $\pm$ 0.002 & \re{0.0172 $\pm$ 0.0131} \\
\textbf{2} & $\pmb{S_3}$ & \gr{0.0244 $\pm$ 0.0062} & \gr{0.0012 $\pm$ 0.0003} & 0.0277 $\pm$ 0.0114 & 0.0146 $\pm$ 0.0066 & 0.0237 $\pm$ 0.0105 & \re{0.0052 $\pm$ 0.002} & 0.0165 $\pm$ 0.0125 \\
\textbf{2} & $\pmb{S_4}$ & 0.0259 $\pm$ 0.0074 & 0.0014 $\pm$ 0.0004 & 0.0274 $\pm$ 0.0107 & \gr{0.0139 $\pm$ 0.0064} & \gr{0.0225 $\pm$ 0.0099} & \gr{0.005 $\pm$ 0.0019} & \gr{0.0164 $\pm$ 0.0123} \\
\midrule
\textbf{3} & $\pmb{S_1}$ & 0.0651 $\pm$ 0.0171 & 0.0019 $\pm$ 0.0004 & 0.0368 $\pm$ 0.0142 & 0.0256 $\pm$ 0.0121 & 0.0399 $\pm$ 0.0193 & 0.0084 $\pm$ 0.0058 & 0.0301 $\pm$ 0.0247 \\
\textbf{3} & $\pmb{S_2}$ & \gr{0.0463 $\pm$ 0.0129} & \gr{0.0018 $\pm$ 0.0005} & \gr{0.0329 $\pm$ 0.0122} & \gr{0.0219 $\pm$ 0.01} & \gr{0.0352 $\pm$ 0.0149} & \gr{0.0069 $\pm$ 0.0036} & \gr{0.0254 $\pm$ 0.0187} \\
\textbf{3} & $\pmb{S_3}$ & 0.0642 $\pm$ 0.016 & 0.0019 $\pm$ 0.0004 & 0.0392 $\pm$ 0.0156 & 0.0252 $\pm$ 0.0127 & 0.0386 $\pm$ 0.0188 & 0.0084 $\pm$ 0.0056 & 0.03 $\pm$ 0.0245 \\
\textbf{3} & $\pmb{S_4}$ & \re{0.0735 $\pm$ 0.0183} & \re{0.002 $\pm$ 0.0005} & \re{0.0395 $\pm$ 0.0151} & \re{0.0266 $\pm$ 0.0127} & \re{0.0411 $\pm$ 0.0189} & \re{0.0088 $\pm$ 0.006} & \re{0.0324 $\pm$ 0.0269} \\
\midrule
\textbf{4} & $\pmb{S_1}$ & 0.0714 $\pm$ 0.0217 & 0.0021 $\pm$ 0.0005 & 0.0401 $\pm$ 0.0152 & 0.0316 $\pm$ 0.0151 & 0.0501 $\pm$ 0.0234 & 0.0139 $\pm$ 0.0051 & 0.0357 $\pm$ 0.0278 \\
\textbf{4} & $\pmb{S_2}$ & \gr{0.0519 $\pm$ 0.0119} & \gr{0.0019 $\pm$ 0.0005} & \gr{0.0347 $\pm$ 0.0132} & \gr{0.022 $\pm$ 0.0103} & \gr{0.0356 $\pm$ 0.0162} & \gr{0.0103 $\pm$ 0.0035} & \gr{0.0261 $\pm$ 0.0204} \\
\textbf{4} & $\pmb{S_3}$ & 0.0744 $\pm$ 0.0233 & 0.0019 $\pm$ 0.0008 & 0.0405 $\pm$ 0.0156 & 0.0317 $\pm$ 0.0146 & 0.0442 $\pm$ 0.0199 & 0.0135 $\pm$ 0.0052 & 0.0349 $\pm$ 0.0276 \\
\textbf{4} & $\pmb{S_4}$ & \re{0.0903 $\pm$ 0.0254} & \re{0.0024 $\pm$ 0.0008} & \re{0.0422 $\pm$ 0.0156} & \re{0.039 $\pm$ 0.0179} & \re{0.0519 $\pm$ 0.0228} & \re{0.015 $\pm$ 0.0054} & \re{0.0408 $\pm$ 0.0325} \\
\midrule
\textbf{5} & $\pmb{S_1}$ & 0.0955 $\pm$ 0.0263 & \gr{0.0031 $\pm$ 0.0006} & 0.0453 $\pm$ 0.0173 & 0.0333 $\pm$ 0.0156 & 0.0461 $\pm$ 0.0195 & 0.0195 $\pm$ 0.0071 & 0.0407 $\pm$ 0.0327 \\
\textbf{5} & $\pmb{S_2}$ & \gr{0.0706 $\pm$ 0.0197} & 0.0031 $\pm$ 0.0007 & \gr{0.0382 $\pm$ 0.0143} & \gr{0.0223 $\pm$ 0.0107} & \gr{0.036 $\pm$ 0.0165} & \gr{0.0154 $\pm$ 0.0057} & \gr{0.031 $\pm$ 0.0244} \\
\textbf{5} & $\pmb{S_3}$ & 0.0953 $\pm$ 0.0265 & \re{0.0034 $\pm$ 0.0007} & 0.0453 $\pm$ 0.0173 & 0.0342 $\pm$ 0.0159 & 0.0504 $\pm$ 0.0226 & 0.0192 $\pm$ 0.0069 & 0.0418 $\pm$ 0.0332 \\
\textbf{5} & $\pmb{S_4}$ & \re{0.1045 $\pm$ 0.0263} & 0.0032 $\pm$ 0.0007 & \re{0.0495 $\pm$ 0.018} & \re{0.0429 $\pm$ 0.019} & \re{0.0552 $\pm$ 0.0233} & \re{0.0217 $\pm$ 0.008} & \re{0.0466 $\pm$ 0.0359} \\
\midrule
\textbf{8} & $\pmb{S_1}$ & 0.1087 $\pm$ 0.0261 & 0.0032 $\pm$ 0.0007 & 0.0501 $\pm$ 0.0196 & 0.0435 $\pm$ 0.0195 & 0.0636 $\pm$ 0.0287 & 0.0301 $\pm$ 0.0112 & 0.0506 $\pm$ 0.0377 \\
\textbf{8} & $\pmb{S_2}$ & \gr{0.0766 $\pm$ 0.0206} & \re{0.0035 $\pm$ 0.0006} & \gr{0.0397 $\pm$ 0.0154} & \gr{0.0263 $\pm$ 0.0126} & \gr{0.0402 $\pm$ 0.017} & \gr{0.0184 $\pm$ 0.0067} & \gr{0.0344 $\pm$ 0.0263} \\
\textbf{8} & $\pmb{S_3}$ & 0.1117 $\pm$ 0.0271 & \gr{0.0031 $\pm$ 0.0007} & 0.0496 $\pm$ 0.0187 & 0.0449 $\pm$ 0.0204 & 0.0593 $\pm$ 0.0259 & 0.0289 $\pm$ 0.0104 & 0.0501 $\pm$ 0.0379 \\
\textbf{8} & $\pmb{S_4}$ & \re{0.1291 $\pm$ 0.026} & 0.0034 $\pm$ 0.0007 & \re{0.0575 $\pm$ 0.0206} & \re{0.0589 $\pm$ 0.0252} & \re{0.0755 $\pm$ 0.0317} & \re{0.0346 $\pm$ 0.0126} & \re{0.0606 $\pm$ 0.0439} \\
\midrule
\textbf{10} & $\pmb{S_1}$ & 0.1369 $\pm$ 0.0366 & \gr{0.0035 $\pm$ 0.0006} & 0.053 $\pm$ 0.02 & 0.0501 $\pm$ 0.0229 & 0.0635 $\pm$ 0.0253 & 0.0354 $\pm$ 0.0129 & 0.0574 $\pm$ 0.0455 \\
\textbf{10} & $\pmb{S_2}$ & \gr{0.1066 $\pm$ 0.0318} & \re{0.0041 $\pm$ 0.001} & \gr{0.0409 $\pm$ 0.0161} & \gr{0.0325 $\pm$ 0.016} & \gr{0.051 $\pm$ 0.0242} & \gr{0.022 $\pm$ 0.0079} & \gr{0.0433 $\pm$ 0.0368} \\
\textbf{10} & $\pmb{S_3}$ & 0.1414 $\pm$ 0.037 & 0.0041 $\pm$ 0.0008 & 0.054 $\pm$ 0.0206 & 0.0558 $\pm$ 0.0254 & 0.0682 $\pm$ 0.0281 & 0.0333 $\pm$ 0.0121 & 0.0599 $\pm$ 0.0474 \\
\textbf{10} & $\pmb{S_4}$ & \re{0.1536 $\pm$ 0.0328} & 0.004 $\pm$ 0.0007 & \re{0.0608 $\pm$ 0.0218} & \re{0.0669 $\pm$ 0.0268} & \re{0.0821 $\pm$ 0.0304} & \re{0.0406 $\pm$ 0.0143} & \re{0.0687 $\pm$ 0.0506} \\
\bottomrule
\end{tabular}
\end{sidewaystable}

\begin{sidewaystable}[htbp]
\renewcommand{\tabcolsep}{0.2cm}
\renewcommand{\arraystretch}{0.96}
\centering
\scriptsize  
\caption{CS results for each combination of parameters.} \label{tbl:app_upppm:results_cs}
\begin{tabular}{ccccccccc} 
\toprule
\multicolumn{2}{c}{\bf Parameters} & \multicolumn{6}{c}{\bf Event logs}  & \multirow{2}{*}{\textbf{Avg.}} \\
\cmidrule(lr){1-2} \cmidrule(lr){3-8}
$\pmb{k}$ & $\pmb{sel}$ & {\bf BPI12} & {\bf BPI13} & {\bf BPI14} & {\bf BPI15} & {\bf CoSeLoG} & {\bf TGN-Hospital}  \\
\midrule
\textbf{2} & $\pmb{S_1}$ & \gr{0.9997 $\pm$ 0.001} & \re{0.9999 $\pm$ 0.0016} & \re{0.8353 $\pm$ 0.2284} & 0.955 $\pm$ 0.088 & \gr{0.9906 $\pm$ 0.0272} & \gr{0.9999 $\pm$ 0.0005} & \gr{0.9634 $\pm$ 0.0578} \\
\textbf{2} & $\pmb{S_2}$ & 0.9996 $\pm$ 0.0019 & \gr{0.9999 $\pm$ 0.0009} & 0.8355 $\pm$ 0.2285 & \re{0.9545 $\pm$ 0.0878} & 0.9902 $\pm$ 0.0301 & 0.9999 $\pm$ 0.0007 & 0.9633 $\pm$ 0.0583 \\
\textbf{2} & $\pmb{S_3}$ & 0.9997 $\pm$ 0.0012 & 0.9999 $\pm$ 0.0011 & \gr{0.8359 $\pm$ 0.2286} & \gr{0.9561 $\pm$ 0.0853} & \re{0.9884 $\pm$ 0.0335} & 0.9999 $\pm$ 0.0006 & 0.9633 $\pm$ 0.0584 \\
\textbf{2} & $\pmb{S_4}$ & \re{0.9992 $\pm$ 0.0037} & 0.9999 $\pm$ 0.0011 & 0.8356 $\pm$ 0.2289 & 0.9556 $\pm$ 0.0877 & 0.9896 $\pm$ 0.0301 & \re{0.9999 $\pm$ 0.0008} & \re{0.9633 $\pm$ 0.0587} \\
\midrule
\textbf{3} & $\pmb{S_1}$ & 0.9962 $\pm$ 0.0184 & \gr{0.9999 $\pm$ 0.0017} & 0.8313 $\pm$ 0.2269 & 0.9269 $\pm$ 0.0975 & \re{0.9613 $\pm$ 0.0912} & 0.9998 $\pm$ 0.0011 & \re{0.9526 $\pm$ 0.0728} \\
\textbf{3} & $\pmb{S_2}$ & \gr{0.9987 $\pm$ 0.0061} & 0.9998 $\pm$ 0.0021 & \gr{0.8341 $\pm$ 0.2272} & \gr{0.944 $\pm$ 0.0915} & \gr{0.9749 $\pm$ 0.0678} & \gr{0.9999 $\pm$ 0.0008} & \gr{0.9586 $\pm$ 0.0659} \\
\textbf{3} & $\pmb{S_3}$ & 0.9946 $\pm$ 0.0288 & \re{0.9998 $\pm$ 0.0026} & \re{0.8304 $\pm$ 0.2279} & 0.9282 $\pm$ 0.0974 & 0.9637 $\pm$ 0.0886 & 0.9998 $\pm$ 0.0012 & 0.9528 $\pm$ 0.0744 \\
\textbf{3} & $\pmb{S_4}$ & \re{0.9945 $\pm$ 0.0307} & 0.9999 $\pm$ 0.0022 & 0.8309 $\pm$ 0.2244 & \re{0.9253 $\pm$ 0.1044} & 0.9709 $\pm$ 0.0688 & \re{0.9998 $\pm$ 0.0013} & 0.9536 $\pm$ 0.072 \\
\midrule
\textbf{4} & $\pmb{S_1}$ & 0.9906 $\pm$ 0.0438 & 0.9998 $\pm$ 0.0025 & 0.8301 $\pm$ 0.2296 & 0.9055 $\pm$ 0.1152 & 0.9578 $\pm$ 0.1083 & 0.9997 $\pm$ 0.0017 & 0.9473 $\pm$ 0.0835 \\
\textbf{4} & $\pmb{S_2}$ & \gr{0.9915 $\pm$ 0.0427} & \gr{0.9998 $\pm$ 0.0022} & \gr{0.8334 $\pm$ 0.229} & \gr{0.9418 $\pm$ 0.094} & \gr{0.9708 $\pm$ 0.091} & \gr{0.9997 $\pm$ 0.0015} & \gr{0.9562 $\pm$ 0.0767} \\
\textbf{4} & $\pmb{S_3}$ & 0.9887 $\pm$ 0.0499 & 0.9998 $\pm$ 0.0026 & 0.8299 $\pm$ 0.2284 & 0.8978 $\pm$ 0.1213 & \re{0.9542 $\pm$ 0.0991} & 0.9996 $\pm$ 0.0018 & 0.945 $\pm$ 0.0839 \\
\textbf{4} & $\pmb{S_4}$ & \re{0.9879 $\pm$ 0.0497} & \re{0.9997 $\pm$ 0.0036} & \re{0.8296 $\pm$ 0.2291} & \re{0.8894 $\pm$ 0.1238} & 0.9553 $\pm$ 0.0894 & \re{0.9996 $\pm$ 0.0019} & \re{0.9436 $\pm$ 0.0829} \\
\midrule
\textbf{5} & $\pmb{S_1}$ & 0.9861 $\pm$ 0.0593 & \re{0.9995 $\pm$ 0.0054} & 0.8281 $\pm$ 0.2292 & 0.8966 $\pm$ 0.1327 & 0.9475 $\pm$ 0.1304 & \re{0.9994 $\pm$ 0.0023} & 0.9429 $\pm$ 0.0932 \\
\textbf{5} & $\pmb{S_2}$ & \gr{0.9886 $\pm$ 0.0508} & \gr{0.9996 $\pm$ 0.0039} & \gr{0.8329 $\pm$ 0.226} & \gr{0.9313 $\pm$ 0.1323} & \gr{0.9648 $\pm$ 0.1105} & \gr{0.9997 $\pm$ 0.002} & \gr{0.9528 $\pm$ 0.0876} \\
\textbf{5} & $\pmb{S_3}$ & 0.9881 $\pm$ 0.0604 & 0.9996 $\pm$ 0.0045 & 0.829 $\pm$ 0.2286 & 0.8872 $\pm$ 0.1375 & \re{0.9415 $\pm$ 0.1407} & 0.9995 $\pm$ 0.0018 & 0.9408 $\pm$ 0.0956 \\
\textbf{5} & $\pmb{S_4}$ & \re{0.9856 $\pm$ 0.0561} & 0.9995 $\pm$ 0.0049 & \re{0.8254 $\pm$ 0.2297} & \re{0.8593 $\pm$ 0.1509} & 0.9433 $\pm$ 0.1239 & 0.9994 $\pm$ 0.0022 & \re{0.9354 $\pm$ 0.0946} \\
\midrule
\textbf{8} & $\pmb{S_1}$ & 0.9737 $\pm$ 0.1118 & \gr{0.9995 $\pm$ 0.0058} & 0.8254 $\pm$ 0.2293 & 0.8683 $\pm$ 0.1403 & 0.931 $\pm$ 0.1467 & 0.9991 $\pm$ 0.0027 & 0.9328 $\pm$ 0.1061 \\
\textbf{8} & $\pmb{S_2}$ & \gr{0.987 $\pm$ 0.0503} & 0.9994 $\pm$ 0.005 & \gr{0.8312 $\pm$ 0.2253} & \gr{0.928 $\pm$ 0.1127} & \gr{0.9569 $\pm$ 0.0538} & \gr{0.9994 $\pm$ 0.0024} & \gr{0.9503 $\pm$ 0.0749} \\
\textbf{8} & $\pmb{S_3}$ & 0.9768 $\pm$ 0.0854 & 0.9994 $\pm$ 0.0055 & 0.8223 $\pm$ 0.2343 & 0.8619 $\pm$ 0.1413 & 0.927 $\pm$ 0.1395 & 0.9992 $\pm$ 0.0025 & 0.9311 $\pm$ 0.1014 \\
\textbf{8} & $\pmb{S_4}$ & \re{0.9653 $\pm$ 0.1162} & \re{0.9992 $\pm$ 0.0071} & \re{0.8205 $\pm$ 0.2277} & \re{0.816 $\pm$ 0.1693} & \re{0.9202 $\pm$ 0.1399} & \re{0.9988 $\pm$ 0.0032} & \re{0.92 $\pm$ 0.1106} \\
\midrule
\textbf{10} & $\pmb{S_1}$ & 0.9731 $\pm$ 0.0942 & 0.9991 $\pm$ 0.0092 & 0.8248 $\pm$ 0.2271 & 0.8418 $\pm$ 0.1784 & 0.9238 $\pm$ 0.1612 & 0.9989 $\pm$ 0.003 & 0.9269 $\pm$ 0.1122 \\
\textbf{10} & $\pmb{S_2}$ & \gr{0.9857 $\pm$ 0.0693} & \gr{0.9992 $\pm$ 0.0107} & \gr{0.8307 $\pm$ 0.2277} & \gr{0.9171 $\pm$ 0.1434} & \gr{0.9473 $\pm$ 0.1607} & \gr{0.9994 $\pm$ 0.0028} & \gr{0.9466 $\pm$ 0.1024} \\
\textbf{10} & $\pmb{S_3}$ & 0.9724 $\pm$ 0.0927 & 0.9988 $\pm$ 0.0103 & 0.822 $\pm$ 0.232 & 0.8211 $\pm$ 0.1892 & 0.9125 $\pm$ 0.1657 & 0.9988 $\pm$ 0.0033 & 0.9209 $\pm$ 0.1155 \\
\textbf{10} & $\pmb{S_4}$ & \re{0.9629 $\pm$ 0.1115} & \re{0.9986 $\pm$ 0.012} & \re{0.8189 $\pm$ 0.228} & \re{0.7655 $\pm$ 0.2174} & \re{0.8982 $\pm$ 0.1854} & \re{0.9983 $\pm$ 0.004} & \re{0.9071 $\pm$ 0.1264} \\
\bottomrule
\end{tabular}
\end{sidewaystable}


\newpage

\begin{figure}[t!]
\centering
\includegraphics[width=0.89\linewidth]{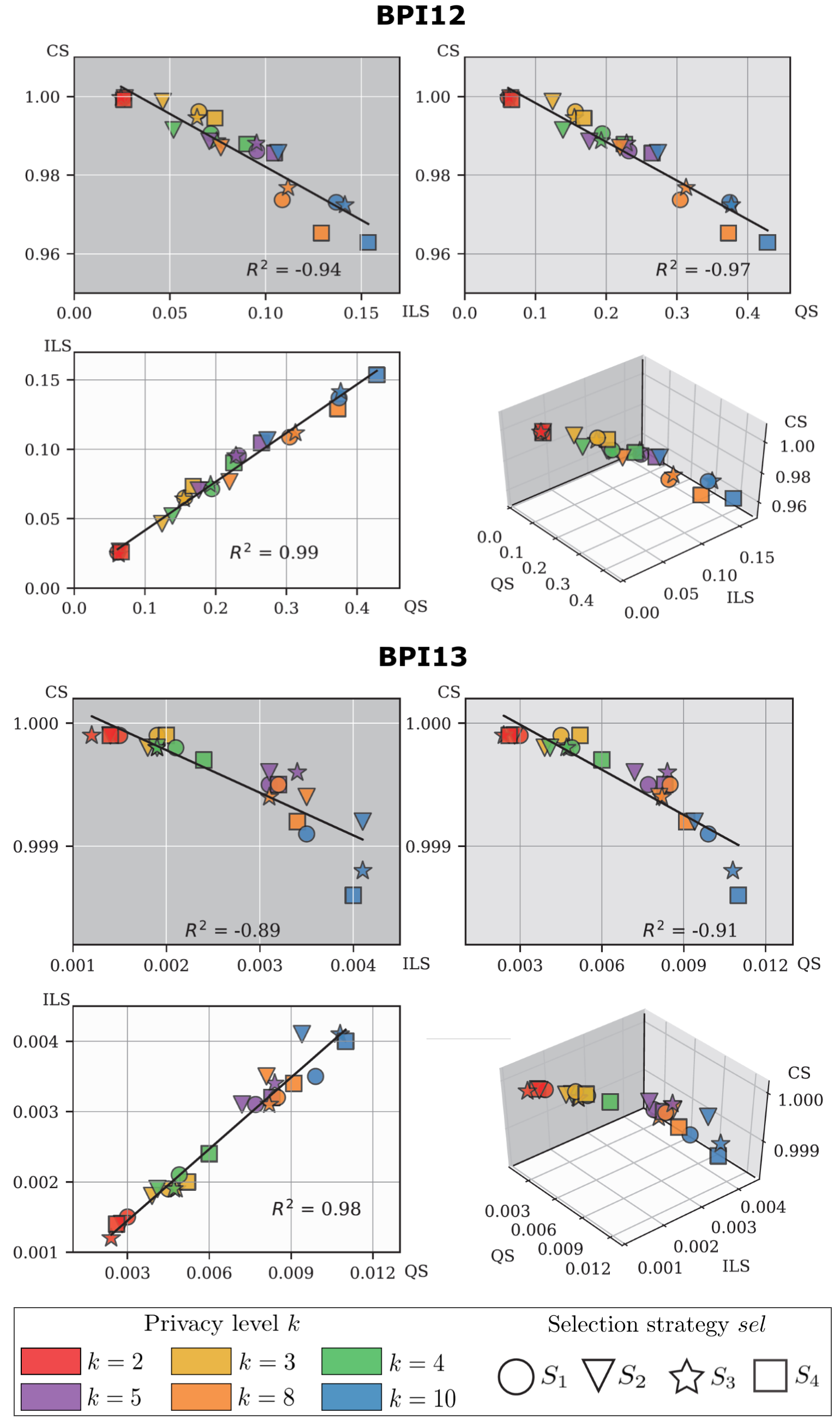}
\caption{Correlation between the QS, ILS and CS results using the BPI12 and BPI13 event logs (adapted from \cite{batista2021uniformization}).}
\label{fig:app_upppm:results_bpi12_bpi13}
\end{figure}

\begin{figure}[t!]
\centering
\includegraphics[width=0.89\linewidth]{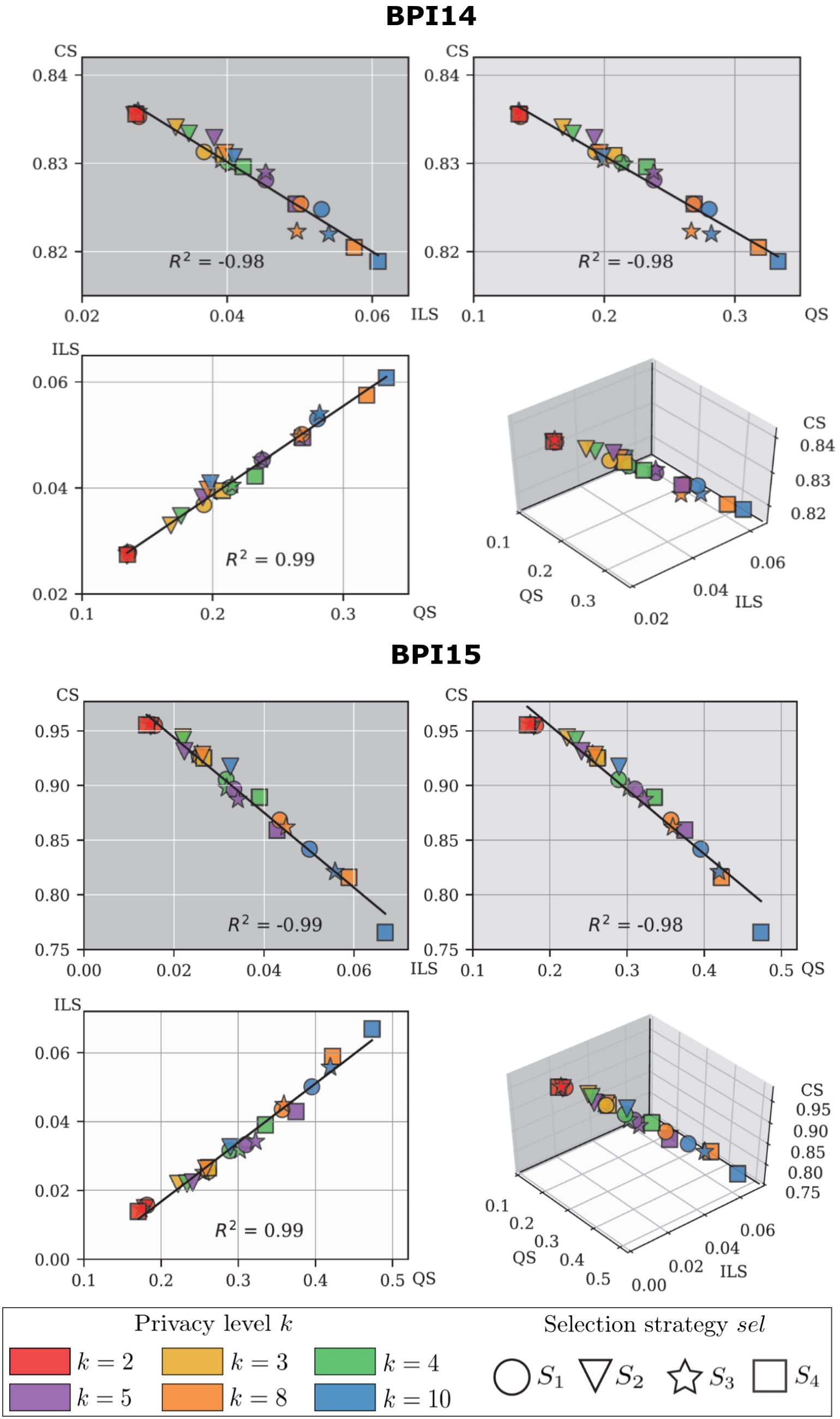}
\caption{Correlation between the QS, ILS and CS results using the BPI14 and BPI15 event logs (adapted from \cite{batista2021uniformization}).}
\label{fig:app_upppm:results_bpi14_bpi15}
\end{figure}

\begin{figure}[t!]
\centering
\includegraphics[width=0.89\linewidth]{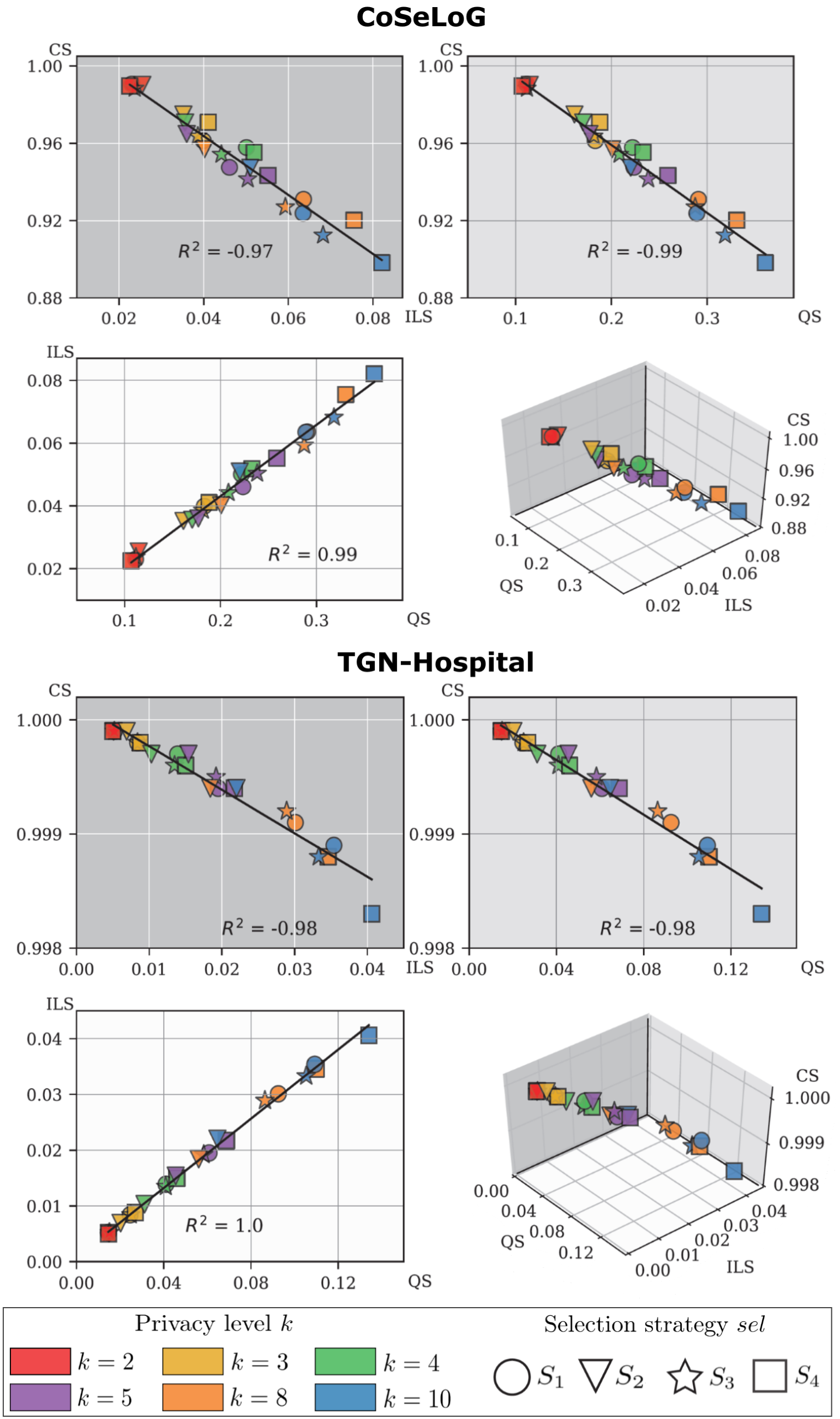}
\caption{Correlation between the QS, ILS and CS results using the CoSeLoG and TGN-Hospital event logs (adapted from \cite{batista2021uniformization}).}
\label{fig:app_upppm:results_coselog_tgn}
\end{figure}


\begin{figure}[t!]
\centering
\includegraphics[width=0.67\linewidth]{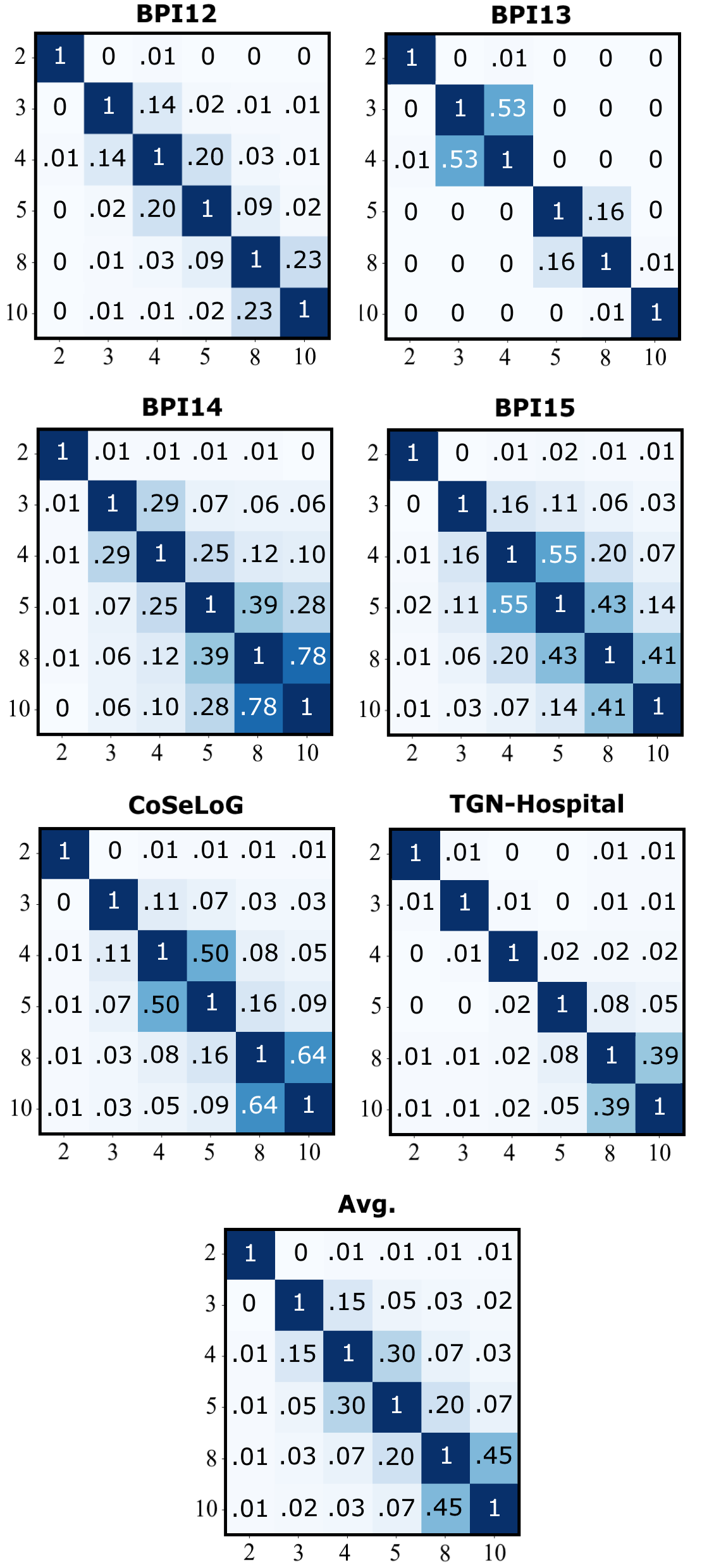}
\caption{Results of the $p$-values from the t-Tests according to the $k$ parameter in the QS results.}
\label{fig:app_upppm:ttest_qs_k}
\end{figure}

\begin{figure}[t!]
\centering
\includegraphics[width=\linewidth]{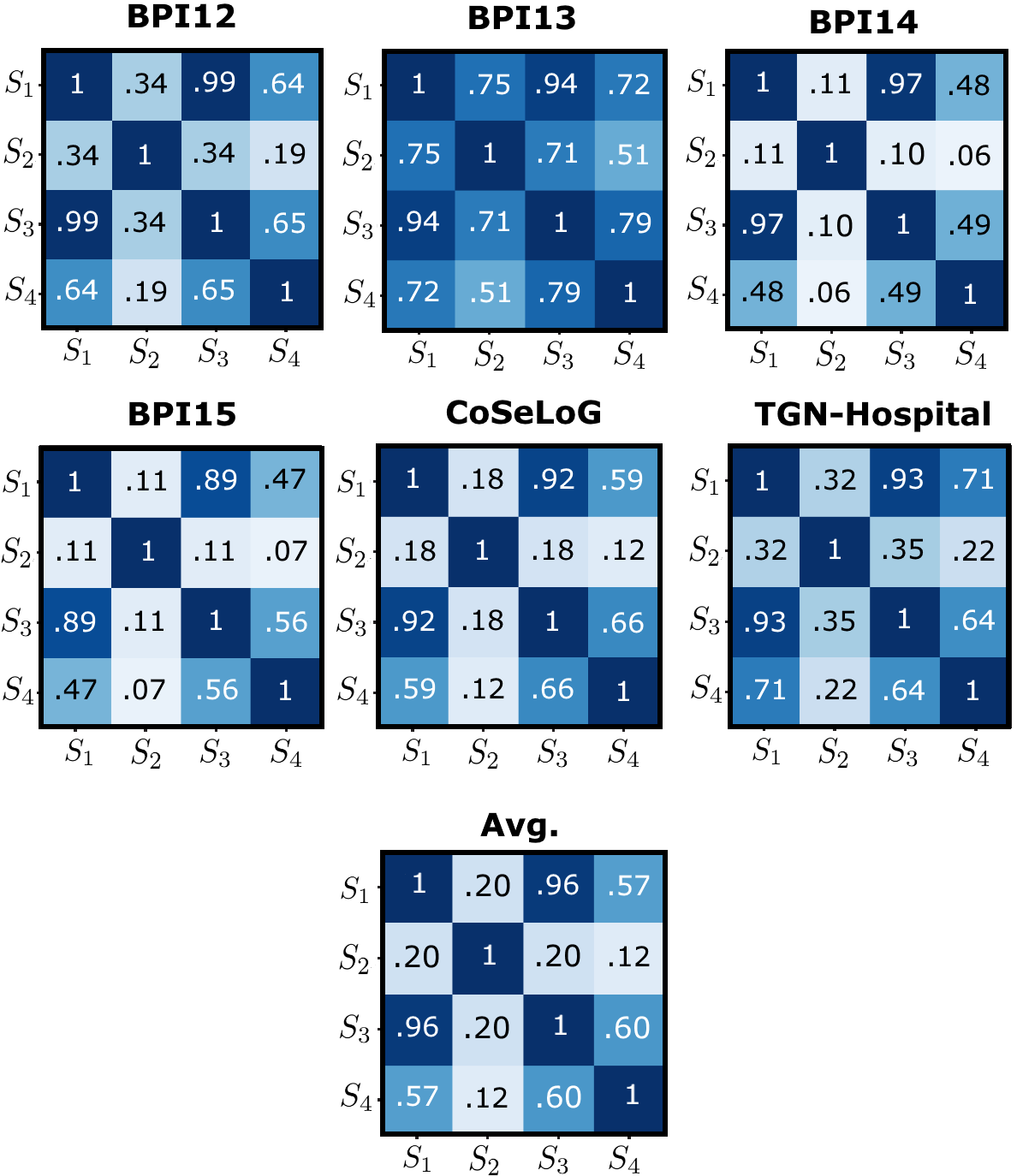}
\caption{Results of the $p$-values from the t-Tests according to the $sel$ parameter in the QS results.}
\label{fig:app_upppm:ttest_qs_strategy}
\end{figure}

\begin{figure}[t!]
\centering
\includegraphics[width=0.67\linewidth]{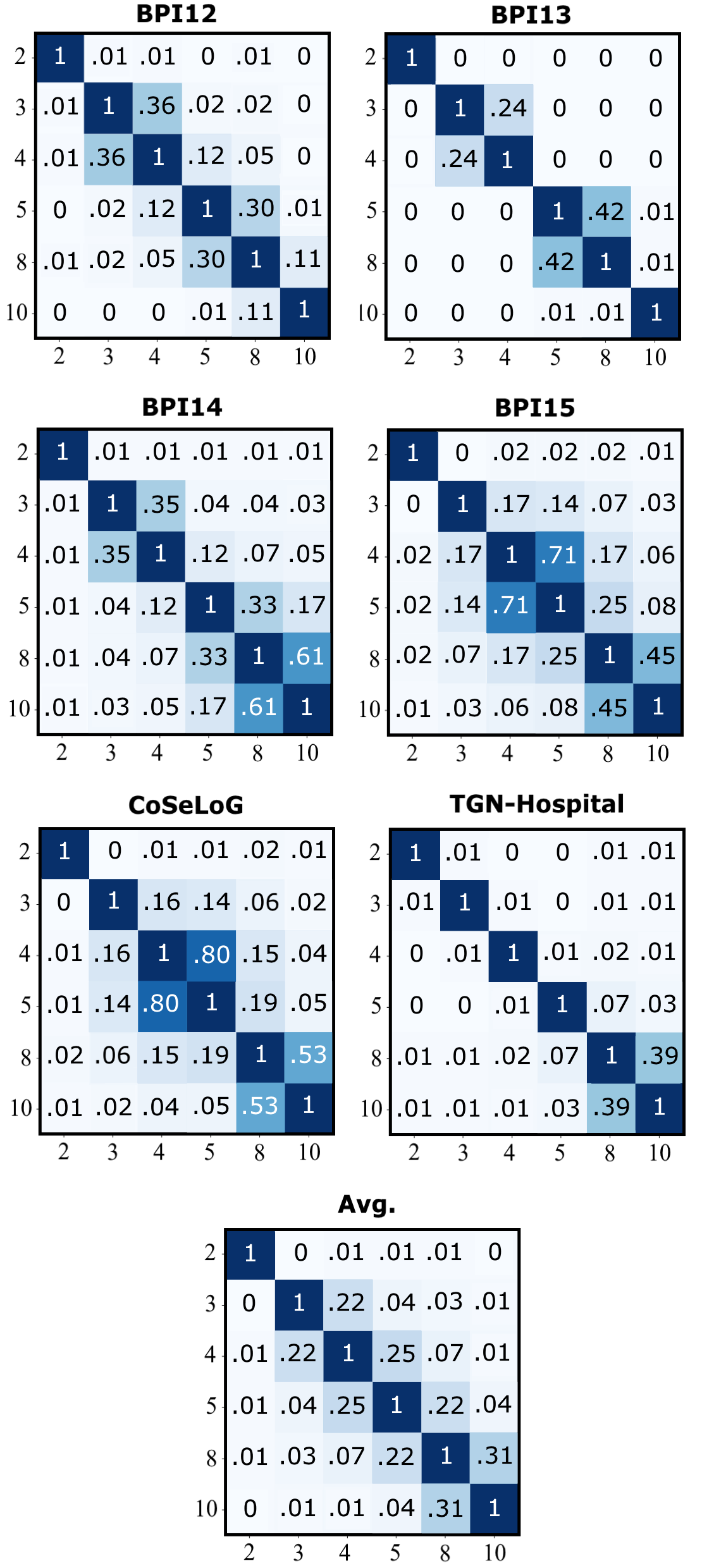}
\caption{Results of the $p$-values from the t-Tests according to the $k$ parameter in the ILS results.}
\label{fig:app_upppm:ttest_ils_k}
\end{figure}

\begin{figure}[t!]
\centering
\includegraphics[width=\linewidth]{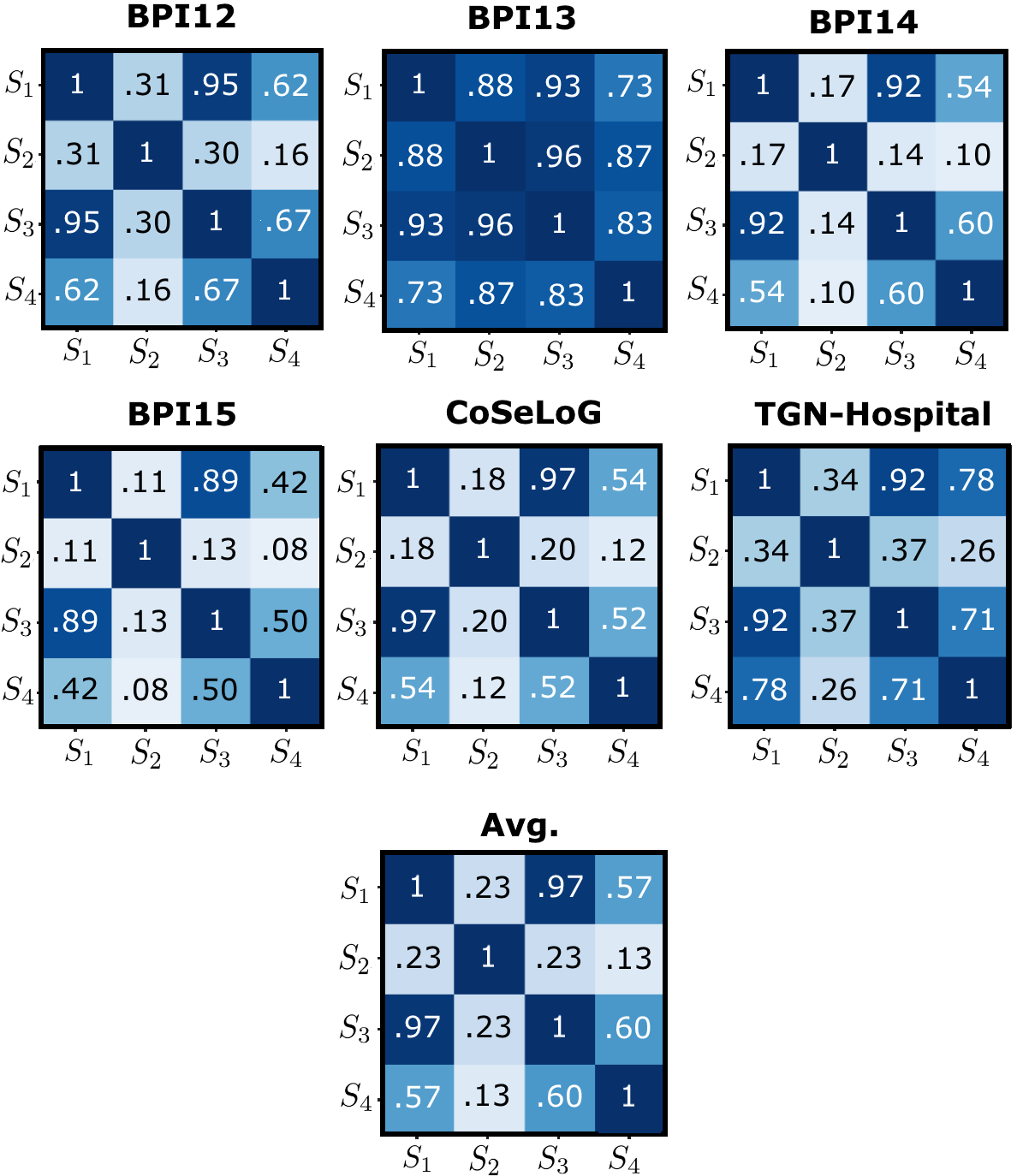}
\caption{Results of the $p$-values from the t-Tests according to the $sel$ parameter in the ILS results.}
\label{fig:app_upppm:ttest_ils_strategy}
\end{figure}

\begin{figure}[t!]
\centering
\includegraphics[width=0.67\linewidth]{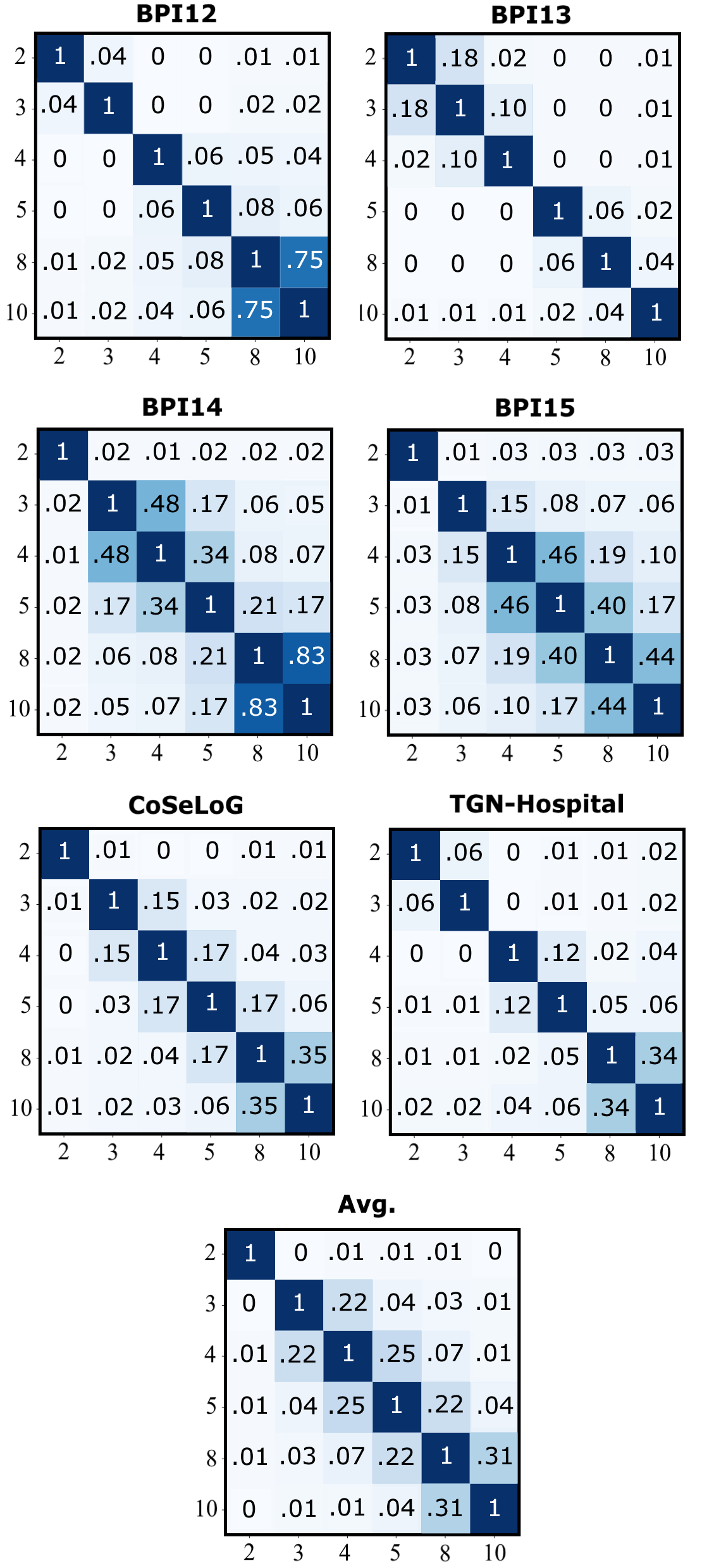}
\caption{Results of the $p$-values from the t-Tests according to the $k$ parameter in the CS results.}
\label{fig:app_upppm:ttest_cs_k}
\end{figure}

\begin{figure}[t!]
\centering
\includegraphics[width=\linewidth]{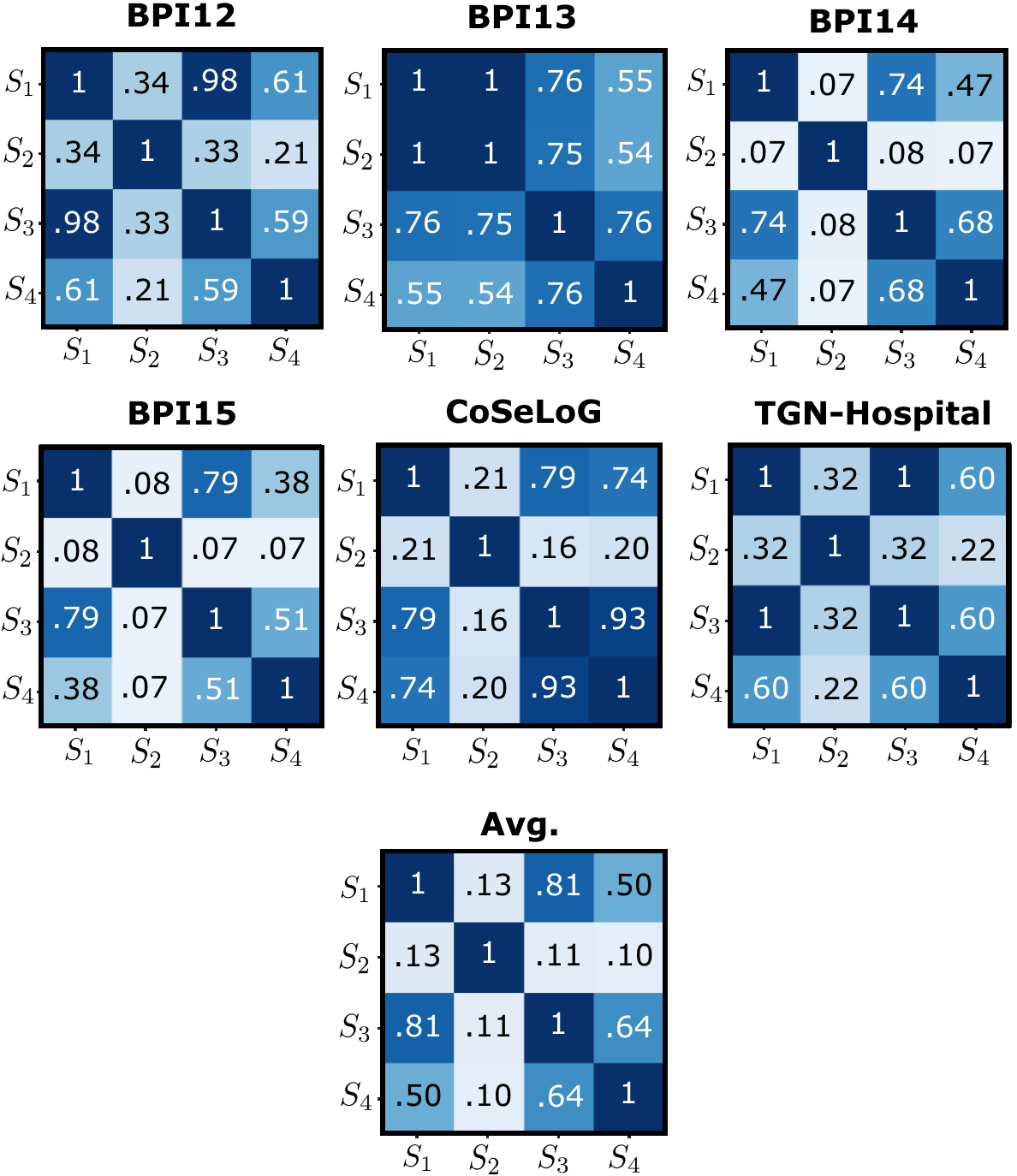}
\caption{Results of the $p$-values from the t-Tests according to the $sel$ parameter in the CS results.}
\label{fig:app_upppm:ttest_cs_strategy}
\end{figure}

\chapter{Results of \textit{k}-PPPM} \label{app:kpppm}

This appendix presents the details of all the results obtained from the execution of the proposed \textit{k}-PPPM method in Chapter \ref{chap:kpppm}.
More specifically, the following information can be found:

\begin{itemize}
\item Tables \ref{tbl:app_kpppm:results_qs_1} to \ref{tbl:app_kpppm:results_qs_4} and Tables \ref{tbl:app_kpppm:results_ils_1} to \ref{tbl:app_kpppm:results_ils_4} list the QS and ILS results obtained, respectively, for each execution of \textit{k}-PPPM with a certain combination of the parameters $k$ (privacy level), $clus$ (clustering algorithm) and $sim$ (similarity measure).
\item Figures \ref{fig:app_kpppm:results_bpi12_bpi13}, \ref{fig:app_kpppm:results_bpi14_bpi15} and \ref{fig:app_kpppm:results_coselog_tgn} depict the correlation between the QS and ILS results, for each execution of \textit{k}-PPPM with a certain combination of parameters.
\item Figures \ref{fig:app_kpppm:ttest_qs_k}, \ref{fig:app_kpppm:ttest_qs_clus} and \ref{fig:app_kpppm:ttest_qs_sim} illustrate the $p$-value results from the t-Tests according to the $k$, $clus$ and $sim$ parameters values, respectively, for the QS results. The $p$-value results for the ILS results are analogously illustrated in Figures \ref{fig:app_kpppm:ttest_ils_k}, \ref{fig:app_kpppm:ttest_ils_clus} and \ref{fig:app_kpppm:ttest_ils_sim}, respectively.
\end{itemize}

\begin{sidewaystable}[htbp]
\renewcommand{\tabcolsep}{0.2cm}
\renewcommand{\arraystretch}{0.96}
\centering
\scriptsize  
\caption{QS results for each combination of parameters (I).} \label{tbl:app_kpppm:results_qs_1}
\begin{tabular}{cccccccccc} 
\toprule
\multicolumn{3}{c}{\textbf{Parameters}} & \multicolumn{6}{c}{\textbf{Event logs}} & \multirow{2}{*}{\textbf{Avg.}} \\
\cmidrule(lr){1-3} \cmidrule(lr){4-9}
$\pmb{k}$ & $\pmb{clus}$ & $\pmb{sim}$ & {\bf BPI12} & {\bf BPI13} & {\bf BPI14} & {\bf BPI15} & {\bf CoSeLoG} & {\bf TGN-Hospital}  \\
\midrule
\textbf{2} & \textbf{MDAV} & \textbf{VEO} & 0.242 $\pm$ 0.145 & 0.12 $\pm$ 0.155 & 0.39 $\pm$ 0.113 & 0.467 $\pm$ 0.174 & 0.302 $\pm$ 0.17 & 0.289 $\pm$ 0.112 & 0.302 $\pm$ 0.145 \\
\textbf{2} & \textbf{MDAV} & \textbf{VR} & 0.248 $\pm$ 0.152 & 0.142 $\pm$ 0.173 & 0.385 $\pm$ 0.12 & 0.478 $\pm$ 0.177 & 0.326 $\pm$ 0.157 & 0.282 $\pm$ 0.11 & 0.31 $\pm$ 0.148 \\
\textbf{2} & \textbf{MDAV} & \textbf{WD} & 0.244 $\pm$ 0.145 & 0.131 $\pm$ 0.166 & 0.386 $\pm$ 0.118 & 0.48 $\pm$ 0.188 & 0.306 $\pm$ 0.173 & 0.291 $\pm$ 0.114 & 0.306 $\pm$ 0.151 \\
\textbf{2} & \textbf{MDAV} & \textbf{DC} & 0.24 $\pm$ 0.151 & \gr{0.113 $\pm$ 0.146} & 0.391 $\pm$ 0.122 & 0.488 $\pm$ 0.212 & 0.315 $\pm$ 0.163 & \gr{0.281 $\pm$ 0.123} & 0.303 $\pm$ 0.153 \\
\textbf{2} & \textbf{KM} & \textbf{VEO} & \gr{0.231 $\pm$ 0.14} & 0.118 $\pm$ 0.151 & 0.386 $\pm$ 0.124 & \gr{0.465 $\pm$ 0.19} & 0.313 $\pm$ 0.157 & 0.29 $\pm$ 0.109 & \gr{0.301 $\pm$ 0.145} \\
\textbf{2} & \textbf{KM} & \textbf{VR} & 0.25 $\pm$ 0.164 & 0.132 $\pm$ 0.167 & 0.386 $\pm$ 0.125 & 0.478 $\pm$ 0.176 & 0.3 $\pm$ 0.177 & 0.291 $\pm$ 0.104 & 0.306 $\pm$ 0.152 \\
\textbf{2} & \textbf{KM} & \textbf{WD} & 0.246 $\pm$ 0.139 & 0.124 $\pm$ 0.157 & \gr{0.382 $\pm$ 0.122} & 0.479 $\pm$ 0.179 & \gr{0.295 $\pm$ 0.162} & 0.289 $\pm$ 0.114 & 0.303 $\pm$ 0.145 \\
\textbf{2} & \textbf{KM} & \textbf{DC} & 0.249 $\pm$ 0.153 & 0.115 $\pm$ 0.151 & 0.403 $\pm$ 0.123 & 0.495 $\pm$ 0.216 & 0.336 $\pm$ 0.182 & 0.289 $\pm$ 0.116 & 0.316 $\pm$ 0.157 \\
\textbf{2} & \textbf{OKA} & \textbf{VEO} & 0.254 $\pm$ 0.163 & 0.139 $\pm$ 0.171 & 0.401 $\pm$ 0.131 & 0.487 $\pm$ 0.19 & 0.35 $\pm$ 0.209 & 0.3 $\pm$ 0.118 & 0.322 $\pm$ 0.164 \\
\textbf{2} & \textbf{OKA} & \textbf{VR} & 0.274 $\pm$ 0.182 & 0.152 $\pm$ 0.175 & 0.405 $\pm$ 0.142 & 0.492 $\pm$ 0.196 & 0.351 $\pm$ 0.215 & 0.307 $\pm$ 0.141 & 0.33 $\pm$ 0.175 \\
\textbf{2} & \textbf{OKA} & \textbf{WD} & 0.264 $\pm$ 0.173 & 0.146 $\pm$ 0.18 & 0.408 $\pm$ 0.147 & 0.498 $\pm$ 0.207 & 0.354 $\pm$ 0.213 & 0.308 $\pm$ 0.15 & 0.33 $\pm$ 0.178 \\
\textbf{2} & \textbf{OKA} & \textbf{DC} & 0.283 $\pm$ 0.181 & 0.152 $\pm$ 0.191 & 0.402 $\pm$ 0.138 & 0.516 $\pm$ 0.199 & \re{0.381 $\pm$ 0.225} & 0.313 $\pm$ 0.142 & 0.341 $\pm$ 0.179 \\
\textbf{2} & \textbf{BL} & \textbf{VEO} & 0.316 $\pm$ 0.169 & 0.279 $\pm$ 0.237 & 0.445 $\pm$ 0.129 & 0.514 $\pm$ 0.196 & 0.366 $\pm$ 0.194 & \re{0.417 $\pm$ 0.147} & 0.389 $\pm$ 0.179 \\
\textbf{2} & \textbf{BL} & \textbf{VR} & \re{0.321 $\pm$ 0.177} & \re{0.284 $\pm$ 0.232} & 0.447 $\pm$ 0.135 & 0.51 $\pm$ 0.197 & 0.37 $\pm$ 0.204 & 0.414 $\pm$ 0.154 & 0.391 $\pm$ 0.183 \\
\textbf{2} & \textbf{BL} & \textbf{WD} & 0.314 $\pm$ 0.171 & 0.283 $\pm$ 0.237 & \re{0.447 $\pm$ 0.137} & 0.518 $\pm$ 0.188 & 0.367 $\pm$ 0.2 & 0.414 $\pm$ 0.15 & 0.39 $\pm$ 0.18 \\
\textbf{2} & \textbf{BL} & \textbf{DC} & 0.319 $\pm$ 0.169 & 0.28 $\pm$ 0.237 & 0.446 $\pm$ 0.135 & \re{0.526 $\pm$ 0.198} & 0.366 $\pm$ 0.21 & 0.413 $\pm$ 0.147 & \re{0.392 $\pm$ 0.183} \\ 
\midrule
\textbf{3} & \textbf{MDAV} & \textbf{VEO} & 0.302 $\pm$ 0.186 & 0.161 $\pm$ 0.174 & \gr{0.431 $\pm$ 0.109} & \gr{0.571 $\pm$ 0.138} & \gr{0.376 $\pm$ 0.168} & 0.348 $\pm$ 0.113 & \gr{0.365 $\pm$ 0.148} \\
\textbf{3} & \textbf{MDAV} & \textbf{VR} & 0.303 $\pm$ 0.178 & 0.166 $\pm$ 0.18 & 0.445 $\pm$ 0.116 & 0.578 $\pm$ 0.169 & 0.415 $\pm$ 0.175 & 0.348 $\pm$ 0.112 & 0.376 $\pm$ 0.155 \\
\textbf{3} & \textbf{MDAV} & \textbf{WD} & 0.313 $\pm$ 0.183 & 0.158 $\pm$ 0.175 & 0.44 $\pm$ 0.124 & 0.574 $\pm$ 0.166 & 0.409 $\pm$ 0.124 & 0.357 $\pm$ 0.125 & 0.375 $\pm$ 0.149 \\
\textbf{3} & \textbf{MDAV} & \textbf{DC} & 0.305 $\pm$ 0.166 & \gr{0.147 $\pm$ 0.168} & 0.434 $\pm$ 0.131 & 0.596 $\pm$ 0.163 & 0.415 $\pm$ 0.154 & \gr{0.34 $\pm$ 0.121} & 0.374 $\pm$ 0.15 \\
\textbf{3} & \textbf{KM} & \textbf{VEO} & 0.308 $\pm$ 0.163 & 0.15 $\pm$ 0.167 & 0.436 $\pm$ 0.118 & 0.574 $\pm$ 0.136 & 0.399 $\pm$ 0.164 & 0.346 $\pm$ 0.114 & 0.369 $\pm$ 0.144 \\
\textbf{3} & \textbf{KM} & \textbf{VR} & 0.308 $\pm$ 0.164 & 0.165 $\pm$ 0.179 & 0.443 $\pm$ 0.108 & 0.563 $\pm$ 0.146 & 0.388 $\pm$ 0.168 & 0.347 $\pm$ 0.114 & 0.369 $\pm$ 0.147 \\
\textbf{3} & \textbf{KM} & \textbf{WD} & 0.321 $\pm$ 0.173 & 0.16 $\pm$ 0.18 & 0.433 $\pm$ 0.117 & 0.574 $\pm$ 0.164 & 0.422 $\pm$ 0.138 & 0.355 $\pm$ 0.114 & 0.378 $\pm$ 0.148 \\
\textbf{3} & \textbf{KM} & \textbf{DC} & 0.311 $\pm$ 0.17 & 0.149 $\pm$ 0.175 & 0.45 $\pm$ 0.12 & 0.573 $\pm$ 0.159 & 0.395 $\pm$ 0.183 & 0.344 $\pm$ 0.117 & 0.37 $\pm$ 0.154 \\
& & \multicolumn{8}{l}{\textit{Continued on next page}}
\end{tabular}
\end{sidewaystable}

\begin{sidewaystable}[htbp]
\renewcommand{\tabcolsep}{0.2cm}
\renewcommand{\arraystretch}{0.96}
\centering
\scriptsize  
\caption{QS results for each combination of parameters (II).} \label{tbl:app_kpppm:results_qs_2}
\begin{tabular}{cccccccccc} 
\toprule
\multicolumn{3}{c}{\textbf{Parameters}} & \multicolumn{6}{c}{\textbf{Event logs}} & \multirow{2}{*}{\textbf{Avg.}} \\
\cmidrule(lr){1-3} \cmidrule(lr){4-9}
$\pmb{k}$ & $\pmb{clus}$ & $\pmb{sim}$ & {\bf BPI12} & {\bf BPI13} & {\bf BPI14} & {\bf BPI15} & {\bf CoSeLoG} & {\bf TGN-Hospital}  \\
\midrule
\textbf{3} & \textbf{OKA} & \textbf{VEO} & 0.371 $\pm$ 0.225 & 0.167 $\pm$ 0.184 & 0.457 $\pm$ 0.125 & 0.58 $\pm$ 0.183 & 0.413 $\pm$ 0.205 & 0.354 $\pm$ 0.128 & 0.39 $\pm$ 0.175 \\
\textbf{3} & \textbf{OKA} & \textbf{VR} & 0.319 $\pm$ 0.189 & 0.192 $\pm$ 0.197 & 0.455 $\pm$ 0.142 & 0.581 $\pm$ 0.191 & 0.462 $\pm$ 0.203 & 0.364 $\pm$ 0.138 & 0.395 $\pm$ 0.176 \\
\textbf{3} & \textbf{OKA} & \textbf{WD} & 0.313 $\pm$ 0.205 & 0.181 $\pm$ 0.2 & 0.457 $\pm$ 0.126 & 0.589 $\pm$ 0.189 & 0.412 $\pm$ 0.189 & 0.373 $\pm$ 0.144 & 0.388 $\pm$ 0.175 \\
\textbf{3} & \textbf{OKA} & \textbf{DC} & \gr{0.294 $\pm$ 0.187} & 0.18 $\pm$ 0.199 & 0.473 $\pm$ 0.131 & 0.577 $\pm$ 0.172 & 0.441 $\pm$ 0.19 & 0.373 $\pm$ 0.145 & 0.39 $\pm$ 0.171 \\
\textbf{3} & \textbf{BL} & \textbf{VEO} & \re{0.401 $\pm$ 0.177} & \re{0.357 $\pm$ 0.232} & 0.499 $\pm$ 0.119 & 0.591 $\pm$ 0.183 & 0.43 $\pm$ 0.187 & 0.491 $\pm$ 0.136 & 0.461 $\pm$ 0.173 \\
\textbf{3} & \textbf{BL} & \textbf{VR} & 0.396 $\pm$ 0.181 & 0.354 $\pm$ 0.233 & \re{0.499 $\pm$ 0.124} & \re{0.605 $\pm$ 0.191} & 0.462 $\pm$ 0.192 & 0.493 $\pm$ 0.132 & 0.468 $\pm$ 0.175 \\
\textbf{3} & \textbf{BL} & \textbf{WD} & 0.396 $\pm$ 0.184 & 0.35 $\pm$ 0.236 & 0.499 $\pm$ 0.119 & 0.605 $\pm$ 0.183 & \re{0.472 $\pm$ 0.168} & \re{0.493 $\pm$ 0.134} & \re{0.469 $\pm$ 0.171} \\
\textbf{3} & \textbf{BL} & \textbf{DC} & 0.398 $\pm$ 0.174 & 0.351 $\pm$ 0.237 & 0.495 $\pm$ 0.127 & 0.589 $\pm$ 0.187 & 0.469 $\pm$ 0.198 & 0.491 $\pm$ 0.131 & 0.465 $\pm$ 0.176 \\ 
\midrule
\textbf{4} & \textbf{MDAV} & \textbf{VEO} & 0.342 $\pm$ 0.187 & 0.171 $\pm$ 0.182 & 0.465 $\pm$ 0.104 & 0.61 $\pm$ 0.131 & 0.461 $\pm$ 0.156 & 0.382 $\pm$ 0.115 & 0.405 $\pm$ 0.146 \\
\textbf{4} & \textbf{MDAV} & \textbf{VR} & 0.325 $\pm$ 0.171 & 0.184 $\pm$ 0.184 & 0.459 $\pm$ 0.119 & 0.601 $\pm$ 0.168 & \gr{0.414 $\pm$ 0.18} & 0.396 $\pm$ 0.119 & \gr{0.396 $\pm$ 0.157} \\
\textbf{4} & \textbf{MDAV} & \textbf{WD} & 0.358 $\pm$ 0.2 & 0.192 $\pm$ 0.192 & 0.463 $\pm$ 0.126 & 0.634 $\pm$ 0.171 & 0.417 $\pm$ 0.164 & 0.388 $\pm$ 0.124 & 0.409 $\pm$ 0.163 \\
\textbf{4} & \textbf{MDAV} & \textbf{DC} & 0.327 $\pm$ 0.178 & 0.168 $\pm$ 0.186 & 0.471 $\pm$ 0.125 & 0.62 $\pm$ 0.156 & 0.47 $\pm$ 0.143 & 0.393 $\pm$ 0.12 & 0.401 $\pm$ 0.151 \\
\textbf{4} & \textbf{KM} & \textbf{VEO} & 0.333 $\pm$ 0.172 & 0.178 $\pm$ 0.178 & \gr{0.458 $\pm$ 0.111} & \gr{0.596 $\pm$ 0.154} & 0.443 $\pm$ 0.158 & \gr{0.377 $\pm$ 0.118} & 0.398 $\pm$ 0.149 \\
\textbf{4} & \textbf{KM} & \textbf{VR} & \gr{0.323 $\pm$ 0.168} & 0.182 $\pm$ 0.181 & 0.467 $\pm$ 0.113 & 0.604 $\pm$ 0.157 & 0.432 $\pm$ 0.174 & 0.388 $\pm$ 0.114 & 0.399 $\pm$ 0.151 \\
\textbf{4} & \textbf{KM} & \textbf{WD} & 0.36 $\pm$ 0.2 & 0.176 $\pm$ 0.183 & 0.472 $\pm$ 0.107 & 0.618 $\pm$ 0.161 & 0.431 $\pm$ 0.166 & 0.392 $\pm$ 0.117 & 0.408 $\pm$ 0.156 \\
\textbf{4} & \textbf{KM} & \textbf{DC} & 0.34 $\pm$ 0.192 & \gr{0.166 $\pm$ 0.181} & 0.48 $\pm$ 0.104 & 0.628 $\pm$ 0.153 & 0.47 $\pm$ 0.193 & 0.382 $\pm$ 0.118 & 0.403 $\pm$ 0.157 \\
\textbf{4} & \textbf{OKA} & \textbf{VEO} & 0.37 $\pm$ 0.208 & 0.244 $\pm$ 0.21 & 0.478 $\pm$ 0.133 & 0.618 $\pm$ 0.173 & 0.475 $\pm$ 0.2 & 0.397 $\pm$ 0.144 & 0.43 $\pm$ 0.178 \\
\textbf{4} & \textbf{OKA} & \textbf{VR} & 0.345 $\pm$ 0.196 & 0.227 $\pm$ 0.214 & 0.485 $\pm$ 0.129 & 0.624 $\pm$ 0.165 & 0.446 $\pm$ 0.196 & 0.401 $\pm$ 0.126 & 0.421 $\pm$ 0.171 \\
\textbf{4} & \textbf{OKA} & \textbf{WD} & 0.352 $\pm$ 0.205 & 0.208 $\pm$ 0.206 & 0.492 $\pm$ 0.135 & 0.639 $\pm$ 0.16 & 0.458 $\pm$ 0.185 & 0.399 $\pm$ 0.142 & 0.425 $\pm$ 0.172 \\
\textbf{4} & \textbf{OKA} & \textbf{DC} & 0.34 $\pm$ 0.201 & 0.229 $\pm$ 0.213 & 0.495 $\pm$ 0.126 & 0.639 $\pm$ 0.132 & 0.489 $\pm$ 0.2 & 0.393 $\pm$ 0.133 & 0.431 $\pm$ 0.168 \\
\textbf{4} & \textbf{BL} & \textbf{VEO} & \re{0.439 $\pm$ 0.171} & 0.389 $\pm$ 0.218 & \re{0.523 $\pm$ 0.113} & 0.635 $\pm$ 0.175 & 0.478 $\pm$ 0.194 & 0.531 $\pm$ 0.126 & 0.499 $\pm$ 0.166 \\
\textbf{4} & \textbf{BL} & \textbf{VR} & 0.433 $\pm$ 0.181 & \re{0.395 $\pm$ 0.22} & 0.521 $\pm$ 0.123 & 0.631 $\pm$ 0.185 & \re{0.495 $\pm$ 0.187} & 0.53 $\pm$ 0.123 & \re{0.501 $\pm$ 0.17} \\
\textbf{4} & \textbf{BL} & \textbf{WD} & 0.432 $\pm$ 0.18 & 0.39 $\pm$ 0.218 & 0.516 $\pm$ 0.129 & 0.633 $\pm$ 0.178 & 0.49 $\pm$ 0.193 & 0.531 $\pm$ 0.128 & 0.499 $\pm$ 0.171 \\
\textbf{4} & \textbf{BL} & \textbf{DC} & 0.427 $\pm$ 0.179 & 0.387 $\pm$ 0.222 & 0.519 $\pm$ 0.126 & \re{0.646 $\pm$ 0.169} & 0.493 $\pm$ 0.192 & \re{0.534 $\pm$ 0.126} & 0.501 $\pm$ 0.169 \\ 
\midrule
& & \multicolumn{8}{l}{\textit{Continued on next page}}
\end{tabular}
\end{sidewaystable}

\begin{sidewaystable}[htbp]
\renewcommand{\tabcolsep}{0.2cm}
\renewcommand{\arraystretch}{0.96}
\centering
\scriptsize  
\caption{QS results for each combination of parameters (III).} \label{tbl:app_kpppm:results_qs_3}
\begin{tabular}{cccccccccc} 
\toprule
\multicolumn{3}{c}{\textbf{Parameters}} & \multicolumn{6}{c}{\textbf{Event logs}} & \multirow{2}{*}{\textbf{Avg.}} \\
\cmidrule(lr){1-3} \cmidrule(lr){4-9}
$\pmb{k}$ & $\pmb{clus}$ & $\pmb{sim}$ & {\bf BPI12} & {\bf BPI13} & {\bf BPI14} & {\bf BPI15} & {\bf CoSeLoG} & {\bf TGN-Hospital}  \\
\midrule
\textbf{5} & \textbf{MDAV} & \textbf{VEO} & 0.364 $\pm$ 0.193 & 0.19 $\pm$ 0.188 & 0.47 $\pm$ 0.107 & 0.644 $\pm$ 0.143 & 0.511 $\pm$ 0.138 & \gr{0.4 $\pm$ 0.114} & 0.43 $\pm$ 0.147 \\
\textbf{5} & \textbf{MDAV} & \textbf{VR} & 0.356 $\pm$ 0.191 & 0.197 $\pm$ 0.193 & 0.477 $\pm$ 0.12 & 0.633 $\pm$ 0.149 & 0.467 $\pm$ 0.16 & 0.402 $\pm$ 0.117 & \gr{0.422 $\pm$ 0.155} \\
\textbf{5} & \textbf{MDAV} & \textbf{WD} & 0.379 $\pm$ 0.197 & 0.204 $\pm$ 0.194 & 0.492 $\pm$ 0.107 & 0.642 $\pm$ 0.158 & 0.482 $\pm$ 0.155 & 0.417 $\pm$ 0.134 & 0.436 $\pm$ 0.157 \\
\textbf{5} & \textbf{MDAV} & \textbf{DC} & 0.349 $\pm$ 0.177 & \gr{0.168 $\pm$ 0.182} & 0.489 $\pm$ 0.126 & 0.642 $\pm$ 0.145 & 0.515 $\pm$ 0.136 & 0.401 $\pm$ 0.129 & 0.436 $\pm$ 0.149 \\
\textbf{5} & \textbf{KM} & \textbf{VEO} & 0.367 $\pm$ 0.178 & 0.183 $\pm$ 0.183 & \gr{0.469 $\pm$ 0.102} & 0.633 $\pm$ 0.132 & 0.487 $\pm$ 0.15 & 0.408 $\pm$ 0.109 & 0.425 $\pm$ 0.142 \\
\textbf{5} & \textbf{KM} & \textbf{VR} & 0.369 $\pm$ 0.18 & 0.199 $\pm$ 0.198 & 0.49 $\pm$ 0.113 & \gr{0.622 $\pm$ 0.158} & 0.489 $\pm$ 0.179 & 0.409 $\pm$ 0.106 & 0.43 $\pm$ 0.156 \\
\textbf{5} & \textbf{KM} & \textbf{WD} & 0.383 $\pm$ 0.198 & 0.206 $\pm$ 0.201 & 0.501 $\pm$ 0.104 & 0.652 $\pm$ 0.135 & 0.491 $\pm$ 0.119 & 0.421 $\pm$ 0.118 & 0.435 $\pm$ 0.146 \\
\textbf{5} & \textbf{KM} & \textbf{DC} & 0.354 $\pm$ 0.18 & 0.184 $\pm$ 0.189 & 0.495 $\pm$ 0.125 & 0.646 $\pm$ 0.151 & 0.505 $\pm$ 0.167 & 0.421 $\pm$ 0.125 & 0.431 $\pm$ 0.156 \\
\textbf{5} & \textbf{OKA} & \textbf{VEO} & 0.372 $\pm$ 0.205 & 0.237 $\pm$ 0.21 & 0.5 $\pm$ 0.124 & 0.633 $\pm$ 0.14 & \gr{0.431 $\pm$ 0.176} & 0.423 $\pm$ 0.157 & 0.433 $\pm$ 0.169 \\
\textbf{5} & \textbf{OKA} & \textbf{VR} & 0.356 $\pm$ 0.201 & 0.268 $\pm$ 0.23 & 0.505 $\pm$ 0.119 & 0.647 $\pm$ 0.135 & 0.469 $\pm$ 0.183 & 0.425 $\pm$ 0.132 & 0.445 $\pm$ 0.167 \\
\textbf{5} & \textbf{OKA} & \textbf{WD} & 0.359 $\pm$ 0.183 & 0.256 $\pm$ 0.217 & 0.488 $\pm$ 0.115 & 0.654 $\pm$ 0.164 & 0.529 $\pm$ 0.152 & 0.435 $\pm$ 0.159 & 0.454 $\pm$ 0.165 \\
\textbf{5} & \textbf{OKA} & \textbf{DC} & \gr{0.34 $\pm$ 0.203} & 0.215 $\pm$ 0.216 & 0.517 $\pm$ 0.128 & 0.651 $\pm$ 0.136 & 0.514 $\pm$ 0.175 & 0.428 $\pm$ 0.139 & 0.448 $\pm$ 0.166 \\
\textbf{5} & \textbf{BL} & \textbf{VEO} & 0.467 $\pm$ 0.17 & 0.404 $\pm$ 0.224 & \re{0.545 $\pm$ 0.108} & 0.667 $\pm$ 0.161 & 0.536 $\pm$ 0.204 & 0.546 $\pm$ 0.122 & \re{0.528 $\pm$ 0.165} \\
\textbf{5} & \textbf{BL} & \textbf{VR} & 0.468 $\pm$ 0.17 & 0.399 $\pm$ 0.224 & 0.537 $\pm$ 0.112 & \re{0.671 $\pm$ 0.154} & 0.534 $\pm$ 0.189 & \re{0.548 $\pm$ 0.123} & 0.526 $\pm$ 0.162 \\
\textbf{5} & \textbf{BL} & \textbf{WD} & 0.465 $\pm$ 0.175 & \re{0.406 $\pm$ 0.217} & 0.525 $\pm$ 0.132 & 0.666 $\pm$ 0.162 & \re{0.537 $\pm$ 0.19} & 0.548 $\pm$ 0.116 & 0.524 $\pm$ 0.165 \\
\textbf{5} & \textbf{BL} & \textbf{DC} & \re{0.471 $\pm$ 0.171} & 0.403 $\pm$ 0.221 & 0.53 $\pm$ 0.123 & 0.661 $\pm$ 0.159 & 0.506 $\pm$ 0.195 & 0.545 $\pm$ 0.123 & 0.519 $\pm$ 0.165 \\ 
\midrule
\textbf{10} & \textbf{MDAV} & \textbf{VEO} & 0.445 $\pm$ 0.192 & 0.239 $\pm$ 0.203 & \gr{0.515 $\pm$ 0.119} & 0.698 $\pm$ 0.126 & 0.542 $\pm$ 0.129 & 0.456 $\pm$ 0.129 & 0.483 $\pm$ 0.15 \\
\textbf{10} & \textbf{MDAV} & \textbf{VR} & 0.463 $\pm$ 0.199 & 0.246 $\pm$ 0.211 & 0.528 $\pm$ 0.101 & \gr{0.677 $\pm$ 0.125} & 0.51 $\pm$ 0.11 & 0.473 $\pm$ 0.113 & 0.483 $\pm$ 0.143 \\
\textbf{10} & \textbf{MDAV} & \textbf{WD} & 0.422 $\pm$ 0.201 & 0.26 $\pm$ 0.208 & 0.545 $\pm$ 0.107 & 0.685 $\pm$ 0.132 & 0.546 $\pm$ 0.177 & 0.468 $\pm$ 0.128 & 0.488 $\pm$ 0.159 \\
\textbf{10} & \textbf{MDAV} & \textbf{DC} & 0.41 $\pm$ 0.199 & \gr{0.207 $\pm$ 0.189} & 0.542 $\pm$ 0.118 & 0.696 $\pm$ 0.133 & 0.546 $\pm$ 0.173 & 0.45 $\pm$ 0.126 & 0.478 $\pm$ 0.156 \\
\textbf{10} & \textbf{KM} & \textbf{VEO} & 0.433 $\pm$ 0.18 & 0.23 $\pm$ 0.195 & 0.53 $\pm$ 0.097 & 0.679 $\pm$ 0.117 & 0.525 $\pm$ 0.122 & 0.459 $\pm$ 0.115 & \gr{0.476 $\pm$ 0.138} \\
\textbf{10} & \textbf{KM} & \textbf{VR} & 0.452 $\pm$ 0.196 & 0.235 $\pm$ 0.2 & 0.532 $\pm$ 0.1 & 0.685 $\pm$ 0.121 & 0.518 $\pm$ 0.105 & 0.474 $\pm$ 0.107 & 0.483 $\pm$ 0.138 \\
\textbf{10} & \textbf{KM} & \textbf{WD} & 0.46 $\pm$ 0.212 & 0.255 $\pm$ 0.207 & 0.537 $\pm$ 0.106 & 0.714 $\pm$ 0.136 & 0.554 $\pm$ 0.148 & 0.48 $\pm$ 0.117 & 0.503 $\pm$ 0.154 \\
\textbf{10} & \textbf{KM} & \textbf{DC} & \gr{0.4 $\pm$ 0.201} & 0.233 $\pm$ 0.202 & 0.535 $\pm$ 0.121 & 0.706 $\pm$ 0.128 & 0.566 $\pm$ 0.133 & 0.464 $\pm$ 0.129 & 0.492 $\pm$ 0.152 \\
& & \multicolumn{8}{l}{\textit{Continued on next page}}
\end{tabular}
\end{sidewaystable}

\begin{sidewaystable}[htbp]
\renewcommand{\tabcolsep}{0.2cm}
\renewcommand{\arraystretch}{0.96}
\centering
\scriptsize
\caption{QS results for each combination of parameters (and IV).} \label{tbl:app_kpppm:results_qs_4}
\begin{tabular}{cccccccccc} 
\toprule
\multicolumn{3}{c}{\textbf{Parameters}} & \multicolumn{6}{c}{\textbf{Event logs}} & \multirow{2}{*}{\textbf{Avg.}} \\
\cmidrule(lr){1-3} \cmidrule(lr){4-9}
$\pmb{k}$ & $\pmb{clus}$ & $\pmb{sim}$ & {\bf BPI12} & {\bf BPI13} & {\bf BPI14} & {\bf BPI15} & {\bf CoSeLoG} & {\bf TGN-Hospital}  \\
\midrule
\textbf{10} & \textbf{OKA} & \textbf{VEO} & 0.477 $\pm$ 0.208 & 0.269 $\pm$ 0.216 & 0.535 $\pm$ 0.131 & 0.709 $\pm$ 0.137 & 0.508 $\pm$ 0.145 & 0.451 $\pm$ 0.144 & 0.493 $\pm$ 0.163 \\
\textbf{10} & \textbf{OKA} & \textbf{VR} & 0.426 $\pm$ 0.229 & 0.304 $\pm$ 0.224 & 0.545 $\pm$ 0.132 & 0.703 $\pm$ 0.135 & 0.522 $\pm$ 0.171 & 0.472 $\pm$ 0.154 & 0.493 $\pm$ 0.174 \\
\textbf{10} & \textbf{OKA} & \textbf{WD} & 0.454 $\pm$ 0.237 & 0.256 $\pm$ 0.228 & 0.546 $\pm$ 0.129 & 0.696 $\pm$ 0.125 & 0.529 $\pm$ 0.159 & \gr{0.449 $\pm$ 0.143} & 0.492 $\pm$ 0.17 \\
\textbf{10} & \textbf{OKA} & \textbf{DC} & 0.447 $\pm$ 0.209 & 0.278 $\pm$ 0.223 & 0.551 $\pm$ 0.111 & 0.695 $\pm$ 0.112 & \gr{0.499 $\pm$ 0.15} & 0.478 $\pm$ 0.146 & 0.491 $\pm$ 0.158 \\
\textbf{10} & \textbf{BL} & \textbf{VEO} & 0.584 $\pm$ 0.159 & 0.431 $\pm$ 0.206 & \re{0.572 $\pm$ 0.098} & \re{0.728 $\pm$ 0.124} & 0.557 $\pm$ 0.162 & \re{0.589 $\pm$ 0.108} & 0.574 $\pm$ 0.143 \\
\textbf{10} & \textbf{BL} & \textbf{VR} & \re{0.586 $\pm$ 0.153} & \re{0.451 $\pm$ 0.206} & 0.557 $\pm$ 0.116 & 0.718 $\pm$ 0.145 & 0.578 $\pm$ 0.162 & 0.579 $\pm$ 0.106 & 0.564 $\pm$ 0.148 \\
\textbf{10} & \textbf{BL} & \textbf{WD} & 0.578 $\pm$ 0.15 & 0.432 $\pm$ 0.201 & 0.571 $\pm$ 0.104 & 0.72 $\pm$ 0.117 & \re{0.583 $\pm$ 0.168} & 0.585 $\pm$ 0.109 & \re{0.585 $\pm$ 0.141} \\
\textbf{10} & \textbf{BL} & \textbf{DC} & 0.585 $\pm$ 0.155 & 0.435 $\pm$ 0.207 & 0.563 $\pm$ 0.098 & 0.723 $\pm$ 0.122 & 0.566 $\pm$ 0.167 & 0.582 $\pm$ 0.111 & 0.582 $\pm$ 0.143 \\
\midrule
\textbf{20} & \textbf{MDAV} & \textbf{VEO} & 0.518 $\pm$ 0.15 & 0.275 $\pm$ 0.215 & \gr{0.552 $\pm$ 0.1} & 0.728 $\pm$ 0.128 & 0.595 $\pm$ 0.14 & 0.495 $\pm$ 0.138 & 0.527 $\pm$ 0.145 \\ 
\textbf{20} & \textbf{MDAV} & \textbf{VR} & 0.497 $\pm$ 0.179 & 0.283 $\pm$ 0.206 & 0.576 $\pm$ 0.091 & 0.725 $\pm$ 0.118 & 0.588 $\pm$ 0.109 & 0.519 $\pm$ 0.096 & 0.531 $\pm$ 0.133 \\
\textbf{20} & \textbf{MDAV} & \textbf{WD} & 0.498 $\pm$ 0.245 & 0.293 $\pm$ 0.216 & 0.577 $\pm$ 0.105 & 0.755 $\pm$ 0.13 & 0.585 $\pm$ 0.131 & 0.501 $\pm$ 0.136 & 0.538 $\pm$ 0.161 \\
\textbf{20} & \textbf{MDAV} & \textbf{DC} & \gr{0.457 $\pm$ 0.199} & \gr{0.269 $\pm$ 0.218} & 0.572 $\pm$ 0.11 & 0.733 $\pm$ 0.123 & 0.587 $\pm$ 0.127 & \gr{0.488 $\pm$ 0.147} & \gr{0.518 $\pm$ 0.154} \\
\textbf{20} & \textbf{KM} & \textbf{VEO} & 0.535 $\pm$ 0.144 & 0.275 $\pm$ 0.216 & 0.571 $\pm$ 0.093 & 0.74 $\pm$ 0.129 & 0.584 $\pm$ 0.137 & 0.5 $\pm$ 0.099 & 0.534 $\pm$ 0.136 \\
\textbf{20} & \textbf{KM} & \textbf{VR} & 0.474 $\pm$ 0.214 & 0.311 $\pm$ 0.208 & 0.56 $\pm$ 0.093 & 0.728 $\pm$ 0.114 & 0.574 $\pm$ 0.133 & 0.523 $\pm$ 0.101 & 0.528 $\pm$ 0.144 \\
\textbf{20} & \textbf{KM} & \textbf{WD} & 0.521 $\pm$ 0.159 & 0.27 $\pm$ 0.194 & 0.572 $\pm$ 0.102 & 0.726 $\pm$ 0.123 & 0.584 $\pm$ 0.136 & 0.515 $\pm$ 0.12 & 0.534 $\pm$ 0.144 \\
\textbf{20} & \textbf{KM} & \textbf{DC} & 0.501 $\pm$ 0.189 & 0.278 $\pm$ 0.228 & 0.567 $\pm$ 0.106 & 0.727 $\pm$ 0.133 & 0.573 $\pm$ 0.137 & 0.519 $\pm$ 0.106 & 0.539 $\pm$ 0.145 \\
\textbf{20} & \textbf{OKA} & \textbf{VEO} & 0.488 $\pm$ 0.233 & 0.304 $\pm$ 0.239 & 0.588 $\pm$ 0.121 & 0.727 $\pm$ 0.104 & 0.593 $\pm$ 0.122 & 0.5 $\pm$ 0.124 & 0.533 $\pm$ 0.157 \\
\textbf{20} & \textbf{OKA} & \textbf{VR} & 0.496 $\pm$ 0.223 & 0.337 $\pm$ 0.209 & 0.582 $\pm$ 0.116 & \gr{0.72 $\pm$ 0.107} & \gr{0.565 $\pm$ 0.111} & 0.519 $\pm$ 0.146 & 0.537 $\pm$ 0.152 \\
\textbf{20} & \textbf{OKA} & \textbf{WD} & 0.485 $\pm$ 0.23 & 0.363 $\pm$ 0.21 & 0.56 $\pm$ 0.109 & 0.751 $\pm$ 0.115 & 0.581 $\pm$ 0.12 & 0.512 $\pm$ 0.168 & 0.542 $\pm$ 0.159 \\
\textbf{20} & \textbf{OKA} & \textbf{DC} & 0.483 $\pm$ 0.22 & 0.33 $\pm$ 0.225 & 0.584 $\pm$ 0.102 & 0.723 $\pm$ 0.107 & 0.595 $\pm$ 0.122 & 0.518 $\pm$ 0.137 & 0.539 $\pm$ 0.152 \\
\textbf{20} & \textbf{BL} & \textbf{VEO} & \re{0.647 $\pm$ 0.139} & 0.447 $\pm$ 0.194 & 0.578 $\pm$ 0.093 & 0.761 $\pm$ 0.123 & 0.603 $\pm$ 0.153 & 0.6 $\pm$ 0.108 & 0.613 $\pm$ 0.135 \\
\textbf{20} & \textbf{BL} & \textbf{VR} & 0.645 $\pm$ 0.158 & 0.433 $\pm$ 0.213 & \re{0.589 $\pm$ 0.094} & 0.764 $\pm$ 0.125 & 0.602 $\pm$ 0.138 & 0.6 $\pm$ 0.102 & 0.608 $\pm$ 0.139 \\
\textbf{20} & \textbf{BL} & \textbf{WD} & 0.646 $\pm$ 0.155 & \re{0.471 $\pm$ 0.199} & 0.585 $\pm$ 0.108 & \re{0.773 $\pm$ 0.122} & 0.607 $\pm$ 0.158 & \re{0.613 $\pm$ 0.1} & \re{0.621 $\pm$ 0.141} \\
\textbf{20} & \textbf{BL} & \textbf{DC} & 0.646 $\pm$ 0.155 & 0.458 $\pm$ 0.187 & 0.585 $\pm$ 0.093 & 0.759 $\pm$ 0.118 & \re{0.617 $\pm$ 0.141} & 0.606 $\pm$ 0.104 & 0.609 $\pm$ 0.133 \\
\bottomrule
\end{tabular}
\end{sidewaystable}


\begin{sidewaystable}[htbp]
\renewcommand{\tabcolsep}{0.2cm}
\renewcommand{\arraystretch}{0.96}
\centering
\scriptsize  
\caption{ILS results for each combination of parameters (I).} \label{tbl:app_kpppm:results_ils_1}
\begin{tabular}{cccccccccc} 
\toprule
\multicolumn{3}{c}{\textbf{Parameters}} & \multicolumn{6}{c}{\textbf{Event logs}} & \multirow{2}{*}{\textbf{Avg.}} \\
\cmidrule(lr){1-3} \cmidrule(lr){4-9}
$\pmb{k}$ & $\pmb{clus}$ & $\pmb{sim}$ & {\bf BPI12} & {\bf BPI13} & {\bf BPI14} & {\bf BPI15} & {\bf CoSeLoG} & {\bf TGN-Hospital}  \\
\midrule
\textbf{2} & \textbf{MDAV} & \textbf{VEO} & 0.055 $\pm$ 0.011 & 0.047 $\pm$ 0.013 & 0.042 $\pm$ 0.018 & 0.037 $\pm$ 0.014 & 0.064 $\pm$ 0.019 & 0.045 $\pm$ 0.018 & 0.048 $\pm$ 0.018 \\
\textbf{2} & \textbf{MDAV} & \textbf{VR} & 0.054 $\pm$ 0.012 & 0.052 $\pm$ 0.017 & \gr{0.042 $\pm$ 0.016} & 0.036 $\pm$ 0.013 & 0.066 $\pm$ 0.019 & \gr{0.04 $\pm$ 0.013} & 0.049 $\pm$ 0.018 \\
\textbf{2} & \textbf{MDAV} & \textbf{WD} & 0.058 $\pm$ 0.014 & 0.055 $\pm$ 0.017 & 0.043 $\pm$ 0.019 & 0.04 $\pm$ 0.018 & 0.063 $\pm$ 0.02 & 0.048 $\pm$ 0.021 & 0.051 $\pm$ 0.02 \\
\textbf{2} & \textbf{MDAV} & \textbf{DC} & 0.052 $\pm$ 0.009 & \gr{0.042 $\pm$ 0.01} & 0.045 $\pm$ 0.019 & 0.038 $\pm$ 0.034 & 0.065 $\pm$ 0.022 & 0.043 $\pm$ 0.016 & 0.051 $\pm$ 0.022 \\
\textbf{2} & \textbf{KM} & \textbf{VEO} & 0.055 $\pm$ 0.011 & 0.044 $\pm$ 0.012 & 0.042 $\pm$ 0.018 & \gr{0.035 $\pm$ 0.012} & 0.066 $\pm$ 0.019 & 0.046 $\pm$ 0.017 & 0.048 $\pm$ 0.018 \\
\textbf{2} & \textbf{KM} & \textbf{VR} & 0.057 $\pm$ 0.014 & 0.047 $\pm$ 0.015 & 0.042 $\pm$ 0.017 & 0.037 $\pm$ 0.012 & \gr{0.056 $\pm$ 0.012} & 0.044 $\pm$ 0.015 & \gr{0.047 $\pm$ 0.016} \\
\textbf{2} & \textbf{KM} & \textbf{WD} & 0.061 $\pm$ 0.016 & 0.051 $\pm$ 0.016 & 0.042 $\pm$ 0.019 & 0.041 $\pm$ 0.017 & 0.059 $\pm$ 0.015 & 0.049 $\pm$ 0.02 & 0.05 $\pm$ 0.019 \\
\textbf{2} & \textbf{KM} & \textbf{DC} & 0.054 $\pm$ 0.011 & 0.043 $\pm$ 0.011 & 0.045 $\pm$ 0.02 & 0.046 $\pm$ 0.041 & 0.078 $\pm$ 0.031 & 0.045 $\pm$ 0.016 & 0.053 $\pm$ 0.027 \\
\textbf{2} & \textbf{OKA} & \textbf{VEO} & \gr{0.051 $\pm$ 0.004} & 0.051 $\pm$ 0.012 & 0.044 $\pm$ 0.022 & 0.039 $\pm$ 0.018 & 0.083 $\pm$ 0.03 & 0.045 $\pm$ 0.016 & 0.052 $\pm$ 0.024 \\
\textbf{2} & \textbf{OKA} & \textbf{VR} & 0.07 $\pm$ 0.024 & 0.054 $\pm$ 0.014 & 0.048 $\pm$ 0.022 & 0.044 $\pm$ 0.021 & 0.071 $\pm$ 0.021 & 0.05 $\pm$ 0.017 & 0.055 $\pm$ 0.026 \\
\textbf{2} & \textbf{OKA} & \textbf{WD} & 0.061 $\pm$ 0.013 & 0.056 $\pm$ 0.015 & 0.049 $\pm$ 0.027 & 0.049 $\pm$ 0.03 & 0.086 $\pm$ 0.031 & 0.052 $\pm$ 0.023 & 0.06 $\pm$ 0.027 \\
\textbf{2} & \textbf{OKA} & \textbf{DC} & 0.079 $\pm$ 0.018 & 0.057 $\pm$ 0.015 & 0.047 $\pm$ 0.022 & 0.05 $\pm$ 0.025 & 0.08 $\pm$ 0.031 & 0.052 $\pm$ 0.021 & 0.061 $\pm$ 0.026 \\
\textbf{2} & \textbf{BL} & \textbf{VEO} & 0.103 $\pm$ 0.024 & 0.124 $\pm$ 0.025 & 0.056 $\pm$ 0.028 & 0.053 $\pm$ 0.028 & 0.087 $\pm$ 0.031 & 0.094 $\pm$ 0.04 & 0.086 $\pm$ 0.039 \\
\textbf{2} & \textbf{BL} & \textbf{VR} & \re{0.105 $\pm$ 0.026} & 0.125 $\pm$ 0.026 & \re{0.058 $\pm$ 0.03} & 0.053 $\pm$ 0.028 & \re{0.094 $\pm$ 0.037} & \re{0.095 $\pm$ 0.041} & \re{0.088 $\pm$ 0.041} \\
\textbf{2} & \textbf{BL} & \textbf{WD} & 0.102 $\pm$ 0.023 & \re{0.126 $\pm$ 0.026} & 0.056 $\pm$ 0.027 & 0.052 $\pm$ 0.028 & 0.086 $\pm$ 0.029 & 0.091 $\pm$ 0.039 & 0.085 $\pm$ 0.039 \\
\textbf{2} & \textbf{BL} & \textbf{DC} & 0.103 $\pm$ 0.025 & 0.124 $\pm$ 0.025 & 0.056 $\pm$ 0.028 & \re{0.054 $\pm$ 0.029} & 0.089 $\pm$ 0.031 & 0.093 $\pm$ 0.04 & 0.086 $\pm$ 0.039 \\
\midrule
\textbf{3} & \textbf{MDAV} & \textbf{VEO} & 0.081 $\pm$ 0.016 & 0.062 $\pm$ 0.017 & 0.053 $\pm$ 0.022 & 0.059 $\pm$ 0.019 & 0.081 $\pm$ 0.02 & 0.056 $\pm$ 0.021 & 0.065 $\pm$ 0.022 \\
\textbf{3} & \textbf{MDAV} & \textbf{VR} & \gr{0.075 $\pm$ 0.01} & 0.058 $\pm$ 0.015 & 0.055 $\pm$ 0.022 & 0.061 $\pm$ 0.018 & 0.088 $\pm$ 0.023 & 0.054 $\pm$ 0.019 & 0.065 $\pm$ 0.022 \\
\textbf{3} & \textbf{MDAV} & \textbf{WD} & 0.091 $\pm$ 0.024 & 0.062 $\pm$ 0.016 & \gr{0.052 $\pm$ 0.023} & 0.065 $\pm$ 0.026 & \gr{0.08 $\pm$ 0.019} & 0.062 $\pm$ 0.027 & 0.069 $\pm$ 0.026 \\
\textbf{3} & \textbf{MDAV} & \textbf{DC} & 0.083 $\pm$ 0.014 & \gr{0.055 $\pm$ 0.012} & 0.056 $\pm$ 0.024 & 0.066 $\pm$ 0.052 & 0.089 $\pm$ 0.026 & 0.053 $\pm$ 0.019 & 0.07 $\pm$ 0.033 \\
\textbf{3} & \textbf{KM} & \textbf{VEO} & 0.083 $\pm$ 0.018 & 0.057 $\pm$ 0.015 & 0.054 $\pm$ 0.022 & 0.057 $\pm$ 0.019 & 0.089 $\pm$ 0.027 & 0.056 $\pm$ 0.02 & 0.066 $\pm$ 0.025 \\
\textbf{3} & \textbf{KM} & \textbf{VR} & 0.081 $\pm$ 0.016 & 0.057 $\pm$ 0.016 & 0.054 $\pm$ 0.021 & \gr{0.055 $\pm$ 0.015} & 0.085 $\pm$ 0.024 & 0.054 $\pm$ 0.018 & \gr{0.064 $\pm$ 0.023} \\
\textbf{3} & \textbf{KM} & \textbf{WD} & 0.094 $\pm$ 0.028 & 0.064 $\pm$ 0.019 & 0.054 $\pm$ 0.023 & 0.062 $\pm$ 0.024 & 0.086 $\pm$ 0.026 & 0.059 $\pm$ 0.023 & 0.07 $\pm$ 0.028 \\
\textbf{3} & \textbf{KM} & \textbf{DC} & 0.078 $\pm$ 0.015 & 0.055 $\pm$ 0.013 & 0.057 $\pm$ 0.024 & 0.064 $\pm$ 0.045 & 0.083 $\pm$ 0.026 & 0.055 $\pm$ 0.019 & 0.069 $\pm$ 0.029 \\
& & \multicolumn{8}{l}{\textit{Continued on next page}}
\end{tabular}
\end{sidewaystable}

\begin{sidewaystable}[htbp]
\renewcommand{\tabcolsep}{0.2cm}
\renewcommand{\arraystretch}{0.96}
\centering
\scriptsize  
\caption{ILS results for each combination of parameters (II).} \label{tbl:app_kpppm:results_ils_2}
\begin{tabular}{cccccccccc} 
\toprule
\multicolumn{3}{c}{\textbf{Parameters}} & \multicolumn{6}{c}{\textbf{Event logs}} & \multirow{2}{*}{\textbf{Avg.}} \\
\cmidrule(lr){1-3} \cmidrule(lr){4-9}
$\pmb{k}$ & $\pmb{clus}$ & $\pmb{sim}$ & {\bf BPI12} & {\bf BPI13} & {\bf BPI14} & {\bf BPI15} & {\bf CoSeLoG} & {\bf TGN-Hospital}  \\
\midrule
\textbf{3} & \textbf{OKA} & \textbf{VEO} & 0.097 $\pm$ 0.067 & 0.059 $\pm$ 0.013 & 0.056 $\pm$ 0.025 & 0.068 $\pm$ 0.03 & 0.097 $\pm$ 0.03 & \gr{0.053 $\pm$ 0.017} & 0.076 $\pm$ 0.051 \\
\textbf{3} & \textbf{OKA} & \textbf{VR} & 0.084 $\pm$ 0.023 & 0.065 $\pm$ 0.017 & 0.057 $\pm$ 0.023 & 0.066 $\pm$ 0.028 & 0.088 $\pm$ 0.043 & 0.059 $\pm$ 0.02 & 0.074 $\pm$ 0.034 \\
\textbf{3} & \textbf{OKA} & \textbf{WD} & 0.086 $\pm$ 0.026 & 0.065 $\pm$ 0.017 & 0.054 $\pm$ 0.027 & 0.065 $\pm$ 0.036 & 0.094 $\pm$ 0.03 & 0.065 $\pm$ 0.028 & 0.073 $\pm$ 0.031 \\
\textbf{3} & \textbf{OKA} & \textbf{DC} & 0.075 $\pm$ 0.017 & 0.069 $\pm$ 0.018 & 0.064 $\pm$ 0.03 & 0.065 $\pm$ 0.027 & 0.098 $\pm$ 0.033 & 0.064 $\pm$ 0.026 & 0.073 $\pm$ 0.028 \\
\textbf{3} & \textbf{BL} & \textbf{VEO} & \re{0.147 $\pm$ 0.034} & 0.143 $\pm$ 0.028 & 0.065 $\pm$ 0.031 & \re{0.07 $\pm$ 0.029} & 0.108 $\pm$ 0.03 & 0.111 $\pm$ 0.047 & 0.107 $\pm$ 0.047 \\
\textbf{3} & \textbf{BL} & \textbf{VR} & 0.145 $\pm$ 0.032 & 0.144 $\pm$ 0.029 & \re{0.067 $\pm$ 0.032} & 0.069 $\pm$ 0.028 & 0.112 $\pm$ 0.034 & \re{0.112 $\pm$ 0.046} & 0.108 $\pm$ 0.047 \\
\textbf{3} & \textbf{BL} & \textbf{WD} & 0.144 $\pm$ 0.031 & 0.142 $\pm$ 0.028 & 0.066 $\pm$ 0.032 & 0.069 $\pm$ 0.03 & 0.109 $\pm$ 0.031 & 0.111 $\pm$ 0.047 & 0.107 $\pm$ 0.046 \\
\textbf{3} & \textbf{BL} & \textbf{DC} & 0.146 $\pm$ 0.033 & \re{0.144 $\pm$ 0.03} & 0.066 $\pm$ 0.032 & 0.068 $\pm$ 0.029 & \re{0.12 $\pm$ 0.036} & 0.111 $\pm$ 0.046 & \re{0.109 $\pm$ 0.048} \\
\midrule
\textbf{4} & \textbf{MDAV} & \textbf{VEO} & 0.105 $\pm$ 0.025 & 0.063 $\pm$ 0.015 & 0.061 $\pm$ 0.024 & 0.072 $\pm$ 0.02 & 0.112 $\pm$ 0.028 & 0.067 $\pm$ 0.023 & 0.08 $\pm$ 0.031 \\
\textbf{4} & \textbf{MDAV} & \textbf{VR} & 0.086 $\pm$ 0.013 & 0.066 $\pm$ 0.018 & \gr{0.059 $\pm$ 0.022} & 0.072 $\pm$ 0.019 & \gr{0.1 $\pm$ 0.021} & 0.069 $\pm$ 0.023 & \gr{0.075 $\pm$ 0.024} \\
\textbf{4} & \textbf{MDAV} & \textbf{WD} & 0.119 $\pm$ 0.038 & 0.077 $\pm$ 0.023 & 0.059 $\pm$ 0.025 & 0.073 $\pm$ 0.037 & 0.106 $\pm$ 0.025 & 0.069 $\pm$ 0.026 & 0.085 $\pm$ 0.036 \\
\textbf{4} & \textbf{MDAV} & \textbf{DC} & 0.101 $\pm$ 0.021 & 0.064 $\pm$ 0.015 & 0.064 $\pm$ 0.026 & 0.074 $\pm$ 0.057 & 0.107 $\pm$ 0.027 & 0.071 $\pm$ 0.026 & 0.082 $\pm$ 0.038 \\
\textbf{4} & \textbf{KM} & \textbf{VEO} & 0.096 $\pm$ 0.017 & 0.064 $\pm$ 0.017 & 0.059 $\pm$ 0.023 & 0.072 $\pm$ 0.019 & 0.117 $\pm$ 0.03 & 0.065 $\pm$ 0.023 & 0.079 $\pm$ 0.03 \\
\textbf{4} & \textbf{KM} & \textbf{VR} & \gr{0.086 $\pm$ 0.012} & 0.063 $\pm$ 0.017 & 0.061 $\pm$ 0.023 & 0.072 $\pm$ 0.018 & 0.105 $\pm$ 0.023 & \gr{0.063 $\pm$ 0.02} & 0.075 $\pm$ 0.025 \\
\textbf{4} & \textbf{KM} & \textbf{WD} & 0.125 $\pm$ 0.039 & 0.068 $\pm$ 0.019 & 0.062 $\pm$ 0.027 & 0.081 $\pm$ 0.033 & 0.106 $\pm$ 0.025 & 0.071 $\pm$ 0.028 & 0.086 $\pm$ 0.037 \\
\textbf{4} & \textbf{KM} & \textbf{DC} & 0.102 $\pm$ 0.022 & \gr{0.062 $\pm$ 0.015} & 0.066 $\pm$ 0.028 & 0.078 $\pm$ 0.057 & 0.113 $\pm$ 0.031 & 0.064 $\pm$ 0.023 & 0.086 $\pm$ 0.039 \\
\textbf{4} & \textbf{OKA} & \textbf{VEO} & 0.132 $\pm$ 0.042 & 0.089 $\pm$ 0.022 & 0.064 $\pm$ 0.028 & 0.082 $\pm$ 0.032 & 0.119 $\pm$ 0.037 & 0.069 $\pm$ 0.028 & 0.094 $\pm$ 0.042 \\
\textbf{4} & \textbf{OKA} & \textbf{VR} & 0.099 $\pm$ 0.022 & 0.08 $\pm$ 0.018 & 0.068 $\pm$ 0.028 & \gr{0.07 $\pm$ 0.021} & 0.114 $\pm$ 0.024 & 0.067 $\pm$ 0.024 & 0.083 $\pm$ 0.029 \\
\textbf{4} & \textbf{OKA} & \textbf{WD} & 0.103 $\pm$ 0.025 & 0.08 $\pm$ 0.017 & 0.072 $\pm$ 0.034 & 0.077 $\pm$ 0.039 & 0.12 $\pm$ 0.032 & 0.069 $\pm$ 0.028 & 0.088 $\pm$ 0.035 \\
\textbf{4} & \textbf{OKA} & \textbf{DC} & 0.103 $\pm$ 0.025 & 0.088 $\pm$ 0.021 & 0.068 $\pm$ 0.03 & 0.077 $\pm$ 0.026 & 0.126 $\pm$ 0.034 & 0.064 $\pm$ 0.024 & 0.088 $\pm$ 0.035 \\
\textbf{4} & \textbf{BL} & \textbf{VEO} & 0.149 $\pm$ 0.03 & 0.154 $\pm$ 0.031 & \re{0.073 $\pm$ 0.035} & 0.084 $\pm$ 0.032 & 0.13 $\pm$ 0.029 & 0.126 $\pm$ 0.053 & 0.117 $\pm$ 0.047 \\
\textbf{4} & \textbf{BL} & \textbf{VR} & 0.15 $\pm$ 0.032 & \re{0.158 $\pm$ 0.033} & 0.072 $\pm$ 0.033 & 0.085 $\pm$ 0.032 & \re{0.131 $\pm$ 0.041} & 0.126 $\pm$ 0.054 & \re{0.12 $\pm$ 0.05} \\
\textbf{4} & \textbf{BL} & \textbf{WD} & \re{0.153 $\pm$ 0.035} & 0.153 $\pm$ 0.03 & 0.073 $\pm$ 0.033 & \re{0.086 $\pm$ 0.033} & 0.126 $\pm$ 0.037 & \re{0.127 $\pm$ 0.053} & 0.12 $\pm$ 0.049 \\
\textbf{4} & \textbf{BL} & \textbf{DC} & 0.151 $\pm$ 0.033 & 0.154 $\pm$ 0.031 & 0.072 $\pm$ 0.032 & 0.084 $\pm$ 0.032 & 0.126 $\pm$ 0.037 & 0.124 $\pm$ 0.052 & 0.118 $\pm$ 0.048 \\
\midrule
& & \multicolumn{8}{l}{\textit{Continued on next page}}
\end{tabular}
\end{sidewaystable}

\begin{sidewaystable}[htbp]
\renewcommand{\tabcolsep}{0.2cm}
\renewcommand{\arraystretch}{0.96}
\centering
\scriptsize  
\caption{ILS results for each combination of parameters (III).} \label{tbl:app_kpppm:results_ils_3}
\begin{tabular}{cccccccccc} 
\toprule
\multicolumn{3}{c}{\textbf{Parameters}} & \multicolumn{6}{c}{\textbf{Event logs}} & \multirow{2}{*}{\textbf{Avg.}} \\
\cmidrule(lr){1-3} \cmidrule(lr){4-9}
$\pmb{k}$ & $\pmb{clus}$ & $\pmb{sim}$ & {\bf BPI12} & {\bf BPI13} & {\bf BPI14} & {\bf BPI15} & {\bf CoSeLoG} & {\bf TGN-Hospital}  \\
\midrule
\textbf{5} & \textbf{MDAV} & \textbf{VEO} & 0.116 $\pm$ 0.025 & 0.066 $\pm$ 0.019 & 0.065 $\pm$ 0.024 & \gr{0.085 $\pm$ 0.021} & 0.117 $\pm$ 0.021 & 0.07 $\pm$ 0.024 & \gr{0.086 $\pm$ 0.032} \\
\textbf{5} & \textbf{MDAV} & \textbf{VR} & 0.108 $\pm$ 0.019 & 0.067 $\pm$ 0.021 & 0.065 $\pm$ 0.023 & 0.089 $\pm$ 0.022 & 0.121 $\pm$ 0.022 & 0.07 $\pm$ 0.023 & 0.087 $\pm$ 0.03 \\
\textbf{5} & \textbf{MDAV} & \textbf{WD} & 0.139 $\pm$ 0.039 & 0.077 $\pm$ 0.026 & 0.068 $\pm$ 0.028 & 0.104 $\pm$ 0.035 & 0.132 $\pm$ 0.042 & 0.079 $\pm$ 0.032 & 0.103 $\pm$ 0.047 \\
\textbf{5} & \textbf{MDAV} & \textbf{DC} & 0.106 $\pm$ 0.018 & \gr{0.064 $\pm$ 0.015} & 0.073 $\pm$ 0.029 & 0.122 $\pm$ 0.051 & 0.129 $\pm$ 0.027 & 0.071 $\pm$ 0.025 & 0.095 $\pm$ 0.04 \\
\textbf{5} & \textbf{KM} & \textbf{VEO} & 0.125 $\pm$ 0.027 & 0.065 $\pm$ 0.019 & \gr{0.062 $\pm$ 0.022} & 0.094 $\pm$ 0.028 & 0.143 $\pm$ 0.035 & \gr{0.07 $\pm$ 0.022} & 0.093 $\pm$ 0.041 \\
\textbf{5} & \textbf{KM} & \textbf{VR} & 0.113 $\pm$ 0.019 & 0.065 $\pm$ 0.02 & 0.067 $\pm$ 0.025 & 0.086 $\pm$ 0.019 & \gr{0.118 $\pm$ 0.019} & 0.072 $\pm$ 0.022 & 0.087 $\pm$ 0.03 \\
\textbf{5} & \textbf{KM} & \textbf{WD} & 0.142 $\pm$ 0.038 & 0.078 $\pm$ 0.027 & 0.072 $\pm$ 0.031 & 0.109 $\pm$ 0.04 & 0.122 $\pm$ 0.023 & 0.082 $\pm$ 0.033 & 0.101 $\pm$ 0.041 \\
\textbf{5} & \textbf{KM} & \textbf{DC} & 0.102 $\pm$ 0.013 & 0.071 $\pm$ 0.017 & 0.075 $\pm$ 0.03 & 0.13 $\pm$ 0.058 & 0.133 $\pm$ 0.031 & 0.078 $\pm$ 0.029 & 0.098 $\pm$ 0.043 \\
\textbf{5} & \textbf{OKA} & \textbf{VEO} & 0.12 $\pm$ 0.034 & 0.085 $\pm$ 0.024 & 0.072 $\pm$ 0.032 & 0.096 $\pm$ 0.03 & 0.121 $\pm$ 0.02 & 0.074 $\pm$ 0.026 & 0.095 $\pm$ 0.034 \\
\textbf{5} & \textbf{OKA} & \textbf{VR} & \gr{0.098 $\pm$ 0.02} & 0.101 $\pm$ 0.033 & 0.066 $\pm$ 0.024 & 0.096 $\pm$ 0.028 & 0.12 $\pm$ 0.02 & 0.077 $\pm$ 0.025 & 0.093 $\pm$ 0.031 \\
\textbf{5} & \textbf{OKA} & \textbf{WD} & 0.105 $\pm$ 0.025 & 0.095 $\pm$ 0.029 & 0.063 $\pm$ 0.026 & 0.106 $\pm$ 0.039 & 0.13 $\pm$ 0.029 & 0.081 $\pm$ 0.033 & 0.097 $\pm$ 0.037 \\
\textbf{5} & \textbf{OKA} & \textbf{DC} & 0.098 $\pm$ 0.021 & 0.079 $\pm$ 0.02 & 0.071 $\pm$ 0.035 & 0.106 $\pm$ 0.037 & 0.134 $\pm$ 0.03 & 0.081 $\pm$ 0.031 & 0.094 $\pm$ 0.035 \\
\textbf{5} & \textbf{BL} & \textbf{VEO} & \re{0.184 $\pm$ 0.043} & 0.157 $\pm$ 0.034 & \re{0.079 $\pm$ 0.036} & 0.167 $\pm$ 0.03 & \re{0.143 $\pm$ 0.034} & \re{0.132 $\pm$ 0.055} & \re{0.166 $\pm$ 0.053} \\
\textbf{5} & \textbf{BL} & \textbf{VR} & 0.181 $\pm$ 0.042 & \re{0.16 $\pm$ 0.036} & 0.078 $\pm$ 0.035 & \re{0.168 $\pm$ 0.031} & 0.135 $\pm$ 0.028 & 0.129 $\pm$ 0.05 & 0.16 $\pm$ 0.052 \\
\textbf{5} & \textbf{BL} & \textbf{WD} & 0.178 $\pm$ 0.038 & 0.154 $\pm$ 0.031 & 0.077 $\pm$ 0.033 & 0.167 $\pm$ 0.031 & 0.137 $\pm$ 0.029 & 0.129 $\pm$ 0.055 & 0.159 $\pm$ 0.05 \\
\textbf{5} & \textbf{BL} & \textbf{DC} & 0.177 $\pm$ 0.039 & 0.154 $\pm$ 0.032 & 0.076 $\pm$ 0.033 & 0.166 $\pm$ 0.03 & 0.139 $\pm$ 0.031 & 0.129 $\pm$ 0.052 & 0.159 $\pm$ 0.05 \\
\midrule
\textbf{10} & \textbf{MDAV} & \textbf{VEO} & 0.191 $\pm$ 0.034 & 0.081 $\pm$ 0.025 & 0.082 $\pm$ 0.026 & 0.154 $\pm$ 0.027 & 0.222 $\pm$ 0.032 & 0.095 $\pm$ 0.03 & 0.138 $\pm$ 0.063 \\
\textbf{10} & \textbf{MDAV} & \textbf{VR} & 0.198 $\pm$ 0.038 & 0.084 $\pm$ 0.027 & \gr{0.082 $\pm$ 0.024} & 0.153 $\pm$ 0.024 & 0.2 $\pm$ 0.029 & 0.098 $\pm$ 0.032 & 0.136 $\pm$ 0.059 \\
\textbf{10} & \textbf{MDAV} & \textbf{WD} & 0.201 $\pm$ 0.039 & 0.094 $\pm$ 0.032 & 0.085 $\pm$ 0.036 & 0.175 $\pm$ 0.074 & 0.225 $\pm$ 0.055 & 0.097 $\pm$ 0.035 & 0.146 $\pm$ 0.084 \\
\textbf{10} & \textbf{MDAV} & \textbf{DC} & 0.161 $\pm$ 0.02 & \gr{0.072 $\pm$ 0.018} & 0.089 $\pm$ 0.036 & 0.183 $\pm$ 0.049 & 0.214 $\pm$ 0.031 & 0.088 $\pm$ 0.028 & 0.136 $\pm$ 0.062 \\
\textbf{10} & \textbf{KM} & \textbf{VEO} & 0.176 $\pm$ 0.024 & 0.082 $\pm$ 0.014 & 0.086 $\pm$ 0.028 & \gr{0.141 $\pm$ 0.018} & 0.201 $\pm$ 0.028 & 0.091 $\pm$ 0.026 & \gr{0.13 $\pm$ 0.052} \\
\textbf{10} & \textbf{KM} & \textbf{VR} & 0.182 $\pm$ 0.026 & 0.085 $\pm$ 0.018 & 0.083 $\pm$ 0.025 & 0.167 $\pm$ 0.035 & 0.194 $\pm$ 0.028 & 0.093 $\pm$ 0.026 & 0.134 $\pm$ 0.055 \\
\textbf{10} & \textbf{KM} & \textbf{WD} & 0.211 $\pm$ 0.044 & 0.095 $\pm$ 0.022 & 0.085 $\pm$ 0.033 & 0.171 $\pm$ 0.063 & 0.255 $\pm$ 0.051 & 0.099 $\pm$ 0.033 & 0.159 $\pm$ 0.078 \\
\textbf{10} & \textbf{KM} & \textbf{DC} & 0.16 $\pm$ 0.022 & 0.092 $\pm$ 0.021 & 0.087 $\pm$ 0.034 & 0.177 $\pm$ 0.058 & 0.211 $\pm$ 0.029 & 0.1 $\pm$ 0.032 & 0.143 $\pm$ 0.06 \\
& & \multicolumn{8}{l}{\textit{Continued on next page}}
\end{tabular}
\end{sidewaystable}

\begin{sidewaystable}[htbp]
\renewcommand{\tabcolsep}{0.2cm}
\renewcommand{\arraystretch}{0.96}
\centering
\scriptsize  
\caption{ILS results for each combination of parameters (and IV).} \label{tbl:app_kpppm:results_ils_4}
\begin{tabular}{cccccccccc} 
\toprule
\multicolumn{3}{c}{\textbf{Parameters}} & \multicolumn{6}{c}{\textbf{Event logs}} & \multirow{2}{*}{\textbf{Avg.}} \\
\cmidrule(lr){1-3} \cmidrule(lr){4-9}
$\pmb{k}$ & $\pmb{clus}$ & $\pmb{sim}$ & {\bf BPI12} & {\bf BPI13} & {\bf BPI14} & {\bf BPI15} & {\bf CoSeLoG} & {\bf TGN-Hospital}  \\
\midrule
\textbf{10} & \textbf{OKA} & \textbf{VEO} & 0.2 $\pm$ 0.044 & 0.101 $\pm$ 0.025 & 0.084 $\pm$ 0.034 & 0.18 $\pm$ 0.087 & 0.201 $\pm$ 0.026 & \gr{0.087 $\pm$ 0.023} & 0.152 $\pm$ 0.075 \\
\textbf{10} & \textbf{OKA} & \textbf{VR} & 0.194 $\pm$ 0.043 & 0.108 $\pm$ 0.03 & 0.084 $\pm$ 0.038 & 0.186 $\pm$ 0.049 & \gr{0.191 $\pm$ 0.029} & 0.102 $\pm$ 0.032 & 0.148 $\pm$ 0.057 \\
\textbf{10} & \textbf{OKA} & \textbf{WD} & 0.202 $\pm$ 0.046 & 0.093 $\pm$ 0.021 & 0.083 $\pm$ 0.037 & 0.16 $\pm$ 0.032 & 0.2 $\pm$ 0.024 & 0.09 $\pm$ 0.027 & 0.14 $\pm$ 0.059 \\
\textbf{10} & \textbf{OKA} & \textbf{DC} & \gr{0.158 $\pm$ 0.019} & 0.1 $\pm$ 0.025 & 0.089 $\pm$ 0.031 & 0.155 $\pm$ 0.03 & 0.204 $\pm$ 0.027 & 0.108 $\pm$ 0.038 & 0.136 $\pm$ 0.049 \\
\textbf{10} & \textbf{BL} & \textbf{VEO} & 0.237 $\pm$ 0.04 & 0.167 $\pm$ 0.031 & 0.091 $\pm$ 0.033 & \re{0.196 $\pm$ 0.029} & \re{0.324 $\pm$ 0.031} & \re{0.165 $\pm$ 0.063} & 0.196 $\pm$ 0.062 \\
\textbf{10} & \textbf{BL} & \textbf{VR} & 0.239 $\pm$ 0.044 & \re{0.172 $\pm$ 0.033} & \re{0.092 $\pm$ 0.033} & 0.193 $\pm$ 0.027 & 0.311 $\pm$ 0.026 & 0.153 $\pm$ 0.056 & 0.193 $\pm$ 0.06 \\
\textbf{10} & \textbf{BL} & \textbf{WD} & 0.242 $\pm$ 0.045 & 0.169 $\pm$ 0.032 & 0.091 $\pm$ 0.032 & 0.191 $\pm$ 0.025 & 0.316 $\pm$ 0.029 & 0.162 $\pm$ 0.06 & 0.195 $\pm$ 0.062 \\
\textbf{10} & \textbf{BL} & \textbf{DC} & \re{0.264 $\pm$ 0.062} & 0.168 $\pm$ 0.032 & 0.092 $\pm$ 0.031 & 0.19 $\pm$ 0.025 & 0.311 $\pm$ 0.025 & 0.159 $\pm$ 0.061 & \re{0.197 $\pm$ 0.068} \\
\midrule
\textbf{20} & \textbf{MDAV} & \textbf{VEO} & 0.278 $\pm$ 0.041 & 0.104 $\pm$ 0.026 & 0.111 $\pm$ 0.025 & 0.307 $\pm$ 0.025 & 0.366 $\pm$ 0.058 & 0.122 $\pm$ 0.034 & 0.215 $\pm$ 0.112 \\
\textbf{20} & \textbf{MDAV} & \textbf{VR} & 0.248 $\pm$ 0.035 & 0.105 $\pm$ 0.029 & \gr{0.107 $\pm$ 0.023} & 0.31 $\pm$ 0.029 & 0.345 $\pm$ 0.053 & 0.116 $\pm$ 0.03 & \gr{0.205 $\pm$ 0.106} \\
\textbf{20} & \textbf{MDAV} & \textbf{WD} & 0.284 $\pm$ 0.039 & 0.112 $\pm$ 0.03 & 0.123 $\pm$ 0.034 & 0.313 $\pm$ 0.075 & 0.362 $\pm$ 0.056 & 0.119 $\pm$ 0.034 & 0.232 $\pm$ 0.128 \\
\textbf{20} & \textbf{MDAV} & \textbf{DC} & 0.264 $\pm$ 0.042 & 0.098 $\pm$ 0.022 & 0.12 $\pm$ 0.031 & 0.318 $\pm$ 0.03 & 0.34 $\pm$ 0.053 & \gr{0.11 $\pm$ 0.028} & 0.208 $\pm$ 0.108 \\
\textbf{20} & \textbf{KM} & \textbf{VEO} & 0.289 $\pm$ 0.044 & 0.103 $\pm$ 0.027 & 0.116 $\pm$ 0.029 & 0.304 $\pm$ 0.029 & 0.365 $\pm$ 0.053 & 0.112 $\pm$ 0.024 & 0.215 $\pm$ 0.113 \\
\textbf{20} & \textbf{KM} & \textbf{VR} & 0.249 $\pm$ 0.034 & 0.111 $\pm$ 0.032 & 0.109 $\pm$ 0.025 & 0.287 $\pm$ 0.028 & 0.365 $\pm$ 0.057 & 0.116 $\pm$ 0.026 & 0.206 $\pm$ 0.106 \\
\textbf{20} & \textbf{KM} & \textbf{WD} & 0.288 $\pm$ 0.044 & 0.111 $\pm$ 0.031 & 0.128 $\pm$ 0.038 & 0.31 $\pm$ 0.092 & 0.36 $\pm$ 0.051 & 0.131 $\pm$ 0.037 & 0.236 $\pm$ 0.129 \\
\textbf{20} & \textbf{KM} & \textbf{DC} & 0.255 $\pm$ 0.03 & \gr{0.095 $\pm$ 0.022} & 0.123 $\pm$ 0.032 & 0.376 $\pm$ 0.075 & 0.387 $\pm$ 0.068 & 0.126 $\pm$ 0.032 & 0.227 $\pm$ 0.13 \\
\textbf{20} & \textbf{OKA} & \textbf{VEO} & \gr{0.239 $\pm$ 0.025} & 0.119 $\pm$ 0.022 & 0.13 $\pm$ 0.038 & 0.288 $\pm$ 0.018 & 0.372 $\pm$ 0.053 & 0.118 $\pm$ 0.03 & 0.211 $\pm$ 0.102 \\
\textbf{20} & \textbf{OKA} & \textbf{VR} & 0.245 $\pm$ 0.029 & 0.124 $\pm$ 0.026 & 0.128 $\pm$ 0.036 & \gr{0.286 $\pm$ 0.024} & \gr{0.338 $\pm$ 0.053} & 0.142 $\pm$ 0.044 & 0.21 $\pm$ 0.092 \\
\textbf{20} & \textbf{OKA} & \textbf{WD} & 0.245 $\pm$ 0.027 & 0.133 $\pm$ 0.03 & 0.111 $\pm$ 0.027 & 0.31 $\pm$ 0.022 & 0.389 $\pm$ 0.052 & 0.137 $\pm$ 0.043 & 0.221 $\pm$ 0.109 \\
\textbf{20} & \textbf{OKA} & \textbf{DC} & 0.248 $\pm$ 0.028 & 0.124 $\pm$ 0.025 & 0.126 $\pm$ 0.035 & 0.296 $\pm$ 0.024 & 0.341 $\pm$ 0.05 & 0.129 $\pm$ 0.036 & 0.211 $\pm$ 0.095 \\
\textbf{20} & \textbf{BL} & \textbf{VEO} & \re{0.323 $\pm$ 0.046} & 0.176 $\pm$ 0.038 & 0.133 $\pm$ 0.028 & 0.387 $\pm$ 0.021 & 0.397 $\pm$ 0.051 & 0.178 $\pm$ 0.053 & 0.266 $\pm$ 0.097 \\
\textbf{20} & \textbf{BL} & \textbf{VR} & 0.313 $\pm$ 0.04 & 0.176 $\pm$ 0.037 & 0.132 $\pm$ 0.026 & 0.385 $\pm$ 0.023 & \re{0.409 $\pm$ 0.059} & 0.177 $\pm$ 0.054 & 0.265 $\pm$ 0.099 \\
\textbf{20} & \textbf{BL} & \textbf{WD} & 0.32 $\pm$ 0.044 & \re{0.179 $\pm$ 0.038} & 0.135 $\pm$ 0.029 & \re{0.392 $\pm$ 0.019} & 0.397 $\pm$ 0.051 & 0.175 $\pm$ 0.051 & \re{0.268 $\pm$ 0.097} \\
\textbf{20} & \textbf{BL} & \textbf{DC} & 0.311 $\pm$ 0.039 & 0.172 $\pm$ 0.035 & \re{0.138 $\pm$ 0.027} & 0.39 $\pm$ 0.021 & 0.4 $\pm$ 0.05 & \re{0.19 $\pm$ 0.06} & 0.266 $\pm$ 0.096 \\
\bottomrule
\end{tabular}
\end{sidewaystable}


\newpage

\begin{figure}[t!]
\centering
\includegraphics[width=0.75\linewidth]{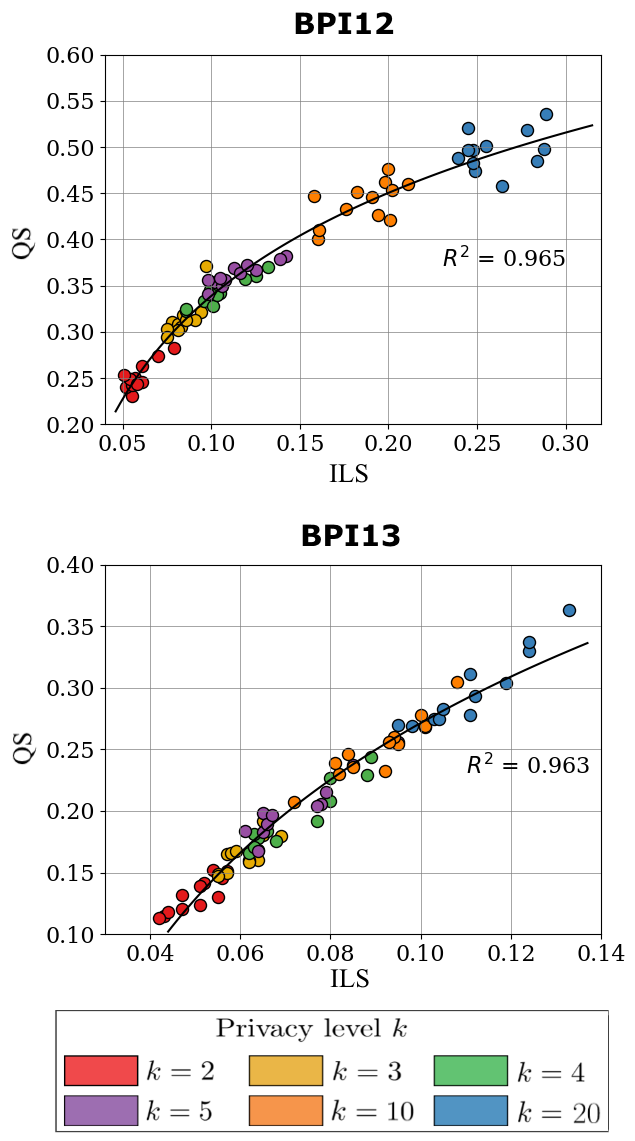}
\caption{Correlation between the QS and ILS results using the BPI12 and BPI13 event logs (reprinted from \cite{kpppm}).}
\label{fig:app_kpppm:results_bpi12_bpi13}
\end{figure}

\begin{figure}[t!]
\centering
\includegraphics[width=0.75\linewidth]{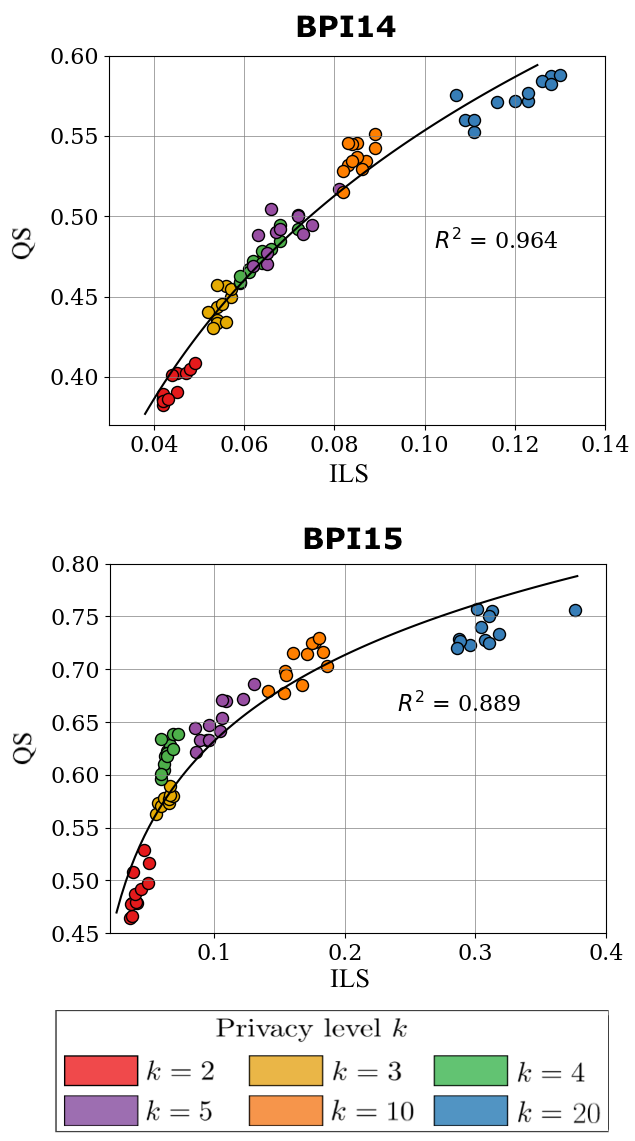}
\caption{Correlation between the QS and ILS results using the BPI14 and BPI15 event logs (reprinted from \cite{kpppm}).}
\label{fig:app_kpppm:results_bpi14_bpi15}
\end{figure}

\begin{figure}[t!]
\centering
\includegraphics[width=0.75\linewidth]{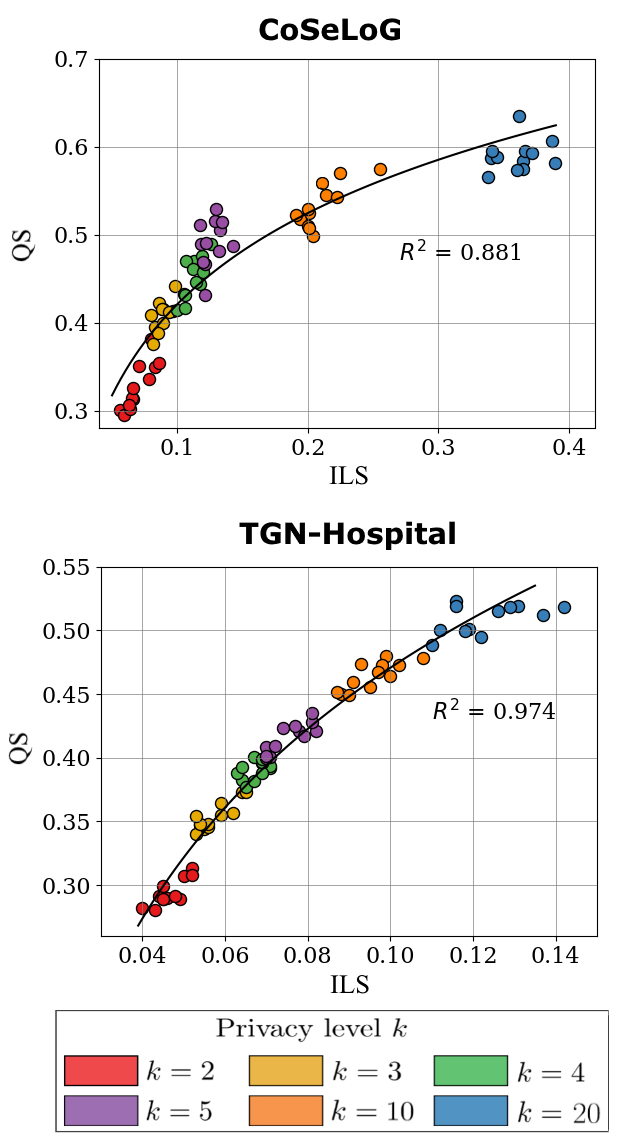}
\caption{Correlation between the QS and ILS results using the CoSeLoG and TGN-Hospital event logs (reprinted from \cite{kpppm}).}
\label{fig:app_kpppm:results_coselog_tgn}
\end{figure}


\begin{figure}[t!]
\centering
\includegraphics[width=0.67\linewidth]{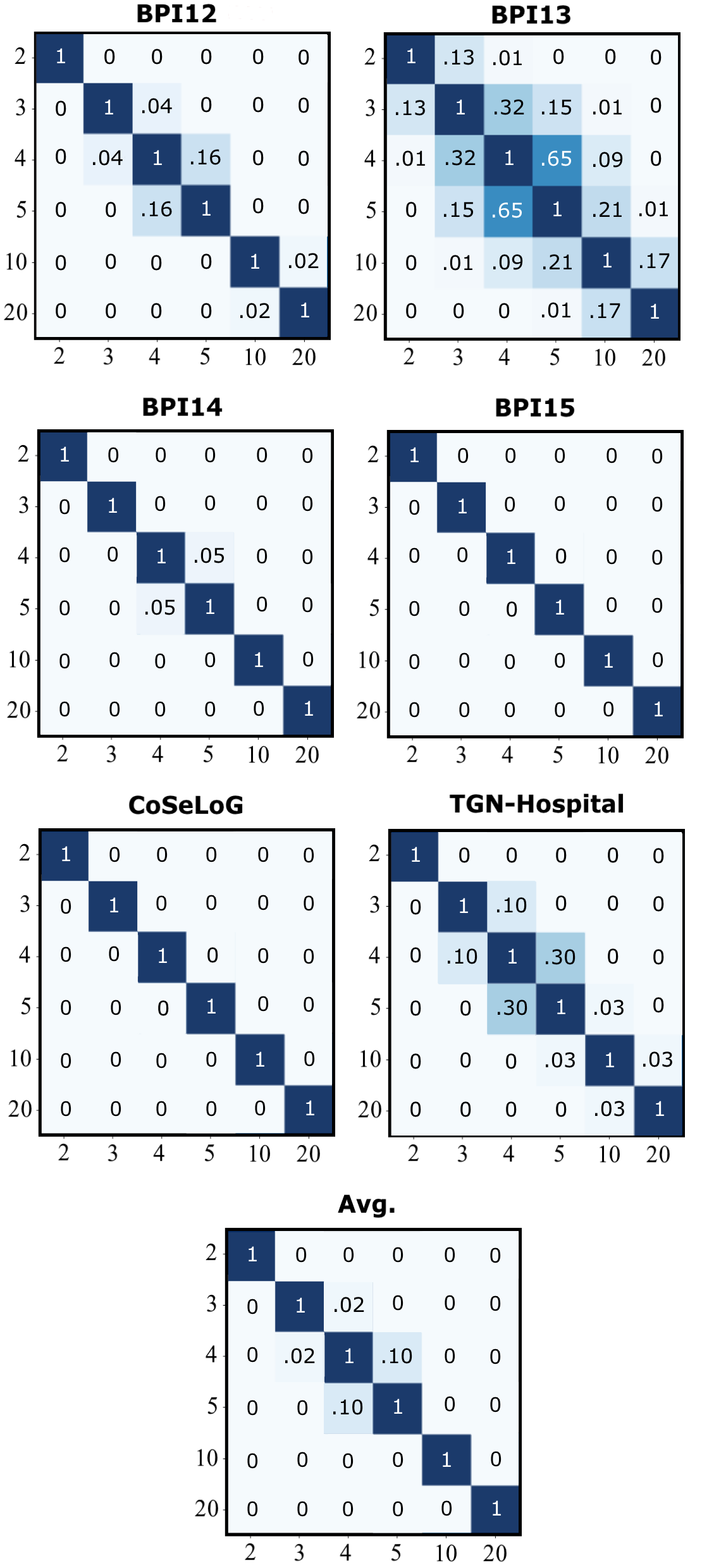}
\caption{Results of the $p$-values from the t-Tests according to the $k$ parameter in the QS results (reprinted from \cite{kpppm}).}
\label{fig:app_kpppm:ttest_qs_k}
\end{figure}

\begin{figure}[t!]
\centering
\includegraphics[width=\linewidth]{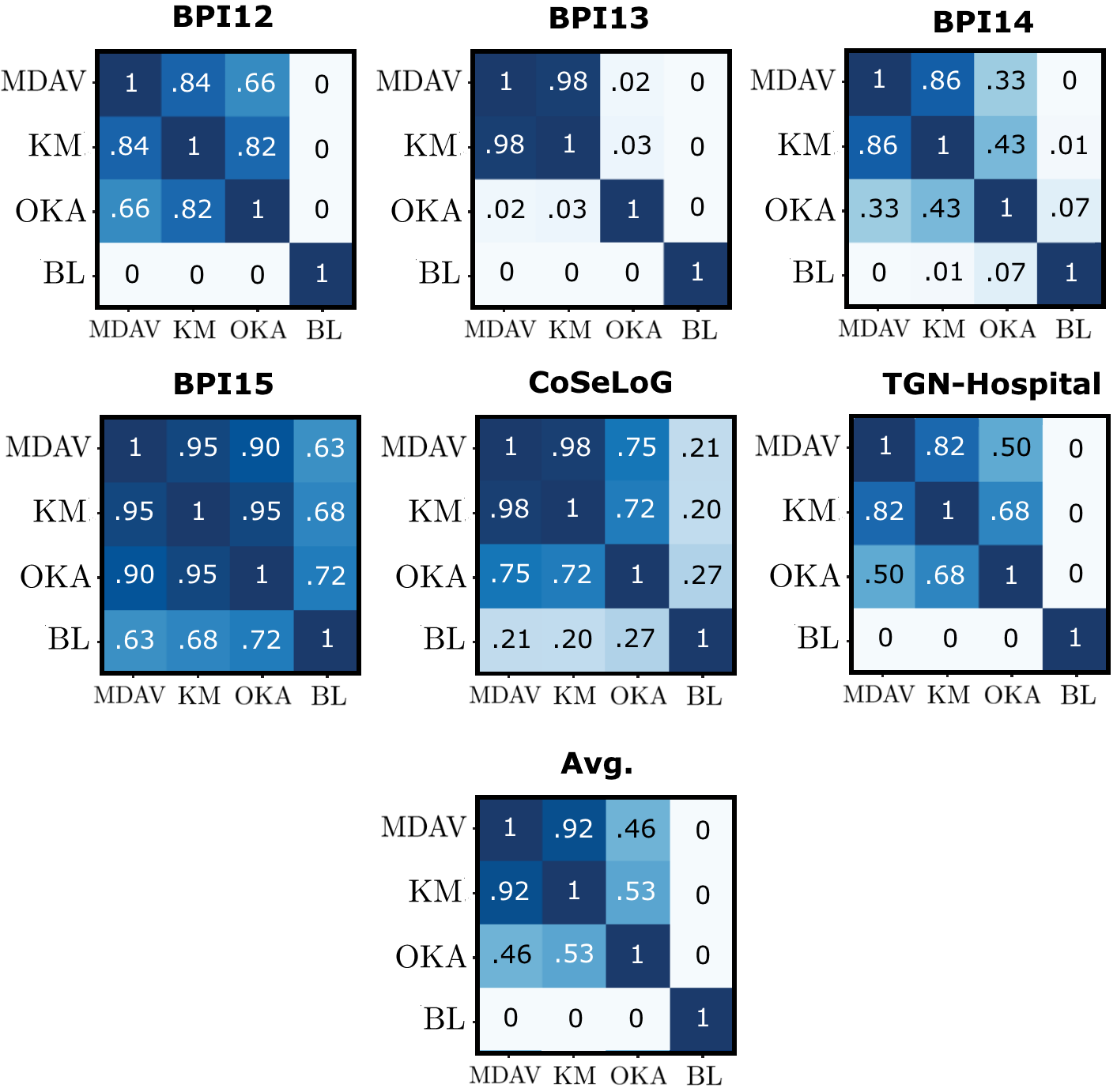}
\caption{Results of the $p$-values from the t-Tests according to the $clus$ parameter in the QS results (reprinted from \cite{kpppm}).}
\label{fig:app_kpppm:ttest_qs_clus}
\end{figure}

\begin{figure}[t!]
\centering
\includegraphics[width=\linewidth]{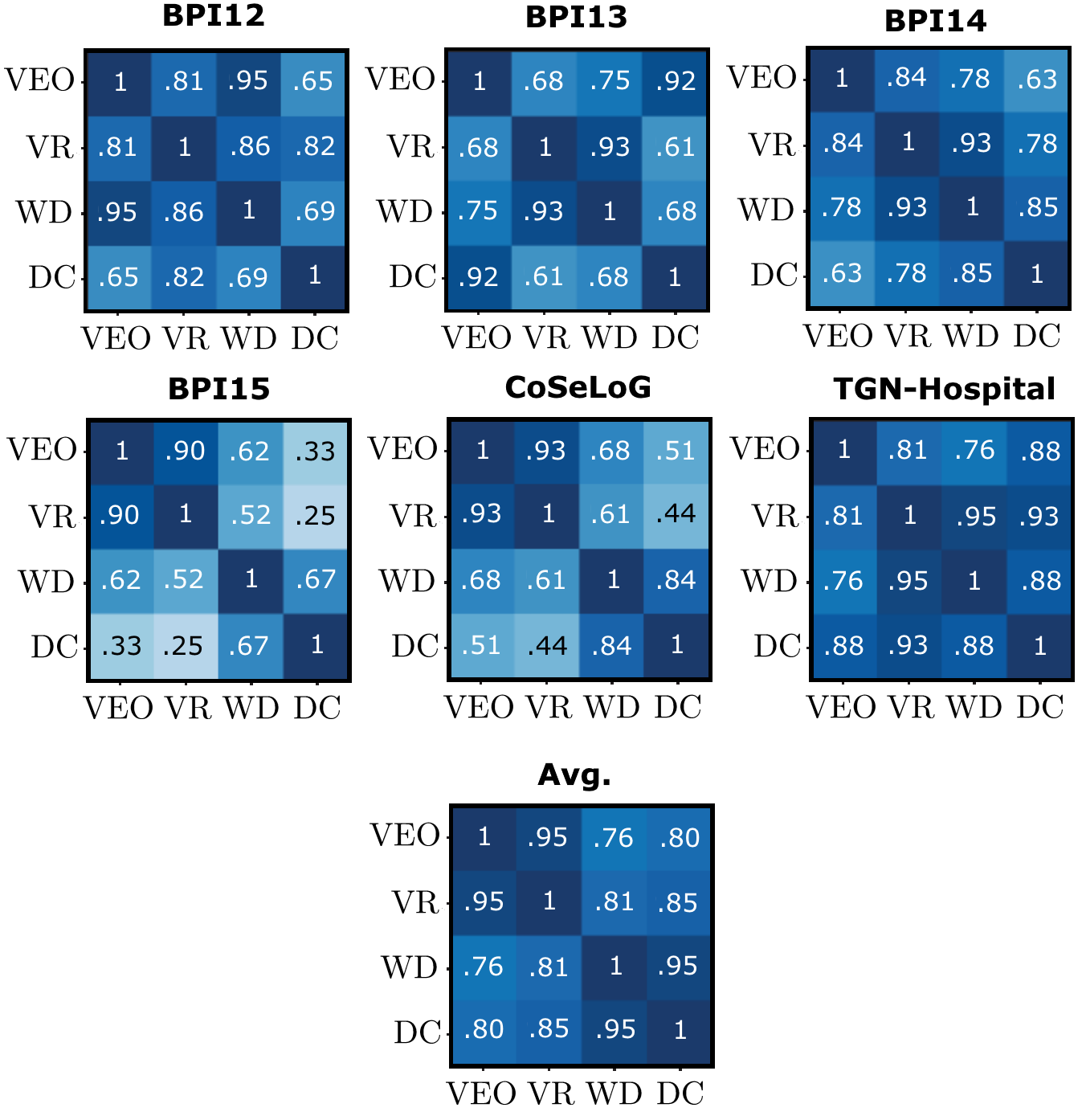}
\caption{Results of the $p$-values from the t-Tests according to the $sim$ parameter in the QS results (reprinted from \cite{kpppm}).}
\label{fig:app_kpppm:ttest_qs_sim}
\end{figure}

\begin{figure}[t!]
\centering
\includegraphics[width=0.67\linewidth]{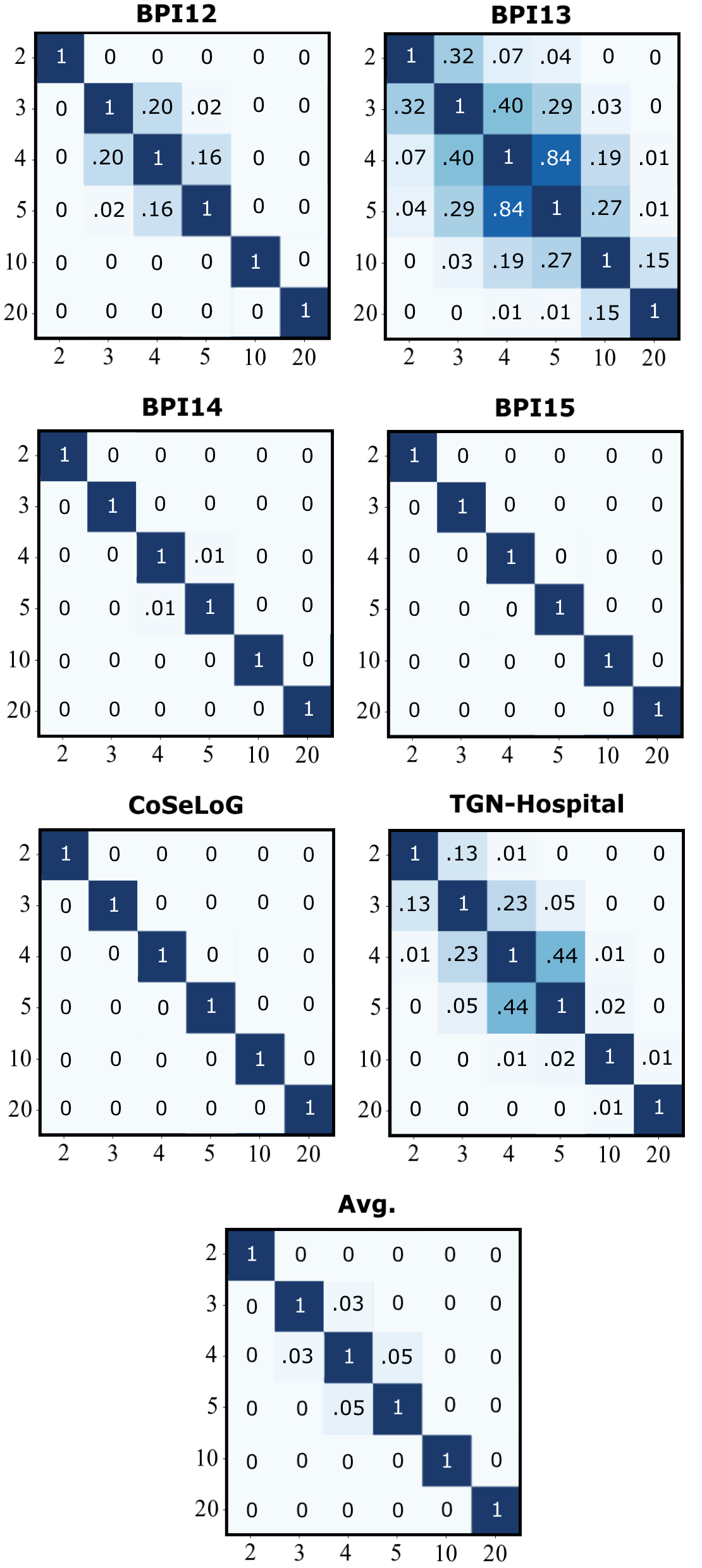}
\caption{Results of the $p$-values from the t-Tests according to the $k$ parameter in the ILS results (reprinted from \cite{kpppm}).}
\label{fig:app_kpppm:ttest_ils_k}
\end{figure}

\begin{figure}[t!]
\centering
\includegraphics[width=\linewidth]{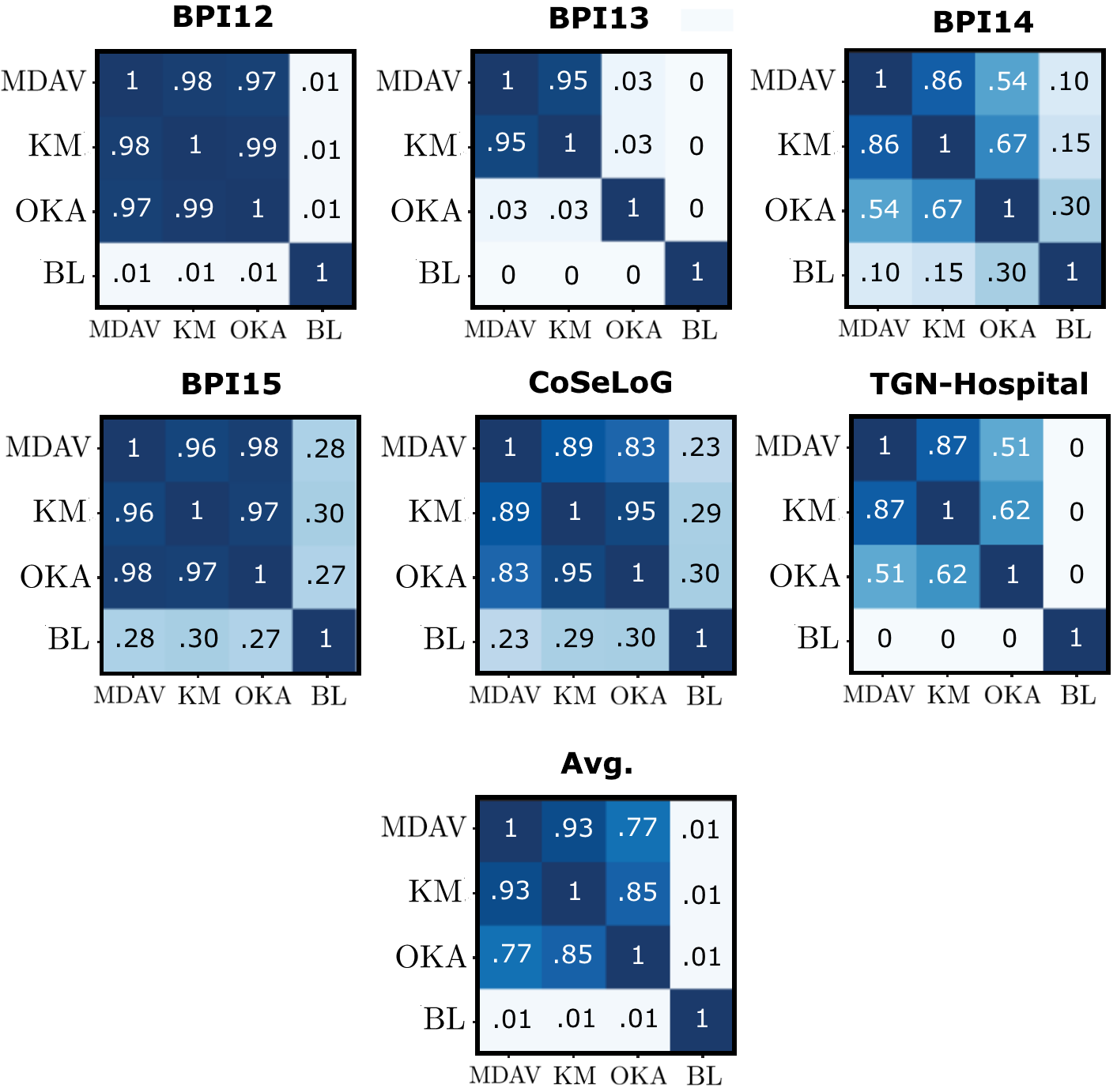}
\caption{Results of the $p$-values from the t-Tests according to the $clus$ parameter in the ILS results (reprinted from \cite{kpppm}).}
\label{fig:app_kpppm:ttest_ils_clus}
\end{figure}

\begin{figure}[t!]
\centering
\includegraphics[width=\linewidth]{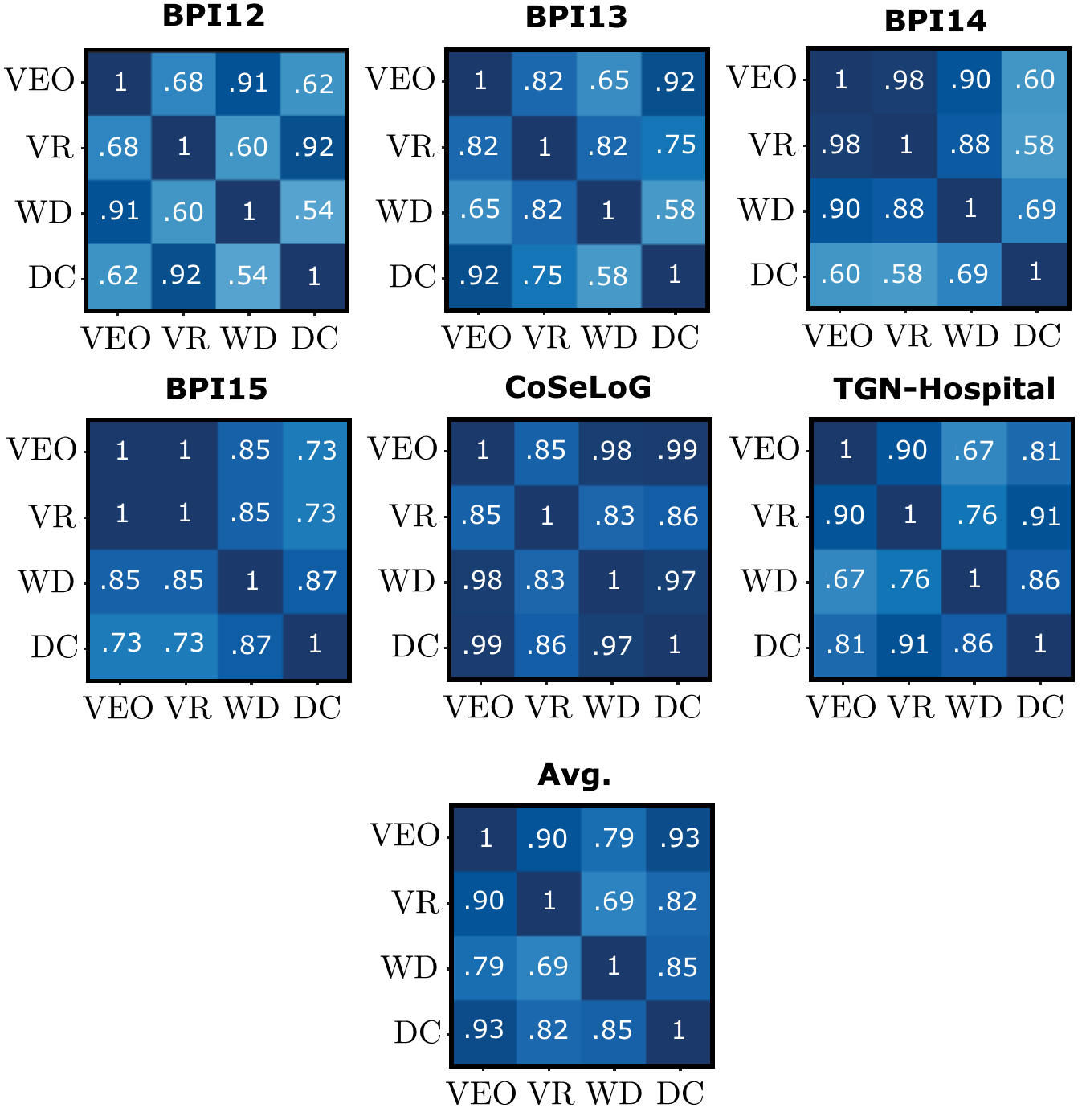}
\caption{Results of the $p$-values from the t-Tests according to the $sim$ parameter in the ILS results (reprinted from \cite{kpppm}).}
\label{fig:app_kpppm:ttest_ils_sim}
\end{figure}

\bibliographystyle{abbrv}
{\footnotesize \bibliography{Thesis}}

\begin{thebibliography}{100}

\bibitem{logrepository}
{4TU.ResearchData Consortium}.
\newblock {4TU.ResearchData -- Science, Engineering, Design}.
\newblock \url{https://data.4tu.nl}, 2013.
\newblock Accessed: 2021-07-11.

\bibitem{adar2007user}
E.~Adar.
\newblock {User 4xxxxx9: Anonymizing Query Logs}.
\newblock In {\em Proceedings of the 16th International Conference on World
  Wide Web -- Query Log Analysis: Social And Technological Challenges
  Workshop}, pages 1--8, Banff, Canada, 2007.

\bibitem{agrawal2000ppdm}
R.~Agrawal and R.~Srikant.
\newblock {Privacy-Preserving Data Mining}.
\newblock In {\em Proceedings of the 3rd ACM SIGMOD International Conference on
  Management of Data}, pages 439--450, Dallas, USA, 2000. ACM.

\bibitem{ahmed2017false}
M.~Ahmed and A.~S. S.~M. Barkat~Ullah.
\newblock {False Data Injection Attacks in Healthcare}.
\newblock In {\em Proceedings of the 15th Australasian Conference on Data
  Mining}, pages 192--202, Melbourne, Australia, 2017. Springer.

\bibitem{al2020intelligence}
F.~Al-Turjman, M.~H. Nawaz, and U.~D. Ulusar.
\newblock {Intelligence in the Internet of Medical Things era: A systematic
  review of current and future trends}.
\newblock {\em Computer Communications}, 150:644--660, 2020.

\bibitem{alvarez2018discovering}
C.~Alvarez, E.~Rojas, M.~Arias, J.~Munoz-Gama, M.~Sep{\'u}lveda, V.~Herskovic,
  and D.~Capurro.
\newblock {Discovering role interaction models in the Emergency Room using
  Process Mining}.
\newblock {\em Journal of Biomedical Informatics}, 78:60--77, 2018.

\bibitem{alves2004alpha1}
A.~K. Alves~de Medeiros, B.~F. van Dongen, W.~M.~P. van~der Aalst, and A.~J.
  M.~M. Weijters.
\newblock Process mining: Extending the $\alpha$-algorithm to mine short loops.
\newblock Technical Report WP 113, BETA Working Paper Series, Technische
  Universiteit Eindhoven, Eindhoven, The Netherlands, 2004.

\bibitem{adi2018alzheimer}
{Alzheimer's Disease International}.
\newblock {World Alzheimer Report 2018 -- The state of the art of dementia
  research: New frontiers}.
\newblock Technical report, Alzheimer's Disease International, London, UK,
  September 2018.

\bibitem{anderson2016mobile}
K.~Anderson, O.~Burford, and L.~Emmerton.
\newblock {Mobile Health Apps to Facilitate Self-Care: A Qualitative Study of
  User Experiences}.
\newblock {\em PLOS ONE}, 11(5):e0156164, 2016.

\bibitem{andrade2019cognitive}
R.~O. Andrade and S.~G. Yoo.
\newblock {Cognitive security: A comprehensive study of cognitive science in
  cybersecurity}.
\newblock {\em Journal of Information Security and Applications}, 48:102352,
  2019.

\bibitem{apwg2020trends}
{Anti-Phishing Working Group}.
\newblock {Phishing Activity Trends Report -- 3rd Quarter 2020}.
\newblock Technical report, Anti-Phishing Working Group, November 2020.

\bibitem{anwar2018green}
M.~Anwar, A.~H. Abdullah, A.~Altameem, K.~N. Qureshi, F.~Masud, M.~Faheem,
  Y.~Cao, and R.~Kharel.
\newblock {Green Communication for Wireless Body Area Networks: Energy Aware
  Link Efficient Routing Approach}.
\newblock {\em Sensors}, 18(10):3237, 2018.

\bibitem{baker2017process}
K.~Baker, E.~Dunwoodie, R.~G. Jones, A.~Newsham, O.~Johnson, C.~P. Price,
  J.~Wolstenholme, J.~Leal, P.~McGinley, C.~Twelves, and G.~Hall.
\newblock {Process mining routinely collected electronic health records to
  define real-life clinical pathways during chemotherapy}.
\newblock {\em International Journal of Medical Informatics}, 103:32--41, 2017.

\bibitem{bakiri2018hardware}
M.~Bakiri, C.~Guyeux, J.-F. Couchot, L.~Marangio, and S.~Galatolo.
\newblock {A Hardware and Secure Pseudorandom Generator for Constrained
  Devices}.
\newblock {\em IEEE Transactions on Industrial Informatics}, 14(8):3754--3765,
  2018.

\bibitem{ball2001health}
M.~J. Ball and J.~Lillis.
\newblock {E-health: transforming the physician/patient relationship}.
\newblock {\em International Journal of Medical Informatics}, 61(1):1--10,
  2001.

\bibitem{basole2015visual}
R.~C. Basole, H.~Park, M.~Gupta, M.~L. Braunstein, D.~H. Chau, and M.~Thompson.
\newblock {A Visual Analytics Approach to Understanding Care Process Variation
  and Conformance}.
\newblock In {\em Proceedings of the 6th Workshop on Visual Analytics in
  Healthcare}, pages 1--8, Chicago, USA, 2015.

\bibitem{batista2015monitoring}
E.~Batista, F.~Borr{\`a}s, and A.~Mart{\'i}nez-Ballest{\'e}.
\newblock {Monitoring People with MCI: Deployment in a Real Scenario for
  Low-Budget Smartphones}.
\newblock In {\em Proceedings of the 6th International Conference on
  Information, Intelligence, Systems and Applications}, pages 1--6, Corf{\'u},
  Greece, 2015. IEEE.

\bibitem{batista2019pmcae}
E.~Batista, F.~Falcone, A.~Mart{\'i}nez-Ballest{\'e}, and A.~Solanas.
\newblock {The Promising Future of Process Mining with the Internet of Events
  in Context-Aware Environments}.
\newblock In {\em Proceedings of the 2nd International Conference on Sensing
  and Instrumentation in IoT Era}, pages 1--6, Lisbon, Portugal, 2019. IEEE.

\bibitem{batista2018hgis}
E.~Batista, A.~Mart{\'i}nez-Ballest{\'e}, M.~Pe{\~n}a, X.~Singla, and
  A.~Solanas.
\newblock {HGIS: A Healthcare-Oriented Approach to Geographic Information
  Systems}.
\newblock In {\em Proceedings of the 6th International Conference on
  Applications in Electronics Pervading Industry, Environment and Society},
  pages 59--65, Pisa, Italy, 2018. Springer.

\bibitem{kpppm}
E.~Batista, A.~Mart{\'i}nez-Ballest{\'e}, and A.~Solanas.
\newblock {Privacy-Preserving Process Mining: A Microaggregation-based
  Approach}.
\newblock {\em Journal of Information Security and Applications}, Submitted.

\bibitem{batista2021sensors}
E.~Batista, M.~A. Moncusi, P.~L{\'o}pez-Aguilar, A.~Mart{\'i}nez-Ballest{\'e},
  and A.~Solanas.
\newblock {Sensors for Context-Aware Smart Healthcare: A Security Perspective}.
\newblock {\em Sensors}, 21(20):6886, 2021.

\bibitem{batista2018process}
E.~Batista and A.~Solanas.
\newblock {Process Mining in Healthcare: A Systematic Review}.
\newblock In {\em Proceedings of the 9th International Conference on
  Information, Intelligence, Systems and Applications}, pages 1--6, Zakynthos,
  Greece, 2018. IEEE.

\bibitem{batista2019skip}
E.~Batista and A.~Solanas.
\newblock {Skip Miner: Towards the Simplification of Spaghetti-like Business
  Process Models}.
\newblock In {\em Proceedings of the 10th International Conference on
  Information, Intelligence, Systems and Applications}, pages 1--6, Patras,
  Greece, 2019. IEEE.

\bibitem{batista2021uniformization}
E.~Batista and A.~Solanas.
\newblock {A uniformization-based approach to preserve individuals' privacy
  during process mining analyses}.
\newblock {\em Peer-to-Peer Networking and Applications}, 14(3):1500--1519,
  2021.

\bibitem{bauer2019elpaas}
M.~Bauer, S.~A. Fahrenkrog-Petersen, A.~Koschmider, F.~Mannhardt, H.~van~der
  Aa, and M.~Weidlich.
\newblock {ELPaaS: Event Log Privacy as a Service}.
\newblock In {\em Proceedings of the 17th International Conference on Business
  Process Management -- Dissertation Award, Doctoral Consortium and
  Demonstration Track}, pages 159--163, Vienna, Austria, 2019. Springer.

\bibitem{bautista2019executive}
C.~A. Bautista, P.~Nydahl, M.~K. Bader, S.~Livesay, A.-K. Cassier-Woidasky, and
  D.~M. Olson.
\newblock {Executive Summary: Post--Intensive Care Syndrome in the
  Neurocritical Intensive Care Unit}.
\newblock {\em Journal of Neuroscience Nursing}, 51(4):158--161, 2019.

\bibitem{belanger2011privacy}
F.~B{\'e}langer and R.~E. Crossler.
\newblock {Privacy in the digital age: A review of information privacy research
  in information systems}.
\newblock {\em MIS Quarterly}, 35(4):1017--1041, 2011.

\bibitem{bilal2017authentication}
M.~Bilal and S.-G. Kang.
\newblock {An Authentication Protocol for Future Sensor Networks}.
\newblock {\em Sensors}, 17(5):979, 2017.

\bibitem{blum2008workflow}
T.~Blum, N.~Padoy, H.~Feu{\ss}ner, and N.~Navab.
\newblock {Workflow mining for visualization and analysis of surgeries}.
\newblock {\em International Journal of Computer Assisted Radiology and
  Surgery}, 3(5):379--386, 2008.

\bibitem{zarpelao2017survey}
B.~Bogaz~Zarpel{\~a}o, R.~Sanches~Miani, C.~Toshio~Kawakani, and S.~C.
  de~Alvarenga.
\newblock {A survey of intrusion detection in Internet of Things}.
\newblock {\em Journal of Network and Computer Applications}, 84:25--37, 2017.

\bibitem{buck2017appip}
C.~Buck and S.~Burster.
\newblock {App Information Privacy Concerns}.
\newblock In {\em Proceedings of the 23rd Americas Conference on Information
  Systems}, pages 2100--2109, Boston, USA, 2017.

\bibitem{coselog}
J.~C. A.~M. Buijs.
\newblock {Receipt phase of an environmental permit application process
  (`WABO'), CoSeLoG project}.
\newblock
  \url{https://doi.org/10.4121/uuid:a07386a5-7be3-4367-9535-70bc9e77dbe6},
  2014.
\newblock Eindhoven University of Technology. Collection. Accessed: 2021-07-13.

\bibitem{burattin2015toward}
A.~Burattin, M.~Conti, and D.~Turato.
\newblock {Toward an Anonymous Process Mining}.
\newblock In {\em Proceedings of the 3rd International Conference on Future
  Internet of Things \& Cloud}, pages 58--63, Rome, Italy, 2015. IEEE.

\bibitem{byun2007kmember}
J.-W. Byun, A.~Kamra, E.~Bertino, and N.~Li.
\newblock {Efficient $k$-Anonymization Using Clustering Techniques}.
\newblock In {\em Proceedings of the 12th International Conference on Database
  Systems for Advanced Applications}, pages 188--200, Bangkok, Thailand, 2007.
  Springer.

\bibitem{cao2012publishing}
J.~Cao and P.~Karras.
\newblock {Publishing Microdata with a Robust Privacy Guarantee}.
\newblock {\em Proceedings of the VLDB Endowment}, 5(11):1388--1399, 2012.

\bibitem{caragliu2011smart}
A.~Caragliu, C.~Del~Bo, and P.~Nijkamp.
\newblock {Smart cities in Europe}.
\newblock {\em Journal of Urban Technology}, 18(2):65--82, 2011.

\bibitem{maita2018systematic}
A.~R. C{\'a}rdenas~Maita, L.~Corr{\^e}a~Martins, C.~R. L{\'o}pez~Paz,
  L.~Rafferty, P.~C.~K. Hung, S.~Marques~Peres, and M.~Fantinato.
\newblock {A systematic mapping study of process mining}.
\newblock {\em Enterprise Information Systems}, 12(5):505--549, 2018.

\bibitem{caron2013healthcare}
F.~Caron, J.~Vanthienen, and B.~Baesens.
\newblock {Healthcare Analytics: Examining the Diagnosis-Treatment Cycle}.
\newblock {\em Procedia Technology}, 9:996--1004, 2013.

\bibitem{caron2014monitoring}
F.~Caron, J.~Vanthienen, K.~Vanhaecht, E.~Van~Limbergen, J.~De~Weerdt, and
  B.~Baesens.
\newblock {Monitoring care processes in the gynecologic oncology department}.
\newblock {\em Computers in Biology and Medicine}, 44:88--96, 2014.

\bibitem{caron2014process}
F.~Caron, J.~Vanthienen, K.~Vanhaecht, E.~Van~Limbergen, J.~Deweerdt, and
  B.~Baesens.
\newblock {A process mining-based investigation of adverse events in care
  processes}.
\newblock {\em Health Information Management Journal}, 43(1):16--25, 2014.

\bibitem{carpenter2020developments}
S.~M. Carpenter, M.~Menictas, I.~Nahum-Shani, D.~W. Wetter, and S.~A. Murphy.
\newblock {Developments in Mobile Health Just-in-Time Adaptive Interventions
  for Addiction Science}.
\newblock {\em Current Addiction Reports}, pages 1--11, 2020.

\bibitem{casino2015context}
F.~Casino, E.~Batista, C.~Patsakis, and A.~Solanas.
\newblock {Context-Aware Recommender for Smart Health}.
\newblock In {\em Proceedings of the IEEE 1st International Smart Cities
  Conference}, pages 1--2, Guadalajara, M{\'e}xico, 2015. IEEE.

\bibitem{casino2017healthy}
F.~Casino, C.~Patsakis, E.~Batista, F.~Borr{\`a}s, and
  A.~Mart{\'i}nez-Ballest{\'e}.
\newblock {Healthy Routes in the Smart City: A Context-Aware Mobile
  Recommender}.
\newblock {\em IEEE Software}, 34(6):42--47, 2017.

\bibitem{casino2018smart}
F.~Casino, C.~Patsakis, E.~Batista, O.~Postolache,
  A.~Mart{\'i}nez-Ballest{\'e}, and A.~Solanas.
\newblock {Smart Healthcare in the IoT Era: A Context-Aware Recommendation
  Example}.
\newblock In {\em Proceedings of the 1st International Symposium on Sensing and
  Instrumentation in IoT Era}, pages 1--4, Shanghai, China, 2018. IEEE.

\bibitem{casson2019wearable}
A.~J. Casson.
\newblock {Wearable EEG and beyond}.
\newblock {\em Biomedical Engineering Letters}, 9(1):53--71, 2019.

\bibitem{cavoukian20107principles}
A.~Cavoukian.
\newblock {Privacy by Design -- The 7 Foundational Principles -- Implementation
  and Mapping of Fair Information Practices}.
\newblock Technical report, Information and Privacy Commissioner of Ontario,
  Ontario, Canada, May 2010.

\bibitem{cechetti2019developing}
N.~P. Cechetti, E.~A. Bellei, D.~Biduski, J.~P.~M. Rodriguez, M.~K. Roman, and
  A.~C.~B. De~Marchi.
\newblock {Developing and implementing a gamification method to improve user
  engagement: A case study with an m-Health application for hypertension
  monitoring}.
\newblock {\em Telematics and Informatics}, 41:126--138, 2019.

\bibitem{chang2020impact}
H.~Chang, J.~Y. Yu, S.~Y. Yoon, S.~Y. Hwang, H.~Yoon, W.~C. Cha, M.~S. Sim,
  I.~J. Jo, and T.~Kim.
\newblock {Impact of COVID-19 Pandemic on the Overall Diagnostic and
  Therapeutic Process for Patients of Emergency Department and Those with Acute
  Cerebrovascular Disease}.
\newblock {\em Journal of Clinical Medicine}, 9(12):3842, 2020.

\bibitem{cheung2021wearable}
C.~C. Cheung, J.~E. Olgin, and B.~K. Lee.
\newblock {Wearable cardioverter-defibrillators: A review of evidence and
  indications}.
\newblock {\em Trends in Cardiovascular Medicine}, 31(3):196--201, 2021.

\bibitem{clarke1997introduction}
R.~Clarke.
\newblock {Introduction to Dataveillance and Information Privacy, and
  Definitions of Terms}.
\newblock Technical report, Xamax Consultancy Pty Ltd., 1997.

\bibitem{classen2018anatomy}
J.~Classen, D.~Wegemer, P.~Patras, T.~Spink, and M.~Hollick.
\newblock {Anatomy of a Vulnerable Fitness Tracking System: Dissecting the
  Fitbit Cloud, App, and Firmware}.
\newblock {\em Proceedings of the ACM on Interactive, Mobile, Wearable and
  Ubiquitous Technologies}, 2(1):1--24, 2018.

\bibitem{cockcroft2016relationship}
S.~Cockcroft and S.~Rekker.
\newblock {The relationship between culture and information privacy policy}.
\newblock {\em Electronic Markets}, 26(1):55--72, 2016.

\bibitem{conforti2017filtering}
R.~Conforti, M.~La~Rosa, and A.~H.~M. ter Hofstede.
\newblock {Filtering Out Infrequent Behavior from Business Process Event Logs}.
\newblock {\em IEEE Transactions on Knowledge and Data Engineering},
  29(2):300--314, 2017.

\bibitem{dasaklis2018blockchain}
T.~K. Dasaklis, F.~Casino, and C.~Patsakis.
\newblock {Blockchain Meets Smart Health: Towards Next Generation Healthcare
  Services}.
\newblock In {\em Proceedings of the 9th International Conference on
  Information, Intelligence, Systems and Applications}, pages 1--8, Zakynthos,
  Greece, 2018.

\bibitem{davies2016data}
S.~Davies.
\newblock {The Data Protection Regulation: A Triumph of Pragmatism over
  Principle?}
\newblock {\em European Data Protection Law Review}, 2:290, 2016.

\bibitem{de2018health}
M.~A. de~Carvalho~Junior and P.~Bandiera-Paiva.
\newblock {Health Information System Role-Based Access Control Current Security
  Trends and Challenges}.
\newblock {\em Journal of Healthcare Engineering}, 2018, 2018.

\bibitem{de2018ddos}
M.~De~Donno, N.~Dragoni, A.~Giaretta, and A.~Spognardi.
\newblock {DDoS-Capable IoT Malwares: Comparative Analysis and Mirai
  Investigation}.
\newblock {\em Security and Communication Networks}, 2018, 2018.

\bibitem{fuentes2018attribute}
J.~M. de~Fuentes, L.~Gonzalez-Manzano, A.~Solanas, and F.~Veseli.
\newblock {Attribute-Based Credentials for Privacy-Aware Smart Health Services
  in IoT-based Smart Cities}.
\newblock {\em Computer}, 51(7):44--53, 2018.

\bibitem{de2016new}
P.~De~Hert and V.~Papakonstantinou.
\newblock {The new General Data Protection Regulation: Still asound system for
  the protection of individuals?}
\newblock {\em Computer Law \& Security Review}, 32(2):179--194, 2016.

\bibitem{vries2017towards}
G.-J. de~Vries, R.~A. Quintano~Neira, G.~Geleijnse, P.~Dixit, and B.~F. Mazza.
\newblock {Towards Process Mining of EMR Data: Case Study for Sepsis
  Management}.
\newblock In {\em Proceedings of the 10th International Joint Conference on
  Biomedical Engineering Systems and Technologies}, volume~5, pages 585--593,
  Porto, Portugal, 2017.

\bibitem{delias2014applying}
P.~Delias, P.~Manolitzas, E.~Grigoroudis, and N.~Matsatsinis.
\newblock {Applying Process Mining to the Emergency Department}.
\newblock In {\em Encyclopedia of Business Analytics and Optimization}, pages
  168--178. IGI Global, 2014.

\bibitem{detro2016enhancing}
S.~P. Detro, D.~Morozov, M.~Lezoche, H.~Panetto, E.~P. Santos, and
  M.~Zdravkovic.
\newblock {Enhancing Semantic Interoperability in Healthcare using Semantic
  Process Mining}.
\newblock In {\em Proceedings of the 6th International Conference on
  Information Society and Techology}, pages 80--85, Kopaonik, Serbia, 2016.

\bibitem{dijkman2008bpmn}
R.~M. Dijkman, M.~Dumas, and C.~Ouyang.
\newblock {Semantics and Analysis of Business Process Models in BPMN}.
\newblock {\em Information and Software technology}, 50(12):1281--1294, 2008.

\bibitem{domingo20073d}
J.~Domingo-Ferrer.
\newblock {A Three-Dimensional Conceptual Framework for Database Privacy}.
\newblock In {\em Proceedings of the 4th Workshop on Secure Data Management},
  pages 193--202, Vienna, Austria, 2007. Springer.

\bibitem{domingo2009measuring}
J.~Domingo-Ferrer and D.~Rebollo-Monedero.
\newblock {Measuring Risk and Utility of Anonymized Data Using Information
  Theory}.
\newblock In {\em Proceedings of the Joint Conference EDBT/ICDT Workshops},
  pages 126--130, Saint-Petersburg, Russia, 2009.

\bibitem{domingo2009h}
J.~Domingo-Ferrer, A.~Solanas, and J.~Castell{\`a}-Roca.
\newblock {\textit{h}(\textit{k})-private information retrieval from
  privacy-uncooperative queryable databases}.
\newblock {\em Online Information Review}, 33(4):720--744, 2009.

\bibitem{domingo2001quantitative}
J.~Domingo-Ferrer and V.~Torra.
\newblock {A Quantitative Comparison of Disclosure Control Methods for
  Microdata}.
\newblock In P.~Doyle, J.~Lane, J.~Theeuwes, and L.~Zayatz, editors, {\em
  Confidentiality, disclosure and data access: theory and practical
  applications for statistical agencies}, pages 111--134. Elsevier, 2001.

\bibitem{domingo2005ordinal}
J.~Domingo-Ferrer and V.~Torra.
\newblock {Ordinal, Continuous and Heterogeneous $k$-Anonymity Through
  Microaggregation}.
\newblock {\em Data Mining and Knowledge Discovery}, 11(2):195--212, 2005.

\bibitem{doshi2018machine}
R.~Doshi, N.~Apthorpe, and N.~Feamster.
\newblock {Machine Learning DDoS Detection for Consumer Internet of Things
  Devices}.
\newblock In {\em Proceedings of the 39th IEEE Security and Privacy Workshops},
  pages 29--35, San Francisco, USA, 2018. IEEE.

\bibitem{duan2004protecting}
Y.~Duan and J.~Canny.
\newblock {Protecting User Data in Ubiquitous Computing: Towards Trustworthy
  Environments}.
\newblock In {\em Proceedings of the 4th International Workshop on Privacy
  Enhancing Technologies}, pages 167--185, Toronto, Canada, 2004. Springer.

\bibitem{duma2020ad}
D.~Duma and R.~Aringhieri.
\newblock {An ad hoc process mining approach to discover patient paths of an
  Emergency Department}.
\newblock {\em Flexible Services and Manufacturing Journal}, 32(1):6--34, 2020.

\bibitem{dumas2013bpm}
M.~Dumas, M.~La~Rosa, J.~Mendling, and H.~A. Reijers.
\newblock {\em {Fundamentals of Business Process Management}}.
\newblock Springer, 2013.

\bibitem{dumas2005pais}
M.~Dumas, W.~M.~P. van~der Aalst, and A.~H.~M. Ter~Hofstede.
\newblock {\em {Process-Aware Information Systems: Bridging People and Software
  through Process Technology}}.
\newblock John Wiley \& Sons, 2005.

\bibitem{dunkl2011assessing}
R.~Dunkl, K.~A. Fr{\"o}schl, W.~Grossmann, and S.~Rinderle-Ma.
\newblock {Assessing Medical Treatment Compliance Based on Formal Process
  Modeling}.
\newblock In {\em Proceedings of the 5th Symposium of the Austrian HCI and
  Usability Engineering Group}, pages 533--546, Graz, Austria, 2011. Springer.

\bibitem{dwork2006differential}
C.~Dwork.
\newblock Differential privacy.
\newblock In {\em Proceedings of the 33rd International Conference on Automata,
  Languages and Programming}, pages 1--12, Venice, Italy, 2006. Springer.

\bibitem{edwards2016privacy}
L.~Edwards.
\newblock {Privacy, security and data protection in smart cities: A critical EU
  law perspective}.
\newblock {\em European Data Protection Law Review}, 2:1--28, 2016.

\bibitem{elkoumy2020secure}
G.~Elkoumy, S.~A. Fahrenkrog-Petersen, M.~Dumas, P.~Laud, A.~Pankova, and
  M.~Weidlich.
\newblock {Secure Multi-Party Computation for Inter-Organizational Process
  Mining}.
\newblock In {\em Enterprise, Business-Process and Information Systems
  Modeling}, pages 166--181. Springer, 2020.

\bibitem{elkoumy2020shareprom}
G.~Elkoumy, S.~A. Fahrenkrog-Petersen, M.~Dumas, P.~Laud, A.~Pankova, and
  M.~Weidlich.
\newblock {Shareprom: A Tool for Privacy-Preserving Inter-Organizational
  Process Mining}.
\newblock In {\em Proceedings of the 18th International Conference on Business
  Process Management -- PhD/Demos}, pages 72--76. Springer, 2020.

\bibitem{elkoumy2021privacy}
G.~Elkoumy, S.~A. Fahrenkrog-Petersen, M.~F. Sani, A.~Koschmider, F.~Mannhardt,
  S.~Nu{\~n}ez~von Voigt, M.~Rafiei, and L.~von Waldthausen.
\newblock {Privacy and Confidentiality in Process Mining -- Threats and
  Research Challenges}.
\newblock {\em ACM Transactions on Management Information System}, 13(1):1--17,
  2021.

\bibitem{elkoumy2020privacy}
G.~Elkoumy, A.~Pankova, and M.~Dumas.
\newblock {Privacy-Preserving Directly-Follows Graphs: Balancing Risk and
  Utility in Process Mining}.
\newblock {\em arXiv preprint arXiv:2012.01119}, pages 1--35, 2020.

\bibitem{elkoumy2021mine}
G.~Elkoumy, A.~Pankova, and M.~Dumas.
\newblock {Mine Me but Don't Single Me Out: Differentially Private Event Logs
  for Process Mining}.
\newblock In {\em Proceedings of the 3rd International Conference on Process
  Mining}, pages 80--87, Eindhoven, The Netherlands, 2021. IEEE.

\bibitem{erdogan2018systematic}
T.~G. Erdogan and A.~Tarhan.
\newblock {Systematic Mapping of Process Mining Studies in Healthcare}.
\newblock {\em IEEE Access}, 6:24543--24567, 2018.

\bibitem{eudpd}
{European Union}.
\newblock {Directive 95/46/EC of the European Parliament and of the Council of
  24 October 1995 on the protection of individuals with regard to the
  processing of personal data and on the free movement of such data}.
\newblock {\em Official Journal of the European Union}, L281:31--50, 1995.

\bibitem{eugdpr}
{European Union}.
\newblock {Regulation (EU) 2016/679 of the European Parliament and of the
  Council of 27 April 2016 on the protection of natural persons with regard to
  the processing of personal data and on the free movement of such data, and
  repealing Directive 95/46/EC (General Data Protection Regulation)}.
\newblock {\em Official Journal of the European Union}, L119:1--88, 2016.

\bibitem{eysenbach2001ehealth}
G.~Eysenbach.
\newblock {What is e-health?}
\newblock {\em Journal of Medical Internet Research}, 3(2):e20, 2001.

\bibitem{fahrenkrog2019providing}
S.~A. Fahrenkrog-Petersen.
\newblock {Providing Privacy Guarantees in Process Mining}.
\newblock In {\em Proceedings of the 31st International Conference on Advanced
  Information Systems Engineering -- Doctoral Consortium}, pages 23--30, Rome,
  Italy, 2019. Springer.

\bibitem{fahrenkrog2019pretsa}
S.~A. Fahrenkrog-Petersen, H.~van~der Aa, and M.~Weidlich.
\newblock {PRETSA: Event Log Sanitization for Privacy-aware Process Discovery}.
\newblock In {\em Proceedings of the 1st International Conference on Process
  Mining}, pages 1--8, Aachen, Germany, 2019. IEEE.

\bibitem{fernandez2013process}
C.~Fern{\'a}ndez-Llatas, J.-M. Benedi, J.~M. Garc{\'i}a-G{\'o}mez, and
  V.~Traver.
\newblock {Process Mining for Individualized Behavior Modeling Using Wireless
  Tracking in Nursing Homes}.
\newblock {\em Sensors}, 13(11):15434--15451, 2013.

\bibitem{fernandez2015process}
C.~Fern{\'a}ndez-Llatas, A.~Lizondo, E.~Monton, J.-M. Benedi, and V.~Traver.
\newblock {Process Mining Methodology for Health Process Tracking Using
  Real-Time Indoor Location Systems}.
\newblock {\em Sensors}, 15(12):29821--29840, 2015.

\bibitem{ferre2021smart}
M.~Ferre, E.~Batista, A.~Solanas, and A.~Mart{\'i}nez-Ballest{\'e}.
\newblock {Smart Health-Enhanced Early Mobilisation in Intensive Care Units}.
\newblock {\em Sensors}, 21(16):5408, 2021.

\bibitem{firs2019respiratory}
{Forum of International Respiratory Societies}.
\newblock {The Global Impact of Respiratory Disease -- Second Edition}.
\newblock Technical report, European Respiratory Society, Sheffield, UK, 2017.

\bibitem{fuchs2017internet}
C.~Fuchs and D.~Trottier.
\newblock {Internet Surveillance after Snowden: A Critical Empirical Study of
  Computer Experts' Attitudes on Commercial and State Surveillance of the
  Internet and Social Media post-Edward Snowden}.
\newblock {\em Journal of Information, Communication and Ethics in Society},
  15(4):412--444, 2017.

\bibitem{dos2019process}
C.~d.~S. Garcia, A.~Meincheim, E.~R. Faria~Junior, M.~R. Dallagassa, D.~M.~V.
  Sato, D.~R. Carvalho, E.~A.~P. Santos, and E.~E. Scalabrin.
\newblock {Process mining techniques and applications -- A systematic mapping
  study}.
\newblock {\em Expert Systems with Applications}, 133:260--295, 2019.

\bibitem{gatta2019opportunities}
R.~Gatta, M.~Vallati, C.~Fernandez-Llatas, A.~Martinez-Millana, S.~Orini,
  L.~Sacchi, J.~Lenkowicz, M.~Marcos, J.~Munoz-Gama, M.~Cuendet, B.~de~Bari,
  L.~Marco-Ruiz, A.~Stefanini, and M.~Castellano.
\newblock {Clinical Guidelines: A Crossroad of Many Research Areas. Challenges
  and Opportunities in Process Mining for Healthcare}.
\newblock In {\em Proceedings of the 17th International Conference on Business
  Process Management Workshops}, pages 545--556, Vienna, Austria, 2019.
  Springer.

\bibitem{meteo_gencat}
{Generalitat de Catalunya, Departament de Territori i Sostenibilitat}.
\newblock {Vols saber què respires?}
\newblock
  \url{http://mediambient.gencat.cat/ca/05_ambits_dactuacio/atmosfera/qualitat_de_laire/vols-saber-que-respires},
  2021.
\newblock In Catalan. Accessed: 2021-07-22.

\bibitem{ghasemi2016review}
M.~Ghasemi and D.~Amyot.
\newblock {Process Mining in Healthcare -- A Systematised Literature Review}.
\newblock {\em International Journal of Electronic Healthcare}, 9(1):60--88,
  2016.

\bibitem{goyal2016mind}
R.~Goyal, N.~Dragoni, and A.~Spognardi.
\newblock {Mind The Tracker You Wear -- A Security Analysis of Wearable Health
  Trackers}.
\newblock In {\em Proceedings of the 31st Annual ACM Symposium on Applied
  Computing}, pages 131--136, Pisa, Italy, 2016. ACM.

\bibitem{grando2011semantic}
M.~A. Grando, M.~H. Schonenberg, and W.~M.~P. van~der Aalst.
\newblock {Semantic-Based Conformance Checking of Computer Interpretable
  Medical Guidelines}.
\newblock In {\em Proceedings of the 4th International Joint Conference on
  Biomedical Engineering Systems and Technologies}, pages 285--300, Rome,
  Italy, 2011. Springer.

\bibitem{graves2008integrative}
B.~A. Graves.
\newblock {Integrative Literature Review: A Review of Literature Related to
  Geographical Information Systems, Healthcare Access, and Health Outcomes}.
\newblock {\em Perspectives in Health Information Management}, 5:1--11, 2008.

\bibitem{leakhiv}
J.~Griffiths.
\newblock {HIV status of over 14,000 people leaked online, Singapore
  authorities say}.
\newblock
  \url{https://edition.cnn.com/2019/01/28/health/hiv-status-data-leak-singapore-intl/index.html},
  January 2019.
\newblock Accessed: 2021-07-14.

\bibitem{gunther2007fuzzy}
C.~W. G{\"u}nther and W.~M.~P. van~der Aalst.
\newblock {Fuzzy Mining -- Adaptive Process Simplification Based on
  Multi-perspective Metrics}.
\newblock In {\em Proceedings of the 5th International Conference on Business
  Process Management}, pages 328--343, Brisbane, Australia, 2007. Springer.

\bibitem{halperin2008pacemakers}
D.~Halperin, T.~S. Heydt-Benjamin, B.~Ransford, S.~S. Clark, B.~Defend,
  W.~Morgan, K.~Fu, T.~Kohno, and W.~H. Maisel.
\newblock {Pacemakers and Implantable Cardiac Defibrillators: Software Radio
  Attacks and Zero-Power Defenses}.
\newblock In {\em Proceedings of the 29th IEEE Symposium on Security and
  Privacy}, pages 129--142, Oakland, USA, 2008. IEEE.

\bibitem{hammer2009reengineering}
M.~Hammer and J.~Champy.
\newblock {\em {Reengineering the Corporation: A Manifesto for Business
  Revolution}}.
\newblock Zondervan, 2009.

\bibitem{havey2005essential}
M.~Havey.
\newblock {\em {Essential Business Process Modeling}}.
\newblock O'Reilly Media, Inc., 2005.

\bibitem{hei2010defending}
X.~Hei, X.~Du, J.~Wu, and F.~Hu.
\newblock {Defending Resource Depletion Attacks on Implantable Medical
  Devices}.
\newblock In {\em Proceedings of the 53th IEEE Global Telecommunications
  Conference}, pages 1--5, Miami, USA, 2010. IEEE.

\bibitem{helfert2009challenges}
M.~Helfert.
\newblock {Challenges of business processes management in healthcare:
  Experience in the Irish healthcare sector}.
\newblock {\em Business Process Management Journal}, 15(6):937--952, 2009.

\bibitem{helm2019adopting}
E.~Helm, A.~M. Lin, D.~Baumgartner, A.~C. Lin, and J.~K{\"u}ng.
\newblock {Adopting Standard Clinical Descriptors for Process Mining Case
  Studies in Healthcare}.
\newblock In {\em Proceedings of the 17th International Conference on Business
  Process Management}, pages 608--619, Vienna, Austria, 2019. Springer.

\bibitem{helm2020towards}
E.~Helm, A.~M. Lin, D.~Baumgartner, A.~C. Lin, and J.~K{\"u}ng.
\newblock {Towards the Use of Standardized Terms in Clinical Case Studies for
  Process Mining in Healthcare}.
\newblock {\em International Journal of Environmental Research and Public
  Health}, 17(4):1348, 2020.

\bibitem{huang2018lbs}
H.~Huang, G.~Gartner, J.~M. Krisp, M.~Raubal, and N.~Van~de Weghe.
\newblock {Location based services: ongoing evolution and research agenda}.
\newblock {\em Journal of Location Based Services}, 12(2):63--93, 2018.

\bibitem{hundepool2012statistical}
A.~Hundepool, J.~Domingo-Ferrer, L.~Franconi, S.~Giessing, E.~S. Nordholt,
  K.~Spicer, and P.-P. De~Wolf.
\newblock {\em {Statistical Disclosure Control}}.
\newblock John Wiley \& Sons, 2012.

\bibitem{ieee1849}
{IEEE Standards Association}.
\newblock {1849-2016 -- IEEE Standard for eXtensible Event Stream (XES) for
  Achieving Interoperability in Event Logs and Event Streams}.
\newblock Technical Report 1849-2016, IEEE Standards Association, 2016.

\bibitem{isaak2018user}
J.~Isaak and M.~J. Hanna.
\newblock {User Data Privacy: Facebook, Cambridge Analytica, and Privacy
  Protection}.
\newblock {\em Computer}, 51(8):56--59, 2018.

\bibitem{istepanian2006mhealth}
R.~Istepanian, S.~Laxminarayan, and C.~S. Pattichis.
\newblock {\em {M-Health: Emerging Mobile Health Systems}}.
\newblock Springer, 2006.

\bibitem{jalali2020employees}
M.~S. Jalali, M.~Bruckes, D.~Westmattelmann, and G.~Schewe.
\newblock {Why Employees (Still) Click on Phishing Links: Investigation in
  Hospitals}.
\newblock {\em Journal of Medical Internet Research}, 22(1):e16775, 2020.

\bibitem{joseph2018adversarial}
A.~D. Joseph, B.~Nelson, B.~I.~P. Rubinstein, and J.~D. Tygar.
\newblock {\em {Adversarial Machine Learning}}.
\newblock Cambridge University Press, 2018.

\bibitem{karagiannis2007sox}
D.~Karagiannis, J.~Mylopoulos, and M.~Schwab.
\newblock {Business Process-Based Regulation Compliance: The Case of the
  Sarbanes-Oxley Act}.
\newblock In {\em Proceedings of the 15th IEEE International Requirements
  Engineering Conference}, pages 315--321, Delhi, India, 2007. IEEE.

\bibitem{kazlouski2021still}
A.~Kazlouski, T.~Marchioro, H.~Manifavas, and E.~P. Markatos.
\newblock {I still See You! Inferring Fitness Data from Encrypted Traffic of
  Wearables}.
\newblock In {\em Proceedings of the 14th International Joint Conference on
  Biomedical Engineering Systems and Technologies}, pages 369--376, Vienna,
  Austrial, 2021.

\bibitem{khanna2015urbanisation}
P.~Khanna.
\newblock {Urbanisation, Technology, and the Growth of Smart Cities}.
\newblock {\em Asian Management Insights}, 2(2):52--59, 2015.

\bibitem{kim2018air}
D.~Kim, Z.~Chen, L.-F. Zhou, and S.-X. Huang.
\newblock {Air pollutants and early origins of respiratory diseases}.
\newblock {\em Chronic Diseases and Translational Medicine}, 4(2):75--94, 2018.

\bibitem{kim2013discovery}
E.~Kim, S.~Kim, M.~Song, S.~Kim, D.~Yoo, H.~Hwang, and S.~Yoo.
\newblock {Discovery of Outpatient Care Process of a Tertiary University
  Hospital Using Process Mining}.
\newblock {\em Healthcare Informatics Research}, 19(1):42--49, 2013.

\bibitem{kontis2017future}
V.~Kontis, J.~E. Bennett, C.~D. Mathers, G.~Li, K.~Foreman, and M.~Ezzati.
\newblock {Future life expectancy in 35 industrialised countries: projections
  with a Bayesian model ensemble}.
\newblock {\em The Lancet}, 389(10076):1323--1335, 2017.

\bibitem{kotzanikolaou2016lightweight}
P.~Kotzanikolaou, C.~Patsakis, E.~Magkos, and M.~Korakakis.
\newblock {Lightweight private proximity testing for geospatial social
  networks}.
\newblock {\em Computer Communications}, 73:263--270, 2016.

\bibitem{koutra2013deltacon}
D.~Koutra, J.~T. Vogelstein, and C.~Faloutsos.
\newblock {DeltaCon: A Principled Massive-Graph Similarity Function}.
\newblock In {\em Proceedings of the SIAM International Conference on Data
  Mining}, pages 162--170, Austin, USA, 2013.

\bibitem{kroll2021enhancing}
J.~A. Kroll, J.~B. Michael, and D.~B. Thaw.
\newblock {Enhancing Cybersecurity via Artificial Intelligence: Risks, Rewards,
  and Frameworks}.
\newblock {\em Computer}, 54(6):64--71, 2021.

\bibitem{krumm2018ubiquitous}
J.~Krumm.
\newblock {\em {Ubiquitous Computing Fundamentals}}.
\newblock CRC Press, 2018.

\bibitem{kuehn2018pacemaker}
B.~M. Kuehn.
\newblock {Pacemaker Recall Highlights Security Concerns for Implantable
  Devices}.
\newblock {\em Circulation}, 138(1):1597--1598, 2018.

\bibitem{kune2013ghost}
D.~F. Kune, J.~Backes, S.~S. Clark, D.~Kramer, M.~Reynolds, K.~Fu, Y.~Kim, and
  W.~Xu.
\newblock {Ghost Talk: Mitigating EMI Signal Injection Attacks against Analog
  Sensors}.
\newblock In {\em Proceedings of the IEEE Symposium on Security and Privacy},
  pages 145--159, Berkeley, USA, 2013. IEEE.

\bibitem{kurniati2018process}
A.~P. Kurniati, G.~Hall, D.~Hogg, and O.~Johnson.
\newblock {Process mining in oncology using the MIMIC-III dataset}.
\newblock {\em Journal of Physics: Conference Series}, 971:012008, 2018.

\bibitem{kurniati2019assessment}
A.~P. Kurniati, E.~Rojas, D.~Hogg, G.~Hall, and O.~A. Johnson.
\newblock {The assessment of data quality issues for process mining in
  healthcare using Medical Information Mart for Intensive Care III, a freely
  available e-health record database}.
\newblock {\em Health Informatics Journal}, 25(4):1878--1893, 2019.

\bibitem{langheinrich2001privacy}
M.~Langheinrich.
\newblock {Privacy by Design -- Principles of Privacy-Aware Ubiquitous
  Systems}.
\newblock In {\em Proceedings of the 1st International Conference on Ubiquitous
  Computing}, pages 273--291, Atlanta, USA, 2001. Springer.

\bibitem{langheinrich2002privacy}
M.~Langheinrich.
\newblock {A Privacy Awareness System for Ubiquitous Computing Environments}.
\newblock In {\em Proceedings of the 2nd International Conference on Ubiquitous
  Computing}, pages 237--245, G{\"o}teborg, Sweden, 2002. Springer.

\bibitem{lara2018elliptic}
C.~A. Lara-Nino, A.~Diaz-Perez, and M.~Morales-Sandoval.
\newblock {Elliptic Curve Lightweight Cryptography: A Survey}.
\newblock {\em IEEE Access}, 6:72514--72550, 2018.

\bibitem{leemans2013inductive}
S.~J.~J. Leemans, D.~Fahland, and W.~M.~P. van~der Aalst.
\newblock {Discovering Block-Structured Process Models From Event Logs -- A
  Constructive Approach}.
\newblock In {\em Proceedings of the 34th International Conference on
  Applications and Theory of Petri Nets and Concurrency}, pages 311--329,
  Milan, Italy, 2013. Springer.

\bibitem{lenz2007support}
R.~Lenz and M.~Reichert.
\newblock {IT support for healthcare processes -- premises, challenges,
  perspectives}.
\newblock {\em Data \& Knowledge Engineering}, 61(1):39--58, 2007.

\bibitem{li2020modeling}
H.~Li and G.~Wen.
\newblock {Modeling reverse thinking for machine learning}.
\newblock {\em Soft Computing}, 24(2):1483--1496, 2020.

\bibitem{li2007process}
J.~Li, D.~Liu, and B.~Yang.
\newblock {Process Mining: Extending $\alpha$-Algorithm to Mine Duplicate Tasks
  in Process Logs}.
\newblock In {\em Advances in Web and Network Technologies, and Information
  Management}, pages 396--407. Springer, 2007.

\bibitem{li2020propagation}
L.~Li, Z.~Yang, Z.~Dang, C.~Meng, J.~Huang, H.~Meng, D.~Wang, G.~Chen,
  J.~Zhang, H.~Peng, and Y.~Shao.
\newblock {Propagation analysis and prediction of the COVID-19}.
\newblock {\em Infectious Disease Modelling}, 5:282--292, 2020.

\bibitem{li2007tcloseness}
N.~Li, T.~Li, and S.~Venkatasubramanian.
\newblock {$t$-Closeness: Privacy Beyond $k$-Anonymity and $l$-Diversity}.
\newblock In {\em Proceedings of the IEEE 23rd International Conference on Data
  Engineering}, pages 106--115, Istanbul, Turkey, 2007. IEEE.

\bibitem{li2009tradeoff}
T.~Li and N.~Li.
\newblock {On the Tradeoff Between Privacy and Utility in Data Publishing}.
\newblock In {\em Proceedings of the 15th ACM SIGKDD International Conference
  on Knowledge Discovery and Data Mining}, pages 517--526, Paris, France, 2009.

\bibitem{liebowitz2015biological}
J.~Liebowitz and R.~Schaller.
\newblock {Biological Warfare: Tampering with implantable medical devices}.
\newblock {\em IT Professional}, 17(5):70--72, 2015.

\bibitem{lin2008oka}
J.-L. Lin and M.-C. Wei.
\newblock {An Efficient Clustering Method for k-Anonymization}.
\newblock In {\em Proceedings of the International Workshop Privacy and
  Anonymity in Information Society}, pages 46--50, Nantes, France, 2008.

\bibitem{liu2016towards}
C.~Liu, H.~Duan, Q.~Zeng, M.~Zhou, F.~Lu, and J.~Cheng.
\newblock {Towards Comprehensive Support for Privacy Preservation
  Cross-Organization Business Process Mining}.
\newblock {\em IEEE Transactions on Services Computing}, 12(4):1--15, 2016.

\bibitem{lopez2021effective}
P.~L{\'o}pez-Aguilar and A.~Solanas.
\newblock {An Effective Approach to the Cross-Border Exchange of Digital
  Evidence Using Blockchain}.
\newblock In {\em Proceedings of the 9th International Conference on
  Applications in Electronics Pervading Industry, Environment and Society},
  pages 1--5, Pisa, Italy, 2021. Springer.

\bibitem{lopez2021human}
P.~L{\'o}pez-Aguilar and A.~Solanas.
\newblock {Human Susceptibility to Phishing Attacks Based on Personality
  Traits: The Role of Neuroticism}.
\newblock In {\em Proceedings of the IEEE 45th Annual Computers, Software, and
  Applications Conference}, pages 1363--1368, Madrid, Spain, 2021. IEEE.

\bibitem{lopez2017challenges}
P.~L{\'o}pez-Iturri, E.~Aguirre, L.~Azpilicueta, J.~J. Astrain,
  J.~Villandangos, and F.~Falcone.
\newblock {Challenges in Wireless System Integration as Enablers for Indoor
  Context Aware Environments}.
\newblock {\em Sensors}, 17(7):1616, 2017.

\bibitem{lopez2018implementation}
P.~Lopez-Iturri, E.~Aguirre, J.~D. Trigo, J.~J. Astrain, L.~Azpilicueta,
  L.~Serrano, J.~Villadangos, and F.~Falcone.
\newblock {Implementation and Operational Analysis of an Interactive Intensive
  Care Unit within a Smart Health Context}.
\newblock {\em Sensors}, 18(2):389, 2018.

\bibitem{machanavajjhala2007ldiversity}
A.~Machanavajjhala, D.~Kifer, J.~Gehrke, and M.~Venkitasubramaniam.
\newblock {$l$-Diversity: Privacy Beyond $k$-Anonymity}.
\newblock {\em ACM Transactions on Knowledge Discovery from Data}, 1(1):1--52,
  2007.

\bibitem{machin2021privacy}
J.~Machin, E.~Batista, A.~Mart{\'i}nez-Ballest{\'e}, and A.~Solanas.
\newblock {Privacy and Security in Cognitive Cities: A Systematic Review}.
\newblock {\em Applied Sciences}, 11(10):4471, 2021.

\bibitem{manisalidis2020environmental}
I.~Manisalidis, E.~Stavropoulou, A.~Stavropoulos, and E.~Bezirtzoglou.
\newblock {Environmental and Health Impacts of Air Pollution: A Review}.
\newblock {\em Frontiers in Public Health}, 8:14, 2020.

\bibitem{mannhardt2017analyzing}
F.~Mannhardt and D.~Blinde.
\newblock {Analyzing the Trajectories of Patients with Sepsis using Process
  Mining}.
\newblock In {\em Proceedings of the 29th International Conference on Advanced
  Information Systems Engineering}, pages 72--80, Essen, Germany, 2017.

\bibitem{mannhardt2019pppm}
F.~Mannhardt, A.~Koschmider, N.~Baracaldo, M.~Weidlich, and J.~Michael.
\newblock {Privacy-Preserving Process Mining}.
\newblock {\em Business \& Information Systems Engineering}, 61(5):595--614,
  2019.

\bibitem{mannhardt2018privacy}
F.~Mannhardt, S.~A. Petersen, and M.~F. Oliveira.
\newblock {Privacy Challenges for Process Mining in Human-Centered Industrial
  Environments}.
\newblock In {\em Proceedings of the 14th International Conference on
  Intelligent Environments}, pages 1--8, Rome, Italy, 2018. IEEE.

\bibitem{mans2008stroke}
R.~S. Mans, M.~H. Schonenberg, G.~Leonardi, S.~Panzarasa, A.~Cavallini,
  S.~Quaglini, and W.~M.~P. van~der Aalst.
\newblock {Process Mining Techniques: an Application to Stroke Care}.
\newblock {\em Studies in Health Technology and Informatics}, 136:573--578,
  2008.

\bibitem{mans2008application}
R.~S. Mans, M.~H. Schonenberg, M.~Song, W.~M.~P. van~der Aalst, and P.~J.~M.
  Bakker.
\newblock {Application of Process Mining in Healthcare -- A Case Study in a
  Dutch Hospital}.
\newblock In {\em Proceedings of the 1st International Joint Conference on
  Biomedical Engineering Systems and Technologies}, volume~25, pages 425--438,
  Funchal, Portugal, 2008. Springer.

\bibitem{mans2015process}
R.~S. Mans, W.~M.~P. van~der Aalst, and R.~J.~B. Vanwersch.
\newblock {\em {Process Mining in Healthcare: Evaluating and Exploiting
  Operational Healthcare Processes}}.
\newblock Springer, 2015.

\bibitem{mans2012process}
R.~S. Mans, W.~M.~P. van~der Aalst, R.~J.~B. Vanwersch, and A.~J. Moleman.
\newblock {Process Mining in Healthcare: Data Challenges when Answering
  Frequently Posed Questions}.
\newblock In {\em Process Support and Knowledge Representation in Health Care},
  pages 140--153, Tallin, Estonia, 2012. Springer.

\bibitem{marin2016security}
E.~Marin, D.~Singelee, F.~D. Garcia, T.~Chothia, R.~Willems, and B.~Preneel.
\newblock {On the (in)security of the Latest Generation Implantable Cardiac
  Defibrillators and How to Secure Them}.
\newblock In {\em Proceedings of the 32nd Annual Conference on Computer
  Security Applications}, pages 226--236, Los Angeles, USA, 2016. ACM.

\bibitem{martin2017role}
K.~D. Martin and P.~E. Murphy.
\newblock {The role of data privacy in marketing}.
\newblock {\em Journal of the Academy of Marketing Science}, 45(2):135--155,
  2017.

\bibitem{martinez2019epemicu}
A.~Mart{\'i}nez-Ballest{\'e}, P.~Gimeno, A.~Marin{\'e}, E.~Batista, and
  A.~Solanas.
\newblock {e-PEMICU: an e-Health Platform to Support Early Mobilisation in
  Intensive Care Units}.
\newblock In {\em Proceedings of the 10th International Conference on
  Information, Intelligence, Systems and Applications}, pages 1--6, Patras,
  Greece, 2019. IEEE.

\bibitem{martinez2013pursuit}
A.~Mart{\'i}nez-Ballest{\'e}, P.~A. P{\'e}rez-Mart{\'i}nez, and A.~Solanas.
\newblock {The Pursuit of Citizens' Privacy: A Privacy-Aware Smart City Is
  Possible}.
\newblock {\em IEEE Communications Magazine}, 51(6):136--141, 2013.

\bibitem{matthews2011data}
G.~J. Matthews and O.~Harel.
\newblock {Data confidentiality: A review of methods for statistical disclosure
  limitation and methods for assessing privacy}.
\newblock {\em Statistics Surveys}, 5:1--29, 2011.

\bibitem{michael2019user}
J.~Michael, A.~Koschmider, F.~Mannhardt, N.~Baracaldo, and B.~Rumpe.
\newblock {User-Centered and Privacy-Driven Process Mining System Design for
  IoT}.
\newblock In {\em Proceedings of the 31st International Conference on Advanced
  Information Systems Engineering}, pages 194--206, Rome, Italy, 2019.

\bibitem{microbit}
{Micro:bit Educational Foundation}.
\newblock {BBC micro:bit}.
\newblock \url{https://microbit.org/}.
\newblock Accessed: 2021-09-01.

\bibitem{miraz2018internet}
M.~H. Miraz, M.~Ali, P.~S. Excell, and R.~Picking.
\newblock {Internet of Nano-Things, Things and Everything: Future Growth
  Trends}.
\newblock {\em Future Internet}, 10(8):68, 2018.

\bibitem{mitnick2003art}
K.~D. Mitnick and W.~L. Simon.
\newblock {\em {The art of deception: Controlling the human element of
  security}}.
\newblock John Wiley \& Sons, 2003.

\bibitem{morgner2018malicious}
P.~Morgner, S.~Pfennig, D.~Salzner, and Z.~Benenson.
\newblock {Malicious IoT Implants: Tampering with Serial Communication over the
  Internet}.
\newblock In {\em Proceedings of the 21st International Symposium on Research
  in Attacks, Intrusions, and Defenses}, pages 535--555, Heraklion, Greece,
  2018. Springer.

\bibitem{mostashari2011cognitive}
A.~Mostashari, F.~Arnold, M.~Mansouri, and M.~Finger.
\newblock {Cognitive cities and intelligent urban governance}.
\newblock {\em Network Industries Quarterly}, 13(3):3--6, 2011.

\bibitem{murata1989petri}
T.~Murata.
\newblock {Petri Nets: Properties, Analysis and Applications}.
\newblock {\em Proceedings of the IEEE}, 77(4):541--580, 1989.

\bibitem{nejatollahi2019post}
H.~Nejatollahi, N.~Dutt, S.~Ray, F.~Regazzoni, I.~Banerjee, and R.~Cammarota.
\newblock {Post-Quantum Lattice-Based Cryptography Implementations: A Survey}.
\newblock {\em ACM Computing Surveys}, 51(6):1--41, 2019.

\bibitem{nergiz2007hiding}
M.~E. Nergiz, M.~Atzori, and C.~Clifton.
\newblock {Hiding the Presence of Individuals from Shared Databases}.
\newblock In {\em Proceedings of the ACM SIGMOD International Conference on
  Management of Data}, pages 665--676, Beijing, China, 2007.

\bibitem{myfitnesspalhack}
A.~Newcomb.
\newblock {Hacked MyFitnessPal Data Goes on Sale on the Dark Web -- One Year
  After the Breach}.
\newblock
  \url{https://fortune.com/2019/02/14/hacked-myfitnesspal-data-sale-dark-web-one-year-breach},
  February 2019.
\newblock Accessed: 2021-07-15.

\bibitem{nidhya2019end}
R.~Nidhya, S.~Karthik, and G.~Smilarubavathy.
\newblock {An End-to-End Secure and Energy-Aware Routing Mechanism for
  IoT-Based Modern Health Care System}.
\newblock In {\em Soft Computing and Signal Processing}, pages 379--388.
  Springer, 2019.

\bibitem{niu2014fine}
B.~Niu, Q.~Li, X.~Zhu, and H.~Li.
\newblock {A Fine-Grained Spatial Cloaking Scheme for Privacy-Aware Users in
  Location-Based Services}.
\newblock In {\em Proceedings of the 23rd International Conference on Computer
  Communication and Networks}, pages 1--8, Shanghai, China, 2014. IEEE.

\bibitem{nivzetic2020iot}
S.~Ni{\v{z}}eti{\'c}, P.~{\v{S}}oli{\'c}, D.~L{\'o}pez-de-Ipi{\~n}a
  Gonz{\'a}lez-de Artaza, and L.~Patrono.
\newblock {Internet of Things (IoT): Opportunities, issues and challenges
  towards a smart and sustainable future}.
\newblock {\em Journal of Cleaner Production}, 274:122877, 2020.

\bibitem{nobles2018human}
C.~Nobles.
\newblock {Botching Human Factors in Cybersecurity in Business Organizations}.
\newblock {\em HOLISTICA -- Journal of Business and Public Administration},
  9(3):71--88, 2018.

\bibitem{von2020quantifying}
S.~Nu\~{n}ez~von Voigt, S.~A. Fahrenkrog-Petersen, D.~Janssen, A.~Koschmider,
  F.~Tschorsch, F.~Mannhardt, O.~Landsiedel, and M.~Weidlich.
\newblock {Quantifying the Re-identification Risk of Event Logs for Process
  Mining}.
\newblock In {\em Proceedings of the 32nd International Conference on Advanced
  Information Systems Engineering}, pages 252--267, Grenoble, France, 2020.
  Springer.

\bibitem{ouaddah2017towards}
A.~Ouaddah, A.~A. Elkalam, and A.~A. Ouahman.
\newblock {Towards a Novel Privacy-Preserving Access Control Model Based on
  Blockchain Technology in IoT}.
\newblock In {\em Proceedings of the Europe, Middle East and North Africa
  Conference on Technology and Security to Support Learning}, pages 523--533,
  Saidia, Morocco, 2017. Springer.

\bibitem{papadimitriou2010web}
P.~Papadimitriou, A.~Dasdan, and H.~Garcia-Molina.
\newblock {Web graph similarity for anomaly detection}.
\newblock {\em Journal of Internet Services and Applications}, 1(1):19--30,
  2010.

\bibitem{papageorgiou2018security}
A.~Papageorgiou, M.~Strigkos, E.~Politou, E.~Alepis, A.~Solanas, and
  C.~Patsakis.
\newblock {Security and Privacy Analysis of Mobile Health Applications: The
  Alarming State of Practice}.
\newblock {\em IEEE Access}, 6:9390--9403, 2018.

\bibitem{park2016ain}
Y.~Park, Y.~Son, H.~Shin, D.~Kim, and Y.~Kim.
\newblock {This ain't your dose: Sensor Spoofing Attack on Medical Infusion
  Pump}.
\newblock In {\em Proceedings of the 10th USENIX Workshop on Offensive
  Technologies}, pages 1--11, Austin, USA, 2016.

\bibitem{partington2015process}
A.~Partington, M.~T. Wynn, S.~Suriadi, C.~Ouyang, and J.~Karnon.
\newblock {Process Mining for Clinical Processes: A Comparative Analysis of
  Four Australian Hospitals}.
\newblock {\em ACM Transactions on Management Information Systems}, 5(4):1--18,
  2015.

\bibitem{patsakis2015emergency}
C.~Patsakis, A.~Papageorgiou, F.~Falcone, and A.~Solanas.
\newblock {s-Health as a driver towards better emergency response systems in
  urban environments}.
\newblock In {\em Proceedings of the 10th IEEE International Symposium on
  Medical Measurements and Applications}, pages 214--218, Torino, Italy, 2015.
  IEEE.

\bibitem{patsakis2014personalized}
C.~Patsakis, R.~Venanzio, P.~Bellavista, A.~Solanas, and M.~Bouroche.
\newblock {Personalized Medical Services using Smart Cities' Infrastructures}.
\newblock In {\em Proceedings of the 9th IEEE International Symposium on
  Medical Measurements and Applications}, pages 1--5, Lisbon, Portugal, 2014.
  IEEE.

\bibitem{pegoraro2021analyzing}
M.~Pegoraro, M.~B. Shankara, W.~M.~P. van~der Aalst, L.~Martin, and G.~Marx.
\newblock {Analyzing Medical Data with Process Mining: a COVID-19 Case Study}.
\newblock In {\em Proceedings of the 12th Workshop on Applications of
  Knowledge-Based Technologies in Business}, pages 1--9, Hannover, Germany,
  2021.

\bibitem{perez2013cities}
P.~A. P{\'e}rez-Mart{\'i}nez, A.~Mart{\'i}nez-Ballest{\'e}, and A.~Solanas.
\newblock {Privacy in Smart Cities -- A Case Study of Smart Public Parking}.
\newblock In {\em Proceedings of the 3rd International Conference on Pervasive
  and Embedded Computing and Communication Systems}, pages 55--59, Barcelona,
  Spain, 2013. INSTICC.

\bibitem{perez2011w3}
P.~A. P{\'e}rez-Mart{\'i}nez and A.~Solanas.
\newblock {W$^3$-Privacy: the Three Dimensions of User Privacy in LBS}.
\newblock In {\em Proceedings of the 12th ACM International Symposium on Mobile
  Ad Hoc Networking and Computing}, pages 1--2, Paris, France, 2011. ACM.

\bibitem{perez2009location}
P.~A. P{\'e}rez-Mart{\'i}nez, A.~Solanas, and A.~Mart{\'i}nez-Ballest{\'e}.
\newblock {Location Privacy Through Users' Collaboration: A Distributed
  Pseudonymizer}.
\newblock In {\em Proceedings of the 3rd International Conference on Mobile
  Ubiquitous Computing, Systems, Services and Technologies}, pages 338--341,
  Silema, Malta, 2009. IEEE.

\bibitem{perimal2014health}
L.~Perimal-Lewis, D.~B. de~Vries, and C.~H. Thompson.
\newblock {Health intelligence: Discovering the process model using process
  mining by constructing Start-to-End patient journeys}.
\newblock In {\em Proceedings of the 7th Australasian Workshop on Health
  Informatics and Knowledge Management}, pages 59--67, Auckland, New Zealand,
  2014. Australian Computer Society.

\bibitem{perimal2012gaining}
L.~Perimal-Lewis, S.~Qin, C.~H. Thompson, and P.~Hakendorf.
\newblock {Gaining Insight from Patient Journey Data using a Process-Oriented
  Analysis Approach}.
\newblock In {\em Proceedings of the 5th Australasian Workshop on Health
  Informatics and Knowledge Management}, pages 59--66, Melbourne, Australia,
  2012. Australian Computer Society.

\bibitem{peterson1977petri}
J.~L. Peterson.
\newblock {Petri Nets}.
\newblock {\em ACM Computing Surveys}, 9(3):223--252, 1977.

\bibitem{phelan2018implementing}
S.~Phelan, F.~Lin, M.~Mitchell, and W.~Chaboyer.
\newblock {Implementing early mobilisation in the intensive care unit: An
  integrative review}.
\newblock {\em International Journal of Nursing Studies}, 77:91--105, 2018.

\bibitem{pika2020privacy}
A.~Pika, M.~T. Wynn, S.~Budiono, A.~H.~M. ter Hofstede, W.~M.~P. van~der Aalst,
  and H.~A. Reijers.
\newblock {Privacy-Preserving Process Mining in Healthcare}.
\newblock {\em International Journal of Environmental Research and Public
  Health}, 17(5):1612, 2020.

\bibitem{poelmans2010combining}
J.~Poelmans, G.~Dedene, G.~Verheyden, H.~Van~der Mussele, S.~Viaene, and
  E.~Peters.
\newblock {Combining business process and data discovery techniques for
  analyzing and improving integrated care pathways}.
\newblock In {\em Proceedings of the 10th Industrial Conference on Data Mining,
  Applications and Theoretical Aspects}, pages 505--517, Berlin, Germany, 2010.
  Springer.

\bibitem{politou2018forgetting}
E.~Politou, E.~Alepis, and C.~Patsakis.
\newblock {Forgetting personal data and revoking consent under the GDPR:
  Challenges and proposed solutions}.
\newblock {\em Journal of Cybersecurity}, 4(1):1--20, 2018.

\bibitem{ponemon2020insider}
{Ponemon Institute LLC}.
\newblock {2020 Cost of Insider Threats Global Report}.
\newblock Technical report, Ponemon Institute, Traverse City, USA, 2020.

\bibitem{priestman2019phishing}
W.~Priestman, T.~Anstis, I.~G. Sebire, S.~Sridharan, and N.~J. Sebire.
\newblock {Phishing in healthcare organisations: threats, mitigation and
  approaches}.
\newblock {\em BMJ Health \& Care Informatics}, 26(1):e100031, 2019.

\bibitem{pyzdek2003six}
T.~Pyzdek and P.~Keller.
\newblock {\em {The Six Sigma}}.
\newblock McGraw-Hill, 2003.

\bibitem{radcliffe2011hacking}
J.~Radcliffe.
\newblock {Hacking Medical Devices for Fun and Insulin: Breaking the Human
  SCADA System}.
\newblock In {\em Black Hat Conference}, volume 2011, Las Vegas, USA, 2011.

\bibitem{rafiei2021pc4pm}
M.~Rafiei, A.~Schnitzler, and W.~M.~P. van~der Aalst.
\newblock {PC4PM: A Tool for Privacy/Confidentiality Preservation in Process
  Mining}.
\newblock {\em arXiv preprint arXiv:2107.14499}, pages 1--5, 2021.

\bibitem{rafiei2019mining}
M.~Rafiei and W.~M.~P. van~der Aalst.
\newblock {Mining Roles From Event Logs While Preserving Privacy}.
\newblock In {\em Proceedings of the 17th International Conference on Business
  Process Management -- Workshop Security and Privacy-enhanced Business Process
  Management}, pages 676--689, Vienna, Austria, 2019. Springer.

\bibitem{rafiei2020practical}
M.~Rafiei and W.~M.~P. van~der Aalst.
\newblock {Practical Aspect of Privacy-Preserving Data Publishing in Process
  Mining}.
\newblock In {\em Proceedings of the 18th International Conference on Business
  Process Management -- Dissertation Award, Doctoral Consortium and
  Demonstration Track}, pages 92--96, Seville, Spain, 2020. Springer.

\bibitem{rafiei2020privacy}
M.~Rafiei and W.~M.~P. van~der Aalst.
\newblock {Privacy-Preserving Data Publishing in Process Mining}.
\newblock In {\em Proceedings of the 18th International Conference on Business
  Process Management Forum}, pages 122--138, Seville, Spain, 2020. Springer.

\bibitem{rafiei2020towards}
M.~Rafiei and W.~M.~P. van~der Aalst.
\newblock {Towards Quantifying Privacy in Process Mining}.
\newblock In {\em Proceedings of the 2nd International Conference on Process
  Mining Workshops}, pages 385--397, Padua, Italy, 2020. Springer.

\bibitem{rafiei2021group}
M.~Rafiei and W.~M.~P. van~der Aalst.
\newblock {Group-based privacy preservation techniques for process mining}.
\newblock {\em Data \& Knowledge Engineering}, 134:101908, 2021.

\bibitem{rafiei2018ensuring}
M.~Rafiei, L.~Von~Waldthausen, and W.~M.~P. van~der Aalst.
\newblock {Ensuring Confidentiality in Process Mining}.
\newblock In {\em Proceedings of the 8th International Symposium on Data-Driven
  Process Discovery and Analysis}, pages 3--17, Seville, Spain, 2018. Springer.

\bibitem{rafiei2018supporting}
M.~Rafiei, L.~von Waldthausen, and W.~M.~P. van~der Aalst.
\newblock {Supporting Confidentiality in Process Mining Using Abstraction and
  Encryption}.
\newblock In {\em Proceedings of the 8th International Symposium on Data-Driven
  Process Discovery and Analysis}, pages 101--123, Seville, Spain, 2018.
  Springer.

\bibitem{rafiei2020tlkc}
M.~Rafiei, M.~Wagner, and W.~M.~P. van~der Aalst.
\newblock {\textit{TLKC}-Privacy Model for Process Mining}.
\newblock In {\em Proceedings of the 14th International Conference on Research
  Challenges in Information Science}, pages 398--416, Limassol, Cyprus, 2020.
  Springer.

\bibitem{rahman2015secure}
M.~Rahman, B.~Carbunar, and U.~Topkara.
\newblock {Secure Management of Low Power Fitness Trackers}.
\newblock {\em IEEE Transactions on Mobile Computing}, 15(2):447--459, 2015.

\bibitem{rebuge2012business}
{\'A}.~Rebuge and D.~R. Ferreira.
\newblock {Business Process Analysis in Healthcare Environments: a Methodology
  based on Process Mining}.
\newblock {\em Information Systems}, 37(2):99--116, 2012.

\bibitem{idc2018data}
D.~Reinsel, J.~Gantz, and J.~Rydning.
\newblock {The Digitization of the World: From Edge to Core}.
\newblock Technical Report US4441331, IDC, Framingham, USA, 2018.

\bibitem{revadigar2017accelerometer}
G.~Revadigar, C.~Javali, W.~Xu, A.~V. Vasilakos, W.~Hu, and S.~Jha.
\newblock {Accelerometer and Fuzzy Vault-Based Secure Group Key Generation and
  Sharing Protocol for Smart Wearables}.
\newblock {\em IEEE Transactions on Information Forensics and Security},
  12(10):2467--2482, 2017.

\bibitem{riano2014copd}
D.~Ria{\~n}o and A.~Solanas.
\newblock {Exploiting the Relation Between Environmental Factors and Diseases:
  A Case Study on Chronic Obstructive Pulmonary Disease}.
\newblock In {\em Proceedings of the 6th International Workshop on Knowledge
  Representation for Health-Care Data, Processes and Guidelines}, pages
  160--173, Vienna, Austria, 2014. Springer.

\bibitem{rojas2016review}
E.~Rojas, J.~Munoz-Gama, M.~Sep{\'u}lveda, and D.~Capurro.
\newblock {Process mining in healthcare: A literature review}.
\newblock {\em Journal of Biomedical Informatics}, 61:224--236, 2016.

\bibitem{rojas2017question}
E.~Rojas, M.~Sep{\'u}lveda, J.~Munoz-Gama, D.~Capurro, V.~Traver, and
  C.~Fern{\'a}ndez-Llatas.
\newblock {Question-Driven Methodology for Analyzing Emergency Room Processes
  Using Process Mining}.
\newblock {\em Applied Sciences}, 7(3):302, 2017.

\bibitem{rose2020zero}
S.~Rose, O.~Borchert, S.~Mitchell, and S.~Connelly.
\newblock {Zero Trust Architecture}.
\newblock Technical Report Special Publication 800-207, National Institute of
  Standards and Technology, 2020.

\bibitem{rostami2013balancing}
M.~Rostami, W.~Burleson, F.~Koushanfar, and A.~Juels.
\newblock {Balancing Security and Utility in Medical Devices?}
\newblock In {\em Proceedings of the 50th Annual Design Automation Conference},
  pages 1--6, Austin, USA, 2013.

\bibitem{roth2013smart}
S.~Roth, J.~Kaivo-Oja, and T.~Hirschmann.
\newblock {Smart regions: Two cases of crowdsourcing for regional development}.
\newblock {\em International Journal of Entrepreneurship and Small Business},
  20(3):272--285, 2013.

\bibitem{rubner2000emd}
Y.~Rubner, C.~Tomasi, and L.~J. Guibas.
\newblock {The Earth Mover's Distance as a Metric for Image Retrieval}.
\newblock {\em International Journal of Computer Vision}, 40(2):99--121, 2000.

\bibitem{samarati1998kanonymity}
P.~Samarati and L.~Sweeney.
\newblock {Protecting Privacy when Disclosing Information: $k$-Anonymity and
  Its Enforcement through Generalization and Suppression}.
\newblock Technical Report CSL-98-04, SRI International, Menlo Park, USA, 1998.

\bibitem{schaub2015context}
F.~Schaub, B.~K{\"o}nings, and M.~Weber.
\newblock {Context-Adaptive Privacy: Leveraging Context Awareness to Support
  Privacy Decision Making }.
\newblock {\em IEEE Pervasive Computing}, 14(1):34--43, 2015.

\bibitem{shahidehpour2018smart}
M.~Shahidehpour, Z.~Li, and M.~Ganji.
\newblock {Smart Cities for a Sustainable urbanization: Illuminating the need
  for establishing smart urban infrastructures}.
\newblock {\em IEEE Electrification Magazine}, 6(2):16--33, 2018.

\bibitem{shlomo2018sdc}
N.~Shlomo.
\newblock {Statistical Disclosure Limitation: New Directions and Challenges}.
\newblock {\em Journal of Privacy and Confidentiality}, 8(1):1--17, 2018.

\bibitem{shoubridge2002detection}
P.~Shoubridge, M.~Kraetzl, W.~D. Wallis, and H.~Bunke.
\newblock {Detection of abnormal change in a time series of graphs}.
\newblock {\em Journal of Interconnection Networks}, 3(01n02):85--101, 2002.

\bibitem{siemens2017connectivism}
G.~Siemens.
\newblock {Connectivism: A Learning Theory for the Digital Age}.
\newblock In R.~E. West, editor, {\em Foundations of Learning and Instructional
  Design Technology}. Pressbooks, 2017.

\bibitem{sleurs2019mobile}
K.~Sleurs, S.~F. Seys, J.~Bousquet, W.~J. Fokkens, S.~Gorris, B.~Pugin, and
  P.~W. Hellings.
\newblock {Mobile health tools for the management of chronic respiratory
  diseases}.
\newblock {\em Allergy}, 74(7):1292--1306, 2019.

\bibitem{solanas2015wandering}
A.~Solanas, E.~Batista, F.~Borras, A.~Mart{\'i}nez-Ballest{\'e}, and
  C.~Patsakis.
\newblock {Wandering Analysis with Mobile Phones: On the relation between
  randomness and wandering}.
\newblock In {\em Proceedings of the 5th International Conference on Pervasive
  and Embedded Computing and Communication Systems}, pages 168--173, Angers,
  France, 2015. INSTICC.

\bibitem{solanas2021privacy}
A.~Solanas, E.~Batista, F.~Casino, A.~Papageorgiou, and C.~Patsakis.
\newblock {Privacy-Oriented Analysis of Ubiquitous Computing Systems: A 5-D
  Approach}.
\newblock In {\em Security of Ubiquitous Computing Systems}, pages 201--213.
  Springer, 2021.

\bibitem{solanas2006vmdav}
A.~Solanas and A.~Martinez-Ballest{\'e}.
\newblock {V-MDAV: A Multivariate Microaggregation With Variable Group Size}.
\newblock In {\em Proceedings of the 17th Computational Statistics Symposium},
  pages 917--925, Rome, Italy, 2006.

\bibitem{solanas2014shealth}
A.~Solanas, C.~Patsakis, M.~Conti, I.~S. Vlachos, V.~Ramos, F.~Falcone,
  O.~Postolache, P.~A. P{\'e}rez-Mart{\'i}nez, R.~Di~Pietro, D.~N. Perrea, and
  A.~Mart{\'i}nez-Ballest{\'e}.
\newblock {Smart Health: A Context-Aware Health Paradigm within Smart Cities}.
\newblock {\em IEEE Communications Magazine}, 52(8):74--81, 2014.

\bibitem{song2008trace}
M.~Song, C.~W. G{\"u}nther, and W.~M.~P. van~der Aalst.
\newblock {Trace Clustering in Process Mining}.
\newblock In {\em Proceedings of the 6th International Conference on Business
  Process Management}, pages 109--120, Milan, Italy, 2008. Springer.

\bibitem{spence2020side}
A.~Spence and S.~Bangay.
\newblock {Side-Channel Sensing: Exploiting Side-Channels to Extract
  Information for Medical Diagnostics and Monitoring}.
\newblock {\em IEEE Journal of Translational Engineering in Health and
  Medicine}, 8:1--13, 2020.

\bibitem{spindler2016encryption}
G.~Spindler and P.~Schmechel.
\newblock {Personal Data and Encryption in the European General Data Protection
  Regulation}.
\newblock {\em Journal of Intellectual Property, Information Technology and
  E-Commerce Law}, 7:163--177, 2016.

\bibitem{bpi13}
W.~Steeman.
\newblock {BPI Challenge 2013, closed problems}.
\newblock
  \url{https://doi.org/10.4121/uuid:c2c3b154-ab26-4b31-a0e8-8f2350ddac11},
  2013.
\newblock Ghent University. Collection. Accessed: 2021-07-13.

\bibitem{steinhubl2013can}
S.~R. Steinhubl, E.~D. Muse, and E.~J. Topol.
\newblock {Can Mobile Health Technologies Transform Health Care?}
\newblock {\em Journal of the American Medical Association},
  310(22):2395--2396, 2013.

\bibitem{stewart2017comment}
D.~W. Stewart.
\newblock {A comment on privacy}.
\newblock {\em Journal of the Academy of Marketing Science}, 45(2):156--159,
  2017.

\bibitem{sun2008enhanced}
X.~Sun, H.~Wang, J.~Li, and T.~M. Truta.
\newblock {Enhanced p-Sensitive k-Anonymity Models for Privacy Preserving Data
  Publishing}.
\newblock {\em Transactions on Data Privacy}, 1(2):53--66, 2008.

\bibitem{sunyaev2015availability}
A.~Sunyaev, T.~Dehling, P.~L. Taylor, and K.~D. Mandl.
\newblock {Availability and quality of mobile health app privacy policies}.
\newblock {\em Journal of the American Medical Informatics Association},
  22(e1):e28--e33, 2015.

\bibitem{suriadi2017event}
S.~Suriadi, R.~Andrews, A.~H.~M. ter Hofstede, and M.~T. Wynn.
\newblock {Event log imperfection patterns for process mining: Towards a
  systematic approach to cleaning event logs}.
\newblock {\em Information Systems}, 64:132--150, 2017.

\bibitem{suriadi2014measuring}
S.~Suriadi, R.~S. Mans, M.~T. Wynn, A.~Partington, and J.~Karnon.
\newblock {Measuring Patient Flow Variations: A Cross-Organisational Process
  Mining Approach}.
\newblock In {\em Proceedings of the 2nd Asia-Pacific Conference on Business
  Process Management}, pages 43--58, Brisbane, Australia, 2014. Springer.

\bibitem{sweeney2002kanonymity}
L.~Sweeney.
\newblock {$k$-anonymity: A model for protecting privacy}.
\newblock {\em International Journal of Uncertainty, Fuzziness and
  Knowledge-Based Systems}, 10(5):557--570, 2002.

\bibitem{tambou2019fines}
O.~Tambou.
\newblock {Lessons from the First Post-GDPR Fines of the CNIL against Google
  LLC}.
\newblock {\em European Data Protection Law Review}, 5(1):80--84, 2019.

\bibitem{temdee2018context}
P.~Temdee and R.~Prasad.
\newblock {\em {Context-Aware Communication and Computing: Applications for
  Smart Environment}}.
\newblock Springer, 2018.

\bibitem{templ2014sdc}
M.~Templ, B.~Meindl, A.~Kowarik, and S.~Chen.
\newblock {Introduction to Statistical Disclosure Control (SDC)}.
\newblock Technical report, IHSN Working Paper 007, August 2014.

\bibitem{yahoohack}
S.~Thielman.
\newblock {Yahoo hack: 1bn accounts compromised by biggest data breach in
  history}.
\newblock
  \url{https://www.theguardian.com/technology/2016/dec/14/yahoo-hack-security-of-one-billion-accounts-breached},
  December 2016.
\newblock Accessed: 2021-07-15.

\bibitem{tillem2016privalpha}
G.~Tillem, Z.~Erkin, and R.~L. Lagendijk.
\newblock {Privacy-Preserving Alpha Algorithm for Software Analysis}.
\newblock In {\em Proceedings of the 37th International Symposium on
  Information Theory and Signal Processing in the Benelux}, pages 136--143,
  Louvain-la-Neuve, Belgium, 2016.

\bibitem{torra2017privacy}
V.~Torra.
\newblock {\em {Data Privacy: Foundations, New Developments and the Big Data
  Challenge}}.
\newblock Springer, 2017.

\bibitem{truta2006privacy}
T.~M. Truta and B.~Vinay.
\newblock {Privacy Protection: p-Sensitive k-Anonymity Property}.
\newblock In {\em Proceedings of the 22nd International Conference on Data
  Engineering Workshops)}, pages 94--94, Atlanta, USA, 2006. IEEE.

\bibitem{un2019population}
{United Nations, Department of Economic and Social Affairs, Population
  Division}.
\newblock {World Population Prospects 2019: Highlights}.
\newblock Technical Report ST/ESA/SER.A/423, United Nations, New York, USA,
  2019.

\bibitem{un2019urba}
{United Nations, Department of Economic and Social Affairs, Population
  Division}.
\newblock {World Urbanization Prospects The 2018 Revision}.
\newblock Technical Report ST/ESA/SER.A/420, United Nations, New York, USA,
  2019.

\bibitem{un2020ageing}
{United Nations, Department of Economic and Social Affairs, Population
  Division}.
\newblock {World Population Ageing 2019}.
\newblock Technical Report ST/ESA/SER.A/444, United Nations, New York, USA,
  2020.

\bibitem{un2017dementia}
{United Nations, Department of Mental Health and Substance Abuse}.
\newblock {Global action plan on the public health response to dementia
  2017-2025}.
\newblock Technical report, United Nations, Geneva, Switzerland, 2017.

\bibitem{urquhart2018realising}
L.~Urquhart, N.~Sailaja, and D.~McAuley.
\newblock {Realising the right to data portability for the domestic Internet of
  things}.
\newblock {\em Personal and Ubiquitous Computing}, 22(2):317--332, 2018.

\bibitem{elshout2008comparing}
S.~van~den Elshout, K.~L{\'e}ger, and F.~Nussio.
\newblock {Comparing urban air quality in Europe in real time: A review of
  existing air quality indices and theproposal of a common alternative}.
\newblock {\em Environment International}, 34(5):720--726, 2008.

\bibitem{aalst2009pais}
W.~M.~P. van~der Aalst.
\newblock {Process-Aware Information Systems: Lessons to be Learned from
  Process Mining}.
\newblock In {\em Transactions on Petri Nets and Other Models of Concurrency
  II}, pages 1--26. Springer, 2009.

\bibitem{aalst2011book}
W.~M.~P. van~der Aalst.
\newblock {\em {Process Mining: Discovery, Conformance and Enhancement of
  Business Processes}}.
\newblock Springer, 2011.

\bibitem{aalst2014scientist}
W.~M.~P. van~der Aalst.
\newblock {Data Scientist: The Engineer of the Future}.
\newblock In {\em Proceedings of the 7th International Conference on
  Interoperability for Enterprises Systems and Applications}, volume~7, pages
  13--26, Albi, France, 2014. Springer.

\bibitem{aalst2016book}
W.~M.~P. van~der Aalst.
\newblock {\em {Process Mining: Data Science in Action}}.
\newblock Springer, 2016.

\bibitem{aalst2016responsible}
W.~M.~P. van~der Aalst.
\newblock {Responsible Data Science: Using Event Data in a ``People Friendly''
  Manner}.
\newblock In {\em Proceedings of the 18th International Conference on
  Enterprise Information Systems}, pages 3--28, Rome, Italy, 2016. SciTePress.

\bibitem{aalst2011manifesto}
W.~M.~P. van~der Aalst, A.~Adriansyah, A.~K. Alves~de Medeiros, F.~Arcieri,
  T.~Baier, T.~Blickle, J.~C. Bose, P.~van~den Brand, R.~Brandtjen, J.~Buijs,
  et~al.
\newblock {Process Mining Manifesto}.
\newblock In {\em Proceedings of the 9th International Conference on Business
  Process Management}, pages 169--194, Clermont-Ferrand, France, 2011.
  Springer.

\bibitem{aalst2011cnets}
W.~M.~P. van~der Aalst, A.~Adriansyah, and B.~van Dongen.
\newblock {Causal Nets: A Modeling Language Tailored towards Process
  Discovery}.
\newblock In {\em Proceedings of the 22nd International Conference on
  Concurrency Theory}, pages 28--42, Aachen, Germany, 2011. Springer.

\bibitem{aalst2005genetic}
W.~M.~P. van~der Aalst, A.~K. Alves~de Medeiros, and A.~J. M.~M. Weijters.
\newblock {Genetic Process Mining}.
\newblock In {\em Proceedings of the 26th International Conference on
  Application and Theory of Petri Nets}, pages 48--69, Miami, USA, 2005.
  Springer.

\bibitem{aalst2007structure}
W.~M.~P. van~der Aalst and C.~W. G{\"u}nther.
\newblock {Finding Structure in Unstructured Processes: The Case for Process
  Mining}.
\newblock In {\em Proceedings of the 7th International Conference on
  Application of Concurrency to System Design}, pages 3--12, Bratislava,
  Slovakia, 2007. IEEE.

\bibitem{aalst2005yawl}
W.~M.~P. van~der Aalst and A.~H.~M. ter Hofstede.
\newblock {YAWL: Yet Another Workflow Language}.
\newblock {\em Information Systems}, 30(4):245--275, 2005.

\bibitem{aalst2004workflow}
W.~M.~P. van~der Aalst and K.~van Hee.
\newblock {\em {Workflow Management: Models, Methods, and Systems}}.
\newblock MIT Press, 2004.

\bibitem{aalst2004alpha}
W.~M.~P. van~der Aalst, A.~J. M.~M. Weijters, and L.~Maruster.
\newblock {Workflow Mining. Discovering Process Models From Event Logs}.
\newblock {\em IEEE Transactions on Knowledge and Data Engineering},
  16(9):1128--1142, 2004.

\bibitem{bpi12}
B.~F. van Dongen.
\newblock {BPI Challenge 2012}.
\newblock
  \url{https://doi.org/10.4121/uuid:3926db30-f712-4394-aebc-75976070e91f},
  2012.
\newblock 4TU. Centre for Research Data. Collection. Accessed: 2021-07-13.

\bibitem{bpi14}
B.~F. van Dongen.
\newblock {BPI Challenge 2014: Activity log for incidents}.
\newblock
  \url{https://doi.org/10.4121/uuid:86977bac-f874-49cf-8337-80f26bf5d2ef},
  2014.
\newblock 4TU. Centre for Research Data. Collection. Accessed: 2021-07-13.

\bibitem{bpi15}
B.~F. van Dongen.
\newblock {BPI Challenge 2015}.
\newblock
  \url{https://doi.org/10.4121/uuid:31a308ef-c844-48da-948c-305d167a0ec1},
  2015.
\newblock 4TU. Centre for Research Data. Collection. Accessed: 2021-07-13.

\bibitem{dongen2005mxml}
B.~F. van Dongen and W.~M.~P. van~der Aalst.
\newblock {A Meta Model for Process Mining Data}.
\newblock In {\em Proceedings of the 17th International Conference on Advanced
  Information Systems Engineering Workshops (EMOI-INTEROP Workshop)}, volume~2,
  pages 309--320, Porto, Portugal, 2005.

\bibitem{zelst2020detection}
S.~J. van Zelst, M.~Fani~Sani, A.~Ostovar, R.~Conforti, and M.~La~Rosa.
\newblock Detection and removal of infrequent behavior from event streams of
  business processes.
\newblock {\em Information Systems}, 90:101451, 2020.

\bibitem{verbeek2010xes}
H.~M.~W. Verbeek, J.~C. A.~M. Buijs, B.~F. van Dongen, and W.~M.~P. van~der
  Aalst.
\newblock {XES, XESame, and ProM 6}.
\newblock In {\em Proceedings of the 22nd International Conference on Advanced
  Information Systems Engineering Forum}, pages 60--75, Hammamet, Tunisia,
  2010. Springer.

\bibitem{vignau201910}
B.~Vignau, R.~Khoury, and S.~Hall{\'e}.
\newblock {10 Years of IoT Malware: a Feature-Based Taxonomy}.
\newblock In {\em Proceedings of the IEEE 19th International Conference on
  Software Quality, Reliability and Security Companion}, pages 458--465, Sofia,
  Bulgaria, 2019. IEEE.

\bibitem{wang2018protecting}
J.~Wang, Z.~Cai, Y.~Li, D.~Yang, J.~Li, and H.~Gao.
\newblock {Protecting query privacy with differentially private
  \textit{k}-anonymity in location-based services}.
\newblock {\em Personal and Ubiquitous Computing}, 22(3):453--469, 2018.

\bibitem{wang2018privacy}
S.~Wang, Q.~Hu, Y.~Sun, and J.~Huang.
\newblock {Privacy Preservation in Location-Based Services}.
\newblock {\em IEEE Communications Magazine}, 56(3):134--140, 2018.

\bibitem{waters2016gis}
N.~Waters.
\newblock {GIS: history}.
\newblock {\em International Encyclopedia of Geography: People, the Earth,
  Environment and Technology}, pages 1--13, 2016.

\bibitem{watson2016uv}
M.~Watson, D.~M. Holman, and M.~Maguire-Eisen.
\newblock {Ultraviolet Radiation Exposure and Its Impact on Skin Cancer Risk}.
\newblock {\em Seminars in Oncology Nursing}, 32(3):241--254, 2016.

\bibitem{weijters2003rediscovering}
A.~J. M.~M. Weijters and W.~M.~P. van~der Aalst.
\newblock {Rediscovering Workflow Models from Event-Based Data using Little
  Thumb}.
\newblock {\em Integrated Computer-Aided Engineering}, 10(2):151--162, 2003.

\bibitem{weijters2006heuristic}
A.~J. M.~M. Weijters, W.~M.~P. van~der Aalst, and A.~K. Alves~de Medeiros.
\newblock {Process Mining with the Heuristics Miner Algorithm}.
\newblock Technical Report WP 166, Technische Universiteit Eindhoven,
  Eindhoven, The Netherlands, 2006.

\bibitem{weiser1991computer}
M.~Weiser.
\newblock {The Computer for the 21st Century}.
\newblock {\em Scientific American}, 265(3):94--104, 1991.

\bibitem{weske2007bpm}
M.~Weske.
\newblock {\em {Business Process Management - Concepts, Languages,
  Architectures}}.
\newblock Springer, 2007.

\bibitem{willenborg2012elements}
L.~Willenborg and T.~De~Waal.
\newblock {\em {Elements of Statistical Disclosure Control}}, volume 155.
\newblock Springer Science \& Business Media, 2012.

\bibitem{wohed2006bpmn}
P.~Wohed, W.~M.~P. van~der Aalst, M.~Dumas, A.~H.~M. ter Hofstede, and
  N.~Russell.
\newblock {On the Suitability of BPMN for Business Process Modelling}.
\newblock In {\em Proceedings of the 4th International Conference on Business
  Process Management}, pages 161--176, Vienna, Austria, 2006. Springer.

\bibitem{who2017environment}
{World Health Organization}.
\newblock {Preventing noncommunicable diseases (NCDs) by reducing environmental
  risk factors}.
\newblock Technical report, World Health Organization, Geneva, Switzerland,
  2017.

\bibitem{who2020decade}
{World Health Organization}.
\newblock {Decade of Healthy Ageing 2020--2030}.
\newblock Technical report, World Health Organization, Geneva, Switzerland,
  April 2020.

\bibitem{who2021asthma}
{World Health Organization}.
\newblock {Asthma}.
\newblock \url{https://www.who.int/news-room/fact-sheets/detail/asthma}, May
  2021.
\newblock Accessed: 2021-08-19.

\bibitem{who2021cvd}
{World Health Organization}.
\newblock {Cardiovascular Diseases}.
\newblock \url{https://www.who.int/health-topics/cardiovascular-diseases}, June
  2021.
\newblock Accessed: 2021-08-19.

\bibitem{who2021copd}
{World Health Organization}.
\newblock {Chronic obstructive pulmonary disease (COPD)}.
\newblock
  \url{https://www.who.int/news-room/fact-sheets/detail/chronic-obstructive-pulmonary-disease-(copd)},
  June 2021.
\newblock Accessed: 2021-08-19.

\bibitem{wright2016big}
A.~Wright, S.~Aaron, and D.~W. Bates.
\newblock {The Big Phish: Cyberattacks Against U.S. Healthcare Systems}.
\newblock {\em Journal of General Internal Medicine}, 31:1115--1118, 2016.

\bibitem{wu2018continuous}
G.~Wu, J.~Wang, Y.~Zhang, and S.~Jiang.
\newblock {A Continuous Identity Authentication Scheme Based on Physiological
  and Behavioral Characteristics}.
\newblock {\em Sensors}, 18(1):179, 2018.

\bibitem{xeswww}
{XES Working Group}.
\newblock {eXtensible Event Stream (XES) Standard}.
\newblock \url{https://www.xes-standard.org}, 2021.
\newblock Accessed: 2021-08-16.

\bibitem{xu2017gait}
W.~Xu, C.~Javali, G.~Revadigar, C.~Luo, N.~Bergmann, and W.~Hu.
\newblock {Gait-Key: A Gait-Based Shared Secret Key Generation Protocol for
  Wearable Devices}.
\newblock {\em ACM Transactions on Sensor Networks}, 13(1):1--27, 2017.

\bibitem{xu2018association}
Z.~Xu, J.~L. Crooks, J.~M. Davies, A.~F. Khan, W.~Hu, and S.~Tong.
\newblock {The association between ambient temperature and childhood asthma: a
  systematic review}.
\newblock {\em International Journal of Biometeorology}, 62(3):471--481, 2018.

\bibitem{blackmarketprice}
M.~Yao.
\newblock {Your Electronic Medical Records Could Be Worth \$1000 To Hackers}.
\newblock
  \url{https://www.forbes.com/sites/mariyayao/2017/04/14/your-electronic-medical-records-can-be-worth-1000-to-hackers/},
  April 2017.
\newblock Accessed: 2021-07-15.

\bibitem{yekhanin2010pir}
S.~Yekhanin.
\newblock {Private Information Retrieval}.
\newblock {\em Communications of the ACM}, 53(4):68--73, 2010.

\bibitem{yetisen2018wearables}
A.~K. Yetisen, J.~L. Martinez-Hurtado, B.~{\"U}nal, A.~Khademhosseini, and
  H.~Butt.
\newblock {Wearables in Medicine}.
\newblock {\em Advanced Materials}, 30(33):1706910, 2018.

\bibitem{yoo2016assessment}
S.~Yoo, M.~Cho, E.~Kim, S.~Kim, Y.~Sim, D.~Yoo, H.~Hwang, and M.~Song.
\newblock {Assessment of hospital processes using a process mining technique:
  Outpatient process analysis at a tertiary hospital}.
\newblock {\em International Journal of Medical Informatics}, 88:34--43, 2016.

\bibitem{yu2016green}
F.~R. Yu, X.~Zhang, and V.~C.~M. Leung.
\newblock {\em {Green Communications and Networking}}.
\newblock CRC Press, 2016.

\bibitem{zarsky2017incompatible}
T.~Z. Zarsky.
\newblock {Incompatible: The GDPR in the Age of Big Data}.
\newblock {\em Seton Hall Law Review}, 47(4):995--1020, 2017.

\bibitem{zhou2014process}
Z.~Zhou, Y.~Wang, and L.~Li.
\newblock {Process Mining Based Modeling and Analysis of Workflows in Clinical
  Care -- A Case Study in a Chicago Outpatient Clinic}.
\newblock In {\em Proceedings of the 11th IEEE International Conference on
  Networking, Sensing and Control}, pages 590--595, Miami, USA, 2014. IEEE.

\bibitem{zigomitros2020survey}
A.~Zigomitros, F.~Casino, A.~Solanas, and C.~Patsakis.
\newblock {A Survey on Privacy Properties for Data Publishing of Relational
  Data}.
\newblock {\em IEEE Access}, 8:51071--51099, 2020.

\end{thebibliography}

\afterpage{\blankpage}
\includepdf[pages=-]{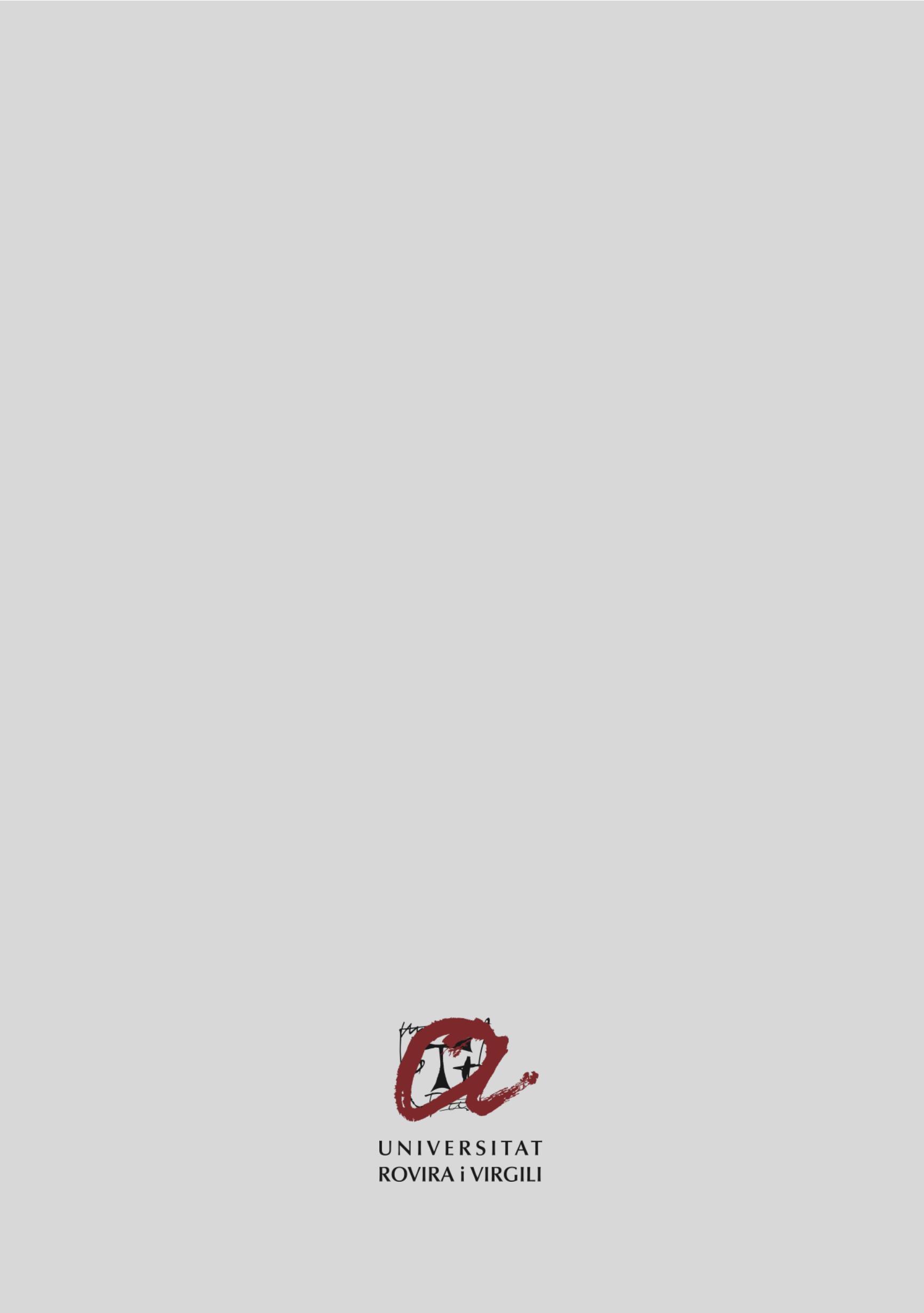}\AtBeginShipout\AtBeginShipoutDiscard

\end{document}